\documentclass[a4paper,12pt,twoside,openright]{report}
\usepackage{cite}                                                                                               

\usepackage[utf8]{inputenc}                                                                                     
\usepackage[T1]{fontenc}                                                                                        
\usepackage[english]{babel}                                                                                     

\usepackage{multirow}                                                                                           
\usepackage[table]{xcolor}                                                                                      
\definecolor{bleuobservatoire}{RGB}{37,39,124}
\definecolor{bleupsl}{RGB}{61,55,133}
\definecolor{mygray}{gray}{0.50}
\definecolor{ddlion}{RGB}{255,245,219}
\usepackage{pifont}                                                                                             
\usepackage{floatrow}                                                                                           

\usepackage[Lenny]{fncychap}                                                                                    
\usepackage{fourier-orns}                                                                                       
\usepackage{adforn}                                                                                             
\usepackage{fancyhdr}                                                                                           

\makeatletter
\def\hrulefill{\leavevmode\leaders\hrule height 0.1pt\hfill\kern\z@}
\makeatother

\fancypagestyle{plain}
{
   \fancyhead[]{}
   \fancyhf{}
   \fancyfoot[C]{\bfseries\color{mygray}\hrulefill\quad\adfhangingleafleft\ \color{black}{\thepage}\ \color{mygray}\adfhangingleafright\quad\hrulefill}
}

\fancyhfoffset{0mm}
\usepackage{enumitem}
\usepackage[section]{placeins}                                                                                  
\usepackage{epigraph}                                                                                           
\usepackage{tocbibind}                                                                                          
\usepackage{minitoc}                                                                                            
\usepackage{thmtools}                                                                                           
\usepackage{bigints}                                                                                            
\usepackage{tensor}                                                                                             
\usepackage{siunitx}                                                                                            
\usepackage{amsmath}                                                                                            
\usepackage{amssymb}                                                                                            
\usepackage{amsfonts}                                                                                           
\usepackage{stmaryrd}                                                                                           

\usepackage{graphicx}                                                                                           
\graphicspath{{Figures/}}
\usepackage[font={small,sl}]{caption}                                                                           
\usepackage{rotating}                                                                                           
\usepackage{eso-pic}                                                                                            

\usepackage{framed}                                                                                             
\usepackage{array}                                                                                              
\usepackage{etoolbox}

\usepackage{nameref}                                                                                            
\usepackage[colorlinks,linkcolor=blue,citecolor=red,pagebackref]{hyperref}                        
\hypersetup
{
   pdfauthor   = {Gr\'egoire Martinon},
   pdfsubject  = {PhD thesis manuscript},
   pdftitle    = {Gravitational systems in asymptotically anti-de Sitter spacetimes},
   pdfkeywords = {Geons, anti-de Sitter, general relativity}
}

\usepackage{setspace}                                                                                           
\usepackage{geometry}                                                                                           
\usepackage[xindy,acronyms,toc]{glossaries}
\newglossary[nlg]{notation}{not}{ntn}{Notations}

\makeglossaries
\newacronym{ds}{dS}{de Sitter}
\newacronym{ads}{AdS}{anti-de Sitter}
\newacronym{sads}{SAdS}{Schwarzschild-anti de-Sitter}
\newacronym{aads}{AAdS}{asymptotically anti de-Sitter}
\newacronym{bk}{BK}{Balasubramanian-Kraus}
\newacronym{amd}{AMD}{Ashtekar-Magnon-Das}
\newacronym{adm}{ADM}{Arnowitt-Deser-Misner}
\newacronym{cft}{CFT}{Conformal Field Theories}
\newacronym{ttf}{TTF}{Two-Time Framework}
\newacronym{gr}{GR}{General Relativity}
\newacronym{jwkb}{JWKB}{Jeffreys-Wentzel-Kramers-Brillouin}
\newacronym{btz}{BTZ}{Ba\~{n}ados-Teitelboim-Zanelli}
\newacronym{amr}{AMR}{Adaptive Mesh Refinement}
\newacronym{gb}{GB}{Gauss-Bonnet}
\newacronym{egb}{GB}{Einstein-Gauss-Bonnet}
\newacronym{kam}{KAM}{Kolmogorov-Arnold-Moser}
\newacronym{itg}{ITG}{Interior Time Gauge}
\newacronym{btg}{BTG}{Boundary Time Gauge}
\newacronym{tov}{TOV}{Tolman-Oppenheimer-Volkoff}
\newacronym{kz}{KZ}{Kolmogorov-Zakharov}
\newacronym{hp}{HP}{Hawking-Page}
\newacronym{qcd}{QCD}{Quantum Chromodynamics}
\newacronym{qed}{QED}{Quantum Electrodynamics}
\newacronym{sym}{SYM}{Super Yang-Mills}
\newacronym{gkp}{GKP}{Gubser-Klebanov-Polyakov}
\newacronym{eh}{EH}{Einstein-Hilbert}
\newacronym{ghy}{GHY}{Gibbons-Hawking-York}
\newacronym{nh}{NH}{Near-Horizon}
\newacronym{kms}{KMS}{Kubo-Martin-Schwinger}
\newacronym{ekg}{EKG}{Einstein-Klein-Gordon}
\newacronym{fg}{FG}{Fefferman-Graham}
\newacronym{lhc}{LHC}{Large Hadron Collider}
\newacronym{qgp}{QGP}{Quark-Gluon Plasma}
\newacronym{ws}{WS}{Weyl-Schouten}
\newacronym{cdm}{CDM}{Cold Dark Matter}
\newacronym{am}{AM}{Andersson-Moncrief}
\newacronym{kis}{KIS}{Kodama-Ishibashi-Seto}
\newacronym{kg}{KG}{Klein-Gordon}
\newacronym{edt}{EDT}{Einstein-De Turck}
\newacronym{eam}{EAM}{Einstein-Andersson-Moncrief}
\newacronym{bssn}{BSSN}{Baumgarte-Shapiro-Shibata-Nakamura}
\newacronym{zamo}{ZAMO}{Zero Angular Momentum Observer}
\newacronym{fput}{FPUT}{Fermi-Pasta-Ulam-Tsingou}
\newacronym{gp}{GP}{Gross-Pitaevskii}
\newacronym{mp}{MP}{Myers-Perry}
\newacronym{ah}{AH}{Abelian-Higgs}
\newacronym{hr}{HR}{Hawking-Reall}

\newglossaryentry{c}
{
  type = notation,
  name = {\ensuremath{c}},
  description = {speed of light in vaccum $\SI{299782458}{m.s^{-1}}$}
}
\newglossaryentry{e}
{
  type = notation,
  name = {\ensuremath{e}},
  description = {elementary electric charge $\SI{1.6021766208(98)e-19}{C}$}
}
\newglossaryentry{epsilon0}
{
  type = notation,
  name = {\ensuremath{\varepsilon_0}},
  description = {vacuum permittivity $\SI{8.854187817620e-12}{A^2.s^4.kg^{-1}.m^{-3}}$},
  sort = {epsilon0}
}
\newglossaryentry{Es}
{
  type = notation,
  name = {\ensuremath{E_S}},
  description = {Schwinger limit $\dfrac{m_e^2 c^3}{\hbar e} = \SI{1.32315305(04)e18}{kg.m.A^{-1}.s^{-3}}$}
}
\newglossaryentry{G}
{
  type = notation,
  name = {\ensuremath{G}},
  description = {gravitational constant $\SI{6.67408(31)e-11}{m^3.kg^{-1}.s^{-2}}$}
}
\newglossaryentry{hbar}
{
  type = notation,
  name = {\ensuremath{\hbar}},
  description = {reduced Planck constant $\SI{1.054571800(13)e-34}{kg.m^2.s^{-1}}$},
  sort = {hbar}
}
\newglossaryentry{kb}
{
  type = notation,
  name = {\ensuremath{k_B}},
  description = {Boltzmann constant $\SI{1.38064852(79)e-23}{kg.m^2.s^{-2}.K^{-1}}$}
}
\newglossaryentry{L}
{
  type = notation,
  name = {\ensuremath{L}},
  description = {AdS length $\sqrt{\dfrac{-3}{\Lambda}}$}
}
\newglossaryentry{Lambda}
{
  type = notation,
  name = {\ensuremath{\Lambda}},
  description = {Cosmological constant},
  sort = {Lambda}
}
\newglossaryentry{lambdac}
{
  type = notation,
  name = {\ensuremath{\lambda_C}},
  description = {Compton wavelength of the electron $\dfrac{\hbar}{m_ec} = \SI{2.4263102367(11)e-12}{m}$},
  sort = {lambdac}
}
\newglossaryentry{me}
{
  type = notation,
  name = {\ensuremath{m_e}},
  description = {electron rest mass $\SI{9.10938356(11)e-31}{kg}$}
}
\newglossaryentry{mpl}
{
  type = notation,
  name = {\ensuremath{m_{Pl}}},
  description = {Planck mass $\sqrt{\frac{\hbar c}{G}} = \SI{2.176470(51)e-8}{kg}$},
  sort = {mpl}
}
\newglossaryentry{mu0}
{
  type = notation,
  name = {\ensuremath{\mu_0}},
  description = {vacuum permeability $\SI{4\pi e-7}{kg.m.s^{-2}.A^{-2}}$},
  sort = {mu0}
}
\newglossaryentry{msun}
{
  type = notation,
  name = {\ensuremath{M_\odot}},
  description = {solar mass $\SI{1.988(55)e30}{kg}$}
}
\newglossaryentry{nsym}
{
  type = notation,
  name = {\ensuremath{\mathcal{N} = 4} SYM},
  description = {Super Yang-Mills theory with four supersymmetries}
}
\newglossaryentry{sigma}
{
  type = notation,
  name = {\ensuremath{\sigma}},
  description = {Stefan-Boltzmann constant $\dfrac{\pi^2k_B^4}{60\hbar^3c^2} = \SI{5.670373(21)e-8}{kg.s^{-3}.K^{-4}}$},
  sort = {sigma}
}

\newcommand{\tn}{\textnormal}
\newcommand{\citationChap}[2]{\epigraph{``\textit{#1}''}{#2}}

\let\oldpagenumbering\pagenumbering
\renewcommand{\pagenumbering}[1]{\cleardoublepage\oldpagenumbering{#1}}

\makeatletter
\newcommand{\clearevenpage}{\clearpage\if@twoside \ifodd\c@page
   \hbox{}\newpage\if@twocolumn\hbox{}\newpage\fi\fi\fi}
\makeatother

\newcommand{\RN}[1]{\textup{\uppercase\expandafter{\romannumeral#1}}}

\newcommand{\cmark}{\ding{51}}
\newcommand{\xmark}{\ding{55}}

\DeclareMathOperator \Tr {Tr}
\DeclareMathOperator \arsinh {arsinh}
\DeclareMathOperator \artanh {artanh}

\declaretheorem[numberwithin=chapter]{theorem}
\declaretheorem[numberwithin=chapter]{definition}
\declaretheorem[numberwithin=chapter]{conjecture}

\newenvironment{myfig}
{
   \begin{figure}[t]
}
{
   \end{figure}
}

\newenvironment{mytab}
{
   \begin{table}[t]
      \rowcolors{1}{white}{ddlion}
}
{
   \end{table}
}

\newenvironment{mystab}
{
   \begin{sidewaystable}[p]
      \rowcolors{1}{white}{ddlion}
}
{
   \end{sidewaystable}
}

\newenvironment{myitem}
{
   \begin{itemize}[label=$\blacktriangleright$,font=\color{bleupsl}]
}
{
   \end{itemize}
}

\newenvironment{myenum}
{
   \begin{enumerate}[label=\protect\ding{\value*},start = 182,font=\large\color{bleupsl}]
}
{
   \end{enumerate}
}

\newenvironment{myenum2}
{
   \begin{enumerate}[label=(\alph*)]
}
{
   \end{enumerate}
}

\newcommand\MyBorder
{
   \put(0,0)
   {
      \parbox[b][\paperheight]{\paperwidth}
      {
         \includegraphics[height=0.45\paperheight]{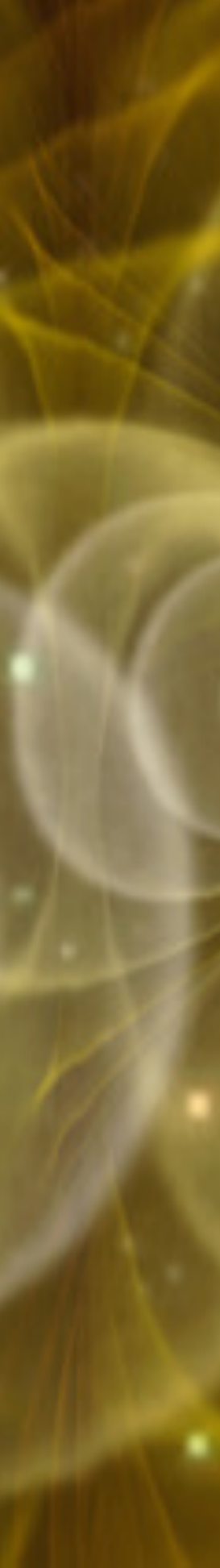}
         \vfill
      }
   }
}

\newcommand\ThesisLabel
{
   \put(0,0)
   {
      \parbox[t][\paperheight]{\paperwidth}
      {
         \hfill
         \colorbox{black}
         {
            \begin{minipage}[b]{3em}
               \centering\Huge\textcolor{orange}{\begin{turn}{-90} T\ \ H\ \ E\ \ S\ \ E \end{turn}\\}
               \vspace{0.2cm}
            \end{minipage}
         }
      }
   }
}

\makeatletter
\newcommand{\frontpage}
{
   \newgeometry{top=2.5cm, bottom=2cm, left=2cm, right=1cm}
   \AddToShipoutPicture*{\MyBorder}
   \AddToShipoutPicture*{\ThesisLabel}
   \begin{titlepage}
   \centering
   \vfill
      \includegraphics[height=1.3cm]{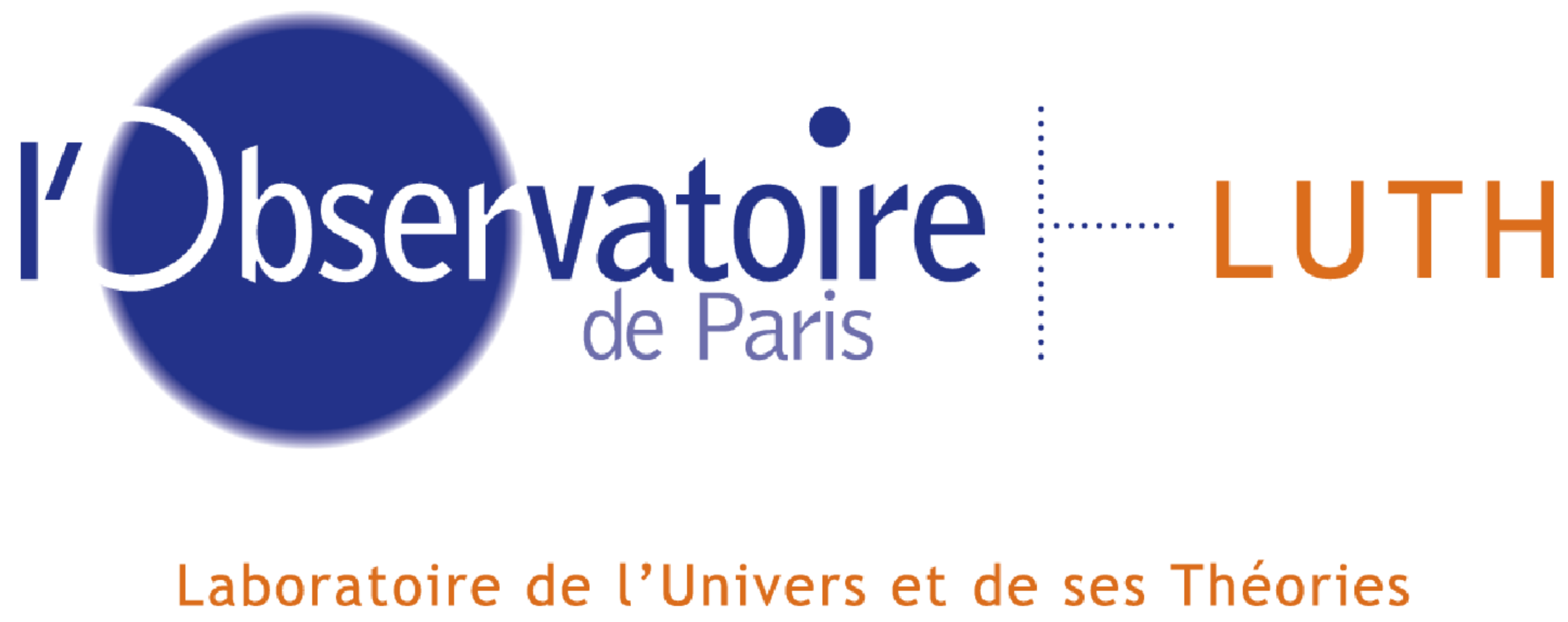}
      \hspace{0.3cm}
      \includegraphics[height=1.3cm]{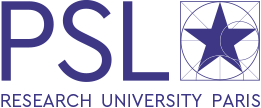}
      \hspace{0.3cm}
      \includegraphics[height=1.3cm]{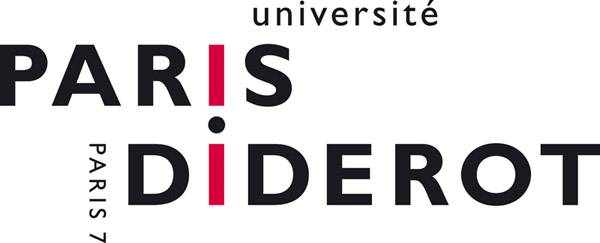}
      \hspace{0.3cm}
      \includegraphics[height=1.3cm]{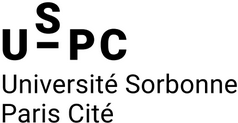}
      \hspace{0.3cm}
      \includegraphics[height=1.3cm]{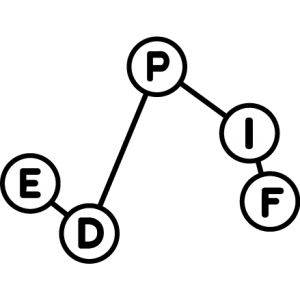}\\
   \vfill
   \vspace{1.0cm}
   {\huge {\bfseries Thèse de doctorat}}\\\vspace{0.2cm}
   {\large de l'Université Sorbonne Paris Cité\\\vspace{0.2cm}
   Préparée à l'Université Paris Diderot}\\\vspace{0.3cm}
    {\large Ecole doctorale n$^\circ$ 564 : Physique en Ile de France}
    \vfill
    {\Large \textbf{Laboratoire de l'Univers et de ses Théories}\\\vspace{0.2cm}
      Equipe Relativité et Objets Compacts}
    \vfill
    {\huge \textcolor{bleuobservatoire}{\bfseries{Gravitational Systems in \\\vspace{0.4cm}
    Asymptotically Anti-de Sitter Space-times}}} \\
    \vfill
       {\LARGE par {\bfseries Grégoire \scshape{Martinon}}}\\\vspace{0.2cm}
    \vfill
    {\Large{\bfseries Thèse de doctorat de Physique}\\\vspace{0.2cm}
       {\large Dirigée par Philippe \scshape{Grandclément}}}\\
    \vfill
       \textit{Presentée et soutenue publiquement à l'Observatoire de Paris, site de Meudon, le 28 juin 2017}\\
    \vfill
    \begin{flushleft}
    \begin{tabular}{lll}
       \textbf{Nathalie Deruelle}     & Université Paris Diderot              & Présidente du jury \\
       \textbf{Vitor Cardoso}         & Instituto Superior Técnico, Lisbonne  & Rapporteur         \\
       \textbf{Andrzej Rostworowski}  & Uniwersytet Jagiellonski, Cracovie    & Rapporteur         \\
       \textbf{Loïc Villain}          & Université François Rabelais, Tours   & Examinateur        \\
       \textbf{Christos Charmousis}   & Université Paris Sud                  & Examinateur        \\
       \textbf{Philippe Grandclément} & Observatoire de Paris                 & Directeur de thèse
    \end{tabular}
    \end{flushleft}
   \vfill
   \end{titlepage}
   \restoregeometry
}
\makeatother

\def\jnl@style{\it}
\def\aaref@jnl#1{{\jnl@style#1}}

\def\aaref@jnl#1{{\jnl@style#1}}

\def\aj{\aaref@jnl{AJ}}                   
\def\araa{\aaref@jnl{ARA\&A}}             
\def\apj{\aaref@jnl{ApJ}}                 
\def\apjl{\aaref@jnl{ApJ}}                
\def\apjs{\aaref@jnl{ApJS}}               
\def\ao{\aaref@jnl{Appl.~Opt.}}           
\def\apss{\aaref@jnl{Ap\&SS}}             
\def\aap{\aaref@jnl{A\&A}}                
\def\aapr{\aaref@jnl{A\&A~Rev.}}          
\def\aaps{\aaref@jnl{A\&AS}}              
\def\azh{\aaref@jnl{AZh}}                 
\def\baas{\aaref@jnl{BAAS}}               
\def\jrasc{\aaref@jnl{JRASC}}             
\def\memras{\aaref@jnl{MmRAS}}            
\def\mnras{\aaref@jnl{MNRAS}}             
\def\pra{\aaref@jnl{Phys.~Rev.~A}}        
\def\prb{\aaref@jnl{Phys.~Rev.~B}}        
\def\prc{\aaref@jnl{Phys.~Rev.~C}}        
\def\prd{\aaref@jnl{Phys.~Rev.~D}}        
\def\pre{\aaref@jnl{Phys.~Rev.~E}}        
\def\prl{\aaref@jnl{Phys.~Rev.~Lett.}}    
\def\pasp{\aaref@jnl{PASP}}               
\def\pasj{\aaref@jnl{PASJ}}               
\def\qjras{\aaref@jnl{QJRAS}}             
\def\skytel{\aaref@jnl{S\&T}}             
\def\solphys{\aaref@jnl{Sol.~Phys.}}      
\def\sovast{\aaref@jnl{Soviet~Ast.}}      
\def\ssr{\aaref@jnl{Space~Sci.~Rev.}}     
\def\zap{\aaref@jnl{ZAp}}                 
\def\nat{\aaref@jnl{Nature}}              
\def\iaucirc{\aaref@jnl{IAU~Circ.}}       
\def\aplett{\aaref@jnl{Astrophys.~Lett.}} 
\def\apspr{\aaref@jnl{Astrophys.~Space~Phys.~Res.}}
\def\bain{\aaref@jnl{Bull.~Astron.~Inst.~Netherlands}} 
\def\fcp{\aaref@jnl{Fund.~Cosmic~Phys.}}  
\def\gca{\aaref@jnl{Geochim.~Cosmochim.~Acta}}   
\def\grl{\aaref@jnl{Geophys.~Res.~Lett.}} 
\def\jcp{\aaref@jnl{J.~Chem.~Phys.}}      
\def\jgr{\aaref@jnl{J.~Geophys.~Res.}}    
\def\jqsrt{\aaref@jnl{J.~Quant.~Spec.~Radiat.~Transf.}}
\def\memsai{\aaref@jnl{Mem.~Soc.~Astron.~Italiana}}
\def\nphysa{\aaref@jnl{Nucl.~Phys.~A}}   
\def\physrep{\aaref@jnl{Phys.~Rep.}}   
\def\physscr{\aaref@jnl{Phys.~Scr}}   
\def\planss{\aaref@jnl{Planet.~Space~Sci.}}   
\def\procspie{\aaref@jnl{Proc.~SPIE}}   

\hypersetup
{
   pdfauthor   = {Grégoire Martinon},
   pdfsubject  = {PhD thesis manuscript},
   pdftitle    = {Gravitational systems in asymptotically anti-de Sitter space-times},
   pdfkeywords = {Geons, anti-de Sitter, general relativity}
}

\begin{document}
\pagenumbering{roman}
\frontpage
\cleardoublepage
\cleardoublepage
\thispagestyle{empty}
\vspace*{\stretch{1}}
\begin{flushright}
   \textit{To Marine.}
\end{flushright}
\vspace*{\stretch{2}}

\chapter*{Acknowledgements}
\addstarredchapter{Acknowledgements}

This PhD thesis would not have been possible without the enthusiastic incentive of Philippe Grandclément. As a PhD advisor, he
was always very attentive to my well-being and my level of advancement. Very often he encouraged my efforts, and did not
spare me a frank and constructive criticism. Without the numerous lively discussions we had, this PhD would not have been as much
successful, and for this I am eternally grateful to him. Such human and scientific qualities are rare and undoubtedly dear, they
are those of an excellent advisor, they are precisely the one of Philippe. Of course, we disagreed many times. While I
am naturally more prone to a picky (but slow) methodology, Philippe was more bold and fearless. I have to confess that
despite my stubbornness, his experienced intuition superseded several times (but not all times) my analysis. I am now deeply
convinced that this complementarity was very successful at the end, and that the present PhD thesis took full advantage of our two
different natures. Without a doubt, most of my work took advantage of the undeniable numerical skills of Philippe, since my daily
framework was rooted in his clever spectral library KADATH. What I learned in numerical relativity, I owe it to him. These three
years of work did not go without many and hard scientific difficulties.  At some point, they were so puzzling that motivation was
fading. Hopefully, Philippe supported me constantly, and without his continuous encouragement, I could have been drowned
away. By now I am aware that many PhDs endure such difficulties, and sometimes with dramatic consequences. Bouncing back, as I did
successfully, requires a strong support of the advisor. This, Philippe did very well, and I am infinitely thankful to him. He is a
model for other advisors, and I hope many other students in the future will benefit from his guidance. Thank you Philippe.

During the first two years of my PhD, I had the great pleasure to work in a close, daily collaboration with Gyula Fodor, who became
in practice my co-supervisor. I shared with him a taste for accuracy, and I soon discovered that he was even more meticulous
and inhabited by the Cartesian doubt than I was. His positive scepticism was of great help in the deep understanding of the
subject, in the development of numerical diagnostics, and in the resolution of our very difficult problems. As an expert of perturbative
techniques, he performed a great deal of an excellent work, that I could not have performed myself in parallel of my numerical
project. In particular, during a large part of the PhD, we were convinced that our numerical results were wrong since they were in
contradiction with previous studies. Multiplying the angles of attack and the numerical diagnostics, the problem remained
irremediably the same. It is only when Gyula presented me his latest sixth order perturbative results that we became aware that,
against all odds, our numerical results were the good ones and independently confirmed by his perturbative analysis. Without Gyula,
the doubt on the correctness of our results would have probably never disappeared. In this sense, he precisely incarnates when
this PhD bounced back\footnote{Would I dare to say ``off the AdS boundary''?} from the depth of anxiety and difficulty to an unprecedented agreement between numerical and analytical
arguments. I owe very much to this fruitful complementarity between Gyula's skills and mines. Even if he had to leave one year
before the end of the PhD, we kept closed mail contacts, and I benefited very much from his sharp eye and high level of
precision. It was a constant pleasure to work with him, and I knew that I could always find a sympathetic ear in his office. For
all this, thank you Gyula.

Of course, this PhD would have never emerged without the original incentive of Peter Forgàcs. He is in some sense the architect of
this PhD thesis since he was the first to propose to Philippe and Gyula a common project about geons in asymptotically anti-de
Sitter space-times. Even if distance did not allow us to meet each other regularly, it was always a great pleasure to discuss with him. His
indestructible enthusiasm has always been a breath of fresh air, even in the heart of difficulties. I will always remember our
``geon team'' meetings in Philippe's office, exchanging during hours and hours our feelings, planning our next to-do list. For
this cheerful atmosphere and all these sound advises, thank you Peter. How can you have so much energy without ever having lunch?

Since a large part of this work was numerical, I was in close contact with all our numerical engineers, among them Fabrice Roy and
Marco Mancini. First, they were invaluable interlocutors and careful colleagues to work with. I have lost the count of the number
of times I ran into their offices to ask for help or advises. Second, I am infinitely obliged to them, since they offered me the
opportunity to use several hundreds of thousands of hours of computing time on national supercomputers. The present research could not
have presented as much results without this (how much timely) numerical support. Fabrice and Marco definitely played a great role
in making this PhD bouncing to success. On top of that, they created a very stimulating participative environment centred around
numerical skills in the laboratory, notably by organising regular training sessions opened to all researchers and students. These
contributed to keep me up-to-date in the technology I used, and much of my workflow was built on their killer tips and advises. I
think the laboratory can sincerely boast about having such an efficient and jovial team of computer scientists, who bring an
overall added value to research. To me they offered not only a great working environment but also a successful PhD. For all that,
thank you.

I have never learned so much that during this PhD in general relativity. In this regard, I had the major opportunity to work with
the best experts of my laboratory. In particular, I wish to warmly thank Alexandre Le Tiec and Eric Gourgoulhon, who shared with
me their invaluable expertise. Many times, I knocked to their door in hope of getting a precise answer, either technical or
conceptual, about the foundations of the theory, and not once was I disappointed. Eric was my very first teacher of general
relativity, and Alexandre, by his continuous hospitality, the second. Furthermore, they were constantly animated by a profusion of
new ideas, some of them I assimilated with success in my research and also in its popularisation. For all your kind
attention and availability, thank you.

If I have learned very much from my teachers, I have even more learned from my peers. Among them, a special thank to Aurélien
Sourie, my close colleague and dear friend. I think it is very important, during the PhD, to not feel alone. By sharing my office
with Aurélien during three years, we were able to share instantaneously our problems, our doubts, our mixed feelings that invariably
populate a PhD student's life. It is so good to have somebody to talk with, to doubt with, to joke with, to
complain with and so on. He always was my first interlocutor at work. Many times, we shared our difficulties in general
relativity and in teaching. Beyond doubt, PhDs have to be shared. For this mutual partnership, thank you Aurélien.

I would also like to thank all my laboratory, especially the Relativity and Compact Objects team: Marion Grould, who became a close
friend this last year, but also Silvano Bonazzola, Jerôme Novak and Micaela Oertel. A big thank to the director of the LUTH:
Stéphane Mazevet. Thanks to Claire Michaut and Martine Mouchet for their listening. It was a pleasure to welcome Zakaria Meliani
everyday in my office and his irreducible sense of humour. He contributed a lot to my mastery of Paraview. The secretaries of
the laboratory did a perfect job in accompanying me during these three years, especially Annie Le Guevellou, but also Omur Avci
and Nathalie Ollivier. Many thanks to Stéphane Méné (with his memorable bursts of laughter) and Jean-Denis Bovas for their careful
handling of our local computing cluster Poincaré. Thanks again to my contemporary PhD students and post-docs: Miguel Marques,
Debarati Chatterjee, Océane Saincir, Guillaume Voisin, Paul de Fromont and Karim Van Aelst. And of course: many thanks to
University Paris Diderot who funded this research project.

All of this would not have been possible without my partner, Marine Cogo, who experienced this PhD as much as I did, and is
supporting me for almost thirteen years by now. Thank you for your kind and ever-present endorsement. A great deal of thanks to my
family: thank you Mum, thank you my two brothers: Quentin and Mathias, with his wife Ilhem and two sons Tamim and Hilal. Thank you
Agathe, my sister in the sky. A great many thanks, naturally, to all my childhood friends, Ben, Toin, Zouz, Alex, Amélie, Soso,
Goomo, but also Elours, Fanny, Vince, Gang-Yun, Gulien, Gulie and Sim. Of course, many thanks to the Cogo's family. Finally, a
great deal of thanks to my regular role-playing table: Loïc (Holy MJ), Charles (Sir Sheltem), Thomas (Kilraillon), Vincent (Leoric
of Redstar) and Romain (Baldric). A new power is rising. Behold and fear my wizardry, for I am Saros the Red and I have just
leveled up.

\chapter*{Résumé}
\addstarredchapter{Résumé}

Ce manuscrit de thèse présente une construction numérique de geons gravitationnels asymptotiquement \gls{ads}. Pourquoi ?

Tout d'abord, le concept de geons a été introduit par Wheeler dans les années 50, dans le contexte de la géométrodynamique. Au
départ, le mot ``geon'' désignait un système auto-gravitant solution des équations d'Einstein-Maxwell en espace-temps
asymptotiquement plat. L'idée sous-jacente était de donner une définition, ou une illustration, du concept de \textit{corps} au sein de la
théorie de la relativité générale. L'intérêt était de construire une description purement géométrique des particules élémentaires.
Au-delà de ses motivations conceptuelles, le problème de la construction des geons était un problème difficile à l'époque, et a
attiré l'attention de nombre de relativistes. Très rapidement, le concept s'est élargi à d'autres champs que les photons, comme
par exemple les scalarons, les neutrinos et les gravitons, autrement dit les ondes gravitationnelles. Cependant, toutes ces
solutions se désagrégeaient lentement avec le temps. En effet, rien ne pouvait empêcher le contenu énergétique initial de se disperser
vers l'infini, donnant lieu à une évaporation progressive.

Les geons ont inspiré nombre de solution auto-gravitantes, en particulier celles constituées de champs bosoniques. En effet, la
matière noire est censée être composée de particules non-collisionnelles interagissant très faiblement avec les champs du modèle
standard. Certains champs scalaires apparaissant naturellement dans plusieurs modèles d'inflations, il se pourrait que la
matière noire soit constituée de champs scalaires fondamentaux, comme les axions. Par ailleurs, si l'on en croit les théories
tenseur-scalaire, qui sont une extension naturelle de la relativité générale, le phénomène de scalarisation spontanée semble
favoriser l'existence de solutions scalaires auto-gravitantes. Comme le principe d'équivalence affirme que toutes les formes de
matière gravitent, les geons et leur cousins scalaires apparaissent comme des outils permettant d'étudier la phénoménologie
gravitationnelle de la matière noire.

Le concept de geon vit son intérêt renouvelé une nouvelle fois très récemment dans un contexte complètement différent, à savoir la conjecture de
l'instabilité d'\gls{ads}. L'espace-temps \gls{ads} a longtemps été considéré comme la plus étrange des solutions maximalement
symétriques de l'équation d'Einstein, qui sont les espace-temps \gls{ds}, Minkowski, et \gls{ads}. Il est défini par une
constante cosmologique négative qui agit comme un puits gravitationnel et empêche les particules massives d'atteindre l'infini du
genre espace. Encore plus particulières sont les trajectoires des photons qui rebondissent littéralement sur l'infini du genre
espace et reviennent à leurs positions initiales en un temps fini (mesuré par un observateur statique). Mais en 1998, si
l'espace-temps \gls{ds} est entré en phase avec la découverte de l'expansion accélérée de l'univers, l'espace-temps \gls{ads} a
été célébré par la communauté des physiciens théoriciens avec l'avènement de la correspondance \gls{ads}-\gls{cft}. Cette percée
de la physique théorique est considérée comme une avancée prometteuse vers l'unification de la gravité et de la théorie quantique des
champs. Un des ingrédients majeurs de cette correspondance est justement l'espace-temps \gls{ads}, qui a été intensivement étudié
par la suite.

Avec l'intérêt croissant pour l'espace-temps \gls{ads}, une question fondamentale est celle de sa stabilité. Ce problème, malgré
sa grande importance, n'a pratiquement pas été abordé pendant plus d'une décennie. C'est seulement depuis 2011 que la communauté
scientifique s'y attèle. Il se trouve que, contrairement à ce qui était auparavant communément accepté, l'espace-temps
\gls{ads} est instable vis-à-vis de la formation de trous noirs, et ce pour de larges classes de perturbations. Autrement dit,
aussi arbitrairement petite soit la perturbation initiale, les non-linéarités peuvent s'accumuler dans le temps au points
concentrer le budget énergétique initiale dans une région infime de l'espace-temps, provoquant la naissance d'une singularité.
Cette instabilité est complètement absente en espace-temps de Minkowski ou \gls{ds}, et a été renommée la conjecture de
l'instabilité \gls{ads}. A l'heure actuelle, cette conjecture est encore ambigüe car la structure même de l'instabilité semble
très compliquée. Notamment, la séparation entre données initiales stables et instables est loin d'être claire. C'est dans ce contexte
que les geons gravitationnels, c'est-à-dire des paquets d'ondes gravitationnelles, ont refait surface récemment dans la
littérature.

En effet, ces geons sont vus pour l'instant comme des îlots de stabilité non-linéaires en espace-temps asymptotiquement
\gls{ads}. En tant que tel, ils sont d'un grand intérêt pour le problème de l'instabilité \gls{ads}, car ils peuvent donner de
précieuses clés de compréhension. Qui plus est, ils représentent un défi technique puisqu'ils vont au-delà de la symétrie
sphérique, une hypothèse difficile à abandonner au vu des limitations numériques. Même si leur construction perturbative a été
initiée en 2012, il a fallu attendre l'année 2015 pour voir apparaître dans la littérature la première construction numérique de
geons. Toutefois, certaines de leurs propriétés sont encore controversées aujourd'hui. Ce manuscrit de thèse est précisément dédié
à l'étude numérique de geons gravitationnels en espace-temps asymptotiquement \gls{ads}.

En première partie, nous insistons sur le contexte scientifique, en particulier la correspondance \gls{ads}-\gls{cft} et
l'instabilité \gls{ads}. Le chapitre \ref{geons} présente un point de vue historique sur la naissance du concept de geons. En
effet, après plusieurs années de recherche, ce concept est devenu de plus en plus confus, et nous tentons d'en donner une
signification rigoureuse. Dans le chapitre \ref{aads}, nous donnons une définition précise de ce que signifie asymptotiquement
\gls{ads}. Ce chapitre permet notamment d'introduire plusieurs concepts utiles, comme la masse et le moment angulaire, qui sont
utilisés tout au long du texte. Le chapitre \ref{adscft} est dédié à la correspondance \gls{ads}-\gls{cft}. C'est une sujet très
vaste, ce qui nous oblige à ne traiter que certains aspects clés et décrire les applications les plus simples susceptibles
d'intéresser des chercheurs non-experts du domaine. Le chapitre \ref{adsinsta} est entièrement dévoué à l'instabilité
\gls{ads}, qui est un sujet densément ramifié. En particulier, nous insistons sur le rôle prépondérant des geons dans ce contexte.
Dans la deuxième partie de ce manuscrit, nous détaillons nos résultats récents, à savoir la construction numérique de geons
gravitationnels asymptotiquement \gls{ads}, ainsi que les techniques qui se sont révélées utiles pour ce problème. Le chapitre
\ref{perturbations} traite le problème de la construction analytique de geons linéaires d'un point de vue perturbatif. C'est le
point de départ de la construction de geons non-linéaires. Le chapitre \ref{gaugefreedom} évoque le problème du choix de jauge.
En particulier, nous discutons en détail la jauge \gls{am}, qui a été introduite initialement dans le contexte du formalisme
$3+1$, et établissons un lien clair avec jauge harmonique et la méthode de De Turck. Finalement, le chapitre \ref{simulations} illustre nos récents
résultats numériques, et dévoile la construction non-linéaire de geons dits excités, dont l'existence était controversée
jusqu'alors dans la littérature. Nous utilisons également tous les concepts introduits précédemment pour établir des diagnostics
d'erreurs numériques précis et originaux.

Les principaux résultats de ce manuscrit sont publiés dans \cite{Martinon17} et sont les suivants. Premièrement, nous donnons une
construction indépendante de geons non-linéaires qui n'ont été construits qu'une seule fois dans la littérature auparavant. Nos
résultats sur les quantités globales sont en tension avec ces précédents travaux, mais fortement soutenus par la convergence de
nos résultats tant analytiques que numériques. Deuxièmement, nous présentons la jauge \gls{am} et discutons ses motivations
théoriques ainsi que son implémentation numérique. Nous clarifions également le lien avec la jauge harmonique combinée avec la
méthode de De Turck. Troisièmement, nous étendons la construction numérique de geons non-linéaires au cas de solutions contenant
un nœud radial (geons dits excités). L'existence de ces geons était au coeur d'un débat jusqu'alors, mais nos résultats
argumentent clairement en défaveur de leur supposée non-existence.

Nous présentons également quatre annexes, dans lesquelles certains points de calculs sont détaillés. L'annexe \ref{grd} est
bref résumé des formules utiles de relativité générale en dimension arbitraire. L'annexe \ref{d+1} donne quelques rappels du
formalisme $d+1$ qui est la fondation théorique de notre procédure numérique. L'annexe \ref{leastaction} décrit le principe de
moindre action de la gravité, qui est d'une importance primordiale dans le contexte des espace-temps asymptotiquement
\gls{ads} et de la correspondance \gls{ads}-\gls{cft}. Enfin, l'annexe \ref{quantumgas} récapitule que résultats du gaz parfait
quantiques qui sont pertinents dans le contexte \gls{ads}-\gls{cft}.

\dominitoc
\tableofcontents
\listoffigures
\listoftables
\listoftheorems
\printglossary[type=notation]
\printglossary[type=\acronymtype]
\pagenumbering{arabic}

\chapter*{Introduction}
\addstarredchapter{Introduction}
\citationChap{We'd stared into the face of Death, and Death blinked first. You'd think that would make us feel brave and invincible. It didn't.}{Rick Yancey}

The present PhD thesis aims at numerically constructing gravitational geons in \gls{aads} space-times. Why?

First of all, the concept of geons finds its origin in the late 50's after Wheeler's interest in geometrodynamics. Initially, the word
``geon'' designated an asymptotically flat and self-gravitating solution of Einstein-Maxwell equations. The underlying motivation
was to give a definition, or an illustration, of the concept of a \textit{body} within \gls{gr}. The hope was to find a purely geometrical
description of elementary particles. Beyond its conceptual motivation, the problem of geon construction was very challenging at
the time and attracted the attention of many researchers in \gls{gr}. Very quickly, the concept broadened to other fields
than photons, like e.g.\ scalarons, neutrinos and even gravitons, or gravitational waves. However, all these solutions were
slowly decaying in time. Indeed, nothing could prevent their initial energy budget to leak out toward infinity, leading to a
progressive evaporation of geons. 

Geons inspired many other self-gravitating solutions, in particular, those made of bosonic fields. Indeed, dark matter is
believed to be mostly composed of cold, collisionless particles that interact very weakly with the ones of the Standard Model. Some scalar
fields naturally appearing in several inflationary scenarios, it could be that dark matter is made of a fundamental
scalar field, such as axions. Furthermore, according to scalar-tensor theories, that are the most natural extension of \gls{gr},
the phenomenon of spontaneous scalarisation argues in favour of the existence of self-gravitating scalar bodies. Since the
equivalence principle ensures that all forms of matter gravitate, geons and there closely related bosonic cousins seem to be
promising channels to look for dark matter gravitational imprints.

The concept of geons got an even larger interest very recently in a completely different context, namely the \gls{ads} instability
conjecture. The \gls{ads} space-time was for a long time considered the weirdest of all maximally symmetric solutions of Einstein
equations, that are the \gls{ds}, Minkowski and \gls{ads} space-times. It has a constant negative curvature that
acts like a gravitational potential and prevents massive particles to ever reach spatial infinity. Even more peculiar are the
trajectories of photons that literally bounce off spatial infinity and come back at their initial position in a finite time (as
measured by a static observer). But in 1998, if its closed counterpart, the \gls{ds} space-time was acknowledged by cosmologists
who have just observed the accelerated expansion of the universe, \gls{ads} space-time was celebrated by a much different
community, namely theoretical physicists trying to unify gravity and quantum field theory. This was the advent of the famous
\gls{ads}-\gls{cft} correspondence. This correspondence was understood as a very significant step toward unification of
some of our modern theories. And as a major component of this achievement, the \gls{ads} space-time was very intensively studied
since then.

With the raising interest in \gls{ads} space-times, one question of fundamental importance was whether it was stable or
not. A problem that, despite its great significance, was almost not explored at all for more than a decade. It is only since 2011 that this
is studied intensively. It turns out that, unlike the previous accepted view, \gls{ads} space-time is unstable against
black hole formation for large classes of arbitrarily small perturbations. Said differently, some perturbations, however small, can build up
non-linearities in time that lead to black hole formation. This feature is completely absent in Minkowski or
\gls{ds} space-times, and is called the \gls{ads} instability conjecture. Presently, this question is far from being fully
answered, since the instability seems to have a very intricate structure. Notably, the demarcation line between non-linearly
stable and non-linearly unstable initial data is far from being clear and distinct. This is the context in which gravitational
geons, i.e.\ self-gravitating gravitational wave packets, recently came back in the literature.

Indeed, these geons are believed to form non-linearly stable islands of stability in \gls{aads} space-times. As such, they are of
great interest in the \gls{ads} instability problem as they could give invaluable clues of understanding. Furthermore, they are a
technical challenge since they are breaking new grounds in this area of research beyond spherical symmetry, an assumption
difficult to abandon in terms of computing time limitations. Even if their perturbative construction was initiated in 2012, it is
only in 2015 that the first numerical geons were obtained, and some of their properties are still controversial today. This
PhD thesis is precisely dedicated to the study of gravitational geons in \gls{aads} space-times.

In the first part of the present manuscript, we aim at giving a comprehensive understanding of the research context, that ranges
from the \gls{ads}-\gls{cft} duality to the instability conjecture. Chapter \ref{geons} presents a historical overview of the concept of geons.
Since after many years of research, the concept got quite confused, we give a precise definition of what a geon is. In
chapter \ref{aads}, we define accurately what is meant by \gls{aads} space-times. This chapter allows us to introduce several concepts, among
them the mass and angular momentum, that are used throughout the manuscript. Chapter \ref{adscft} is dedicated to the celebrated
\gls{ads}-\gls{cft} correspondence. This is a very large area of research and we only aim at giving some key concepts as well as
simple real-world applications for non-experts. Chapter \ref{adsinsta} is entirely devoted to the \gls{ads} instability conjecture, which is a
densely ramified topic. In particular, we stress the special role played by geons in this context. In the second part
of this manuscript, we detail our recent research results, namely the numerical construction of \gls{aads} gravitational geons, as
well as the techniques that proved useful in that matter. Chapter \ref{perturbations} treats the problem of the construction of geons from a
perturbative point of view. This can be seen as the starting point of the construction of fully non-linear geons. Chapter
\ref{gaugefreedom} deals the problem of gauge freedom. In particular, we discuss in detail the so-called \gls{am} gauge, which
was first introduced in the context of 3+1 formalism, and makes clear the link with the De Turck method. Finally, chapter
\ref{simulations} exhibits our most recent results about numerical geons. We also take advantage of all previously discussed
materials to monitor precisely the numerical errors of our solutions.

The main results of the present manuscript are published in \cite{Martinon17} and are the followings. First, we give an
independent construction of fully non-linear geons that were constructed previously in the literature. Our results on global
quantities are in tension with these previous works, but are strongly supported by convergent analytical and numerical arguments.
Secondly, we present the so-called \gls{am} gauge and discuss its theoretical motivations as well as its numerical implementation.
We also make clear the link between this gauge and the harmonic gauge enforced by the De-Turck method. Thirdly, we
extend the numerical constructions of fully non-linear geons to solutions with more than two angular nodes, as well as to the
three excited families exhibiting one radial node. The existence of these excited geons is actually at the heart of a lively
debate in the literature. However, our results clearly argue in favour of the existence of such solutions.

We also present four appendices. Appendix \ref{grd} is a quick summary of useful formulas in \gls{gr} in an arbitrary number of
dimensions. Appendix \ref{d+1} deals with the $d+1$ formalism, that is at the theoretical foundation of our numerical procedure.
Appendix \ref{leastaction} describes the least action principle of gravity, which is of major significance in the context of
\gls{aads} space-times and of the \gls{ads}-\gls{cft} correspondence. Finally, Appendix \ref{quantumgas} very briefly recaps some
quantum perfect gas results, that are relevant to the \gls{ads}-\gls{cft} context.

\setcounter{mtc}{9}       
\part{Geons and Anti-de Sitter space-time}
\chapter{``Anything that can radiate, does radiate''}
\label{geons}
\addcontentsline{lof}{chapter}{\nameref{geons}}
\addcontentsline{lot}{chapter}{\nameref{geons}}
\citationChap{The fact that we live at the bottom of a deep gravity well, on the surface of a gas covered planet going around a nuclear fireball 90 million miles away and think this to be normal is obviously some indication of how skewed our perspective tends to be.}{Douglas Adams}
\minitoc

In 1955, Wheeler made the following analysis \cite{Wheeler55}. The geodesic equations of \gls{gr} determine
the motion of test particles in a known gravitational field. This is probably the most important prediction of the theory and
certainly the one that has received the most thorough observational confirmations. The geodesic equation was even considered a
consequence of Einstein's equation in \cite{Infeld49}, by a limiting procedure bringing a localised mass to a test particle of vanishing
mass and radius. However the notion of a test particle is an idealisation: it assumes that the matter field is not a smooth function
but a distribution, a mathematical tool that is forbidden by the non-linearities of Einstein's equation \cite{Geroch87}. The
notion of test particle should then be promoted to the more appropriate concept of a body. But what is the nature of a body? Is it a
singularity of the metric? Or should we postulate that the metric is regular and count on quantum mechanics to explain how this
can be so near an elementary particle?

Wheeler advocated for a third option, that was purely classical, everywhere regular and entirely describable within the framework
of \gls{gr}: geons, or Gravitational-Electromagnetic Entities. Their simplest realisation is most easily
visualised as a standing electromagnetic wave bent into a closed circular torus of high energy concentration, held together by its
own gravitational field. Such an object has all the intrinsic properties of a body such as mass and angular momentum, it exerts a
gravitational pull on other bodies and moves through the space-time according to the field equations and the background geometry.
It was even an incentive toward the so-called geometrodynamics paradigm, according to which elementary particles could be
represented as geons. This idea aimed at unifying quantum mechanics and \gls{gr}.

At first, the literature focused on electromagnetic geons, as originally designed by Wheeler, but progressively, the concept
broadened and finally encompassed all kinds of self-gravitating fields, be they massive, massless, with or without intrinsic
spins. In this chapter, we focus on asymptotically flat space-times, as they were the historical playground for this kind of
solutions. Since the concept of geon broadened largely with time, we hereafter review the different kinds of geons and give a
precise definition compatible with the evolution of the concept over the years.

\section{Gravitational-electromagnetic entities (GEONs)}

Originally, geons were made only of electromagnetic field, hence their name. Even restricted to electromagnetic field, there is a
rich variety of geons and different geometries, among them the toroidal, spherical or even cylindrical geons. Hereafter, we derive rough
estimates of some properties of these objects, especially the toroidal and thermal ones that are the simplest families. It turns out that
dimensional analysis and a handful of physical concepts are sufficient to describe them and give results very close to
accurate numerical simulations.

\subsection{Toroidal geons}

\begin{figure}[t]
   \includegraphics[width = 0.49\textwidth]{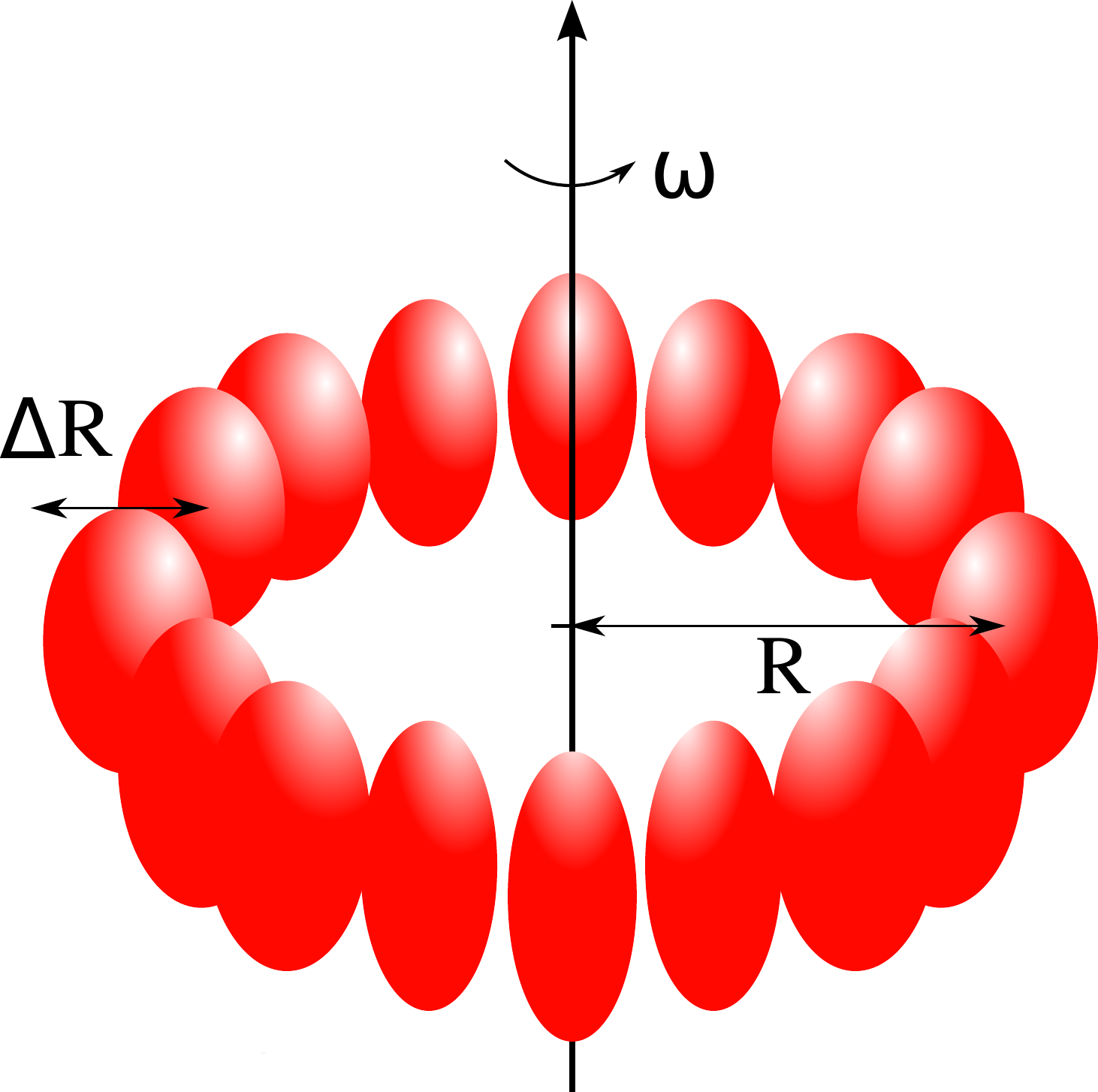}
   \caption[Toroidal geon]{Picture of a toroidal geon consisting of an electromagnetic standing wave trapped by its own gravitational potential
   in a circular orbit. Regions of large electromagnetic energy densities are pictured in red. The electromagnetic energy is
   confined to a torus, called the active region. Credits: G. Martinon.}
   \label{torgeon}
\end{figure}

Toroidal geons, pictured in figure \ref{torgeon}, have properties that can be given on the basis of heuristic arguments. In
particular, the deflection of a pencil of light in such a geon is not much different from the deflection of a pencil of light by
the Sun. Equating the kinematic and gravitational accelerations of a photon in circular orbit at a radius $R$ from a massive
object of mass $M$, we get
\begin{equation}
   \frac{\gls{c}^2}{R} \sim \frac{\gls{G}M}{R^2} \iff R \sim \frac{\gls{G}M}{\gls{c}^2}.
   \label{compact}
\end{equation}
This equation makes it clear that geons are highly relativistic objects, since their compactness is of order one.
If the toroidal wave is unidirectional, the angular momentum $J$ of the system is given by
\begin{equation}
   J \sim M\gls{c}R \sim \frac{\gls{c}^3}{\gls{G}}R^2,
\end{equation}
and scales like the square of the geon radius. As a point of comparison, recall that particles in a Keplerian disk have an
angular momentum scaling like the square root of the radial distance from the centre.

A natural question arising in the toroidal geometry is how thin is the torus compared to its major radius $R$. Actually, the thickness of the torus, or the active region,
depends on the wavelength. Denoting $a$ the number of nodes (or the azimuthal index number) and requiring it to be a large
integer, the wavelength $\lambda$ is a small fraction of the circumference, which allows us to deduce the frequency $\omega$ of
the electromagnetic wave:
\begin{equation}
   \lambda \sim \frac{2\pi R}{a} \quad \tn{and} \quad \omega = \frac{2\pi\gls{c}}{\lambda} \sim \frac{a\gls{c}}{R}.
   \label{wavegeon}
\end{equation}
Close to the toroidal active region, the gravitational field is approximately that of an infinite cylinder. This induces a logarithmic
decrease of the gravitational potential\footnote{This can be shown in Newtonian gravity as a standard application of Gauss's
theorem.}. Besides, the wave cannot be confined to a region smaller than $\lambda$ because of diffraction effects, so that the lateral
extension $\Delta R$ of the geon is proportional to the wavelength multiplied by a logarithmic factor:
\begin{equation}
   \Delta R \sim \ln\left( \frac{R}{\lambda}\right)\lambda.
\end{equation}
Ernst \cite{Ernst57b} made precise analytical calculations for the infinitely long cylinder (the so-called linear geon) and the infinitely thin
ring cases, being appropriate limits of the toroidal one as seen respectively from a very close or very far away observer.

In order to give size limit estimations for a geon, it is useful to compute how its mass scales with its action.
According to Hamilton's variational principle\footnote{See the excellent review \cite{Gray04} as well as \cite{Gray96}. The book of
Goldstein \cite{Goldstein02}, section 8.6, provides a very pedagogical derivation of equation \eqref{hamilton}. This is just the
consequence of a variational principle without Dirichlet boundary conditions for the initial and end state: the $E\delta t$ plays
the role of non-vanishing boundary terms.}, the average energy $E$ of the system is related to the variation of its action $S$ by
\begin{equation}
   E\delta t + \delta S = 0,
   \label{hamilton}
\end{equation}
where $\delta t$ is the duration time along a trial trajectory. This relation looks very much like the Hamilton-Jacobi equation,
but holds also for dissipative systems, which is our case of interest\footnote{See \cite{Galley13} for a generalisation of Hamiltonian
mechanics in dissipative systems.}. If we consider a simple monochromatic geon that leaks very slowly an infinitesimal amount of
radiation energy $dE$, we can thus argue, in light of \eqref{hamilton}, that its reduced action $\mathcal{S} = S/2\pi$ decreases by
an amount $d \mathcal{S}$ obeying
\begin{equation}
   \gls{c}^2 dM = dE = \omega d \mathcal{S}.
   \label{hamjacobi}
\end{equation}
This tends to increase the frequency $\omega$, since we have (equations \eqref{compact} and \eqref{wavegeon})
\begin{equation}
   \frac{\gls{G}M}{\gls{c}^2}\sim R \sim \frac{a\gls{c}}{\omega}.
   \label{omega}
\end{equation}
Multiplying together \eqref{hamjacobi} and \eqref{omega}, we obtain
\begin{equation}
   \gls{G}d(M^2) \sim a\gls{c} d \mathcal{S} \iff M \sim \sqrt{\frac{a\gls{c} \mathcal{S}}{\gls{G}}}.
   \label{scaling}
\end{equation}
We have thus recovered that the mass of a geon scales like the square root of its action.

What are the size limits of geons? The mass-action scaling relation \eqref{scaling} allows to answer this question. Let us
consider the lower limit first. The smaller the geon, the larger its electromagnetic energy density has to be to balance the
gravitational force. So the electromagnetic field is very large for small geons. How intense can it be? When an electric field
$E$ working on an elementary charge $\gls{e}$ over the localisability distance, the Compton wavelength $\gls{lambdac}$, can impart
to an electron an energy $\gls{e}E\gls{lambdac}$ of the order of its rest mass $\gls{me}\gls{c}^2$, it brings forth from empty
space pairs of positrons and electrons and the purely classical description is lost. Classical geons should then obey
\begin{equation}
   E < \gls{Es} \equiv \frac{\gls{me}^2 \gls{c}^3}{\gls{hbar} \gls{e}} \simeq \SI{1.3e18}{V.m^{-1}},
\end{equation}
which is called the Schwinger limit. Moreover, by dimensional analysis, it can be inferred that the action of the geon scales like
\begin{equation}
   S \sim \frac{a\gls{c}^7}{\gls{G}^2 E^2 \gls{epsilon0}}.
\end{equation}
With \eqref{compact} and \eqref{scaling}, we deduce that the mass and radius of this limiting classical geon are
\begin{equation}
   M \sim \frac{a\gls{c}^4}{\sqrt{\gls{G}\gls{epsilon0}}\gls{G}\gls{Es}} = a \SI{e36}{kg}\quad \tn{and} \quad R \sim \frac{a\gls{c}^2}{\sqrt{\gls{G}\gls{epsilon0}}\gls{Es}} = a\SI{e9}{m}.
\end{equation}
The smallest geon has then a size of $\sim 1\%$ of an astronomical unit and a mass of $\sim 10^6 \gls{msun}$. It is not impossible to
build geons below this limit, but the object would become quantum in nature and the particle pair production mechanism should be taken
into account. So strictly speaking, this lower bound is only a limit on the classical regime of geons. The intuition suggests that
the additional amount of radiation provided by the quantum pair creation process could possibly accelerate the decay of such small
geons. On the other hand, there is no upper size limit. A geon of Hubble radius $\sim \SI{14}{Gly}$ would have a mass of $\sim
\SI{e56}{kg}$. Geons can then have masses between $10^{36}$ and $\SI{e56}{kg}$.

Given the large compactness parameter and large mass of the geon, we can reasonably expect the electromagnetic field to be very
large too in geons. To have an idea of how intense the electromagnetic field in the active region is, we can consider that, essentially, the whole mass
is due to its electromagnetic field. We can roughly equate the electromagnetic energy with the mass of the geon. Denoting
$E$ the electric field, the electromagnetic energy density is $\sim \gls{epsilon0} E^2$. The volume of the active region is $\sim R \Delta
R^2$ where $\Delta R$ is the torus thickness. It then follows
\begin{equation}
   \gls{epsilon0} E^2 R \Delta R^2 \sim M\gls{c}^2,
\end{equation}
so that the potential difference $E \Delta R$ is universal to all toroidal geons:
\begin{equation}
   E \Delta R \sim \sqrt{\frac{M\gls{c}^2}{\gls{epsilon0} R}} \sim \frac{\gls{c}^2}{\sqrt{\gls{epsilon0} \gls{G}}} \sim \SI{e27}{V}.
\end{equation}
This can impart to an electron in the neighbourhood of the active region an energy greater than its rest mass by a factor
\begin{equation}
   \frac{\gls{e}E\Delta R}{\gls{me}\gls{c}^2} \sim \frac{\gls{e}}{\gls{me}\sqrt{\gls{epsilon0}\gls{G}}} \sim 10^{21}.
\end{equation}
This colossal number is actually the square root of the ratio of the electrostatic force to the gravitational force exerted by two
identical electrons on each other.

Geons are thus extreme self-gravitating objects, but do they well illustrate the concept of a body ? Far from the active region,
the field decreases exponentially with a characteristic length of order $\sim \lambda$. Strictly speaking, a geon is thus not in
principle an isolated entity, but the field outside is so extremely small in comparison with the field inside that for most
purposes, the geon has the character of a well-defined localised body.

Of fundamental interest is the question of stability in time of geons. Because of photon-photon interactions, some of the energy
leaks toward infinity. The field outside the geon is then a continual transport of energy outward. As the geon slowly loses mass,
it shrinks in size and the frequency $\omega$ goes up in inverse proportion according to \eqref{compact} and \eqref{wavegeon}:
\begin{equation}
   \frac{dM}{M} = \frac{dR}{R} = -\frac{d\omega}{\omega} = -\alpha \omega dt,
   \label{decaygeon}
\end{equation}
where $\alpha$ is a dimensionless factor called attrition. The solution is $t = t_0 - 1/\alpha\omega$ so that the mass decreases linearly in
time. A rather involved computation can estimate the attrition to be
\begin{equation}
   \alpha \simeq \exp(-1.52 a),
   \label{attrition}
\end{equation}
so that geons with short wavelengths are radiating much more efficiently. Said differently, their lifetime grows exponentially with
the azimuthal number, and arbitrary long living geons can be constructed. Different mechanisms of decay include potential barrier
penetration and photon-photon collisions, also called secondary waves excitations.

\subsection{Spherical geons}

Beyond toroidal geons, it is natural to investigate the structure of geons in a simpler geometry. A substantial part of
\cite{Wheeler55} is dedicated to the mathematical treatment of spherical geons. They consist in an incoherent sum of identical
monochromatic toroidal geons with a uniform spherically symmetric distribution of their angular momenta. The spherical geon thus
constructed has a shell-shaped active region.

\begin{figure}[t]
   \includegraphics[width = 0.49\textwidth]{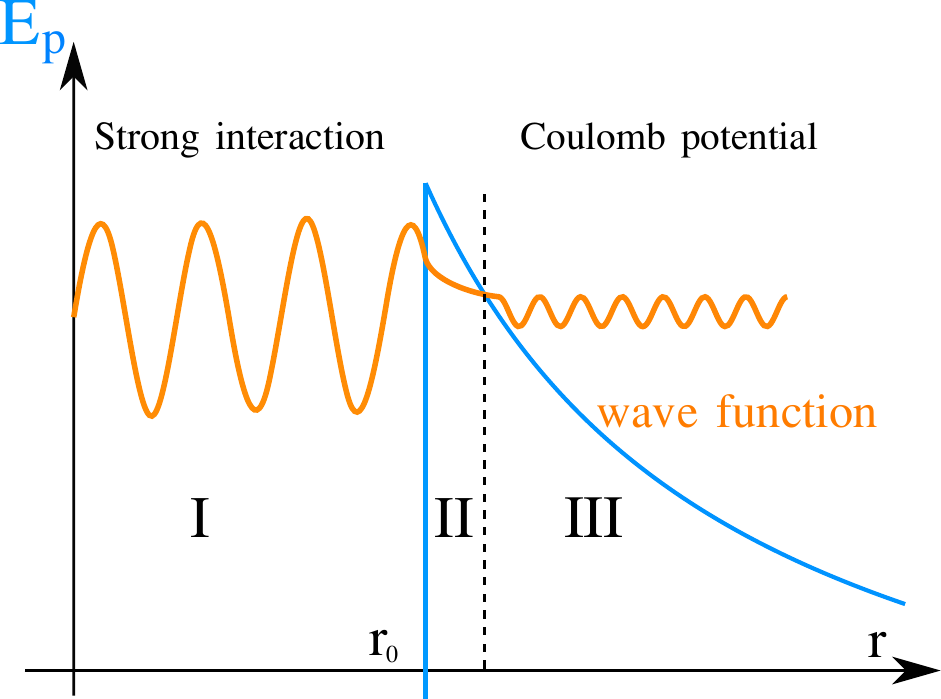}
   \includegraphics[width = 0.49\textwidth]{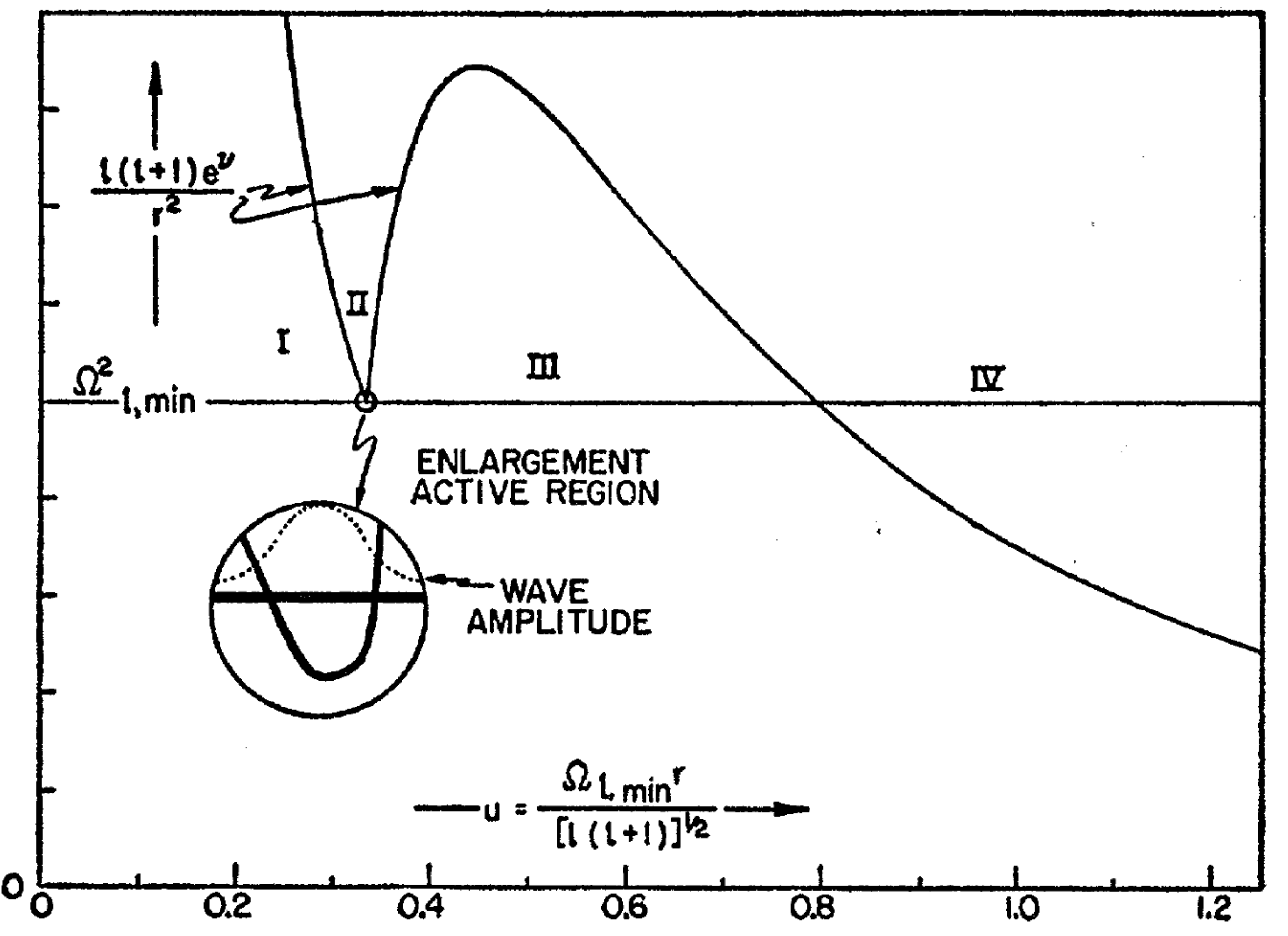}
   \caption[Radioactive alpha decay and geons]{In the radioactive alpha decay problem (left), the wave function of the $\alpha$ particle is
   submitted to a potential combining the strong interaction in the nucleus of the atom ($r \leq r_0$) and Coulomb's interaction when it
   is outside ($r > r_0$), since both the $\alpha$ particle and the nucleus are positively charged. From a classical point of
   view, the escape is impossible, but quantum mechanics allows the wave function to have an exponential tail that leaks outward
   the potential barrier. Thus, a small part of the wave function can propagate outside the nucleus and the probability of
   measuring the position of the particle outside the nucleus is non-zero. This is qualitatively the same picture for photons
   trapped in a geon (right). The active region corresponds to region II where photons are trapped in the minimum of the effective
   potential, and region IV is the leaking region. Credits: G. Martinon and \cite{Wheeler55}.}
   \label{alpha}
\end{figure}

Einstein's equations need sophisticated averaging techniques in near spherical symmetry to be tractable. In \cite{Wheeler55}, the metric is chosen
to be spherically symmetric with coordinates $(t,r,\theta,\varphi)$ but the potential vector for an individual toroidal mode is
given by the following ansatz:
\begin{subequations}
\begin{align}
   ds^2 &= -e^\nu \gls{c}^2dt^2 + e^\lambda dr^2 + r^2(d\theta^2 + \sin^2\theta d\varphi^2),\\
   A_\varphi &= \sin(\omega t)R(r)\sin\theta \frac{d}{d\theta}P_l(\cos\theta),
\end{align}
\end{subequations}
where $\omega$ is the frequency of the toroidal wave and $P_l$ is a Legendre polynomial. The radial structure of this geon, encoded
in the functions $R(r)$, $\nu(r)$ and $\lambda(r)$, obeys a radial wave equation (master equation) revealing an effective
potential. Such an equation is mathematically closed to the alpha decay problem where a quantum wave
function can escape a confining potential by quantum tunnelling, illustrated in figure \ref{alpha}. The problem can then be solved
via \gls{jwkb} techniques\footnote{It consists in solving approximately a differential equation in asymptotic regions where some
   terms can be simplified, and then match the solutions with continuity arguments. For example, in the alpha decay problem, the
   wave function obeys an equation of the form
\begin{equation*}
   \frac{d^2 \psi}{dx^2} = \left(\frac{1}{x} - \frac{1}{x_0} \right)\psi,
\end{equation*}
where $x$ is an adimensional radius and $x_0$ is the so-called turning point, corresponding to the minimal distance between an
alpha particle and the nucleus of an atom in a frontal Coulomb collision. There are then three asymptotic regions \RN{1}, \RN{2}, \RN{3} that
corresponds to $x \ll x_0$, $x \simeq x_0$ and $x \gg x_0$ respectively. In regions \RN{1} and \RN{3}, a slowly varying envelope
approximation can be used while in region \RN{2} an expansion around $x_0$ makes the computation tractable. Solving the equation in
each region and then matching the solutions preserving the continuity of $\psi$ and its first derivative gives an approximate
solution of the problem.}.
Here, photons are in a sense trapped in a potential combining the geon self-gravitation and the centrifugal force, but some of
them can nonetheless leak through the potential barrier with a non-vanishing amplitude.

One important numerical result obtained by Wheeler is that a clock at the centre of the spherical geon ticks at about 33\%
of the rate of an identical clock far away. Ernst \cite{Ernst57a} demonstrated analytically that this rate was precisely $1/3$
using a Ritz variational principle. He also gave analytical results for the mass and radius that were numerically guessed by Wheeler:
\begin{equation}
   M = \frac{4a\gls{c}^3}{27\gls{G}\omega} \quad \tn{and}\quad R = \frac{a\gls{c}}{3\omega}.
\end{equation}
The compactness of spherical geons is then universal and take the exact value $4/9$. Note that this compactness
matches precisely the limit obtained by Buchdahl four years later in 1959 for incompressible fluid spheres and fluid
governed by barotropic equations of state. This is so because the stress-energy tensor of a massless tensor field is traceless and
because of the thin-shell spherical symmetry \cite{Buchdahl59a,Buchdahl59b,Buchdahl59c}. Furthermore, the peak value of the
root-mean-square (rms) value of the electrical field is
\begin{equation}
   E_{max}^{rms} = 0.46\frac{a^{1/3}\gls{c}^4}{\gls{G}^{3/2}M}.
\end{equation}
All numerical factors are very close to 1, emphasising the power of dimensional analysis.

\begin{figure}[t]
   \includegraphics[width = 0.49\textwidth]{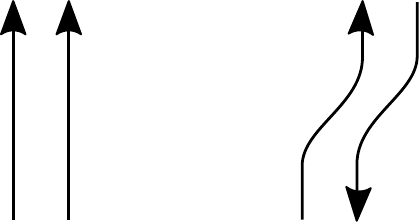}
   \caption[Self-attraction of photons]{Left: two photons with parallel propagation vectors do not interact gravitationally.
   Right: two photons with antiparallel propagation vector attract each other. Credits: G. Martinon.}
   \label{photons}
\end{figure}

Even if mathematically tractable, spherical geons are less stable than toroidal ones. Indeed, two nearly parallel pencils of light
attract gravitationally each other when their propagation vectors are oppositely directed, and not at all when similarly directed
\cite{Tolman34}. This is illustrated in figure \ref{photons}. Consequently a system of randomly oriented circular rays of light
drops to a state of greater stability when half of the angular momentum vectors orient themselves parallel, half antiparallel, to
a certain direction in space. The number of attractive bonds between orbits is then maximised. Spherical geons thus constitute an
unstable equilibrium, that spontaneously tends toward a toroidal configuration.

About the physical interest of geon configurations, we borrow these words from Wheeler \cite{Wheeler55}: ``\textit{geons make only this visible
contribution to science: it completes the scheme of classical physics by providing for the first time an acceptable classical
theory of the concept of body. One's interest in following geon theory down into the quantum domain will depend upon one's
considered view of the relation between very small geons and elementary particles}''. This is reminiscent of the now deprecated
geometrodynamics paradigm within which elementary particles were thought to be different manifestations of gravitational
bodies, including geons.

\subsection{Thermal geons}

Thermal geons \cite{Power57} are a generalisation of spherical geons where all confined modes with all possible azimuthal numbers are distributed
according to the most natural statistical law. A mode having frequency $\omega$ has then an idealised excitation
\begin{equation}
   m_{\omega}\gls{c}^2 \equiv E_{\omega} = \frac{\gls{hbar} \omega}{\exp\left( \frac{\gls{hbar} \omega}{\gls{kb}T} \right) - 1}.
   \label{bb}
\end{equation}
These objects are static, spherically symmetric and parametrised by a single parameter, the temperature $T$. They thus form the
simplest family of geons, and some of there properties can be estimated with very few arguments and calculations.

Thermal geon are merely a statistical self-gravitating system. Since the photons are supposed to be trapped in the gravitational
potential, it is reasonable to approximate a thermal geon by a self-gravitating black body obeying \eqref{bb}. For a black-body radiation, the energy
density in the active region should be approximately
\begin{equation}
   \frac{M\gls{c}^2}{R^3} \sim \frac{\pi^2\gls{kb}^4T^4}{15\gls{hbar}^3 \gls{c}^3},
\end{equation}
and on the other hand, the relativistic configuration imposes $GM/R\gls{c}^2 \sim 1$ so that
\begin{equation}
   M \sim M_T \equiv \sqrt{\frac{\gls{hbar}^3 \gls{c}^{11}}{\gls{G}^3 \gls{kb}^4T^4}} \quad \tn{and} \quad R \sim R_T \equiv \sqrt{\frac{\gls{hbar}^3 \gls{c}^7}{\gls{G} \gls{kb}^4T^4}}.
   \label{roughMR}
\end{equation}
Mass and radius therefore scale like $O(T^{-2})$, so that the heat capacity $dM/dT$ is negative. This indicates thermodynamical
instability. However, we can notice that this is also the case for black holes when the Hawking radiation is taken into account
(see section \ref{beyondflat}). This is a simple manifestation of the spontaneous evaporation of these objects. Hotter geons correspond to
higher energy densities with smaller mass and size. Numerical integration of the equations gives more precise results for the mass
and the outer boundary, namely
\begin{equation}
   M \simeq 0.12 M_T \quad \tn{and} \quad R \simeq 0.36 R_T.
   \label{mrthermal}
\end{equation}
This confirms again the rough dimensional analysis of \eqref{roughMR}. The compactness is then $1/3$, which is a lower value than
for the previous spherical case, in particular because the thin-shell approximation does not hold any more.

Just like there toroidal counterparts, thermal geons feature a maximum temperature and a inferior size limit determining the
limit beyond which the electron pair creation process becomes significant. Naively, we expect that this process is shut down
if the kinetic energy is small compared to the rest mass of a pair of electrons, namely $\gls{kb} T < 2
\gls{me}\gls{c}^2$. But a slightly more precise argument consists in assuming that the black-body energy density should be smaller than the
Schwinger limit. This yields
\begin{equation}
   \frac{\pi^2\gls{kb}^4T^4}{15\gls{hbar}^3 \gls{c}^3} < \frac{\gls{epsilon0}\gls{Es}^2}{2} = \frac{\gls{epsilon0}}{2}\left(\frac{\gls{me}^2\gls{c}^3}{\gls{e} \gls{hbar}}\right)^2 \iff \gls{kb}T \lesssim 1.7\ \gls{me}\gls{c}^2,
\end{equation}
where we used the fine structure constant $\alpha \equiv \gls{e}^2/4\pi\gls{epsilon0}\gls{hbar} \gls{c} \simeq 1/137$. The
limiting size of a thermal geon is thus $\sim \SI{e9}{m}$ and its corresponding mass $\sim \SI{e36}{kg}$. These values are very similar to the toroidal case results.

Thermal geons exhibit an intricate deformation of space-time. For instance, geodesic motions are very rich in such configurations.
For a ray which can escape to infinity, the impact parameter $P$ is defined as the asymptotic separation of the ray and a
parallel ray which comes straight through the centre of the geon without reflection.  Bound rays\footnote{that can nonetheless
escape on long time scales via potential barrier penetration.} can have impact parameters only between the two limits
\begin{equation}
   P_1 = 0.51 R_T \quad \tn{and} \quad P_2 = 0.62 R_T,
\end{equation}
and move always between radii
\begin{equation}
   R_{min} = 0.14 R_T \quad \tn{and} \quad R_{max} = 0.30 R_T.
\end{equation}

\begin{figure}[t]
   \includegraphics[width = 0.49\textwidth]{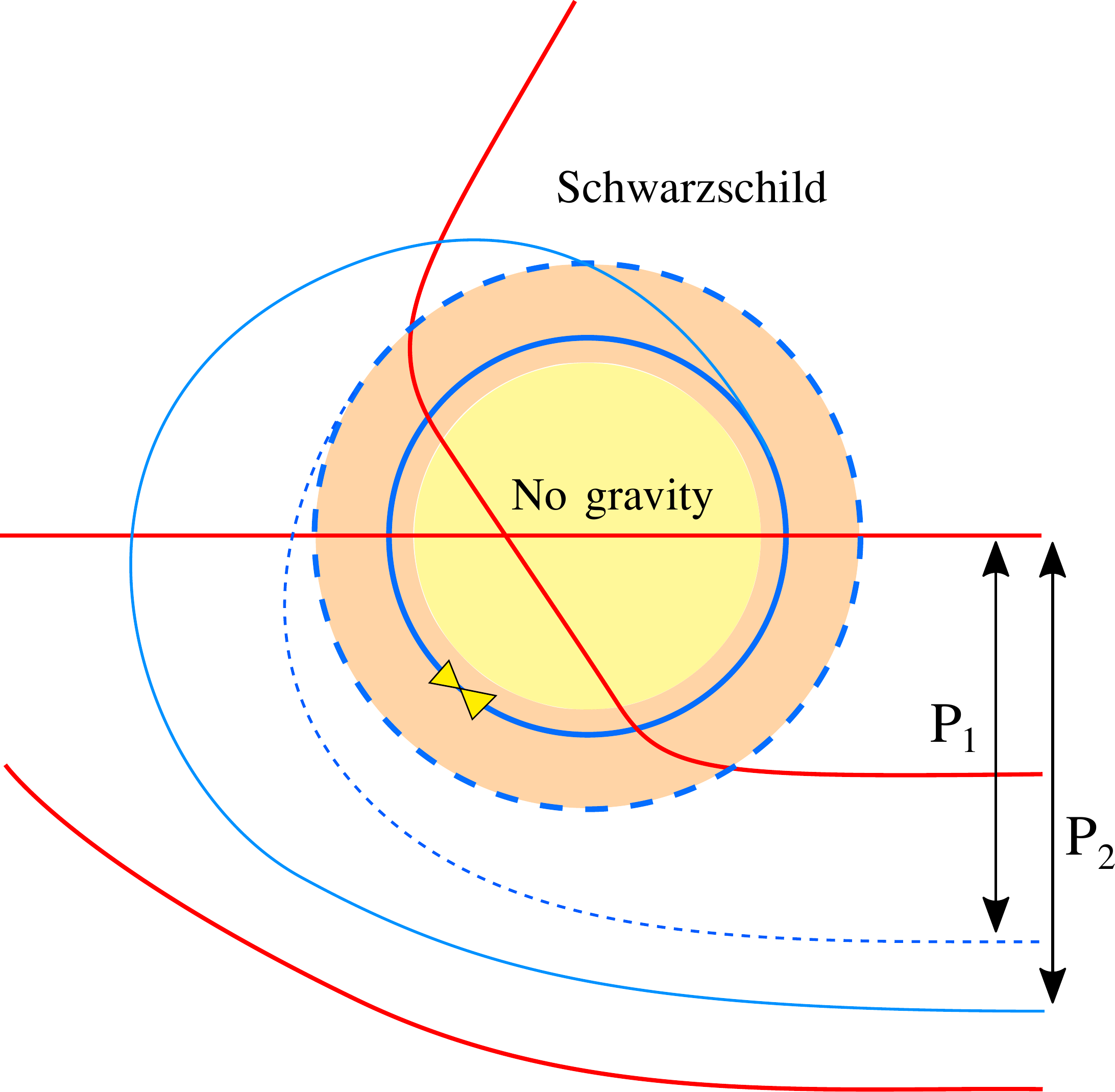}
   \caption[Null geodesics in a thermal geon]{Null geodesics behaviour in a thermal geon. The active region of the geon is
      pictured in shaded orange and is limited by $R_{min}$ and $R_{max}$. The interior region in pictured in shaded yellow and is
      a region where the gravitational potential is constant, with a flat geometry. Outside the active region, Birkhoff's
      theorem implies that the geometry matches the Schwarzschild metric. In red are pictured three typical null geodesics.
      Any geodesic intersecting the centre of the geon is not deflected. The solid blue line depicts a limiting null geodesics
      that could enter or leave the single stable circular orbit with impact parameter $P_2$. The dashed blue line depicts a
      limiting null geodesics that could enter or leave the single unstable circular orbit with impact parameter $P_1$. The
      two solid yellow triangles represents the allowed solid angle for bound rays intersecting the stable circular orbit. Any ray
      intersecting this point and having a propagation vector outside the pictured solid angle leaks from the geon sooner or later.
      Credits: G. Martinon, adapted from \cite{Power57}.}
   \label{thergeon}
\end{figure}

Among these null geodesics, there is exactly one which is circular and stable, at radius and impact parameter
\begin{equation}
   R_{circle} = 0.19 R_T \quad \tn{and} \quad P_{circle} = P_2 = 0.62 R_T.
\end{equation}
Figure \ref{thergeon} illustrates the behaviour of photons in a thermal geon. The circular orbit having $R = R_{max}$ and $P = P_1$ is
unstable but marks the boundary between orbits that can be trapped and those that cannot. Photons emitted toward the
centre of the geon are not deflected and the trapped ones follow infinitely precessing elliptic orbits. Furthermore, in the active
region, bound rays fill only a fraction of the entire solid angle, so that the actual energy density is less than the usual black
body radiation. Near the stable circular orbit, the solid angle occupied by bound rays is the largest and takes the value $\sim
0.59\times4\pi$.

Like their toroidal and spherical cousins, thermal geons are doomed to decay in time. When a ray escapes from the active region,
there is less pressure to sustain the gravitational collapse and the geon contracts. The electromagnetic energy content decreases
slowly not only by the monomolecular process of potential barrier penetration but also by bimolecular processes in which two photons
collide, either to produce a pair of electrons, or to go off as photons in new directions. For temperatures small compared to
$\gls{me}\gls{c}^2$, the number of electron pair production process is exponentially small and proportional to $\sim \exp(-\alpha
\gls{me}\gls{c}^2/\gls{kb}T)$ where $\alpha$ is a numerical factor of order unity. Electron pair production processes can then be
neglected in comparison to elastic collisions. Not all photon collisions result in a loss of energy for the system, since sometimes
the new quanta are emitted in trapped regions. A naive application of the Stefan-Boltzmann law gives
\begin{equation}
   \gls{c}^2 \frac{dM}{dt} = -4\pi R^2 \gls{sigma} T^4,
\end{equation}
where $\gls{sigma}$ denotes the Stefan-Boltzmann constant. With \eqref{roughMR}, we
recover that the mass decreases linearly in time, similarly to the toroidal case. This decay has a characteristic timescale
\begin{equation}
   \tau^{naive} = 0.058\sqrt{\frac{\gls{hbar}^3 \gls{c}^5}{\gls{G}\gls{kb}^4T^4}}\si{s}.
\end{equation}
However, the authors of \cite{Power57} performed the computation of the characteristic depletion time $\tau$ taking into account
the temperature dependence of the photon-photon collision section. They relied on \gls{qed} results, and gave the
more precise estimate
\begin{equation}
   \tau^{QED} \simeq 10^{-12}\left( \frac{\gls{me}\gls{c}^2}{\gls{kb}T} \right)^9 \si{s}.
\end{equation}
For instance, imposing $\tau = \SI{1}{yr}$ implies $\gls{kb}T\sim 0.07 \gls{me}\gls{c}^2$, $M_T \sim \SI{e39}{kg}$ and $R_T \sim
\SI{e12}{m}$. As the energy loss continues, the thermal geon shrinks in size, grows denser and hotter and loses energy at a rapidly
increasing rate. As the temperature rises to the neighbourhood of $\gls{me}\gls{c}^2$, pair production processes rapidly increase
in importance and quantum effects need to be taken into account.

\subsection{Pure magnetic or electric geons: Melvin's universes}

In 1964, Melvin, in a quest for unveiling new geons, found a solution of Einstein-Maxwell equations that was static, cylindrically symmetric and sustained by a pure
magnetic field (or equivalently a pure electric field) pointing parallel to the direction of the symmetry axis \cite{Melvin64}. Denoting by
$B_0$ the magnetic field on the polar axis and introducing the range radius
\begin{equation}
   \overline a = \frac{2\gls{c}^2}{B_0}\sqrt{\frac{\gls{mu0}}{4\pi \gls{G}}} \simeq 10^{19}\left( \frac{\SI{1}{T}}{B_0} \right)\si{m},
\end{equation}
the solution for pure magnetic geons reads
\begin{equation}
   ds^2 = (1+\rho^2)^2(-\gls{c}^2dt^2 + d\rho^2 + dz^2) + \frac{\rho^2}{(1+\rho^2)^2}d\varphi^2,
\end{equation}
where $z$ is the coordinate along the symmetry axis, $\varphi$ the azimuthal angle, and $\rho = r/\overline a$.

Their stability was investigated by Melvin himself in \cite{Melvin65} and Thorne in \cite{Thorne65}, who introduced the
concept of cylindrical \textit{C} energy, which was minimised by Melvin's solution. The author found that pure magnetic geons
were indeed stable against radial perturbations, which was in deep contrast to the toroidal case. Melvin's solution can be seen as a
limiting case of an average toroidal geon as seen by an observer very close to the active region, reminiscent of the linear case of
Ernst \cite{Ernst57b}. The stability of Melvin's solution means that toroidal geons are probably stable against collapse of their
minor radius $\Delta R$, but not to their major radius $R$. This is an example of the Faraday flux resistance to gravitational
collapse: the flux is held together by the balance between its outwardly directed pressure and its inwardly directed gravitational
pull.

Strictly speaking, these pure magnetic solutions are not bodies. They were appropriately renamed \textit{Melvin's universes} after
several studies of their geodesics \cite{Thorne65,Melvin65,Melvin66}. Indeed, no photon can have a circular
orbit beyond $r = \overline a/\sqrt{3}$, the gravitational attraction toward the symmetry axis being too large; while a photon
moving parallel to the symmetry axis is not deflected at all.  In other words, the escape velocity is infinite. So not only these
pure magnetic geons cannot collapse but nor can they send information to infinity. They thus constitute complete and closed universe:
``no news can enter or leave out''. More precisely, all particles, no matter what their initial positions and velocities are
(except those moving parallel to the symmetry axis), must have their orbits lying wholly or partially within the cylindrical
region $r < \overline a /\sqrt{3}$. This is mainly because such a universe is not asymptotically flat.

\section{Other types of geons}

If geons were first confined to Einstein-Maxwell equations, the concept rapidly extended to other types of fields. The basic
ingredient of geons, which is the photon, was thus progressively replaced by neutrinos, gravitational waves, scalar fields and
even spin-1 fields. The initial denomination of geon (gravitational-electromagnetic entity) was thus largely corrupted. However,
the name geon was still, in any case, deeply rooted to the concept of a self-gravitating and localised body. Hereafter we review
the different context in which a self-gravitating solution has been dubbed geon, and discuss their main properties.

\subsection{Neutrino geons}

In \cite{Brill57}, geons where the electromagnetic field was replaced by a massless Dirac field were studied. They were still called
geons though. Calculations were very similar but a new feature came into play in the effective potential: the spin-orbit coupling.

In analogy to photons, two neutrinos attract each other when their propagation vectors are antiparallel and not at all when they are parallel.
Therefore, neutrinos in a toroidal neutrino geon are in their most stable configuration when half go around one way and
half the other way.

In essence, geons can be constructed out of neutrinos in much the same way as their electromagnetic counterparts, and they
share the same qualitative features. One notable difference when going from Bose-Einstein to Fermi-Dirac statistics is the
well-known numerical factor of $7/8$ in the black-body energy density (see equation \eqref{78} of appendix \ref{quantumgas}).

\subsection{Gravitational geons}

After the study of electromagnetic and neutrino solutions, Brill and Hartle tackled the problem of gravitational geons,
i.e.\ made exclusively of gravitational waves \cite{Brill64}. The authors restricted their work to thin-shell spherically symmetric configurations.

In a gravitational geon, there is a background geometry which is strongly curved but which is treated as static or changing only
slowly with time. Superimposed on this background, is a ripple of space-time which has an amplitude very small
compared to unity and which obeys a linear wave equation. However, the reduced wavelength is so short compared to the dimensions
of the geon that the effective energy density is substantial. This effective energy density averaged in time serves
as a source of energy that produces the curvature of the background geometry. Averaging techniques similar to the Hartree-Fock self-consistent
approach for electrons in atoms are relevant for this particular problem.

The metric can be split into
\begin{equation}
   g_{\alpha\beta} = \overline{g}_{\alpha\beta} + h_{\alpha\beta}\quad \tn{with} \quad |h| \ll |g|,
\end{equation}
where, unlike usual perturbative calculations, the background $\overline{g}$ is unknown while the perturbation $h$ is decomposed
onto spherical harmonics. Einstein's equation can be put into the form
\begin{equation}
   G_{\alpha\beta}(g_{\mu\nu}) = 0 = G_{\alpha\beta}(\overline{g}_{\mu\nu}) + \langle \Delta G_{\alpha\beta}(\overline{g}_{\mu\nu},h_{\mu\nu})\rangle,
\end{equation}
where $G_{\alpha\beta}$ denotes Einstein's tensor and brackets denote an average over a time long compared to the
fluctuation period of $h_{\mu\nu}$ but short compared to the time needed by light to cross the gravitational geon. This can be rewritten
\begin{equation}
   G_{\alpha\beta}(\overline{g}_{\mu\nu}) = - \langle \Delta G_{\alpha\beta}(\overline{g}_{\mu\nu},h_{\mu\nu})\rangle,
\end{equation}
where the right-hand side is at least second order in $h_{\mu\nu}$ (the time average being zero at first order). This clearly translates the fact that the background metric is sourced
by small ripples of space-time. Such a framework has motivated Isaacson in 1968 \cite{Isaacson68a,Isaacson68b} to
develop the formalism of geometrical optics and effective energy-momentum tensor for gravitational waves.

For each frequency and radial dependence there are numerous solutions differing only in angular dependence and related to each
other by rotations about the centre of the gravitational geon. After averaging over many modes, only the time and radial
dependence remain, so that the calculation can be performed for a purely radial wave equation on the one mode with the simplest
feature.

It turns out that the radial structure of the gravitational geon obeys a master equation very similar to the one of the
electromagnetic and neutrino cases, so that many of their features are recovered, in particular the compactness value of $4/9$.
This result seems to be generalisable to any massless field confined to thin-shell spherical symmetry, obeying a linear
field equation, be it Maxwell, Dirac or the linearised Einstein's equation.

This problem generated a renewed interest in \cite{Anderson97,Perry99}, where the averaging formalism was
questioned and improved, and the effective energy-momentum tensor of the ripples was proven to be gauge-invariant. The problem of
size limits was never investigated for gravitational geons and, to the best of our knowledge, no quantum effects is supposed to appear before the Planck limit.

\subsection{Complex scalar geons: boson stars}

Geons made of a complex scalar field were first constructed by Kaup in 1968 \cite{Kaup68}, closely followed by Ruffini and
Bonazzola in 1969 \cite{Ruffini69}. The former dubbed them \gls{kg} geons while the latter gave birth to the concept of boson
stars (see \cite{Schunk03,Liebling12} for a review). Boson stars are very famous in the literature because they are completely
stable against perturbations and do not decay.

Boson stars consist of a complex scalar field solving the \gls{ekg} equations. Because of the $U(1)$ symmetry, all
the solutions have a conserved current in the complex plane. They are stationary, since only the phase of the scalar field is
oscillating. Furthermore, one important feature of boson stars is that the pressure is not isotropic.

In spherical symmetry, the radial structure of these objects is again determined by a master equation very similar to the
electromagnetic geon case. Because the field is massive, these kinds of geons have a maximum mass. In
\cite{Ruffini69}, it is found that the maximum mass of spherically symmetric boson stars is given by
\begin{equation}
   M_{max} \sim 10^{-15}\left(\frac{\SI{1}{kg}}{m}\right) \si{kg},
\end{equation}
where $m$ is the mass of the bosonic particle. The authors applied the formalism of second quantisation and treated also the case of
self-gravitating spin-1/2 fermions obeying Dirac's equation and recovered the equations of structure of a self-gravitating
spherically symmetric degenerated Fermi gas, obtained previously by Oppenheimer and Volkoff in 1939
\cite{Oppenheimer39}.

Similarly to electromagnetic geons, the scalar field around a boson star decreases exponentially, but is nowhere zero. Thus, in contrast
with neutron stars, the radius of a boson star is ill-defined, and the convention is to take the radius containing 99\% of
the total mass.

\begin{figure}[t]
   \includegraphics[width = 0.49\textwidth]{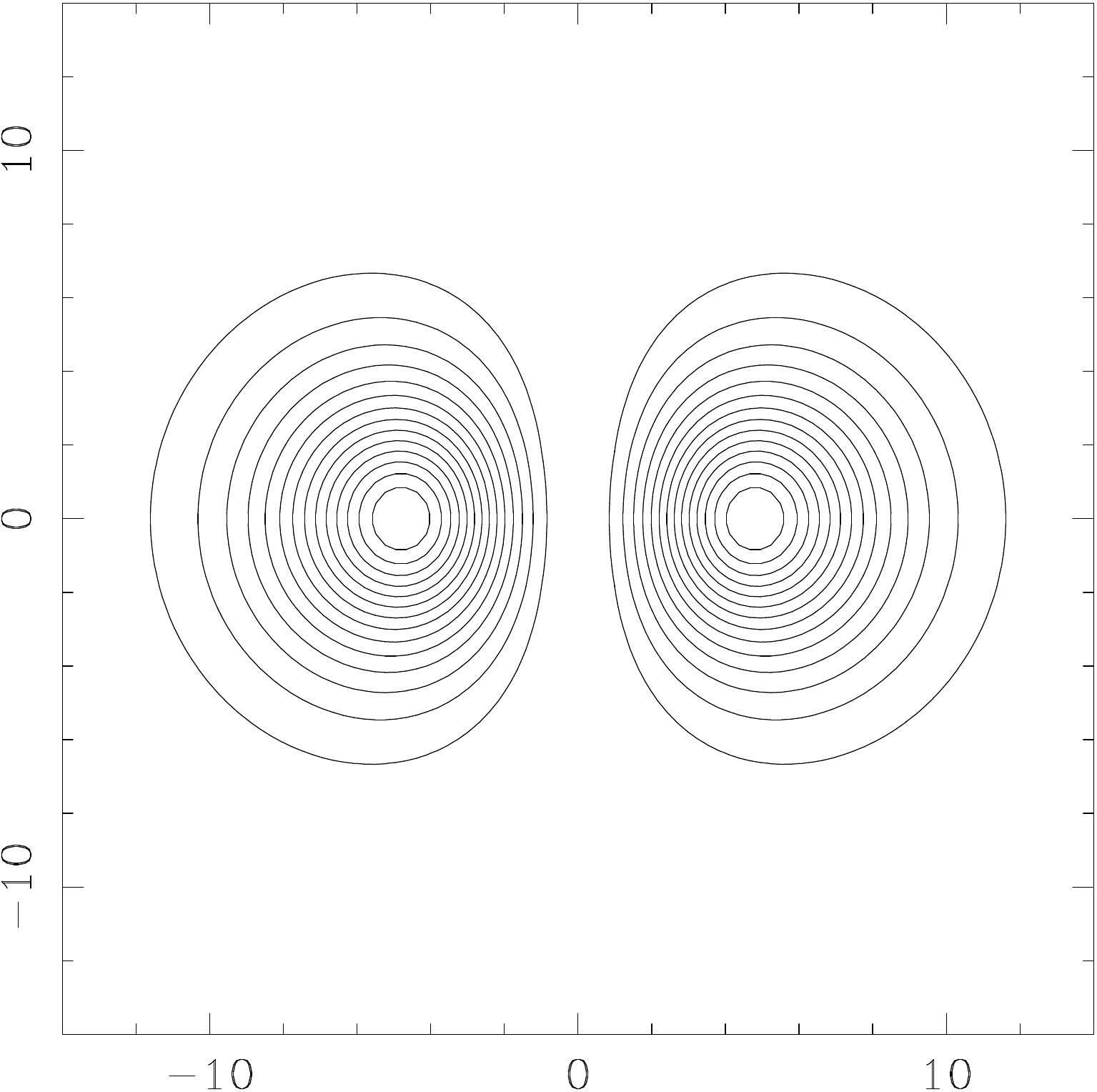}
   \caption[Isocontours of a toroidal rotating boson star]{Isocontours of the scalar field modulus in a meridian plane of constant
      $(t,\varphi)$ for a rotating toroidal boson star having $k=2$ and $\gls{hbar}\omega = 0.9m\gls{c}^2$. The length unit is
      $\gls{hbar}/m\gls{c}$ where $m$ is the mass of the boson.
      Credits: \cite{Grandclement14}.}
   \label{bosoncontour}
\end{figure}

Rotating boson stars were constructed in \cite{Schunck96,Schunk98,Ryan97,Yoshida97,Kleihaus05,Hartmann10,Brihaye14,Grandclement14}
assuming the following ansatz for the scalar field in spherical coordinates $(t,r,\theta,\varphi)$
\begin{equation}
   \phi = \phi_0(r,\theta)\exp[i(\omega t - k\varphi)].
\end{equation}
Boson stars have then quantified modes of rotations labelled by an integer $k$ and a scalar field frequency $\omega$. On figure
\ref{bosoncontour}, it is clear that, when rotating, they lose the spherical symmetry. Instead, they adopt the form of a torus,
which is reminiscent of toroidal electromagnetic geons. Boson stars with no repulsive terms in their interaction potential were
shown to be generically unstable in five dimensions in \cite{Brihaye16}, an unexpected feature very different from the
4-dimensional case.  This illustrates how important can be the number of space-time dimensions in stability problems.

\begin{figure}[t]
   \includegraphics[width = 0.49\textwidth]{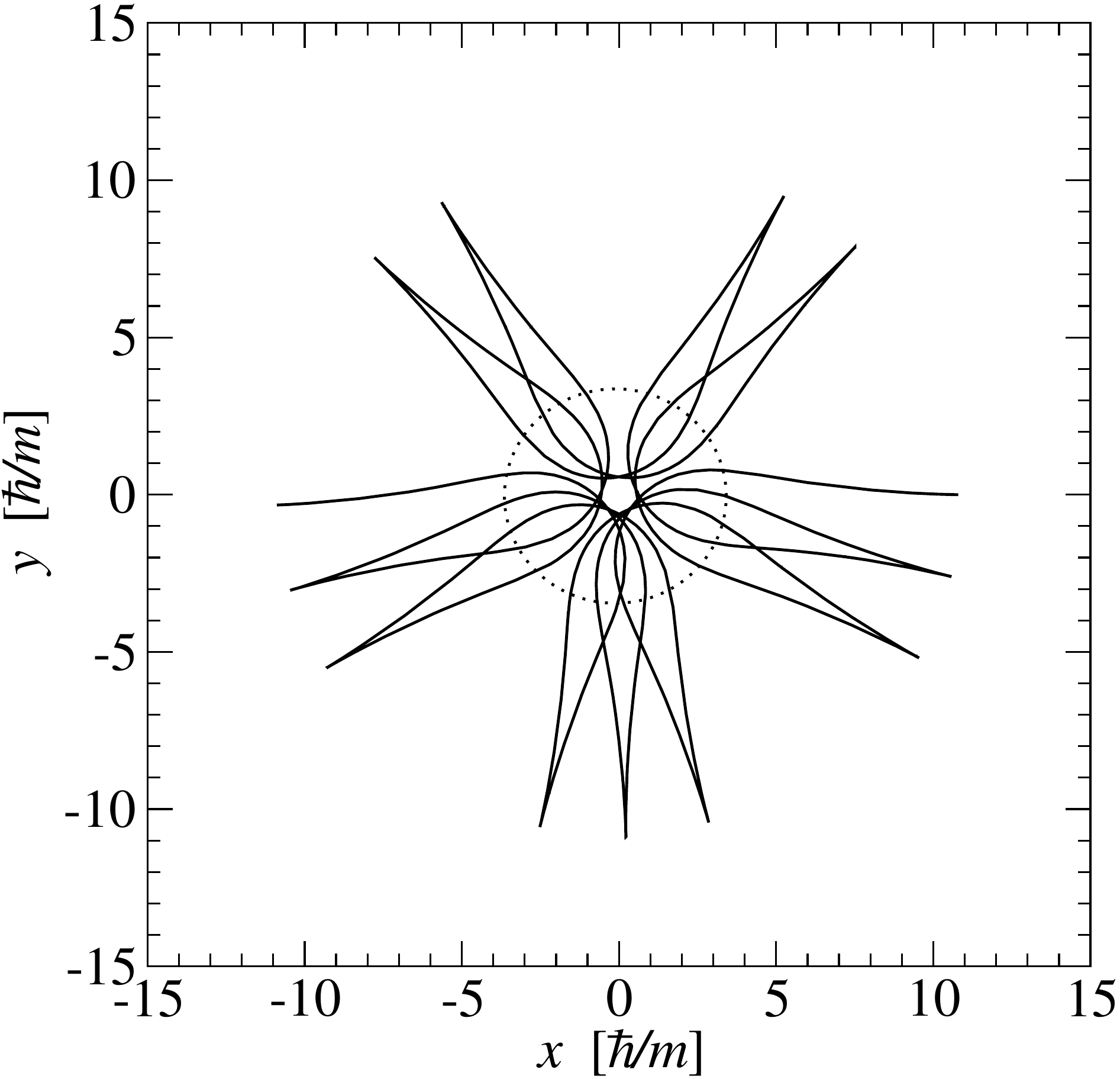}
   \caption[Orbits around rotating boson stars]{Orbit of an initially static test particle in the equatorial plane of a rotating
   boson star having $k = 2$ and $\gls{hbar}\omega = 0.75m\gls{c}^2$. The particle is initially at rest but builds up an angular
   velocity because of frame-dragging effects. The dotted circle marks the maximum of the scalar field modulus, where the boson
   star is the densest. Credits: \cite{Grandclement14}.}
   \label{petal}
\end{figure}

The orbits of test particles in a boson star geometry are very rich in nature. If an initially static test particle falls
toward a rotating boson star, it progressively builds up an angular velocity with respect to a static observer at infinity because of
frame-dragging effects. Once the particle has crossed the bosonic torus, the gravitational field drops very quickly to a near flat
geometry and the particle follows an almost straight line before re-entering the bosonic torus where frame-dragging starts again.
This gives the so-called \textit{pointy petals orbits} of figure \ref{petal}. In \cite{Grandclement17}, even stranger orbits were
found: light points, or stationary photons with respect to a static observer at infinity.

Charged boson star were studied for the first time in the late 80's in \cite{Jetzer89a,Jetzer89b}, and more recently in
\cite{Pugliese13}. Since identically charged particles repel each other, there can be no attraction between two bosons if their
individual charge is too high. Typically, the boson charge $q$ is restricted to be
\begin{equation}
   q < q_{crit} \equiv \sqrt{4\pi \gls{epsilon0}\gls{G}}m,
\end{equation}
where $m$ is the boson mass. This equation simply results from equating the Coulomb and Newton gravitational forces. However, for
relativistic self-gravitating systems, this does not take into account the gravitational binding energy per particle. In some
sense, it is expected that the curvature of space-time induced by the collective effect could allow charged boson
stars with boson charge $q = q_{crit}$. And indeed, in \cite{Pugliese13}, such configurations were obtained numerically in
spherical symmetry and proved to be stable.

\subsection{Real scalar geons: oscillatons}

It was in 1991 that Seidel and Suen considered objects similar to boson stars but this time with a real scalar field: oscillating
soliton stars, or oscillatons \cite{Seidel91}. Because the field has no imaginary part, no Noether current can be conserved, so
that the solutions of the \gls{ekg} equations feature a time-dependent metric. The authors succeeded to build numerically
spherically symmetric oscillatons.

In \cite{Seidel94}, the authors presented numerical simulations of scalar cloud collapse with different types of scalar fields in
spherical symmetry. They observed that in general, massless scalar fields dispersed and never formed compact object while massive
complex scalar fields collapsed to stationary boson stars and massive real scalar fields collapsed to time-periodic oscillatons.
The mass ejection ensures sometimes that no black hole is formed, a phenomenon called gravitational cooling. Because of spherical
symmetry, the question of fragmentation and of the Jean's instability was not addressed.

\begin{figure}[t]
   \includegraphics[width = 0.49\textwidth]{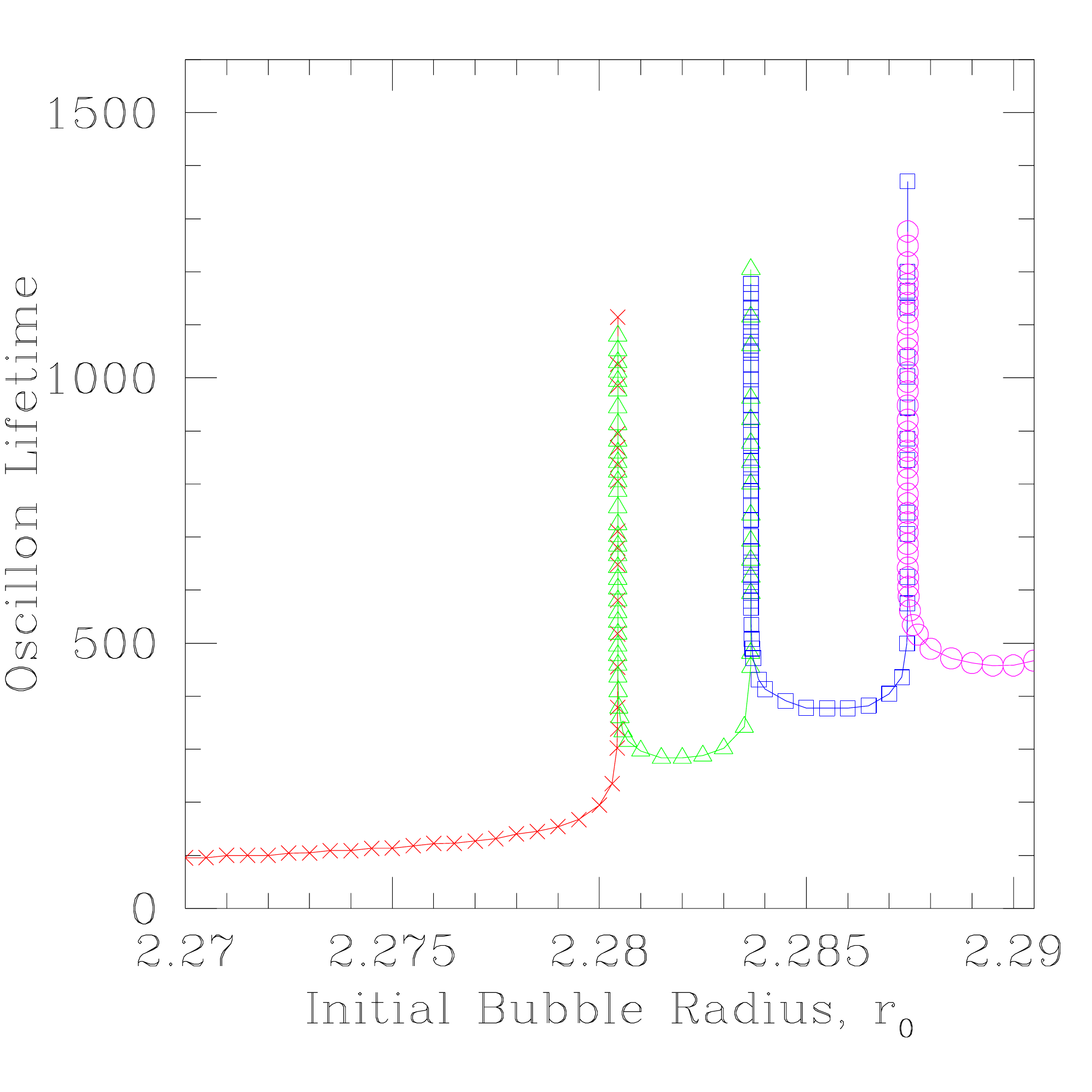}
   \caption[Resonant lifetimes of oscillons]{Oscillon lifetimes as a function of the width $r_0$ of the initial scalar wave packet.
      The lifetime seems to diverge logarithmically for three resonant parameters but is everywhere finite. Different colours and
      symbols correspond to families with different modulations of their oscillations. Credits: \cite{Honda02}.}
   \label{oscillon}
\end{figure}

Oscillatons are actually distant cousins of oscillons (see e.g.\
\cite{Bogolyubskii77,Segur87,Gleiser94,Copeland95,Honda02,Fodor06,Arodz06,Koutvitsky06,Arodz08,Fodor09a,Fodor09b,Amir12}), which
are localised, time-dependent, spherically symmetric solution of the \gls{kg} equation in a fixed everywhere flat space-time with a
double-well shaped potential
\begin{equation}
   \partial_t^2 \phi - \gls{c}^2\partial_x^2 \phi + \phi(\phi^2 - 1) = 0.
\end{equation}
Oscillons feature a lifetime much longer than their dynamical time, and can last sometimes for times longer than the age of the
universe\footnote{
Let us briefly mention the case of the sine-Gordon equation
\begin{equation}
   \partial_t^2 \phi - \gls{c}^2\partial_x^2 \phi + \sin\phi = 0,
\end{equation}
which is a true exception to the maxim ``anything that can radiate does radiate''. Indeed, the so-called breather solutions of
this equation do not radiate and are exactly periodic.}.
A scalar potential with a double-well shape is enough to trigger non-linearities that are responsible for their very long
lifetime. In particular, resonances emerge for some values of the parameter $r_0$ representing the width of the initial
oscillon wave packet, as illustrated in figure \ref{oscillon}. Fine-tuning can thus lead to oscillons with very long lifetimes. However,
\cite{Fodor06} demonstrated numerically that, in the case of a double-well shaped potential, no oscillon with finite mass could be
exactly periodic. They showed that there was necessarily an energy leak and a secular change in periodicity, so that even if not
obvious at first sight on figure \ref{oscillon}, lifetimes were always finite.

Changing the boundary conditions in order to let waves coming from infinity can turn the radiation into a standing wave tail. Such a
configuration is a reasonable approximation of the core part of the oscillon, but has an infinite mass. The authors proposed to
rename such configurations quasi-breathers. In \cite{Segur87,Fodor09a,Fodor09b}, it was shown that the energy emission rate of
small-amplitude oscillons was exponentially small in terms of their central amplitude.

In the case of oscillatons, even if the scalar field potential is just a mass term, gravity brings enough non-linearities to allow
for very long-living solutions. Even if this point was originally missed by Seidel and Suen, oscillatons, despite their
non-linearly stability\footnote{Two distinct notions have to be distinguished here: non-linear stability and decay in time. Actually, it is possible
to compute oscillaton configurations that are stable against perturbations, even at higher-than-first order levels. In other
words, no perturbation of an oscillaton can grow unboundedly in time, this is called non-linear stability. However, these oscillatons are not asymptotically flat, since they
feature a $O(r^{-1})$ tail at space-like infinity. The only way for oscillatons to be still asymptotically flat is to decay
slowly in time, but this does not conflict with their non-linear stability.}, radiate and decay in time because of the quantum
decay of scalarons into gravitons and because of the emission of gravitational and scalar waves. Indeed, the oscillations of the
metric tend to excite scalar waves that escape to infinity. In \cite{Page04}, denoting $\mu =
Mm/\gls{mpl}^2$ the dimensionless product of the oscillaton mass $M$ and the scalar mass $m$, Page estimated that the mass of the
oscillaton was decreasing due to the classical contribution $C(\mu)$ and the quantum contribution $Q(\mu)$ according to
\begin{equation}
   \frac{\gls{G}}{\gls{c}^3}\frac{dM}{dt} = -C(\mu) - Q(\mu) \quad \tn{with} \quad C(\mu) = \frac{\gamma}{\mu^2}e^{-\frac{\alpha}{\mu^2}} \quad \tn{and} \quad Q(\mu) = q \frac{m^2}{\gls{mpl}^2} \mu^5,
\end{equation}
where
\begin{equation}
   \alpha \simeq 40, \quad \gamma \simeq \num{3.8e6}, \quad q \simeq \num{8.5e-3}.
   \label{alphacq}
\end{equation}
This decay was accompanied by a secular change in periodicity. However, the computations of \cite{Page04} where flawed by the
divergence of the perturbative expansion the author considered. Furthermore, in spherical symmetry, due to Birkhoff's theorem, no
gravitational waves can be emitted. The classical mass loss of spherically symmetric oscillatons is thus entirely due to scalar radiation. In
\cite{Fodor10b}, using Borel resummation techniques, it was shown that the proportionality constant $\gamma$ in \eqref{alphacq} was
$\sim 200$ times smaller. 

Oscillatons are thus adiabatically radiating, just like electromagnetic geons, but with a mass
loss rate so low that it was missed by previous numerical studies. The more massive the boson, the shortest the lifetime.
Furthermore, oscillatons cannot radiate away their mass in a finite time, which make them very nearly stable for astronomical
purposes.

As many compact objects in \gls{gr}, oscillatons have a maximum mass beyond which they collapse to a black hole. In
table \ref{oscillatonlifetime}, we give the mass loss of oscillatons in a Hubble time $\sim \SI{14}{Gly}$ as a function of the
scalar field mass. All configurations are supposed to be maximally massive initially.

\begin{mytab}
\begin{tabular}{cccc}
\hline
                    $m\gls{c}^2/\SI{}{eV}$  & $M_{max}/\gls{msun}$ & $M/\gls{msun}$   & loss ($\%$)      \\
\hline
$\num{e-35}$ & $\num{8.20e24}$      & $\num{8.20e24}$  & $\num{2.91e-17}$ \\
$\num{e-30}$ & $\num{8.20e19}$      & $\num{8.20e19}$  & $\num{2.91e-12}$ \\
$\num{e-25}$ & $\num{8.20e14}$      & $\num{8.20e14}$  & $\num{2.90e-7}$  \\
$\num{e-20}$ & $\num{8.20e9}$       & $\num{8.20e9}$   & $\num{2.09e-2}$  \\
$\num{e-15}$ & $\num{8.20e4}$       & $\num{7.87e4}$   & $4.00$           \\
$\num{e-10}$ & $\num{8.20e-1}$      & $\num{7.06e-1}$  & $13.9$           \\
$\num{e-5}$  & $\num{8.20e-6}$      & $\num{6.24e-6}$  & $23.8$           \\
     1       & $\num{8.20e-11}$     & $\num{5.56e-11}$ & $32.2$           \\
$\num{e5}$   & $\num{8.20e-16}$     & $\num{4.98e-16}$ & $39.2$           \\
$\num{e10}$  & $\num{8.20e-21}$     & $\num{4.51e-21}$ & $45.0$           \\
$\num{e15}$  & $\num{8.20e-26}$     & $\num{4.11e-26}$ & $49.9$           \\
\hline
\end{tabular}
\caption[Lifetimes of oscillatons]{Mass $M$ of an oscillaton made of a scalar field of mass $m$. Each configuration is assumed to
   be initially maximally massive. The mass $M$ corresponds to the mass of this configuration after a Hubble
   time $\SI{14}{Gly}$ of decay. Credits: \cite{Grandclement11}.}
\label{oscillatonlifetime}
\end{mytab}

Deeper studies of oscillatons and their stability can be found in
\cite{Alcubierre02,Alcubierre03,Urena02,Balakrishna08,Matos09,Fodor10a,Fodor10b,Grandclement11,Urena12,Marsh15}. The authors of
\cite{Fodor10a} found that a positive cosmological constant is an additional source of decay of oscillatons. In
\cite{Grandclement11}, it was demonstrated numerically that no oscillaton with finite mass could be exactly time-periodic. Akin to
oscillons, oscillatons have then always a finite lifetime.

\subsection{Spin-1 geons: Proca stars and vector oscillatons}

\begin{figure}[t]
   \includegraphics[width = 0.49\textwidth]{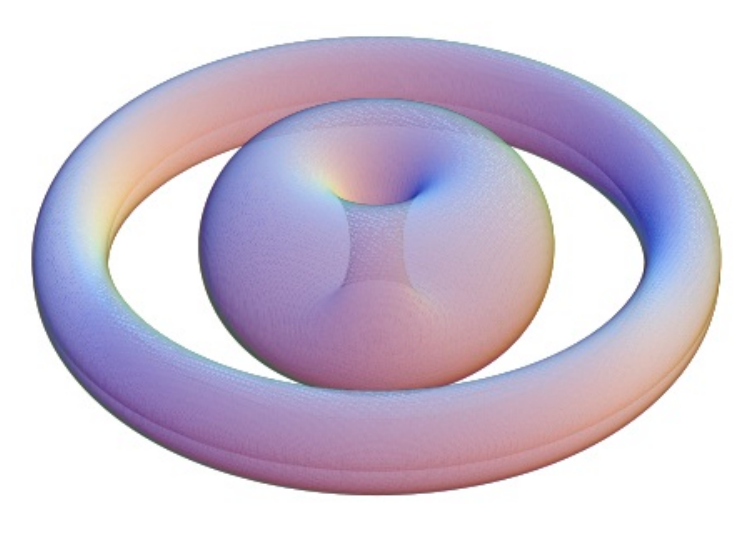}
   \caption[Shape of rotating Proca stars]{Surfaces of constant energy density for a rotating Proca star with $m=2$ and
      $\gls{hbar}\omega=0.8\mu\gls{c}^2$, where $\mu$ is the mass of Proca field.
      The topology is di-ring-like. Credits: \cite{Herdeiro16b}.}
   \label{procatorus}
\end{figure}

Spin-1 bosonic stars are the most recent geon-type solutions. On the one hand, Proca stars were obtained in \cite{Brito16a} and
can be regarded as Abelian spin-1 field geons, i.e.\ made of complex and massive ``photons''. Because of the $U(1)$ symmetry,
there is a conserved complex current. The authors looked for spherically symmetric as well as axisymmetric rotating solutions.
Similarly to boson stars, Proca stars exhibit a maximum mass connected, at least in spherical symmetry, to a stable and an
unstable branch of solutions. Unlike boson stars, the topology of rotating Proca star is not limited to tori, but can exhibit
Saturn-like or di-ring-like isocontours, as illustrated in figure \ref{procatorus}. Their stability was investigated in
\cite{Gual17} in spherical symmetry. The authors found that the stable and unstable branch connected each other at the maximum
\gls{adm} mass.

On the other hand, vector oscillatons are the real vector field counterparts of Proca stars. They were obtained for the first time
in \cite{Brito16c} in spherical symmetry. The metric and vector fields were Fourier expanded in time with coefficients displaying a
radial dependence. It turned out that the Fourier series converged extremely rapidly and only a handful of modes were needed. The
vector oscillatons thus obtained were constituted of a real massive vector field. Due to the time dependence of such solutions, studying
their linear stability was a very challenging problem that was not addressed yet. The intuition is that, due to their close
similarity with boson and Proca stars, vector oscillatons (apart from their well-known secular decay) display a stable and an
unstable branch that are connected at the maximum mass configuration. Unstable configurations are thought to collapse to a black
hole or to migrate back to the stable branch via mass ejection, the so-called gravitational cooling mechanism originally observed
by Seidel and Suen \cite{Seidel94}.

\section{Geons in the sky}

The applications of geons are very far from what Wheeler originally expected. The actual main motivation for studying these
objects is to find dark matter candidates, since cosmology teaches us that near a third of the mass of the universe is not
baryonic. Be it a scalar, electromagnetic, Proca, neutrino or gravitational, all geon-type solutions provide new ingredients to
add to Einstein's theory that could be distinguished by present and future astronomical observations.

\subsection{Radio astronomical detection}

\begin{figure}[t]
   \includegraphics[width = 0.49\textwidth]{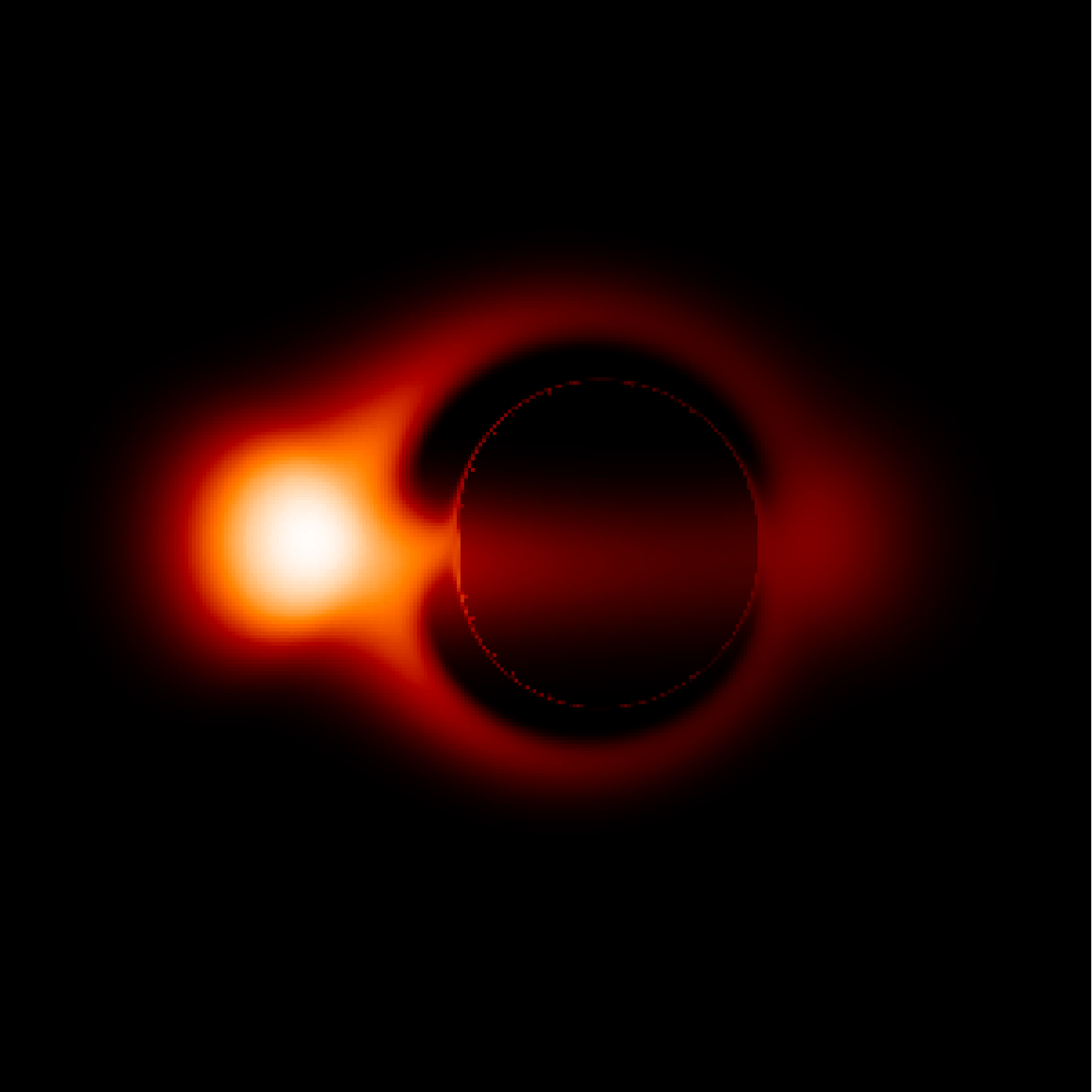}
   \includegraphics[width = 0.49\textwidth]{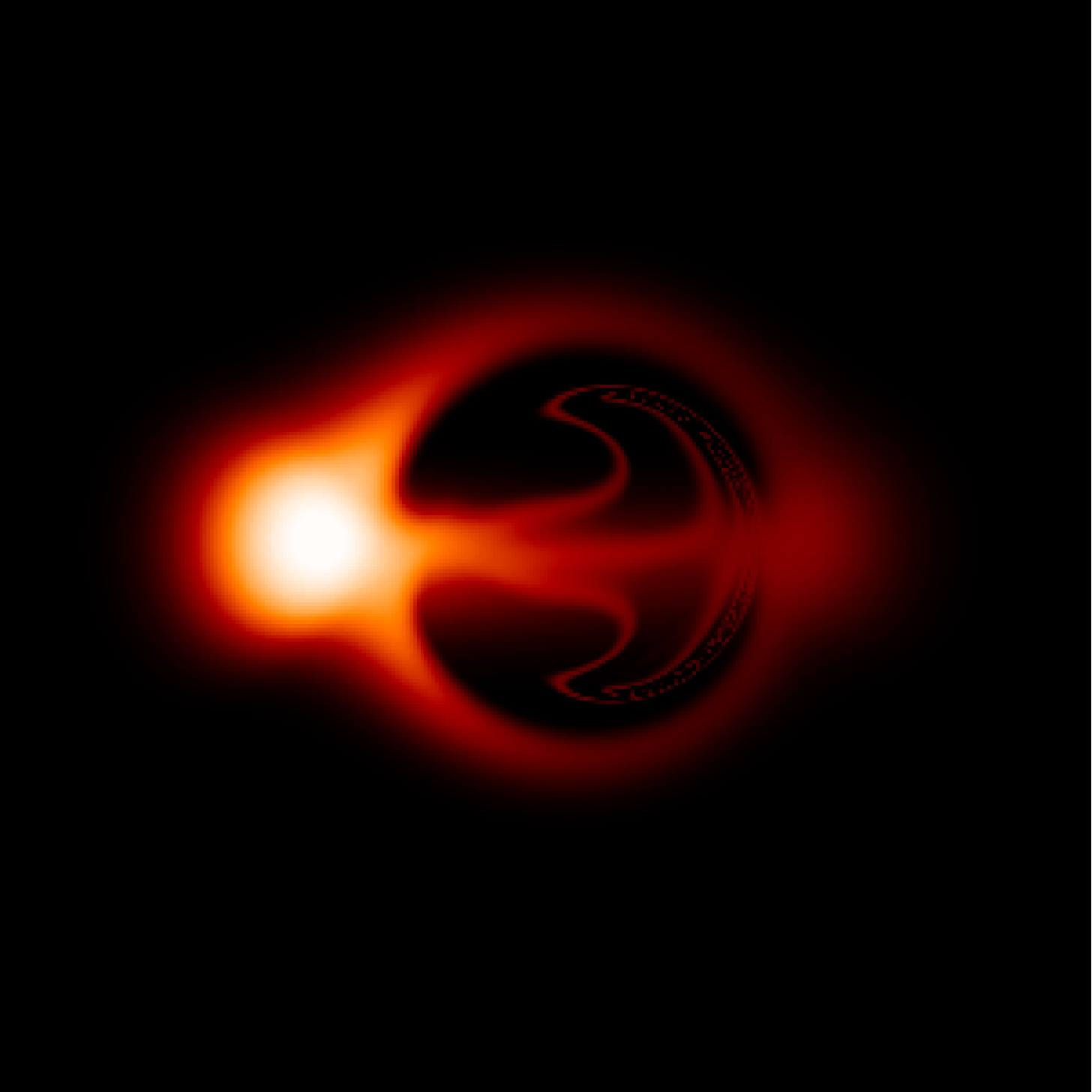}
   \caption[Images of a boson star and a black hole]{Left panel: image of a Kerr black hole with mass identical to Sgr A$^*$, the
      compact object at the centre of the Milky Way, and a reduced angular momentum $a \equiv \gls{c}J/\gls{G}M^2 = 0.9$. Right panel: image of a
      rotating boson star with $k=1$ and same mass and angular momentum. The images are computed at a wavelength $\lambda =
      \SI{1.3}{mm}$. Credits: \cite{Vincent16a}.}
   \label{bosonimage}
\end{figure}

Of particular astrophysical relevance are boson and Proca stars since they share an absolute stability in time with no decay at
all. A key challenge in present astronomy is precisely to detect potential evidence in favour of (or against) the existence of
boson stars in the universe. Notably, in \cite{Vincent16a}, simulated images of a Kerr black hole are compared to simulated images
of a boson star sharing the same mass and angular momentum. The difference can be seen on figure \ref{bosonimage}, as it would
look like in radio astronomical observations. It turns out that, even if boson stars do not have an event horizon, their images
are very much alike those of a Kerr black hole. In particular, they exhibit a very large shadow region. Thus, imagery does not
seem to be the most relevant criterion to distinguish between these two types of compact objects.

Let us also mention that the GRAVITY instrument is currently harvesting data about the galactic centre in order to discriminate
between Kerr and boson star geodesics by observing the orbits of the closest stars. The project of the Event Horizon Telescope (EHT)
should also contribute to this promising research area in the future.

\subsection{Interaction with black holes}

When a Kerr black hole is surrounded by a scalar or Proca field, it is said to have respectively scalar of Proca hairs. The
problem of scalar hairy black hole detection with radio astronomical observations was investigated in \cite{Vincent16b}. The authors
concluded that, provided that the deviation from Kerr space-time is large enough, scalar hairs could be inferred from several
properties of the photon ring, the projection of the innermost photon orbit.

\begin{figure}[t]
   \includegraphics[width = 0.49\textwidth]{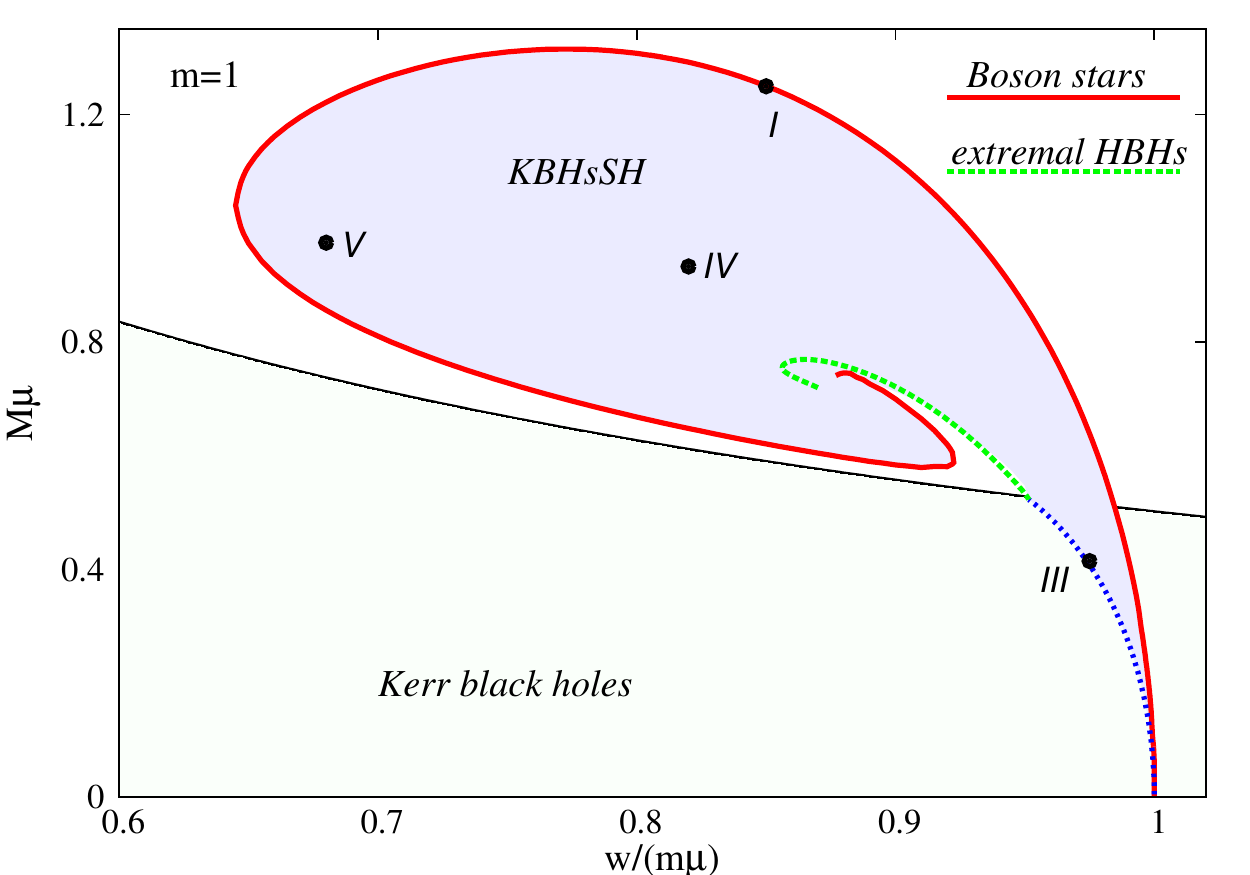}
   \includegraphics[width = 0.49\textwidth]{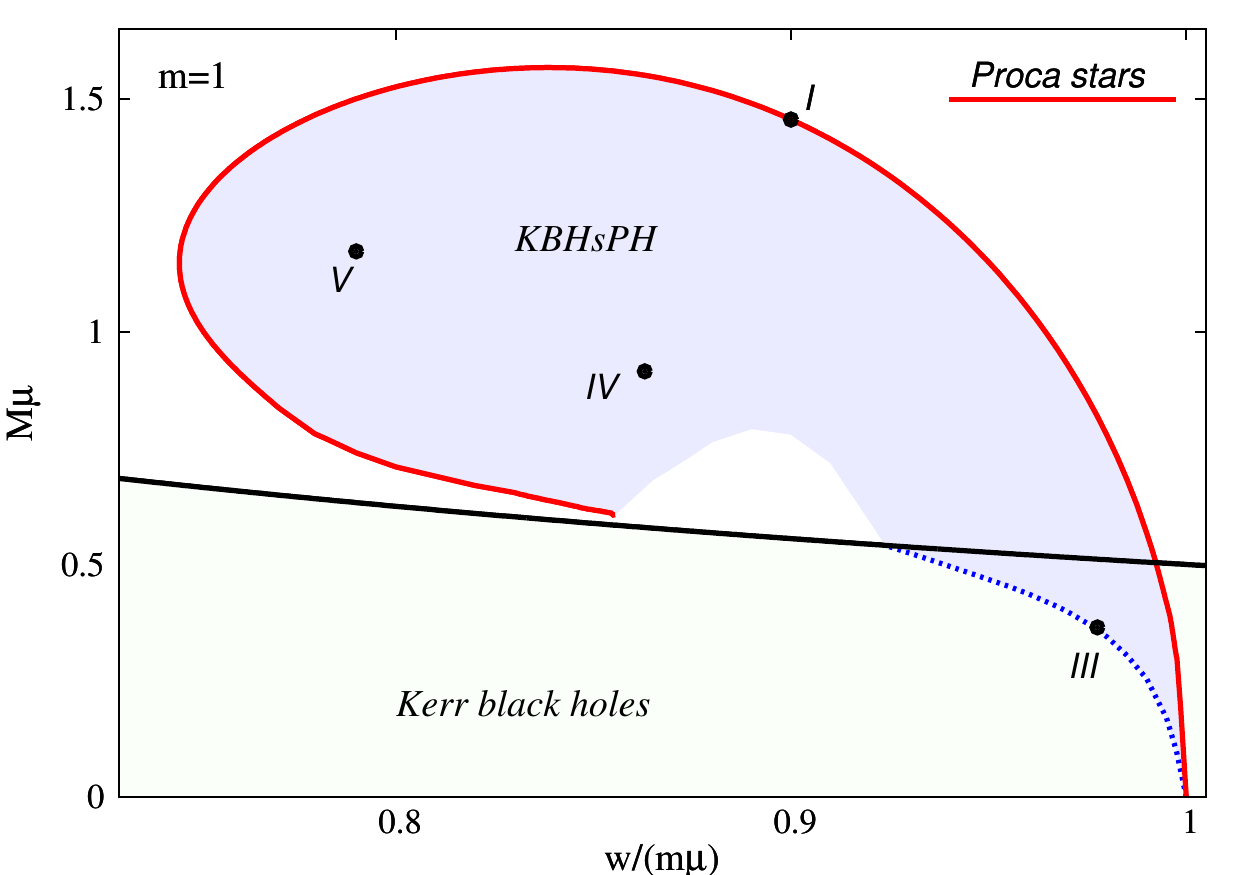}
   \caption[Kerr black holes with scalar and Proca hairs]{Mass-frequency diagram for an azimuthal number $m=1$ Kerr
      black hole with scalar hairs (left) and Proca hairs (right). Boson and Proca stars existence curves are pictured in solid
      red. Extremal Kerr solutions lie on the solid black curve. Hairy black holes exist in the blue shaded regions. Credits: \cite{Herdeiro16b}.}
   \label{kerrbosonproca}
\end{figure}

Interactions between black holes and Proca fields were studied in \cite{Zilhao15,Herdeiro16b}. Proca stars and boson stars are
close counterparts and are both connected to Kerr black holes. Such configurations can spread much larger regions in
parameter space than Kerr black holes or boson/Proca stars alone, as pictured in figure \ref{kerrbosonproca}.

Regarding oscillatons as an alternative to the black hole hypothesis for Sgr A$^*$ at the centre of the galaxy, they were
studied in \cite{Alcubierre02}, and as dark matter candidates in \cite{Matos09,Marsh15}. In these studies, it appeared that the
collapse of a real scalar cloud could lead to a viable oscillaton at the centre of galaxies, with a density profile that could be
distinguished with future measurements of cosmological structure formation.

\subsection{Interaction with standard model stars}

Not only black holes can have real scalar of vector hairs, but ordinary stars can also interact with bosonic fields. In
\cite{Brito16c,Brito15b}, the interaction between standard model stars with potential dark matter solitonic cores were studied. The
authors advocated that ordinary stars could host a scalar or vector core during their formation, due to the collapse of a cloud of
gas in a bosonic environment. They also argued that such hybrid stars with scalar pulsating cores could be formed via accretion by lumps of
oscillatons. In particular, the gravitational cooling mechanism originally discovered by Seidel and Suen \cite{Seidel94} seemed to be very
efficient to avoid black hole formation in these scenarios. This thus contradicted the accepted view according to which a star
would accrete more and more bosonic matter until a critical mass is reached and a black hole is formed. Actually in
\cite{Brito16c}, the unstable branch was never reached: all accretion scenarios seemed to efficiently migrate to the stable branch
via gravitational cooling.

One of the main results of these papers is that stars featuring an oscillaton core oscillate, driven by the scalar field, at a
frequency
\begin{equation}
   f = \SI{2.5e14}{}\left( \frac{m \gls{c}^2}{\SI{}{eV}} \right)\SI{}{Hz},
\end{equation}
where $m$ is the mass of the boson. For axion-like particles with masses $mc^2 \sim \SI{1e-5}{eV}$, these stars would emit in the
microwave band. This suggests that, with precise asteroseismology measurements, stars could work as dark matter detectors.

\section{Beyond asymptotically flat geons}
\label{beyondflat}

Geons are in essence strongly relativistic and highly non-linear systems. In some way, they allow us to probe
the very deep nature of Einstein's general theory of gravitation. They display a very intricate geodesic structure which is a open
window onto their space-time richness.

Such self-gravitating bodies have attracted a lot of attention in the literature, and our list of references is not exhaustive.
Table \ref{geonrecap} summarises the properties of geons covered in this chapter. 

\begin{mystab}
\begin{small}
\begin{tabular}{llllccccc}
\hline
geon               & field                          & geometry                 & asymptotically    & $M_{min}$ & $M_{max}$ & decay & year & first reference     \\
\hline
toroidal           & photons                        & toroidal                 & flat              & \cmark    & \xmark    & \cmark & 1955 & \cite{Wheeler55}   \\
spherical          & photons                        & spherical                & flat              & \cmark    & \xmark    & \cmark & 1955 & \cite{Wheeler55}   \\
thermal            & photons                        & spherical                & flat              & \cmark    & \xmark    & \cmark & 1957 & \cite{Power57}     \\
Melvin universe    & electric or magnetic           & spherical                & Melvin's universe & \xmark    & \xmark    & \xmark & 1964 & \cite{Melvin64}    \\
neutrino           & neutrinos                      & spherical                & flat              & \xmark    & \xmark    & \cmark & 1957 & \cite{Brill57}     \\
gravitational      & gravitons                      & spherical                & flat              & \xmark    & \xmark    & \cmark & 1957 & \cite{Brill64}     \\
boson stars        & complex massive scalar         & spherical/toroidal       & flat              & \xmark    & \cmark    & \xmark & 1968 & \cite{Kaup68}      \\
oscillatons        & real massive scalar            & spherical/toroidal       & flat              & \xmark    & \cmark    & \cmark & 1991 & \cite{Seidel91}    \\
Proca stars        & complex Abelian massive vector & spherical/Saturn/di-ring & flat              & \xmark    & \cmark    & \xmark & 2016 & \cite{Brito16a}    \\
vector oscillatons & real massive vector            & spherical                & flat              & \xmark    & \cmark    & \cmark & 2016 & \cite{Brito16c}    \\
\hline
\end{tabular}
\end{small}
\caption[Summary of geon properties]{Summary of geon properties in asymptotically flat space-times. $M_{min}$ and $M_{max}$ stand
for minimum and maximum mass respectively. For the electromagnetic geons, the minimum mass is not a physical limit, but a boundary
between quantum and classical treatment. Credits: G. Martinon.}
\label{geonrecap}
\end{mystab}

\subsection{Definition of geons}

Even if the concept of geons was crystal clear at the beginning, namely a gravitational-electromagnetic entity, it became more and
more confuse as several ingredients replaced the electromagnetic fields in subsequent realisations of geons (like scalar field,
gravitational waves, neutrinos\ldots). So at this point, we find it useful to give a definitive definition to the concept of geon:
\begin{definition}[Geons]
A geon is a regular self-gravitating localised body made of fundamental bosonic or massless fermionic fields that has a finite mass
and is periodic or quasi-periodic in time.
\label{defgeon}
\end{definition}
By quasi-periodicity, we mean that the eigen frequencies of these geons can potentially suffer from secular changes on time-scales
that are much larger than the crossing time, which is typically the time it takes for a photon to go through or circle around the
geon. With this definition, white dwarfs and neutron stars cannot be geons since they are made of several kinds of elementary
fermions (at least protons and neutrons). Black holes are not geons either since they are singular and Melvin's universes have a
priori an infinite mass with respect to a Minkowski background. Thus, most of the self-gravitating systems described in this
chapter are geons. An interesting theorem by Gibbons and Stuart \cite{Gibbons84} has established the absence of asymptotically
flat solutions of Einstein's equation which are time-periodic and empty near infinity. This implies that if asymptotically flat
geons with an oscillating field exist, they cannot be strictly time-periodic but they should necessarily radiate and suffer a
secular change in their frequency.

\subsection{Analogy with Hawking radiation}

As all the above geon cases have shown, there seems to be few exceptions to the rule ``anything that can radiate does radiate''.
Since the works of Bekenstein in 1973 and Hawking in 1975 \cite{Bekenstein73,Hawking75}, it is well known that even black holes do
radiate away their mass and angular momentum via particle pair production. The mechanism is completely different, because it is
quantum in nature, but still illustrates the above maxim. For example, a Schwarzschild black hole of mass $M$ emits a black-body
radiation at a temperature $T_H$ given by
\begin{equation}
   \gls{kb}T_H = \frac{\gls{hbar} \gls{c}^3}{8\pi \gls{G}M}.
\end{equation}
This is to be compared to the thermal geon result for which $M\propto T^{-2}$. A crude estimate of the mass loss
\cite{Page76} gives, according to the Stefan-Boltzmann law:
\begin{equation}
   \gls{c}^2\frac{dM}{dt} = -4\pi r_S^2 \gls{sigma} T_H^4,
\end{equation}
where $r_S = 2\gls{G}M/\gls{c}^2$ is the Schwarzschild radius. It then follows that
\begin{equation}
   \frac{dM}{dt} = -\frac{\alpha}{M^2} \quad \tn{with} \quad \alpha = \frac{\gls{hbar} \gls{c}^4}{15360\pi \gls{G}^2} \simeq \SI{3.84e15}{kg^3.s^{-1}},
\end{equation}
which can be solved, denoting $M_0 = M(t=0)$, by
\begin{equation}
   M(t) = M_0\left( 1 - \frac{3\alpha t}{M_0^3} \right)^{1/3}.
\end{equation}
The characteristic depletion time is thus given by
\begin{equation}
   \tau = \frac{M_0^3}{3\alpha}.
\end{equation}
In comparison, toroidal geons decay linearly in time. With \eqref{decaygeon} and \eqref{attrition}, we find that a toroidal geon with
azimuthal number $a = 143$ and a Schwarzschild black hole both of initial mass $M = 10^6 \gls{msun}$ decay with the same
characteristic time, which is much larger than the age of the universe. So the decay property of geons is in some sense similar to
black hole evaporation.

\subsection{How to suppress geons decay}

Finally, be it geons or black holes, many solutions of Einstein's equations tend to evaporate, even if they
can be very nearly stable for astronomical purposes. A legitimate theoretical question is then: how to get rid of this inexorable
decay that flaws most of these objects? \gls{ads} space-time brings a simple answer. With its negative
cosmological constant, it prevents radiation to leak toward infinity and acts like a confining reflecting box.
Furthermore, black holes and geons are thermodynamically unstable in asymptotically flat space-times, having a negative heat capacity
$dM/dT < 0$, but this problem is completely cured in \gls{aads} ones \cite{Hawking82}. In theory, geon solutions could thus survive when
settled into such space-times. This topic is examined further in chapter \ref{adsinsta}. The next chapter is attached to the
definition of \gls{aads} space-times and chapter \ref{adscft} explains why they are of physical interest.

\chapter{Asymptotically anti de-Sitter space-times}
\label{aads}
\addcontentsline{lof}{chapter}{\nameref{aads}}
\addcontentsline{lot}{chapter}{\nameref{aads}}
\citationChap{Trips to fairly unknown regions should be made twice; once to make mistakes and once to correct them.}{John Steinbeck}
\minitoc

\gls{ads} space-time is the maximally symmetric solution of Einstein's equation in vacuum with negative cosmological constant. As a
consequence, it has a constant negative curvature. In this regard, it is a close counterpart of Minkowski and \gls{ds} space-times which are the two
other maximally symmetric solutions of Einstein's equation, with vanishing and positive cosmological constant respectively. The
concept of a cosmological constant was introduced by Einstein in 1917 \cite{Einstein17} in order to get static solutions of the
universe. This was the accepted view at the time. However, de Sitter was convinced that Einstein's theory was a prediction of the
expansion of the universe. He studied \gls{gr} in a series of three articles published between 1916 and
1917 \cite{deSitter16a,deSitter16b,deSitter17a,deSitter17b}, where he notably introduced the \gls{ds} space-time. This was interpreted as
a definitive argument according to which one of \gls{gr} predictions was a putative universe in expansion. Even if Einstein referred to the cosmological
constant as its ``biggest blunder'' soon after Hubble discovered that the universe was truly expanding \cite{Hubble29}, it turns
out that since 1998, we know that the (positive) cosmological constant is the best fit of the cosmological data and lead to the
famous $\gls{Lambda}$-\gls{cdm} model, whose \gls{cdm} part was already discussed in 1932 by Einstein and de Sitter himself
\cite{Einstein32}.

The interest of the \gls{ads} space-time does not lie in cosmology but in theoretical physics and the unification of quantum
theories with \gls{gr}. Another interest of this space-time is that it allows us to better understand \gls{gr} itself, since it
exhibits a lot more features and instabilities that its Minkowski and \gls{ds} counterparts. We discuss in detail these
aspects in chapters \ref{adscft} and \ref{adsinsta}. In this chapter, we aim at giving a rigorous definition of what is precisely
an \gls{aads} space-time. The common feature of such space-times is the presence of a reflective boundary.

Many technicalities of this chapter are delegated to appendices about \gls{gr} (appendix \ref{grd}), $d+1$ formalism (appendix
\ref{d+1}) and principle of least action (appendix \ref{leastaction}), whose reading is strongly encouraged at this point of the
manuscript. From now on and hereafter, the space-time is of dimension $n$. For convenience, we also introduce $d = n-1$ which is
the number of dimensions of hypersurfaces and $p = n-2$ which is the number of dimensions of the spheres. We set the speed of
light to $\gls{c} = 1$.

\section{Anti-de Sitter space-time}

We start by discussing the simplest \gls{aads} space-time, \gls{ads} itself. We define it as the only maximally symmetric solution of Einstein's
equation in vacuum with negative cosmological constant.

\subsection{Maximally symmetric space-times}

There are two equivalent definitions of a maximally symmetric space-time:
\begin{myenum}
   \item A space-time is said to be maximally symmetric if and only if it has the maximal number $n(n+1)/2$ of Killing
      vectors.
   \item A space-time is said to be maximally symmetric if and only if it is spatially homogeneous and isotropic.
\end{myenum}
It is shown in \cite{Wald84} that definition 2 has the following consequence: any maximally symmetric space-time has constant
scalar curvature. \gls{ads} is precisely the maximally symmetric space-time with constant \textit{negative} curvature. It is
solution of the vacuum Einstein's equation with negative cosmological constant
\begin{equation}
   G_{\alpha\beta} + \gls{Lambda}g_{\alpha\beta} = 0, \quad \gls{Lambda} < 0.
   \label{einmaxsym}
\end{equation}
The constant curvature property implies that its Riemann tensor is of the form
\begin{equation}
   R_{\alpha\beta\gamma\delta} = -\frac{1}{\gls{L}^2}(g_{\alpha\gamma}g_{\beta\delta} - g_{\alpha\delta}g_{\beta\gamma}),
   \label{Rmaxsym}
\end{equation}
where $\gls{L}$ is called the \gls{ads} length, or \gls{ads} radius. The contractions of this equation are
\begin{equation}
   R_{\alpha\beta} = -\frac{n-1}{\gls{L}^2}g_{\alpha\beta} \quad \tn{and} \quad R = -\frac{n(n-1)}{\gls{L}^2}.
   \label{Rmaxsym2}
\end{equation}
Combining \eqref{einmaxsym}, \eqref{Rmaxsym} and \eqref{Rmaxsym2}, we get a relation between the cosmological
constant $\gls{Lambda}$ and the \gls{ads} length $\gls{L}$. They are related to each other by
\begin{equation}
   \gls{Lambda} = -\frac{(n-1)(n-2)}{2 \gls{L}^2}.
   \label{lambdaL}
\end{equation}
This equation is used extensively in the remainder of this chapter.

\subsection{Usual coordinates system}
\label{coordinates}

How to give an expression of the metric of \gls{ads} space-time? A systematic way of building maximally symmetric space-times
with constant negative curvature is to consider the isometric immersion of an hyperboloid in an $n+1$ flat space-time with
two time-like directions and whose metric reads
\begin{equation}
   ds^2_{n+1} = -dU^2 - dV^2 + \sum_{i=1}^{n-1}dX_i^2.
   \label{ds2n+1}
\end{equation}
The \gls{ads} space-time is then identified with the quadric of equation
\begin{equation}
   -U^2 - V^2 + \sum_{i=1}^{n-1}X_i^2 = -\gls{L}^2.
   \label{quadric}
\end{equation}
Equations \eqref{ds2n+1} and \eqref{quadric} ensure that \gls{ads} space-time has $n(n+1)/2$ Killing vectors that generate the
group $SO(n-1,2)$ \cite{Bengtsson98}. This construction is thus preserving the maximal symmetry. There are many ways of parametrising the
quadric \eqref{quadric}, some of them giving suitable coordinate systems of \gls{ads}.

\paragraph{Poincaré coordinates} For example, the constraint \eqref{quadric} can be satisfied by parametrising
\begin{subequations}
\begin{align}
   U &= \frac{\gls{L}^2}{2r}\left[ 1 + \frac{r^2}{\gls{L}^4}\left( \gls{L}^2 + \vec{x}^2 - t^2 \right) \right],\\
   V &= \frac{r}{\gls{L}}t,\\
   X_i &= \frac{r}{\gls{L}}x_i, \quad i = 1,\ldots,n-2,\\
   X_{n-1} &= \frac{\gls{L}^2}{2r}\left[ 1 - \frac{r^2}{\gls{L}^4}\left( \gls{L}^2 - \vec{x}^2 + t^2 \right) \right],
\end{align}
\end{subequations}
where $\vec{x}$ is a vector of Cartesian-type coordinates $(x_1,\ldots,x_{n-2})$ of Euclidean norm $\vec{x}^2$ and $r \geq 0$.
Combining with \eqref{ds2n+1}, the metric restricted to the quadric is the one of the \gls{ads} space-time itself, namely
\begin{equation}
   ds^2 = -\frac{r^2}{\gls{L}^2}dt^2 + \frac{\gls{L}^2}{r^2}dr^2 + \frac{r^2}{\gls{L}^2}d \vec{x}^2.
   \label{adspoincare}
\end{equation}
These coordinates are the so-called Poincaré coordinates. They do not cover the entire space-time, but only its so-called Poincaré
patch. They are mainly used in the context of the \gls{ads}-\gls{cft} correspondence (chapter
\ref{adscft}).

\paragraph{Global coordinates} Another way of satisfying the constraint \eqref{quadric} is to parametrise
\begin{subequations}
\begin{align}
   U &= \gls{L}\cosh\rho\cos \tau,\\
   V &= \gls{L}\sinh\rho\sin \tau,\\
   X_i &= \gls{L}\sinh\rho \widehat{x}_i,
\end{align}
\end{subequations}
where $\tau \in [0,2\pi[$, $\rho \geq 0$ and $\widehat{x}_i$ parametrise the unit $p$-sphere:
\begin{subequations}
\begin{align}
   \sum_{i=1}^p \widehat{x}_i^2 &= 1,\\
   \widehat{x}_1 &= \cos\varphi_1,\\
   \widehat{x}_2 &= \sin\varphi_1\cos\varphi_2,\\
   \widehat{x}_3 &= \sin\varphi_1\sin\varphi_2\cos\varphi_3,\\
             &\ \ \vdots\\
   \widehat{x}_{p} &= \sin\varphi_1\ldots\sin\varphi_{p-1}\cos\varphi_{p},\\
   \widehat{x}_{p+1} &= \sin\varphi_1\ldots\sin\varphi_{p-1}\sin\varphi_{p},
\end{align}
\end{subequations}
In this equation, $\varphi_i \in [0,\pi]$ for $i = 1,\ldots,p-1$ and $\varphi_{p} \in [0,2\pi[$. Combining with \eqref{ds2n+1} gives
\begin{equation}
   ds^2 = \gls{L}^2(-\cosh^2\rho d\tau^2 + d\rho^2 + \sinh^2\rho d\Omega_p^2),
   \label{ds2hyp}
\end{equation}
where
\begin{equation}
   d \Omega_p^2 = d\varphi_1^2 + \sin^2\varphi_1 d\varphi_2^2 + \cdots + \sin^2\varphi_1\ldots\sin^2\varphi_{p-1}d\varphi_p^2.
\end{equation}
In equation \eqref{ds2hyp}, $\tau$ is adimensional and $2\pi$-periodic. The isometric immersion in $\mathbb{R}^{n+1}$ in these
coordinates is pictured on figure \ref{adsimmersion}. We usually unwrap the time-like direction and consider the so-called
covering space of \gls{ads} space-time where $t \equiv L\tau \in \mathbb{R}$, which finally gives
\begin{equation}
   ds^2 = -\cosh^2\rho dt^2 + \gls{L}^2( d\rho^2 + \sinh^2\rho d\Omega_p^2).
   \label{adsglobal}
\end{equation}
These coordinates are the so-called global (or hyperbolic) coordinates.

\begin{myfig}
   \includegraphics[width = 0.49\textwidth]{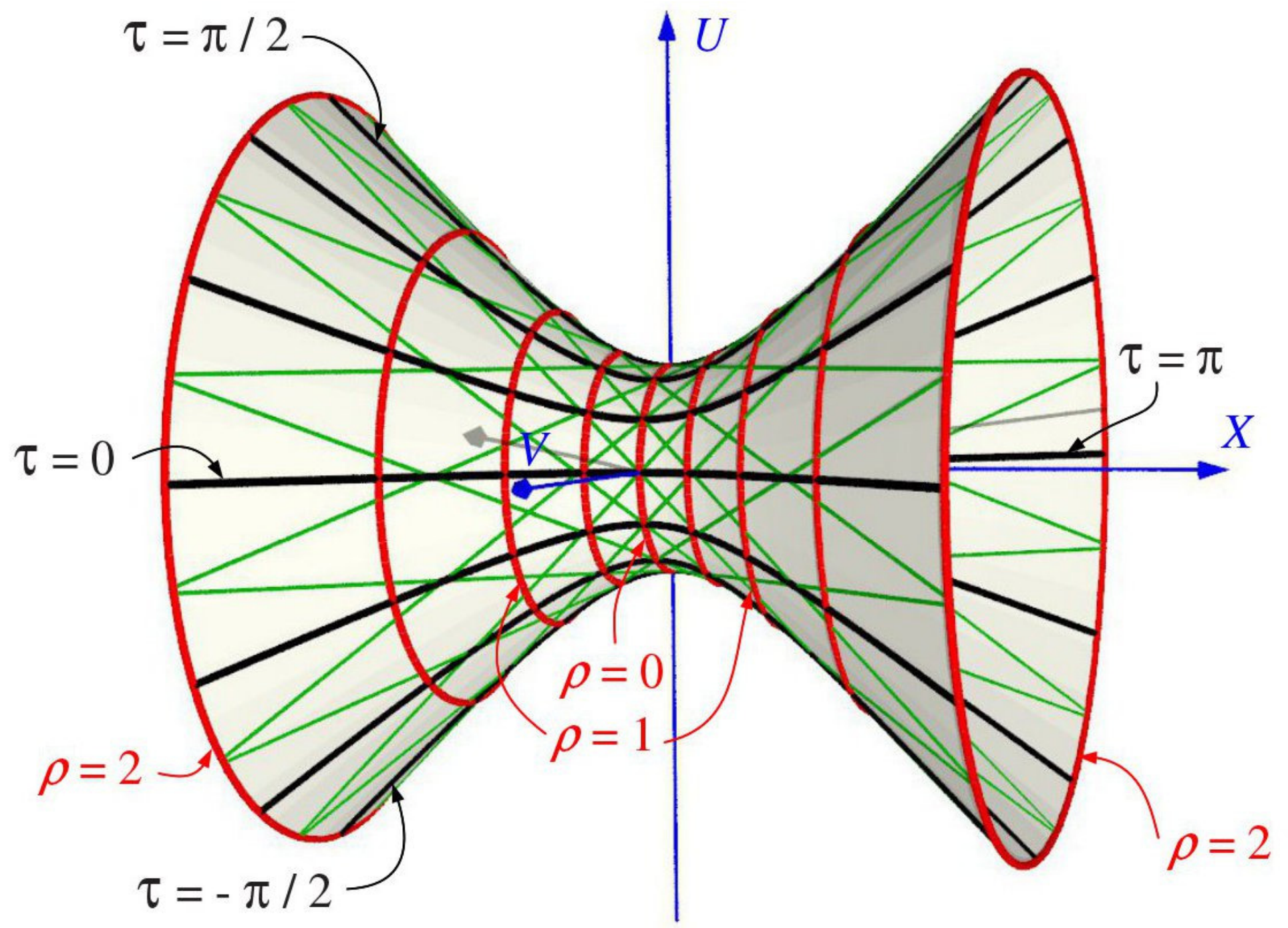}
   \caption[Isometric immersion of anti-de Sitter in $\mathbb{R}^{n+1}$]{Isometric immersion of the $n$-dimensional \gls{ads}
   space-time in $\mathbb{R}^{n+1}$ with two time-like directions $U$ and $V$. $n-2$ dimensions are suppressed on this plot. Red
   curves are constant $\rho$ lines, black curves are constant $\tau$ lines, and green curves are null geodesics, be it for the
   metric \eqref{ds2n+1} in $\mathbb{R}^{n+1}$ or the metric \eqref{ds2hyp} in $n$-dimensional \gls{ads} space-time. Notice that
   in $\mathbb{R}^{n+1}$, these geodesics are mere straight lines. Even if not obvious on the figure, \gls{ads} is homogeneous and
   the centre of the figure is not particular (recall for example the similar case of the origin in spherical coordinates). Credits: E. Gourgoulhon.}
   \label{adsimmersion}
\end{myfig}

\paragraph{Static coordinates} Starting from \eqref{adsglobal} and letting
\begin{subequations}
\begin{align}
   \rho &= \gls{L}\arsinh\left( \frac{r}{\gls{L}} \right),\\
   d\rho &= \frac{dr}{\gls{L}\sqrt{1 + \dfrac{r^2}{\gls{L}^2}}},
\end{align}
\end{subequations}
another form of the \gls{ads} metric can be obtained, namely
\begin{equation}
   ds^2 = -\left( 1 + \frac{r^2}{\gls{L}^2} \right)dt^2 + \frac{dr^2}{1+\dfrac{r^2}{\gls{L}^2}} + r^2 d\Omega_p^2,
   \label{adsstatic}
\end{equation}
where $r \geq 0$. These coordinates are the so-called static\footnote{This adjective is inherited from the positively curved
\gls{ds} space-time. Indeed, trading $1+r^2/\gls{L}^2$ for $1 - r^2/\gls{L}^2$, we get the \gls{ds} metric in static
coordinates, in opposition to \gls{ds} global coordinates in which the metric is time-dependent.}
(or Schwarzschild) coordinates.

\paragraph{Conformal coordinates} From \eqref{adsstatic}, defining
\begin{subequations}
\begin{align}
   r &= \gls{L}\tan x,\\
   dr &= \gls{L}(1 + \tan^2 x)dx,
\end{align}
\end{subequations}
leads to
\begin{equation}
   ds^2 = \frac{1}{\cos^2x}(-dt^2 + \gls{L}^2(dx^2 + \sin^2xd\Omega_p^2)),
   \label{adsconformal}
\end{equation}
where $x \in [0,\pi/2[$. These coordinates are the so-called conformal coordinates. Transforming $t \to t/\gls{L}$ allows to
factorise $\gls{L}$, thus making clear that \gls{ads} space-time is conformal to half of Einstein's universe whose
metric is $ds^2 = -dt^2 + d\Omega_{p+1}^2$ with $x \in [0,\pi[$.

\paragraph{Isotropic coordinates} If this time we consider \eqref{adsglobal} and let
\begin{subequations}
\begin{align}
   \rho &= 2 \artanh\left( \frac{r}{\gls{L}} \right),\\
   d\rho &= \frac{2dr}{\gls{L}}\left( 1 - \frac{r^2}{\gls{L}^2} \right),
\end{align}
\end{subequations}
it comes
\begin{equation}
   ds^2 = -\left( \frac{1+\dfrac{r^2}{\gls{L}^2}}{1 - \dfrac{r^2}{\gls{L}^2}} \right)dt^2 + \left(
   \frac{2}{1-\dfrac{r^2}{\gls{L}^2}} \right)^2(dr^2 + r^2 d\Omega_p^2),
   \label{adsisotropic}
\end{equation}
where $r \in [0,\gls{L} [$. These coordinates are the so-called isotropic (or spatially conformally flat) coordinates. Indeed, the
last parenthesis is the metric of $\mathbb{R}^{p+1}$. They are represented on the tessellation of figure \ref{tessellation}, which
illustrates the diverging behaviour of the metric components in the neighbourhood of $r = \gls{L}$. Such tessellations where
popularised by the Dutch artist Escher\footnote{The web is full of different artistic patterns when looking for ``hyperbolic Escher''.}.

\begin{myfig}
   \includegraphics[width = 0.49\textwidth]{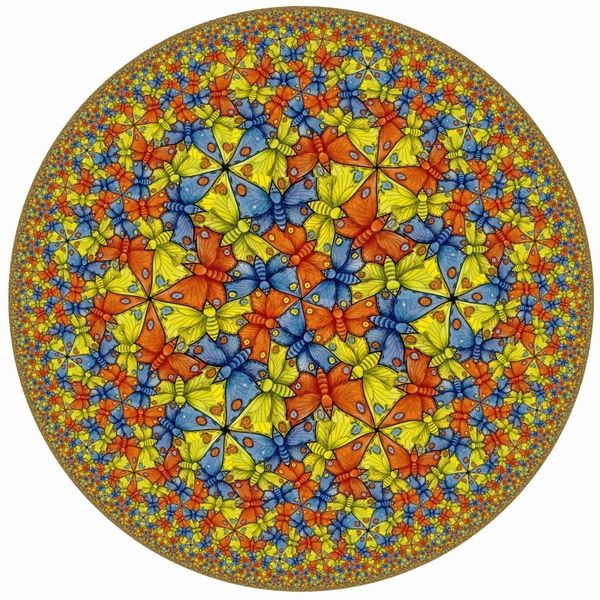}
   \caption[Tessellation of the hyperbolic plane]{Tessellation of the hyperbolic plane, which is a $t = cst$ slice of
      \gls{ads} space-time in isotropic coordinates whose metric is \eqref{adsisotropic}. The density of butterflies increases
      drastically near $r = \gls{L}$. This is because the proper area is proportional to $g_{rr}$ and diverges like
      $O\left( (\gls{L}-r)^{-2} \right)$ in this neighbourhood. Credits: \cite{Escher}.}
   \label{tessellation}
\end{myfig}

It is remarkable on this metric that if we multiply it by $(1-r^2/\gls{L}^2)^2/(1+r^2/\gls{L}^2)^2$, it becomes
\begin{equation}
   -dt^2 + \left( \frac{2}{1+\dfrac{r^2}{\gls{L}^2}} \right)^2(dr^2 + r^2 d\Omega_p^2).
\end{equation}
In particular, at $r = \gls{L}$, we recover the geometry of the Minkowski space-time. This is called the conformal flatness of the
\gls{ads} boundary. It is at the heart of the definition of \gls{aads} space-times (see section \ref{defaads}).

\subsection{Radial geodesics}

Now that we have defined the \gls{ads} space-time and given several coordinates adapted to its geometry, we would like to
understand the physics at play. In order to probe its properties, an enlightening tool is the study of
radial geodesics. In this section, we fix once and for all the coordinate system to be static (equation \eqref{adsstatic}).

\paragraph{Null radial geodesics} As a first probe, let us consider the motion of a photon in \gls{ads} space-time. The null radial
geodesics describing the motion of photons obeys $ds^2 = 0$ (with $d\Omega_p^2 = 0$), which in light of \eqref{adsstatic} gives
\begin{equation}
  dt = \pm \frac{dr}{1+\dfrac{r^2}{\gls{L}^2}} \iff t = t_0 \pm \gls{L}\arctan \left( \frac{r}{\gls{L}} \right),
\end{equation}
where $t_0$ is a constant. The striking feature of this equation is that $r = +\infty$ can be reached in a finite coordinate time $t_0
\pm \pi \gls{L}/2$. If we believe in energy conservation, the mass of the whole space-time mass should be continuous at all time.
Then the photon should bounce off spatial infinity and come back to its initial position. Other possibilities are the dissipation
of energy (no photons comes back at all) or the injection of energy at the boundary (more than one photon come back). The energy
conservation is naturally called the reflective boundary condition of the \gls{ads} boundary (see below section
\ref{adsboundary}). This feature is counter-intuitive since it is forbidden in Minkowski space-time. In the flat case, the photon
reaches infinity only in a infinite time (as measured by a static observer).

So the time coordinate remains finite even for a photon reaching spatial infinity. But what about the affine parameter? Since the
metric is independent of the time coordinate, the vector $\partial_t^\alpha$ is a Killing vector. We can thus introduce the
conserved energy of the photon
\begin{equation}
   E \equiv -g_{\mu\nu}p^\mu \partial_t^\nu = -g_{tt}p^t = -p_t,
   \label{defEgeo}
\end{equation}
where $p^\mu$ is its 4-momentum. The equation $p_\mu p^\mu = 0$ that holds for null geodesics then reduces to
\begin{equation}
   \frac{E^2}{g_{tt}} + g_{rr}(p^r)^2 = 0 \iff \frac{dr}{d\lambda} = \pm E \iff r = r_0 \pm E\lambda,
\end{equation}
where $\lambda$ is an affine parameter of the photon. The photon thus reaches $r = +\infty$ in a finite coordinate time, but with
an infinite affine parameter.

\paragraph{Time-like radial geodesics} A second probe of the physics at play is the study of time-like radial geodesics that describe the motion
of massive particles. Sticking to the definition of energy \eqref{defEgeo}, and denoting by $v^\mu$ the 4-velocity of the test
particle, the normalisation condition is now $v_\mu v^\mu = -1$, so that, with $d\Omega_p^2 = 0$
\begin{equation}
   \frac{E^2}{g_{tt}} + g_{rr}(v^r)^2 = -1 \iff \frac{dr}{d\tau} = \pm \sqrt{E^2 - 1 - \frac{r^2}{\gls{L}^2}},
   \label{drdtau}
\end{equation}
where $\tau$ is the proper time of the massive test particle. Since $E^2 - 1 = (dr/d\tau)^2 + r^2/\gls{L}^2$, the quantity $E^2 -
1$ is always positive. However, in the square root of \eqref{drdtau}, the argument is not always positive, especially for large $r$
coordinates. There is thus a turning point beyond which the motion is forbidden, namely
\begin{equation}
   r_\star = \gls{L}\sqrt{E^2 - 1}.
\end{equation}
This is the equivalent to spatial infinity for photons, but this time, since the particle has a non-zero mass, it cannot go that
far and is constrained to turn back at a finite radius.

The differential equation \eqref{drdtau} can be integrated as follows:
\begin{equation}
   \tau = cst \pm \frac{1}{\sqrt{E^2-1}}\bigints_r^{r_\star}\frac{dr}{\sqrt{1 - \dfrac{r^2}{r_\star^2}}}.
\end{equation}
By the change of variables
\begin{subequations}
\begin{align}
   r &= r_\star \sin\varphi,\\
   dr &= r_\star \cos\varphi d\varphi,
\end{align}
\label{varphichange}
\end{subequations}
we get
\begin{equation}
   \tau = \tau_0 \pm \gls{L}\arcsin\left( \frac{r}{r_\star} \right).
   \label{tautimelike}
\end{equation}
The test particle thus reaches the turning point $r_\star$ in a finite proper time $\tau_0 \pm \pi \gls{L}/2$.
In order to express the result as a function of the time coordinate, we notice that
\begin{equation}
   \frac{dt}{d\tau} = v^t = -\frac{E}{g_{tt}} = \frac{E}{1+\dfrac{r^2}{\gls{L}^2}} \iff \frac{dt}{d\varphi} =
   \pm\frac{E \gls{L}}{1 + (E^2 - 1)\sin^2\varphi},
\end{equation}
where we have used \eqref{varphichange}, \eqref{tautimelike} and $d\tau = \pm\gls{L}d\varphi$. Introducing a last change of variables
\begin{subequations}
\begin{align}
   \varphi &= \arctan u,\\
   d\varphi &= \frac{du}{1+u^2},
\end{align}
\end{subequations}
it comes
\begin{align}
\nonumber   t &= cst \pm \int_\varphi^{\pi/2} \frac{E \gls{L}d\varphi}{\cos^2\varphi + E^2 \sin^2\varphi} = cst \pm E
   \gls{L}\int_\varphi^{\pi/2}\frac{(1 + \tan^2\varphi)d\varphi}{1 + E^2\tan^2\varphi}\\
   &= cst \pm E \gls{L}\int_u^\infty \frac{du}{1 + E^2u^2} = t_0 \pm \arctan(uE).
\end{align}
Recovering $r$ through $u$ and $\varphi$, we end with
\begin{equation}
   t = t_0 \pm \gls{L}\arctan\left( \frac{Er}{r_\star\sqrt{1-\dfrac{r^2}{r_\star^2}}} \right).
\end{equation}
On this equation, and exactly like the photon case, the turning point is reached in a finite coordinate time $t_0 \pm \pi
\gls{L}/2$. Thus, a round trip between the origin of \gls{ads} space-time and the \gls{ads} boundary always takes a coordinate
time interval $\pi \gls{L}$, be the test particle massive or not. The notable difference between time-like and null geodesics is
that the affine coordinate of photons is infinite at the turning point, while the proper time of massive test particles remains
finite and is independent of the initial conditions.

\paragraph{Summary} On figure \ref{geodesics}, both null and time-like geodesics are pictured. It is clear on that plot that the
motion of radial geodesics is $2\pi \gls{L}$-periodic. The negative curvature of the space-time acts like an additional source of attraction that
refocuses the bunch of world-lines toward the centre. Furthermore, since the turning point is further for higher initial energies,
it is clear that the null geodesics match the limiting $E \to \infty$ time-like geodesics, i.e.\ the limit where the rest mass
is negligible compared to the kinetic energy. Chapter \ref{adsinsta} describes in detail how this very particular boomerang behaviour
can have dramatic consequences on the gravitational dynamics in \gls{aads} space-times.

\begin{myfig}
   \includegraphics[width=0.49\textwidth]{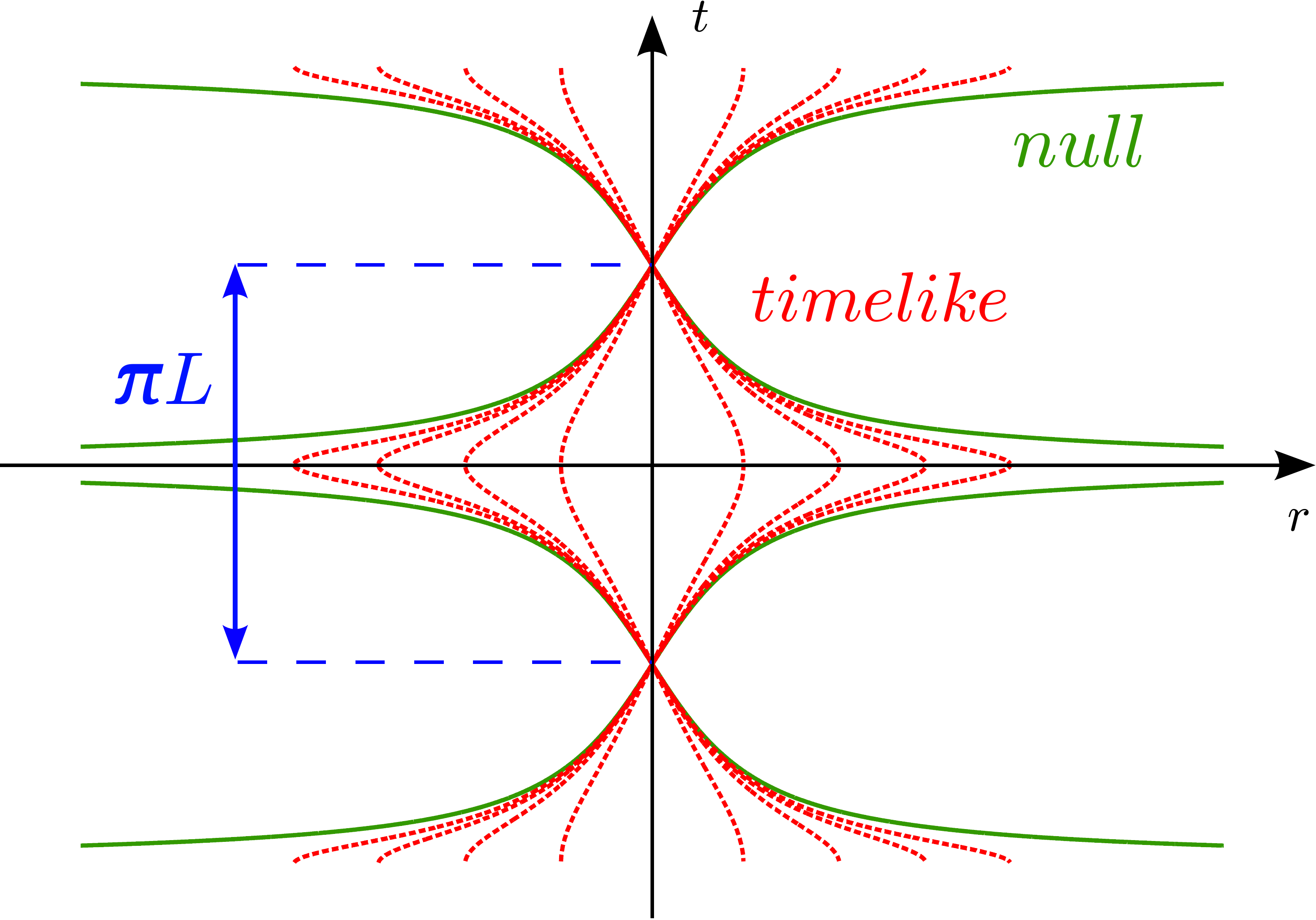}
   \caption[Radial geodesics in anti-de Sitter]{Radial geodesics in vacuum \gls{ads} space-time. Time-like geodesics are pictured in dotted red
   lines and null ones in solid green. A bunch of time-like geodesics exhibiting an initial kinetic energy spread comes
   back to its initial position in a unique and finite amount of time, $\pi \gls{L}$, where $\gls{L}$ is the \gls{ads} radius
   directly linked to the cosmological constant by \eqref{lambdaL}. Credits: G. Martinon.}
   \label{geodesics}
\end{myfig}

\subsection{The AdS boundary}
\label{adsboundary}

As we have seen in the study of geodesics, at some point in the demonstration we were forced to prescribe the boundary conditions at space-like
infinity. In particular for null geodesics, energy conservation lead us to choose reflective boundary conditions. This is not
mandatory though. Indeed, one characteristic of \gls{ads} space-time is the absence of a global Cauchy surface. Unlike Minkowski space-time,
specification of initial data on a complete space-like hypersurface does not lead to a unique prediction of the future state of a
dynamical system. This is because radiation is not specified by the initial data and, in absence of a boundary conditions
prescription, can propagate in (or out) from infinity. In particular, there is no notion of null infinity in \gls{aads}
space-times: photons are doomed to bounce off an infinite number of times on spatial infinity, which is a time-like hypersurface.
All possible boundary conditions for scalar, electromagnetic and gravitational perturbations of \gls{ads} were determined in
\cite{Ishibashi04}. In this paper, the linearised Einstein's equations are reduced to a single master wave equation (discussed in
details in chapter \ref{perturbations}) for a master function $\Phi$, on which can be imposed Dirichlet (vanishing of the field),
Neumann (vanishing of its derivative), or Robin (a linear combination of the latter) type conditions at space-like infinity.

What is called the \gls{ads} boundary is precisely spatial infinity where boundary conditions have to be prescribed. It is the
$n-1$-submanifold defined by a constant radius that reaches its upper bound. Table \ref{tabbound} illustrates the definition of the
\gls{ads} boundary in the different systems of coordinates of section \ref{coordinates}. Note that the metric components are always diverging
near the \gls{ads} boundary in these coordinates.

\begin{table}[t]
\rowcolors{1}{white}{ddlion}
\begin{tabular}{lll}
\hline
Coordinates & length element & \gls{ads} boundary\\
\hline
Poincaré  & $ds^2 = -\frac{r^2}{\gls{L}^2}dt^2 + \frac{\gls{L}^2}{r^2}dr^2 + \frac{r^2}{\gls{L}^2}d \vec{x}^2$ & $r = +\infty$\\
global    & $ds^2 = -\cosh^2\rho dt^2 + \gls{L}^2( d\rho^2 + \sinh^2\rho d\Omega_p^2)$ & $\rho = +\infty$\\
static    & $ds^2 = -\left( 1 + \frac{r^2}{\gls{L}^2} \right)dt^2 + \frac{\gls{L}^2 dr^2}{\gls{L}^2+r^2} + r^2 d\Omega_p^2$ & $r = +\infty$\\
conformal & $ds^2 = \frac{1}{\cos^2x}(-dt^2 + \gls{L}^2(dx^2 + \sin^2xd\Omega_p^2))$ & $x = \pi/2$\\
isotropic & $ds^2 = -\left( \frac{\gls{L}^2+r^2}{\gls{L}^2 - r^2} \right)dt^2 + \left( \frac{2 \gls{L}^2}{\gls{L}^2-r^2} \right)^2(dr^2 + r^2 d\Omega_p^2)$ & $r = \gls{L}$\\
\hline
\end{tabular}
\caption[AdS boundary in different coordiantes systems]{The \gls{ads} boundary in the coordinates systems studied above.}
\label{tabbound}
\end{table}

The \gls{ads} boundary is a time-like hypersurface and plays a central role in gravitational dynamics in \gls{gr} as well as in the context of the
celebrated \gls{ads}-\gls{cft} correspondence (chapter \ref{adscft}). It can be viewed as the edge of an $n$-dimensional billiard
on which everything bounces off. From now on and hereafter, we denote it by $\partial \mathcal{M}$. By definition, it  has the
topology of $\mathbb{R}\times \mathcal{S}^p$. This is not obvious in some Poincaré coordinates (equation \eqref{adspoincare}), where the
boundary seems to have the topology of $\mathbb{R}\times \mathbb{R}^p$. This is simply because Poincaré coordinates cover only part of
the \gls{ads} space-time and thus part of its boundary.

\section{Asymptotically anti-de Sitter space-times}

\gls{ads} space-time is maximally symmetric because it describes a zero-mass vacuum in a negatively curved geometry.
However, we are mainly interested in configurations where fields or even black holes can propagate in such a background geometry. Such
configurations cannot be \gls{ads}, and are called \gls{aads}. It is well-known that any solution of Einstein's equation with vanishing
cosmological constant is not always asymptotically flat (recall for example Melvin's universes in chapter \ref{geons}). Similarly,
any solution of Einstein's equation with negative cosmological constant is not a priori \gls{aads}. In this section, we provide a
rigorous definition of the concept of \gls{aads} space-times. The first attempts of definition started in the 80's, notably in
\cite{Abbott82,Henneaux85} where the argument was based on first order perturbation of the background \gls{ads} or
Kerr-\gls{ads} geometry. However in this section, we rely mainly on the conformal formalism, whose foundations lie in Penrose's
conformal treatment of infinity, which was first proposed in \cite{Hawking83} in the case of \gls{aads} space-times, and then
investigated in more details in \cite{Ashtekar84,Ashtekar00,Wald00,Hollands05a}.

\subsection{The Weyl-Schouten theorem}

The notion of conformal flatness plays an important role in the definition of \gls{aads} space-times, as we have glimpsed in the
case of isotropic coordinates \eqref{adsisotropic}. We thus use the following definition:
\begin{definition}[Conformal flatness]
A Riemannian manifold is said to be conformally flat if and only if for each point $x$ there exists a neighbourhood
$U$ of $x$ and a smooth function $\Omega$ such that the conformal metric $\widehat{g}_{\alpha\beta} = \Omega^2 g_{\alpha\beta}$
has a vanishing Riemann curvature tensor on $U$.
\end{definition}
The example of the isotropic coordinates \eqref{adsisotropic} already gave us an illustration of this definition for the
\gls{ads} boundary. We can thus conclude that the \gls{ads} boundary is conformally flat. This is the common denominator between
all \gls{aads} space-times. Furthermore the notion of conformal flatness can be characterised by the so-called \gls{ws} theorem:
\begin{theorem}[Characterisation of conformal flatness]
A Riemannian manifold of dimension $n$ is conformally flat if and only if
\begin{myitem}
   \item The Weyl tensor vanishes (case $n \geq 4$)
   \item The Cotton tensor vanishes (case $n = 3$)
\end{myitem}
\label{wstheorem}
\end{theorem}
This theorem is the cornerstone of the definition of \gls{aads} space-times.

\subsection{Definition of asymptotically anti-de Sitter space-times}
\label{defaads}

We do not consider the very special 3-dimensional case and define an \gls{aads} $n$-dimensional (with $n \geq 4$) space-time
($\mathcal{M},g_{\alpha\beta}$) as follows\footnote{Beware that in our notations, $g_{\alpha\beta}$ denotes the physical metric
and $\widehat{g}_{\alpha\beta}$ denotes the conformal one while in \cite{Ashtekar84,Ashtekar00} the opposite convention is
taken.} \cite{Ashtekar84,Ashtekar00}.
\begin{definition}[Asymptotically anti-de Sitter space-times]
The space-time ($\mathcal{M},g_{\alpha\beta}$) is said to be \gls{aads} if there exists a manifold $\widehat{\mathcal{M}}$ with boundary $\partial
\widehat{\mathcal{M}}$ equipped with a metric $\widehat{g}_{\alpha\beta}$ and a diffeomorphism from $\mathcal{M}$ onto
$\widehat{\mathcal{M}} - \partial \widehat{\mathcal{M}}$ such that
\begin{myenum}
   \item there exists a function $\Omega$ on $\mathcal{M}$ for which on $\widehat{\mathcal{M}}$
      \begin{equation}
         \widehat{g}_{\alpha\beta} = \Omega^2 g_{\alpha\beta},
         \label{confg}
      \end{equation}
   \item $\partial \widehat{\mathcal{M}}$ is topologically $\mathbb{R}\times \mathcal{S}^p$, $\Omega$ vanishes on $\partial \widehat{\mathcal{M}}$ but its gradient
      $\widehat{\nabla}_\alpha \Omega$ is nowhere vanishing on $\partial \widehat{\mathcal{M}}$,
   \item on $\mathcal{M}$, $g_{\alpha\beta}$ is solution of Einstein's equation with negative cosmological constant (see equation
      \eqref{eineq} of appendix \ref{grd}), and the energy-momentum tensor is such that $\Omega^{2-n}T_{\alpha\beta}$ admits a
      smooth limit to $\partial \widehat{\mathcal{M}}$,
   \item the Weyl tensor $\widehat{C}_{\alpha\beta\gamma\delta}$ of $\widehat{g}_{\alpha\beta}$ is such that
      $\Omega^{4-n}\widehat{C}_{\alpha\beta\gamma\delta}$ is smooth on $\widehat{\mathcal{M}}$ and vanishes at $\partial \widehat{\mathcal{M}}$,
   \item for the $n=4$ case only, the pseudo-magnetic part of $\Omega^{3-n}\widehat{C}_{\alpha\beta\gamma\delta}$ vanishes at
      $\partial \widehat{\mathcal{M}}$.
\end{myenum}
\label{aadsdefinition}
\end{definition}
These properties ensure that the asymptotic symmetry group of the physical space-time is the \gls{ads} symmetry group
\cite{Ashtekar00}. Several remarks are relevant at this point.

The metric $g_{\alpha\beta}$ and $\widehat{g}_{\alpha\beta}$ are two distinct metrics on two distinct manifolds. Thus, the manifolds
$\mathcal{M}$ and $\widehat{\mathcal{M}} - \partial \widehat{\mathcal{M}}$ need to be identified by a diffeomorphism, otherwise
equation \eqref{confg} makes no sense. As we have seen in section \ref{coordinates}, the \gls{ads} metric components are always diverging near
$\partial \mathcal{M}$. The conformal metric $\widehat{g}_{\alpha\beta}$ of condition 1 is precisely the multiplication of the
physical metric by a factor $\Omega^2$ that suppresses the divergence. For example, in the case of isotropic coordinates (equation
\eqref{adsisotropic}), we have seen that we could choose $\Omega = (1-r^2/\gls{L}^2)/(1+r^2/\gls{L}^2)$. The conformal metric is
thus more convenient as it is regular everywhere. We can thus write near the \gls{ads} boundary
\begin{subequations}
\begin{align}
   g_{\alpha\beta} \underset{\partial\mathcal{M}}{=} O(\Omega^{-2}) \quad &\tn{and} \quad g^{\alpha\beta} \underset{\partial\mathcal{M}}{=} O(\Omega^2),\\
   \widehat{g}_{\alpha\beta} \underset{\partial\widehat{\mathcal{M}}}{=} O(1) \quad &\tn{and} \quad \widehat{g}^{\alpha\beta} \underset{\partial\widehat{\mathcal{M}}}{=} O(1).
\end{align}
\end{subequations}
The conformal geometry is then perfectly regular.

Condition 2 ensures that $\Omega$ can be chosen as a suitable radial coordinate of the conformal completion
$(\widehat{\mathcal{M}},\widehat{g}_{\alpha\beta})$, at least in the neighbourhood of the \gls{ads} boundary. It has only one
first order zero that signals the \gls{ads} boundary. The topology of the boundary seems to exclude the Poincaré patch of
\gls{ads} (equation \eqref{adspoincare}). This is an artefact due to the partial covering of the whole \gls{ads} space-time in
these coordinates (see for example \cite{Ashtekar14} that overcome this point).

Condition 3 merely requires the metric to be solution of Einstein's equation and that the energy-momentum tensor vanishes
sufficiently rapidly near the \gls{ads} boundary. In particular, it means that
\begin{equation}
   T_{\alpha\beta} \underset{\partial\mathcal{M}}{=} O(\Omega^{n-2}) \quad \tn{and} \quad T \underset{\partial\mathcal{M}}{=} O(\Omega^n),
   \label{OT}
\end{equation}
near the \gls{ads} boundary.

Condition 4 is of most importance. In particular, it implies that
\begin{equation}
   \widehat{C}_{\alpha\beta\gamma\delta} \underset{\partial\widehat{\mathcal{M}}}{=} O(\Omega^{n-3}),
   \label{OC}
\end{equation}
near the \gls{ads} boundary. Thus, $\widehat{C}\indices{^\alpha_{\beta\gamma\delta}}$ vanishes on $\partial \widehat{\mathcal{M}}$ and
$C\indices{^\alpha_{\beta\gamma\delta}}$ vanishes on $\partial \mathcal{M}$ since the Weyl tensor is a conformal invariant (see appendix
\ref{conftransform}). We show in section \ref{demoaads} that conditions 4 and 5 simply mean that the Weyl tensor of
$\widehat{q}_{\alpha\beta}$ the induced metric on $\partial \widehat{\mathcal{M}}$ vanishes. According to the \gls{ws} theorem
\ref{wstheorem}, this is equivalent to the conformal flatness of the \gls{ads} boundary. Combined with condition 1, we can thus sum up the above conditions
in the following sentence \cite{Papadimitriou05}:

\begin{framed}
   A space-time is said to be \gls{aads} if and only if it is solution of Einstein's equation and exhibits a conformally flat boundary on
   which the metric has a second order pole.
\end{framed}

We have already seen in section \ref{coordinates} that the conformal flatness of the \gls{ads} boundary was manifest in the
isotropic coordinates. It is actually a generic feature of \gls{ads} space-times and at the heart of the definition of
\gls{aads} space-times. Hereafter, we make clear the link between this property and conditions 4 and 5, and we use the
formalism of conformal transformations to define global charges. In order to make clearer the argument, we summarise our notations
in table \ref{confrecap}.

\begin{mystab}
\begin{tabular}{lll}
\hline
Quantity & Physical space-time & Conformal space-time \\[0.1cm]
\hline
manifold & $\mathcal{M}$ & $\widehat{\mathcal{M}}$ \\[0.1cm]
boundary & $\partial \mathcal{M}$ & $\partial \widehat{\mathcal{M}}$ \\[0.1cm]
normal to $\Omega = cst$ hypersurfaces & $\widehat{n}_\alpha = \nabla_\alpha \Omega$ & $\widehat{n}_\alpha = \widehat{\nabla}_\alpha \Omega$ \\[0.1cm]
metric & $g_{\alpha\beta}$ & $\widehat{g}_{\alpha\beta} = \Omega^2 g_{\alpha\beta}$ \\[0.1cm]
covariant derivative & $\nabla$ & $\widehat{\nabla}$ \\[0.1cm]
Riemann tensor & $R\indices{^\alpha_{\beta\gamma\delta}}$ & $\widehat{R}\indices{^\alpha_{\beta\gamma\delta}}$ \\[0.1cm]
Schouten tensor & $S_{\alpha\beta}$ & $\widehat{S}_{\alpha\beta}$ \\[0.1cm]
Cotton tensor & $C_{\alpha\beta\gamma} = \nabla_{[\alpha}S_{\beta]\gamma}$ & $\widehat{C}_{\alpha\beta\gamma} = \widehat{\nabla}_{[\alpha}\widehat{S}_{\beta]\gamma}$ \\[0.1cm]
Weyl tensor & $C\indices{^\alpha_{\beta\gamma\delta}}$ & $\widehat{C}\indices{^\alpha_{\beta\gamma\delta}} = C\indices{^\alpha_{\beta\gamma\delta}}$ \\[0.1cm]
electric and pseudo-electric Weyl tensors & $E_{\alpha\beta} = C_{\alpha\mu\beta\nu}r^\mu r^\nu$ & $\widehat{\varepsilon}_{\alpha\beta} = \Omega^{3-n}\widehat{C}_{\alpha\mu\beta\nu}\widehat{n}^\mu \widehat{n}^\nu$ \\[0.1cm]
unit normal to $Sigma_r$ & $r_\alpha$ & $\widehat{r}_\alpha = \Omega r_\alpha$ \\[0.1cm]
induced metric onto $Sigma_r$ & $q_{\alpha\beta}$ & $\widehat{q}_{\alpha\beta} = \Omega^2 q_{\alpha\beta}$ \\[0.1cm]
covariant derivative onto $Sigma_r$ & $D$ & $\widehat{D}$ \\[0.1cm]
extrinsic curvature tensor of $Sigma_r$ & $\Theta_{\alpha\beta} = -q \indices{^\mu_\alpha}q \indices{^\nu_\beta}\nabla_\alpha r_\beta$ & $\widehat{\Theta}_{\alpha\beta} = -\widehat{q} \indices{^\mu_\alpha}\widehat{q} \indices{^\nu_\beta}\widehat{\nabla}_\alpha \widehat{r}_\beta$ \\[0.1cm]
intrinsic Riemann tensor of $Sigma_r$ & $\mathcal{R}_{\alpha\beta\gamma\delta}$ & $\widehat{\mathcal{R}}_{\alpha\beta\gamma\delta}$ \\[0.1cm]
intrinsic Schouten tensor of $Sigma_r$ & $\mathcal{S}_{\alpha\beta}$ & $\widehat{\mathcal{S}}_{\alpha\beta}$ \\[0.1cm]
intrinsic Cotton tensor of $Sigma_r$ & $\mathcal{C}_{\alpha\beta\gamma} = D_{[\alpha}\mathcal{S}_{\beta]\gamma}$ & $\widehat{\mathcal{C}}_{\alpha\beta\gamma} = \widehat{D}_{[\alpha}\widehat{\mathcal{S}}_{\beta]\gamma}$ \\[0.1cm]
intrinsic Weyl tensor of $Sigma_r$ & $\mathcal{C}\indices{^\alpha_{\beta\gamma\delta}}$ & $\widehat{\mathcal{C}}\indices{^\alpha_{\beta\gamma\delta}}$ \\[0.1cm]
Killing vector of the \gls{ads} boundary & $\xi^\alpha$ & $\widehat{\xi}^\alpha = \xi^\alpha$ \\[0.1cm]
Killing equation on the \gls{ads} boundary & $D^{(\alpha}\xi^{\beta)} = 0$ & $(n-1)\widehat{D}^{(\alpha}\widehat{\xi}^{\beta)} = \widehat{D}_{\mu}\widehat{\xi}^{\mu}\widehat{q}^{\alpha\beta}$ \\[0.1cm]
normal to $\Sigma_t$ & $u_\alpha$ & $\widehat{u}_\alpha = \Omega u_\alpha$ \\[0.1cm]
a $t = cst$ slice of the \gls{ads} boundary & $\Sigma$ & $\widehat{\Sigma}$ \\[0.1cm]
induced metric onto $\Sigma$ & $\sigma_{\alpha\beta}$ & $\widehat{\sigma}_{\alpha\beta} = \Omega^2 \sigma_{\alpha\beta}$ \\[0.1cm]
charge associated to $\Sigma$ & $Q_\xi^{BK}[\Sigma]$ & $Q_{\widehat{\xi}}^{AMD}[\widehat{\Sigma}]$ \\[0.1cm]
\hline
\end{tabular}
\caption[Notations in the conformal formalism]{Summary of all notations related to the conformal formalism used in this chapter.
   As in appendix \ref{d+1}, $\Sigma_x$ denotes an $x = cst$ hypersurface. Credits: G. Martinon.}
\label{confrecap}
\end{mystab}

\subsection{Some properties of conformal transformations}

Conformal transformations being the key ingredient of the definition of \gls{aads} space-times, we discuss here some of their
properties (see also appendix \ref{conftransform}). Since $\Omega$ can be interpreted as a good radial coordinate (at least in a
neighbourhood of the \gls{ads} boundary) according to condition 2, we introduce its gradient
\begin{equation}
   \widehat{n}_{\alpha} \equiv \widehat{\nabla}_\alpha \Omega \quad \tn{and} \quad \widehat{n}^\alpha \equiv
   \widehat{g}^{\alpha\mu}\widehat{n}_\mu.
\end{equation}
This vector is by construction normal to the $\Omega = cst$ hypersurfaces. How does this vector behaves near the \gls{ads}
boundary? The answer can be obtained with the help of the Schouten tensor \cite{Ashtekar00 }. In appendix \ref{conftransform}, we
have seen that the Schouten tensor of $\widehat{g}_{\alpha\beta}$ and $g_{\alpha\beta}$ are related to each other by
\begin{equation}
   \widehat{S}_{\alpha\beta} = S_{\alpha\beta} - (n-2)\nabla_\alpha\nabla_\beta\ln\Omega + (n-2)\nabla_\alpha\ln\Omega\nabla_\beta\ln\Omega - \frac{n-2}{2}g_{\alpha\beta}\nabla^\mu\ln\Omega\nabla_\mu\ln\Omega.
\end{equation}
We would like to trade the $\nabla$ for $\widehat{\nabla}$ associated to $\widehat{g}_{\alpha\beta}$. For any tensor
$T\indices{^\alpha_\beta}$, we have
\begin{equation}
   \nabla_\gamma T\indices{^\alpha_\beta} = \widehat{\nabla}_\gamma T\indices{^\alpha_\beta} + (\Gamma\indices{^\alpha_{\gamma\mu}} - \widehat{\Gamma}\indices{^\alpha_{\gamma\mu}}) T\indices{^\mu_\beta} - (\Gamma\indices{^\mu_{\gamma\beta}} - \widehat{\Gamma}\indices{^\mu_{\gamma\beta}}) T\indices{^\alpha_\mu}.
   \label{nablahatT}
\end{equation}
Thus, with the help of \eqref{confGamma}, it comes directly
\begin{equation}
   S_{\alpha\beta} = \widehat{S}_{\alpha\beta} + \frac{n-2}{\Omega}\widehat{\nabla}_\alpha \widehat{n}_\beta -
   \frac{n-2}{2\Omega^2}\widehat{n}_\mu \widehat{n}^\mu \widehat{g}_{\alpha\beta}.
   \label{sab}
\end{equation}
If we multiply this equation by $\Omega^2$ and combine with the Einstein's equation in the form \eqref{eineq4}, it comes
\begin{equation}
   \frac{\gls{Lambda}}{n-1}\widehat{g}_{\alpha\beta} + 8\pi \gls{G}\Omega^2 \widetilde{T}_{\alpha\beta} = \Omega^2
   \widehat{S}_{\alpha\beta} + (n-2)\Omega\widehat{\nabla}_\alpha \widehat{n}_\beta -
   \frac{n-2}{2}\widehat{n}_\mu \widehat{n}^\mu \widehat{g}_{\alpha\beta},
   \label{boundaryeval}
\end{equation}
where we have used the shortcut notation
\begin{equation}
   \widetilde{T}_{\alpha\beta} \equiv T_{\alpha\beta} - \frac{T}{n-1}g_{\alpha\beta}.
\end{equation}
Evaluating \eqref{boundaryeval} at the \gls{ads} boundary, i.e.\ at $\Omega = 0$, using the near-boundary development of
$T_{\alpha\beta}$ \eqref{OT} and the \gls{Lambda}-\gls{L} relation \eqref{lambdaL}, it comes
\begin{equation}
   \widehat{n}_\mu \widehat{n}^\mu \mathrel{\widehat{=}} -\frac{2 \gls{Lambda}}{(n-1)(n-2)} = \frac{1}{\gls{L}^2},
   \label{nmunmu}
\end{equation}
where the symbol $\widehat{=}$ denotes equality restricted to the \gls{ads} boundary $\Omega = 0$. The vector $\widehat{n}^\alpha$ is
thus space-like near the \gls{ads} boundary, and its norm is directly related to the \gls{ads} radius. If instead of $\Omega^2$ we
multiply \eqref{sab} by $\Omega$, it comes (with again the help of \eqref{lambdaL})
\begin{equation}
   \frac{\gls{Lambda}}{n-1}\frac{\widehat{g}_{\alpha\beta}}{\Omega} + 8\pi \gls{G}\Omega \widetilde{T}_{\alpha\beta} = \Omega
   \widehat{S}_{\alpha\beta} + (n-2)\widehat{\nabla}_\alpha \widehat{n}_\beta - \frac{n-2}{2\Omega}\widehat{n}_\mu \widehat{n}^\mu \widehat{g}_{\alpha\beta}.
\end{equation}
Evaluating this equation at $\Omega = 0$, it comes
\begin{equation}
   \lim_{\Omega\to 0}\frac{1}{\Omega}\left( \widehat{n}_\mu \widehat{n}^\mu - \frac{1}{\gls{L}^2} \right)\widehat{g}_{\alpha\beta}
   \mathrel{\widehat{=}} 2 \widehat{\nabla}_\alpha \widehat{n}_\beta.
   \label{lim1}
\end{equation}
The limit in the left-hand side is well-defined and finite according to \eqref{nmunmu}. Taking the trace with $\widehat{g}^{\alpha\beta}$ yields
\begin{equation}
   \lim_{\Omega\to 0}\frac{1}{\Omega}\left( \widehat{n}_\mu \widehat{n}^\mu - \frac{1}{\gls{L}^2} \right) \mathrel{\widehat{=}}
   \frac{2}{n} \widehat{\nabla}_\mu \widehat{n}^\mu.
   \label{lim2}
\end{equation}
Combining \eqref{lim1} and \eqref{lim2}, it comes directly
\begin{equation}
   \widehat{\nabla}_\alpha \widehat{n}_\beta \mathrel{\widehat{=}} \frac{1}{n} \widehat{\nabla}_\mu \widehat{n}^\mu
   \widehat{g}_{\alpha\beta}.
   \label{nablan}
\end{equation}
This relation is of great help for defining a gauge that makes the computations simpler.

\subsection{Conformal gauge freedom}

Of course, the conformal factor $\Omega$ is not unique, and can be chosen arbitrarily provided that condition 2 is satisfied. This is in some sense a gauge choice. It
turns out that a convenient gauge is the one for which \eqref{nablan} is identically zero. How can we impose such a condition? Let
us consider a distinct conformal factor defined by
\begin{equation}
   \overline{\Omega} \equiv \omega \Omega, \quad \overline{n}_{\alpha} \equiv \overline{\nabla}_\alpha \overline{\Omega}, \quad
   \overline{g}_{\alpha\beta} \equiv \overline{\Omega}^2 g_{\alpha\beta} = \omega^2 \widehat{g}_{\alpha\beta},
\end{equation}
where $\omega$ is a function of the coordinates. We can thus compute explicitly
\begin{equation}
   \overline{\nabla}_\alpha \overline{n}_\beta = \widehat{n}_\alpha \widehat{\nabla}_\beta \omega + \widehat{n}_\beta \widehat{\nabla}_\alpha \omega +
   \omega(\widehat{\nabla}_\alpha \widehat{n}_\beta - [\overline{\Gamma} \indices{^\mu_{\alpha\beta}} - \widehat{\Gamma}
   \indices{^\mu_{\alpha\beta}}]\widehat{n}_\mu) +
   \Omega(\widehat{\nabla}_\alpha \widehat{\nabla}_\beta \omega - [\overline{\Gamma} \indices{^\mu_{\alpha\beta}} - \widehat{\Gamma}
   \indices{^\mu_{\alpha\beta}}]\widehat{\nabla}_\mu \omega),
\end{equation}
where we have forced the emergence of $\widehat{\nabla}$ with \eqref{nablahatT}. Evaluating this equation on the \gls{ads} boundary
($\Omega = 0$), and combining with \eqref{confGamma}, it comes
\begin{equation}
   \overline{\nabla}_\alpha \overline{n}_\beta \mathrel{\widehat{=}} \omega \widehat{\nabla}_\alpha \widehat{n}_\beta + \widehat{g}_{\alpha\beta} \widehat{n}^\mu
   \widehat{\nabla}_\mu \omega.
\end{equation}
Recalling \eqref{nablan}, we have thus recovered that
\begin{equation}
   \overline{\nabla}_\alpha \overline{n}_\beta \mathrel{\widehat{=}} \left(\frac{\omega}{n} \widehat{\nabla}^\mu \widehat{n}_\mu +
   \widehat{n}^\mu \widehat{\nabla}_\mu \omega\right)\widehat{g}_{\alpha\beta}.
\end{equation}
In view of this result, it appears that we have the freedom to chose $\omega$ such that $\overline{\nabla}_\alpha \overline{n}_\beta
\mathrel{\widehat{=}} 0$ just by solving a first order differential equation for $\omega$. From now on and hereafter, we work in
a conformal gauge such that
\begin{equation}
   \widehat{\nabla}_\alpha \widehat{n}_\beta \mathrel{\widehat{=}} 0.
   \label{confgauge}
\end{equation}
This choice does not completely fix the conformal gauge freedom though. Indeed, once this particular choice of conformal
factor is done, any transformation $\Omega \to \omega \Omega$ satisfying $\widehat{n}^\mu \widehat{\nabla}_\mu \omega
\mathrel{\widehat{=}} 0$ leaves unchanged the above property. All important results that are derived hereafter are conformally
invariant, and thus independent of this gauge choice. However, in this gauge, the argument is simpler.

\subsection{Conformal flatness of the AdS boundary}
\label{demoaads}

Even if not obvious at first sight, conditions 4 and 5 of definition \ref{aadsdefinition} are equivalent to the conformal
flatness of \gls{ads} boundary. The argument is the following \cite{Ashtekar00}.

\paragraph{Case $n \geq 5$} We first introduce the following conformal $d+1$
decomposition (appendix \ref{d+1}) perpendicular to the \gls{ads} boundary with unit normal vector
\begin{equation}
   \widehat{r}_\alpha \equiv -\frac{\partial_\alpha\Omega}{\sqrt{\widehat{g}^{\mu\nu}\partial_\mu \Omega \partial_\nu\Omega}} =
   -\frac{\widehat{n}_\alpha}{\sqrt{\widehat{n}_\mu\widehat{n}^\mu}}.
   \label{defra}
\end{equation}
The vector $\widehat{r}_\alpha$ is oriented toward decreasing $\Omega$, which corresponds to increasing radii in the neighbourhood
of $\partial \widehat{\mathcal{M}}$. Note that equation \eqref{nmunmu} implies
\begin{equation}
   \widehat{r}_{\alpha} \mathrel{\widehat{=}} - \gls{L}\widehat{n}_\alpha.
   \label{rn}
\end{equation}
The conformal induced metric on $\Omega = cst$ hypersurfaces is
\begin{equation}
   \widehat{q}_{\alpha\beta} = \widehat{g}_{\alpha\beta} - \widehat{r}_\alpha \widehat{r}_\beta.
\end{equation}
From the orthogonality condition $\widehat{q}_{\alpha\mu}\widehat{r}^\mu = 0$, we deduce also that
\begin{equation}
   \widehat{q}_{\alpha\mu} \widehat{n}^\mu = 0.
   \label{confortho}
\end{equation}

Secondly, in order to use the simplifications provided by our conformal gauge choice (equation \eqref{confgauge}), it is
convenient to introduce $\widehat{\Theta}_{\alpha\beta}$, the extrinsic curvature tensor of constant $\Omega$ hypersurfaces associated to
the conformal induced metric $\widehat{q}_{\alpha\beta} = \Omega^2 q_{\alpha\beta}$. By definition (see equation \eqref{Kdef2}), we have
\begin{equation}
   \widehat{\Theta}_{\alpha\beta} = -\widehat{q}\indices{^\mu_\alpha}\widehat{q}\indices{^\nu_\beta}\widehat{\nabla}_\mu \widehat{r}_\nu.
   \label{defthetahat}
\end{equation}
Combining with our the definition \eqref{defra}, the orthogonality condition \eqref{confortho} the conformal gauge condition
\eqref{confgauge}, it comes
\begin{equation}
   \widehat{\Theta}_{\alpha\beta} \mathrel{\widehat{=}} 0.
\end{equation}
This allows us to study the geometry of the \gls{ads} boundary. Indeed, the Gauss relation \eqref{gauss} becomes
\begin{equation}
   \widehat{\mathcal{R}}_{\alpha\beta\gamma\delta} \mathrel{\widehat{=}}
   \widehat{q}\indices{^\mu_\alpha}\widehat{q}\indices{^\nu_\beta}\widehat{q}\indices{^\rho_\gamma}\widehat{q}\indices{^\sigma_\delta}\widehat{R}_{\mu\nu\rho\sigma},
\end{equation}
where $\widehat{\mathcal{R}}_{\alpha\beta\gamma\delta}$ is the Riemann tensor of $\widehat{q}_{\alpha\beta}$.
Given that $\widehat{C}_{\alpha\beta\gamma\delta} \mathrel{\widehat{=}} 0$ in virtue of \eqref{OC}, we get by successive
contractions (the relation between the Riemann and Weyl tensors being \eqref{weyldef})
\begin{subequations}
\begin{align}
   \widehat{\mathcal{R}}_{\alpha\beta\gamma\delta} &\mathrel{\widehat{=}} \frac{1}{n-2}(\widehat{q}_{\alpha\gamma} \widehat{q}\indices{^\mu_\beta}\widehat{q}\indices{^\nu_\delta} - \widehat{q}_{\alpha\delta} \widehat{q}\indices{^\mu_\beta}\widehat{q}\indices{^\nu_\gamma} - \widehat{q}_{\beta\gamma} \widehat{q}\indices{^\mu_\alpha}\widehat{q}\indices{^\nu_\delta} + \widehat{q}_{\beta\delta} \widehat{q}\indices{^\mu_\alpha}\widehat{q}\indices{^\nu_\gamma})\widehat{S}_{\mu\nu},\\
   \widehat{\mathcal{R}}_{\alpha\beta} &\mathrel{\widehat{=}} \frac{1}{n-2}\left( (n-3)\widehat{q}\indices{^\mu_\alpha}\widehat{q}\indices{^\nu_\beta} + \widehat{q}_{\alpha\beta}\widehat{q}^{\mu\nu} \right)\widehat{S}_{\mu\nu},\\
   \widehat{\mathcal{R}} &\mathrel{\widehat{=}} 2 \widehat{q}^{\mu\nu}\widehat{S}_{\mu\nu},\\
   \label{confS}
   \widehat{\mathcal{S}}_{\alpha\beta} &\mathrel{\widehat{=}} \frac{n-3}{n-2}\widehat{q}\indices{^\mu_\alpha}\widehat{q}\indices{^\nu_\beta}\widehat{S}_{\mu\nu},\\
   \label{confC0}
   \widehat{\mathcal{C}}_{\alpha\beta\gamma\delta} &\mathrel{\widehat{=}} 0,
\end{align}
\end{subequations}
where $\widehat{\mathcal{S}}_{\alpha\beta}$ and $\widehat{\mathcal{C}}_{\alpha\beta\gamma\delta}$ are respectively the Schouten and Weyl
tensors of $\widehat{q}_{\alpha\beta}$, and $\widehat{S}_{\alpha\beta}$ is the Schouten tensor of $\widehat{g}_{\alpha\beta}$. According to
the \gls{ws} theorem \ref{wstheorem}, equation \eqref{confC0} indicates that for $n \geq 5$, the \gls{ads} boundary is conformally
flat. Condition 4 of definition \ref{aadsdefinition} thus appears as a mere sufficient condition for the conformal flatness of the \gls{ads} boundary in dimensions
$n \geq 5$.

\paragraph{Conformal Einstein-Bianchi identity} Before moving to the $n = 4$ case, we need to discuss the conformal Bianchi identity. We
start from the Einstein-Bianchi equation \eqref{einbianchi}, that we repeat here for convenience
\begin{equation}
   \nabla_\mu C \indices{^\mu_{\alpha\beta\gamma}} = 16\pi\frac{n-3}{n-2}\nabla_{[\beta}\widetilde{T}_{\gamma]\alpha}.
\end{equation}
In this equation, $C \indices{^\mu_{\alpha\beta\gamma}} = \widehat{C} \indices{^\mu_{\alpha\beta\gamma}}$ by conformal invariance
(equation \eqref{weylconf2}). Now trading the $\nabla$ for $\widehat{\nabla}$ with \eqref{nablahatT} and using \eqref{confGamma} as
well as the Bianchi identity \eqref{bianchi12}, it comes
\begin{equation}
   \widehat{\nabla}_\mu \widehat{C}\indices{^\mu_{\alpha\beta\gamma}} -
   \frac{n-3}{\Omega}\widehat{C}\indices{^\mu_{\alpha\beta\gamma}}\widehat{n}_{\mu} =
   \frac{16\pi \gls{G}}{\Omega}\frac{n-3}{n-2}[\widehat{\nabla}_{[\beta}(\Omega \widetilde{T}_{\gamma]\alpha}) +
   \widehat{g}_{\alpha[\gamma}\widetilde{T}_{\beta]\mu}\widehat{n}^\mu].
   \label{nablamuC}
\end{equation}
Thanks to the near-boundary behaviour of the Weyl tensor (equation \eqref{OC}), we can introduce
\begin{equation}
   \widehat{K}_{\alpha\beta\gamma\delta} \equiv \frac{1}{\Omega^{n-3}}\widehat{C}_{\alpha\beta\gamma\delta},
\end{equation}
which has a smooth $O(1)$ limit near the \gls{ads} boundary. Equation \eqref{nablamuC} thus boils down to
\begin{equation}
   \widehat{\nabla}^\mu \widehat{K}_{\alpha\beta\gamma\mu} \mathrel{\widehat{=}} \lim_{\Omega\to 0} - \frac{16\pi
      \gls{G}}{\Omega^{n-2}}\frac{n-3}{n-2} [\widetilde{T}_{\alpha[\mu}\widehat{g}_{\beta]\gamma}\widehat{n}^\mu +
      \widehat{\nabla}_{[\alpha}(\Omega \widetilde{T}_{\beta]\gamma})].
      \label{dK}
\end{equation}
The limit on the right-hand side is well-defined and finite according to \eqref{OT}. This is the conformal Einstein-Bianchi identity. It is a
kind of conservation equation for the conformal Weyl tensor.

\paragraph{Case $n = 4$} The case $n=4$ is particular, since $\widehat{\mathcal{C}}_{\alpha\beta\gamma\delta}$ lives on a 3-dimensional manifold and is
thus identically zero (see equation \eqref{weylcompind} of appendix \ref{grd}). According to the \gls{ws} theorem \ref{wstheorem}, the \gls{ads} boundary is conformally flat if and only if its
conformal Cotton tensor (see definition \eqref{cottondef})
\begin{equation}
   \widehat{\mathcal{C}}_{\alpha\beta\gamma} = \widehat{D}_{[\alpha}\widehat{\mathcal{S}}_{\beta]\gamma},
\end{equation}
is zero. With \eqref{confS} and the definition of the connection $\widehat{D}$ associated to $\widehat{q}_{\alpha\beta}$ (see
equation \eqref{covdefd+1}), it can be shown that
\begin{equation}
   \widehat{\mathcal{C}}_{\alpha\beta\gamma} =
   \frac{1}{2}\widehat{q}\indices{^\mu_\alpha}\widehat{q}\indices{^\nu_\beta}\widehat{q}\indices{^\rho_\gamma}\widehat{C}_{\mu\nu\rho},
   \label{cotton1}
\end{equation}
where $\widehat{C}_{\alpha\beta\gamma}$ is the Cotton tensor of $\widehat{g}_{\alpha\beta}$. In \eqref{nablamuC}, if we evaluate
the terms precisely at the \gls{ads} boundary, we notice that all matter terms vanish in view of \eqref{OT}. Thus combining with
the Bianchi identity \eqref{bianchi22} for the conformal metric, we can eliminate the $\widehat{\nabla}_\mu
\widehat{C}\indices{^\mu_{\alpha\beta\gamma}}$ term and get a simple expression for the Cotton tensor, namely
\begin{equation}
   \widehat{C}_{\alpha\beta\gamma} \mathrel{\widehat{=}} -\widehat{K}_{\alpha\beta\gamma\mu}\widehat{n}^\mu.
   \label{cotton2}
\end{equation}
We thus conclude that equation \eqref{cotton1} and \eqref{cotton2} imply
\begin{equation}
   \widehat{\mathcal{C}}_{\alpha\beta\gamma} \mathrel{\widehat{=}}
   -\frac{1}{2}\widehat{q}\indices{^\mu_\alpha}\widehat{q}\indices{^\nu_\beta}\widehat{q}\indices{^\rho_\gamma}\widehat{n}^\sigma
   \widehat{K}_{\mu\nu\rho\sigma}.
   \label{CK}
\end{equation}
This equation can be simplified further by applying the $d+1$-decomposition of the Weyl tensor (equation \eqref{d+1C}) to
$\widehat{K}_{\alpha\beta\gamma\delta}$. Defining the pseudo-electric and pseudo-magnetic conformal Weyl tensors as
\begin{subequations}
\begin{align}
\label{defelec}
   \widehat{\varepsilon}_{\alpha\beta} \equiv \gls{L}^2 \widehat{K}_{\alpha\mu\beta\nu}\widehat{n}^\mu \widehat{n}^\nu,\\
\label{defmagn}
   \widehat{B}_{\alpha\beta} \equiv \gls{L}^2 \star \widehat{K}_{\alpha\mu\beta\nu}\widehat{n}^\mu \widehat{n}^\nu,
\end{align}
\end{subequations}
and using \eqref{rn}, it can be shown that
\begin{equation}
   \widehat{K}_{\alpha\beta\gamma\mu}\widehat{n}^\mu \mathrel{\widehat{=}} \widehat{\varepsilon}_{\alpha\gamma}\widehat{n}_\beta -
   \widehat{\varepsilon}_{\beta\gamma}\widehat{n}_\alpha + \widehat{\epsilon}_{\alpha\beta\mu}\widehat{B}\indices{^\mu_\gamma},
   \label{Keps}
\end{equation}
where we have defined the conformal 3-volume form of $\partial \widehat{\mathcal{M}}$
\begin{equation}
   \widehat{\epsilon}_{\alpha\beta\gamma} \equiv \widehat{n}^\mu \widehat{\epsilon}_{\mu\alpha\beta\gamma}.
\end{equation}
The triple projection onto $\partial \widehat{\mathcal{M}}$ of \eqref{Keps}, in combination with \eqref{CK} gives
\begin{equation}
   \widehat{\mathcal{C}}_{\alpha\beta\gamma} \mathrel{\widehat{=}}
   \frac{1}{2}\widehat{q}\indices{^\mu_\alpha}\widehat{q}\indices{^\nu_\beta}\widehat{q}\indices{^\rho_\gamma}\widehat{\epsilon}_{\mu\nu\sigma}\widehat{B}\indices{^\sigma_\rho}.
\end{equation}
Thus we conclude that for $n=4$, the 3-dimensional \gls{ads} boundary is conformally flat if and only if the pseudo-magnetic Weyl
tensor vanishes, namely
\begin{equation}
   \widehat{B}_{\alpha\beta} \mathrel{\widehat{=}} 0.
\end{equation}
This is the so-called reflective boundary condition. Condition 4 of definition \ref{aadsdefinition} is thus a mere sufficient
condition for the conformal flatness of the \gls{ads} boundary in the special case $n = 4$. We can thus conclude that an
\gls{aads} space-time is characterised by the conformal flatness character of its boundary $\partial \mathcal{M}$, provided it
constitutes a second order pole of the physical metric (condition 1).

\section{Global charges}
\label{globcharge}

Now that we have precisely discussed the meaning of the \gls{aads} conditions, we can continue the argument in order to unveil
conserved charges \cite{Ashtekar84,Ashtekar00}. In general, these charges are associated to a Killing vector, and can thus be a
valuable tool to compute the mass and the angular momentum of \gls{aads} solutions.

\subsection{Ashtekar-Magnon-Das charges}

As usual when defining global charges, we aim at deriving a conservation equation \cite{Brown93}. The global charge is then
obtained by integrating this conservation equation (like in classical hydrodynamics). If we project \eqref{dK} onto
$\widehat{n}^\alpha \widehat{n}^\gamma \widehat{q} \indices{^\beta_\nu}$, the left-hand side becomes
\begin{equation}
   \widehat{n}^\mu \widehat{n}^\nu \widehat{q}\indices{^\rho_\alpha}\widehat{\nabla}^\sigma \widehat{K}_{\mu\rho\nu\sigma}
   \mathrel{\widehat{=}} \widehat{q}\indices{^\rho_\alpha}\widehat{\nabla}^\sigma (\widehat{K}_{\mu\rho\nu\sigma}\widehat{n}^\mu
   \widehat{n}^\nu ),
   \label{leftdK}
\end{equation}
where we have used our gauge condition \eqref{confgauge}, and the right-hand side becomes
\begin{equation}
   -\frac{8\pi \gls{G}(n-3)}{\gls{L}^2}\widehat{n}^\mu \widehat{q}\indices{^\nu_\alpha}\widehat{T}_{\mu\nu},
   \label{rightdK}
\end{equation}
where we have used the orthogonality condition \eqref{confortho} and have introduced the regularised energy-momentum tensor
\begin{equation}
   \widehat{T}_{\alpha\beta} \equiv \frac{1}{\Omega^{n-2}}\widetilde{T}_{\alpha\beta}.
   \label{thatdef}
\end{equation}
Note that thanks to \eqref{OT}, this tensor is $O(1)$ near $\partial\mathcal{M}$. We define the momentum carried by the matter
field as
\begin{equation}
   \widehat{p}_\alpha \equiv - \widehat{q}\indices{^\mu_\alpha}\widehat{n}^\nu \widehat{T}_{\mu\nu},
   \label{phatdef}
\end{equation}
such that the above projection of equation \eqref{dK} (with left-hand side \eqref{leftdK} and right-hand side \eqref{rightdK}) finally reduces to
\begin{equation}
   \widehat{D}^\mu \widehat{\varepsilon}_{\alpha\mu} \mathrel{\widehat{=}} 8\pi \gls{G}(n-3) \widehat{p}_\alpha.
   \label{depsilon}
\end{equation}
This reveals the sought-after conservation equation, which deal with $\widehat{\varepsilon}_{\alpha\beta}$, the pseudo-electric part of the conformal Weyl tensor
defined in \eqref{defelec}. Equation \eqref{depsilon} is closely related to the so-called \gls{amd} charges.

Indeed, the charge definition directly results from the integration of \eqref{depsilon}. Let us consider $\widehat{\xi}^\alpha$ a
conformal Killing vector of the \gls{ads} boundary. It then obeys the conformal Killing equation
\begin{equation}
   \widehat{D}^{(\alpha}\widehat{\xi}^{\beta)} \mathrel{\widehat{=}} \frac{1}{n-1}\widehat{q}^{\alpha\beta}\widehat{D}_\mu \widehat{\xi}^\mu.
\end{equation}
Contracting \eqref{depsilon} with $\widehat{\xi}^\alpha$ gives
\begin{equation}
   \widehat{D}^\mu(\widehat{\varepsilon}_{\mu\nu}\widehat{\xi}^\nu) \mathrel{\widehat{=}} 8\pi \gls{G}(n-3) \widehat{p}_\mu \widehat{\xi}^\mu,
\end{equation}
where we have used the traceless character of the pseudo-electric Weyl tensor, $\widehat{q}^{\mu\nu}
\widehat{\varepsilon}_{\mu\nu} = 0$ (see section \ref{elecpartweyl} of appendix \ref{d+1}). Integrating this relation on the
conformal \gls{ads} boundary $\partial \widehat{\mathcal{M}}$ between times $t'$ and $t''$ gives
\begin{equation}
   \int_{\partial \widehat{\mathcal{M}}(t',t'')} \widehat{D}^\mu(\widehat{\varepsilon}_{\mu\nu}\widehat{\varepsilon}^\nu) \sqrt{|q|}d^dy =
   \int_{\partial \widehat{\mathcal{M}}(t',t'')} 8\pi \gls{G}(n-3) \widehat{p}_\mu \widehat{\xi}^\mu \sqrt{|\widehat{q}|}d^dy.
\end{equation}
Using Stokes theorem it comes \cite{Ashtekar84,Ashtekar00}
\begin{equation}
   \int_{t' \bigcap \partial \widehat{\mathcal{M}}}^{t'' \bigcap \partial \widehat{\mathcal{M}}} \widehat{\varepsilon}_{\mu\nu} \widehat{u}^\mu \widehat{\xi}^\nu
   \sqrt{\widehat{\sigma}}d^{p} z = 8\pi \gls{G}(n-3)\int_{\partial \widehat{\mathcal{M}}(t',t'')} \widehat{p}_\mu\widehat{\xi}^\mu
   \sqrt{|\widehat{q}|}d^dy,
   \label{consAMD}
\end{equation}
where $\widehat{u}^\alpha$ is the conformal unit normal vector to $t = cst$ slices, $\widehat{\sigma}_{\alpha\beta} =
\widehat{q}_{\alpha\beta} + \widehat{u}^\alpha \widehat{u}^\beta$ is the
conformal induced metric on the $t = cst$ slices of the \gls{ads} boundary, and the notation $\int_{t' \bigcap \partial \mathcal{M}}^{t''
\bigcap \partial \mathcal{M}}$ denotes the difference between the integrals over the $t = t' = cst$ and $t = t'' = cst$ slices
of $\partial \widehat{\mathcal{M}}$ (see figure \ref{global}).

This motivates the following charge definition attached to the boundary conformal Killing vector
$\widehat{\xi}^\alpha$:
\begin{equation}
   Q^{AMD}_{\widehat{\xi}} [\widehat{\Sigma}] = -\frac{\gls{L}}{8\pi \gls{G}(n-3)}\oint_{\widehat{\Sigma}} \widehat{\varepsilon}_{\mu\nu} \widehat{u}^\mu \widehat{\xi}^\nu
   \sqrt{\widehat{\sigma}}d^p z,
   \label{QAMD}
\end{equation}
where $\widehat{\Sigma}$ is a $t = cst$ hypersurface of the conformal \gls{ads} boundary $\partial \widehat{\mathcal{M}}$. Indeed, this charge is
conserved in time according to \eqref{consAMD}, provided the energy-momentum tensor vanishes fast enough (i.e.\ like
$O(\Omega^{n-3})$ at least, according to \eqref{thatdef} and \eqref{phatdef}) near the boundary. This is the so-called \gls{amd}
charge.

Note that this charge definition is invariant under conformal transformations, and thus independent of the conformal gauge
freedom \eqref{confgauge}. Indeed, under the transformation $\Omega \to \omega \Omega$, it comes (see table \ref{confrecap} and section \ref{conftransform}
of the appendix)
\begin{equation}
   \widehat{K}_{\alpha\beta\gamma\delta} \to \frac{1}{\omega^{n-1}}\widehat{K}_{\alpha\beta\gamma\delta}, \quad \widehat{r}^\alpha \to
   \frac{1}{\omega}\widehat{r}^\alpha, \quad \widehat{\varepsilon}_{\alpha\beta} \to
   \frac{1}{\omega^{n-3}}\widehat{\varepsilon}_{\alpha\beta}, \quad \widehat{u}^\alpha \to \frac{1}{\omega}\widehat{u}^\alpha, \quad
   \sqrt{\widehat{\sigma}} \to \omega^{n-2} \sqrt{\widehat{\sigma}}.
\end{equation}
The conformal Killing vector $\widehat{\xi}^\alpha$ being unchanged under such a transformation, it is clear that the \gls{amd}
charge \eqref{QAMD} is independent of our conformal gauge. Furthermore, the mass and angular momentum of \gls{aads} space-times are closely linked to the charges
attached to the boundary conformal Killing vectors $\partial_t^{\alpha}$ and $\partial_\varphi^{\beta}$ (see chapter \ref{simulations}).

\subsection{Balasubramanian-Kraus charges}
\label{bkcharge}

Even if the conformal transformations were essential for the definition of \gls{aads} space-times, and very useful for the
definition of the \gls{amd} charges, there are other ways of defining global charges. Among them, the \gls{bk} charges only rely on
the least action principle. Again, our goal is to get a conservation equation whose integration yields a suitable charge
definition. If $n = 3,4,5$, the action of \gls{gr} with negative cosmological constant is (see \cite{Balasubramanian99} and appendix \ref{leastaction}, equations \eqref{ghyct}
and \eqref{lctd})
\begin{equation}
   S = \frac{1}{16\pi \gls{G}}\int_\mathcal{M} (R - 2 \gls{Lambda}) \sqrt{-g}d^{d+1}x - \frac{1}{8\pi \gls{G}}\oint_{\partial \mathcal{M}}
   \left[\Theta + \frac{n-2}{\gls{L}} + \frac{\gls{L}}{2(n-3)}\mathcal{R}\right] \sqrt{|q|}d^dy,
\end{equation}
where $\Theta$ and $\mathcal{R}$ refer respectively to the mean extrinsic and intrinsic curvatures of $r = cst$ slices. Introducing
$r_\alpha$ the unit normal vector to these hypersurfaces, oriented toward the \gls{ads} boundary, the induced metric is
$q_{\alpha\beta} = g_{\alpha\beta} - r_\alpha r_\beta$. In appendix \ref{leastaction} equation \eqref{taudef}, it is shown that
the associated quasi-local stress tensor is defined by
\begin{equation}
   \tau_{\alpha\beta} \equiv -\frac{2}{\sqrt{|q|}}\frac{\delta S}{\delta q^{\alpha\beta}} = \frac{1}{8\pi \gls{G}}\left(\Theta_{\alpha\beta} - \Theta q_{\alpha\beta} - \frac{n-2}{\gls{L}}q_{\alpha\beta} + \frac{\gls{L}}{n-3} \mathcal{G}_{\alpha\beta}\right),
   \label{quasilocalst}
\end{equation}
where $\Theta_{\alpha\beta}$ is the extrinsic curvature tensor of the $r = cst$ hypersurfaces and $\mathcal{G}_{\alpha\beta}$ is
the Einstein tensor of the induced metric $q_{\alpha\beta}$. Denoting by $D$ the connection of $q_{\alpha\beta}$, and using the
well-known result according to which the metric and the Einstein tensor are divergence-free (equations \eqref{divmetric} and
\eqref{divG}), it can be shown that the momentum constraint of the $d+1$ Einstein's equation (equation \eqref{mom}) implies
\begin{equation}
   D^\mu \tau_{\alpha\mu} = p_\alpha,
   \label{dtau}
\end{equation}
where we have defined the mixed projection of the energy-momentum tensor (or simply momentum)
\begin{equation}
   p_\alpha = - q^\mu_\alpha r^\nu T_{\mu\nu}.
\end{equation}
In order to compute the charge associated to $\tau_{\alpha\beta}$, we assume that there exist a Killing vector $\xi^\alpha$ on the
\gls{ads} boundary $\partial \mathcal{M}$, such that
\begin{equation}
   D^{(\alpha} \xi^{\beta)} \mathrel{\widehat{=}} 0,
   \label{killing}
\end{equation}
where indices in parentheses are symmetrised. Contracting \eqref{dtau} with $\xi^\alpha$, it comes
\begin{equation}
   D^\mu (\tau_{\mu\nu} \xi^\nu) =  p_\mu\xi^\mu,
   \label{dmutau}
\end{equation}
where we have used the symmetry $\tau_{\alpha\beta}$ and the antisymmetry of the Killing equation \eqref{killing}. This is the
sought-after conservation equation, and it deals with the quasi-local stress tensor. The charge definition is directly related to
this equation. Integrating \eqref{dmutau} on the \gls{ads} boundary between times $t'$ and $t''$, it comes
\begin{equation}
   \int_{t'}^{t''}\int_{\partial \mathcal{M}} D^{\mu}(\tau_{\mu\nu}\xi^\nu)\sqrt{|q|}d^d y = \int_{t'}^{t''}\int_{\partial
      \mathcal{M}} p_\mu\xi^\mu \sqrt{|q|}d^dy
   \sqrt{|q|} d^dy.
\end{equation}
Using Stokes theorem \cite{Brown93}, we get
\begin{equation}
   \int_{t' \bigcap \partial \mathcal{M}}^{t'' \bigcap \partial \mathcal{M}} \tau_{\mu\nu} u^\mu \xi^\nu
   \sqrt{\sigma}d^{p} z = \int_{t'}^{t''}\int_{\partial \mathcal{M}} p_\mu\xi^\mu \sqrt{|q|}d^dy,
   \label{consBK}
\end{equation}
where $u^\alpha$ is the unit normal vector to $t = cst$ slices, $\sigma_{\alpha\beta} = q_{\alpha\beta} + u^\alpha u^\beta$ is the
induced metric on the $t = cst$ slices of the \gls{ads} boundary, and the notation $\int_{t' \bigcap \partial \mathcal{M}}^{t''
\bigcap \partial \mathcal{M}}$ denotes the difference between the integrals over the $t = t' = cst$ and $t = t'' = cst$ slices
of $\partial \mathcal{M}$ (see figure \ref{global}).

\begin{myfig}
   \includegraphics[width = 0.49\textwidth]{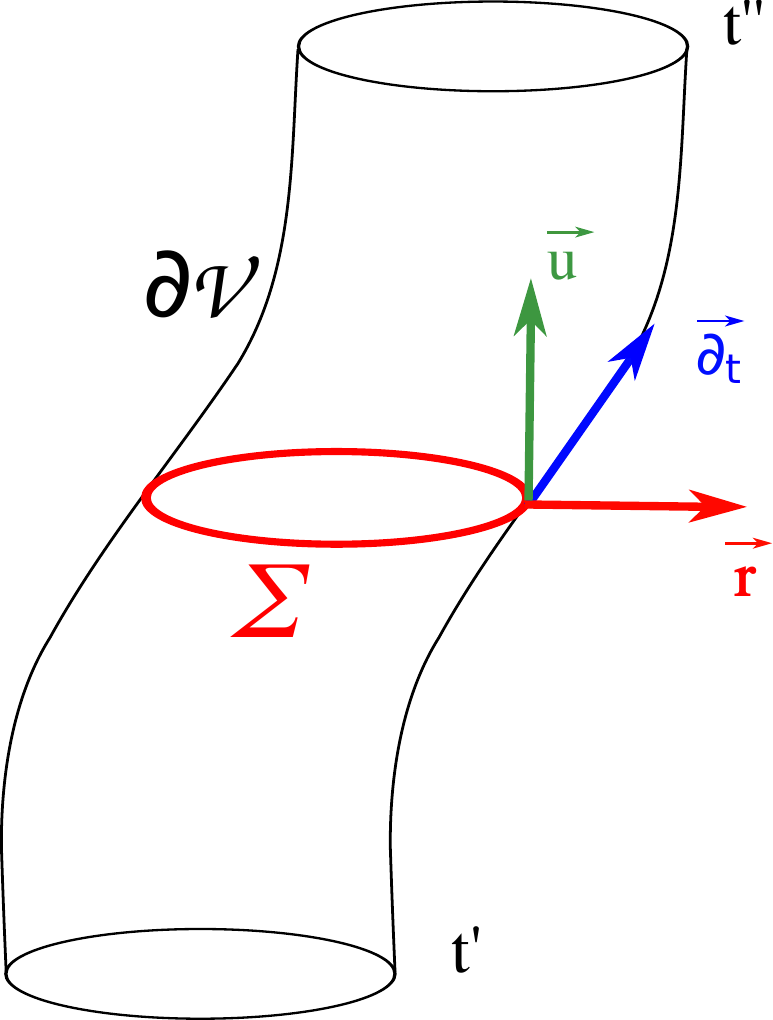}
   \caption[Global representation of the AdS boundary]{The \gls{ads} boundary is the tube of world-lines parallel to $\partial_t$
   at spatial infinity. The unit normal vector to $t = cst$ slices is $u^\alpha$ and the unit normal vector to $r = cst$ slices
   is $r^\alpha$. The times $t'$ and $t''$ are merely the limits of the interval of integration on $\partial\mathcal{M}$. Credits:
G. Martinon.}
   \label{global}
\end{myfig}

This motivates the following charge definition attached to the boundary Killing vector
$\xi^\alpha$:
\begin{equation}
   Q^{BK}_\xi [\Sigma] = \oint_\Sigma \tau_{\mu\nu}u^\mu \xi^\nu \sqrt{\sigma}d^{p}z,
   \label{QBK}
\end{equation}
where $\Sigma$ is a $t = cst$ hypersurface of the \gls{ads} boundary. Indeed, this charge is conserved in time according to
\eqref{consBK} provided the energy-momentum tensor vanishes fast enough (i.e.\ like $O(\Omega^{n-3})$ at least, since
$\sqrt{|q|} \underset{\partial \mathcal{M}}{=} O(\Omega^{1-n})$ and $r^\alpha \underset{\partial \mathcal{M}}{=} O(\Omega)$) near
the boundary. This is the so-called \gls{bk} charge.

At first sight, eyeballing equation \eqref{quasilocalst}, one may think that
$\tau_{\alpha\beta}$ behaves as $O(\Omega^{-2})$ near the \gls{ads} boundary, but it is actually a $O(\Omega^{n-2})$ for the pure
\gls{ads} space-time, and a $O(\Omega^{n-3})$ for \gls{aads} solutions\footnote{The physical energy-momentum tensor of the dual
\gls{cft} (see chapter \ref{adscft}) is actually given by $\tau_{\alpha\beta}/\Omega^{n-3}$ on the \gls{ads} boundary.}. Since
$\sqrt{\sigma} \underset{\partial \mathcal{M}}{=} O(\Omega^{2-n})$ and $u^\mu \underset{\partial \mathcal{M}}{=} O(\Omega)$ near
the \gls{ads} boundary, it is clear that the charge is non-zero if and only if $\tau_{\alpha\beta} \underset{\partial
\mathcal{M}}{=} O(\Omega^{n-3})$.

As well as for the \gls{amd} case, the mass and angular momentum of \gls{aads} space-times are closely linked to the charges
attached to the boundary Killing vectors $\partial_t^{\alpha}$ and $\partial_\varphi^{\beta}$ (see chapter \ref{simulations}).

\subsection{Relation between AMD and BK charges}

We have defined two independent sets of charges: the \gls{amd} and \gls{bk} charges. How do they relate to
each other? In order to compare them, we need to relate \eqref{QBK} and \eqref{QAMD}. The former involves the quasi-local stress
tensor and the latter the pseudo-electric part of the Weyl tensor. However, since $\widehat{n}_\alpha$ is not a unit vector, and
$\widehat{K}_{\alpha\beta\gamma\delta}$ is the rescaled Weyl tensor, $\widehat{\varepsilon}_{\alpha\beta}$ is not exactly the
electric Weyl tensor. In this section, we thus consider the quasi-local stress tensor $\tau_{\alpha\beta}$ and the proper electric
Weyl tensor
\begin{align}
   \tau_{\alpha\beta} &= \frac{1}{8\pi \gls{G}}\left(\Theta_{\alpha\beta} - \Theta q_{\alpha\beta} + \frac{\gls{L}}{n-3} - \frac{n-2}{\gls{L}}q_{\alpha\beta}\right),\\
   E_{\alpha\beta} &\equiv C_{\alpha\mu\beta\nu}r^\mu r^\nu = -\mathcal{R}_{\alpha\beta} + \Theta\Theta_{\alpha\beta} - \Theta_{\alpha\mu}\Theta \indices{^\mu_\beta} +
   \frac{2 \gls{Lambda}}{n-1}q_{\alpha\beta},
   \label{trueE}
\end{align}
where we have used the $d+1$-decomposition of $E_{\alpha\beta}$ tensor given in the appendix (equation \eqref{Ed+1}) and dropped
the matter terms. Indeed we assume that the energy-momentum tensor is vanishing sufficiently slowly
(at least like $O(\Omega^{n-1})$, see equation $\eqref{OT}$) so that it its contribution to the surface integrals is zero for both
charges. With the help of the \gls{Lambda}-\gls{L} relation \eqref{lambdaL} and the Hamiltonian constraint of Einstein's
equation (equation \eqref{ham}), it can be shown that
\begin{align}
\nonumber   8\pi \gls{G}\tau_{\alpha\beta} + \frac{\gls{L}}{n-3}E_{\alpha\beta} &= \frac{\gls{L}}{n-3}\left( \Theta\Theta_{\alpha\beta} -
   \Theta_{\alpha\mu}\Theta \indices{^\mu_\beta} + \frac{1}{2}[\Theta_{\mu\nu}\Theta^{\mu\nu} - \Theta^2]q_{\alpha\beta} \right)
   \\ &- \frac{n-2}{2 \gls{L}}q_{\alpha\beta} + \Theta_{\alpha\beta} - \Theta q_{\alpha\beta}.
   \label{difint}
\end{align}
From the definition of the conformal extrinsic curvature tensor \eqref{defthetahat}, using $\widehat{r}_\alpha = \Omega r_\alpha$,
$\widehat{q}\indices{^\alpha_\beta} = q \indices{^\alpha_\beta}$, the orthogonality condition \eqref{confortho}, and trading the
$\widehat{\nabla}$ for $\nabla$ with \eqref{nablahatT} and \eqref{confGamma}, it comes
\begin{subequations}
\begin{align}
   \Theta_{\alpha\beta} &= \frac{1}{\Omega}\widehat{\Theta}_{\alpha\beta} + \frac{1}{\Omega^2}\widehat{q}_{\alpha\beta} \widehat{n}\cdot \widehat{r},\\
   \Theta \indices{^\alpha_\beta} &= \Omega \widehat{\Theta}\indices{^\alpha_\beta} + \delta \indices{^\alpha_\beta}\widehat{n}\cdot \widehat{r},\\
   \Theta^{\alpha\beta} &= \Omega^3 \widehat{\theta}^{\alpha\beta} + \Omega^2 \widehat{q}^{\alpha\beta} \widehat{n}\cdot \widehat{r},\\
   \Theta &= \Omega \widehat{\Omega} + (n-1) \widehat{n}\cdot \widehat{r},
\end{align}
\label{thetas}
\end{subequations}
where we have defined
\begin{equation}
   \widehat{n}\cdot \widehat{r} \mathrel{\equiv} \widehat{g}^{\mu\nu}\widehat{n}_\mu \widehat{r}_\nu.
\end{equation}
The relations \eqref{thetas} allow us to write equation \eqref{difint} in terms of hatted tensors,
namely
\begin{align}
\nonumber   8\pi \gls{G}\tau_{\alpha\beta} + \frac{\gls{L}}{n-3}E_{\alpha\beta} &= \frac{\gls{L}}{n-3}\left( \widehat{\Theta}\widehat{\Theta}_{\alpha\beta} -
   \widehat{\Theta}_{\alpha\mu}\widehat{\Theta} \indices{^\mu_\beta} + \frac{1}{2}[\widehat{\Theta}_{\mu\nu}\widehat{\Theta}^{\mu\nu} -
   \widehat{\Theta}^2]\widehat{q}_{\alpha\beta} \right)\\ &
   + \frac{\gls{L}\widehat{n}\cdot \widehat{r} + 1}{\Omega}(\widehat{\Theta}_{\alpha\beta} - \widehat{\Theta}\widehat{q}_{\alpha\beta}) -
   \frac{n-2}{2}\frac{(\gls{L}\widehat{n}\cdot \widehat{r} + 1)^2}{\gls{L}\Omega^2}\widehat{q}_{\alpha\beta}.
   \label{difint2}
\end{align}
At this point we can notice that our conformal gauge condition \eqref{confgauge} combined with the definition \eqref{defthetahat} implies
$\widehat{\Theta}_{\alpha\beta} \mathrel{\widehat{=}} 0$, so that
\begin{equation}
   \widehat{\Theta}_{\alpha\beta} \underset{\partial \widehat{\mathcal{M}}}{=} O(\Omega),
   \label{Otheta}
\end{equation}
near the \gls{ads} boundary. Moreover, recycling the results of \eqref{rn} and \eqref{lim2}, it comes
\begin{equation}
   \lim_{\Omega\to 0}\frac{1}{\Omega}(\gls{L}\widehat{n}\cdot \widehat{r} + 1) = \lim_{\Omega\to 0}\frac{1}{\Omega}(-\gls{L}^2 \widehat{n}\cdot \widehat{n} + 1)
   \mathrel{\widehat{=}} - \frac{2}{n}\widehat{\nabla}^\mu\widehat{n}_\mu \mathrel{\widehat{=}} 0,
\end{equation}
where the last equality stems from the conformal gauge choice \eqref{confgauge}. Thus,
\begin{equation}
   \gls{L}\widehat{n}\cdot \widehat{r} + 1 \underset{\partial \mathcal{M}}{=} O(\Omega^2),
   \label{Olnn}
\end{equation}
in the neighbourhood of the \gls{ads} boundary. In light of \eqref{Otheta} and \eqref{Olnn}, we define
\begin{equation}
   \widehat{\vartheta}_{\alpha\beta} \equiv \frac{1}{\Omega}\widehat{\Theta}_{\alpha\beta} \quad \tn{and} \quad \widehat{f} \equiv
   \frac{1}{\Omega^2}(\gls{L}\widehat{n}\cdot \widehat{r} + 1).
\end{equation}
By construction, both $\widehat{\vartheta}_{\alpha\beta}$ and $\widehat{f}$ are of order $O(1)$ near the \gls{ads} boundary. Equation
\eqref{difint2} thus takes the form
\begin{align}
   & 8\pi \gls{G}\tau_{\alpha\beta} + \frac{\gls{L}}{n-3}E_{\alpha\beta} = \Omega^2 \widehat{\Delta}_{\alpha\beta} \quad \tn{with}\\
   \nonumber\widehat{\Delta}_{\alpha\beta} &= \frac{\gls{L}}{n-3}\left( \widehat{\vartheta}\widehat{\vartheta}_{\alpha\beta} -
   \widehat{\vartheta}_{\alpha\mu}\widehat{\vartheta} \indices{^\mu_\beta} + \frac{1}{2}[\widehat{\vartheta}_{\mu\nu}\widehat{\vartheta}^{\mu\nu} -
   \widehat{\vartheta}^2]\widehat{q}_{\alpha\beta} \right)
   + \widehat{f}(\widehat{\vartheta}_{\alpha\beta} - \widehat{\vartheta}\widehat{q}_{\alpha\beta}) -
   \frac{n-2}{2}\widehat{f}^2\widehat{q}_{\alpha\beta}.
\end{align}
The advantage of this equation is that $\widehat{\Delta}_{\alpha\beta} \underset{\partial \mathcal{M}}{=} O(1)$ near the \gls{ads} boundary. It is thus clear that the
left-hand side is $O(\Omega^2)$.

We are now able to examine the difference between the \gls{bk} and \gls{amd} charges. Given that
\begin{equation}
   u^\alpha = \frac{1}{\Omega} \widehat{u}^\alpha,\quad \sqrt{\sigma} = \frac{1}{\Omega^{n-2}} \sqrt{\widehat{\sigma}}, \quad
   \widehat{\varepsilon}_{\alpha\beta} \mathrel{\widehat{=}} \frac{1}{\Omega^{n-3}} E_{\alpha\beta},
\end{equation}
where we have used \eqref{rn}, \eqref{trueE} and \eqref{weylconf}, it comes
\begin{align}
\nonumber   Q^{BK}_{\widehat{\xi}}[\widehat{\Sigma}] - Q^{AMD}_{\xi}[\Sigma] &= \oint_\Sigma \tau_{\mu\nu}\xi^\mu u^\nu \sqrt{\sigma}d^pz +
   \frac{\gls{L}}{8\pi \gls{G}(n-3)}\oint_{\widehat{\Sigma}} \widehat{\varepsilon}_{\mu\nu}\widehat{\xi}^\mu \widehat{u}^\nu
   \sqrt{\widehat{\sigma}}d^pz\\
\nonumber   &= \oint_{\widehat{\Sigma}} \tau_{\mu\nu}\widehat{\xi}^\mu \frac{\widehat{u}^\nu}{\Omega}\frac{\sqrt{\widehat{\sigma}}}{\Omega^{n-2}}d^pz +
   \frac{\gls{L}}{8\pi \gls{G}(n-3)}\oint_{\widehat{\Sigma}} \frac{E_{\mu\nu}}{\Omega^{n-3}}\widehat{\xi}^\mu \widehat{u}^\nu
   \sqrt{\widehat{\sigma}}d^pz\\
   &= \oint_{\widehat{\Sigma}}\Omega^{5-n}\widehat{\Delta}_{\mu\nu}\widehat{\xi}^\mu \widehat{u}^\nu \sqrt{\widehat{\sigma}}d^pz,
\end{align}
where we have identified $\Sigma = \widehat{\Sigma}$ and $\xi^\mu = \widehat{\xi}^\mu$. Indeed, by virtue of \eqref{killingconfkilling}, a Killing
vector for $q_{\alpha\beta}$ is a conformal Killing vector for $\widehat{q}_{\alpha\beta}$. The integrand is thus
$O(\Omega^{5-n})$ on the surface of integration, so that the two charges agree each other for $n = 4$. In \cite{Ashtekar00}, the
authors have shown that for $n = 5$ and for the charges associated to the Killing vector $\partial_t^\alpha$, the result is
\begin{equation}
   Q^{BK}_{\partial_t}[\widehat{\Sigma}] - Q^{AMD}_{\partial_t}[\Sigma] = \frac{3\pi \gls{L}^2}{32 \gls{G}},
   \label{casimir1}
\end{equation}
which corresponds to a Casimir energy (discussed in chapter \ref{adscft} section \ref{casimirsec}). The equivalence between \gls{amd} and
\gls{bk} charges is thus dependent of the number of dimensions. This issue is also discussed in details in
\cite{Papadimitriou05,Hollands05a}, but only in the $n \leq 5$ case. In dimensions $n \geq 6$, the expression for the quasi-local
stress tensor $\tau_{ij}$ is much more complicated and so is the expression of $\widehat{\Delta}_{\alpha\beta}$. It is thus not clear that
the shift between \gls{amd} and \gls{bk} is independent of the dimension. To the best of our knowledge, there is no
published work trying to compare these two definitions of charges in dimensions $n \geq 6$. Furthermore, it is argued in
\cite{Ashtekar00} that the \gls{bk} charges are slightly less covariant than the \gls{amd} ones since they could potentially depend on the
cross section $\widehat{\Sigma}$. A background term subtraction technique is sometimes employed to cure this problem (see e.g.\
\cite{Ashtekar14,Hollands05b}). The equivalence between the \gls{amd} and \gls{bk} masses and the first historical definitions of
\cite{Abbott82,Henneaux85} in 4-dimensional \gls{aads} space-times is performed in \cite{Hollands05a}.

Finally, let us mention that the positivity of the charges is established in \cite{Gibbons83,Townsend84,Ashtekar14}, provided
matter fields obey the dominant energy condition and decay sufficiently rapidly near the \gls{ads} boundary. This is the so-called
positive-energy theorem.

\section{Boundary conditions in AAdS space-times}

We have already mentioned the work of \cite{Ishibashi04} where all possible boundary conditions in \gls{aads} space-times were
enumerated for first order perturbations. The global \gls{aads} definition discussed in this chapter does not provide sufficient boundary
conditions for matter fields in such space-times. Indeed, conditions 4 and 5 of definition \ref{aadsdefinition} are not boundary
conditions, since they fix a number of conditions that is in general different from the number of degrees of freedom of the gravitational dynamics.
However, they can be viewed as potential a posteriori checks that the boundary conditions used in a non-linear problem (whatever they are)
do indeed preserve the \gls{aads} asymptotics.

These conditions are particularly relevant in the field of numerical relativity. Indeed, it is common to use boundary conditions
that are not overwhelmingly complicated and that are adapted to the numerical problem at hand. However, in highly non-linear
configurations, Dirichlet or Neumann boundary conditions are not a priori sufficient to preserve the \gls{aads} asymptotics. The
examination of the conformal Weyl tensor as well as its magnetic part are thus invaluable criteria that can help to determine if a
numerical solution is valid or not. This kind of numerical test was performed notably in \cite{Martinon17}, and are described in
detail in chapter \ref{simulations}.

Finally, the \gls{amd} and \gls{bk} charges are two independent sets of charges that can be computed independently in either
analytical or numerical investigations. As such, in chapter \ref{simulations} and \cite{Martinon17}, we use both of them to determine the mass and
angular momentum of numerical geons, combining appropriately the charges associated to the boundary Killing fields $\partial_t^\alpha$
and $\partial_\varphi^\alpha$ (in spherical coordinates). The agreement between the two sets of charges in four dimensions can
thus be used as a strong argument in favour of the correctness of the \gls{aads} asymptotics.

\chapter{The AdS-CFT correspondence}
\label{adscft}
\addcontentsline{lof}{chapter}{\nameref{adscft}}
\addcontentsline{lot}{chapter}{\nameref{adscft}}
\citationChap{Physics is like sex: sure, it may give some practical results, but that's not why we do it.}{Richard Phillips Feynman}
\minitoc

Of the three maximally symmetric space-times (Minkowski, \gls{ds} and \gls{ads}), \gls{ads} has long been considered the
weirdest. Indeed, Minkowski space-time is the simplest and can be used as a practical background for the dynamics of bodies in a
galaxy, while the \gls{ds} space-time is associated to many cosmological models since we live in a world with presumably a positive
cosmological constant $\Lambda \sim \SI{e-52}{m}$.

It is not before 1998 that \gls{ads} space-time was promoted to ``real-world''
physics with the advance of the \gls{ads}-\gls{cft} conjecture (named also correspondence or duality). The seminal paper of Maldacena
\cite{Maldacena99} is by now the most cited ever article in high energy physics, harvesting a total of $\sim 13 000$ citations at
the date of the present manuscript, which represents roughly two citations a day for the last 19 years. This translates the huge
interest of the high-energy physics community in the correspondence. And indeed, it is undeniable that it shed new
light on the challenging problem of the unification of gravity with quantum mechanics. The reason of such a large success is that
computations that were deemed impossible in quantum field theories became trivial when the duality was uncovered. The
correspondence can be seen as a kind of sophisticated change of variables, but along with variables, we also change theories,
switching constantly between string (gravity) and quantum field theory.

As its name suggests, the \gls{ads}-\gls{cft} conjecture is not a theorem and is not demonstrated mathematically. However, it
strengthens as the number of its validations grows at day in and day out. Since the author of the present manuscript is
not an expert in quantum theories, several arguments below are discussed but admitted without demonstration.

In this chapter, we try and give an very brief overview of the concepts involved in this area, and discuss several
applications. In particular, even if not obvious at first sight, we argue that gravitational dynamics in \gls{ads}
space-times are dual to some properties of \glspl{qgp}, which are produced in hadron colliders but are very tough to study both
theoretically and experimentally.

The conjecture is the subject of an abundant literature. The interested reader is referred to the seminal articles
\cite{Maldacena99,Witten98} as well as the different reviews \cite{Aharony00,Hubeny15} and references therein. We would also like
to stress that the book \cite{Natsuume15}, even if sometimes imprecise, was of great help to write this chapter. Discussing the
correspondence is easier and relevant for 5-dimensional \gls{ads} space-time. Hereafter, we use capital Latin letters $A,B,M,N$
for indices in bulk 5-dimensional \gls{ads} volume, Greek letters $\alpha,\beta,\mu,\nu$ for indices corresponding to its
4-dimensional boundary, and Latin letters $i,j,k,l$ for 3-space indices on the boundary. We set $\gls{hbar} = \gls{c} = \gls{kb} =
1$. We also denote by $\partial \mathcal{V}$ the \gls{ads} boundary.

\section{The quantum side of the duality}

As chapter \ref{aads} entirely dealt with one side of the correspondence, namely the \gls{ads} space-time, let us describe briefly the
other side of the duality. The main motivation for studying \glspl{cft} is that they are close to \gls{qcd}, the theory
describing the strong interaction.

\subsection{Overview of quantum chromodynamics}

In \gls{qcd}, the following picture holds.

\begin{myitem}
   \item Hadrons, in particular protons and neutrons, are made of quarks, which are the fundamental representation of the $SU(3)$ gauge symmetry.
   \item Quarks interact with each other via the strong interaction.
   \item The strong interaction is propagated by gluons, that belong to the adjoint representation of $SU(3)$.
   \item Quarks have six flavours: u, d, t, b, c, s.
   \item Quarks have a colour charge red, green or blue, and antiquarks have an anticolour charge antired, antigreen or antiblue.
   \item Gluons carry one colour and one anticolour charges, so that they interact with themselves (a feature absent of \gls{qed} since photons carry no charges).
   \item At high energies, the interaction is weaker because the creation of many virtual particles implies an important screening of the charges (like the Debye screening in plasma physics). This is called the asymptotic freedom.
   \item At low energies, the interaction is stronger and quarks and gluons are confined in colour neutral states, namely hadrons. This is called the confined phase.
\end{myitem}

The \gls{qgp} is the state describing the asymptotic freedom, i.e.\ when there is a profusion of virtual particles created by the
strong interaction screening the colour charges of the quarks. The plasma is said to be ionised, in the sense of colour charge. When
the \gls{qgp} cools down, it ``crystallises'' into hadrons. In a simplified view, we can thus roughly say that quarks and
gluons exist in two phases: a ``solid'' phase, or hadron phase at low temperatures, and a ``liquid'' phase, namely the \gls{qgp}
at high temperatures. A simplified picture of the quark-gluon phase diagram is depicted on figure \ref{qcdphase}. The critical
temperature between the two phases is
\begin{equation}
   T_c \sim \SI{2e12}{K} \sim \SI{200}{MeV} \sim \SI{1}{fm},
\end{equation}
in the usual set of units used in literature. As a point of comparison, proto-neutron stars formed in supernovas reach
temperatures of $\sim 0.1 T_c$, but the \gls{lhc} can probe temperatures of order $\sim 5 T_c$.

\begin{myfig}
   \includegraphics[width = 0.49\textwidth]{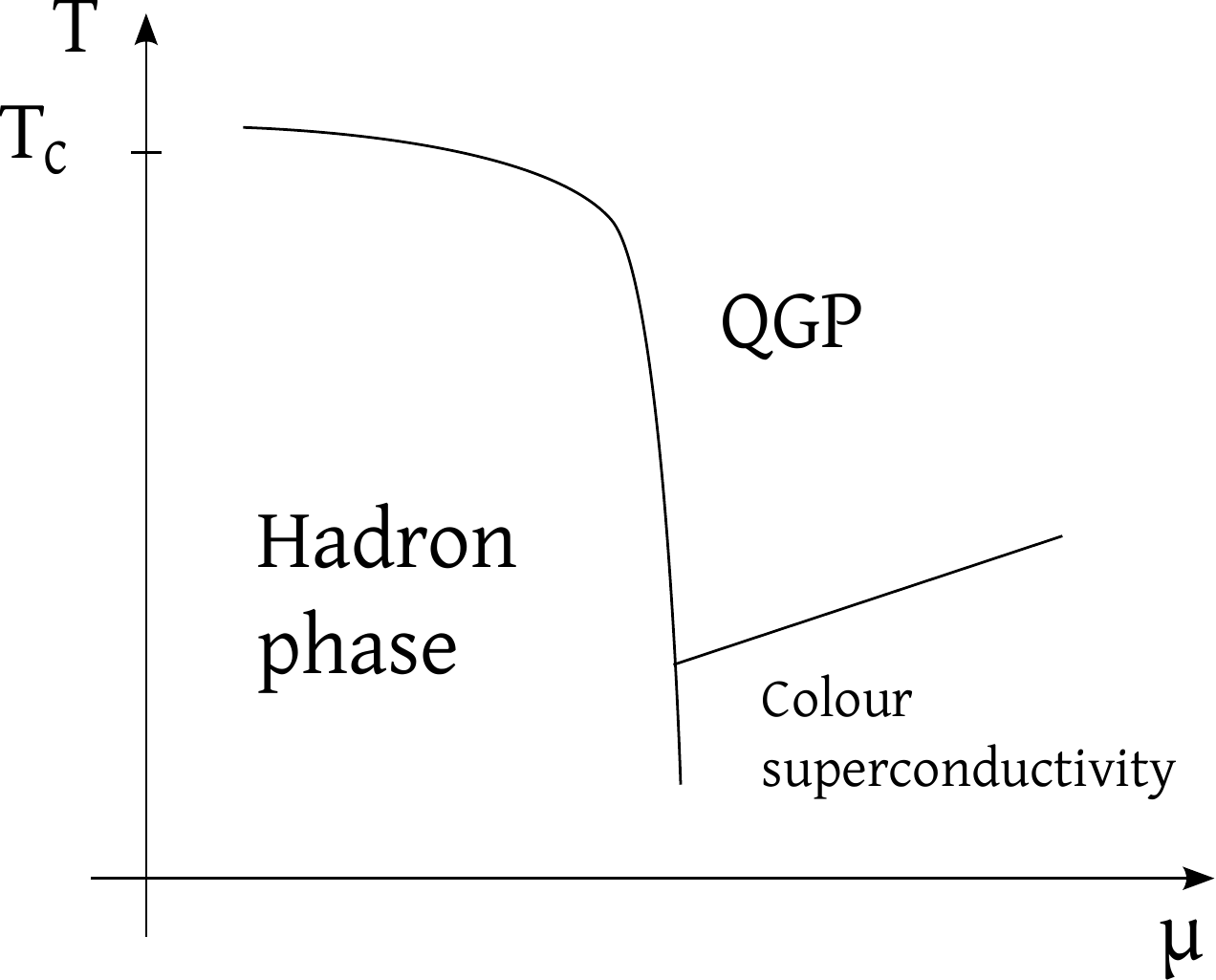}
   \caption[Phase diagram of quantum chromodynamics]{Simplified phase diagram of \gls{qcd}, in the $T$-$\mu$ plane, where $T$ is
   the temperature and $\mu$ the baryonic chemical potential. Below the critical temperature $T_c$, the hadron (or confined phase)
   is dominating while above $T_c$, the asymptotic freedom gives rise to a \gls{qgp}. Credits: \cite{Natsuume15}.}
   \label{qcdphase}
\end{myfig}

\subsection{The quark-gluon plasma}
\label{qgpsec}

The \gls{qgp} are thus present in the \gls{lhc}: when two hadrons collide each other, a \gls{qgp} is formed in the
overlapping region, as pictured in figure \ref{qgpcoll}. However, they are not directly observable. What is observed in colliders
is the hail of hadrons that form when the \gls{qgp} cools down. Since the \gls{qgp} has a non-zero viscosity and is almond-shaped
the flow of emitted particle is anisotropic, as can be seen on figure \ref{qgpcoll}. The number of particles emitted by the \gls{qgp}
between angles $\varphi$ and $\varphi + d\varphi$ is $dN$ and can be parametrised by
\begin{equation}
   \frac{dN}{d\varphi} = N(1 + 2 \nu_2 \cos(2\varphi) + \ldots).
\end{equation}
What is observable is the so-called elliptic flow $\nu_2$. Once it is measured, we can infer the properties of the \gls{qgp} via numerical
investigations. For example, lattice \gls{qcd} simulations helped by experimental data give the ratio of the shear viscosity $\eta$ to the entropy density $s$
in \gls{qgp}:
\begin{equation}
   \frac{\eta}{s} \simeq \frac{1.68}{4\pi} \quad \tn{at} \quad T = 1.65 T_c.
\end{equation}
This is typically the temperatures reached in the \gls{lhc}.

\begin{myfig}
   \includegraphics[width = 0.44\textwidth]{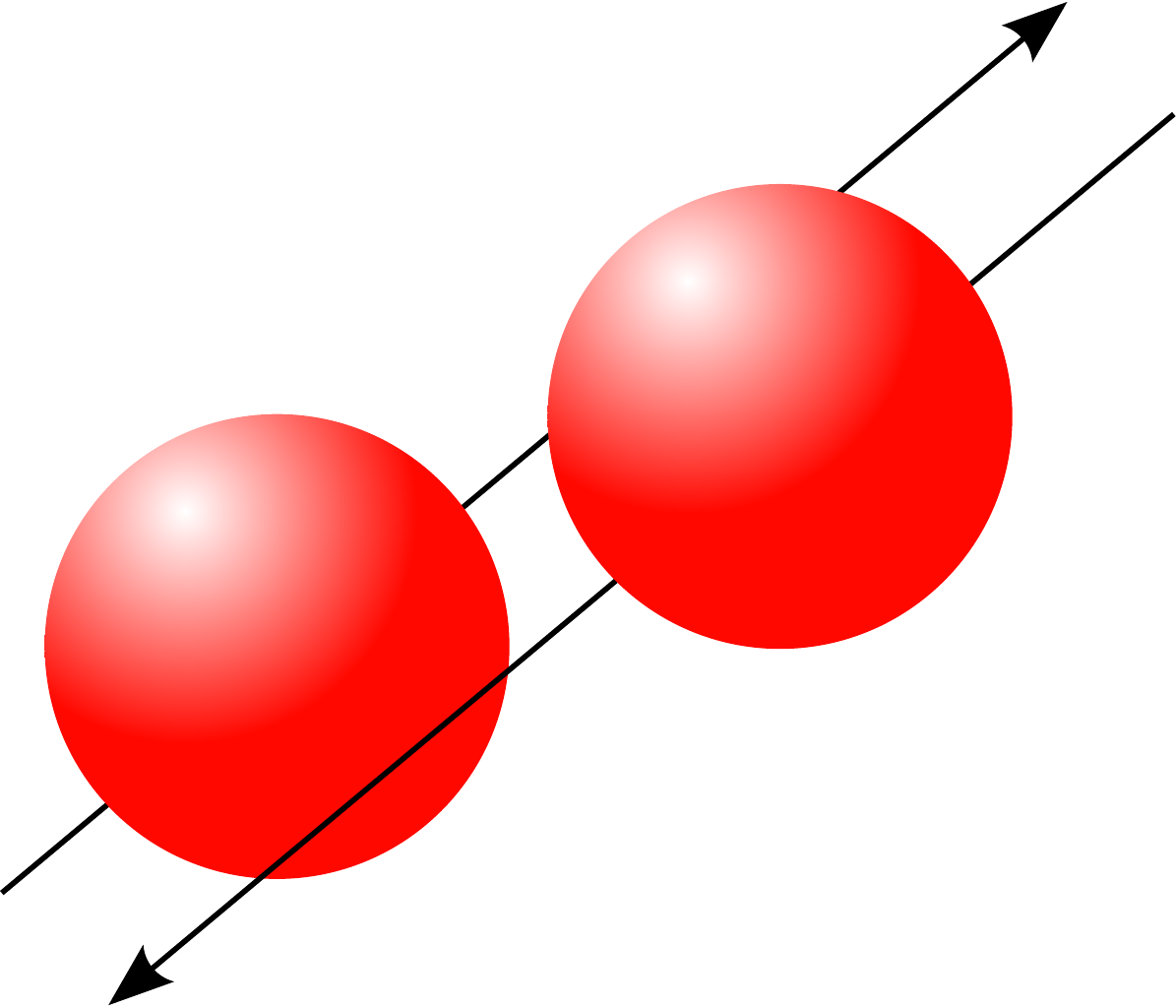}
   \includegraphics[width = 0.53\textwidth]{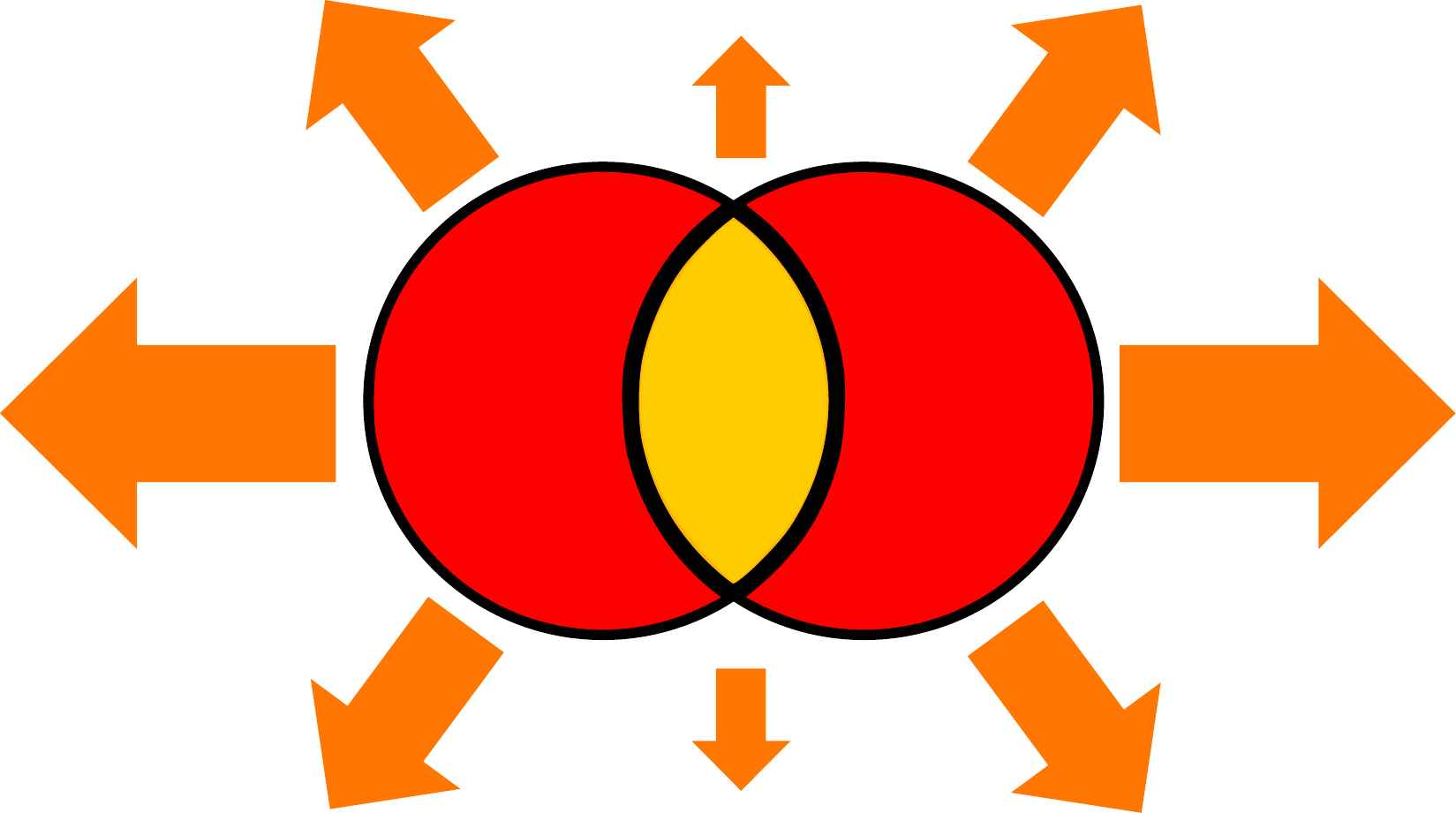}
   \caption[Quark-gluon plasma in hadron colliders]{Two hadrons (in red) collide off-centre. In the overlapping region (yellow), a
      \gls{qgp} is formed. The emission of particles ensuing is anisotropic (orange arrows) because of the non-zero viscosity of
      the plasma and its almond shape. Credits: \cite{Natsuume15}.}
   \label{qgpcoll}
\end{myfig}

The objective of the \gls{ads}-\gls{cft} correspondence is to provide a theoretical framework that can explain
such experimental values. Indeed, the \gls{qgp} is strongly coupled, a regime in which \gls{qcd} is untractable. However, in the
regime $T \gtrsim T_c$, \gls{qcd} is close to a supersymmetric conformal theory. The correspondence precisely deals with
this kind of simpler theories rather than \gls{qcd} directly.

\subsection{The large-$N_c$ limit}
\label{largeNc}

A standard approximation of \gls{qcd} consists in considering not three (red, green, blue) but $N_c$ different colours in
superconformal theories. There are now two parameters which are the coupling constant\footnote{The letters $YM$ stand for
Yang-Mills.} $g_{YM}$ and the number of colours $N_c$. The coupling constant appears as a factor $1/g_{YM}$ in front of gauge field
terms in the action. A useful parameter is the so-called 't Hooft coupling which is defined by
\begin{equation}
   \lambda \equiv g_{YM}^2 N_c.
\end{equation}
The pair of independent parameters ($\lambda$,$N_c$) is as much valid as the pair ($g_{YM}$,$N_c$). We detail in section
\ref{keyfeats} that the \gls{ads}-\gls{cft} duality deals with weak coupling on the gravity (string) theory side and strong coupling in the
gauge theory side. We are thus only interested in the strong coupling limit of superconformal theories. More precisely, we define
the large-$N_c$ limit as
\begin{equation}
   N_c \to \infty, \quad g_{YM} \to 0 \quad \tn{while keeping fixed} \quad \lambda \gg 1.
\end{equation}
The fact that $\lambda$ is fixed but very large is precisely a strong coupling limit. The large-$N_c$ limit of superconformal
theories is the relevant limit in the \gls{ads}-\gls{cft} correspondence.

\subsection{$\mathcal{N} = 4$ super Yang-Mills theory}
\label{n4sym}

Among conformal theories, \gls{nsym} plays a central role as it is present in the simplest manifestation of the duality,
namely the \gls{ads}$_5$-\gls{nsym} duality. For the purposes of this chapter, we only need a minimal knowledge of this
theory. Namely, it has $\mathcal{N} = 4$ supersymmetries, and  contains several bosons and fermions, among them:
\begin{myitem}
   \item 1 massless gauge field $A_\mu$,
   \item 6 massless scalar fields $\phi_i$,
   \item 4 massless Weyl fermions $\lambda_I$.
\end{myitem}

Since \gls{nsym} is supersymmetric, the theory features as many bosons as adjoint fermions.  Furthermore the gauge field has two
polarisations $A_\mu^+$ and $A_\mu^-$. There are thus 8 independent bosonic fields.  Each of these fields is an $N_c\times N_c$
matrix with a normalised trace. Thus, the number of bosonic degrees of freedom in the theory is
\begin{equation}
   N_B = 8(N_c^2 - 1).
   \label{NBF}
\end{equation}
Because of supersymmetry, each bosonic field has also an adjoint fermionic counterpart, so that the number of fermionic degrees of
freedom is $N_F = N_B$. Nothing much has to be known about this theory for the purposes of the present chapter.

\section{The AdS-CFT conjecture}

Basically, the conjecture relates to types of theories: quantum gravity and quantum field theories. The former denotes essentially
string theory, while the latter encompass several non-gravitational superconformal theories (said gauge theories). One of its most
famous illustration is between string theory in 5-dimensional \gls{ads} space-time and the \gls{nsym} theory in the large-$N_c$ limit
(defined earlier in section \ref{largeNc}). This chapter deals precisely with this duality. Conformal theories like \gls{nsym} can be used as
simple proxies of \gls{qcd}.

\subsection{Key features}
\label{keyfeats}

The \gls{ads}-\gls{cft} conjecture can be summed up as follows \cite{Hubeny15}:
\begin{myenum}
   \item there is a mapping between a quantum theory of gravity (string theory) and an ordinary non-gravitational quantum field
      theory (or gauge theory),
   \item \gls{ads}-\gls{cft} is a strong/weak coupling duality, i.e.\ when the gauge theory is strongly coupled and thus hard
      to solve, it is dual to a weakly coupled string theory, whose dominant contribution is \gls{gr} and easy to solve,
   \item the mapping is holographic, i.e.\ there is one more spatial dimension in the \gls{ads} side than in the \gls{cft} side.
\end{myenum}

Feature 1 motivates the equivalent ``gauge-gravity'' denomination. The idea of mapping is quite vague for now. We will see in
sections \ref{gkpwitt} and \ref{gkpwitt2} that this concept can be translated into one single equation, the so-called \gls{gkp}-Witten
relation.

Feature 2 is the reason why the correspondence is so much successful and interesting for the high-energy physics
community, as it allows to do complicated calculations in the \gls{cft} side by solving a simple problem in \gls{ads} space-time.
In this chapter, we never consider string theory beyond its leading order approximation that is \gls{gr}, and thus neglect the
so-called $\alpha'$ (also known as post-\gls{gr} or string) corrections. At this order of approximation, classical \gls{gr} is
(almost) all what is needed to perform \gls{ads}-\gls{cft} computations.

Feature 3 interprets the link between the two theories. Namely, the
\gls{cft} dual space-time is said to ``live'' on the \gls{ads} boundary, and the \gls{ads} space-time itself is often referred as the
bulk theory. This point is made clearer if we use the Fefferman-Graham coordinates (section \ref{fgcoor}).

The class of field theories that are conformal (\gls{cft}) do not describe reality, but can be seen as approximations of \gls{qcd}
in the strong coupling limit. They are an invaluable tool to understand the physics of strong interaction. In this sense,
\gls{ads}-\gls{cft} is like ideal hydrodynamics: even if it is an idealisation, it lays down the fundamentals of our comprehension
of the physics at hand.

\subsection{Fefferman-Graham coordinates}
\label{fgcoor}

The idea behind \gls{fg} coordinates is to put the bulk \gls{ads} metric into a standard form which makes the link with the
\gls{cft} straightforward. Namely, in the \gls{fg} coordinates, the metric reads
\begin{equation}
   ds^2 = \frac{1}{u^2}(g_{\mu\nu}dx^\mu dx^\nu + du^2),
   \label{fgequ}
\end{equation}
where $\mu$ and $\nu$ are 4-dimensional indices in our conventions and $u$ is a radial coordinate that is precisely zero on the
\gls{ads} boundary. This metric has a second order pole on the \gls{ads} boundary, consistently with definition
\ref{aadsdefinition} of chapter \ref{aads}. The 4-dimensional metric $g_{\alpha\beta}$ appearing in the 5-dimensional element is
precisely the metric in the \gls{cft} dual theory that ``lives'' on the boundary, i.e.\ in the limit $u\to 0$.

Since the dual \gls{cft} side is non-gravitational, the 4-dimensional metric has usually a background term
$\overline{g}_{\alpha\beta}$ such that
\begin{equation}
   g_{\alpha\beta} = \overline{g}_{\alpha\beta} + \ldots,
   \label{dots}
\end{equation}
and that just translates the geometry of the \gls{ads} boundary. It can be the metric of $\mathbb{R}\times
\mathcal{S}^3$ if the bulk \gls{ads} space-time is spherically symmetric, or the one of $\mathbb{R}\times\mathbb{R}^3$ in the case of planar \gls{ads}
black holes in Poincaré coordinates, to give a few examples. The dots in \eqref{dots} host information about the energy-momentum tensor of the dual quantum
system, as is demonstrated in \cite{Haro01}.

\subsection{Hawking temperature}
\label{HawkingT}

So far we have discussed the mapping between the gravitational theory in \gls{ads} and the gauge theory. But in practice, how does
it work? An intuitive answer is that, thanks to the laws of black hole thermodynamics that were uncovered in
\cite{Bekenstein73,Bardeen73,Hawking75}, we can simply identify the thermodynamical quantities of black holes in \gls{ads} with
those of the dual quantum system that lives on the boundary. This is actually a consequence of the \gls{gkp}-Witten relation that
we discuss later in section \ref{gkpwitt}. In section \ref{qgpeq}, we illustrate and demonstrate this statement on a practical example. For
now let us focus on temperature.

Since the discovery of Hawking radiation in \cite{Hawking75}, we know that black holes, because of the causal nature of the
horizon, can separate virtual particles resulting from quantum fluctuations of vacuum. Black holes thus emit a black-body
radiation with a certain temperature, the so-called Hawking temperature. A consequence of this phenomenon in the context of the
\gls{ads}-\gls{cft} conjecture is that if the bulk \gls{ads} space-time hosts a black hole with temperature $T$, the dual quantum
system is also at temperature $T$.  On the other hand, a bulk space-time with no black hole has an undefined, or zero temperature,
and the dual system can only be at zero temperature as well.

A simple procedure for determining a black hole temperature is the following. Suppose the black hole metric can be written in the
following form
\begin{equation}
   ds^2 = -f(r) dt^2 + \frac{dr^2}{f(r)} + \ldots,
   \label{fr}
\end{equation}
where $t$ is a time coordinate and $r$ a radial coordinate. If now we go to Euclidean time\footnote{This is also called a Wick
rotation of $\pi/2$ in the complex plane. The formalism of Euclidean (or imaginary) time is often used in quantum statistical
mechanics, where it can be shown that Green's function are periodic in the Euclidean time with a period that scales like the
inverse of the temperature. Euclidean time is a way of looking at the time dimension as if it were a space dimension: the
motion is possible forward or backward along Euclidean time, just like it is possible to move left and right in space.} $t_E =
it$, it comes
\begin{equation}
   ds^2 = f(r) dt_E^2 + \frac{dr^2}{f(r)} + \ldots.
\end{equation}
The horizon lies at the coordinate $r_0$ such that $f(r_0) = 0$. Near the horizon, we can thus approximate $f(r)$ by
\begin{equation}
   f(r) \simeq f'(r_0)(r - r_0),
\end{equation}
such that the \gls{nh} metric reads
\begin{equation}
   ds^2|_{NH} = \frac{dr^2}{f'(r_0)(r-r_0)} + f'(r_0)(r-r_0)dt_E^2 + \ldots.
\end{equation}
We can make this expression cleaner by a change of variables. Namely if we let
\begin{equation}
   \rho = 2\sqrt{\frac{(r-r_0)}{f'(r_0)}} \quad \tn{and} \quad d\rho = \frac{1}{\sqrt{(r-r_0)f'(r_0)}},
\end{equation}
the \gls{nh} metric becomes
\begin{equation}
   ds^2|_{NH} = d\rho^2 + \rho^2 d\left( \frac{f'(r_0)}{2}t_E \right)^2 + \ldots.
   \label{ds2nh}
\end{equation}
This now looks like a regular flat geometry in polar coordinates $d\rho^2 + \rho^2 d\theta^2$. However the polar angle
$\theta$ is $2\pi$-periodic with identification of the points $\theta = 0$ and $\theta = 2\pi$. If the parenthesis $f'(r_0)t_E/2$
of \eqref{ds2nh} is not $2\pi$-periodic, the metric exhibits what is called a conical singularity on the
horizon $\rho = 0$. Since it is well-known that in general the metric is not singular on the horizon, it means that the Euclidean time $t_E$ is
necessarily $\beta$-periodic with
\begin{equation}
   \beta = \frac{4\pi}{f'(r_0)}.
\end{equation}
In statistical mechanics of quantum systems, the periodicity of the Euclidean time is known to be related to the temperature $T$
in the canonical ensemble, this is the so-called \gls{kms} condition \cite{Kubo57,Martin59}. We thus define appropriately the proper temperature
of the black hole by
\begin{equation}
   T \equiv \frac{1}{\beta} = \frac{f'(r_0)}{4\pi}.
   \label{T}
\end{equation}
This definition agrees with the results of Hawking \cite{Hawking75} and can be identified with the temperature of the dual system.

\subsection{The GKP-Witten relation at equilibrium}
\label{gkpwitt}

The \gls{gkp}-Witten relation is the cornerstone of the \gls{ads}-\gls{cft} duality. Namely, it provides a relation between two
quantities: one deals only with the gravitational theory in \gls{ads}, while another characterises the conformal
theory. The correspondence is entirely encoded in this equal sign.

Let us denote by $Z$ the generating functional of a quantum field theory. It is  analogous to the partition function in classical
statistical mechanics and determines fully the theory. The \gls{gkp}-Witten relation basically states that, in the large-$N_c$
limit and at thermodynamical equilibrium, the generating functional of a string theory in bulk \gls{ads} is equal to the
generating functional of a \gls{cft} living on the boundary. Namely
\begin{equation}
   Z_{AdS} = Z_{CFT}.
\end{equation}
We admit this result because its discussion is much beyond the scope of the present manuscript. Let us just mention that the argument is based on the
similarity of the topology of Feynman diagrams in both string theory and \gls{qcd}, and that both \gls{ads} and \gls{cft} share
the conformal invariance $SO(2,4)$. The fact that the string theory is embedded in \gls{ads} space-time stems from conformal
invariance requirements on both side of the correspondence.

Looking for a quantum theory of gravity, it can be expected to be formulated in terms of a path integral
\begin{equation}
   Z = \int \mathcal{D}g e^{iS_L},
\end{equation}
where $g$ is the metric, $S_L$ is the Lorentzian action, and $\mathcal{D} g$ is the measure of the path integral. It is a concise
way to represent the infinite-dimensional integral over all possible configurations on all of space-time. We would like to make
the integral convergent, which is most easily achieved with an real exponential factor. We thus go to Euclidean time $t_E = it$
such that
\begin{equation}
   Z = \int \mathcal{D}g e^{-S_E},
\end{equation}
where $S_E$ is the Euclidean action, namely the gravitational action in the Euclidean formalism that we already used in section
\ref{HawkingT}. However, this integral is still diverging, which is one of the problems in unifying gravity with quantum field
theory. Nonetheless, even if not strictly legal, we want to approximate this integral by the saddle-point method. Namely, of all
the possible configurations of the metric $g$ in the path integral, we retain only the major contribution which is the one that
makes the action stationary. Thus
\begin{equation}
   Z \simeq e^{-S_E^\star},
   \label{pathintegral}
\end{equation}
where $S_E^\star$ is the on-shell Euclidean action, i.e.\ the Euclidean action evaluated precisely for solutions of the equations of
motion. This equation is only valid in the large-$N_c$ limit of section \ref{largeNc}. Thus we end up finally
with the very useful formula
\begin{equation}
   \lim_{N_c \gg \lambda \gg 1} Z_{CFT} = e^{-S_E^\star}.
   \label{gkpwitten}
\end{equation}
The main advantage of this relation is that the left-hand side is very difficult (if not impossible) to compute in the
conformal quantum field theory, while the right-hand side just needs evaluating the classical action of \gls{gr} (see
appendix \ref{leastaction}), plus potential $\alpha'$ corrections of string theory.

One remaining question is how to relate the fundamental constants on both sides of the duality. The gravitational side necessarily
contains the gravitational constant $\gls{G}$ and the \gls{ads} length \gls{L}, while the quantum side necessarily deals with
the number of colours $N_c$ and the 't Hooft coupling $\lambda$ (section \ref{largeNc}). If physical quantities like
temperature and entropy has to be identified on both sides of the correspondence, there must exist a dictionary that identify the two sets of fundamental
constants. This is called the \gls{ads}-\gls{cft} dictionary. In the particular case of the \gls{ads}$_5$-\gls{nsym} duality, it reads
\begin{equation}
   N_c^2 = \frac{\pi}{2}\frac{\gls{L}^3}{\gls{G}_5} \quad \tn{and} \quad \lambda = \left( \frac{\gls{L}}{l_s} \right)^4,
   \label{dictionary}
\end{equation}
where $\gls{G}_5$ is the gravitational constant in 5-dimensional gravity, and $l_s$ is the length of strings in string theory. Again,
we admit these relations. In this chapter, we only need the first relation for $N_c$.

\section{Equilibrium in the strong coupling limit}
\label{qgpeq}

As a first approach to \gls{ads}-\gls{cft} computations, the thermodynamical equilibrium in \gls{nsym} is among the simplest
computations that we can make. We thus consider a planar black hole in 5-dimensional \gls{ads}. The advantage of this example is
that the boundary geometry is $\mathbb{R}\times\mathbb{R}^3$, and is thus directly related to our real-world Minkowski space-time.
Furthermore, we can validate several concepts of the correspondence, like the identification of black hole thermodynamics with
thermodynamics of the dual system, as well as the \gls{gkp}-Witten relation \eqref{gkpwitten}.

\subsection{Planar horizons}
\label{planar}

Consider the 5-dimensional \gls{aads} planar black hole metric on the Poincaré patch of \gls{ads} (see equation
\eqref{adspoincare} of chapter \ref{aads})
\begin{equation}
   ds^2 = -\frac{r^2}{\gls{L}^2}\left( 1 - \frac{r_0^4}{r^4} \right)dt^2 + \frac{\gls{L}^2}{r^2}\frac{dr^2}{1-\dfrac{r_0^4}{r^4}} + \frac{r^2}{\gls{L}^2}(dx^2 + dy^2 + dz^2).
   \label{pbh}
\end{equation}
The radius $r_0$ denotes the limit of the planar horizon, that have an infinite extent in all three directions $(x,y,z)$. Notice
that this metric is invariant under the scaling symmetry ($\mu \in \{t,x,y,z\}$)
\begin{equation}
   x^\mu \to a x^\mu, \quad r \to \frac{r}{a}, \quad r_0 \to \frac{r_0}{a},
   \label{scalepbh}
\end{equation}
for any positive real factor $a$. This means that all planar horizons are equivalent under this rescaling. In particular, there is
no characteristic temperature. This means that the dual \gls{nsym} theory does not allow phase transitions in $\mathbb{R}^4$ and
describes the asymptotic freedom limit only.

\subsection{Thermodynamics of planar black holes}

The length element \eqref{pbh} is clearly of the form \eqref{fr}, so we can compute the planar black hole temperature with \eqref{T}. The
temperature of the planar black hole is thus
\begin{equation}
   T = \frac{r_0}{\pi \gls{L}^2}.
   \label{tpbh}
\end{equation}

According to the four laws of black hole thermodynamics \cite{Bekenstein73,Bardeen73,Hawking75}, the entropy of the black hole is
given by the area $A$ of the horizon via
\begin{equation}
   S = \frac{A}{4\gls{G}_5}.
\end{equation}
In the 5-dimensional planar black hole case, the area is a 3-dimensional volume that is infinite. It is thus more appropriate to
deal with a portion of the horizon $V_3 = L_x L_y L_z$ and to compute the entropy density, i.e.\ the entropy per horizon
area unit. Because the volume element on the horizon is $(r_0/\gls{L})^2(dx^2 + dy^2 + dz^2)$, the associated area is $A = (r_0/\gls{L})^3 V_3$,
so that the entropy density is
\begin{equation}
   s \equiv \frac{S}{V_3} = \frac{1}{4 \gls{G}_5}\left( \frac{r_0}{\gls{L}} \right)^3.
\end{equation}
If we use the \gls{ads}-\gls{cft} dictionary \eqref{dictionary} and make explicit the temperature dependence with \eqref{tpbh}, it
comes
\begin{equation}
   s = \frac{\pi^2}{2}N_c^2 T^3.
   \label{spbh}
\end{equation}
The $O(N_c^2)$ dependence of the entropy characterises a deconfined phase (confined phases have $s = O(N_c)$). Due to its scale
invariance, the dual \gls{nsym} cannot describe a confined phase.

From the second law of thermodynamics, we get the energy density
\begin{equation}
   de = Tds = \frac{3\pi^2}{2}N_c^2 T^3 dT \iff e = \frac{3\pi^2}{8}N_c^2 T^4.
\end{equation}
This formula reveals a $T^4$ dependence of the energy density. It is also known as the Stefan-Boltzmann law and is standard for
statistical systems at equilibrium. From Euler's law\footnote{It can be demonstrated as follows. From the extensivity of the
energy $E$, we get that for any real number $\lambda$, $E(\lambda S, \lambda V) = \lambda E(S,V)$, where $S$ is the entropy and
$V$ the thermodynamical volume. Derivating this equation with respect to $\lambda$, we get
\begin{equation}
   S\frac{\partial E}{\partial S}\bigg|_V + V \frac{\partial E}{\partial V}\bigg|_S = E \iff TS - PV = E,
\end{equation}
where we have used the thermodynamical definitions $T = \frac{\partial E}{\partial S}\big|_V$ and $P = -\frac{\partial
E}{\partial V}\big|_S$.} $e = Ts - P$, we get
\begin{equation}
   P = \frac{\pi^2}{8}N_c^2 T^4 \iff e = 3 P.
   \label{e3p}
\end{equation}
The $e = 3P$ equation translates the traceless character of the dual energy-momentum tensor. This is a property that is common to
all dual \gls{cft} and is due to the so-called local Weyl invariance. Finally, the free energy is by definition
\begin{equation}
   F = E - TS \iff F = -\frac{\pi^2}{8}V_3 N_c^2 T^4.
\end{equation}

\subsection{Comparison with the quantum perfect gas}

How far from the perfect gas case are these results? If the dual system were a perfect gas, its entropy density should be (see
appendix \ref{quantumgas})
\begin{equation}
   s_{PG}^{SYM} = \left(N_B + \frac{7}{8}N_F\right)\frac{2\pi^2}{45}T^3 = \frac{2\pi^2}{3}(N_c^2 - 1)T^3,
\end{equation}
where we have used the bosonic and fermionic degrees of freedom of the \gls{nsym} theory derived in section \ref{n4sym} equation
\eqref{NBF}. We thus find that in the large-$N_c$ limit, the entropy density of the dual system of the 5-dimensional \gls{aads}
planar black hole (equation \eqref{spbh}) is $(3/4)^{th}$ of the \gls{nsym} perfect gas limit:
\begin{equation}
   s = \frac{3}{4}s_{PG}^{SYM}.
\end{equation}
At this point, it should be reminded that the large-$N_c$ limit is a strong coupling limit, not a perfect gas limit. The above
formula is thus a \textit{prediction} of the behaviour of the \gls{nsym} theory in a strong coupling limit. This result was obtained rather
simply with the \gls{ads}-\gls{cft} correspondence, whereas it would have been untractable in the \gls{nsym} theory. It
is also in agreement with numerical simulations. The simplicity of the above calculations is one of the reasons why the
correspondence is so much popular.

\subsection{Euclidean action of planar black holes}

So far, we have identified the thermodynamical quantities of black holes thermodynamics in \gls{ads} with those of the \gls{cft}
dual, without demonstration. In this section, we want to truly justify this identification from first principles, namely from the
\gls{gkp}-Witten relation \eqref{gkpwitten} of the \gls{ads}-\gls{cft} correspondence. We switch to Euclidean formalism $t_E = it$
and use the variables
\begin{equation}
   u = \frac{r_0}{r} \quad \tn{and} \quad h = 1 - u^4,
\end{equation}
so that the Euclidean metric reads
\begin{equation}
   ds^2 = \frac{r_0^2}{\gls{L}^2 u^2}(hdt_E^2 + dx^2 + dy^2 + dz^2) + \frac{\gls{L}^2}{hu^2}du^2.
   \label{pbhu}
\end{equation}
Note that the \gls{ads} boundary lies now at $u=0$ and the horizon at $u = 1$. Moreover, as in Euclidean formalism the signature
of the metric must be $(+++++)$, the region where $h < 0$ is forbidden: the Euclidean space-time simply terminates at $u = 1$.

Our goal is to compute the Euclidean action that is required by the \gls{gkp}-Witten relation. As explained in appendix
\ref{leastaction}, the action has three contributions: the \gls{eh} term, the \gls{ghy} term, and the counter-term. From the
change of variable $t_E = i t$, the link between the Lorentzian $S_L$ and the Euclidean action $S_E$ is
\begin{equation}
   i S_L = - S_E,
\end{equation}
or equivalently $\sqrt{-g}\to -\sqrt{g}$.

\paragraph{The Einstein-Hilbert term} It is given by
\begin{equation}
   S_{EH} = -\frac{1}{16\pi \gls{G}_5}\int (R - 2\gls{Lambda})\sqrt{g}d^5x.
\end{equation}
With the formulas \eqref{Rmaxsym2} and \eqref{lambdaL} of chapter \ref{aads}, it comes $R - 2\Lambda = -8/\gls{L}^2$. Furthermore,
from the metric \eqref{pbhu}, we find $\sqrt{g} = r_0^4/\gls{L}^3 u^5$. Since the Euclidean time is $\beta$-periodic, we integrate
from $t = 0$ to $t = \beta$. The horizon having an infinite spatial extension, we restrict the integral to a finite volume $V_3$
in the $(x,y,z)$ directions. Finally, we learn in appendix \ref{leastaction} that the integral may temporarily diverge near the
\gls{ads} boundary, so that we run the integration down to $u = \varepsilon \ll 1$ but not 0, at least momentarily. These
divergences cancel each other in the total action. All these arguments taken into account, we end up with the on-shell result
\begin{equation}
   S_{EH}^\star = \frac{1}{2\pi \gls{G}_5}\frac{r_0^4}{\gls{L}^5}\int_0^\beta dt \int_{V_3} d^3x \int_\varepsilon^1
   \frac{du}{u^5} = \frac{\beta V_3}{8\pi \gls{G}_5}\frac{r_0^4}{\gls{L}^5}\left( \frac{1}{\varepsilon^4}-1 \right).
   \label{seh}
\end{equation}

\paragraph{The \gls{ghy} term} It is given by (see appendix \ref{leastaction} equation \eqref{sghy})
\begin{equation}
   S_{GHY} = \frac{1}{8\pi \gls{G}_5}\int_{\partial \mathcal{V}} \Theta \sqrt{q}d^4x,
\end{equation}
where $\partial \mathcal{V}$ is the \gls{ads} boundary $u = \varepsilon \to 0$, $q$ is the determinant of the induced metric
$q_{AB} = g_{AB} - n_{A}n_{B}$ on $\partial \mathcal{V}$, $n_A$ is the unit normal vector to $\partial \mathcal{V}$ and $\Theta$
is the mean extrinsic curvature of $\partial \mathcal{V}$. The expression of $n_A$ is very simple in the coordinates $(t,x,y,z,u)$:
\begin{equation}
   n^A = \left( 0,0,0,0,-\frac{1}{\sqrt{g_{uu}}} \right).
   \label{nM}
\end{equation}
This ensures the normalisation $g_{MN}n^Mn^N = 1$. The minus sign in \eqref{nM} is chosen so as to make $n^A$ pointing toward
increasing $r$ coordinates and thus decreasing $u$. The mean extrinsic curvature $\Theta$ is the trace of the extrinsic curvature
tensor $\Theta_{\alpha\beta}$ which is half of the Lie derivative of $q_{\alpha\beta}$ along $n^A$ (see appendix \ref{d+1}
equation \eqref{evolgamma}):
\begin{equation}
   \Theta_{\alpha\beta} = -\frac{1}{2}\left( n^M \partial_M q_{\alpha\beta} + q_{\alpha M}\partial^M n_\beta + q_{\beta M} \partial^M n_\alpha\right).
\end{equation}
In this equation, the last two terms are zero since the metric is diagonal and $n^\mu = 0$. Taking the trace, we find
\begin{equation}
   \Theta = -\frac{1}{2}n^u q^{\mu\nu}\partial_u q_{\mu\nu} = -\frac{1}{2}n^u \partial_u \ln(q) = -\frac{n^u}{\sqrt{q}}\partial_u \sqrt{|q|}.
   \label{Thetanu}
\end{equation}
According to the Euclidean metric \eqref{pbhu}, we find
\begin{equation}
   n^u = -\frac{u \sqrt{h}}{\gls{L}}, \quad \sqrt{q} = \frac{r_0^4 \sqrt{h}}{\gls{L}^4 u^4}, \quad \Theta = \frac{2u^4-4}{\gls{L}\sqrt{h}}.
   \label{nusqth}
\end{equation}
Collecting all these results, we get the on-shell result (note that the integral just brings a factor $\beta V_3$)
\begin{equation}
   S_{GHY}^{\varepsilon\star} = \frac{\beta V_3}{8\pi \gls{G}_5}\lim_{u \to \varepsilon}\left(\sqrt{q}\Theta\right) = \frac{\beta
   V_3}{8\pi \gls{G}_5}\frac{r_0^4}{\gls{L}^5}\left( 2 - \frac{4}{\varepsilon^4} \right).
   \label{sghyeps}
\end{equation}

\paragraph{The counter-term} In the \gls{ads}-\gls{cft} context, adding a counter-term to the action (see appendix
\ref{leastaction} equation \eqref{lctd}) is called holographic renormalisation. In five dimensions, it is
\begin{equation}
   S_{CT} = \frac{1}{8\pi \gls{G}_5}\int_{\partial \mathcal{V}} \frac{3}{\gls{L}}\left(1 + \frac{\gls{L}^2}{12}\mathcal{R} \right)\sqrt{q}d^4x,
   \label{footn}
\end{equation}
where $\mathcal{R}$ is the Ricci scalar curvature of $q_{\alpha\beta}$. However, for planar black holes, a calculation shows that
$\mathcal{R}$ is zero\footnote{Indeed, from \eqref{pbh}, the induced metric on the $r = cst$ hypersurfaces is $ds^2|_{r = cst}
\sim -\tn{cst}_1 dt^2 +
\tn{cst}_2 (dx^2 + dy^2 + dz^2)$. By a suitable rescaling of the variables, it can be brought to the Minkowski metric which has zero
curvature.}. Thus, with the help of \eqref{nusqth}, follows the on-shell result
\begin{equation}
   S_{CT}^{\varepsilon\star} = \frac{3\beta V_3}{8\pi \gls{G}_5 \gls{L}}\lim_{u \to \varepsilon}\sqrt{q} = \frac{3\beta V_3}{8\pi \gls{G}_5}\frac{r_0^4}{\gls{L}^5} \left( \frac{1}{\varepsilon^4} - \frac{1}{2}\right).
   \label{scteps}
\end{equation}

\paragraph{Thermodynamical quantities} Combining \eqref{seh}, \eqref{sghyeps}, \eqref{scteps}, all diverging terms in the
$\varepsilon \to 0$ limit vanish, and the total on-shell Euclidean action is
\begin{equation}
   S_E^\star = -\frac{\beta V_3}{16\pi \gls{G}_5}\frac{r_0^4}{\gls{L}^5}.
\end{equation}
From the \gls{gkp}-Witten relation \eqref{gkpwitten}, we obtain the partition function of the dual system by
\begin{equation}
   Z = e^{-S_E^\star}.
\end{equation}
Regarding the quantum dual system, statistical mechanics teaches that the free energy is nothing but
\begin{equation}
   F \equiv -\frac{1}{\beta}\ln Z = \frac{1}{\beta}S_E^\star = -\frac{V_3}{16\pi \gls{G}_5}\frac{r_0^4}{\gls{L}^5}.
   \label{F}
\end{equation}
Switching from $r_0$ to the Hawking temperature $T$ with \eqref{tpbh}, and using the \gls{ads}-\gls{cft} dictionary \eqref{dictionary}, it comes
\begin{equation}
   F = -\frac{\pi^2}{8}V_3 N_c^2 T^4.
\end{equation}
At this point, the temperature is the Hawking temperature of the black hole, not the one of the dual system. However, the $T^4$ dependence of the free energy is
so much reminiscent of the Stefan-Boltzmann law that it is very natural to identify the black hole temperature with the one of the dual
system. Then, with the usual thermodynamical relations
\begin{equation}
   s = -\frac{1}{V_3}\frac{\partial F}{\partial T}, \quad P = -\frac{\partial F}{\partial V_3}, \quad e = \frac{F}{V_3} + Ts,
\end{equation}
we recover the previous results of section \ref{planar}
\begin{equation}
   s = \frac{\pi^2}{2} N_c^2 T^3, \quad P = \frac{\pi^2}{8}N_c^2 T^4, \quad e = \frac{3\pi^2}{8}N_c^2 T^4.
\end{equation}
Thus we were perfectly right when identifying the black hole entropy with the dual system entropy in section \ref{planar}: it is a
consequence of the \gls{gkp}-Witten relation, that is the backbone of the \gls{ads}-\gls{cft} correspondence.

\section{Hawking-Page phase transition}
\label{hawkingpage}

The second application of \gls{ads}-\gls{cft} we consider is the \gls{hp} phase transition. Namely, it states that equilibrium
configurations are not always dual to black holes in the gravity side, but can also be described by the thermal \gls{ads} metric if the
temperature is below some critical value. This time, the geometry of the boundary is not $\mathbb{R}\times\mathbb{R}^3$ as in the \gls{qgp} case
but $\mathbb{R}\times \mathcal{S}^3$, where $\mathcal{S}^3$ is the 3-sphere. The computation proceeds as usual, namely with the
computation of the on-shell action and with identification of the thermodynamical quantities.

\subsection{The Schwarzschild-AdS black hole}

We consider the Euclidean 5-dimensional \gls{sads} metric in static coordinates (defined in equation \eqref{adsstatic} of chapter
\ref{aads})
\begin{equation}
   ds^2 = f(r)dt^2 + \frac{dr^2}{f(r)} + r^2 d\Omega_3 \quad \tn{with} \quad f(r) = 1 + \frac{r^2}{\gls{L}^2} - \frac{r_0^4}{\gls{L}^2r^2},
\end{equation}
where $d\Omega_3 = d\psi^2 + \sin^2\psi(d\theta^2 + \sin^2\theta d\varphi^2)$ is the unit 3-sphere length element ($\psi,\theta
\in [0,\pi]$ and $\varphi \in [0,2\pi[$). The horizon
lies at $r_+$ such that
\begin{equation}
   f(r_+) = 0 \iff r_+^4 + \gls{L}^2 r_+^2 - r_0^4 = 0.
   \label{r0}
\end{equation}

Using the definition \eqref{T} of Hawking temperature of the black hole, we find that it obeys
\begin{equation}
   T = \frac{2r_+^2 + \gls{L}^2}{2\pi r_+ \gls{L}^2}.
\end{equation}
This function is actually bounded below and a computation shows that
\begin{equation}
   \frac{d T}{dr_+} = 0 \iff r_+ = \frac{\gls{L}}{\sqrt{2}} \iff T = T_1 \equiv \frac{\sqrt{2}}{\pi \gls{L}}.
   \label{defT1}
\end{equation}
Thus the temperature of the black hole is constrained by $T \geq T_1$. No black hole with smaller than $T_1$ temperature can exist.
Furthermore, for $T \geq T_1$, two black holes solutions are possible, since
\begin{equation}
   r_+ = \frac{\pi \gls{L}^2 T}{2}\left( 1 \pm \sqrt{1 - \frac{T_1^2}{T^2}} \right).
   \label{bigsmall}
\end{equation}
Depending on the sign choice in this equation, there are two distinct solutions for a given temperature $T$: a small black hole or
a large black hole.

\subsection{Free energy}

In order to compute thermodynamical quantities, we first compute the Euclidean on-shell action and use the \gls{gkp}-Witten
relation at equilibrium \eqref{gkpwitten}. As usual, the action is made of three terms

\paragraph{The bulk action} The bulk (or \gls{eh}) action is

\begin{equation}
   S_{bulk} = -\frac{1}{16\pi \gls{G}_5}\int (R - 2\gls{Lambda})\sqrt{g}d^5x.
\end{equation}
In 5-dimensional vacuum \gls{ads} space-time, $R - 2\gls{Lambda} = -8/\gls{L}^2$, and $\sqrt{g} = r^3\sin^2\psi\sin\theta$, such that
the on-shell result is
\begin{equation}
   S_{bulk}^\star = \lim_{r\to\infty}\frac{\pi \beta}{\gls{G}_5 \gls{L}^2}\int_{r_+}^r r^3dr = \frac{\pi\beta}{4 \gls{G}_5
   \gls{L}^2}(r^4 - r_+^4).
   \label{sbulksads}
\end{equation}

\paragraph{The \gls{ghy} term} It is given by
\begin{equation}
   S_{GHY} = \frac{1}{8\pi \gls{G}_5}\int_{\partial \mathcal{V}} \sqrt{q}\Theta d^4x.
\end{equation}
In the same spirit of section \ref{qgpeq}, we have
\begin{equation}
   n^r = \sqrt{f}, \quad \sqrt{q} = \sqrt{f}r^3\sin^2\psi\sin\theta, \quad \Theta = -\frac{n^r}{\sqrt{q}}\partial_r\sqrt{q},
\end{equation}
and the on-shell computation gives
\begin{equation}
   S_{GHY}^\star = -\lim_{r\to\infty}\frac{\pi\beta}{8 \gls{G}_5}(r^3 f' + 6r^2f) = -\lim_{r\to\infty}\frac{\pi\beta}{4 \gls{G}_5\gls{L}^2}(4r^4 + 3r^2 \gls{L}^2 - 2r_0^4).
   \label{sghysads}
\end{equation}

\paragraph{The counter-term} It is given by (see equation \eqref{lctd} of the appendix \ref{leastaction})
\begin{equation}
   S_{CT} = \frac{3}{8\pi \gls{G}_5 \gls{L}}\int_{\partial \mathcal{V}} \left( 1 + \frac{\gls{L}^2}{12}\mathcal{R} \right)\sqrt{q}d^4x.
\end{equation}
The Ricci tensor $\mathcal{R}$ of the induced metric $q_{\alpha\beta}$ on $\mathcal{\partial \mathcal{V}}$ is nothing but the Ricci tensor of the 3-sphere
$\mathcal{R} = 6/r^2$, such that the on-shell counter-term is
\begin{equation}
   S_{CT}^\star = \lim_{r\to\infty}\frac{3\pi\beta r^4}{4 \gls{G}_5 \gls{L}^2}\sqrt{1 + \frac{\gls{L}^2}{r^2} - \frac{r_0^4}{r^4}}\left( 1 +
   \frac{\gls{L}^2}{2r^2} \right) = \lim_{r\to\infty}\frac{3\pi\beta}{4 \gls{G}_5 \gls{L}^2}\left( r^4 + r^2 \gls{L}^2 +
   \frac{\gls{L}^4}{8} - \frac{r_0^4}{2} \right).
   \label{sctsads}
\end{equation}

\paragraph{The total action} Combining \eqref{sbulksads}, \eqref{sghysads} and \eqref{sctsads}, and trading $r_0$ for $r_+$ with
\eqref{r0}, we get for the total on-shell Euclidean action
\begin{equation}
   S_E^\star = \frac{\pi\beta}{8 \gls{G}_5 \gls{L}^2}\left[ \frac{3 \gls{L}^4}{4} + \gls{L}^2 r_+^2 - r_+^4 \right].
\end{equation}
Notice that all diverging terms have vanished, as explained in appendix \ref{leastaction}. The free energy, given by
$S_E^\star/\beta$ as in section \ref{qgpeq} equation \eqref{F}, is thus
\begin{equation}
   F = \frac{\pi}{8 \gls{G}_5 \gls{L}^2}\left[ \frac{3 \gls{L}^4}{4} + \gls{L}^2 r_+^2 - r_+^4 \right].
   \label{Fsads}
\end{equation}
Rigorously speaking, we should have two different contributions for the small and the large black hole (see equation \eqref{bigsmall}).
However, it can be demonstrated that the free energy of the small black hole is always the largest. Indeed, the difference
of free energies between a large black hole (LBH) and a small black hole (SBH) at the same temperature $T$ is
\begin{equation}
   F^{LBH} - F^{SBH} = \pi^2 \gls{L}^6 T^2 \sqrt{1-\frac{T_1^2}{T^2}}\left[ 1 - \pi^2 \gls{L}^2\left( T^2 -
      \frac{T_1^2}{2} \right) \right],
\end{equation}
but $T \geq T_1$ and $T_1 = \sqrt{2}/\pi \gls{L}$, so that
\begin{equation}
   T^2 - \frac{T_1^2}{2} \geq \frac{T_1^2}{2} = \frac{1}{\pi^2 \gls{L}^2}.
\end{equation}
This establishes that $F^{LBH}$ is always smaller than $F^{SBH}$. According to the semi-classical approximation of the
gravitational path integral \eqref{pathintegral}, the free energy is dominated by whichever term is bigger. So from now on, we
neglect the small black hole contribution and take \eqref{Fsads} for granted, where $r_+$ is considered a function of the
temperature given by the choice of the plus sign in \eqref{bigsmall}.

\subsection{Casimir energy}
\label{casimirsec}

One striking feature of \eqref{Fsads} is that the free energy is not vanishing when the radius of the black hole shrinks to zero.
In the $r_+\to0$ limit, the thermodynamical energy of pure 5-dimensional \gls{ads} is thus
\begin{equation}
   E = \frac{3\pi \gls{L}^2}{32 \gls{G}_5} = \frac{3 N_c^2}{16 \gls{L}},
\end{equation}
where we have used the \gls{ads}-\gls{cft} dictionary \eqref{dictionary} to get the second equality. This is precisely the
discrepancy between the \gls{bk} and \gls{amd} mass in five dimensions (equation \eqref{casimir1}). Note that in the free fields
limit of \gls{nsym} on $\mathbb{S}\times \mathbb{R}^3$, the Casimir energy is
\begin{equation}
   E_{Casimir} = \frac{3(N_c^2 - 1)}{16 \gls{L}}.
\end{equation}
Thus, the energy (or mass) of the vacuum 5-dimensional \gls{ads} space-time is not zero but corresponds precisely to this Casimir energy in
the large-$N_c$ limit, revealing another striking connection between gravity in \gls{ads} and conformal theories. Notice that this
time, it is a result coming from quantum field theory (the Casimir energy) that sheds light on a very counter-intuitive aspect of
pure gravity (the non-vanishing mass of 5-dimensional vacuum \gls{ads} space-time).

\subsection{Thermodynamical stability}

In order to discuss the thermodynamical stability of \gls{sads} black holes, let us consider the difference between the
free energies of the \gls{sads} black hole and the pure vacuum \gls{ads} metric (see equation \eqref{Fsads})
\begin{equation}
   \Delta F \equiv F_{SAdS} - F_{AdS} = \frac{\pi r_+^2}{8 \gls{G}_5 \gls{L}^2}(\gls{L}^2  - r_+^2).
\end{equation}
There is a critical radius $r_+$ for which $\Delta F = 0$. This corresponds to
\begin{equation}
   r_+ = \gls{L} \iff T = T_2 \equiv \frac{3}{2\pi \gls{L}},
\end{equation}
where in light of equation \eqref{defT1}, $T_2 > T_1$. We can thus discuss the so-called \gls{hp} phase transition originally
uncovered in \cite{Hawking82}:
\begin{myitem}
   \item If $T < T_1$, no black hole can exist, the equilibrium is a thermal gas in \gls{ads}.
   \item If $T_1 < T < T_2$, then $\Delta F > 0$, this means that a meta-stable black hole can form but it is not the minimum of the free
      energy and thus evaporates via Hawking radiation to reach the thermal \gls{ads} state of equilibrium and thus $\Delta F
      = 0$. Said differently, the black hole evaporates too quickly and is gone before its radiation had the time to bounce off the
      \gls{ads} boundary and maintain a non-vanishing horizon.
   \item If $T > T_2$, then $\Delta F \leq 0$ the black hole solution does minimise the free energy and is a thermodynamical
      equilibrium.
\end{myitem}
The \gls{hp} phase transition happens for $T = T_2$ when $\Delta F$ flips sign. It is a first order phase transition since $F$
is continuous but not its first derivative across the transition.

We have thus recovered that, unlike the planar black holes of section \ref{planar}, spherical black holes in \gls{ads} can describe phase transitions. This
is very reminiscent of the hadron-\gls{qgp} phase transition of \gls{qcd}. However, the analogy is not complete. In the case of
\gls{sads} black holes,
the phase transition in the \gls{nsym} theory takes place on the 3-sphere $\mathcal{S}^3$ whereas it occurs in $\mathbb{R}^3$ in
\gls{qcd} and comes from the dynamics in a very complicated way. \glspl{cft} cannot have confining phases in $\mathbb{R}^3$, so that
\gls{nsym} is not a confining gauge theory in the same sense as \gls{qcd}.

\section{Viscosity of quark-gluon plasma}
\label{viscoqgp}

Last but not least, the third application of the correspondence we study is the shear viscosity of \gls{qgp}. As we have seen in section
\ref{qgpsec}, the ratio of the shear viscosity $\eta$ to the entropy density $s$ of \glspl{qgp} can be inferred by combining
measurements of elliptic flows in hadron colliders with heavy numerical simulations. As strange as it may seem, the \gls{ads}-\gls{cft}
duality gives a theoretical framework that allows to explain and predict this value of $\eta/s$ with the help of \gls{gr}
in \gls{ads} space-time. Basically, since quasi-normal modes of an event horizon are always damped, black holes can be said
dissipative. Within the correspondence, this feature is dual to the viscosity of a \gls{qgp}, as illustrated on figure \ref{dualvisco}. In
this section, we show how the correspondence can be used to deduce the ratio $\eta/s$ of the shear viscosity to the entropy
density. The technicality of the argument is more involved than in the equilibrium case of section \ref{qgpeq}, so let us sketch
briefly the computational strategy.

\begin{myfig}
   \includegraphics[width = 0.49\textwidth]{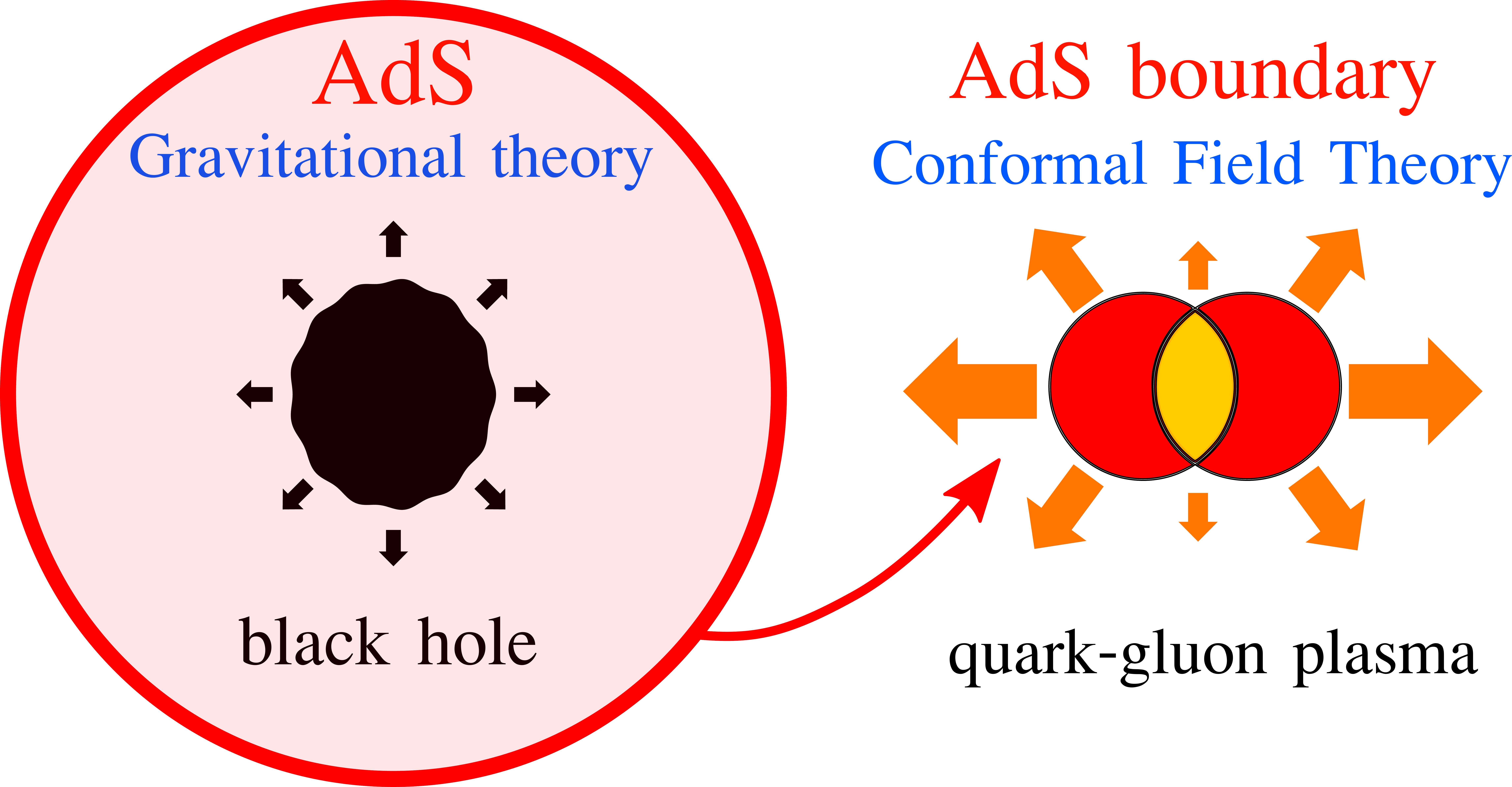}
   \caption[Viscosity of QGPs from black hole quasi-normal modes in AdS space-times]{Studying the quasi-normal modes of a black
      hole in \gls{aads} space-times (left) amounts to computing the viscosity of a \gls{qgp} described by a \gls{cft} on the
      \gls{ads} boundary (right). Credits: G. Martinon.}
   \label{dualvisco}
\end{myfig}

We first generalise the \gls{gkp}-Witten relation to non-equilibrium situations. Indeed, the viscosity of a fluid is essentially a
dynamical quantity and the equilibrium computations of the previous sections are not powerful enough to study it. Secondly, we focus on
the expression of the energy-momentum tensor of a viscous fluid, that is relevant for the \gls{qgp}. Thirdly, we derive the Kubo
formula that links a perturbation in the bulk \gls{ads} space-time to the shear viscosity of a dual quantum system on the \gls{ads}
boundary.

After these quite general considerations, we focus on a particular application. Namely, we perturb the planar black hole
background with a metric perturbation $h_{xy}$. We then solve the equations of \gls{gr} for this perturbation. The boundary
conditions on the horizon must translate that nothing can escape out of it, and the boundary value of the perturbation is left
undetermined and parametrised by $h_{xy}^{(0)}$. This function $h_{xy}^{(0)}$ is precisely needed in both the out-of-equilibrium
\gls{gkp}-Witten relation and the Kubo formula. Combining these two equations together allows to find $T^{xy}$, the $xy$ component of
the energy-momentum tensor of the perturbed \gls{qgp} and therefore its shear viscosity.

The striking result of this procedure is that it gives results that are very close to the combined experimental-numerical
determination described in section \ref{qgpsec}, whereas obtaining this result in \gls{qcd} is untractable. We discuss the
implications of such a result at the end of this section.

\subsection{GKP-Witten relation out of equilibrium}
\label{gkpwitt2}

When the dual system is perturbed, the \gls{gkp}-Witten relation \eqref{gkpwitten} is not valid any more
as the partition function is undefined out-of-equilibrium. However, the perturbation can be described by an operator
$\mathcal{O}$ on the quantum side, and by a metric (or field) perturbation in the gravity side. Thus, the \gls{gkp}-Witten
relation generalises to
\begin{equation}
   \left\langle \exp\left(i\int \phi^{(0)}\mathcal{O}\right)\right\rangle = \exp(i S^\star[\phi|_{\partial \mathcal{V}} = \phi^{(0)}]).
\end{equation}
The left-hand side is the generating functional of the dual theory with a source term $\phi^{(0)}$ generated by an operator
$\mathcal{O}$. The right-hand side is just the evaluation of the on-shell action when the bulk perturbation has a boundary value
$\phi^{(0)}$. We are not interested directly this relation but rather in its consequence, that is
\begin{equation}
   \langle \mathcal{O} \rangle = \frac{\delta S[\phi^{(0)}]}{\delta \phi^{(0)}}.
   \label{angleO}
\end{equation}
The left-hand side $\langle \mathcal{O}\rangle$ is the expectation value of an operator on the \gls{cft} side, for example one component
of the energy-momentum tensor. The right-hand side is merely the functional derivative of the boundary action with respect to the
boundary perturbation $\phi^{(0)}$. From the 4-dimensional point of view, $\phi^{(0)}$ is an external source, but from the 5-dimensional
point of view, $\phi$ is a field propagating in the bulk space-time. Namely, the \gls{ads}-\gls{cft} correspondence claims that an
external source of the gauge theory can have a 5-dimensional origin. Said differently, a bulk field acts as an external source for
a boundary operator.

This gives the following \gls{ads}-\gls{cft} recipe:
\begin{myenum}
   \item consider a perturbation $\phi$ of \gls{ads} with boundary value $\phi^{(0)}$,
   \item solve the equations of motion for $\phi$ in the bulk \gls{ads} space-time,
   \item compute the on-shell action as a functional of $\phi^{(0)}$,
   \item deduce the expectation value of the associated operator $\mathcal{O}$ with \eqref{angleO}.
\end{myenum}
Namely, hereafter we consider a perturbation $\phi = h_{xy}$ and deduce the expectation value of the boundary energy-momentum tensor $\langle
T^{xy}\rangle$. Combined with the Kubo formula discussed in section \ref{kuboformula}, it allows us to compute the shear viscosity of the \gls{qgp}. The main advantage
of this framework is that all first three steps just require \gls{gr}. The last step, even if motivated by
quantum considerations, needs no more knowledge than equation \eqref{angleO}.

At this point, let us mention that the out-of-equilibrium \gls{gkp}-Witten relation should better deal with the Euclidean action,
as in section \ref{gkpwitt} discussing equilibrium. Hereafter, we keep using the Lorentzian action as it allows us to make computations simpler by using
Fourier transforms. It turns out that, for the humble purposes of the present manuscript, the results obtained are the right ones, thanks
to the Lorentzian prescription \cite{Natsuume15}.

\subsection{Energy-momentum tensor of a viscous fluid}

A standard approach to describe the energy-momentum tensor of the \gls{qgp} is to start from the perfect fluid
approximation and perform an expansion in terms of the derivatives of the fluid 4-velocity $u^\alpha$. The first order
hydrodynamics are thus described by an energy-momentum tensor of the form
\begin{equation}
   T^{\alpha\beta} = (e + P)u^\alpha u^\beta + P g^{\alpha\beta} + \tau^{\alpha\beta},
   \label{Talphabeta}
\end{equation}
where the last term is a linear combination of first order derivatives of $u^\alpha$:
\begin{equation}
   \tau^{\alpha\beta} = t^{\alpha\beta\mu\nu}\nabla_\mu u_\nu.
\end{equation}
The tensor $t^{\alpha\beta\mu\nu}$ is unknown for the moment. However, assuming isotropy of space-time, we can reduce its number of degrees
of freedom. Typically, take four vectors $A^\alpha$, $B^\alpha$, $C^\alpha$, $D^\alpha$ and construct the scalar
\begin{equation}
   S = t^{\mu\nu\rho\sigma}A_\mu B_\nu C_\rho D_\sigma.
   \label{Siso}
\end{equation}
Requiring isotropy means that this scalar should not depend on the particular orientation of the vectors, but only on their norms
and relative orientations. This implies that it can only be a linear combination of their scalar products, namely
\begin{equation}
   S = \alpha (A\cdot B)(C \cdot D) + \beta (C \cdot A)(B \cdot D) + \gamma (A \cdot D)(B \cdot C),
\end{equation}
where $\alpha$, $\beta$, $\gamma$ are unknown parameters. The scalar products are computed with the metric so we get
\begin{equation}
   S = \underbrace{(\alpha g^{\mu\nu}g^{\rho\sigma} + \beta g^{\mu\rho}g^{\nu\sigma} + \gamma
   g^{\mu\sigma}g^{\nu\rho})}_{t^{\mu\nu\rho\sigma}}A_\mu B_\nu C_\rho D_\sigma.
\end{equation}
By identification with \eqref{Siso}, we get a 3-parameter expression for the tensor $t^{\alpha\beta\mu\nu}$. Requiring the
energy-momentum tensor to be symmetric, namely $\tau^{\alpha\beta} = \tau^{\beta\alpha}$, we infer that $\beta = \gamma$. We thus
have shown that isotropy of space-time implies the following form of $\tau^{\alpha\beta}$:
\begin{equation}
   \tau^{\alpha\beta} = \alpha g^{\alpha\beta} \nabla_\mu u^\mu + \beta (\nabla^\alpha u^\beta + \nabla^\beta u^\alpha).
   \label{tau0}
\end{equation}
The two parameters are closely linked to the bulk and shear viscosity. Indeed, let us define
\begin{subequations}
\begin{align}
   \eta &= -\beta &\quad \tn{shear viscosity},\\
   \zeta &= -\alpha - \frac{2}{3}\beta &\quad \tn{bulk viscosity}.
\end{align}
\label{viscosity}%
\end{subequations}
In the frame attached to the worldline $u^\alpha$, namely the fluid rest-frame, it is natural to require the viscous forces to not
contribute to the energy density nor to the momentum of the fluid. This would require $\tau^{0\alpha} = 0$ in the fluid
rest-frame. The covariant formulation of this argument is
\begin{equation}
   \tau^{\alpha\mu}u_\mu = 0.
   \label{restframe}
\end{equation}
Denoting $P^{\alpha\beta} = g^{\alpha\beta} + u^\alpha u^\beta$ the projector on the space-like hypersurface orthogonal to
$u^\alpha$, and combining \eqref{tau0}, \eqref{viscosity}, \eqref{restframe}, we end up with
\begin{equation}
   \tau^{\alpha\beta} = -P^{\alpha\mu}P^{\beta\nu}\left[ \eta \left( \nabla_\alpha u_\beta + \nabla_\beta u_\alpha -
   \frac{2}{3}g_{\alpha\beta}\nabla_\mu u^\mu \right) + \zeta g_{\alpha\beta} \nabla_\mu u^\mu \right].
   \label{tau}
\end{equation}
This final expression of the viscous part of the energy-momentum tensor can be used in the context of the \gls{ads}-\gls{cft} correspondence to link
the shear viscosity $\eta$ with a perturbation of the bulk \gls{ads} space-time. This is the purpose of the Kubo formula for viscosity.

\subsection{Kubo formula for viscosity}
\label{kuboformula}

Consider a first order perturbation $h_{\alpha\beta}^{(0)}$ of the 4-dimensional quantum space-time living on the \gls{ads}
boundary
\begin{equation}
   ds^2 = g^{(0)}_{\mu\nu}dx^\mu dx^\nu = -dt^2 + dx^2 + dy^2 + dz^2 + 2h_{xy}^{(0)}(t)dxdy.
   \label{ds2xy}
\end{equation}
The metric $g^{(0)}_{\mu\nu}$ is nothing but the boundary value of $g_{\mu\nu}$ defined in Fefferman-Graham coordinates, equation
\eqref{fgequ} of section \ref{fgcoor}. For a fluid at rest, we also have
\begin{equation}
   u^\mu = (1,0,0,0) \quad \tn{and} \quad u_{\mu} = (-1,0,0,0).
\end{equation}
We want to compute the expectation value of the  energy-momentum tensor due to the metric perturbation at first order. Let us
denote it by $\langle T^{xy}\rangle$. We also assume that the gravitational perturbation does not perturb\footnote{This is also the
case for a gravitational wave perturbation in the transverse-traceless gauge.} the 4-velocity at first order. From \eqref{ds2xy}, the only
non-zero Christoffel symbols are $\Gamma\indices{^t_{xy}}, \Gamma^x_{tx}, \Gamma\indices{^y_{ty}}, \Gamma\indices{^x_{ty}},
\Gamma\indices{^y_{tx}}$ and the only non-zero
components of $\nabla_\alpha u_\beta$ are
\begin{equation}
   \nabla_x u_y = \nabla_y u_x = \Gamma\indices{^t_{xy}} = \frac{1}{2}\partial_t h_{xy}^{(0)}.
\end{equation}
With these results and \eqref{tau}, it comes at first order
\begin{equation}
   \langle T^{xy}\rangle = -Ph_{xy}^{(0)} - 2\eta \Gamma\indices{^t_{xy}} = -Ph_{xy}^{(0)} - \eta\partial_t h_{xy}^{(0)}.
\end{equation}
Taking the Fourier transform of this equation yields the Kubo formula for the shear viscosity
\begin{equation}
   \langle T^{xy}(\omega, \vec q = \vec 0)\rangle = (-P + i\omega \eta) h_{xy}^{(0)},
   \label{kubo}
\end{equation}
where we have taken into account that our perturbation was only time-dependent and thus that our result was only valid in the
limit of zero propagation vector $\vec q = \vec 0$. We apply this formula to the \gls{qgp} case in section \ref{viscotoentropy}.

Finally, let us mention that repeating the above calculations with a perturbation
\begin{equation}
   ds^2 = -dt^2 + (1 + h^{(0)}(t))(dx^2 + dy^2 + dz^2),
\end{equation}
yields another Kubo formula for the bulk viscosity:
\begin{equation}
   \langle T^i_i(\omega,\vec q = \vec 0)\rangle = \frac{9}{2}\zeta i\omega h^{(0)},
\end{equation}
Depending on the metric perturbation, we can thus determine either the shear or bulk viscosities.

\subsection{Framework}

We now consider the particular case of the planar black hole to compute the shear viscosity of the \gls{qgp}. We use the following
ingredients.
\begin{myitem}
   \item The background metric of the planar black hole (section \ref{planar})
\begin{equation}
   ds^2 = \frac{r_0^2}{\gls{L}^2 u^2}(-hdt^2 + dx^2 + dy^2 + dz^2) + \frac{\gls{L}^2}{hu^2}du^2,
   \label{i1}
\end{equation}
where $u = r_0/r$, $h = 1- u^4$ and whose thermodynamical quantities verify
\begin{equation}
   T = \frac{r_0}{\pi \gls{L}^2}, \quad e = 3P = \frac{3\pi^2}{8}N_c^2 T^4, \quad s = \frac{1}{4 \gls{G}_5}\left( \frac{r_0}{\gls{L}} \right)^3.
   \label{i11}
\end{equation}
\item The \gls{ads}-\gls{cft} dictionary \eqref{dictionary}
\begin{equation}
   N_c^2 = \frac{\pi}{2}\frac{\gls{L}^3}{\gls{G}_5} \quad \tn{and} \quad \lambda = \left( \frac{\gls{L}}{l_s} \right)^4.
   \label{i2}
\end{equation}
\item The energy-momentum tensor of the dual system (equations \eqref{Talphabeta} and \eqref{tau})
\begin{equation}
   T^{\alpha\beta} = (e + P)u^\alpha u^\beta + P g_{\alpha\beta} - P^{\alpha\mu}P^{\beta\nu}\left[ \eta \left( \nabla_\alpha u_\beta + \nabla_\beta u_\alpha -
   \frac{2}{3}g_{\alpha\beta}\nabla_\mu u^\mu \right) + \zeta g_{\alpha\beta} \nabla_\mu u^\mu \right].
   \label{i3}
\end{equation}
Note that from the local Weyl invariance ($T^\mu_\mu = 0$), it comes $\zeta = 0$.
\item The Kubo formula \eqref{kubo}
   \begin{equation}
      \langle T^{xy}\rangle = (-P + i\omega \eta) h_{xy}^{(0)},
      \label{i4}
   \end{equation}
   where $\omega$ is the frequency of the bulk perturbation $h_{xy}$ and $h_{xy}^{(0)}$ its boundary value.
\end{myitem}

We now add the following bulk perturbation
\begin{equation}
   \phi \equiv \frac{u^2 \gls{L}^2}{r_0^2}h_{xy},
   \label{i5}
\end{equation}
so that the metric take the following matrix representation in the frame $(t,x,y,z,u)$
\begin{equation}
   g_{\alpha\beta} = \frac{r_0^2}{u^2 \gls{L}^2} \left(
   \begin{array}{ccccc}
      -h & 0 & 0 & 0 & 0\\
      0  & 1 & \phi & 0 & 0\\
      0  & \phi & 1 & 0 & 0\\
      0 & 0 & 0 & 1 & 0 \\
      0 & 0 & 0 & 0 & \frac{\gls{L}^4}{r_0^2 h}
   \end{array}
   \right ).
   \label{i6}
\end{equation}

We now have to compute the action which comprises three terms: the \gls{eh} (or bulk) term, the \gls{ghy} term, and the
counter-term (appendix \ref{leastaction}).

\subsection{Action}

\paragraph{The bulk action} Its expression is
\begin{equation}
   S_{bulk} = \frac{1}{16\pi \gls{G}_5}\int (R - 2 \gls{Lambda}) \sqrt{-g}d^5x.
\end{equation}
As the perturbation $\phi$ is supposed to be small, we retain only terms that are at most second order in the perturbation.
Computing the Ricci scalar\footnote{A symbolic computation software might be of help.} of the metric \eqref{i1}, it comes (setting
$\phi' = \partial_u \phi$)
\begin{subequations}
\begin{align}
            S_{bulk} &= S_0 + S_1 + S_2,\\
            S_0 &= -\frac{r_0^4}{16\pi \gls{G}_5 \gls{L}^5}\int \left( \frac{8}{u^5} \right)d^5x,\\
            S_1 &=\frac{r_0^2}{8\pi \gls{G}_5 \gls{L}} \int \frac{1}{u^3}\partial_x\partial_y \phi d^5x,\\
\nonumber   S_2 &=\frac{r_0^4}{16\pi \gls{G}_5 \gls{L}^5}\int \bigg[ \frac{2 h}{u^3}\phi \phi'' + \frac{2 \gls{L}^4}{r_0^2 u^3}\phi \partial_z^2 \phi - \frac{2 \gls{L}^4}{r_0^2 hu^3}\phi \partial_t^2 \phi + \frac{3h}{2u^3}\phi'^2 - \frac{8}{u^4}\phi\phi'\\
                &+ \frac{3 \gls{L}^4}{2 r_0^2 u^3}(\partial_z\phi)^2 - \frac{3 \gls{L}^4}{2 r_0^2 hu^3}(\partial_t \phi)^2 + \frac{4}{u^5}\phi^2 \bigg]d^5x.
\end{align}
\end{subequations}
The term $S_0$ is the background (equilibrium) part that we already have discussed in equation \eqref{seh}.
In order to simplify these expressions, we perform the 4-dimensional Fourier transform of $\phi$. Denoting $k = (-\omega,\vec q)$
with $\vec q$ a 3-dimensional wave vector, we set
\begin{equation}
   \phi(t,x,y,z,u) = \int \frac{d^4k}{(2\pi)^4} e^{-i\omega t + i q\cdot x}\phi_k(u),
   \label{fourier}
\end{equation}
where $q\cdot x = q_x x + q_y y + q_z z$. It is useful to recall that
\begin{equation}
   \int d^4x e^{-i(\omega + \omega')t + i(q + q')\cdot x} = (2\pi)^4 \delta(\omega + \omega')\delta(q + q').
   \label{deltafourier}
\end{equation}
Let us also define a scalar product for functions:
\begin{equation}
   f\cdot g = \int \frac{d^4k}{(2\pi)^4}fg.
\end{equation}
We also trade the $r_0$ for temperature $T$ with \eqref{i11} and denote
\begin{equation}
   \mathfrak{w} = \frac{\omega}{T} \quad \tn{and} \quad \mathfrak{q} = \frac{q_z}{T}.
\end{equation}
With all these notations at hand, it is (tedious but) straightforward to show that
\begin{subequations}
\begin{align}
          S_0 &= -\frac{r_0^4V_4}{2\pi \gls{G}_5 \gls{L}^5}\int_0^1 \left( \frac{du}{u^5} \right),\\
          S_1 &= 0,\\
\nonumber S_2 &=\frac{r_0^4}{16\pi \gls{G}_5 \gls{L}^5}\int_0^1 \bigg[ \frac{h}{u^3}\left( \frac{3}{2}\phi_k'\cdot\phi_{-k}' + 2\phi_k''\cdot \phi_{-k}\right) - \frac{8}{u^4}\phi_k'\cdot\phi_{-k} \\
              &\qquad\qquad\qquad + \left( \frac{\mathfrak{w}^2 - \mathfrak{q}^2h}{2\pi^2hu^3} + \frac{4}{u^5} \right)\phi_k\cdot\phi_{-k}\bigg]du.
\end{align}
\end{subequations}
The $S_2$ term is nothing but a 1-dimensional second order action, which is discussed in section \ref{1daction} of appendix
\ref{leastaction}. We can thus proceed to a cascade of integrations by parts and get
\begin{align}
\nonumber S_2 &= \frac{r_0^4}{32\pi \gls{G}_5 \gls{L}^5} \int_0^1 \left[ \left( \frac{h}{u^3}\phi_k' \right)' + \frac{\mathfrak{w}^2 - \mathfrak{q}^2h}{\pi^2hu^3}\phi_k \right]\cdot\phi_{-k}du\\
&+ \frac{r_0^4}{16\pi \gls{G}_5 \gls{L}^5}\left[ \frac{3h}{2u^3}\phi_k\cdot\phi_{-k}' - \frac{4}{u^4}\phi_k\cdot\phi_{-k} - \left( \frac{h}{u^3} \right)'\phi_k\cdot \phi_{-k}
\right]_{u=0}^{u=1}.
\label{S2}
\end{align}
On the first line can be read the equation of motion for $\phi_k$ while the second line contains all boundary terms. The
divergence at $u = 0$ is only temporary and is fixed with the \gls{ghy} term and the counter-term. After a minimal rewriting,
we thus get for the total on-shell action
\begin{equation}
   S_{bulk}^\star = \frac{r_0^4}{16\pi \gls{G}_5 \gls{L}^5}\left[ \frac{2V_4}{u^4} + \left(1-\frac{1}{u^4}\right)\phi_k\cdot \phi_{-k} + \frac{3}{2}\left(\frac{1}{u^3}-u\right)\phi_k\cdot\phi_{-k}' \right]_{u=0}^{u=1}.
   \label{Sbulkstar}
\end{equation}
At this point, we should fix the boundary conditions for $\phi$ at $u=1$. We delay the argument up to section \ref{resolution} and
admit momentarily that $\phi \underset{u\to 1}{\propto} \ln(1 - u^4)$, such that all $\phi$ terms in \eqref{Sbulkstar} vanish
when they are evaluated at $u=1$. We thus conclude that
\begin{equation}
   S_{bulk}^\star = \lim_{u\to 0} \frac{r_0^4}{16\pi \gls{G}_5 \gls{L}^5}\left[ V_4\left(-\frac{2}{u^4} + 2 \right) + \left(\frac{1}{u^4}-1\right)\phi_k\cdot \phi_{-k} - \frac{3}{2u^3}\phi_k\cdot\phi_{-k}'\right].
   \label{Sbulkstar2}
\end{equation}

\paragraph{The Gibbons-Hawking-York term} It is defined by
\begin{equation}
   S_{GHY} = -\frac{1}{8\pi \gls{G}_5}\int_{u=0}\Theta\sqrt{-q}d^4x.
\end{equation}
We have already seen (equation \eqref{Thetanu}) that we could compute the mean extrinsic curvature of the \gls{ads} boundary as
\begin{equation}
   \Theta = -\frac{n^u}{\sqrt{-q}}\partial_u \sqrt{-q} \quad \tn{with} \quad n^u = - \frac{1}{\sqrt{g_{uu}}}.
\end{equation}
With \eqref{i1}, it thus comes at second order in $\phi$:
\begin{equation}
   \sqrt{-q} = \frac{r_0^4}{\gls{L}^4}\frac{\sqrt{h}}{u^4}\left( 1 - \frac{\phi^2}{2} \right), \quad n^u = -\frac{u\sqrt{h}}{\gls{L}}.
\end{equation}
The \gls{ghy} term thus reads at second order in $\phi$
\begin{equation}
   S_{GHY} = -\frac{r_0^4}{8\pi \gls{G}_5 \gls{L}^5}\int_{u=0}\left[u\sqrt{h}\left( \frac{\sqrt{h}}{u^4} \right)'\left( 1-
   \frac{\phi^2}{2} \right) - \frac{h}{u^3}\phi\phi'\right]d^4x.
\end{equation}
Using the Fourier representation \eqref{fourier}, it comes
\begin{equation}
   S_{GHY} = \lim_{u\to 0}\frac{r_0^4}{8\pi \gls{G}_5}\gls{L}^5 \left[ V_4\left( \frac{4}{u^4} - 2 \right) + \left(
   -\frac{2}{u^4} + 1 \right)\phi_k\cdot \phi_{-k} + \frac{1}{u^3}\phi_k\cdot \phi_{-k}' \right].
   \label{sghyqp}
\end{equation}

\paragraph{The counter-term} Knowing that planar black holes have zero intrinsic curvature on hypersurfaces $u = cst$ (see
footnote below \eqref{footn}), the on-shell counter-term is
\begin{equation}
   S_{CT}^\star = -\frac{3}{8\pi \gls{G}_5 \gls{L}}\int_{u=0}\sqrt{-q}d^4x = -\frac{3}{8\pi \gls{G}_5
   \gls{L}}\int_{u=0}\frac{\sqrt{h}}{u^4}\left( 1 - \frac{\phi^2}{2} \right)d^4x.
\end{equation}
In Fourier representation, this amounts to
\begin{equation}
   S_{CT}^\star = \lim_{u\to 0}\frac{3}{8\pi \gls{G}_5 \gls{L}}\left[ V_4\left( -\frac{1}{u^4} + \frac{1}{2} \right) + \left( \frac{1}{2u^4} -
   \frac{1}{4} \right)\phi_k\cdot\phi_{-k} \right].
   \label{sctqp}
\end{equation}

\paragraph{Total action} Combining \eqref{Sbulkstar2}, \eqref{sghyqp} and \eqref{sctqp}, it comes for the total on-shell action
\begin{equation}
   S^\star = \lim_{u\to 0}\frac{r_0^4}{16\pi \gls{G}_5 \gls{L}^5}\left[ V_4 - \frac{1}{2}\phi_k\cdot\phi_{-k} + \frac{1}{2u^3}\phi_{-k}\cdot\phi_k' \right].
   \label{Svisco}
\end{equation}
Even if not obvious at first sight, the third term in the hook is not divergent. This becomes clear when solving the equation of
motion near $u = 0$.

\subsection{Resolution of the equations of motion}
\label{resolution}

The equation of motion appears clearly on the first line of \eqref{S2}. In the following, we solve it in the case of a small
$\mathfrak{w}$ and a vanishing $\mathfrak{q}$. It thus reads
\begin{equation}
   \left( \frac{h}{u^3}\phi_k' \right)' + \frac{\mathfrak{w}^2}{\pi^2hu^3}\phi_k = 0.
   \label{motionphik}
\end{equation}
To make the resolution easier, let us expand the solution in powers of $\mathfrak{w}$ as
\begin{equation}
   \phi_k(u) = F_0(u) + \mathfrak{w}F_1(u) + \mathfrak{w}^2 F_2(u) + \ldots.
   \label{wexpansion}
\end{equation}
At orders 0 and 1, $F_0$ and $F_1$ both satisfy
\begin{equation}
   \left( \frac{h}{u^3}F' \right)' = 0 \iff F(u) = A + B\ln(1-u^4),
\end{equation}
so that
\begin{equation}
   \phi_k = A_0 + \mathfrak{w}A_1 + (B_0 + \mathfrak{w}B_1)\ln(1-u^4) + O(\mathfrak{w}^2),
\end{equation}
where the $A$ and $B$ are the integration constants of orders 0 and 1. We introduce the equivalent but more standard notation
\begin{equation}
   \phi_k = \phi_k^{(0)}(1 + \phi_k^{(1)}\ln(1-u^4)) + O(\mathfrak{w}^2).
   \label{phik0}
\end{equation}
In this equation, $\phi_k^{(0)}$ has the clear meaning of the boundary value of $\phi_k$. In order to get an expression for
$\phi_k^{(1)}$, we need to impose boundary conditions on the horizon $u = 1$. We thus develop the equation of motion
\eqref{motionphik} near $u=1$ (for example $h = 1 - u^4 \underset{u\to1}{\sim} 4(1 - u)$). We thus get
\begin{equation}
   \phi_k'' - \frac{\phi_k'}{1-u} + \left( \frac{\mathfrak{w}}{4\pi} \right)^2 \frac{\phi_k}{(1-u)^2} = 0 \qquad (\tn{when } u \to 1).
\end{equation}
Looking for a regular power-law solution $\phi_k = (1-u)^\lambda$, the dispersion relation $\lambda^2 +
(\mathfrak{w}/4\pi)^2 = 0$ comes directly. Hence
\begin{equation}
   \phi_k \underset{u\to1}{\sim} C (1-u)^{\frac{i \mathfrak{w}}{4\pi}} + D(1-u)^{\frac{-i \mathfrak{w}}{4\pi}},
   \label{phikcomb}
\end{equation}
where $C$ and $D$ are constants of integration. In order to better understand the significance of these two independent solutions,
let us move to tortoise coordinates $u_\star$ defined by $g_{tt} = - g_{u_\star u_\star}$, or equivalently, $ds^2 \propto -dt^2 +
du_\star^2 + \ldots$. Given the metric \eqref{i1}, we find that $du_\star^2 = \gls{L}^4 du^2/r_0^2 h^2$, or equivalently
\begin{equation}
   u_\star = -\frac{\gls{L}^2}{r_0}\int \frac{du}{h} \underset{u\to 1}{\sim} \frac{1}{4\pi T}\ln(1-u) + cst,
\end{equation}
where we have used $h \underset{u\to1}{\sim} 4(1-u)$ and the temperature result \eqref{i11}. In these coordinates, equation
\eqref{phikcomb} becomes
\begin{equation}
   \phi_k \underset{u\star\to0}{\sim} C e^{i\omega u_\star} + De^{-i\omega u_\star}.
   \label{phikint}
\end{equation}
We can get more insight of what this means if we consider the simple example for a monochromatic wave. It would imply, after taking the inverse Fourier transform
\begin{equation}
   \phi \underset{u\star\to0}{\sim} C \underbrace{e^{-i\omega(t - u_\star)}}_{\tn{outgoing wave}} + D\underbrace{e^{-i\omega(t + u_\star)}}_{\tn{ingoing wave}}.
\end{equation}
This is nothing but a linear combination of monochromatic ingoing and outgoing waves. The ingoing wave is the one moving toward
increasing $u$ (decreasing radius $r = r_0/u$) and thus decreasing $u_\star$. We thus naturally set $C = 0$ as a boundary
condition, in order to forbid waves outgoing from the horizon. Moreover, since we are looking for an expansion in $\mathfrak{w}$
(see equation \eqref{wexpansion}), we keep only the first order in $\mathfrak{w}$, so that from \eqref{phikint} it remains only
\begin{equation}
   \phi_k \underset{\substack{u\to1\\\mathfrak{w}\to0}}{\propto} 1 - \frac{i \mathfrak{w}}{4\pi}\ln(1 - u^4) + O(\mathfrak{w}^2).
\end{equation}
Comparing to \eqref{phik0}, we are thus able to conclude that the solution of \eqref{motionphik} is unambiguously
\begin{equation}
   \phi_k = \phi_k^{(0)}(1 + \phi_k^{(1)}\ln(1-u^4)) + O(\mathfrak{w}^2) \quad \tn{with} \quad \phi_k^{(1)} = - \frac{i \mathfrak{w}}{4\pi}.
   \label{phik1}
\end{equation}
The coefficient $\phi_k^{(0)}$ is left undetermined, as explained in section \ref{gkpwitt2}.

\subsection{Viscosity to entropy density ratio}
\label{viscotoentropy}

We are now able to evaluate precisely the on-shell action \eqref{Svisco}. With the result \eqref{phik1}, the limit $u \to 0$ is no more
divergent and it comes
\begin{equation}
   S^\star = \frac{r_0^4}{16\pi \gls{G}_5 \gls{L}^5}\left[ V_4 - \frac{1}{2}\phi_{-k}^{(0)}\cdot\phi_k^{(0)} +
   2\phi_{-k}^{(0)}\cdot\left(\phi_k^{(1)}\phi_k^{(0)}\right) \right].
   \label{sstar}
\end{equation}
In order to get the metric on the \gls{cft} side, we introduce the \gls{fg} coordinates, discussed in section \ref{fgcoor}. Introducing
the variable $\widetilde{u}$ defined by
\begin{equation}
   u^2 = \frac{2\widetilde{u}^2}{\widetilde{u}^2 + \widetilde{u}_0^2}\quad \tn{with} \quad \widetilde{u}_0 = \frac{\sqrt{2}}{r_0},
\end{equation}
the metric \eqref{i6} in \gls{fg} coordinates takes the following form
\begin{equation}
   ds^2 = \frac{1}{\widetilde{u}^2}\left[ -\frac{\widetilde{u}_0^2 - \widetilde{u}^2}{\widetilde{u}_0^2 + \widetilde{u}^2}dt^2 + \left( 1 +
   \frac{\widetilde{u}^4}{\widetilde{u}_0^4} \right)(dx^2 + dy^2 + dz^2 + 2\phi dxdy) + d \widetilde{u}^2 \right].
\end{equation}
On the \gls{ads} boundary, i.e.\ taking the limit $\widetilde{u}\to 0$, we can thus read the \gls{cft} metric as in section
\ref{fgcoor} equation \eqref{fgequ}:
\begin{equation}
   ds^2|_{CFT} = g_{\mu\nu}^{(0)}dx^\mu dx^\nu = -dt^2 + dx^2 + dy^2 + dz^2 + 2\phi^{(0)}dxdy.
\end{equation}
At this point, we use a key concept of the \gls{ads}-\gls{cft} correspondence already discussed in section \ref{gkpwitt2}: the
bulk gravitational perturbations are considered as boundary source terms for the \gls{cft}. This allows us to define the \gls{cft}
energy-momentum tensor with equation \eqref{Tdef} of the appendix. On the \gls{cft} side, it thus comes, at second order in
$h_{xy}^{(0)}$
\begin{equation}
   T^{\alpha\beta} = \frac{2}{\sqrt{-g^{(0)}}}\frac{\delta S}{\delta g^{(0)}_{\alpha\beta}} \iff \delta S = \frac{1}{2}\int
   T^{\mu\nu}h_{\mu\nu}^{(0)}\sqrt{-g^{(0)}}d^4x = \int T^{xy}\phi^{(0)}d^4x,
\end{equation}
or equivalently, in Fourier space (recall \eqref{deltafourier})
\begin{equation}
   \langle T^{xy} \rangle = \frac{\delta S}{\delta \phi^{(0)}} \iff \langle T^{xy}_k \rangle = \frac{\delta S}{\delta \phi_{-k}^{(0)}}.
\end{equation}
Combining with \eqref{sstar}, we finally get the expression of the expectation value for the energy-momentum tensor on the
\gls{cft} side, namely
\begin{equation}
   \langle T^{xy}_k \rangle = \frac{r_0^4}{16\pi \gls{G}_5 \gls{L}^5}\left(-1 + 4\phi_k^{(1)}\right)\phi_k^{(0)}.
\end{equation}
By identification with the Kubo formula \eqref{kubo}, and using the thermodynamical relations between temperature and entropy density
\eqref{i11}, it follows
\begin{equation}
   P = \frac{r_0^4}{16\pi \gls{G}_5 \gls{L}^5} = \frac{\pi^2}{8}N_c^2 T^4 \quad \tn{and} \quad \phi_k^{(1)} = i \mathfrak{w}\frac{\eta}{s}.
\end{equation}
The expression of the pressure $P$ obtained with the \gls{ads}-\gls{cft} dictionary \eqref{dictionary} is nothing but the background (or
equilibrium) result of section \ref{qgpeq}, equation \eqref{e3p}. As for $\phi_k^{(1)}$, by identification with our previous
determination \eqref{phik1}, we immediately recover the desired and fundamental result
\begin{equation}
   \frac{\eta}{s} = \frac{1}{4\pi}.
\end{equation}

\subsection{Interpretation}

If we reintroduce the fundamental constants of physics, we have found that in the large-$N_c$ limit of the \gls{nsym} theory
\begin{equation}
   \frac{\eta}{s} = \frac{\gls{hbar}}{4\pi \gls{kb}} \simeq \SI{6.1e-13}{K.s}.
\end{equation}
As a point of comparison, this ratio is 3300 larger for a typical nitrogen gas in the atmosphere, so that \glspl{qgp} can be said to
have a very small ratio $\eta/s$ compared to ordinary matter. Of course we made the calculation in the ideal setup of
the \gls{ads}-\gls{cft} correspondence. For real-world \glspl{qgp}, the $1/4\pi$ value is a lower limit, namely at finite coupling
we expect
\begin{equation}
   \frac{\eta}{s} \gtrsim \frac{1}{4\pi}.
\end{equation}
This is in very good agreement with the combined experimental-numerical determination of this ratio (see section \ref{qgpsec}),
which is around $\sim 1.5/4\pi$. This is all the power of the correspondence: $\eta/s$ is impossible to compute in
\gls{qcd}, but the \gls{ads}-\gls{cft} correspondence allows to give a value that is very close to the combined
numerical-experimental measurements by simply looking at the perturbation of a black hole in \gls{ads} space-time. It has long been known that quasi-normal modes of black holes are by
essence dissipative since part of the energy is absorbed by the horizon. This is the so-called ringdown of black hole
perturbations. \gls{ads}-\gls{cft} gives a striking connection between the ``viscosity'' of black holes and the viscosity of
\glspl{qgp}.

This lower limit on $\eta/s$ can be conjectured to be a fundamental limit valid for \textit{all} fluids. Heuristically, for a fluid of mass
density $\rho$, whose particles have mass $m$, a thermal average speed $\overline{v}$ and a mean free path $\overline{l}$, the shear viscosity
and entropy density are roughly given by
\begin{equation}
   \eta \simeq \rho \overline{v} \overline{l} \quad \tn{and} \quad s \simeq \frac{\rho}{m}\gls{kb},
\end{equation}
where $\rho/m$ is the number of particles considered. Using the Heisenberg uncertainty principle
\begin{equation}
   m \overline{v} \overline{l} \gtrsim \gls{hbar},
\end{equation}
(or equivalently, $\overline{l}$ slightly superior to the de Broglie wavelength) it comes
\begin{equation}
   \frac{\eta}{s} \gtrsim \frac{\gls{hbar}}{\gls{kb}}.
   \label{roughetas}
\end{equation}
It could thus be conjectured that the numerical factor $1/4\pi$ is universal and is a fundamental limit of quantum field
theories. The \gls{ads}-\gls{cft} computation gives the precise value of the undetermined multiplicative factor in \eqref{roughetas}, namely
$1/4\pi$.

However, we should keep in mind that our calculation is an idealisation. Table \ref{qgpadscft} emphasizes the
differences between the assumptions we have used and the real properties of the \gls{qgp} of \gls{qcd}. There are two types of
corrections we can thus conceive, namely $1/\lambda$ and $1/N_c$ corrections. For instance, the $1/\lambda$ corrections correspond
to $\alpha'$ (i.e.\ string) corrections. The computation gets much more involved and the result is of the order of
\begin{equation}
   \frac{\eta}{s} \simeq \frac{1.2}{4\pi},
\end{equation}
which is even closest to the experimental value. However, as a trade-off between accuracy and complexity, our simple computation with
classical \gls{gr} is already very efficient. On conceptual grounds, it is very surprising and at the same time very exciting that
the properties of \glspl{qgp} that are described by \gls{qcd} can be computed considering black holes in a purely \gls{gr}
setting.

\begin{mytab}
\begin{tabular}{ccc}
\hline
parameter                             & \gls{qgp} & large-$N_c$ limit \\
\hline
$\lambda$                             & $\sim 10$ & $\infty$            \\
$N_c$                                 & $\sim 3$  & $\infty$            \\
$g_{YM}$ & $\sim 1$  & 0                   \\
\hline
\end{tabular}
\caption[Comparison between real QGP and AdS-CFT]{Parameters for the real \gls{qgp} and the \gls{ads}-\gls{cft} computation in the large-$N_c$
limit. $N_c$ is the number of colours and $\lambda$ the 't Hooft coupling. Credits: \cite{Natsuume15}.}
\label{qgpadscft}
\end{mytab}

\section{AdS-CFT and beyond}

This chapter aimed at presenting simple applications of the correspondence that are accessible to non-experts of quantum field
theories. Notably, we have studied the thermodynamical equilibrium of \glspl{qgp}, the \gls{hp} phase transition, and the shear viscosity
of \glspl{qgp}. The \gls{hp} phase transition can be extended to several other kinds of phase transition, among them the very
popular holographic superconductors, whose study is much beyond the scope of the present manuscript.

The point is that, for the first time in the history of \gls{gr}, \gls{ads} space-time touches real-world
physics in a very unexpected way, namely the physics of \glspl{qgp} produced in the \gls{lhc}. The duality was then extended, giving
rise to a profusion of dualities, which are listed in \cite{Hubeny15}. The domains of applications of the gauge-gravity
correspondence now reaches the field of condensed matter physics, in particular with the advance of the holographic
superconductors. For many people, the correspondence is a promising route to follow in order to unify quantum field theory and gravity.

Given the rising importance of the correspondence in the literature, it comes as no surprise that more and more relativists tried
to probe deeper the gravitational dynamics of the \gls{ads} space-time, that were almost unexplored before 1998. Despite the
rising interest in this space-time, the question of its non-linear stability was not really addressed before 2011. And actually, it turns out that
\gls{ads} space-time does not only gives precious informations about superconformal theories, but also provides an unexpected
richness of features in purely gravitational physics.

\chapter{Anti-de Sitter instability}
\label{adsinsta}
\addcontentsline{lof}{chapter}{\nameref{adsinsta}}
\addcontentsline{lot}{chapter}{\nameref{adsinsta}}
\citationChap{Perfection is achieved only on the point of collapse.}{Cyril Northcote Parkinson}
\minitoc

Einstein's equations in vacuum admit three maximally symmetric solutions: Minkowski (flat), \gls{ds} and \gls{ads} space-times.
They correspond respectively to zero, positive and negative curvature and cosmological constant \gls{Lambda}. Both Minkowski and
\gls{ds} space-times are non-linearly stable, which means that no small perturbation can grow unbounded. Mathematical demonstrations
can be found respectively in \cite{Christodoulou93} and \cite{Friedrich86}. Both proofs rely on a dispersion mechanism: any
perturbation should decay one way or another, and this is possible when waves can propagate freely toward infinity without
back-reacting substantially on the metric. The decay rate of the perturbation is exponential in \gls{ds} and the non-linear
stability can be inferred relatively easily. In Minkowski, the decay rate is borderline and the non-linear stability proof is much
more subtle.

\gls{ads} space-time is much different. The negative cosmological constant acts like a gravitational potential that
prevents time-like geodesics from ever reaching infinity. Instead, test particles, whatever their initial speeds, come back to their
starting point in a finite proper time. This is even true for null geodesics. In this case, a photon can reach infinity in a
finite time (as measured by a static observer). Assuming that the total energy (or mass) of space-time is
conserved\footnote{This is a natural assumption, but it is not mandatory though.}, this photon has to bounce back on space-like
infinity, at the so-called \gls{ads} boundary. In other words, this space-time acts like a reflective confining box: any test
particle, be it massive or not, follows a straight line in vacuum and comes back at its starting point in a finite time. It thus
oscillates perpetually, as if it were trapped in a 4-dimensional billiard table, as illustrated in figure \ref{geodesics} of chapter
\ref{aads}. This confining mechanism is clearly incompatible with the decay argument needed for establishing a non-linear
stability proof. And actually, no such proof exists so far.

It has long been known that \gls{ads} was \textit{linearly} stable, since the seminal works of
\cite{Avis78,Breitenlohner82,Abbott82}. It means that no small perturbation can have a mode that is unbounded in time at first
order in amplitude. Nevertheless, linear stability does not imply non-linear stability. Very curiously, the emergence of the
\gls{ads}-\gls{cft} correspondence in 1998 (see \cite{Maldacena98,Maldacena99,Witten98,Aharony00,Hubeny15} and chapter
\ref{adscft}) triggered an avalanche of papers, but very few of them were concerned about the non-linear stability of \gls{ads}.
The first questioning of this point appeared in 2006 with the works of \cite{Dafermos06,Anderson06}. Notably in \cite{Anderson06}
the following ``rigidity'' theorem was demonstrated:

\begin{theorem}[Anti-de Sitter rigidity theorem]
The only globally regular \gls{aads} space-time that tends to \gls{ads} at arbitrarily long times is \gls{ads} itself.
\end{theorem}

This is clearly the opposite of the decay condition for perturbations and it is quite natural for a space-time that cannot radiate its
energy to infinity. However, the author conjectured that \gls{ads} was probably non-linearly stable, and this was the global
consensus at that time, with the notable exception of \cite{Dafermos06}.

The number of researchers working on the \gls{ads} non-linear stability problem drastically increased in 2011, after the seminal
paper of Bizo\'n and Rostworowski \cite{Bizon11}. The more numerous the outcoming papers in the field, the more intricate
and subtle the problem seemed to be. This gave rise to lively debates in the community, but also to an abundance of new numerical experiments
and formalisms. In this chapter, we give an overview of the state of the art of the \gls{ads} non-linear stability
problem and hope to disentangle the information harvested in the literature. We set the speed of light to unity, $\gls{c} = 1$.

\section{Weakly turbulent instability of anti-de Sitter}

In 2011, Bizo\'n and Rostworowski studied the collapse of a spherically symmetric scalar wave packet in \gls{ads}. The setup of their
experiment was reused so often that we deem useful to reproduce it here.

\subsection{Einstein-Klein-Gordon equations in spherical symmetry}

Consider the \gls{ekg} system of equations with cosmological constant in four dimensions:
\begin{subequations}
\begin{align}
   G_{\alpha\beta} + \gls{Lambda} g_{\alpha\beta} &= 8\pi \gls{G}\left(\nabla_\alpha\phi\nabla_\beta \phi - \frac{1}{2}g_{\alpha\beta} \nabla_\mu \phi \nabla^\mu \phi\right),\\
   \nabla_\mu \nabla^\mu \phi &= 0,
\end{align}
\label{einsteinscalar}%
\end{subequations}
where $g_{\alpha\beta}$ is the metric, $G_{\alpha\beta}$ is Einstein's tensor, $\nabla$ is the connection associated to the
metric, and $\phi$ is a real massless scalar field. In spherical symmetry and in conformal coordinates (see section
\ref{coordinates} equation \eqref{adsconformal}), the metric can be put into the form\footnote{Note that both $t$ and $x$ coordinates have no dimensions. The
physical time is thus $\gls{L}t$.}
\begin{equation}
   ds^2 = \frac{\gls{L}^2}{\cos^2x}(-Ae^{-2\delta}dt^2 + A^{-1}dx^2 + \sin^2xd\Omega^2), \quad x \in \left[0, \frac{\pi}{2}\right[,
   \label{ansatzscalar}
\end{equation}
where $d\Omega^2 = d\theta^2 + \sin^2\theta d\varphi^2$ is the angular part of the length element. The metric functions $A$, $\delta$ and
scalar field $\phi$ are supposed to depend only on $(t,x)$ and the \gls{ads} boundary lies at $x = \pi/2$. Accordingly, we choose notations in
which overdots and primes indicate time and radial derivatives respectively. Defining
\begin{equation}
   \Phi = \phi' \quad \tn{and} \quad \Pi = A^{-1}e^\delta \dot{\phi},
\end{equation}
the system of equations \eqref{einsteinscalar} boils down to evolution equations for the dynamical variables (in units $4\pi
\gls{G} = 1$)
\begin{subequations}
\begin{align}
   \dot{\Phi} &= (Ae^{-\delta}\Pi)',\\
   \dot{\Pi} &= \frac{1}{\tan^2x}(\tan^2x A e^{-\delta}\Phi)',
\end{align}
\label{time}
\end{subequations}
and constraint equations
\begin{subequations}
\begin{align}
   \label{time3}
   \dot{A} &= -2\sin x \cos x A^2 e^{-\delta} \Phi\Pi,\\
   \label{cons1}
   A' &= \frac{1 + 2\sin^2x}{\sin x \cos x}(1 - A) - \sin x \cos x A (\Phi^2 + \Pi^2),\\
   \label{cons2}
   \delta' &= -\sin x\cos x(\Phi^2 + \Pi^2).
\end{align}
\end{subequations}
Because the scalar field is massless, the \gls{ads} length scale $\gls{L}$ drops out of the equations. The variables $\Phi$ and
$\Pi$ are evolved in time with \eqref{time} while the constraints \eqref{cons1}-\eqref{cons2} are used to update the metric
at each time step. The equation \eqref{time3} is used as a monitor of code precision. This system of equations supplied with
Dirichlet boundary conditions and compatible initial data is locally well-posed\cite{Holzegel12,Holzegel13a}.

\subsection{Numerical observations}

The initial data is chosen to be a localised wave packet with Gaussian shape:
\begin{equation}
   \Phi(0,x) = 0 \quad \tn{and} \quad \Pi(0,x) = \varepsilon \exp\left( -\frac{\tan^2x}{\sigma^2} \right),
\end{equation}
where $\varepsilon$ denotes the amplitude and $\sigma$ the width of initial data. Apparent horizon formation is signalled by the
vanishing of the blackening factor $A$ in the metric \eqref{ansatzscalar}, such that the apparent horizon radius lies at its
largest root. More numerical details can be found in \cite{Maliborski13c}.

Letting the system evolve in time leads to the following observations. If the amplitude is large, the scalar wave packet directly
collapses to a black hole. Lowering the amplitude and repeating the experiment, the horizon radius $x_H$ decreases and eventually
tends to zero when a critical amplitude $\varepsilon_0$ is reached. This would have been the end of the story in asymptotically
flat space-times, as was noted by Choptuik and collaborators in 1993 \cite{Abraham93,Choptuik93}.

If the simulation is run with an amplitude $\varepsilon < \varepsilon_0$, the scalar field starts to contract but does not
form a black hole. It then spreads and reaches infinity in a finite time slightly larger than the null geodesic\footnote{It depends
also on how large the initial data is, i.e.\ how large is the parameter $\sigma$. The wider the initial data, the shorter the
time to reach the boundary.} one $t \gtrsim \pi \gls{L}/2$, as was noted later in \cite{Garfinkle12}. Because of the reflective
boundary conditions, the field bounces off the \gls{ads} boundary. When it comes back to the origin, self-gravitation had the time
to build up a more peaked scalar field profile, so much as to collapse to a black hole when approaching the origin $x = 0$. This leads to a
second branch of collapsing solutions that undergoes one reflection.

When the amplitude is lowered down to a second critical value $\varepsilon_1 < \varepsilon_0$, the resulting black hole has a
horizon radius going to zero. Initial data with amplitudes smaller that $\varepsilon_1$ have to bounce off the \gls{ads}
boundary twice before collapsing. And so on and so forth. A sequence of critical amplitude $\varepsilon_n$ can then be
constructed, where $n$ is the number of bounces of the initial data having amplitudes $\varepsilon_n < \varepsilon <
\varepsilon_{n-1}$. This behaviour is illustrated in figure \ref{adsinstability}. Note that on this plot, $x_H$ denotes the horizon radius
right at the point of collapse. In the long term evolution, after several partial absorptions and reflections of the scalar field
on the horizon, all the scalar field falls into the black hole and the metric settles down to the \gls{sads} solution with a mass
parameter in agreement with the mass of the initial data \cite{Garfinkle12}. 

Furthermore, it was mathematically proved in \cite{Holzegel13a} that the Schwarzschild-\gls{ads} solution was non-linearly stable
in spherical symmetry. Soon after black hole formation, the space-time thus settles down to a stable and stationary Schwarzschild
family of solution. The striking feature unveiled by \cite{Bizon11} is that however small the amplitude of the initial data is, a
black hole is formed all the same. This suggests to look at a perturbative approach and see if some indications of collapse can
be inferred.

\begin{myfig}
   \includegraphics[width = 0.49\textwidth]{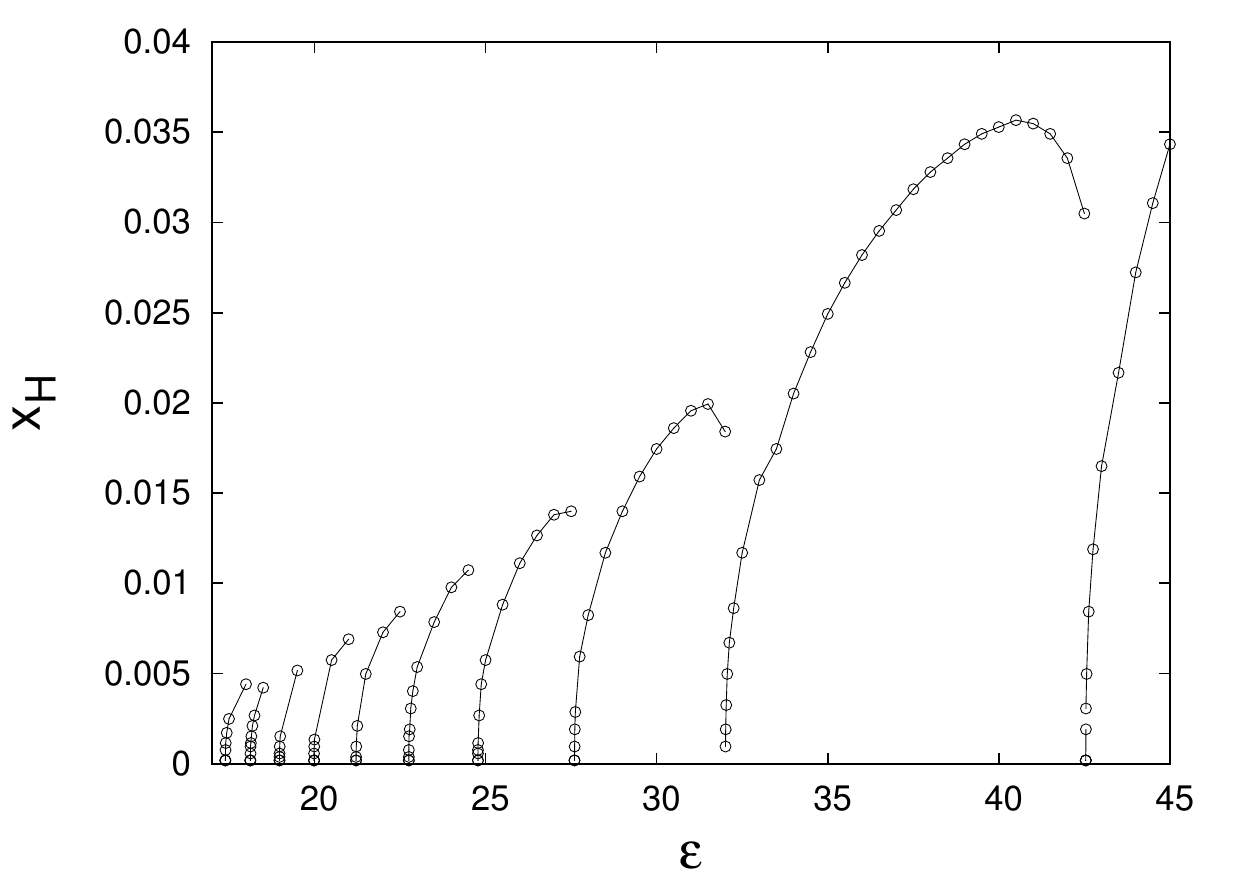}
   \includegraphics[width = 0.49\textwidth]{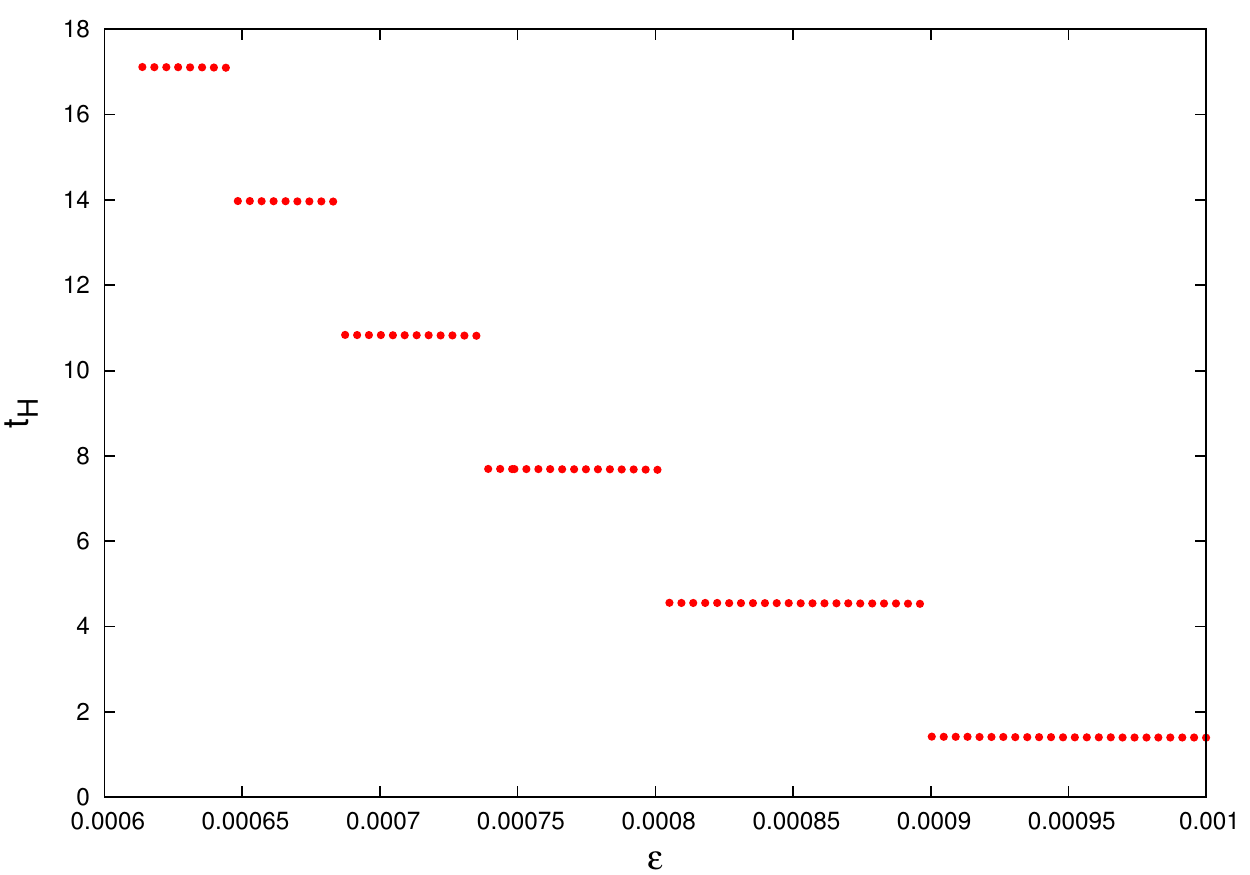}
   \caption[Anti-de Sitter instability]{Left: horizon radius $x_H$ as a function of the amplitude $\varepsilon$ of the initial
   wave packet. The far right curve describes the prompt collapse of the scalar wave packet, the other curves on the left are
   respectively for 1 to 9 reflections off the \gls{ads} boundary. Right: time of black hole formation $t_H$ as a function of the
   initial amplitude of the scalar wave packet. Each step corresponds to an additional reflection off the \gls{ads} boundary. The
   steps are separated by an amount of time $\Delta t \gtrsim \pi \gls{L}$, i.e.\ slightly larger than the time of a null round
   trip in \gls{ads}. Credits: \cite{Bizon11,Jalmuzna11}.}
   \label{adsinstability}
\end{myfig}

\subsection{Perturbative approach}
\label{pertscal}

We denote by $\varepsilon$ the small parameter encoding the initial data amplitude. The \gls{ekg} system of equations
\eqref{einsteinscalar} is invariant under the transformation $\phi \to -\phi$. Thus, under the transformation $\varepsilon \to
-\varepsilon$, the scalar field should just change sign and the metric change not at all. This is why the three functions
admits the following even-odd expansion:
\begin{subequations}
\begin{align}
\phi &= \varepsilon\phi_1 + \varepsilon^3 \phi_3 + \ldots,\\
A &= 1 - \varepsilon^2 A_2 - \ldots,\\
\delta &= \varepsilon^2\delta_2 + \ldots.
\end{align}
\end{subequations}
At first order $O(\varepsilon)$, the \gls{ekg} system boils down to
\begin{equation}
   \ddot{\phi}_1 + \widehat{L}\phi_1 = 0 \quad \tn{with} \quad \widehat{L} = -\frac{1}{\tan^2x}\partial_x(\tan^2x\partial_x).
   \label{pert1}
\end{equation}
This equation can be solved by diagonalising the operator $\widehat{L}$. Defining an inner product on the Hilbert space of solutions as
\begin{equation}
   (f,g) = \int_0^{\pi/2}f(x)g(x)\tan^2x dx,
\end{equation}
an orthonormal basis of solutions is \cite{Maliborski13c}
\begin{subequations}
\begin{align}
   \widehat{L}e_j &= \omega_j^2 e_j \quad \tn{where} \quad \omega_j^2 = (3 + 2j)^2\quad \tn{and} \quad j \in \mathbb{N},\\
   e_j(x) &= d_j \cos^3 x P_j^{\left(\frac{1}{2},\frac{3}{2}\right)}(\cos(2x)) \quad \tn{with} \quad d_j = \frac{2\sqrt{j!(j+2)!}}{\Gamma(j+3/2)},
\end{align}
\label{eigenmode}%
\end{subequations}
where $\Gamma$ is Euler's Gamma function and $P_j^{\left(\frac{1}{2},\frac{3}{2}\right)}$ are Jacobi polynomials. All eigenvalues
are real and positive because the operator $\widehat{L}$ is self-adjoint. This means that no eigenmode is unstable,
which is consistent with the linear stability of \gls{ads} space-time \cite{Avis78,Breitenlohner82,Abbott82}. In addition, all frequencies
are equidistant: the spectrum is said to be resonant. We can now solve \eqref{pert1} through decomposition on the basis
$(e_j)_{j \in \mathbb{N}}$, which gives:
\begin{equation}
   \phi_1(t,x) = \sum_{j=0}^\infty a_j \cos(\omega_j t + \beta_j)e_j(x),
   \label{phi1}
\end{equation}
where $a_j$ and $b_j$ are real constants. In other words, at first order the solution is merely oscillating in time with a spatial
dependence governed by the $(e_j)_{j \in \mathbb{N}}$ functions.

At second order, the equations relative to $A_2$ and $\delta_2$ admit the following solutions:
\begin{subequations}
\begin{align}
   A_2(t,x) &= \frac{\cos^3x}{\sin x}\int_0^x[\dot{\phi_1}^2(t,y) + {\phi_1'}^2(t,y)]\tan^2ydy,\\
   \delta_2(t,x) &= -\int_0^x[\dot{\phi_1}^2(t,y) + {\phi_1'}^2(t,y)]\sin y \cos ydy.
\end{align}
\end{subequations}
At third order, it comes
\begin{equation}
   \ddot{\phi_3} + \widehat{L}\phi_3 = S(\phi_1,A_2,\delta_2) \equiv -2(A_2 + \delta_2)\ddot{\phi}_1 - (\dot{A}_2 + \dot{\delta}_2)\dot{\phi}_1 - (A_2' + \delta_2')\phi_1'.
   \label{phi3}
\end{equation}
Projecting this equation on the basis $(e_j)_{j \in \mathbb{N}}$ yields, denoting $c_j^{(3)} = (\phi_3,e_j)$ and $S_j = (S,e_j)$:
\begin{equation}
   \forall j, \quad \ddot{c}_j^{(3)} + \omega_j^2 c_j^{(3)} = S_j.
   \label{cj3}
\end{equation}
A straightforward (but tedious) look at the right-hand side of \eqref{cj3} makes it clear that it contains some resonant terms
$\cos(\omega_jt)$ or $\sin(\omega_j t)$ every time there exists a triplet $(j_1,j_2,j_3)$ such that
\begin{equation}
   a_{j_1} \neq 0, a_{j_2} \neq 0, a_{j_3} \neq 0 \quad \tn{and} \quad \omega_j = \omega_{j_1} + \omega_{j_2} - \omega_{j_3}.
\end{equation}
This is due to the resonant character of the spectrum of the operator $\widehat{L}$ and gives rise to secular resonances, i.e.\ solutions of the form
\begin{equation}
   \phi_3 \sim t\sin(\omega_jt) + \ldots .
\end{equation}
These solutions are thus diverging linearly in time. We infer that when $\phi_3$ and $\phi_1$ are of the same order of magnitude
(namely after a time $t = O(\varepsilon^{-2})$), the perturbative scheme breaks down, as we expect higher orders terms to be smaller than leading terms
in convergent series. Such resonances are quite common in perturbative expansions and they can sometimes be cured by redefining the
expansion parametrisation. For example the Poincaré-Lindstedt method consists in expanding the frequencies $\omega_j$ as
\begin{equation}
   \omega_j = \omega_j^{(0)} + \varepsilon^2 \omega_j^{(2)} + \ldots .
\end{equation}
Substituting this expression into \eqref{phi1} and \eqref{phi3} leads to the suppression of many secular resonances if the
constants $\omega_j^{(2)}$ are chosen astutely. However, in the case under study, if some resonances of \eqref{cj3} are indeed
removable, others are not and the expansion is truly diverging on time-scales $t= O(\varepsilon^{-2})$ \cite{Bizon11}.

The perturbative approach thus provides an analytical argument in favour of the black hole formation: any small perturbation
cannot remain small and it takes a time $t= O(\varepsilon^{-2})$ to reach the fully-non linear regime. This is in very good
agreement with numerical results, for which the time of collapse does indeed scales as the inverse square of the amplitude (figure
\ref{adsinstability}). In analogy with quantum mechanics, this statement describes the instability in position-space. In order to get
the momentum-space picture, we can take advantage of the perturbative approach and define the energy per mode of a solution.

\subsection{Energy per mode}

For convenience, we introduce
\begin{equation}
   \quad \Pi_j = (\sqrt{A}\Pi,e_j) \quad \tn{and} \quad \Phi_j = (\sqrt{A}\Phi,e_j'),
\end{equation}
the projections of $\Pi$ and $\Phi$ on the bases $(e_j)_{j \in \mathbb{N}}$ and $(e_j')_{j \in \mathbb{N}}$
respectively\footnote{Recall that $e_j' = \frac{d e_j}{dx}$.}. Note that $(e_i,e_j) = \delta_{ij}$ but
$(e_i',e_j') = \omega_j^2 \delta_{ij}$. The total mass of the system can be expressed as\footnote{Recall that the energy density
   $\rho$ measured by a static observer at infinity is $T_{tt}$ where $T_{\alpha\beta} = \left(\nabla_\alpha\phi\nabla_\beta\phi -
   \frac{1}{2}g_{\alpha\beta} \nabla_\mu \phi\nabla^\mu \phi\right)$, so that $\rho = \frac{A^2e^{-\delta}}{2}(\Phi^2 + \Pi^2)$. Due
to spherical symmetry, the mass can be computed by integrating the energy density, but with a rescaled volume form. This is
reminiscent of the famous \gls{tov} equations and mass function for spherically symmetric neutron stars.}
\begin{equation}
   M = \frac{1}{2}\int_0^{\pi/2}(A\Phi^2 + A\Pi^2)\tan^2x dx.
\end{equation}
Applying Parseval's identity, it comes
\begin{equation}
   M = \sum_{j=0}^{\infty} E_j(t) \quad \tn{with} \quad E_j = \Pi_j^2 + \frac{\Phi_j^2}{\omega_j^2}.
\end{equation}
We can then reasonably interpret $E_j$ as a proxy for the energy contained in the $j^{th}$ mode (recall however that energy is not local).
It turns out that the signature of the instability is very clear when considering the energy per mode: it is characterised by a
weakly turbulent cascade.

\subsection{Weakly turbulent cascade}

In order to better understand the mechanism of black hole formation, the authors of \cite{Bizon11} tried to evolve
initial data obtained with a single eigenmode of the linear operator $\widehat{L}$, namely
\begin{equation}
   (\phi,\dot{\phi})_{t=0} = \varepsilon(e_0(x),0).
\end{equation}
From a perturbative point of view, this single-mode initial data displays a single resonant term in the expression of $S_0$, and
this term can indeed be removed by Poincaré-Lindstedt regularisation. It suggests that such initial data should not collapse into a
black hole. This was actually checked via numerical evolution in time: a single-mode excitation is indeed non-linearly stable.

On the other hand, the so-called two-mode initial data (coefficients are defined in equation \eqref{eigenmode})
\begin{equation}
   (\phi,\dot{\phi})_{t=0} = \varepsilon\left(\frac{e_0(x)}{d_0} + \frac{e_1(x)}{d_1},0\right),
   \label{twomode}
\end{equation}
displays an irremovable secular resonance and does lead to black hole formation in a time $t = O(\varepsilon^{-2})$. It was then
numerically observed in \cite{Bizon11} that the energy of the system was cascading to higher order modes and hence higher spatial frequencies. This
cascade behaviour is illustrated on figure \ref{turbu}. Black hole formation then provides a natural cut-off that eventually
forbids the transfer of energy to smaller and smaller scales indefinitely.

\begin{myfig}
   \includegraphics[width = 0.49\textwidth]{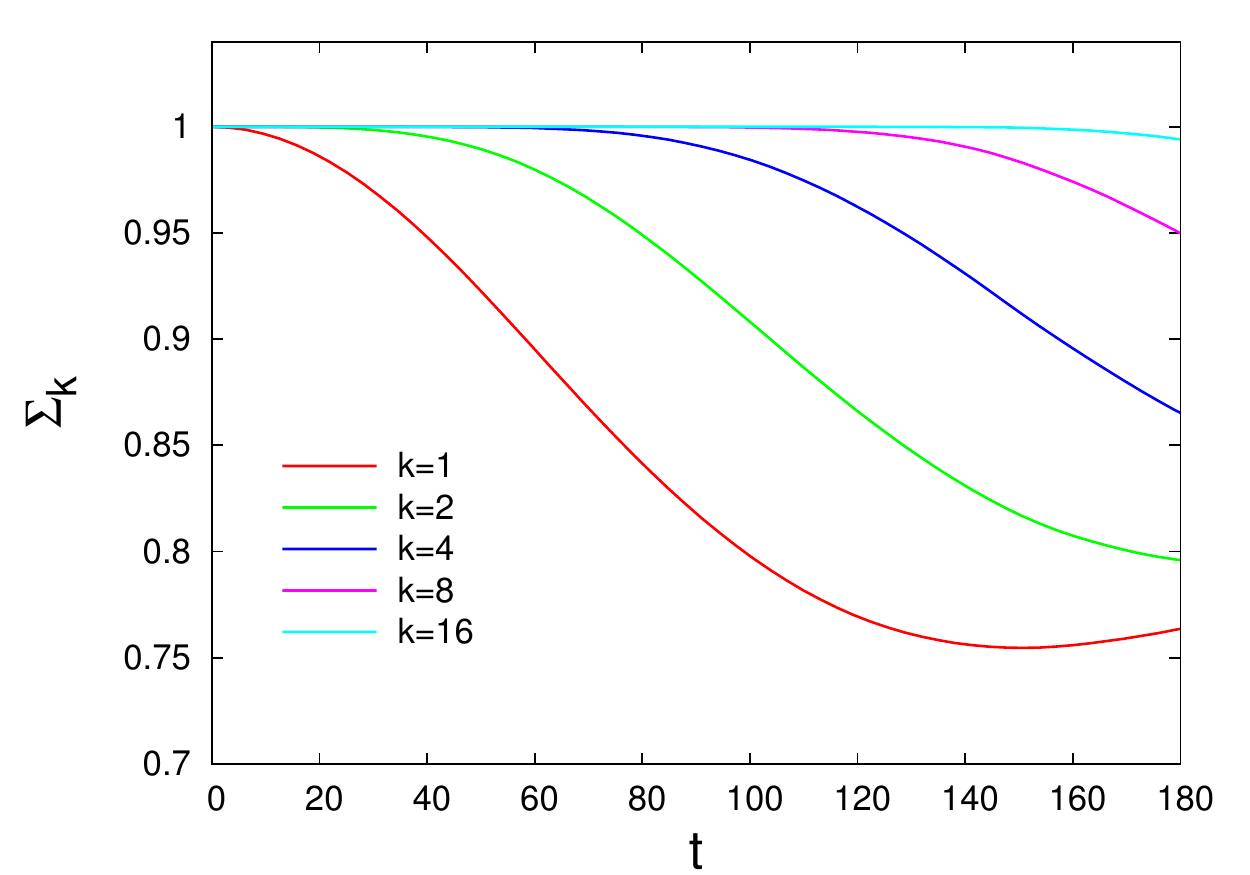}
   \caption[Turbulent cascade]{Fraction of the total energy contained in the first $k+1$ modes $\Sigma_k =
      \sum_{j=0}^k E_j$, for an initial excitation of the $j=0$ and $j=1$ modes (equation \eqref{twomode}). The energy in the first sixteen modes
      $\Sigma_{16}$ is almost constant while the energy in the first modes $\Sigma_1$, $\Sigma_2$, etc., decreases, which means
      that the energy is flowing to higher modes. Credits: \cite{Bizon11}.}
      \label{turbu}
\end{myfig}

\subsection{The anti-de Sitter instability conjecture}

The important features of the simulations are the following:
\begin{myitem}
   \item a black hole is formed however small the initial amplitude of the perturbation is,
   \item the formation time $t_H$ of the apparent horizon scales like $O(\varepsilon^{-2})$,
   \item energy is flowing from low to high spatial frequencies during evolution.
\end{myitem}

The first and third properties gave rise to the so-called weakly turbulent behaviour. This lead to the following instability
conjecture \cite{Bizon14}:

\begin{conjecture}[Anti-de Sitter instability]
\gls{ads} is unstable against black hole formation for a large class of arbitrarily small perturbations.
\end{conjecture}

In \cite{Friedrich14}, it was argued that the reflective boundary conditions were a key ingredient of the
instability precisely because of the absence of a decay condition for perturbations. Let us also mention that a diverging growth
of the frequency of fluctuations was already observed in \cite{Greenwood10} in a \gls{ads} space with a big crunch scenario. 

\subsection{Other features of the instability}

From the geodesic analysis (figure \ref{geodesics} of chapter \ref{aads}), it is expected that the crossing time of a massless scalar wave packet
is $\sim \pi \gls{L}$. However, the authors of \cite{Garfinkle12} observed that the time to form a black
hole after one reflection was slightly larger than the time needed for null geodesics to do one round trip in \gls{ads} space-time.
Non-linearities thus tend to slow down the massless scalar field. A snapshot of the time-radial plane of the evolution of the
scalar field is pictured in figure \ref{timeradial} for both direct and delayed collapse.

\begin{myfig}
   \includegraphics[width = 0.49\textwidth]{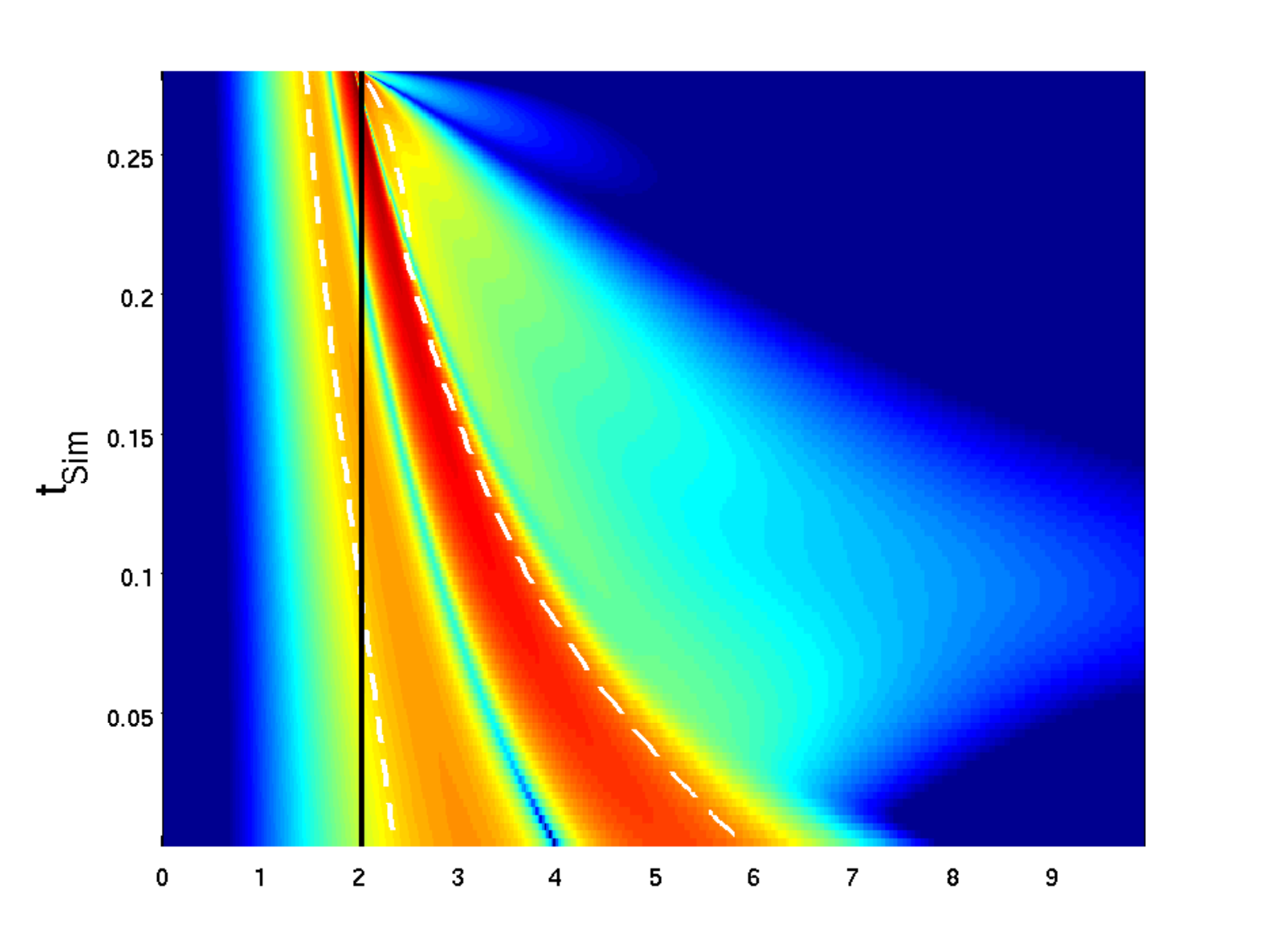}
   \includegraphics[width = 0.49\textwidth]{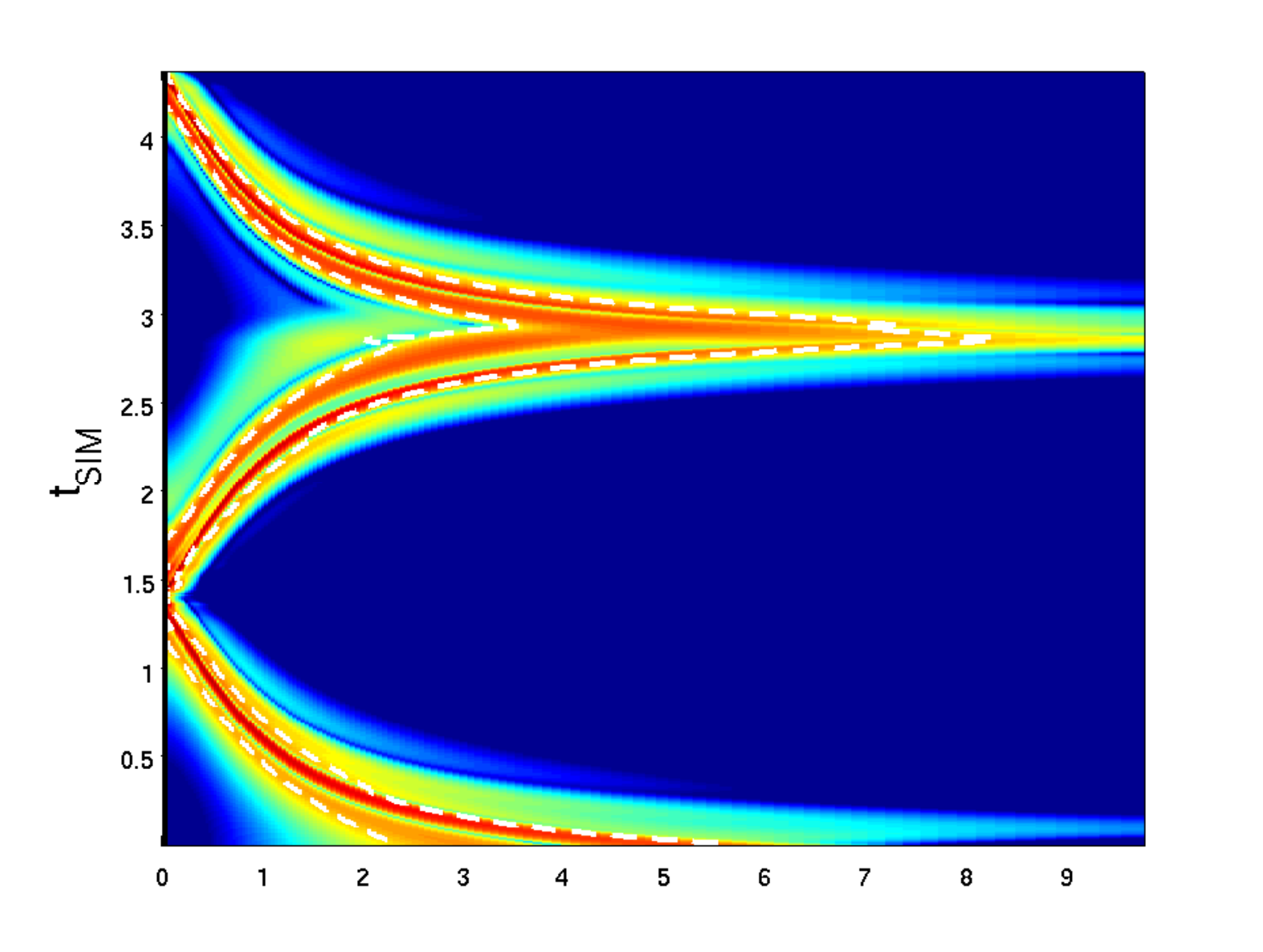}
   \caption[Space-time diagrams of massless scalar field collapse]{Time-radial planes during the evolution of a
      Gaussian massless scalar field initial data. The radii are in units of $\gls{L}$, and colour encodes the scalar energy density.
      Vertical black lines indicate the apparent horizon $r_H$ appearing at the end of the simulation. The initial amplitude is
      150 times smaller on the right panel and the corresponding apparent horizon lies at $r_H = 0.0049 \gls{L}$ after one bounce
      off the \gls{ads} boundary. This figure is highly reminiscent of figure \ref{geodesics} for null geodesics. Credits: \cite{Garfinkle12}.}
   \label{timeradial}
\end{myfig}

The instability in five dimensions was first investigated in \cite{Garfinkle11}. In contradiction with \cite{Bizon11}, they did
not observe black hole formation below a certain amplitude threshold. However, this was only due to their spatial resolution (6400
points): the apparent horizon $x_H$ could not be resolved with such a few number of points. This was demonstrated in
\cite{Jalmuzna11} whose authors used $2^{17} \sim 130 000$ spatial points.

The instability is not only present in four and five dimensions, but in all dimensions. Indeed, in \cite{Jalmuzna11}, it was
demonstrated that the spectrum of the linear operator $\widehat{L}$ was resonant in all dimensions, with equally spaced
eigen frequencies. The perturbative approach always gives rise to secular resonant terms at third order, indicating a
breakdown at time $t = O(\varepsilon^{-2})$, independent of the number of dimensions.

More and more arguments were gathered suggesting that the instability was systematic, i.e.\ independent of the initial data. For
example, the instability was recovered for complex scalar fields \cite{Buchel12,Liebling13}. Let us also mention the Vaidya setup
of \cite{Abajo14,Silva15} where the Gaussian wave-packet is replaced by a Gaussian shell with initial data:
\begin{equation}
   \Phi(0,x) = 0 \quad \tn{and} \quad \Pi(0,x) = \varepsilon \exp\left( -\frac{\tan^2\left(\frac{\pi}{2} - x\right)}{\sigma^2} \right),
\end{equation}
where the scalar field is initially concentrated close to the \gls{ads} boundary.

The weakly turbulent behaviour is not solely an intrinsic property of \gls{ads}, but it was also observed in flat space-time enclosed in a
cavity. The underlying idea is that the reflective boundary conditions of \gls{ads} space-times can be mimicked by a flat
space-time with appropriate boundary conditions at a finite radius. However, this analogy holds only for spherically
symmetric distributions. Indeed, the bouncing time is not isotropic in such a cavity if the matter distribution is not spherically
symmetric, whereas \gls{ads} space-time is strictly isotropic, whatever is the initial distribution. In \cite{Maliborski12},
Dirichlet boundary conditions were imposed at a finite radius $R$ of Minkowski space-time, with Gaussian initial data. Again,
arbitrarily small amplitudes lead to black hole formation after potentially several bounces on the $r = R$ wall.  Figure
\ref{energy_spec} shows the spectrum $E_j$ versus $j$ of the data at different times of the evolution. The turbulent cascade
toward high-$j$ modes is visible and just before black hole formation, the spectrum approaches a power law of exponent $\alpha
\sim -1.2$. This value seems universal since it is independent of the functional form of the initial data and is also observed in
the 4-dimensional \gls{aads} case.

\begin{myfig}
   \includegraphics[width = 0.49\textwidth]{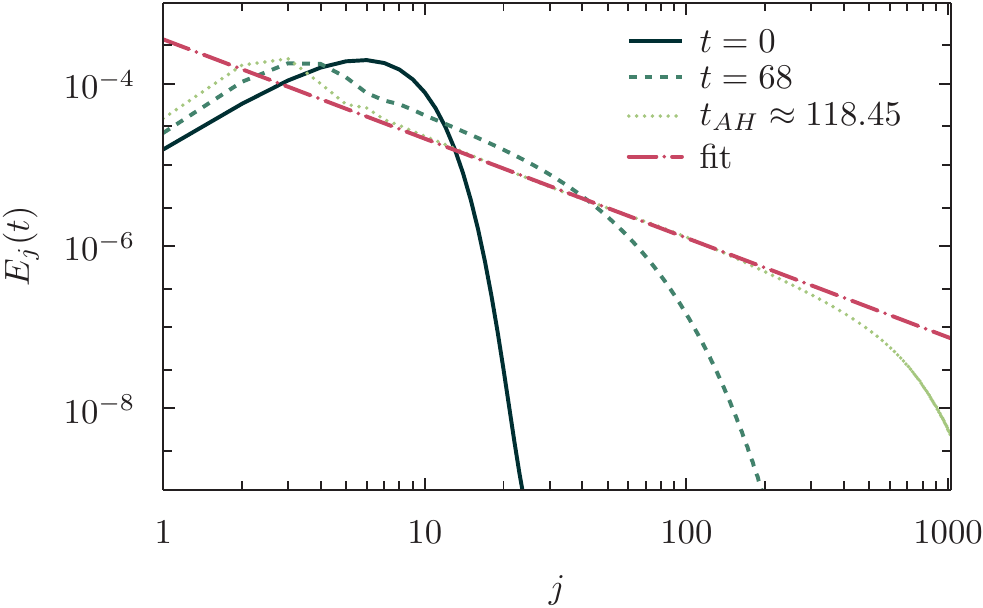}
   \caption[Energy spectrum before collapse]{Energy spectrum at different times before collapse of a solution undergoing several
      dozens of reflections in flat space-time enclosed in a cavity. Just before the apparent horizon formation at time $t_{AH}$,
      the spectrum is approximately a power law with a universal exponent $\alpha \sim -1.2$. Credits: \cite{Maliborski12}.}
   \label{energy_spec}
\end{myfig}

Finally, let us observe that the vocabulary of turbulence interfered with the \gls{ads} instability field mainly because of the cascade
of energy to higher spatial frequencies. The analogy with fluid turbulence was even pushed further with the study of time frequencies
in \cite{Oliveira13}. In particular, it was shown that the power spectrum of the Ricci scalar at the origin was characterised by a
\gls{kz} power spectrum, i.e.\
\begin{equation}
   P(\omega) = \omega^{-s}\quad \tn{with} \quad s = 1.7 \pm 0.1,
\end{equation}
$\omega$ being the time frequency. The numerical value of $s$ seems to be universal as it holds in both four and five dimensions with
Gaussian or 20-mode initial data.

\section{Black hole formation in AdS space-times}

The details of black hole formation in \gls{gr} are in striking analogy with phase transitions. This analogy was uncovered by the
seminal work of Choptuik and collaborators \cite{Choptuik93,Abraham93} in asymptotically flat space-times. Since black holes form
more easily in \gls{aads} space-times, and given that there are several branches of black hole formations, it is legitimate to ask how
close to the asymptotically flat case the formation of black hole with a negative cosmological constant is.

\subsection{Critical phenomena in AAdS space-times}
\label{critical}

From an historical perspective, it is important to mention the numerical simulations by Pretorius and Choptuik
in 2000 \cite{Pretorius00} and Husain and collaborators in 2003 \cite{Husain03} in \gls{aads} space-times. In these papers, the
authors evolved in time a spherically symmetric Gaussian wave packet initial data made of a massless scalar field. Like in the asymptotically
flat case, if the scalar wave packet amplitude $\varepsilon$ is larger than a critical value $\varepsilon_\star$, then the scalar
field collapses directly to a black hole with apparent horizon $r_H$. This corresponds to the far right curve of figure
\ref{adsinstability} (left panel). The goal of these papers, however, was to characterise the critical behaviour of black hole
formation. Namely, on the point of collapse, for amplitudes $\varepsilon \gtrsim \varepsilon_\star$, the authors observed that
the apparent horizon radius $r_H$ was governed by
\begin{equation}
   \ln r_H \sim \gamma_r \ln (\varepsilon - \varepsilon_\star) + r_0 + F_r(\ln(\varepsilon - \varepsilon_\star)),
\end{equation}
where $r_0$ is a constant and $F_r$ a sinusoidal function of period (or echoing period) $\Delta_r$. This is a typical feature of
critical phenomena and phase transitions, illustrated on the left panel of figure \ref{critic}. From the numerical simulations, it
was measured that
\begin{subequations}
\begin{align}
   \gamma_r &\simeq 1.2 \quad \tn{in 3 dimensions},\\
   \gamma_r &\simeq 0.37 \quad \tn{and} \quad \Delta_r \simeq 3.44 \quad \tn{in 4 dimensions.}
   \label{criticscalar}
\end{align}
\end{subequations}
These values are universal, in the sense that they are independent of the value of the cosmological constant \gls{Lambda} and of the
functional form of the initial scalar wave packet. They also match the values found earlier in asymptotically 4-dimensional flat
space-times. This is to be expected since the physics probed is quite local (and thus independent of the asymptotics) for the formation
of arbitrarily small black holes. Unfortunately, at the time, the authors of \cite{Husain03} were not interested in values of
$\varepsilon$ smaller than $\varepsilon_\star$, so they missed the breakthrough of the so-called \gls{ads} weakly turbulent
instability.

The three dimensional case of \cite{Pretorius00} is particular because black holes have a minimum mass below which their formation is strictly
impossible (this is the so-called \gls{btz} metric \cite{Banados92}). It was thus observed in \cite{Pretorius00} that even if
non-linearities build up in such a setting, the scalar field does not collapse even after several bounces off the \gls{ads}
boundary. Still, non-linearities build up and a sub-pulse structure emerges, i.e.\ the initial Gaussian profile breaks off into
several distinct wave packets, as shown on the right panel of figure \ref{critic}.

\begin{myfig}
   \includegraphics[width = 0.49\textwidth]{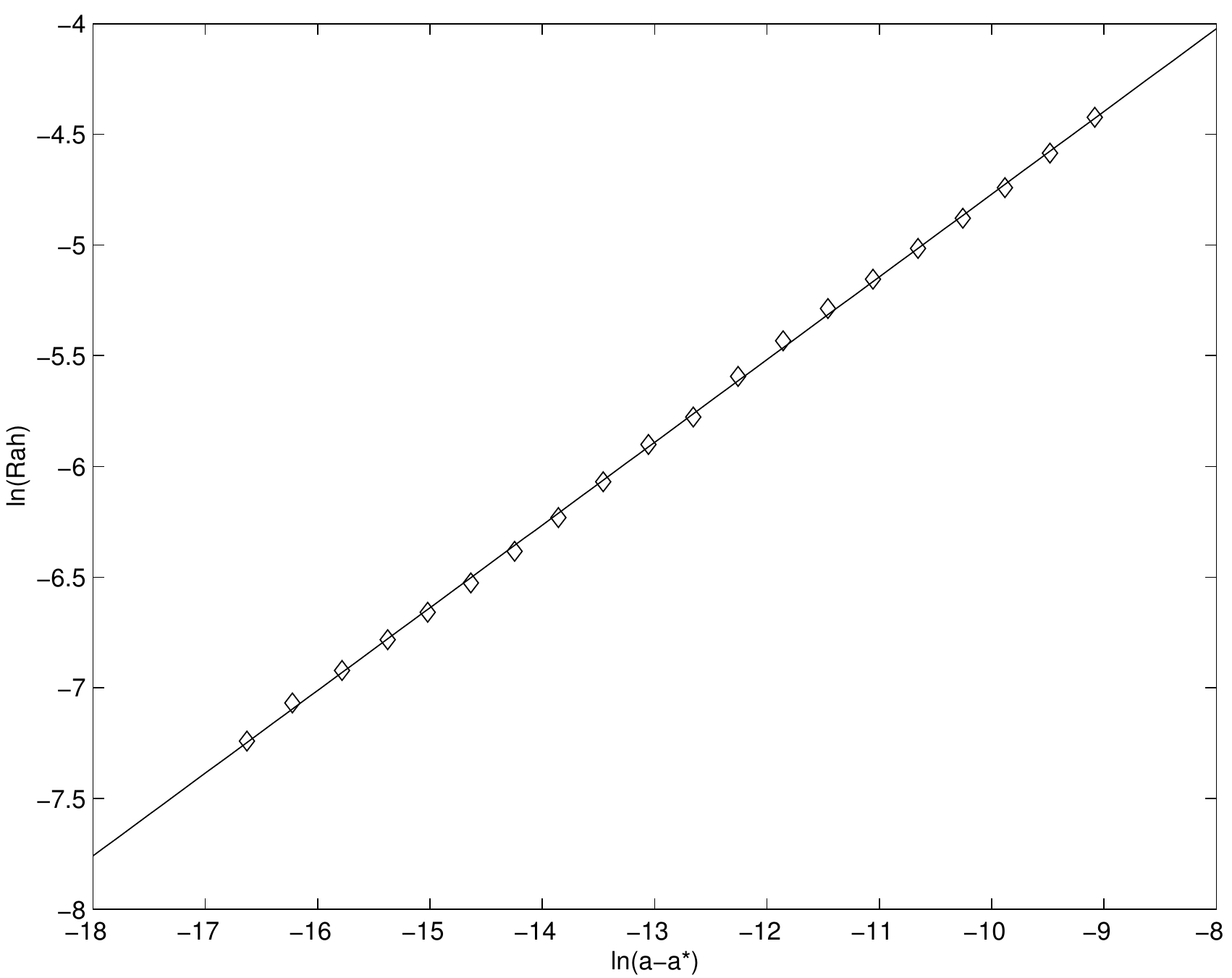}
   \includegraphics[width = 0.26\textwidth]{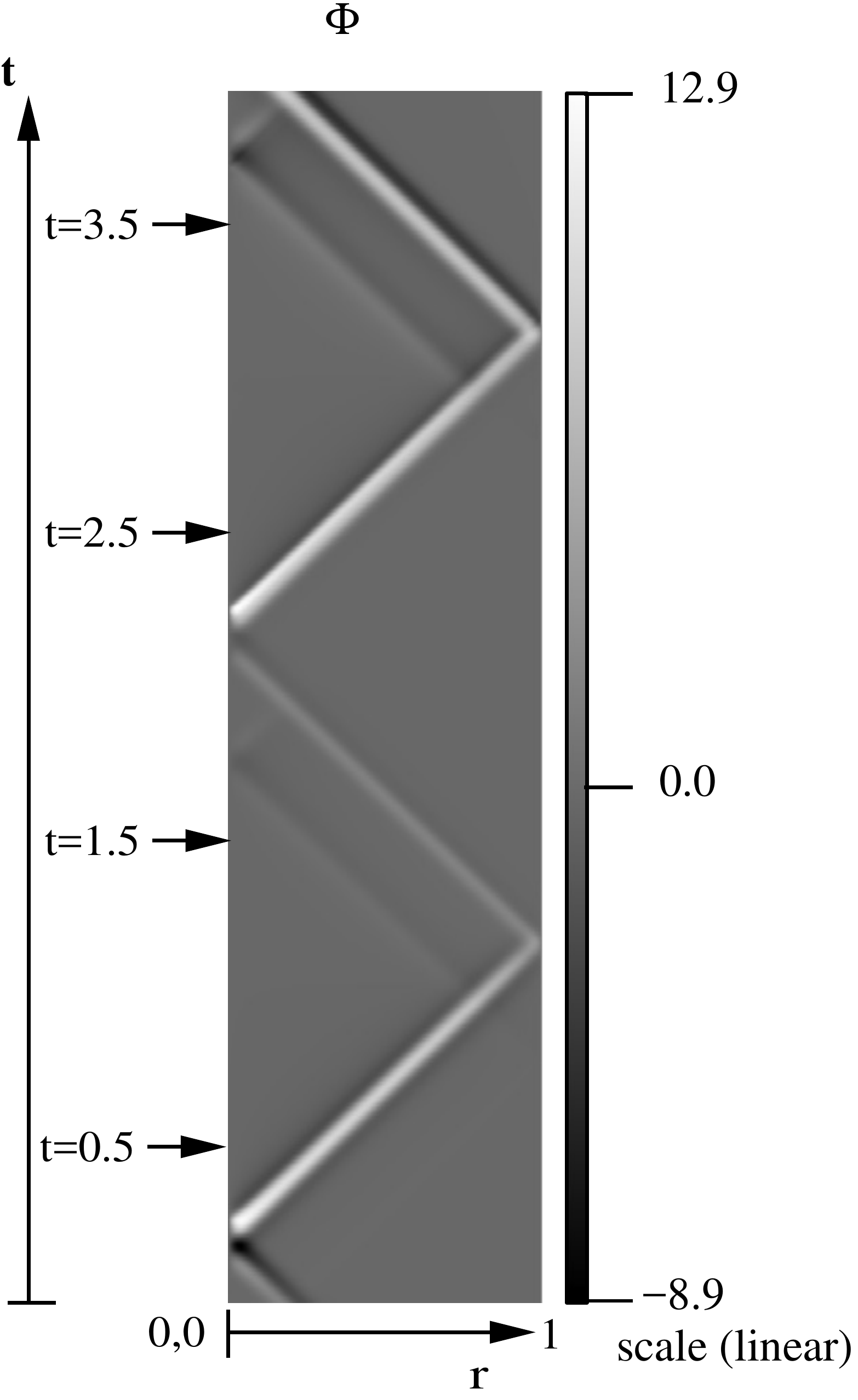}
   \caption[Critical phenomena]{Left: log-log plot of the apparent horizon $r_H$ as a function of the amplitude
      $\varepsilon -\varepsilon_\star$ (denoted by $a$ in the label) of the initial data. Small oscillations of echoing period
      $\Delta$ around the straight line of slope $\gamma$ can be spotted. Right: space-time diagram of the scalar field profile in 3-dimensional
      \gls{ads} with Gaussian initial data. The amplitude is smaller than the threshold of black hole formation, so no black hole
      can appear. Credits: \cite{Husain03,Pretorius00}.}
      \label{critic}
\end{myfig}

In the \gls{ads} instability context, the authors of \cite{Bizon11} checked that in the right neighbourhood of each critical
amplitude $\varepsilon_n$, i.e.\ for $\varepsilon \gtrsim \varepsilon_n$ (see figure \ref{adsinstability}), the power-law
behaviour of \cite{Choptuik93,Abraham93,Husain03} was recovered, namely for $\varepsilon \gtrsim \varepsilon_n$:
\begin{equation}
   x_H \sim (\varepsilon - \varepsilon_n)^{\gamma_r} \quad \tn{with} \quad \gamma_r \sim 0.37.
\end{equation}

These critical phenomena associated to the \gls{ads} instability were refined further in \cite{Olivan16a,Olivan16b}. The authors
were able to resolve precisely the apparent horizon formation and looked at the fine structure of critical collapse. Unlike
previous studies, they focused on the left neighbourhood of critical points that only exist in \gls{ads} space-time. Denoting by
$M_g^{n+1}$ the mass gap at which starts the left branch, and $\varepsilon_n$ the corresponding critical amplitude of initial data
undergoing $n$ bounces, they have shown that in the left neighbourhood of critical points ($\varepsilon \lesssim \varepsilon_n$)
the black hole mass $M_H$ was obeying
\begin{equation}
   M_H - M_g^{n+1} \propto (\varepsilon_n - \varepsilon)^{\xi},
   \label{critleft}
\end{equation}
with $\xi \sim 0.7$. This value of $\xi$ is universal, i.e.\ independent of the number of bounces $n$, and of the functional form
of the initial data. Moreover, looking at the maximal value of Ricci scalar at the origin $R_{max}(x=0)$ on the point of collapse,
they observed that in the left neighbourhood of a critical point ($\varepsilon \lesssim \varepsilon_n$)
\begin{equation}
   \ln R_{max}(x=0) = -2\gamma_l \ln (\varepsilon_n - \varepsilon) + b_0 + F_l(\ln(\varepsilon_n - \varepsilon)),
\end{equation}
where again $F_l$ is a sinusoidal function with echoing period equal to $\Delta_l$. The parameters $\gamma_l$ and
$\Delta_l$ are related to their right branch counterparts by (see equation \eqref{criticscalar})
\begin{equation}
   \gamma_l = \gamma_r \quad \tn{and} \quad \Delta_l = \frac{\Delta_r}{2\gamma_r}.
   \label{leftright}
\end{equation}
This feature is called the discrete self-similarity near the critical points in \gls{ads} space-time and is best illustrated in figure
\ref{selfsim}.

\begin{myfig}
   \includegraphics[width = 0.55\textwidth]{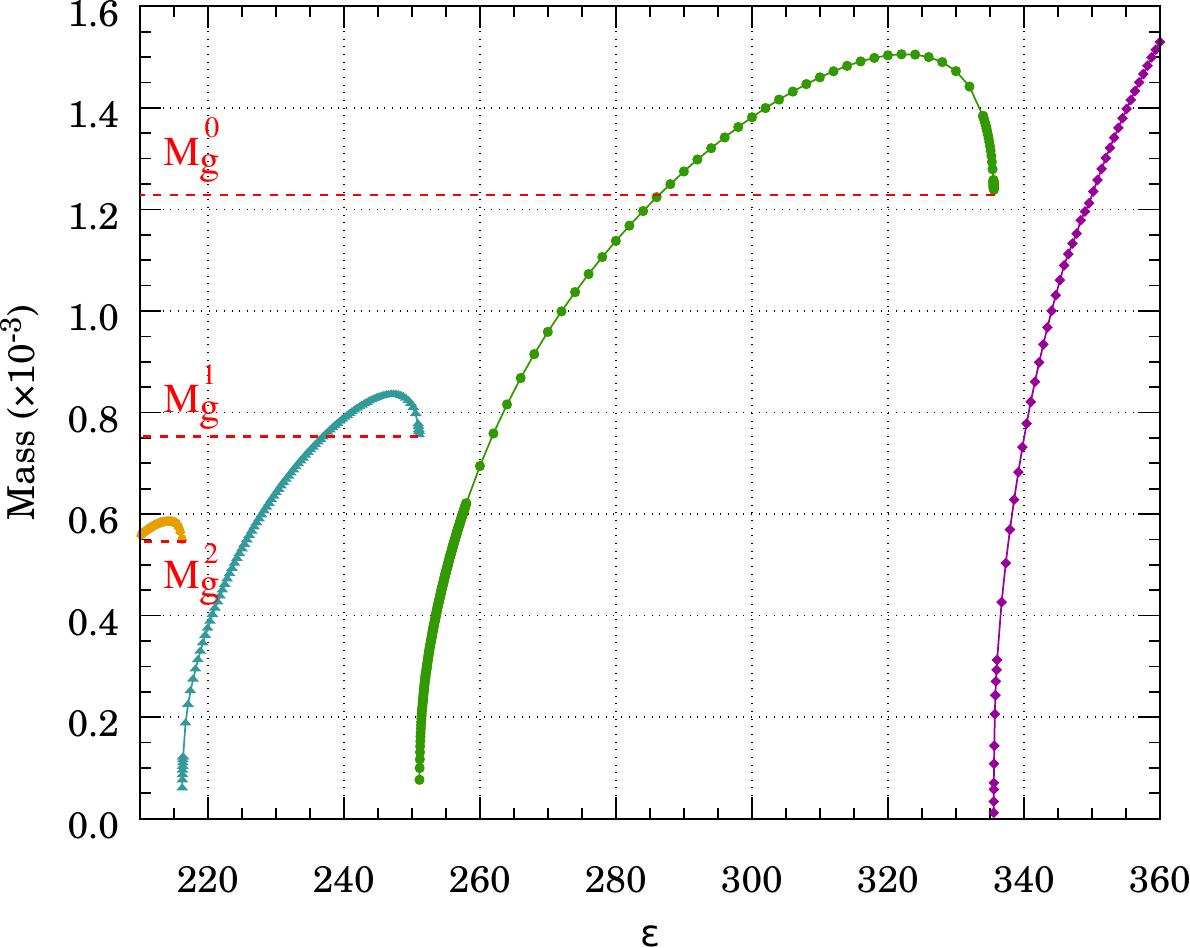}
   \includegraphics[width = 0.44\textwidth]{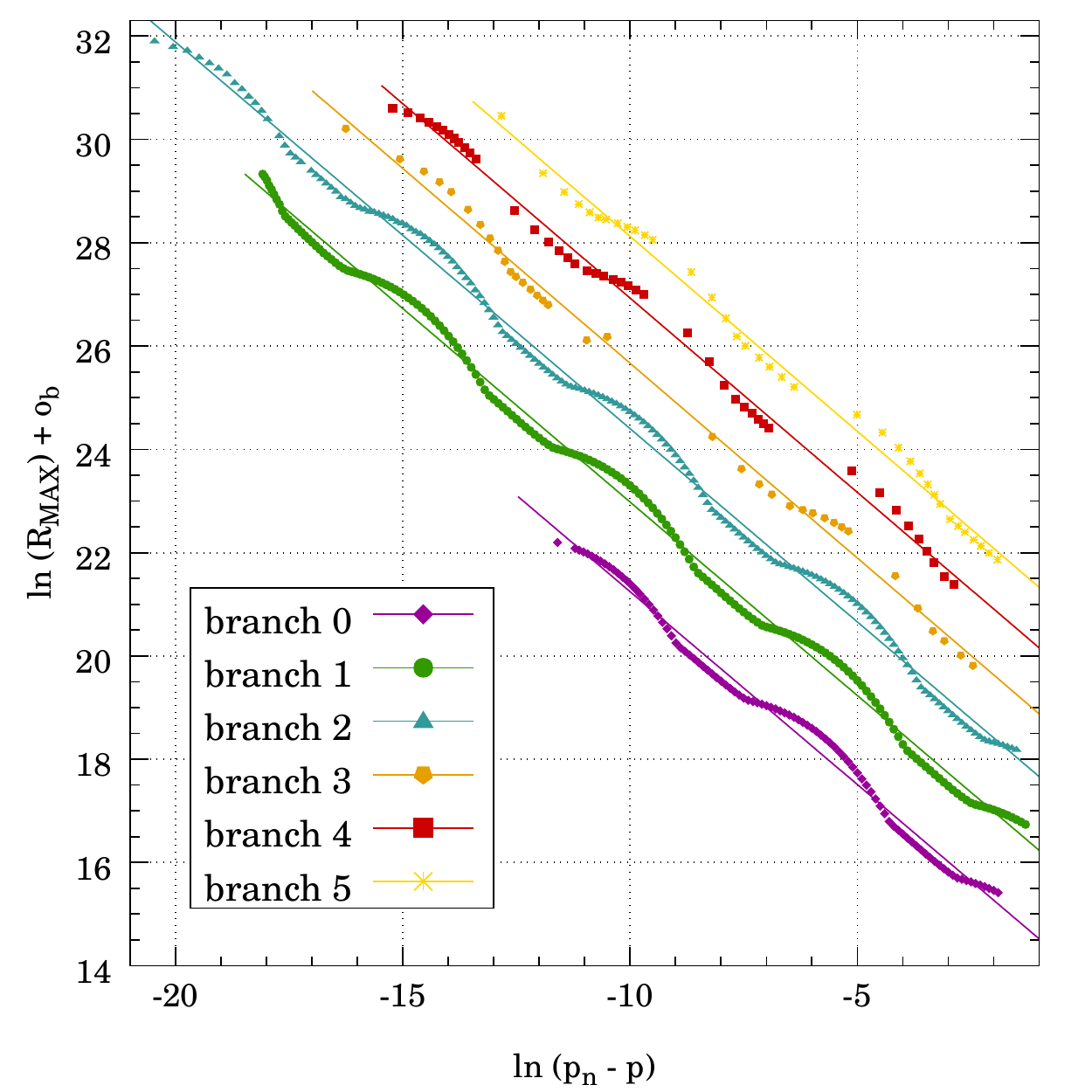}
   \caption[Discrete self-similarity near a critical point]{Left panel: mass of the apparent horizon as a
      function of the amplitude of the initial data. In the right neighbourhood of a critical point,
      $M_H \sim (\varepsilon - \varepsilon_n)^{\gamma_r}$ while in the left neighbourhood of a critical point $M_H - M_g^{n+1}
      \sim (\varepsilon_n - \varepsilon)^\xi$. Right panel: critical behaviour of the Ricci scalar at the origin $R_{max}(x=0)$
      for fixed width $\sigma$ of the initial data and for different branches: from direct collapse ($b = 0$) to five-bounce
      collapse ($b=5$). An offset $o_b$ has been added to distinguish between the curves. In the labels, $p$ stands for $\varepsilon$.
      Credits: \cite{Olivan16b}.}
   \label{selfsim}
\end{myfig}

Thus, critical phenomena in \gls{aads} space-times are much richer than in the asymptotically flat case. First, there is an
infinity of black holes formation channels indexed by the number of bounces. Second, each critical point has not only a right
branch but also a left branch (attached to a mass gap), which are related to each other by \eqref{leftright}.

\subsection{Singularity theorems in AAdS space-times}

Given the strength of the instability conjecture, a natural question is to know whether we can prove it via a singularity
theorem. In the asymptotically flat case, singularity theorems were proved by Hawking and Penrose \cite{Hawking73}. In a
simplified formulation, the theorems imply that if
\begin{myenum}
   \item the null energy conditions holds, i.e.\ $\forall v, v_{\mu}v^{\mu} \geq 0, R_{\mu\nu}v^\mu v^\nu \geq 0$,
   \item the strong causality or chronology conditions holds, i.e.\ there exists no closed time-like curve,
   \item there exists a region of strong gravity, i.e.\ a closed trapped surface,
\end{myenum}
then the space-time is not time-like nor null geodesically complete, i.e.\ there exist some geodesics that never reach infinity.
Said differently, the space-time is singular and can exhibits a black hole or a naked singularity.

These conditions are discussed in the \gls{aads} case in \cite{Ishibashi12}. In view of the instability conjecture,
condition (c) has to be removed, as it was numerically observed that even weak gravity initial data leads to black hole formation.
However, eliminating this condition rises difficulties that were discussed in details in \cite{Ishibashi12} but not
overcome. Nonetheless, the authors managed to give sufficient conditions for a singularity to form in the simplified case of a
perfect fluid in spherical symmetry, by examining carefully the Raychaudhuri's equation. Notably, the naked singularity formation
was not excluded.

Very recently, the author of \cite{Moschidis17a,Moschidis17b} mathematically and rigorously proved that the spherically symmetric
Einstein-radial massless Vlasov system was non-linearly unstable against black hole formation. This setup, also called the
Einstein-null dust system, is a simplified model of the \gls{ekg} equations, and can be seen as a high frequency limit of the
latter (some non-linear terms being dropped out). This tour de force can be interpreted as the very first proof of the \gls{ads}
instability conjecture in the simplest possible setting, and as such as a specific singularity theorem.

No other attempt of singularity theorem demonstrations in \gls{ads} has been attempted to the best of our knowledge. And indeed,
it seems that the formation of black holes is not universal. Several islands of stability\footnote{This denomination was
originally coined in \cite{Dias12b}.} were found in the literature, namely
families of non-linearly stable initial data that never collapse.

\section{Quenching the turbulent cascade}

Black holes form for arbitrarily small amplitudes in a very large number of cases (as we have seen in the previous sections),
but is it mandatory? If the instability conjecture was confirmed many times, an even more challenging problem was to find
solutions that resisted black hole formation and circumvented the instability.

\subsection{The hard wall model}

Black hole formation is expected to occur when the energy get concentrated in such a small region that an apparent horizon can form.
The region where such a focus of energy is favoured is obviously the origin in spherical symmetry. What happens then to the
instability if we prevent the scalar field to ever reach the origin? This question was tackled in \cite{Craps14c,Silva16} with
the so-called hard wall implementation. Namely, the authors placed a wall at a radial coordinate $z = z_0$ while the \gls{ads}
boundary lied at $z = 0$. The scalar field could only move between these two boundaries, as pictured in figure \ref{hwall}. Furthermore, on
the hard wall, Dirichlet or Neumann boundary conditions were imposed, and both the 3 and 4-dimensional cases were studied.

\begin{myfig}
   \includegraphics[width = 0.49\textwidth]{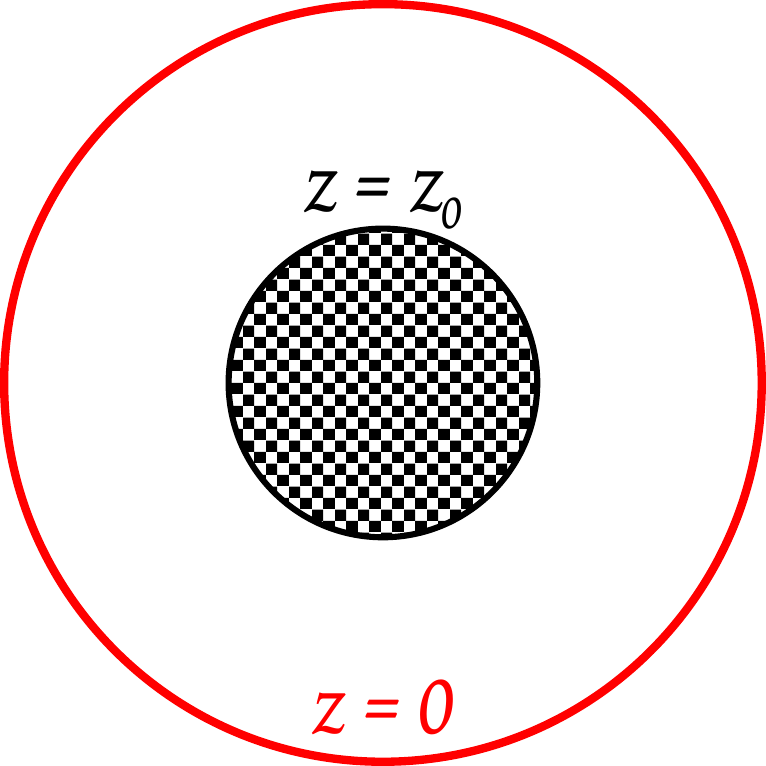}
   \caption[The hard wall model]{In the hard wall model, a wall, i.e.\ Dirichlet or Neumann boundary conditions, is enforced at a
      coordinate $z_0$. The coordinate $z$ is an inverse radius, so that the \gls{ads} boundary lies at $z=0$. The chessboard
      patterned region is forbidden, so that the scalar field is restricted to move only between the wall and the boundary,
   bouncing back and forth between the two. Credits: G. Martinon.}
   \label{hwall}
\end{myfig}

The only input of data was performed via time-dependent energy injection on the \gls{ads} boundary\footnote{For example in
\cite{Krishnan16} the authors advocate that a more natural boundary condition for \gls{ads} is to hold the renormalised
boundary stress tensor fixed, instead of the boundary metric.} imposing
\begin{equation}
   \phi(z=0,t) = \varepsilon e^{-\frac{t^2}{\delta t^2}}.
\end{equation}
The authors observed that for small enough amplitudes $\varepsilon$, the scalar pulse generated by the energy injection bounced
forever back and forth between the \gls{ads} boundary and the hard wall. For amplitudes larger than some threshold
$\varepsilon_0$, a black hole formed with a horizon smaller than $z_0$ (i.e.\ larger than the radius of the wall). The intuition is
that a black hole is formed if the black hole that would be formed in ordinary \gls{ads} space-time (without a hard wall) has
its event horizon outside the hard wall. Otherwise the infalling shell is scattered back by the hard wall before it reaches its
Schwarzschild radius.

This behaviour was observed in all dimensions, with all boundary conditions considered, and is summarised in figure \ref{phasewall}.
These results were sustained by both analytical and numerical arguments. In particular, the frequencies of the linearised modes
did not display obvious resonances, except in the case of Neumann boundary conditions in four dimensions. The hard wall model thus provides a
defocusing mechanism that can suppress the turbulent cascade below a certain amplitude threshold. This is in deep contrast to the
instability conjecture that precisely deny the existence of such a threshold.

\begin{myfig}
   \includegraphics[width = \textwidth]{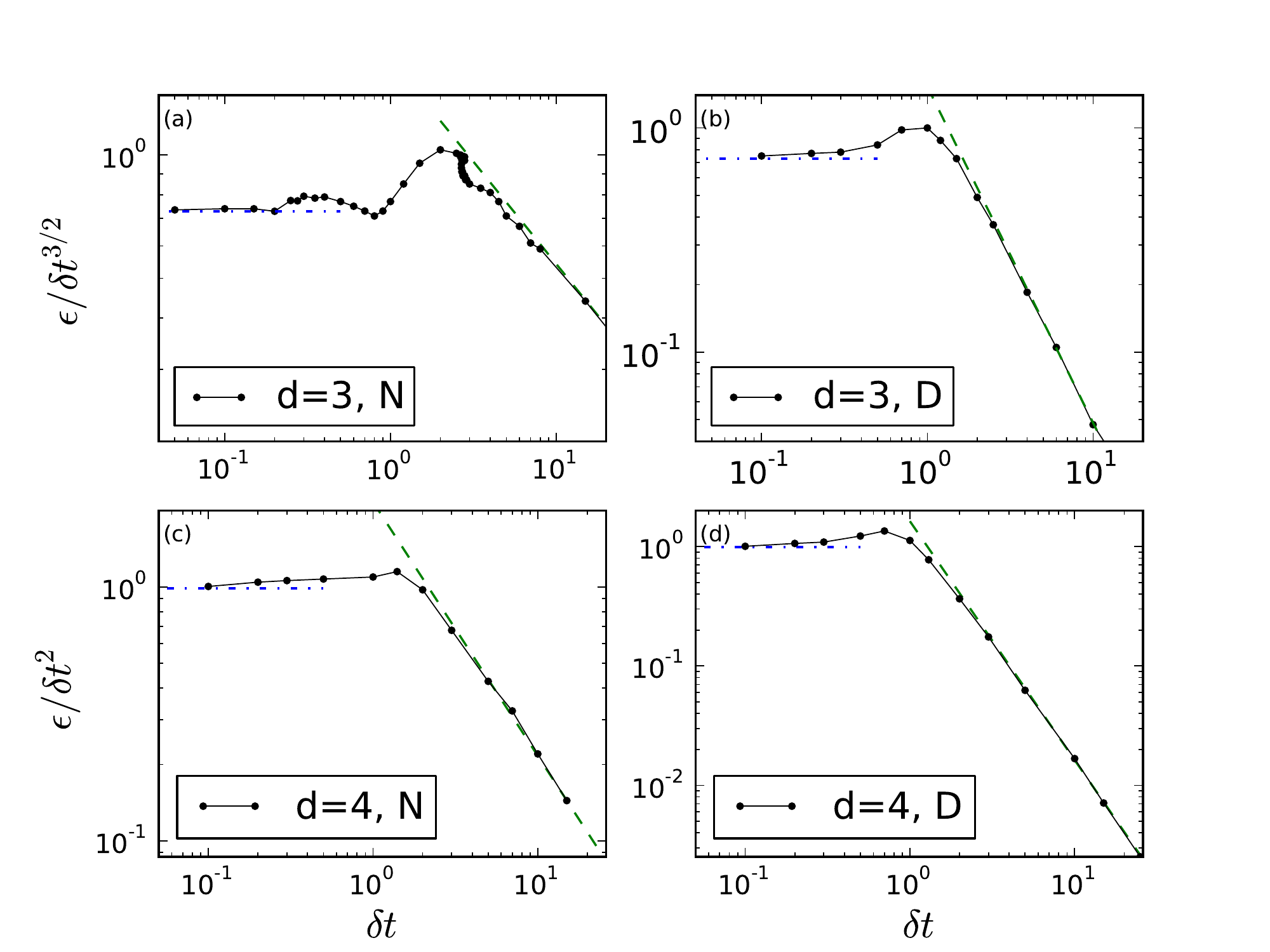}
   \caption[Dynamical phase space diagram of the hard wall model]{Dynamical phase space diagram for black hole formation in
      the hard wall model. The dimension is $d = 3$ or $4$ and Dirichlet (D) or Neumann (N) boundary conditions are imposed on the
      wall. Black holes are formed for parameters above the dots computed numerically, while the scattering phase occurs below. Straight lines
      correspond to analytical limits. Credits: \cite{Craps14c}.}
   \label{phasewall}
\end{myfig}

\subsection{Time-periodic solutions}
\label{tpsol}

A few years before the instability conjecture was established, the quest for building black holes with scalar hairs in \gls{ads}
was triggered by \cite{Basu10}. In this work and its extensions (see e.g.\ \cite{Gentle12,Dias12c,Dias17b}), Reissner-Nordström
black holes surrounded by a spherically symmetric charged scalar field, in what is called the \gls{ah} model, were built either
perturbatively or numerically. These configurations have at least two parameters: one drives the size of the horizon and another
drives the amplitude of the scalar cloud. In this formalism, taking the zero-size limit of the horizon leads naturally to boson
stars (when the scalar field is complex and massive) or charged scalar periodic solutions, dubbed solitons. These solutions obtained in
\cite{Astefanesei03,Basu10,Gentle12,Dias12c,Dias17b} were the very first \gls{aads} time-periodic solutions to emerge in the
literature, even before the instability conjecture was formulated. Furthermore, there were clues that these solutions were stable
against linear perturbations. Apart from boson stars (see \cite{Dias12b} and section \ref{beyondspher} below), the non-linear
stability of these solitons was never investigated though.

The first numerical evolutions of non-collapsing solutions in full \gls{aads} space-times appeared concomitantly in
\cite{Maliborski13b,Buchel13}, in the spherically symmetric \gls{ekg} setup. In \cite{Buchel13}, the authors called them
``boson stars'', but since their scalar field was massless, this denomination was not strictly correct. The authors found
initial conditions that were immune to the non-linear instability below some amplitude threshold $\varepsilon_0$. These
solutions exhibited a power-law spectrum (in terms of pseudo-spectral coefficients) for collapsing solutions while non-collapsing
configurations featured an exponential spectrum. In particular, the authors observed that Gaussian initial data with large width
$\sigma \gtrsim 0.4$ did belong to these non-collapsing solutions, i.e.\ their collapsing time diverged to infinity at a finite
amplitude. This is illustrated in figure \ref{largesigma}. The argument was that widely distributed mass energy prevented the energy to flow to
smaller and smaller scales. Instead, the energy was perpetually exchanged between the first modes. However, this claim was tempered
by \cite{Maliborski13a} whose authors confirmed the results but did reignite instability for even larger values of the width parameter
$\sigma \gtrsim 8$. A similar behaviour was uncovered in the Einstein-Maxwell setup in \cite{Arias16}.

\begin{myfig}
   \includegraphics[width = 0.49\textwidth]{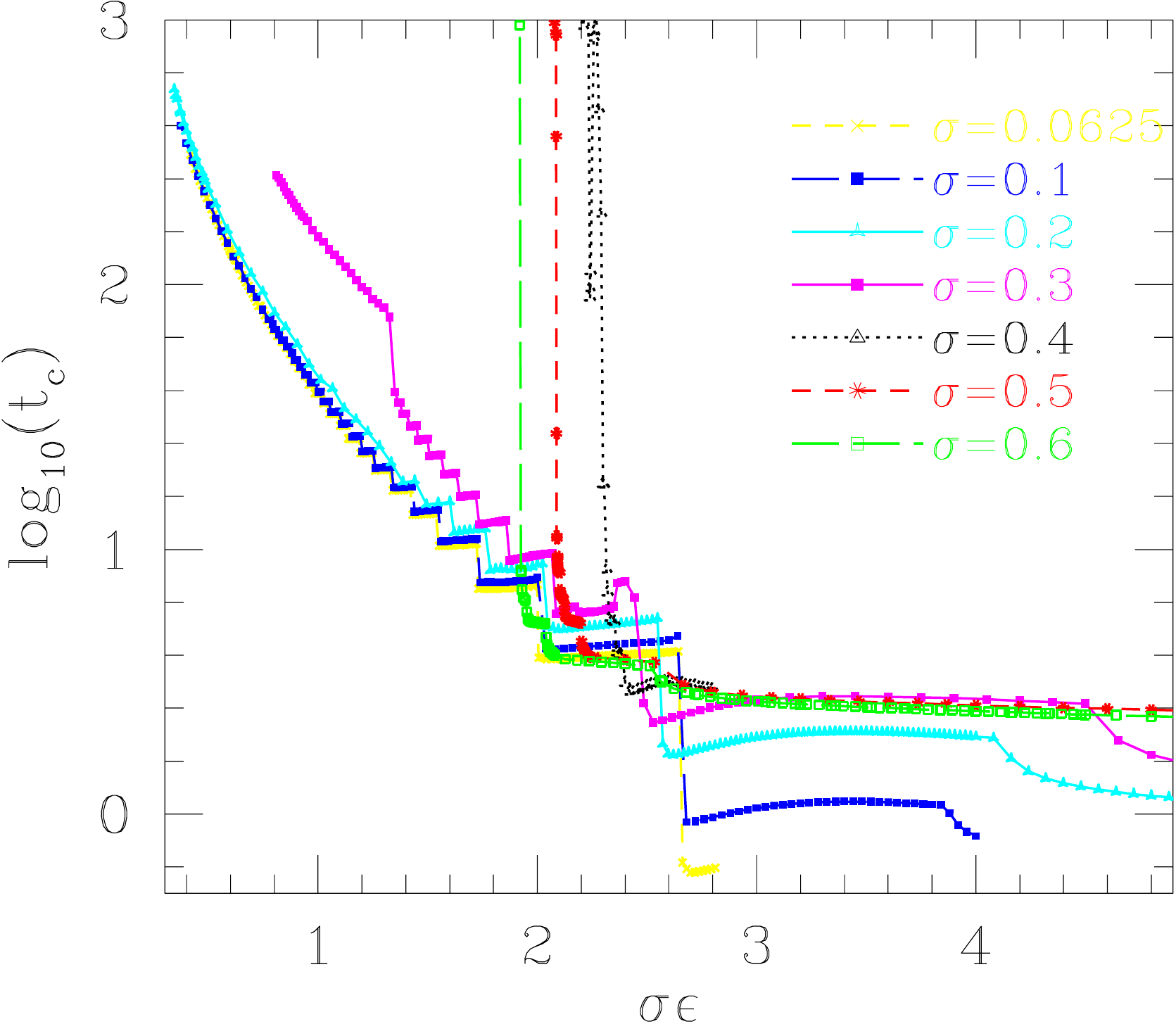}
   \caption[Gaussian initial data with large width]{Collapse times for Gaussian initial data of amplitude $\varepsilon$ and
      width $\sigma$. For large $\sigma$, no collapse is observed, i.e.\ the time of collapse $t_c$ diverges to infinity at a
      small but non-zero amplitude. For small $\sigma$, the instability is recovered. In comparison, Bizo\'n and Rostworowski
      \cite{Bizon11} originally used $\sigma = 1/16 = 0.0625$. Credits: \cite{Buchel13}.}
   \label{largesigma}
\end{myfig}

Almost simultaneously in \cite{Maliborski13b}, time-periodic solutions were constructed and shown to be non-linearly stable with
the help of spectral methods. The authors started by constructing perturbatively periodic solutions, that they used as a seed for a
Newton-Raphson solver that could find fully non-linear generalisations. They then plugged the result into an evolution code and
monitored the phase space of spectral coefficients. In particular, they proved that high coefficients remained bounded, as shown
in figure \ref{phasespace}, demonstrating that no turbulent cascade was at play. The perturbed solution was not periodic any more
but quasi-periodic with orbits close to the perturbative periodic solution. This clearly highlighted the existence of a stable periodic
attractor immune to the non-linear stability.

\begin{myfig}
   \includegraphics[width = \textwidth]{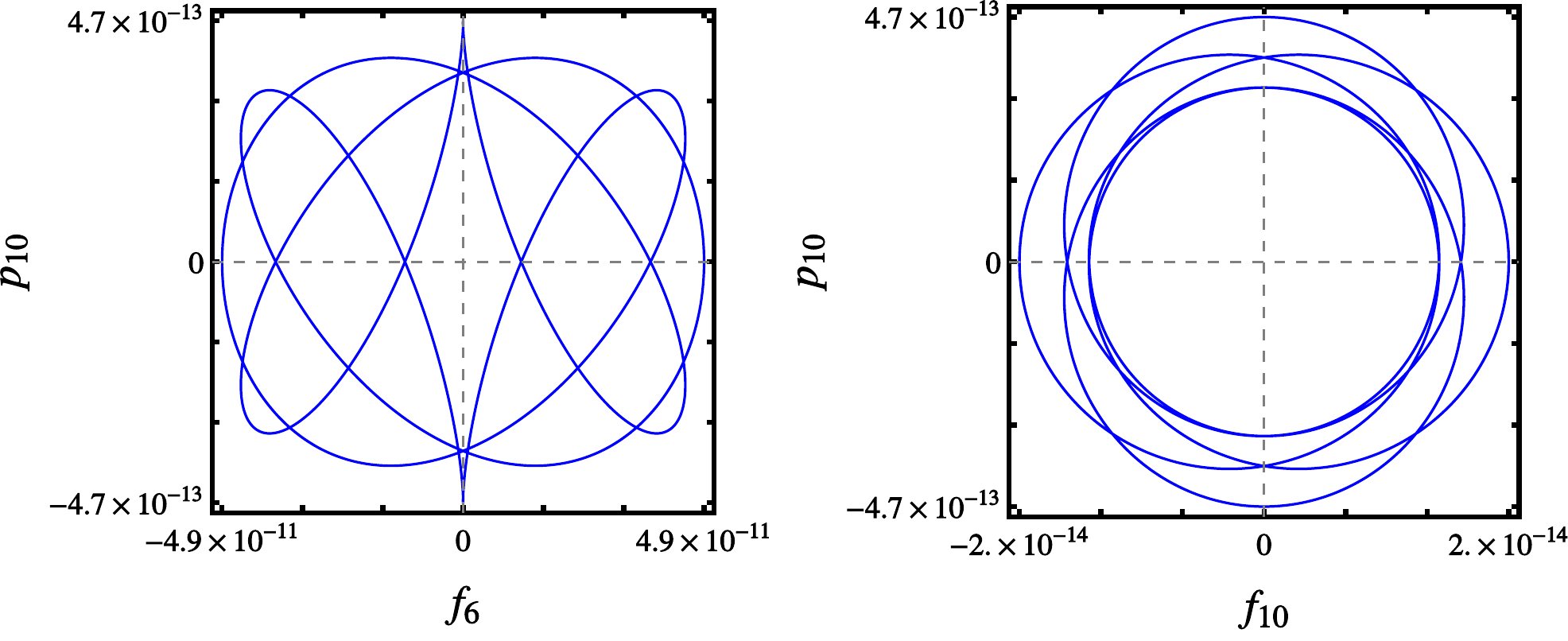}
   \caption[Phase space of stable periodic solution]{Slices of the coefficients space for a stable and
      near-periodic solution in \gls{ads} space-time over 500 periods of time. Not only the coefficients remain bounded and no
      weakly turbulent instability is observed, but the trajectory in phase space turns out to be quasi-periodic. Credits: \cite{Maliborski13b}.}
      \label{phasespace}
\end{myfig}

The perturbative construction of time-periodic solutions consists in choosing an ansatz (see equations \eqref{einsteinscalar} and
\eqref{ansatzscalar})
\begin{equation}
   \phi(t,x) = e^{i\Omega t}f(x), \quad \delta(t,x) = d(x), \quad A(t,x) = \mathcal{A}(x),
\end{equation}
for the three dynamical functions at play. The differential equations for $f$, $d$ and $\mathcal{A}$ are \cite{Maliborski13b}
\begin{subequations}
\begin{align}
   -\Omega^2 \frac{e^d}{\mathcal{A}}f &= \frac{1}{\tan^2 x}(\tan^2x \mathcal{A} e^{-d} f')',\\
   d' &= -\sin x \cos x\left[ {f'}^2 + \left( \frac{\Omega e^d}{\mathcal{A}}f \right)^2 \right],\\
   \mathcal{A}' &= \frac{1 + 2\sin^2 x}{\sin x \cos x} (1 - \mathcal{A}) + \mathcal{A}d'.
\end{align}
\end{subequations}
These equations can be solved order by order by expanding $\Omega$, $f$, $d$ and $\mathcal{A}$ in a small amplitude parameter $\varepsilon$.
Choosing a dominant mode $e_\gamma$ in the zero-amplitude limit, it comes
\begin{subequations}
\begin{align}
   f &= \varepsilon f_1 + \varepsilon^3 f_3 + \ldots \quad \tn{with} \quad f_1(x) \propto e_\gamma(x),\\
   \mathcal{A} &= 1 - \varepsilon^2 \mathcal{A}_2 - \ldots,\\
   d &= \varepsilon^2 d_2 + \ldots,\\
   \Omega &= \omega_\gamma + \varepsilon^2 \Omega_2 + \ldots,
\end{align}
\end{subequations}
where $\Omega$ is expanded around the dominant frequency $\omega_\gamma$ according to the Poincaré-Lindstedt method. After projection of the
unknown functions on the eigen basis $(e_j)_{j \in \mathbb{N}}$ all secular resonances are removable by fine-tuning the $\Omega_i$ coefficients. The
perturbative algorithm for the construction of time-periodic solutions was extended in \cite{Kim15} to tachyonic fields and pushed
to $20^{th}$ order in the amplitude.

The formalism of the time-periodic solution \cite{Maliborski13b} was extended to odd spatial dimensions in \cite{Fodor15}. The
authors explored the parameter space further and found bifurcations and resonances in the non-linear solutions that were missed by
\cite{Maliborski13b}. The stability branch of these solutions are shown in figure \ref{fodor}, as well as the agreement between
numerical and perturbative techniques. Stable solutions remain close to the initial data for ever, while unstable ones quickly
collapse to black holes. This is very reminiscent of usual self-gravitating systems that exhibit a maximum mass that is the boundary
between stable and unstable behaviours.

\begin{myfig}
   \includegraphics[width = 0.49\textwidth]{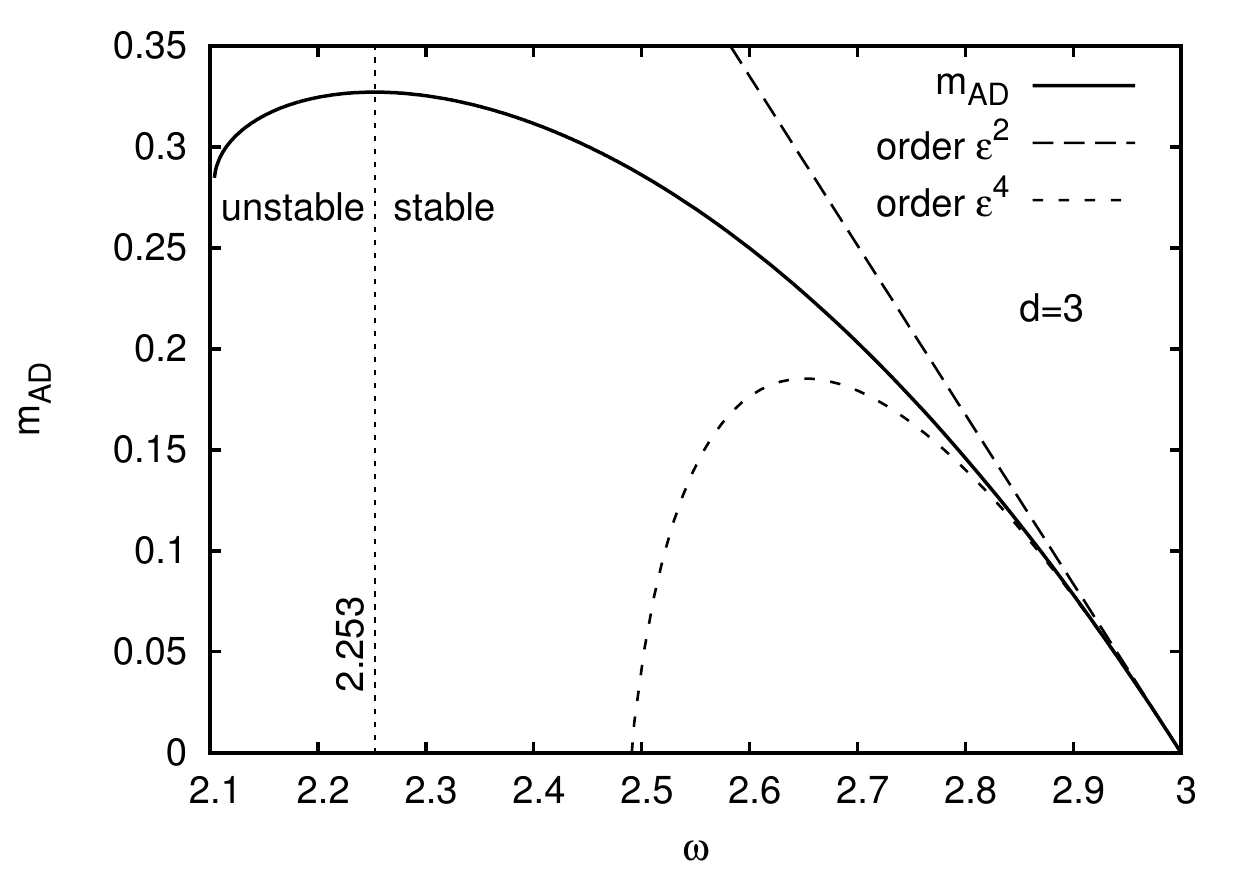}
   \caption[Stability branch of time-peridic solutions]{Mass of a time-periodic solution as a function of the
      oscillation frequency $\Omega$. At low amplitudes, $\Omega = 3$. Below a certain critical value $\Omega = 2.253$, the solutions become
      unstable against black hole formation. Perturbative (dashed) and numerical results (solid) agree well each other in the low amplitude limit.
      Credits: \cite{Fodor15}.}
      \label{fodor}
\end{myfig}

Why are time-periodic solutions non-linearly stable? One explanation was given in \cite{Maliborski14} where the authors studied
the spectrum of a linear perturbation superimposed on a time-periodic background. Unlike the vacuum \gls{ads} background case,
they observed that the spectrum of the corresponding linear operator $\widehat{L}$ was now only asymptotically resonant:
\begin{equation}
   \omega_j = Cj + D + O\left( \frac{1}{j} \right),
\end{equation}
such that eigen frequencies were equidistant only in the large-$j$ limit. The idea was the following: a resonant spectrum leads to
non-linear instability while an only asymptotically resonant spectrum could lead to an amplitude threshold below which the
instability is suppressed (this point is further discussed in section \ref{roleres} below). The argument gathered
momentum with the time evolution of a Gaussian perturbation around this time-periodic background. This did suppress the turbulent
cascade for sufficiently low amplitudes of the perturbation, highlighting that the time-periodic solution could be an attractor immune
to non-linear instability. This point is illustrated in figure \ref{timepattractor}. For stable solutions, the energy spectrum settles
down to an exponential form at long times, adding weight to the stability and regularity statement.

\begin{myfig}
   \includegraphics[width = 0.49\textwidth]{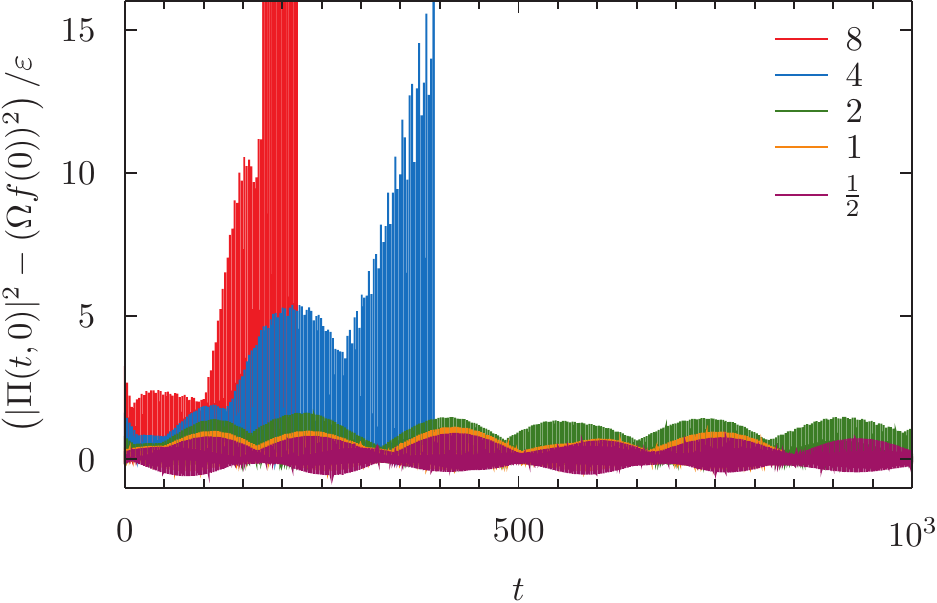}
   \caption[Stability of time-periodic solutions]{Time-periodic background perturbed by a Gaussian profile. The
      plot shows the time evolution of the scalar field amplitude $\Pi(t,0)$ (with time-periodic background subtracted). The
      amplitude of the perturbation is labelled by different colours. For large amplitudes, the instability is triggered, while
      for low amplitudes, it is suppressed. Credits: \cite{Maliborski14}.}
   \label{timepattractor}
\end{myfig}

The time-periodic and large-width Gaussian solutions were the first non-trivial dynamical examples of non-linearly stable
solutions in \gls{aads} space-times, suggesting that the instability conjecture had a much richer structure than was previously thought. It was
noticed in \cite{Abajo14} that the scalar field profiles of large Gaussian initial data of \cite{Buchel13} and of the time-periodic
solutions of \cite{Maliborski13b} were in fact very close to each other and probably belonged to the same island of stability.
Notice that these time-periodic solutions could well be called geons, according to our definition \ref{defgeon} in chapter \ref{geons}.

\subsection{The two-time framework (TTF)}
\label{ttf}

We have seen in section \ref{pertscal} that the Poincaré-Lindstedt method could remove secular resonances arising at third order in the
expansion, at least in some cases. This method works very well and at all orders in the case of time-periodic solutions
(section \ref{tpsol}). However, in the general case, some irremovable resonances appear and are responsible for the
\gls{ads} instability. 

The perturbative equations at third order \eqref{cj3} are nothing but a system of non-linearly coupled oscillators. Such systems
belong to the class of non-integrable Hamiltonian systems. For a long time, they were believed to obey the ergodic hypothesis. In
1955, in order to test this statement, Fermi and his collaborators performed a numerical simulation of a chain of 64
non-linearly coupled oscillators \cite{Fermi55}. At the time, the authors expected the system would exhibit thermalisation, an
ergodic behaviour in which the system becomes random with all modes excited more or less equally. Instead the system displayed a
very intricate quasi-periodic solution. This is the so-called \gls{fput} paradox. This result showed that non-integrable
Hamiltonian systems were not always ergodic. In subsequent works (see \cite{Benettin08} for a review), it was shown that the
quasi-periodic behaviour on certain time scales could be studied within the \gls{ttf}.

Transposed to the \gls{ads} instability problem, the ergodicity hypothesis advocates for a systematic instability, since if very
small scales are substantially excited, black hole formation becomes very likely. Thus, the \gls{ads} instability would appear as
an ergodic thermalisation process, echoed in the dual \gls{cft} (see section \ref{cftinterp}). However, given the similarity between the
\gls{fput} problem and equations \eqref{cj3}, it can be reasonably expected that gravitational dynamics in \gls{ads} space-time
are not ergodic. There could exist quasi-periodic solutions that do not explore the whole phase space and thus avoid black hole
formation. As for the \gls{fput} problem, \gls{ttf} might be of great help in finding such solutions.

The \gls{ttf} aims at providing a systematic way of removing secular resonances and thus building
non-linearly stable solutions. It was first introduced in \cite{Balasubramanian14}, motivated by the \gls{fput} analogy, and refined in
\cite{Craps14a,Craps15a,Buchel15}. Recycling the results of section \ref{pertscal}, we have already seen that at first order, the
solutions could be written
\begin{equation}
   \phi_1(t,x) = \sum_{j=0}^\infty (\alpha_j e^{-i\omega_j t} + \overline{\alpha}_j e^{i\omega_j t})e_j(x),
\end{equation}
where a bar means complex conjugation and $\alpha_j$ are constant complex amplitudes. The idea is the following: given that both
perturbative results and numerical simulations have proved that some initial data becomes singular in a time $t =
O(\varepsilon^{-2})$, let us introduce a new time-scale, or slow-time
\begin{equation}
   \tau \equiv \varepsilon^2 t.
\end{equation}
The intuition is that if the dynamics involves rapid oscillations superimposed on a slow drift behaviour, there should be some
sort of simplified effective description of the slow motion, in which the fast oscillations have been averaged out. This is
at the heart of multiple-scale analysis. We thus expect the small amplitude scalar field to undergo large variations in a time-scale
$\tau = O(1)$, while it oscillates on a much shorter time-scale (i.e.\ it bounces many times before collapsing). This suggests to make
the envelope of oscillations vary slowly, on a time-scale $\tau = O(1)$. We thus write\footnote{This is nothing but a
variation of the constants method.}:
\begin{equation}
   \phi_1(t,\tau,x) = \sum_{j=0}^\infty [\alpha_j(\tau) e^{-i\omega_j t} + \overline{\alpha}_j(\tau) e^{i\omega_j t}]e_j(x).
\end{equation}
Paying attention that now $\partial_t \rightarrow \partial_t + \varepsilon^2 \partial_\tau$, the second order equations for $A_2$
and $\delta_2$ are unchanged but at third order a new term $\partial_t \partial_\tau \phi_1$ appears (compare with
\eqref{phi3}), namely
\begin{equation}
   \partial_t^2{\phi_3} + \widehat{L}\phi_3 + 2\partial_t \partial_\tau \phi_1 = S(\phi_1,A_2,\delta_2).
   \label{phi3slow}
\end{equation}
Projecting onto the basis $(e_j)_{j \in \mathbb{N}}$, it comes (in contrast with \eqref{cj3})
\begin{equation}
   \forall j, \quad \ddot{c}_j^{(3)} + \omega_j^2 c_j^{(3)} - 2i\omega_j(\partial_\tau \alpha_j e^{-i\omega_j t} - \partial_\tau \overline{\alpha}_j e^{i\omega_j t}) = S_j.
\end{equation}
Of course, the introduction of the slow-time did not remove the resonant terms $e^{\pm i \omega_j t}$ on the right-hand side, but
we are now free to enforce them to vanish by imposing
\begin{equation}
   -2i\omega_j \partial_\tau\alpha_j = (\tn{component } e^{-i\omega_j t} \tn{ of } S_j) = \sum_{klm} S_{jklm}\overline{\alpha}_k \alpha_l \alpha_m,
   \label{ttf1}
\end{equation}
where $S_{jklm}$ are real constants representing all the possible resonant channels $\omega_j + \omega_k = \omega_l +
\omega_m$. Other channels like e.g.\ $\omega_j = \omega_k + \omega_l + \omega_m$ can be shown to vanish and a
brute-force calculation of the $S_{jklm}$ coefficients is presented in \cite{Craps14a}. At this point, all functions $c_j^{(3)}$ remain
bounded in time by construction and hence are of little interest, so that we now focus on the $\alpha_j$. Moving to the
exponential complex representation
\begin{equation}
   \alpha_j(\tau) = A_j(\tau)e^{iB_j(\tau)},
   \label{complexalpha}
\end{equation}
with $A_j$ the real amplitude and $B_j$ the real phase, equation \eqref{ttf1} becomes
\begin{subequations}
\begin{align}
   2\omega_j \frac{d A_j}{d \tau} &= \sum_{\substack{j+k=l+m \\ \{j,k\} \neq \{l,m\}}} S_{jklm}A_k A_l A_m \sin(B_j + B_k - B_l - B_m),\\
   2\omega_j \frac{d B_j}{d \tau} &= T_j A_j^2 + \sum_{i\neq j} R_{ij}A_j^2 + \frac{1}{A_j}\sum_{\substack{j+k=l+m \\ \{j,k\} \neq \{l,m\}}} S_{jklm}A_k A_l A_m \cos(B_j + B_k - B_l - B_m),
\end{align}
\label{ttfequations}%
\end{subequations}
where $\{j,k\} \neq \{l,m\}$ means that neither $j$ nor $k$ is equal to either $l$ or $m$, $T_j = S_{jjjj}$ and $R_{ij} =
S_{ijji} + S_{jiji}$. These equations are called the \gls{ttf} equations. Alternative denominations are resonant \cite{Bizon15b},
renormalisation flow \cite{Craps14a} or time-averaged equations \cite{Evnin16}.

Defining
\begin{equation}
   N = \sum_{j=0}^\infty \omega_j A_j^2 \quad \tn{and} \quad E = \sum_{j=0}^\infty \omega_j^2 A_j^2,
   \label{NE}
\end{equation}
it was shown in \cite{Craps15a,Buchel15,Yang15} that $N$ and $E$ were conserved quantities, namely
\begin{equation}
   \frac{dN}{d\tau} = 0 \quad \tn{and} \quad \frac{dE}{d\tau} = 0.
\end{equation}
These quantities are then interpreted as $N$ the total particle number (in analogy to quantum mechanics) and $E$ the total energy
of the system. There exists a third conserved quantity, that was uncovered in \cite{Craps15a} and which represents the total
interaction energy between modes. In \cite{Buchel15}, it was argued that if there was a cascade of energy toward high-$j$ modes,
the simultaneous conservation of $E$ and $N$ induced that there was an inverse cascade of particle number toward low-$j$ modes, and
vice versa. This is called the dual cascade phenomenon.

The \gls{ttf} equations are third order in amplitudes and are valid for durations up to $\tau = O(1)$. It would be possible to
introduce other time-scales like $\varepsilon^4 t$ to go further in time. But usually, evolving the \gls{ttf} equations up to
$\tau = O(1)$ brings enough information to conclude about the stability.

In general, the \gls{ttf} equations are evolved numerically. As the number of coupled equations is infinite, a cut-off (or
truncation number) $j_{max}$ has to be enforced. The advantage of these equations is that they are ordinary differential equations
in time, whereas the full \gls{ekg} system \eqref{einsteinscalar} is a system of partial differential equations. Within the \gls{ttf}, the spatial
dependence is entirely encoded on the basis functions $(e_j)_{j \in \mathbb{N}}$, which is somewhat reminiscent of spectral methods in
numerical analysis. The equations are thus less computationally demanding, which contributed to the democratisation of the
\gls{ads} non-linear instability study. Incidentally, the number of publications in the field drastically increased after 2014 and
\cite{Balasubramanian14}.

The \gls{ttf} equations and conservation laws were generalised to non-spherically symmetric scalar field collapse in
\cite{Yang15} with the help of spherical harmonic decomposition. The resonant channels depend on the number of dimensions and the authors
expected the system of equation to display an underlying symmetry that would simplify their systematic determination. An $SU(d)$
symmetry was precisely demonstrated in \cite{Evnin15} and refined in \cite{Evnin16}.

Last but not least, the \gls{ttf} equations are invariant under the transformation
\begin{equation}
   \alpha_j(\tau) \rightarrow \varepsilon \alpha_j(\tau/\varepsilon^2),
   \label{scalingsym}
\end{equation}
which means that if a solution of amplitude one does something at slow-time $\tau$, then the same solution with amplitude
$\varepsilon$ does the same thing at slow-time $\tau/\varepsilon^2$. This feature was observed in the non-linear case as early as
\cite{Bizon11}. It actually comes out naturally from the \gls{ttf} equations.

\subsection{The analyticity strip method}
\label{analstrip}

A standard approach is to evolve in time the \gls{ttf} equations and to determine if the energy spectrum (equation \eqref{NE}) becomes
singular, namely if there is an energy flow toward high-$j$ modes so as to get a power-law spectrum instead of an exponential one.
The underlying concept is the one of analyticity strip, described in \cite{Sulem83} and used for the first time in the
\gls{ads} instability context in \cite{Bizon13}. It consists in fitting the coefficients $A_j$ (or the energy per mode $E_j =
\omega_j^2 A_j^2$ according to \eqref{NE}) as a function of $j$ at each slow-time step by
\begin{equation}
   A_j(\tau) = C(\tau) j^{-\gamma(\tau)}e^{-\rho(\tau)j}.
   \label{analstripeq}
\end{equation}
In practice, the fit is performed on a reduced set of modes that is away from $j = 0$ and $j = j_{max}$
in order to minimise truncation errors. The function $\rho(\tau)$ is the analyticity radius of the solutions. If $\rho$ stays
strictly positive, it means that the solution is regular at all times. If the radius of analyticity hits zero, it is a strong hint that
the solution blows up in a finite time. The existence of a time $\tau_0$ at which $\rho(\tau_0) = 0$ is a necessary but not
sufficient condition for black hole formation (see section \ref{necsuf} below).

The \gls{ttf} equations coupled to the analyticity strip method are thus less effective than fully non-linear evolutions to find
unstable solutions, but they are very good (and computationally cheap) at exploring the \gls{ads} sea to find islands of
stability. They also bring a kit of analytical tools to better understand the deep structure of the problem.

\subsection{The two-mode controversy}

One of the first playgrounds of the \gls{ttf} framework was the two-mode initial data (equation \eqref{twomode}). It was first
studied in \cite{Bizon11} as a minimal setting for triggering the instability (recall that a single mode initial data is
non-linearly stable). Subsequently, the authors of \cite{Balasubramanian14} tried to evolve the \gls{ttf} equations for the
equal-energy two-mode initial data with a cut-off $j_{max} = 47$. Surprisingly they found that their code was showing no sign of
collapse. After a lively debate \cite{Bizon15a,Balasubramanian15,Buchel15,Bizon15b,Green15}, the simulations of
\cite{Bizon15b,Deppe15b} confirmed that the two-mode initial data was really collapsing and that the \gls{ttf} as
well as the full \gls{gr} simulations of \cite{Balasubramanian14} suffered from resolution problems and a too small truncation
number.

In particular in \cite{Bizon15b}, the authors carefully scrutinised the two-mode initial data both in full \gls{gr} and with the \gls{ttf}
equations with a cut-off $j_{max} = 172$. Not only did they confirm that the two-mode initial data was collapsing in the full theory,
but they also showed the agreement with \gls{ttf} via the analyticity strip method. Namely, they observed that the analyticity radius
was dropping to zero in a finite slow-time $\tau_\star \simeq 0.509$ and that the exponent $\gamma$ in \eqref{analstripeq} was
tending to $2$ at this date. They inferred via the \gls{ttf} equations \eqref{ttfequations} that such behaviours implied a
logarithmic divergence of the phase derivatives $d B_j/d\tau$ ($B_j$ being defined in \eqref{complexalpha}) and checked that this
was indeed the case numerically, as pictured in figure \ref{blowup}.

\begin{myfig}
   \includegraphics[width = 0.48\textwidth]{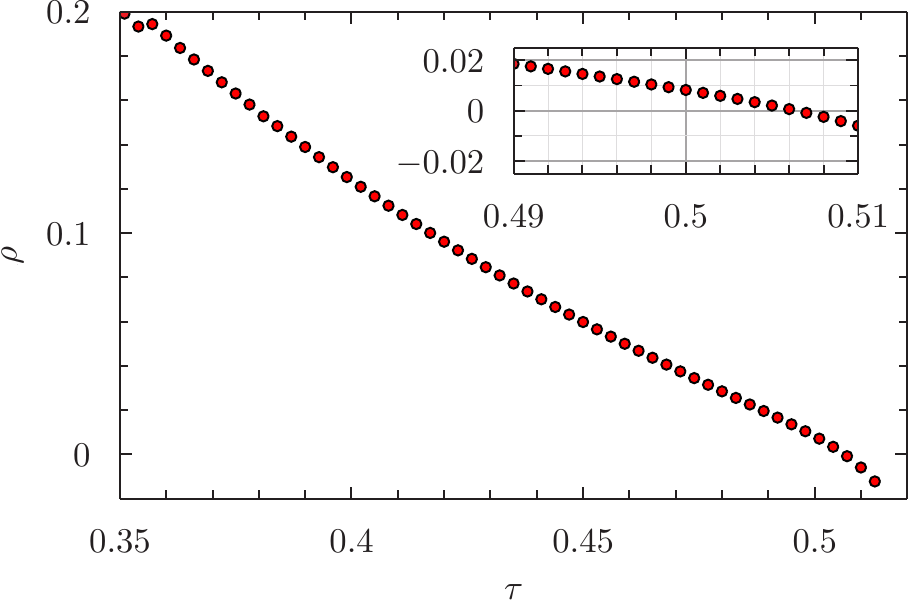}
   \includegraphics[width = 0.49\textwidth]{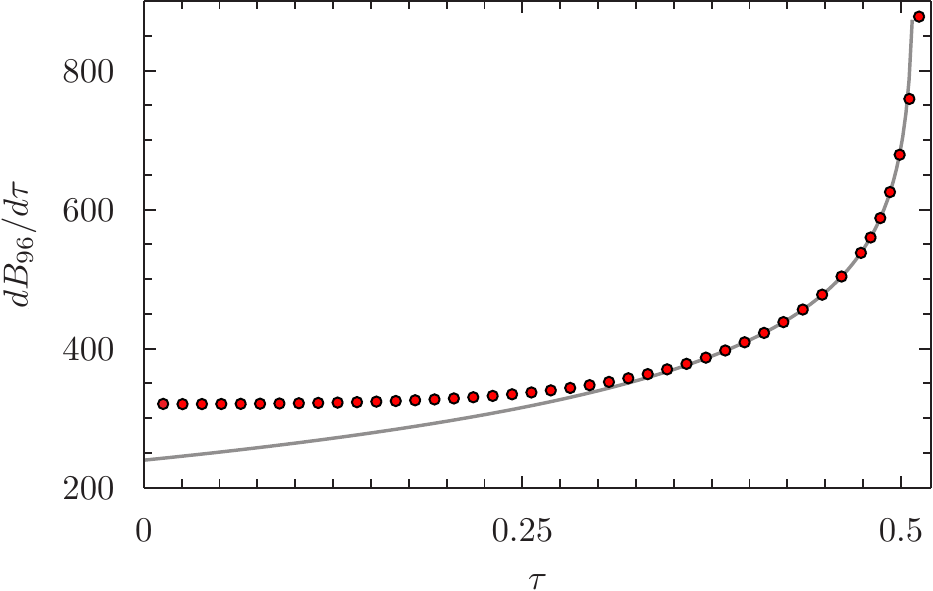}
   \caption[Instability of the two-mode initial data]{Left: analyticity radius for the \gls{ttf} evolution of the two-mode initial
   data. It hits zero in a finite slow-time, suggesting a potential consecutive blow-up. Right: slow-time derivative of the $n=96$
   phase $B_n$. A logarithmic blow-up $a \ln(\tau_\star - \tau) + b$ is fitted and is observed for all $n > 20$. Credits:
   \cite{Bizon15b}.}
   \label{blowup}
\end{myfig}

This behaviour was supported by analytical calculations in \cite{Craps15b,Craps15c}, with a discussion about the gauge-dependence
of the result, in particular for different choices of boundary conditions $\delta(t,x = \pi/2)$. Indeed, the two gauges used
in the literature are
\begin{subequations}
\begin{align}
   \delta(t,x=0) &= 0, \quad \tn{\gls{itg}},\\
   \delta(t,x=\pi/2) &= 0, \quad \tn{\gls{btg}},
\end{align}
\end{subequations}
where $\delta$ is the metric coefficient in \eqref{ansatzscalar}. The authors of \cite{Craps15b,Craps15c} have shown analytically
that the logarithmic divergence of $dB_j/d\tau$ was suppressed in the \gls{btg}. This was confirmed numerically by \cite{Deppe16b}
whose author demonstrated that these features existed only in the \gls{itg} gauge. This was given more support in
\cite{Dimitrakopoulos16} whose authors studied precisely the gauge dependence of the \gls{ttf} equations. They concluded that the gauge
was impacting only the phases $B_j$ but not the amplitudes $A_j$. Furthermore, if one gauge gives singular results and not the
other, it reveals that there is a infinite redshift between the two gauges, so that \gls{ttf} does become invalid as a
perturbation theory and an instability is triggered. Finally, the results of \cite{Deppe16b} indicated that the logarithmic
blow-up of the phase derivative was completely suppressed in both \gls{itg} and \gls{btg} gauges in higher dimensions than four.

Based on these results, full \gls{gr} and \gls{ttf} finally agreed each other about the two-mode initial data: it is unstable, it
collapses to a black hole, and both frameworks can detect it, either by the vanishing of the blackening factor $A$ of by the
vanishing of the analyticity radius $\rho$. Thanks to the scaling symmetry \eqref{scalingsym}, it is established that the two-mode
initial data does collapse for arbitrarily small amplitudes, a regime usually out of reach of numerical simulations.

\begin{myfig}
   \includegraphics[width = 0.47\textwidth]{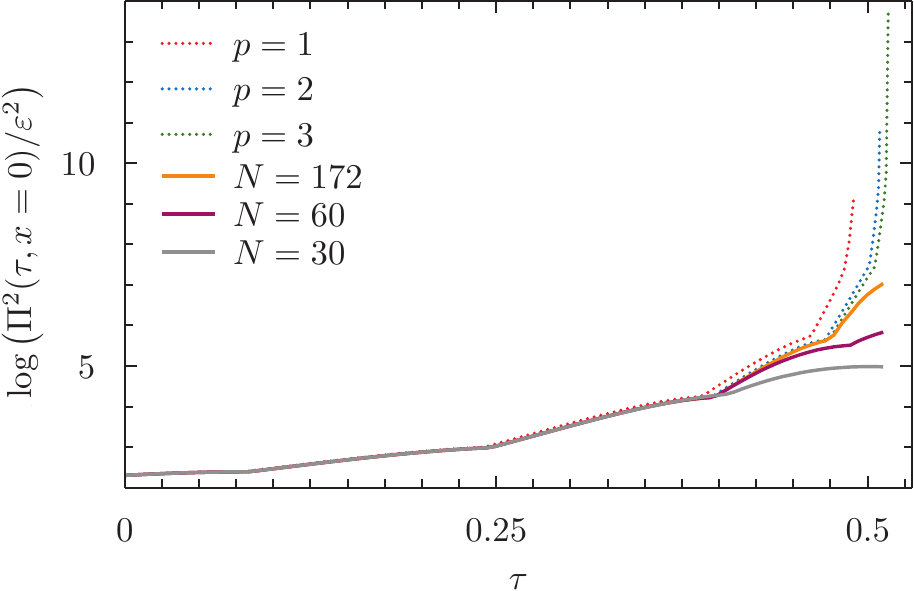}
   \includegraphics[width = 0.51\textwidth]{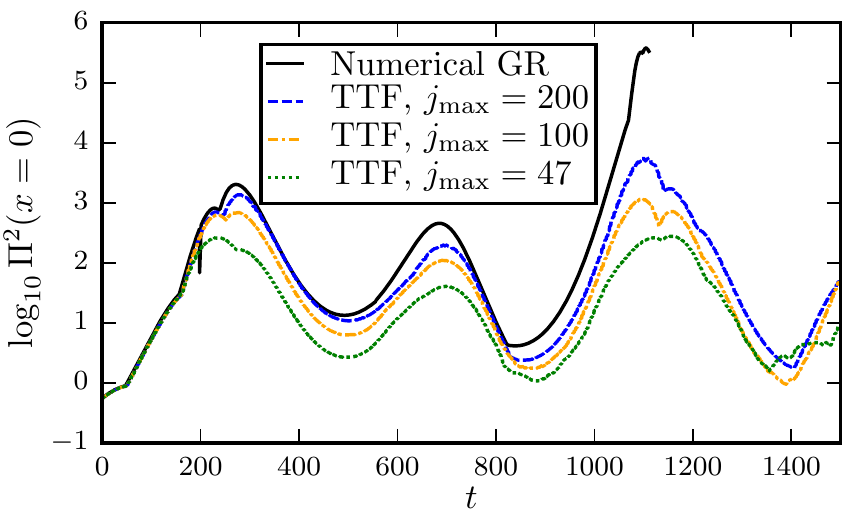}
   \caption[Comparison of TTF framework with full GR]{Left: upper envelope of $\Pi^2(t,0)$ in the full \gls{gr} evolution of
   two-mode initial data with $\varepsilon = (2\pi)^{-3/2}2^{-p}$ for $p = 1,2,3$. The corresponding solutions of the
   truncated \gls{ttf} system are shown in solid as the truncation number $N$ increases. Right: idem but with a larger
   amplitude $\varepsilon = 0.09$. In this case, three cascades of energy (increasing curve) and two inverse cascade (decreasing
   curve) are visible before collapse. Credits: \cite{Bizon15b,Green15}.}
   \label{agreementttf}
\end{myfig}

The agreement between the two kinds of time evolution can be clearly seen in figure \ref{agreementttf}. For small amplitudes (left
panel), the convergence of the \gls{ttf} solution to the full \gls{gr} with increasing truncation number is unambiguous. The case
of amplitude $\varepsilon = 0.09$ (right panel) taken from \cite{Green15} was subject to lively debates. Indeed, if full \gls{gr}
simulations indicate black hole formation at $t \sim 1080$, \gls{ttf} disagrees and undergoes an inverse cascade after this date.

The author of \cite{Deppe16b} studied more deeply this point with the help of the analyticity-strip method. He first observed that
the two-mode initial data was unambiguously unstable in 9-dimensional \gls{ads}, as can be seen on the left panel of figure
\ref{analtrunc}. This suggested that the instability was favoured in higher dimensions. Back to the 4-dimensional problem, he
observed that the truncation number was slightly impacting the location of the first root of the analyticity radius, as can be
seen on the right panel of figure \ref{analtrunc}. The analyticity radius hits zero at time $t \sim 800$, i.e.\ before the
non-linear collapse at time $t\sim 1080$. The point is: nothing prevents the \gls{ttf} equations to be evolved in time after the
analyticity radius hits zero, like in the right panel of figure \ref{agreementttf}. However, the physical meaning of the evolution
after this stage should not be considered too seriously. Even if \gls{ttf} solutions do not blow up at this stage, the only
reliable criterion for instability is the vanishing of the analyticity radius. The fact that it hits zero at time $t\sim 800$ is
already a strong indication of a near subsequent singularity formation.
\begin{myfig}
   \includegraphics[width = 0.49\textwidth]{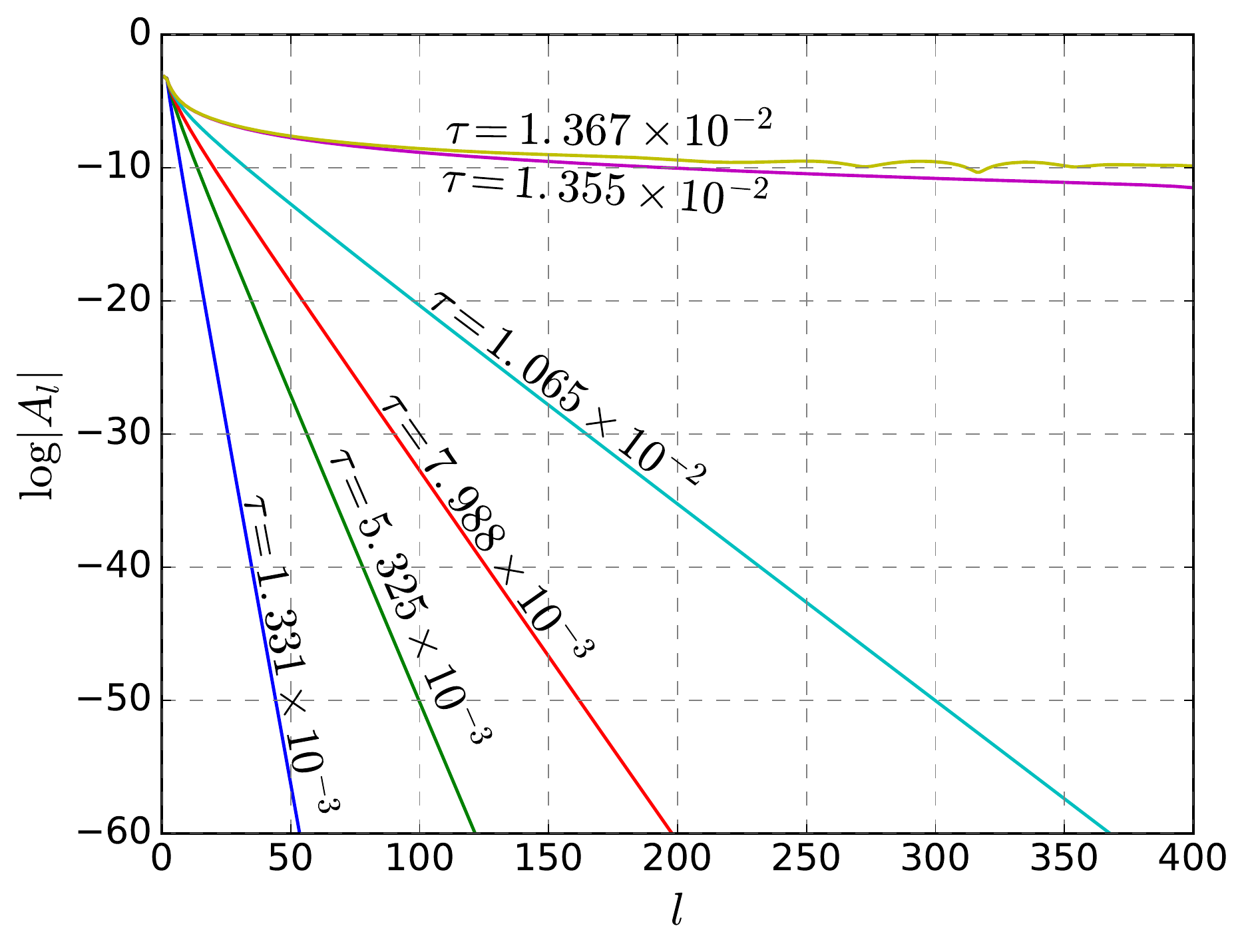}
   \includegraphics[width = 0.49\textwidth]{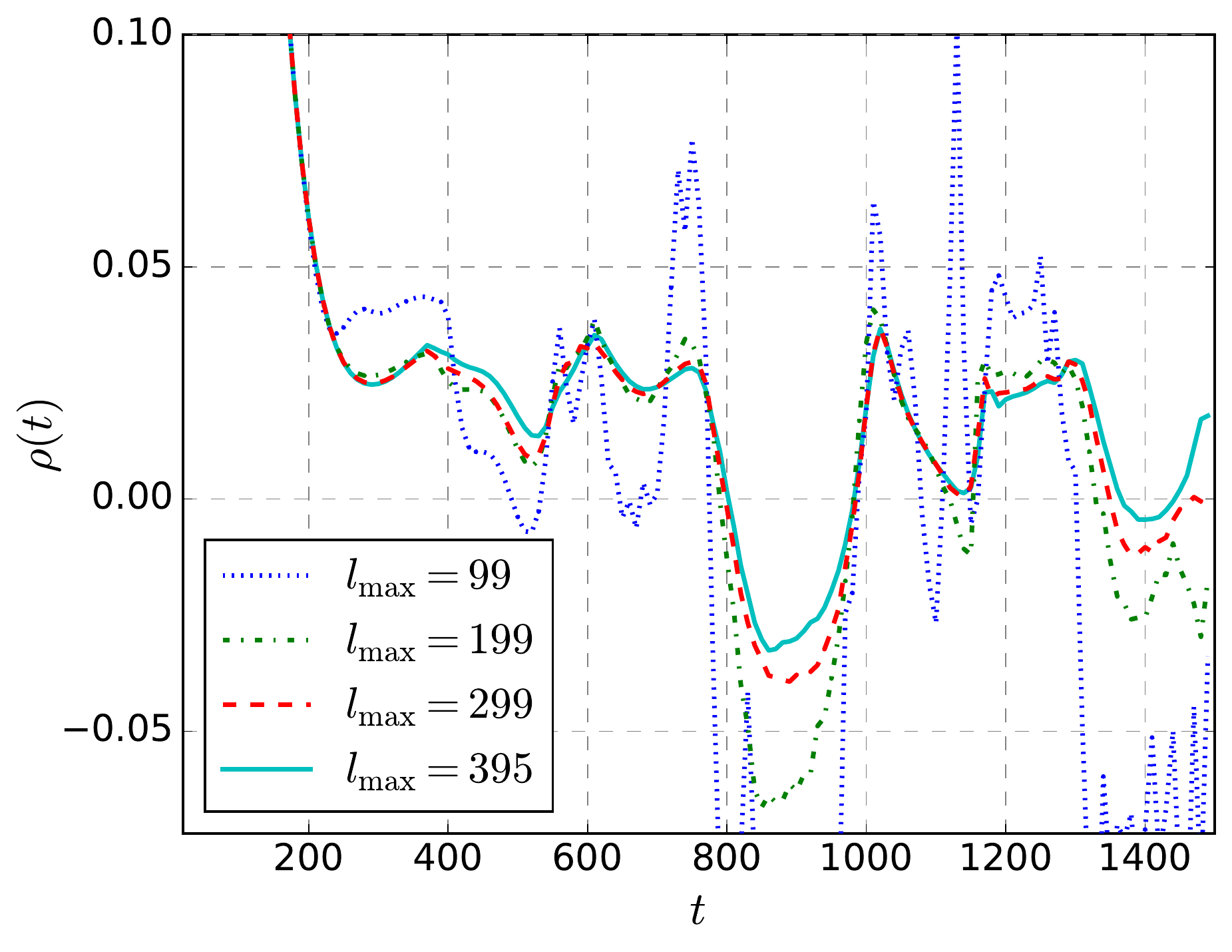}
   \caption[Singular spectrum for two-mode initial data in the TTF formalism]{Left: spectra at different slow-times $\tau$ for the
      two-mode initial data in 9-dimensional \gls{ads}. The spectrum becomes singular unambiguously at $\tau = \num{1.356e-2}$.
      Right: analyticity radius for the 4-dimensional case. The variability of the roots of $\rho$ with truncation number
      $l_{max}$ indicates that the evolutions suffer from truncation errors to some extent. The time axis is the same as in
      figure \ref{agreementttf} where collapse occurs at $t \sim 1080$. Credits: \cite{Deppe16b}.}
   \label{analtrunc}
\end{myfig}

\subsection{Generation of islands of stability with TTF}

Besides being an alternative and cheaper numerical method to study the instability conjecture, the \gls{ttf} is a valuable tool to
build islands of stability, namely initial data that are non-linearly stable. The first paper employing \gls{ttf} in the \gls{ads}
instability problem was \cite{Balasubramanian14}, where the authors uncovered a new family of periodic and stable solutions by
imposing real amplitudes (equation \eqref{complexalpha})
\begin{equation}
   A_j = \varepsilon \frac{e^{-\mu j}}{\omega_j},
   \label{ttfinit}
\end{equation}
with a truncation number $j_{max} = 47$. In figure \ref{qpttf}, it is clearly visible that such initial data remained stable with a
bounded and quasi-periodic departure from the initial conditions. In particular, no turbulent cascade was observed. This was
confirmed in \cite{Basu15} whose authors pushed the truncation number to $j_{max} = 150$ and observed the same stabilisation.

\begin{myfig}
   \includegraphics[width = 0.49\textwidth]{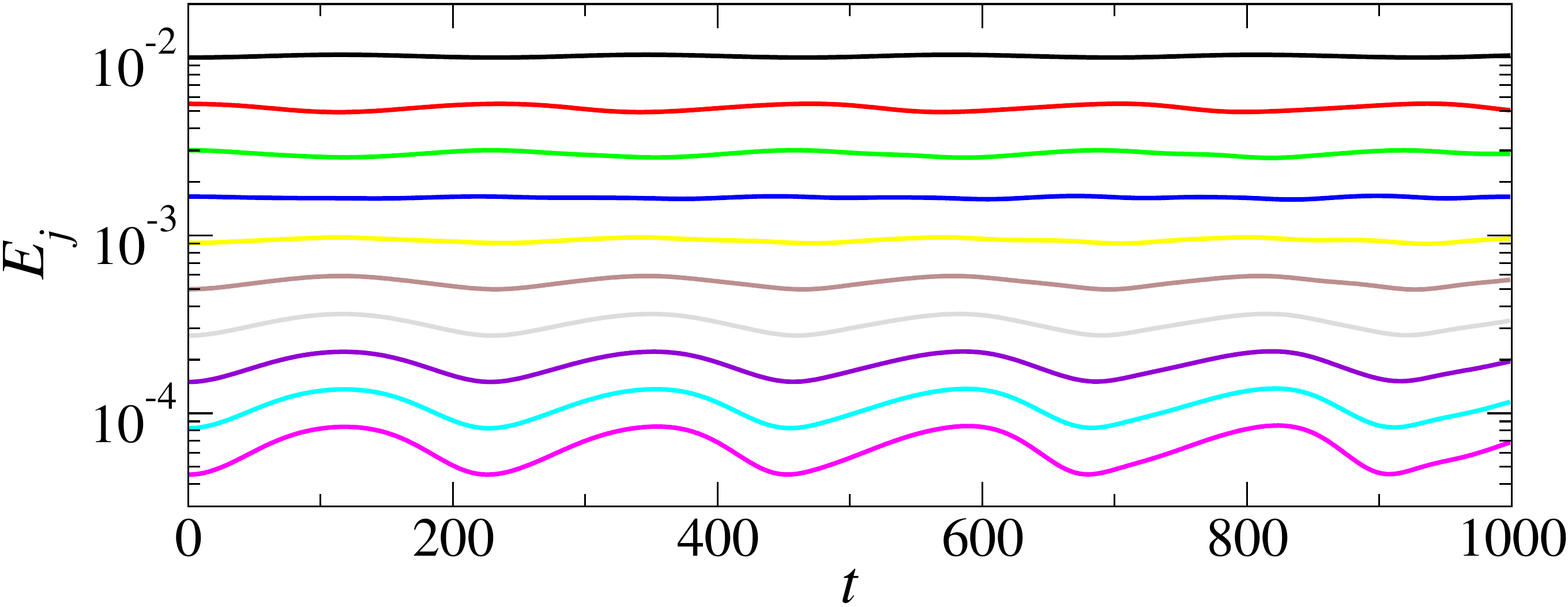}
   \caption[Quasi-periodic solutions with two-time framework]{Energy per mode for initial data \eqref{ttfinit} with $\mu = 0.3$
   and an initial random perturbation. The energies in the first modes remain always close to their initial values, signalling a
   stable solution within the duration of the simulation. Credits: \cite{Balasubramanian14}.}
   \label{qpttf}
\end{myfig}

Similarly in \cite{Green15}, a whole two-parameter family of stable quasi-periodic solutions was found, inspired by \cite{Balasubramanian14}
and equation \eqref{ttfinit}, namely
\begin{equation}
   \alpha_j(\tau) = a_je^{-ib_j\tau},
\end{equation}
where the parameters $a_j$ and $b_j$ could be finely tuned so that energy flows between modes were perfectly balanced. In figure
\ref{circleqp}, a quasi-periodic solution was perturbed with a small two-mode contribution. For moderate amplitudes of the
perturbation, the phase space of the \gls{ttf} exhibited a quasi-periodic behaviour, which was reminiscent of the quasi-periodic
solutions of section \ref{tpsol}. The authors were also able to recover the stable/unstable behaviour of Gaussian initial data
(recall that stability is recovered for widths $0.5 \lesssim \sigma \lesssim 8$ ), as depicted in figure \ref{gaussianttf}: stable
initial data featured an exponential spectrum while unstable initial data presented a power-law spectrum.

\begin{myfig}
   \includegraphics[width = 0.49\textwidth]{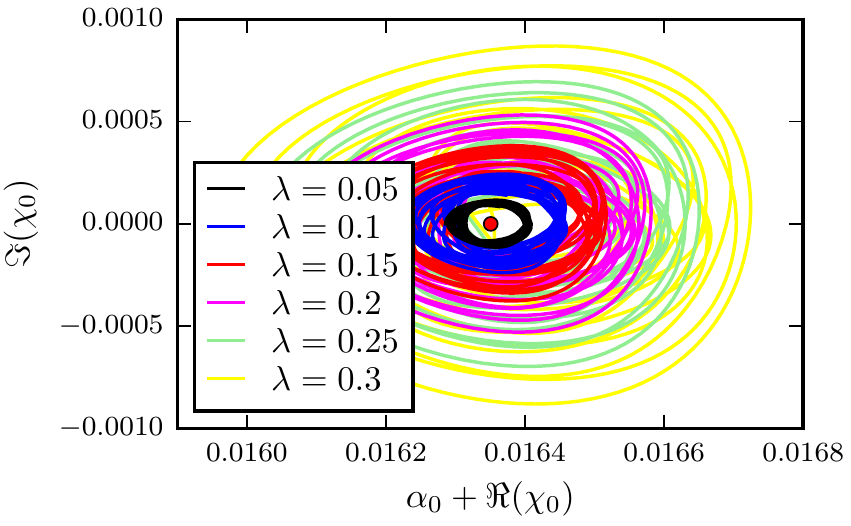}
   \caption[Oscillation around a quasi-periodic solution]{Phase space of initial data interpolating a quasi-periodic solution with the two-mode
      initial data such that $E_j = (1-\lambda)E_j^{QP} + \lambda E_j^{two-mode}$. The quantities $\alpha$ and $\chi$ represents the
      amplitude of the solution. The red dot at the centre is a pure time-periodic solution. When $\lambda \neq 0$, the
      solution merely oscillates in phase space around the periodic solution with no growth of the amplitude.
      Credits: \cite{Green15}.}
   \label{circleqp}
\end{myfig}

\begin{myfig}
   \includegraphics[width = 0.46\textwidth]{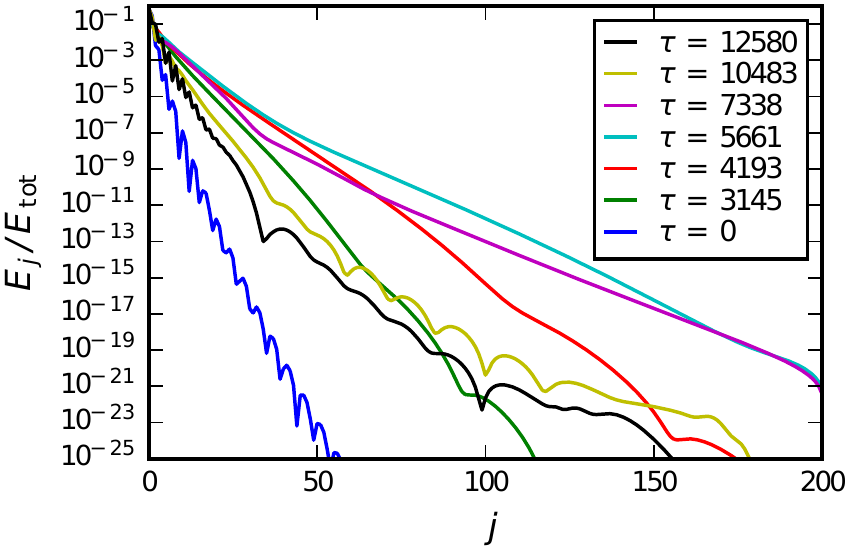}
   \includegraphics[width = 0.49\textwidth]{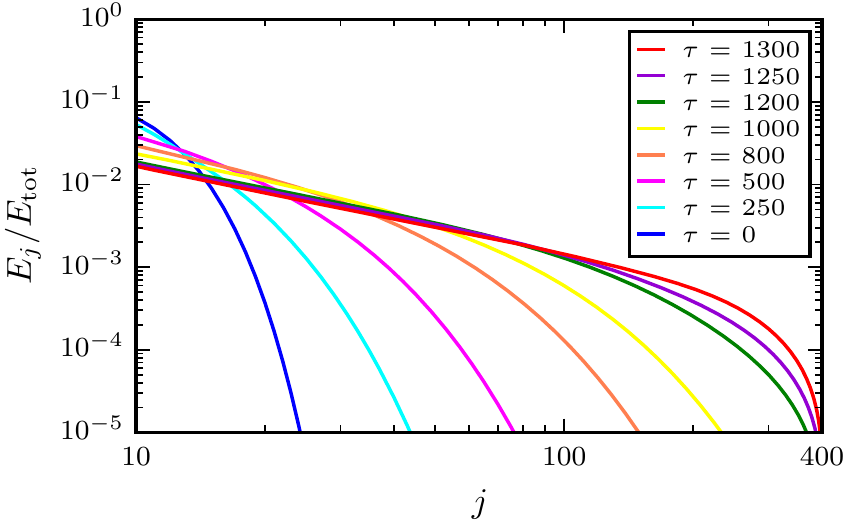}
   \caption[Turbulent cascade detection with two-time framework]{Left: energy spectrum of stable Gaussian initial data $\sigma = 0.4$
      with exponential tail at all slow-times. Right: energy spectrum of unstable Gaussian initial data $\sigma = 1/16$ with a
      power-law tail at large slow-time indicating turbulent cascade and potential black hole formation. Credits: \cite{Green15}.}
   \label{gaussianttf}
\end{myfig}

\subsection{Beyond spherical symmetry}
\label{beyondspher}

Surprisingly, from a historical perspective, the very first island of stability was uncovered before the instability conjecture itself!
Fully non-linear spherically symmetric solutions of \gls{ekg} equation were built as early as 2003, in \cite{Astefanesei03}, where
the authors obtained the first boson stars in \gls{aads} space-times. Their solutions were also proved to be linearly stable against
perturbations and exhibited a maximum mass that was smaller than their asymptotically flat counterparts. According to our definition
of chapter \ref{geons}, this was the very first construction of an \gls{aads} geon.

In the literature, the large majority of studies in the field of the instability conjecture assumed
spherical symmetry. Birkhoff's theorem states that at the exterior of a spherically symmetric non-rotating body the metric
should match the one of \gls{sads}. Thus no gravitational waves can be emitted in such space-times. Gravitational dynamics were thus
induced with a scalar field. This is the simplest setting we can imagine, and also the cheapest, computationally speaking.

The first study beyond spherical symmetry was that of \cite{Dias12a}, which was also the first attempt to build solutions
that resisted black hole formation. The authors solved the vacuum Einstein's equation with a negative cosmological constant
perturbatively with the \gls{kis} formalism (see \cite{Kodama00,Kodama03,Kodama04,Ishibashi04} and chapter
\ref{perturbations}). The solutions, called geons, could be described by four parameters which were the amplitude and three quantum
numbers $(n,l,m)$ that corresponds to the number of radial nodes $n$ and the spherical harmonic seed $Y_m^l$.

Recalling the scalar field knowledge, and in the same spirit of time-periodic solutions (section \ref{tpsol}), it is known that
single-mode excitations do not suffer from secular resonances However, in the
gravitational sector, the Poincaré-Lindstedt method was also successful in removing all secular
resonances appearing at third and fifth order in the case of an $(l,m,n) = (2,2,0)$ seed. This suggested that fully non-linear
$(2,2,0)$ geons could be constructed at arbitrary orders and could thus provide the first island of stability that was not spherically symmetric.

In these configurations, the angular momentum provides a natural barrier against black hole formation, and the rotation is
described by a helical Killing vector. Geons thus play the role of the non-linear time-periodic functions of
\cite{Maliborski13b,Fodor15,Kim15} in the gravitational sector, i.e.\ fundamental non-linearly stable modes of vibrations of
\gls{ads} space-time. The fully non-linear numerical construction of gravitational $(l,m,n) = (2,2,0)$ geons was initiated in \cite{Horowitz15} in the
harmonic gauge with the help of the de-Turck method and spectral discretisation\footnote{See \cite{Dias16b} for an excellent
review on these two popular numerical methods applied to stationary solutions.}. Chapter \ref{gaugefreedom} of the present
manuscript is dedicated to the connection between the de-Turck method and the 3+1 gauge used in \cite{Martinon17}.

If the existence of geons stands on firm arguments, their non-linear stability is still a partially unanswered question. One
expects that, akin to the time-periodic solutions of the scalar sector (section \ref{tpsol}), geons are non-linearly stable
attractors. In \cite{Dias12b}, some perturbative arguments in favour of their non-linear stability were given. The conclusion of
this work was that geons and boson stars were non-linearly stable unless they were embedded in very high dimensional space-times
or if perturbations had low differentiability, i.e.\ were far from being $\mathcal{C}^\infty$. However, the proof was missing some
theorems that were proven only in the analogous case of the non-linear Schrödinger equation. So this result was half-way between a
demonstration and a conjecture. The equivalent of two-mode initial data for geons was also considered perturbatively, but this
time irremovable secular resonances did emerge, like in the scalar case.

Not much work has been done beyond the $(l,m,n) = (2,2,0)$ geon before \cite{Dias16a,Dias17a} (and more recently \cite{Fodor17}),
whose authors greatly extended the perturbative results of \cite{Dias12a}. They worked at fixed quantum numbers $(l,m,n)$ and
obtained the allowed linear frequencies
\begin{subequations}
\begin{align}
   \omega_S &= l + 1 + 2n,\\
   \omega_V &= l + 2 + 2n,
\end{align}
\label{omegasv}%
\end{subequations}
$n$ being a positive integer, and the labels $S$ and $V$ corresponding to scalar-type or vector-type perturbations (see chapter
\ref{perturbations}). For these single-mode geon excitations, the authors found many configurations with irremovable secular
resonances, indicating that non-spherically symmetric systems could be even more unstable than spherically symmetric ones. This
claim was tempered in \cite{Rostworowski16} whose author argued that these resonances
were indeed removable if, instead of a single mode, a linear combination of modes sharing the same frequency $\omega$ and
azimuthal number $m$ were considered. This amounted to work at fixed frequency and azimuthal number $(\omega,m)$ instead of fixed
$(l,m,n)$, $\omega$ being a degenerate function of $l$ and $n$ according to \eqref{omegasv}. This claim
was made stronger in \cite{Rostworowski17a,Rostworowski17b} where the author found that the number of possible geon excitations
with a given frequency $\omega$ was precisely equal to the multiplicity of the frequency, suggesting that all secular resonances could be
removed. An illustration of these arguments is given in figure \ref{geoncontroversy}. The fully non-linear numerical study
\cite{Martinon17} put an end to the debate by explicitly constructing non-linear \textit{excited} geon solutions, thus confirming
the arguments of both \cite{Dias16a,Dias17a} (non-existence of excited $(l,m,n) = (2,2,1)$ geons) and
\cite{Rostworowski16,Rostworowski17a,Rostworowski17b} (existence of three excited families $(\omega,m) = (5,2)$). The authors of
\cite{Martinon17} also  exhibited a phase space diagram that was supported by both analytical and numerical results but was in
contradiction with the previous study \cite{Horowitz15}. Part \ref{part2} of the present manuscript is precisely devoted to the
numerical construction of these geons.

\begin{myfig}
   \includegraphics[width = 0.49\textwidth]{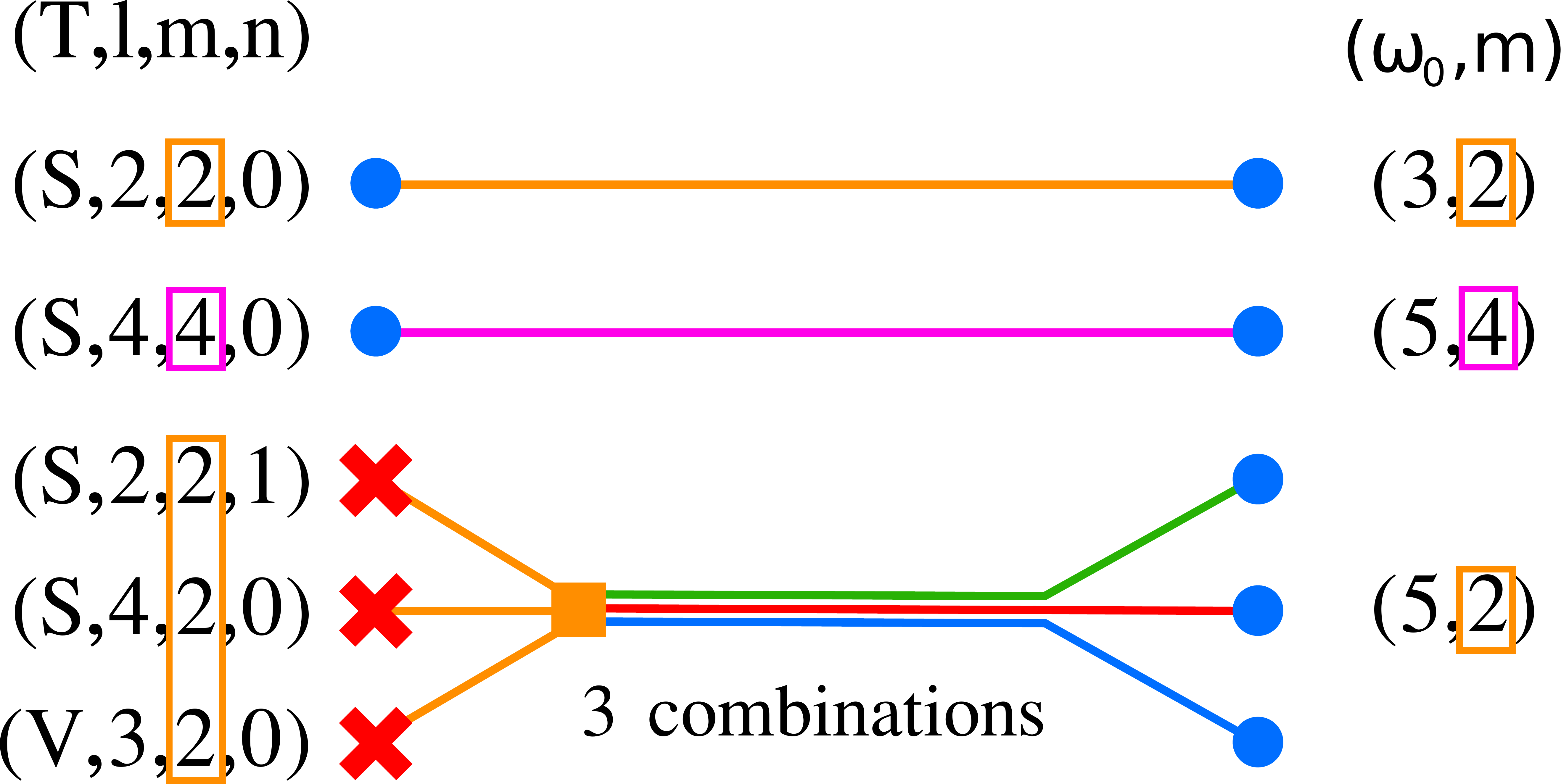}
   \caption[Two approaches for building geons perturbatively]{Two approaches for building geons perturbatively. $S$ and $V$ stand for scalar-type and vector-type perturbations respectively.
      The first point of view consists in looking for non-linear extensions of fixed $(l,m,n)$ single-mode geon seeds. This is
      represented by the left part of the picture. In this case, many single-modes appear to suffer from irremovable secular
      resonances (red crosses). The second point of view, on the right, considers non-linear extensions of fixed $(\omega,m)$ geon seeds,
      $\omega$ being a degenerated function of $l$ and $n$ (equation \eqref{omegasv}). In this case, it turns out that all secular
      resonances can be removed (blue spheres). The non-linear geon extensions thus obtained are merely linear combinations of
      single-modes geons. The number of allowed combinations matches precisely the degeneracy of the frequency $\omega$ at fixed
   $m$. Credits: G. Martinon.}
   \label{geoncontroversy}
\end{myfig}

A question of fundamental interest is how geons are linked to black holes in \gls{aads} space-times. Precisely, the black
resonators of \cite{Dias15} were configurations where a rotating black hole lied at the centre of a $l=m=2$ geon. However, they were shown
to be unstable against superradiance (see also\footnote{The main result of \cite{Green16} was that black holes with ergoregions in
\gls{ads} were linearly unstable for perturbations of rotation speeds above the \gls{hr} bound of superradiance
\cite{Hawking00} (see \cite{Brito15a} for a review on superradiance). To make a long story short, superradiant instability, or
black hole bomb, happens when a rotating black hole absorbs a wave and radiates away another one with much higher amplitude via a
Penrose process. If a mirror is placed around the black hole so as to send this energetic wave back into the black hole, the
process can repeat indefinitely and back-react substantially on the metric even if the initial perturbation was small
\cite{Press72}. This mechanism is of course at play in \gls{ads} space-time because of the reflective boundary conditions (see e.g.\
\cite{Dias13,Cardoso14} for dedicated reviews).} \cite{Green16}). The reason is that black resonators have a geon as their
zero-horizon radius limit and then merge with Kerr (precisely at the onset of superradiance) when the gravitational hair vanishes.
This is illustrated on figure \ref{blackres}. Black resonators were candidates for a putative stationary endpoint of the
superradiant instability, but the authors \cite{Niehoff16} demonstrated that it was impossible, and that maybe the stationary
endpoint of superradiance may not exist at all. These peculiar black holes exhibit only one single Killing vector, and thus
constitute the generalisation the scalar hairy black holes obtained previously in \cite{Dias11}. These single Killing field black
holes with scalar hair displayed a boson star as a zero-horizon radius limit. They then merged with \gls{mp} black holes (see
\cite{Emparan08} for a review) at the onset of superradiance, i.e.\  when the scalar hair vanishes.

\begin{myfig}
   \includegraphics[width = 0.49\textwidth]{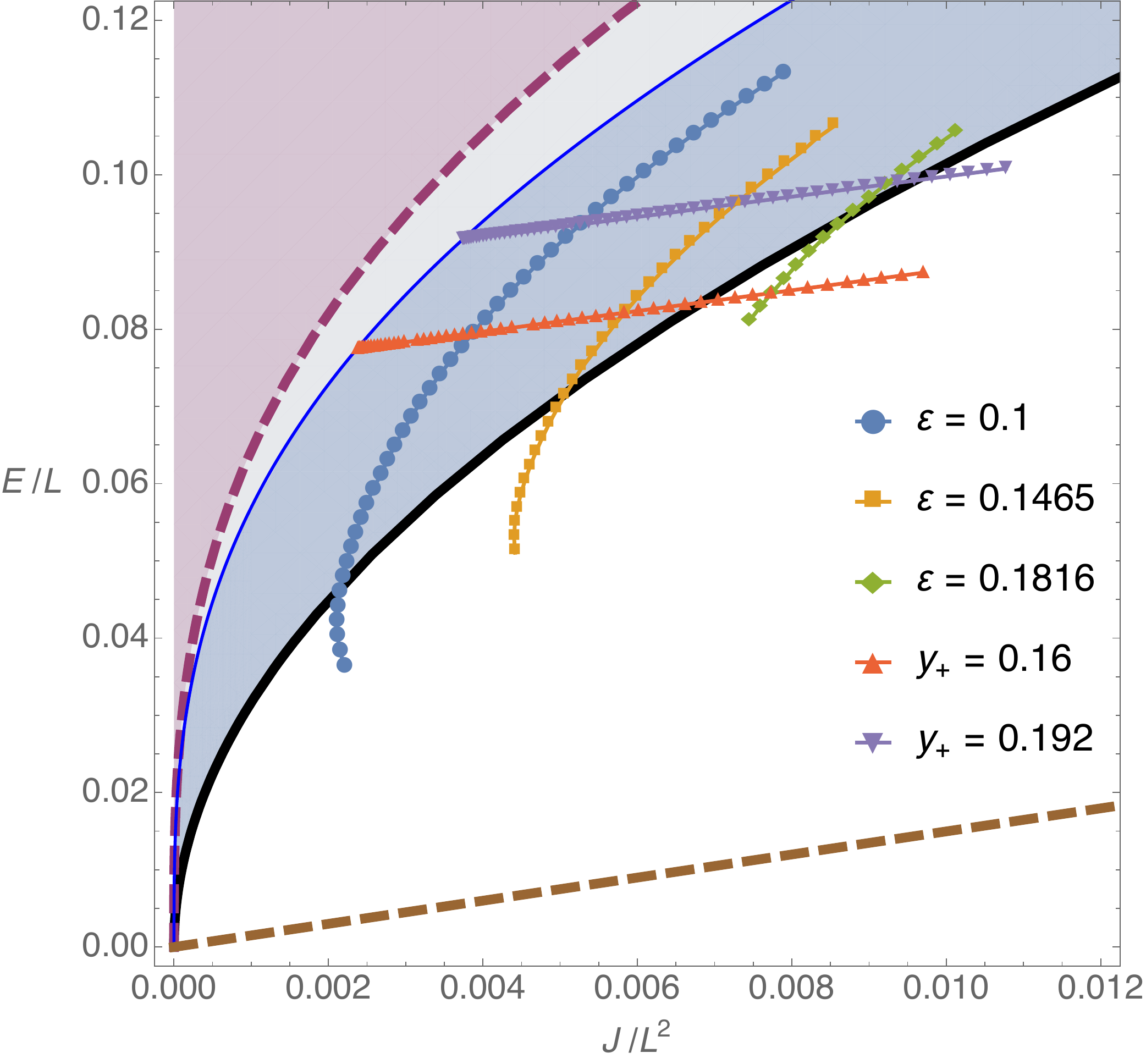}
   \caption[Phase diagram of phase resonators]{Mass-angular momentum diagram of black resonators. The thick black line depicts
      extremal Kerr-\gls{ads} such that non-extremal ones lie above. The dashed purple region delimits black holes with rotation
      frequency $\Omega_H L \leq 1$. The onset of superradiant instability for the $l = m = 2$ mode is denoted by the thin blue
      line. The bottom dashed line represent $l = m = 2$ geons. Data points represent black resonators obtained numerically. Credits: \cite{Dias15}.}
   \label{blackres}
\end{myfig}

As close counterparts of \gls{aads} gravitational geons, let us mention the Einstein-Maxwell-\gls{ads} spinning
solitons obtained in \cite{Herdeiro16a,Herdeiro16c} and the \gls{aads} Proca stars made of a massive Abelian field of
\cite{Duarte16}. Proca stars were spherically symmetric and shown to be stable against radial perturbations. As for the Einstein-Maxwell
solitons, they could also be dressed with a black hole at their centre, in much the same way as the scalar hairy black holes of
\cite{Dias11}. Since no symmetry was assumed, the obtained black holes could have a multipolar structure, depicted in figure
\ref{bhmultipole}, unlike in Minkowski space-time. The stability of these solutions was not investigated though.

\begin{myfig}
   \includegraphics[width = 0.32\textwidth]{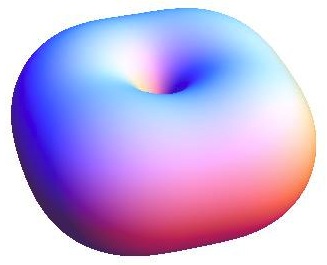}
   \includegraphics[width = 0.32\textwidth]{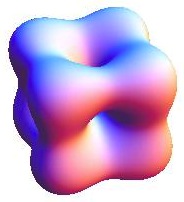}
   \includegraphics[width = 0.32\textwidth]{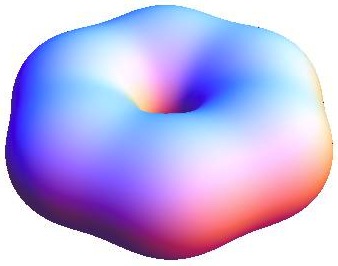}\\
   \includegraphics[width = 0.32\textwidth]{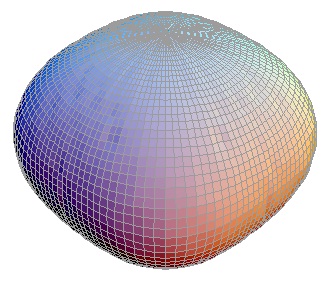}
   \includegraphics[width = 0.32\textwidth]{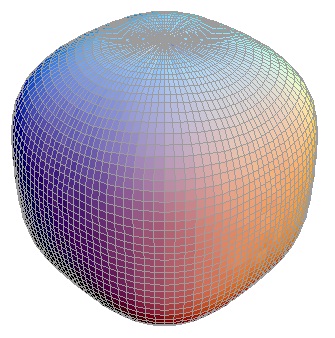}
   \includegraphics[width = 0.32\textwidth]{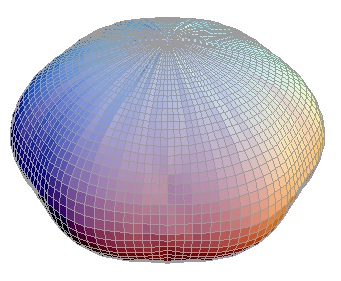}
   \caption[Multipolar structure of AdS-electrovacuum solitons and black holes]{Top: surface of constant energy density for the
      Einstein-Maxwell-\gls{ads} solitons with regular electric multipoles. Bottom: isometrics embedding for the horizon of
      \gls{ads}-electro-vacuum black holes. The same multipole (from left to right) are used: $(l,m) = (2,2),(3,2),(3,3)$. The
      topology of horizon matches that of the energy isocontours of the surrounding soliton. Credits: \cite{Herdeiro16c}.}
   \label{bhmultipole}
\end{myfig}

As a final remark on black holes, it is important to mention that the authors of
\cite{Holzegel13b} have studied analytically the non-linear stability of \gls{sads} black holes with no spherical symmetry
assumption and conjectured that they were non-linearly unstable. More generally, what was granted in spherical symmetry seems to
break down when no symmetry is involved.

Very recently, the first ever numerical evolution of a massless scalar field beyond spherical symmetry was investigated in
\cite{Bantilan17}. The authors chose to work in 5-dimensional \gls{ads} space-time with an $SO(3)$ symmetry, such that dynamics
occurred only in the $(t,x,y)$ directions (compared to $(t,r)$ for spherical symmetry). The numerical evolution was also restricted
to zero angular momentum initial data, which was parametrised by two amplitudes $A$ and $B$, $A$ being a spherically symmetric
component and $B$ a first harmonic excitation. The scalar field $\phi$ initial data then read
\begin{equation}
   \phi(\rho,\chi) = A f(\rho) + B g(\gamma) \cos \chi,
\end{equation}
where $\rho$ was a radial coordinate, $\chi$ an angular coordinate, and functions $f$ and $g$ were piecewise $C^2$ functions.
The spherically symmetric case is recovered whenever $B$ is zero. The main results of this study are summarised in figure
\ref{nonspherical}. At fixed mass, non-spherically symmetric initial data undergo fewer bounces before black hole formation. At
fixed number of bounces, the non-spherically symmetric initial data need less mass to collapse. This lead the authors to
conjecture that the instability was even more prominent beyond spherical symmetry, as already suggested by \cite{Dias16a,Dias17a}.
These results are somewhat difficult to compare to the spherically symmetric knowledge since no low amplitude limit is probed
numerically. 

\begin{myfig}
   \includegraphics[width = 0.49\textwidth]{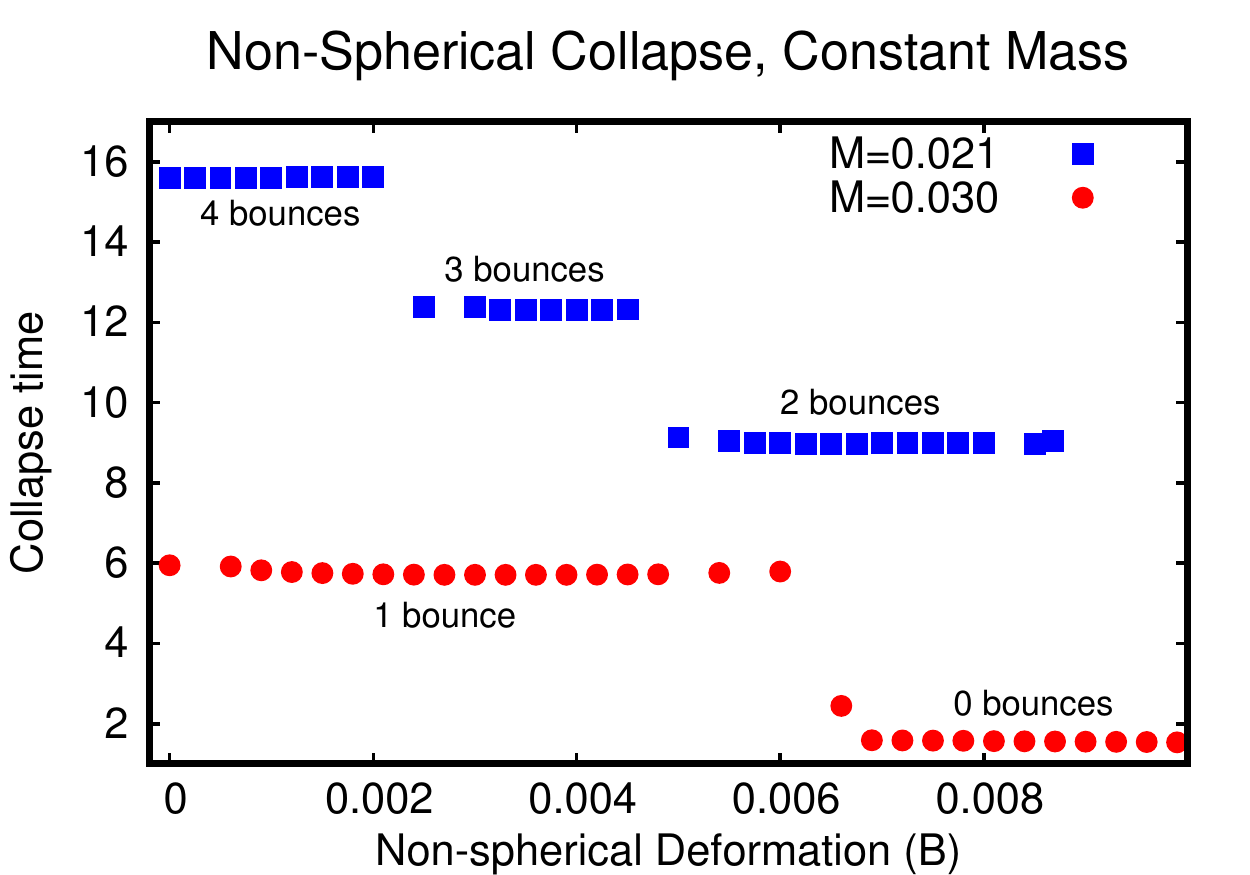}
   \includegraphics[width = 0.49\textwidth]{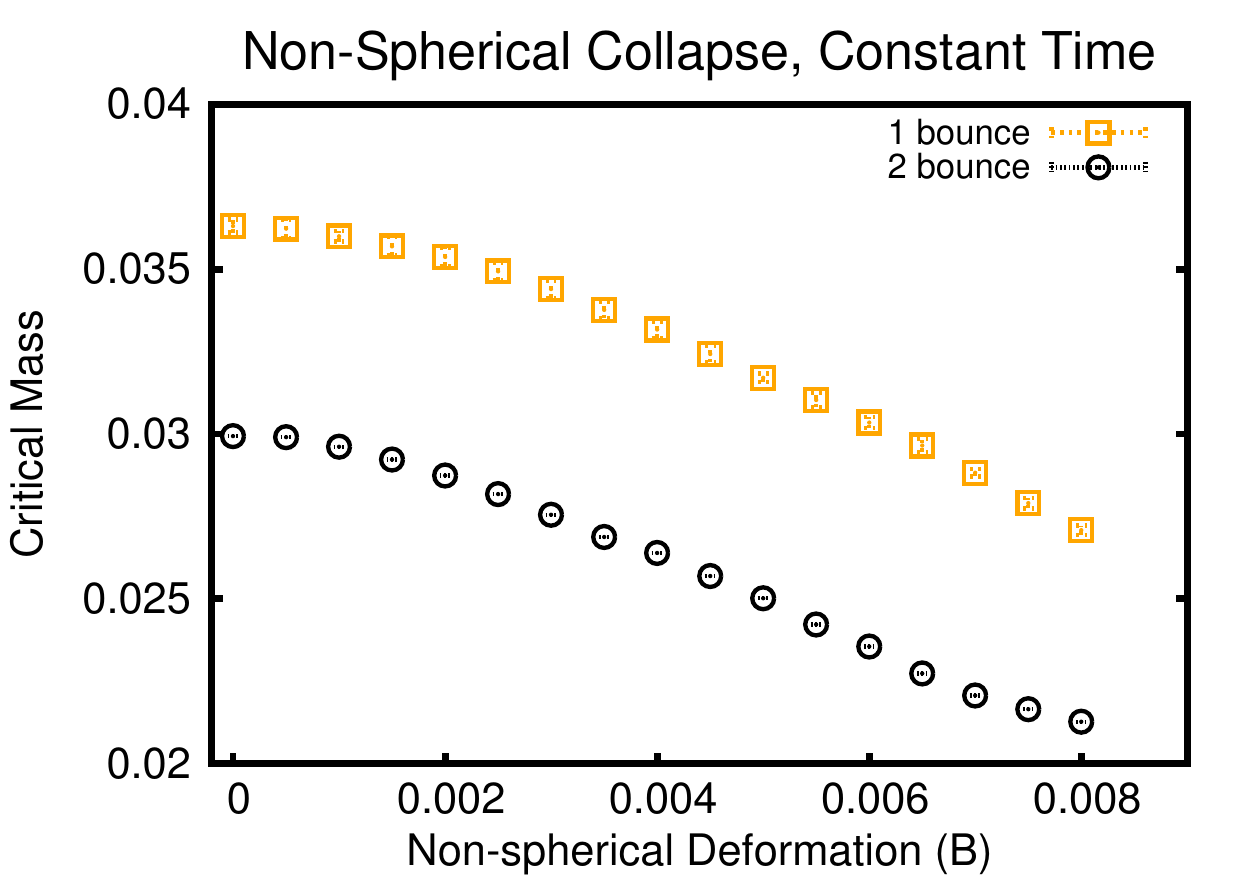}
   \caption[Earlier collapse without spherical symmetry]{Left: Collapse time as a function of the non-spherically symmetric
      deformation of the initial scalar profile, for fixed total mass $M = 0.021$ (blue squares) and $M = 0.030$ (red circles).
      Right: Maximum mass for which a black hole is formed after $N$ bounces, $N$ being 1 (yellow squares) or 2 (black circles),
      as a function of the non-spherical deformation. Credits: \cite{Bantilan17}.}
   \label{nonspherical}
\end{myfig}

Almost simultaneously, the authors of \cite{Choptuik17} performed simulations of a massless scalar field with an azimuthal number $m=1$
ansatz. They compared two families of initial data, one with zero angular momentum $J$ and one with $J \simeq 0.155 E$, $E$ being
the total mass. They observed numerically that non-zero angular momentum simulations were still unstable at low amplitudes.
However, it did take more time to form a black hole compared to the zero angular momentum case. This is illustrated in figure
\ref{twoinit}. The end state of the stability could be a \gls{mp} black hole or the hairy black holes of \cite{Dias11}.

\begin{myfig}
   \includegraphics[width = 0.49\textwidth]{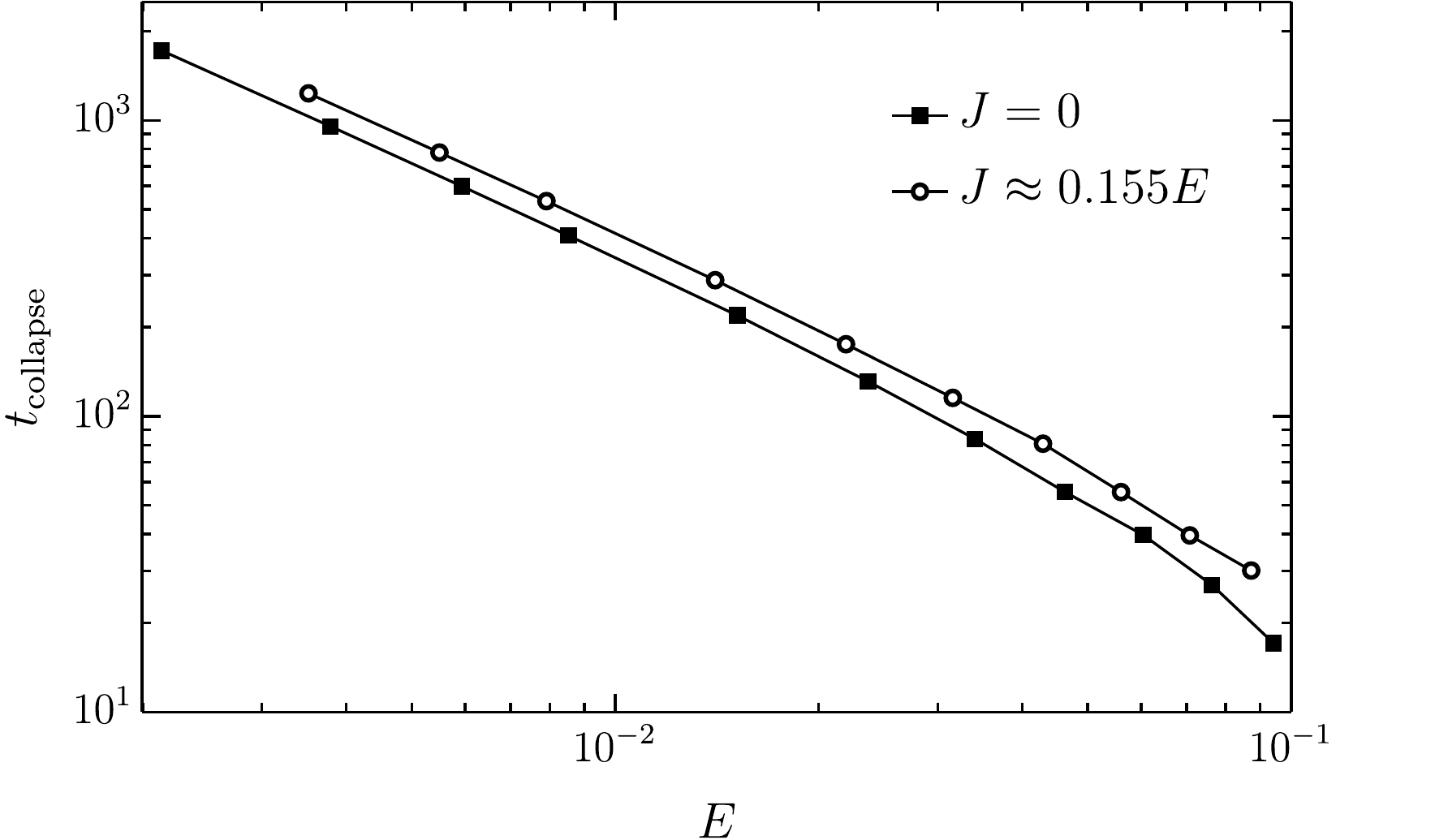}
   \caption[Later collapse with angular momentum]{Time of collapse as a function of the initial energy content $E$ for two
      kinds of initial data: one with zero angular momentum $J=0$ and one with $J \simeq 0.155 E$. The latter simulations do not
      suppress the instability but delay it. Credits: \cite{Choptuik17}.}
   \label{twoinit}
\end{myfig}

Thus, \cite{Bantilan17} argues that no spherical symmetry speed up the instability, while \cite{Choptuik17} advocates that angular
momentum delays black hole formation. This suggests that the structure of the instability beyond spherical symmetry is quite
involved, and not very well-known for the moment. These results are very promising for the future. For the time being, the
qualitative understanding of the \gls{ads} instability is not challenged by angular momentum or asymmetric considerations, but it
is much too soon to consider it a definitive statement. The freedom of initial data being much larger in higher dynamical
dimensions, and much more computationally expansive, we expect that progress in this area of research, however exciting, will be
much slower than in the spherically symmetric case (which literally boomed in about a few years).

\section{Structure of the instability}

So far, we have discussed configurations that were either non-linearly stable or non-linearly unstable. However, the boundary
between these two behaviours is not so clean, and the same family of solution can switch between stability and instability. We
already spotted this phenomenon for the Gaussian initial data, that is generally unstable except for width $0.4 \lesssim \sigma
\lesssim 8$ (section \ref{tpsol}). However, this transition can be much more intricate and chaotic.

\subsection{Chaotic footprints}
\label{chaosfoot}

Apart from Gaussian or two-mode initial data, another family of experiments in \gls{aads} space-times consists in the interaction
of two concentric thin shells of a perfect fluid with a linear equation of state $p = w \rho$ where $p$ is the pressure, $\rho$
the energy density and $w$ a free parameter, that can be constrained by energy conditions (weak, null or strong). This problem is
close to the collapse of a scalar wave packet as several authors \cite{Pretorius00,Abajo14,Silva15,Deppe15a,Deppe15b,Deppe16a}
noticed that broad pulse initial data had the tendency to break off into several sub-pulses interacting with each other.
Furthermore, this setup was the first to exhibit chaotic behaviours in the \gls{ads} instability context.

The first study of this problem was initiated in \cite{Mas15}. The authors patched several Schwarzschild metrics together with
Israel junction conditions (see for example \cite{Poisson04}) between each shell and studied the effective potential of
interaction. They concluded that periodic solutions should then exist.

The numerical experiment was performed in \cite{Cardoso16} in a flat space-time enclosed in a cavity and in \cite{Brito16b} in
\gls{ads} space-time. This problem could be seen as the simplest two-body problem as it was one-dimensional. Several kinds of
behaviours were observed: prompt collapse, delayed collapse and perpetual oscillatory motion. The confinement seems crucial to
ensure that the shells collide each other repeatedly, thus allowing small effects to build up in time.

Examining the number of crossings of the shells during evolution and before collapse, several striking results emerged. First a
fractal-like structure was clearly visible, as depicted in figure \ref{fractal} (left panel) for initial conditions where the two
shells started at the same positions $R_i$. Like in many chaotic systems, the authors unveiled windows of stability, i.e.\ some
ranges of $R_i$ in which no collapse occurred.

Moreover, for a different set of initial conditions, the mass of the black hole exhibited a critical behaviour in the left
neighbourhood of critical points, namely
\begin{equation}
   M_H - M_{n+1} \propto ( \delta_n - \delta )^\gamma,
\end{equation}
where $\delta$ encodes the mass-energy content of the initial data and $M_n$ was the black hole mass at the critical amplitude
$\delta_n$. The coefficient $\gamma$ was found to be $\sim 0.95$ but actually depended on the equation of state parameter $w$. This
critical behaviour is illustrated in figure \ref{fractal} (right panel).

\begin{myfig}
   \includegraphics[width = 0.44\textwidth]{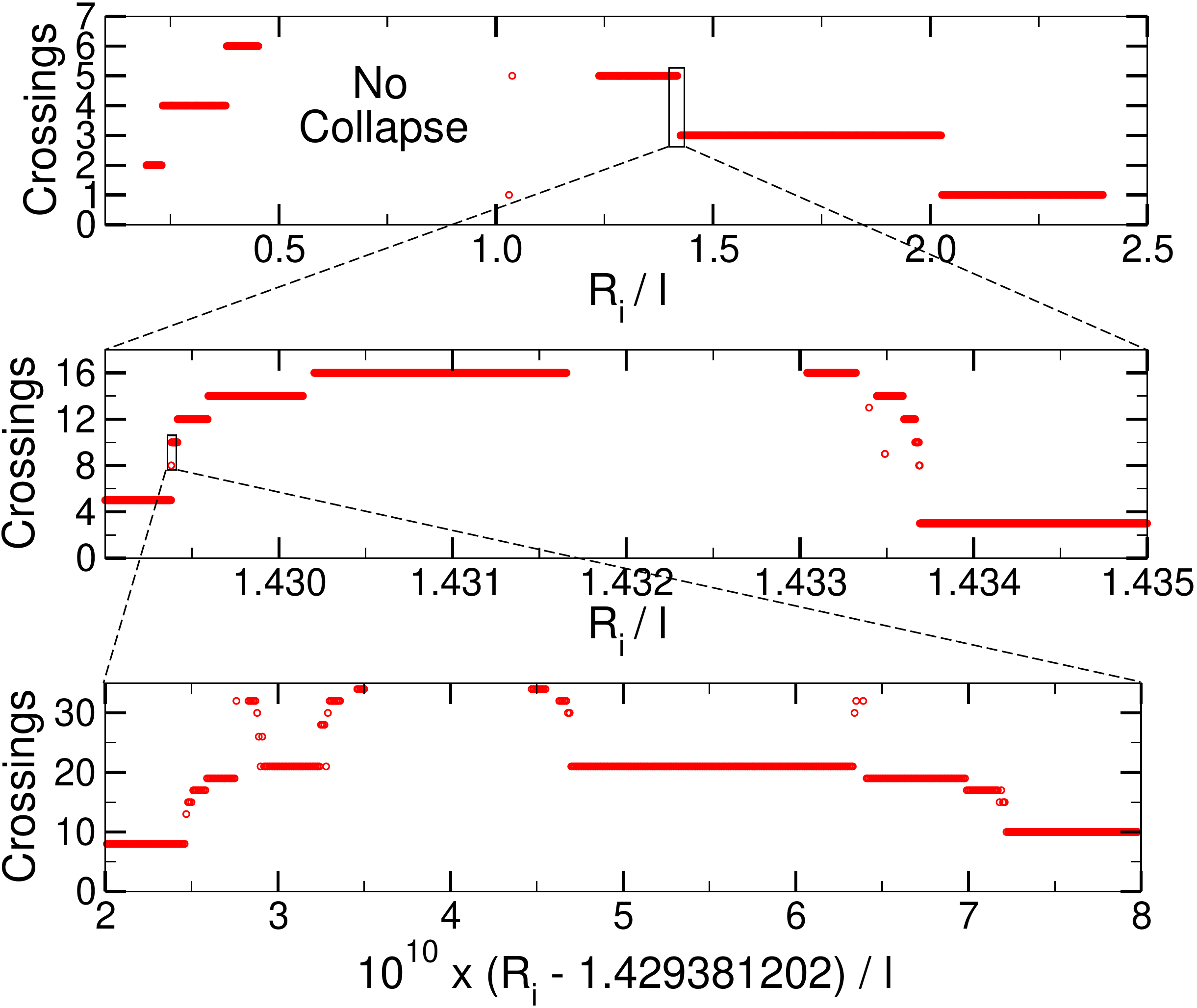}
   \includegraphics[width = 0.54\textwidth]{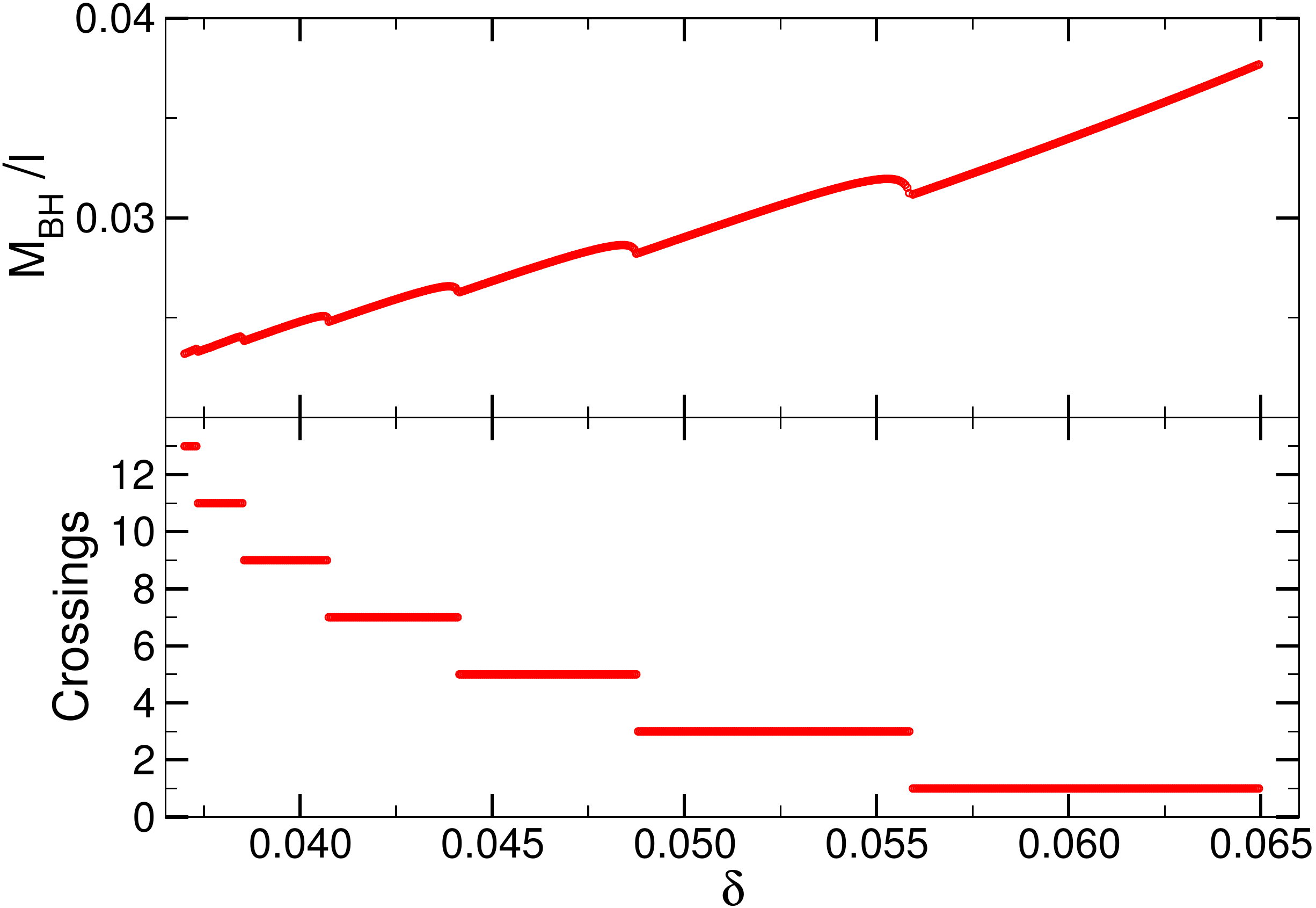}
   \caption[Fractal and critical behaviours in the interacting shells configuration]{Left: number of crossings between the two
      shells before collapse as a function of $R_i/\gls{L}$, their common initial position. There are regions where no collapse
      occurs, and a fractal-like structure emerges where an arbitrarily large number of crossings takes place. Right: black hole
      mass and number of crossings as a function of the initial mass-energy content encoded in the $\delta$ parameter for
      a different set of initial conditions. In the left neighbourhood of the $n^{th}$ critical point starting at a mass
      $M_{n+1}$, the mass of the black hole behaves as $M_{H} - M_{n+1} \sim (\delta_n - \delta)^\gamma$. Credits:
      \cite{Brito16b}.}
   \label{fractal}
\end{myfig}

Finally, non-collapsing solutions explored the space of parameters in a chaotic way, as can be seen on figure \ref{chaos}, where
the phase space of the inner and outer shell displayed fractal curves. The thin shell model thus unfolded a very rich structure as
well as simple examples of chaotic islands of stability that were highly sensitive to initial conditions.

\begin{myfig}
   \includegraphics[width = 0.49\textwidth]{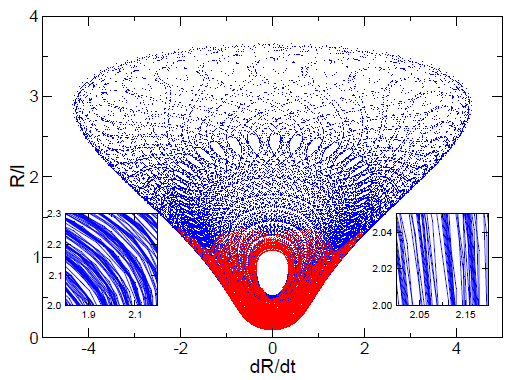}
   \caption[Chaos in non-collapsing interacting shells]{Phase space position-speed of a non-collapsing solution in the thin-shell
      problem. Red depicts the innermost shell's orbit and blue the outermost. Insets show zooms in phase space. Credits: \cite{Brito16b}.}
   \label{chaos}
\end{myfig}

\subsection{Structure of islands of stability}
\label{topoinsta}

As more and more islands of stability were uncovered, a legitimate question was: how large were these islands in the instability sea
? In order to better understand the structure of the instability, let us momentarily consider the simpler case of Minkowski
space-time. This space-time is non-linearly stable when it is slightly perturbed \cite{Christodoulou93}. Thus no black
hole can be formed when the amplitude of the initial data is arbitrarily small. On the other hand, it is well known that many asymptotically
flat self-gravitating objects like neutron stars, white dwarves, boson stars or Proca stars have a maximum mass, i.e.\ 
collapse to a black hole if their mass is too large. We have also seen in chapter \ref{geons} that electromagnetic or
gravitational geons have no maximum mass.

All these features can be summed up in an abstract picture, depicted in figure \ref{islandpictureflat}. In a polar
representation, we can insert different families of self-gravitating solutions at different angles and use the radial direction as
a measure of the amplitude (or equivalently the mass) of a particular solution within a given family. On the one hand, at large
amplitudes, many solutions collapse to a black hole, with the notable exception of geons. On the other hand, since Minkowski
space-time is non-linearly stable against small perturbations, no black hole can be formed around the Minkowski background, such
that this space-time lies at the centre of an ``island'' of stability.

\begin{myfig}
   \includegraphics[width = 0.49\textwidth]{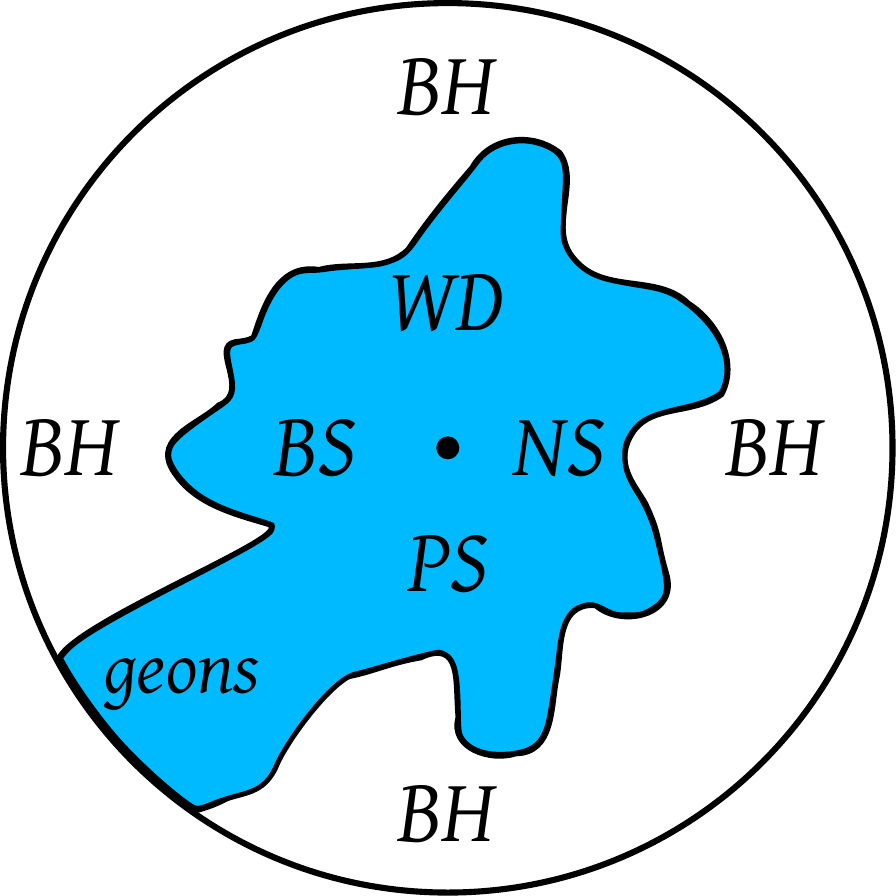}
   \caption[Generic stability of asymptotically flat space-times]{Generic stability diagram of asymptotically flat space-times.
      The radial direction denotes amplitudes of the initial data, so that Minkowski space-time lies at the centre. Angular
      direction is an abstraction that denotes the space of parameters. For example (non-rotating) neutron stars (NS) lie at an
      angle $\theta = 0$, white dwarves (WD) at $\theta = \pi/2$, boson stars (BS) at $\theta = \pi$ and Proca stars (PS) at
      $\theta = 3\pi/2$. We have also inserted electromagnetic geons at $\theta = 5\pi/4$. Since all these objects have a maximum mass,
      they all form a black hole (BH) for sufficiently large amplitudes, which corresponds to the white region. Electromagnetic
      geons on the other hand can have arbitrarily large masses, so they never form a black hole. Since Minkowski space-time is
      non-linearly stable for small perturbations, no black hole can be formed in the neighbourhood of the central Minkowski
      point. Credits: G. Martinon.}
   \label{islandpictureflat}
\end{myfig}

This kind of diagram is very instructive for studying the structure of the \gls{ads} instability. Let us first introduce the following
set of definitions\footnote{The adjective ``generic'' used in this section should not be counfounded with the mathematical
property of genericity, and is used as a synonym of ``almost always''. The distinction with the mathematical property of genericity
is important to perform, though it lacks a rigorous definition of the topology and of a measure in the space of initial data (thanks
to Piotr Bizo\'n for pointing this out).}, taken from \cite{Dimitrakopoulos15a}, and all illustrated in figure \ref{islandpicture}:
\begin{myitem}
   \item ``Generic instability'' means that the set of stable initial conditions (not forming a black hole) shrinks to measure zero
      in the zero-amplitude limit $\varepsilon \rightarrow 0$.
   \item ``Generic stability'' means that the set of unstable initial conditions (forming a black hole) shrinks to measure zero
      in the zero-amplitude limit $\varepsilon \rightarrow 0$.
   \item ``Mixed instability'' means that both sets of initial conditions have non-zero measures in the zero-amplitude limit $\varepsilon \rightarrow 0$.
\end{myitem}
In order to grasp the meaning of these definitions, let us consider a mass isocontour, pictured by a dashed circle in figure
\ref{islandpicture}. If we mentally try to progressively reduce its radius down to zero (zero-amplitude limit), we see
that the circle tends to become completely white on the left diagram (a black hole is always formed), rainbow-like on the central
one (a black hole is never formed), and half-white half-coloured on the right one (black hole may or may not form depending on the
initial data). The three different concepts listed above thus correspond to different colour end states for a limiting isomass circle
whose radius is shrinking to zero.

\begin{myfig}
   \includegraphics[width = 0.98\textwidth]{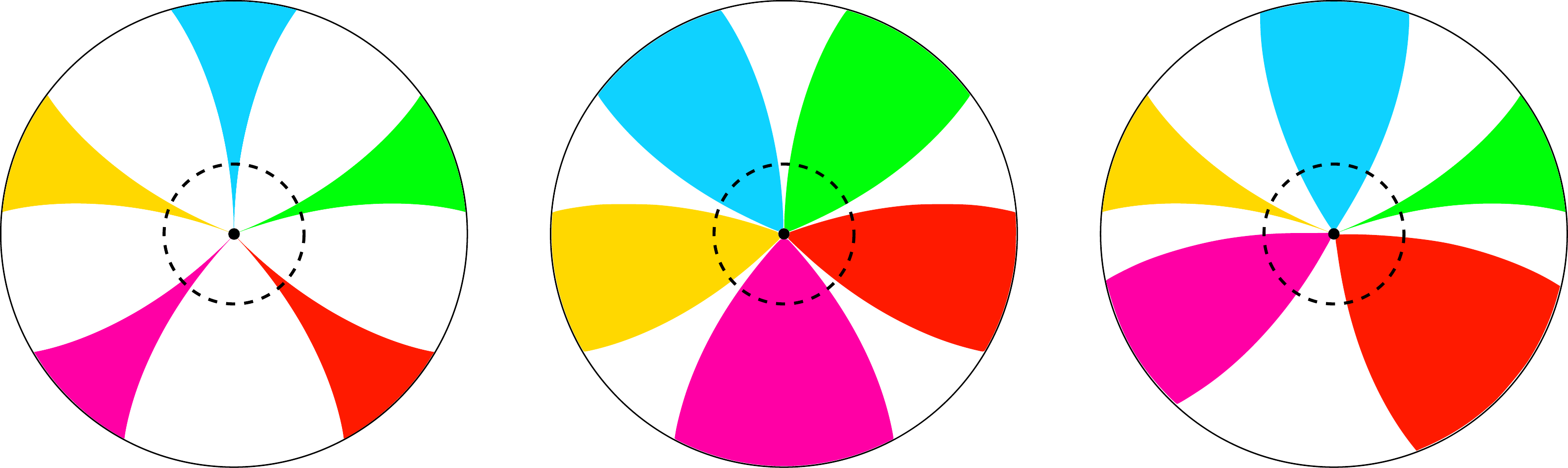}
   \caption[Generic stability in AAdS space-times]{Schematic representation of the generic stability
      definitions. The radial direction denotes amplitudes of the initial data, so that pure vacuum \gls{ads} space-time lies at
      the centre. Angular direction is an abstraction that denotes the space of parameters, like the quantum numbers of geons, or
      more generally any functional dependence. Coloured regions correspond to initial data that are non-linearly stable and never
      collapse, different colours corresponding to different families of stable solutions. White regions indicate initial data
      that are non-linearly unstable and collapse to a black hole. The left panel describes the ``generic instability'' picture,
      the middle panel describes the ``generic stability'' picture, and the right panel describes a possible ``mixed stability''
   picture. On each diagram, we show a mass isocontour with a dashed circle. Credits: G. Martinon.}
   \label{islandpicture}
\end{myfig}

After the discovery of the weakly turbulent instability \cite{Bizon11} and the perturbative construction of geons \cite{Dias12a}, the
general idea was that \gls{ads} space-time was generically unstable \cite{Dias12b}, as can be seen on the left panel of figure
\ref{islands} reproducing that of \cite{Dias12b}. However, the work of \cite{Dimitrakopoulos15b} took full advantage
of the scaling symmetry \eqref{scalingsym} of the \gls{ttf} equations to demonstrate that any non-collapsing solution of amplitude
$\varepsilon$ remained stable in the $\varepsilon \rightarrow 0$ limit in the fully non-linear theory. This forbids the cuspy shape
of the diagram and argues in favour of instability \textit{corners}. In particular, if non-collapsing solutions form a set of
measure non-zero at finite amplitudes, then they persist to be a set of measure non-zero when the amplitude tends to zero.
Figure \ref{islands} illustrates the tension between the original statement of \cite{Dias12b} and the theorems demonstrated in
\cite{Dimitrakopoulos15b}. The former advocates for the ``generic instability'' picture while the latter argues in favour of the
``mixed stability'' hypothesis.

\begin{myfig}
   \includegraphics[width = 0.37\textwidth]{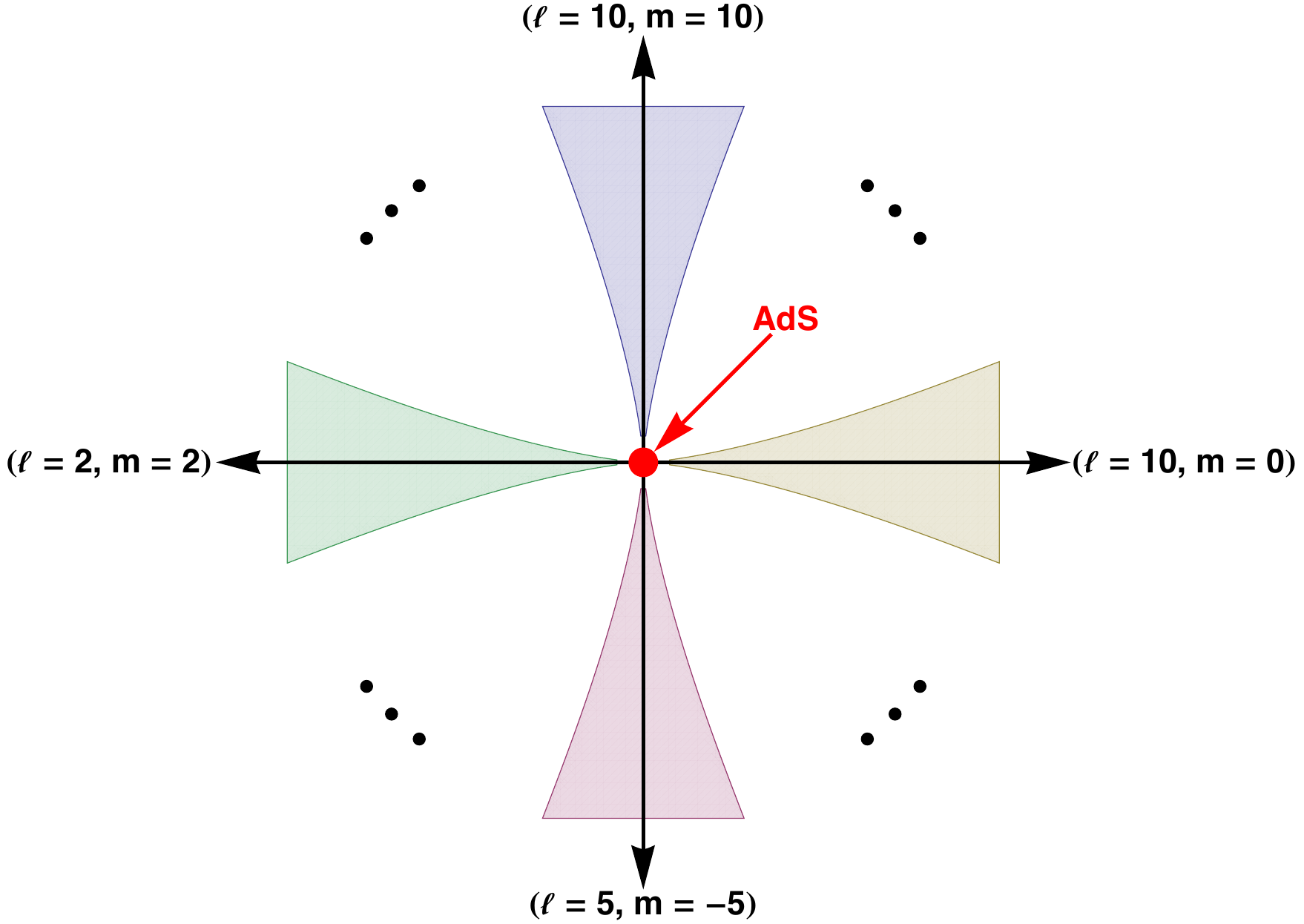}
   \includegraphics[width = 0.28\textwidth]{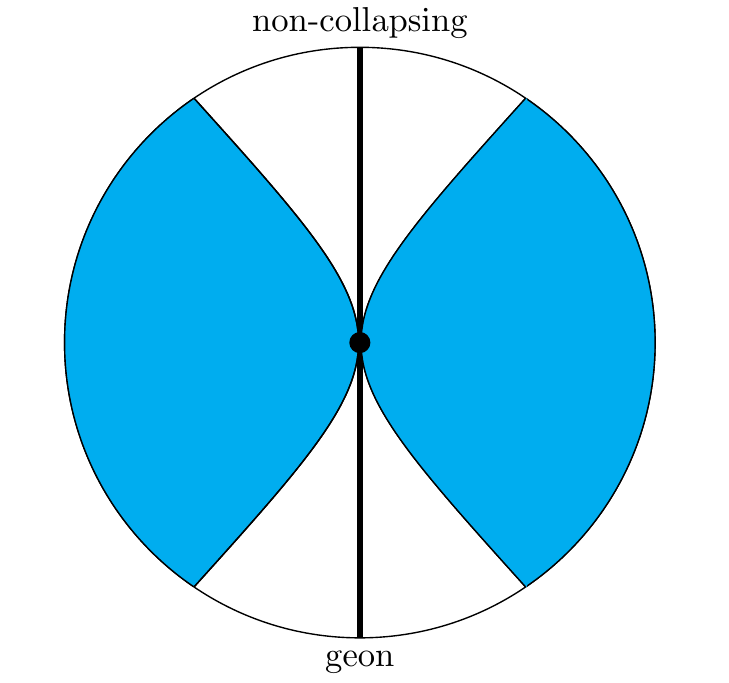}
   \includegraphics[width = 0.28\textwidth]{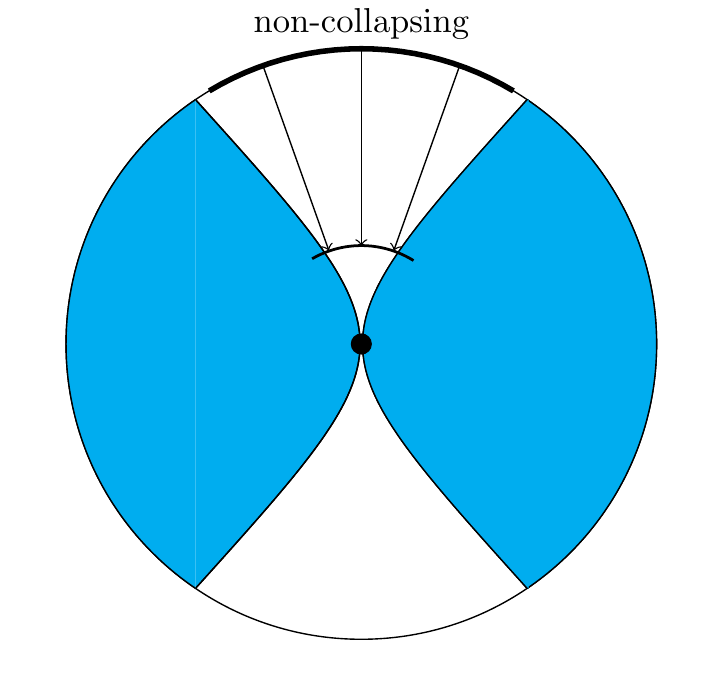}
   \caption[Islands of stability]{Left panel: diagrammatic picture of islands of stability. Geons solutions lie on the black
      arrows. Shaded regions denote islands of stability around geon solutions. Because perturbation theory about empty \gls{ads}
      leads to geons only for a measure zero set of seed solutions in the generic instability picture, each such region has been
      drawn so that empty \gls{ads} lies at a cusp. Middle and right panels: phase-space diagrams of the stability island
      conjecture. Initial perturbations in the blue region collapse while the unshaded region represents islands of stability. The
      theorems of \cite{Dimitrakopoulos15b} show that one can transport non-collapsing solutions radially without triggering
      instability. This is in direct contradiction with the cuspy nature of stable regions, that would be better drawn as
      corners. Credits:
      \cite{Dias12b,Dimitrakopoulos15b}.}
   \label{islands}
\end{myfig}

To summarise, the weakly turbulent instability of \gls{ads} \cite{Bizon11} ruled out the ``generic stability'' hypothesis. This
lead some people to adopt the ``generic instability'' picture \cite{Dias12b}. But actually, the emergence of numerical islands of
stability in combination with the \gls{ttf} framework revealed a more appropriate ``mixed stability'' representation, where unlike the
right panel of figure \ref{islands}, all stability regions are stability \textit{corners} \cite{Dimitrakopoulos15b}. If we also
take into account the chaotic behaviour exhibited in section \ref{chaosfoot}, the boundaries between stable and unstable solutions in
the mixed stability picture of figure \ref{islandpicture} might well be fractals.

\section{Conditions for collapse}
\label{necsuf}

So far, we have discussed several examples of non-linearly stable or unstable solutions. But given the mixture of possibilities
and the intricate structure of the \gls{ads} instability, a legitimate question is: can we give a set of a few necessary or
sufficient conditions for collapse?

\subsection{Competition between focusing and defocusing}
\label{competition}

In hope of giving a mechanism of the instability, intuition argues that self-gravitation of a scalar wave packet tends to
always contract the field every times its typical size is small and self-interaction is large, i.e.\ every times it crosses the
origin. And so on so forth until black hole formation. However, it is a mistake to believe that self-gravitation only contributes to
contracting the scalar field profile. Incidentally, the existence of non-linearly stable solutions suggests that there is at least
one other effect that counterbalances the contraction. Islands of stability should thus result from an endless competition between contraction and
dilatation.

Indeed, in order to understand the islands of stability found in the literature, the authors of \cite{Dimitrakopoulos15a}
developed a perturbative and heuristic argumentation: gravitational self-interaction leads to tidal
deformations which are equally likely to focus or defocus energy. This stresses the potential repulsive nature of gravitation.
A daily illustration of this statement lies in the tidal effects of the Moon onto the Earth that induce a stretching and not a
compacting. The idea is that stable solutions oscillate between focusing and defocusing dynamics. On figure \ref{focusing}, the
simple example of a scalar pulse with one peak region and one extended tail region is carried out (analytical arguments can be
found in \cite{Dimitrakopoulos15a}). On the one hand, if the peak enters first the central region near the origin, the dominant
behaviour is that of contraction. On the other hand, if the tail reaches the origin first, it flattens the peaked region via tidal
interactions. Thus, there is clearly a competition between contraction and dilatation. These two behaviours can act alternatively
in similar proportions like for stable solutions, or fluctuate so much as to finally form a black hole, as depicted in figure
\ref{flowenergy}. It is important to note that these considerations rely on position-space analysis instead of the very popular
energy spectrum analysis.

\begin{myfig}
   \includegraphics[width = 0.49\textwidth]{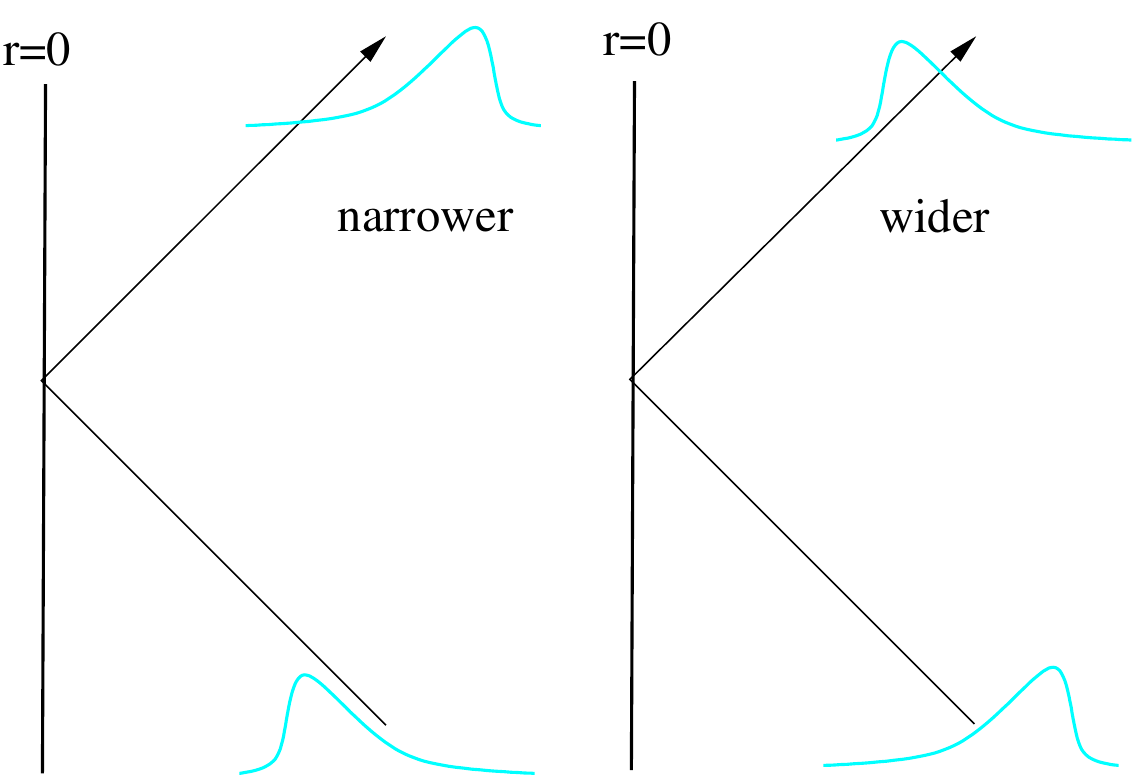}
   \caption[Focusing and defocusing of an asymmetric shell]{A thin shell with higher energy density in its front comes out
      narrower as it crosses the origin, because of self-gravitating contraction effects. However, the flipped configuration with high
      energy in the tail comes out wider because of tidal effects inducing a dilatation of the profile. Credits: \cite{Dimitrakopoulos15a}.}
   \label{focusing}
\end{myfig}
\begin{myfig}
   \includegraphics[width = 0.49\textwidth]{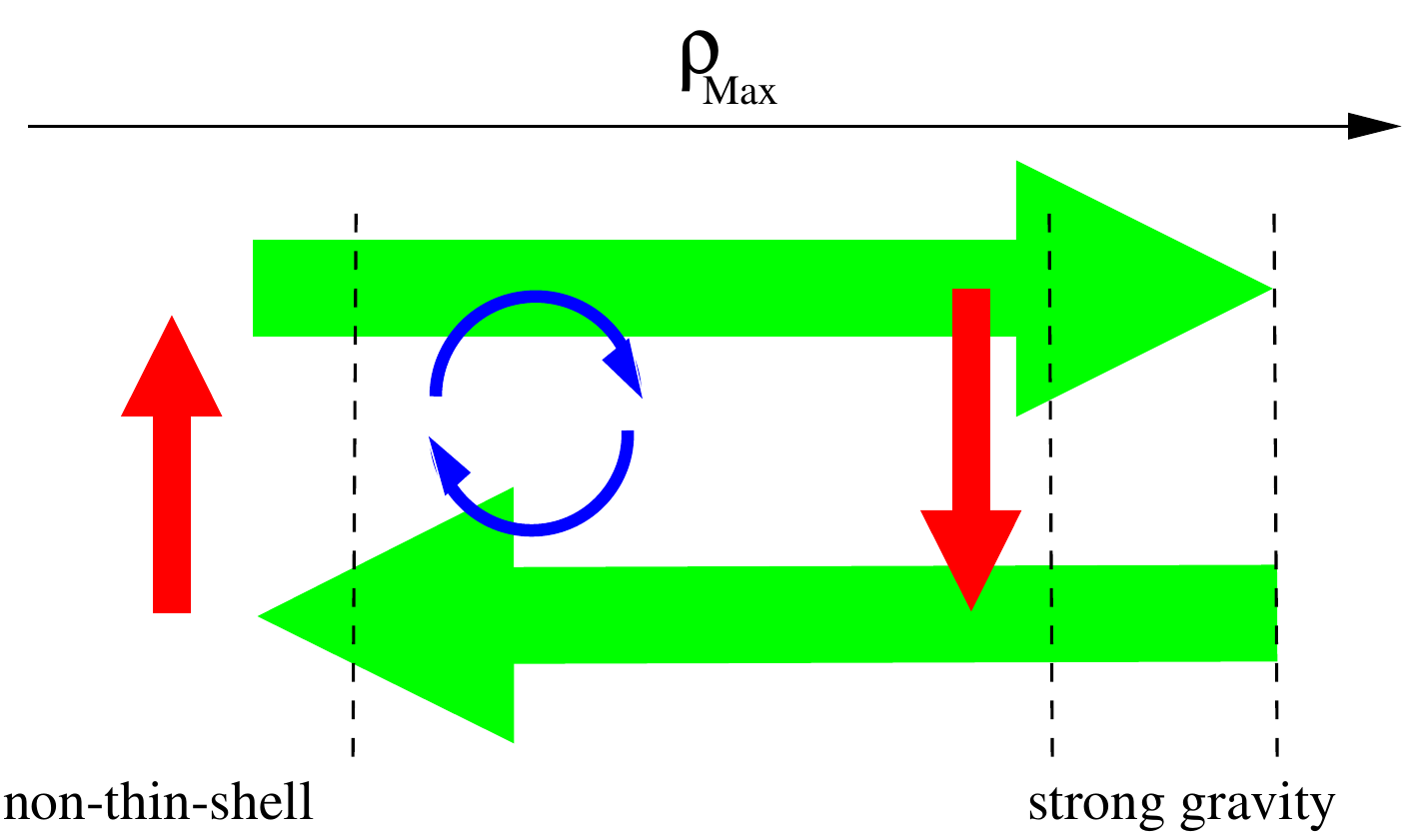}
   \includegraphics[width = 0.49\textwidth]{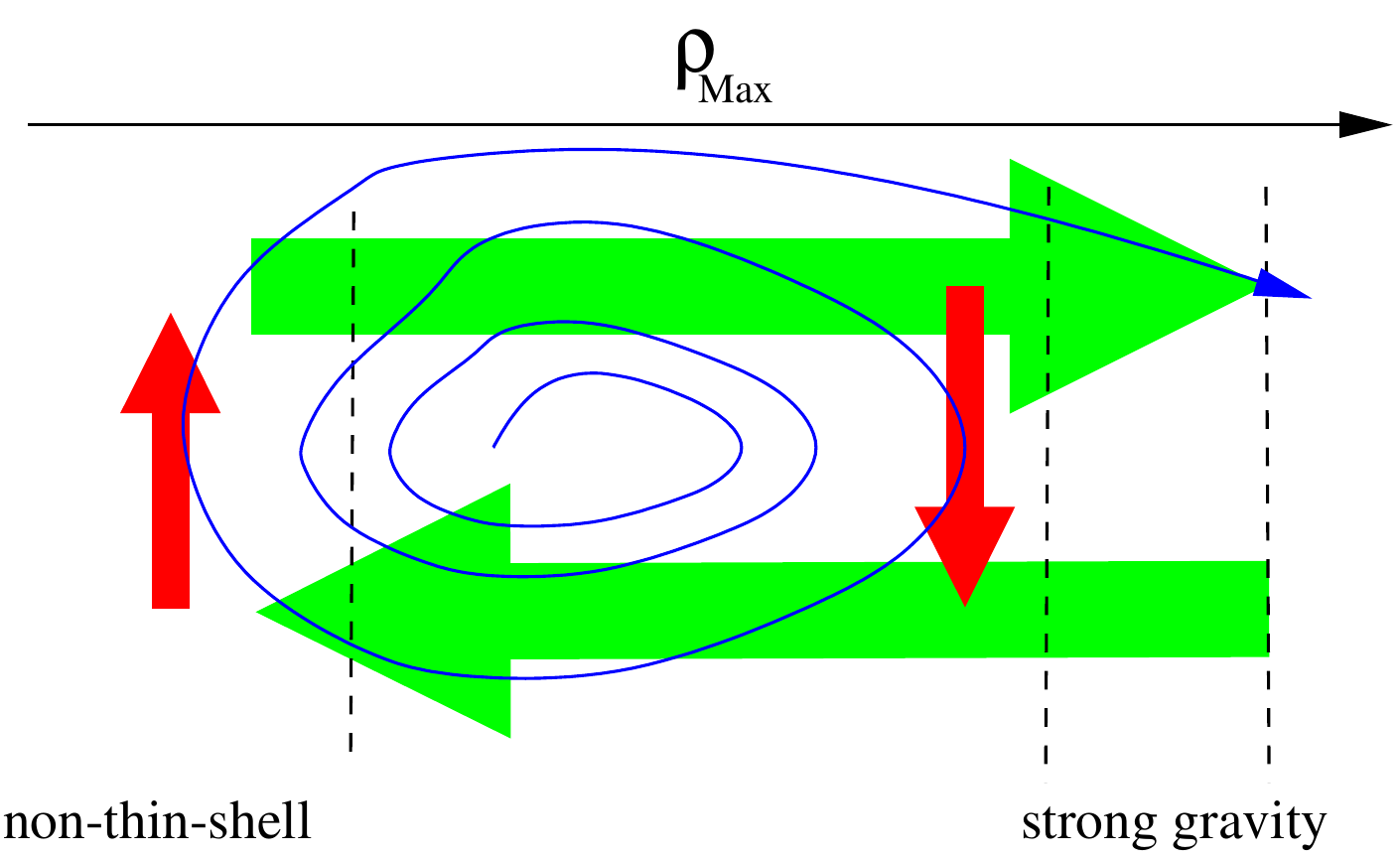}
   \caption[Flow of energy in stable and unstable configurations]{The horizontal axis is the peak energy density of the scalar
      pulse. The green left-oriented arrow represents defocusing effects while the right-oriented one represents focusing effects. There can be
      stable solutions (left panel) that never collapse because of a circular flow pattern (exchange) between these two competitive
      behaviours, while unstable solutions (right panel) fluctuate so much as to form eventually a black hole. Credits:
      \cite{Dimitrakopoulos15a}.}
   \label{flowenergy}
\end{myfig}

\subsection{Phase coherence}

Because of the competition between contraction and tidal dilatation, it can be guessed that collapse
occurs only when the successive contraction stages of the wave packet motion add up coherently during the motion without being
inhibited by the dilatation stages. This is the idea of phase coherence: small effects can build up in time if they are summed
coherently, like a child's swing gathering momentum each time it is pushed forward with the right timing.

The authors of \cite{Freivogel16} took over this argument and showed that a power-law spectrum, usually classified as unstable,
was not a sufficient condition for black hole formation. They uncovered that such a spectrum could indeed belong to a stable
solution if the phases $B_j$ were incoherent. They formulated the condition for phase coherence as (see equation
\eqref{ttfequations})
\begin{equation}
   B_j(\tau) = j\gamma(\tau) + \theta(\tau) + \ldots,
   \label{coherent}
\end{equation}
where dots represents anything that goes to zero in the large-$j$ limit. Said differently, the phases between modes are said to be
coherent if they are (asymptotically) equidistant. For example in 4-dimensional \gls{aads} space-times, if the phase coherence
condition \eqref{coherent} is satisfied, any spectrum $A_n \sim n^{-\alpha}$ with $\alpha > 3$ remains regular at all times while
black holes form for $\alpha < 2$. Between these two limits, substantial back-reaction on the metric is at work and would need
numerical simulations to conclude about black hole formation. On the other hand, whatever is the value of $\alpha$, any
phase-incoherent initial data never collapses to a black hole in a time $O(\varepsilon^{-2})$. Phase
coherence thus appears as an additional necessary condition for collapse. To prove that this condition was not sufficient,
the authors, by fine-tuning the phases, were able to build power-law spectrum solutions where there was no energy transfer among
the modes and hence non-linear stability. Furthermore they demonstrated that the two-mode initial data were particularly prone to
provide coherent phases and hence to lead to black hole formation. This was also demonstrated for unstable Gaussian initial data
\cite{Dimitrakopoulos17}.

The lesson from \cite{Freivogel16} is thus the following: when employing the \gls{ttf} equations to probe unstable configurations,
the solution can be reasonably declared unstable if (i) the analyticity radius hits zero in a finite slow-time (section
\ref{analstrip}), (ii) the phase coherence condition \eqref{coherent} holds and (iii) the exponent of the power-law
spectrum\footnote{Recall that an analyticity radius of zero implies a power-law spectrum.} should be in some interval (superior to
two in the 4-dimensional case), given in \cite{Freivogel16}.

\subsection{The role of a resonant spectrum}
\label{roleres}

A large part of the literature focused on the massless scalar field in \gls{ads} with spherical symmetry. This setting has a
linearised Einstein's operator that is resonant, i.e.\ with equidistant eigenvalues (see section \ref{pertscal}). In order to
investigate if this is a sufficient condition for the instability to be triggered, it is interesting to study different situations
where the resonant character of the spectrum is broken in order to see if the instability is suppressed or maintained. Evolving a
massive scalar field or imposing Neumann boundary conditions in an enclosed cavity are two possible ways of breaking the resonant
spectrum.

For example, massive scalar field initial data in spherical symmetry were evolved in time in the following articles. In \cite{Okawa14a},
initial data were evolved in an asymptotically flat space-time, mimicking a confining mechanism with a $\phi^4$
potential\footnote{For another confining mechanism mimicking that of \gls{ads} space-time, see also \cite{Biasi17} that studies
the Gross-Pitaevskii equation with attractive non-linearity in a harmonic potential. In this non-gravitational study, the authors
found that turning off the resonant spectrum always gave rise to a minimum threshold amplitude below which the wave function never
becomes singular.}. Even if delayed collapse (i.e.\ after several ``bounces'' off the potential barrier) was observed, there was always a
finite threshold below which no collapse occurred. In \cite{Okawa14b}, the confining mechanism was enforced within a flat enclosed
cavity with Dirichlet or Neumann boundary conditions. In the case of Dirichlet boundary conditions, and denoting by $\mu$ the mass
of the scalar field, the spectrum of the linear operator was
\begin{equation}
   \omega_j = \sqrt{\mu^2 + \frac{j^2\pi^2}{R^2}},
\end{equation}
and thus only asymptotically resonant, i.e.\ resonant only for infinite wave-numbers. Finally in \cite{Okawa15}, the standard setup was studied with a massive scalar
field. One important result of these works was that time evolutions with non-resonant spectrum collapsed earlier than in the fully
resonant case, as illustrated in figure \ref{nonresonant}. Furthermore, the authors recovered that in space-time, be the
spectrum resonant or not, all the original features of the massless scalar case were present (instability for very small
amplitudes and finely-tuned islands of stability). These results thus suggested that a resonant spectrum was not a necessary
condition for collapse.

\begin{myfig}
   \includegraphics[width = 0.49\textwidth]{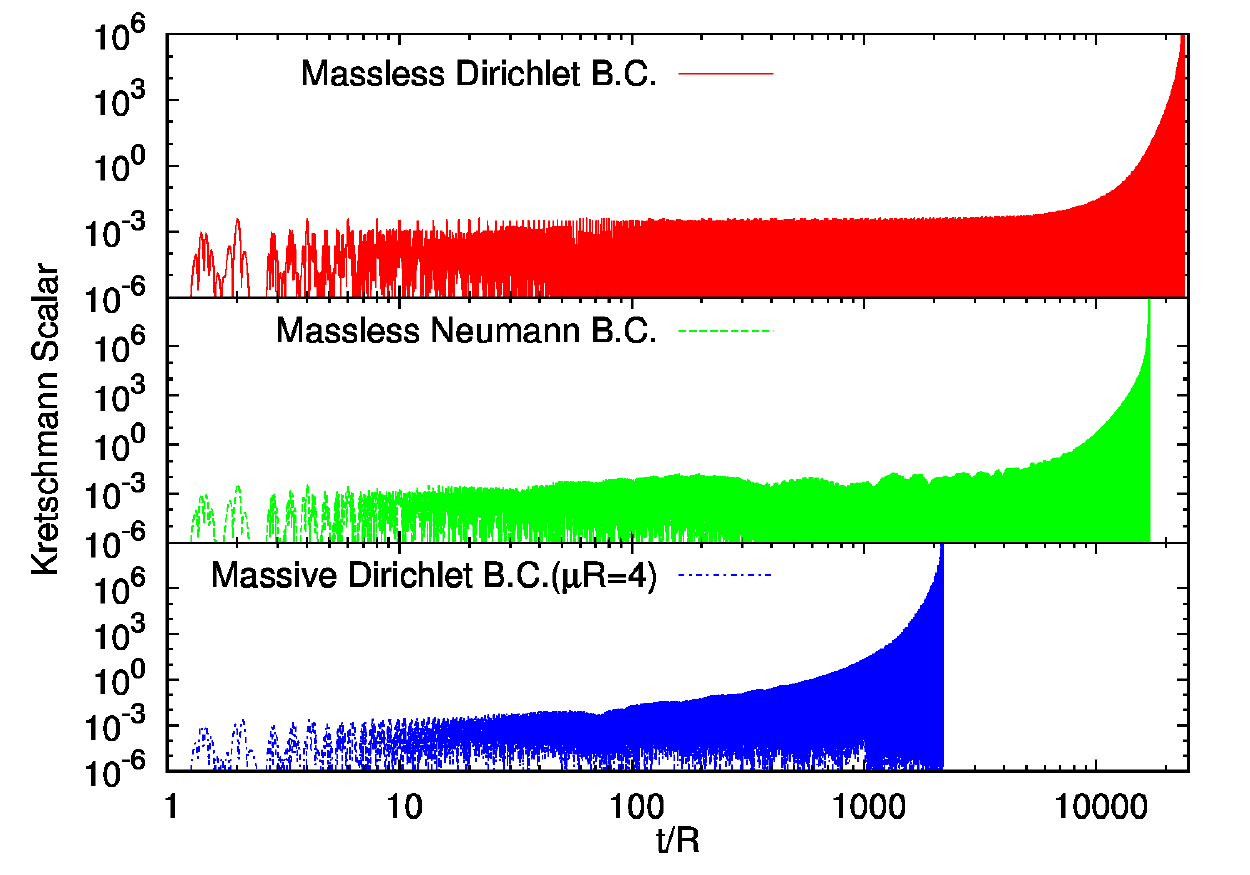}
   \caption{Evolution in time of the Kretschmann scalar for three identical Gaussian
      initial data in a flat enclosed cavity. Top: massless scalar field with Dirichlet boundary conditions, the spectrum is fully
      resonant. Middle: massless scalar field with Neumann boundary conditions, the spectrum is asymptotically resonant. Bottom:
      massive scalar field with Dirichlet boundary conditions, the spectrum is non-resonant. Credits: \cite{Okawa14b}.}
   \label{nonresonant}
\end{myfig}

In order to further investigate this point, it is also possible to investigate the dynamics of a massless scalar field in a flat
enclosed cavity, as initially suggested in \cite{Maliborski12}. At the time, the authors shared the opinion that the resonant
spectrum was not a necessary condition for the instability. However, the careful analysis beyond spherical symmetry of
\cite{Dias12b} (see section \ref{beyondspher}) came out soon after and established an opposite conclusion. In order to solve
this contradiction between analytical and numerical studies, the flat enclosed cavity was scrutinised once more in
\cite{Maliborski14}, whose authors pursued their initial work \cite{Maliborski12}. Decreasing the amplitude of the initial data,
they did find this time that, with Neumann boundary conditions, there existed a threshold $\varepsilon_0$ below which no collapse
occurred. The case of Neumann boundary conditions was particular precisely because the spectrum of the linear operator was only
asymptotically resonant. Indeed, the eigen frequencies obey $\tan(\omega_j R) = \omega_j R$, where $R$ is the radius of the
cavity. In the large-$j$ limit, it comes
\begin{equation}
   \omega_j = \frac{\pi}{R}\left( j + \frac{1}{2} \right) + O\left( \frac{1}{j} \right),
\end{equation}
so that these frequencies are only asymptotically equidistant. This setup was generalised to the charged scalar field in
\cite{Ponglertsakul16} with qualitatively identical results: no collapse below a certain threshold of amplitude. This was also analytically supported by arguments within the
\gls{kam} theory in \cite{Menon16}. Indeed, non-linear dynamics theorems argue that when the spectrum is not
perfectly resonant, there always exists a finite (but possibly arbitrarily small and difficult to spot numerically) threshold
$\varepsilon_0$ below which the instability is suppressed.

Thus the three numerical studies \cite{Okawa14a,Okawa14b,Okawa15} dealing with massive scalar fields had opposite conclusions to the ones of
\cite{Maliborski14,Ponglertsakul16} whose authors used a massless scalar field. However, thanks to the analytical studies
\cite{Dias12b,Menon16}, the role of a resonant spectrum is now well understood to be that of a necessary (but not sufficient)
condition for the instability.

At this point, it is wise to point out two numerical limitations: (i) the amplitude of numerical initial data
cannot be arbitrarily small and (ii) numerical simulations run for a finite amount of time.  Thus, if an unstable solution is
spotted in a numerical simulation, it is rigorously impossible to state that it remains unstable in the zero-amplitude
limit\footnote{In contrast with stable solutions whose stability is preserved by decreasing the amplitude, as explained in section
\ref{topoinsta} via the \gls{ttf} equations.}. In the same vein, if a stable solution is observed to be stable for a finite
(even if very long) time in numerical simulations, it is rigorously impossible to state that it remains stable in the
infinite-time limit. The different numerical setups and precisions used in the literature may explain the apparent disputes
between several authors. This is why analytical frameworks like \gls{kam} theory, perturbations, analyticity strip and
\gls{ttf} are invaluable tools to get more insight in the structure of the instability.

For the sake of completeness, let us mention that a lot more new classes of islands of stability were uncovered in the massive
scalar case. For example in \cite{Fodor14,Fodor15}, time-periodic solutions (or breathers) were constructed perturbatively and
numerically. Moreover, by fine-tuning the mass of the scalar field, whole families of non-linearly stable Gaussian initial data
were found in \cite{Deppe15a}. Finally in \cite{Okawa15}, initial data made of three Gaussian wave packets were observed to remain
non-linearly stable by continuously exchanging energy between its constitutive parts. This is in full agreement with the
focusing/defocusing mechanism discussed in section \ref{competition}.

\section{Instability when no black hole is allowed}

So far, we always confounded non-linear instability and black hole (or apparent horizon) formation. How would the instability be
expressed if, by one way or another, black hole formation was simply forbidden? There are at least two simple cases when black
hole formation is not allowed: 3-dimensional \gls{aads} space-times and \gls{egb} gravity.

\subsection{3-dimensional AAdS space-times}

The particular case of 3-dimensional \gls{aads} space-times was studied in \cite{Bizon13}. As mentioned
in section \ref{critical}, an \gls{aads} 3-dimensional (or \gls{btz}) black hole has a minimum mass $M_0$, so that initial data with total mass $M < M_0$ can
never form this kind of singularity. Thus, the natural turbulent cascade cut-off of black hole formation does not
prevail any more, and what is observed is an exponential growth of the Sobolev norm $H_2(t) = \lVert \phi''(t,x) \rVert_2$ as well
as an exponential decrease of the analyticity radius, shown in figure \ref{sobolev}. Unlike the higher dimensional cases, the
analyticity radius never reaches zero.

\begin{myfig}
   \includegraphics[width = 0.49\textwidth]{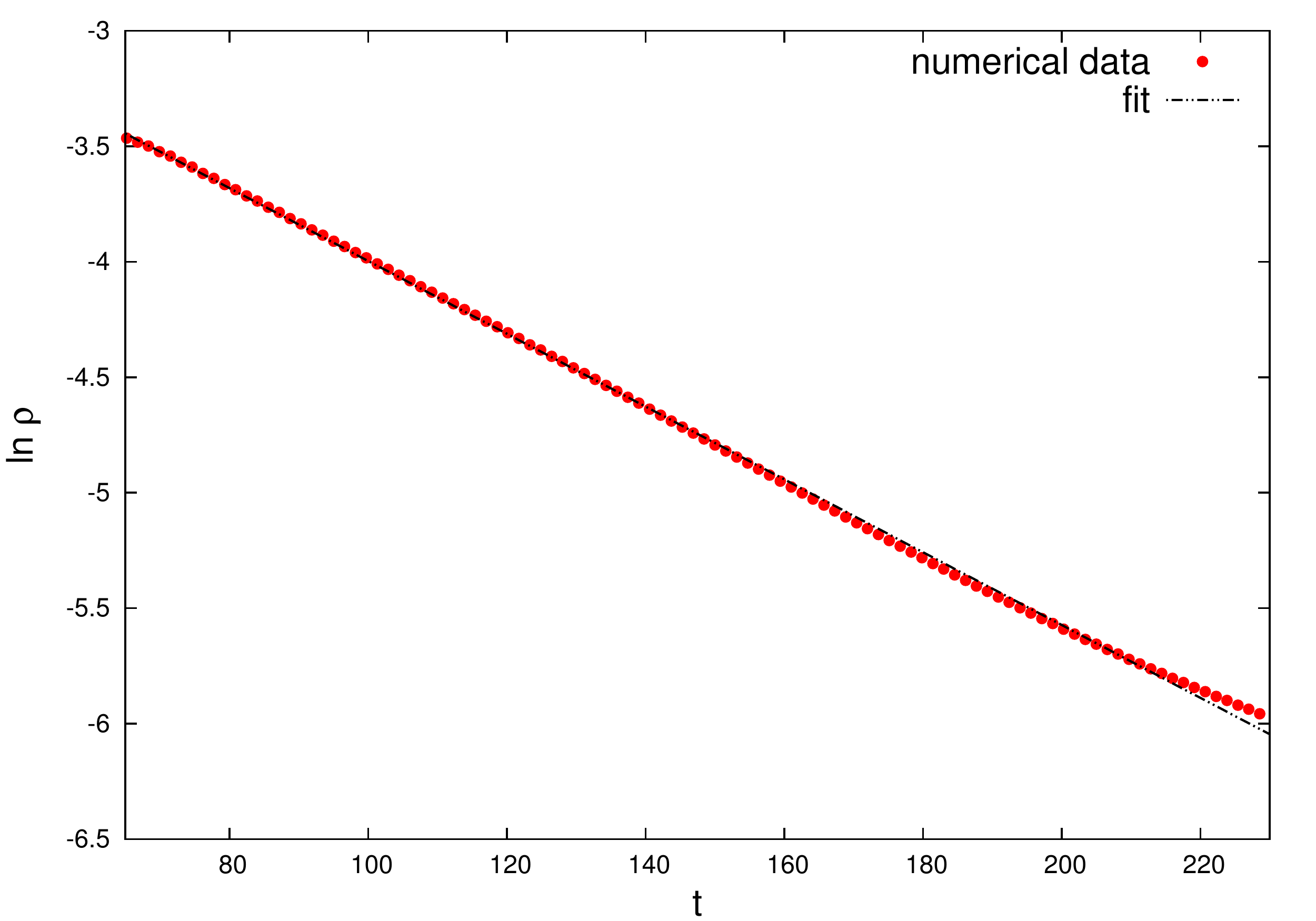}
   \includegraphics[width = 0.49\textwidth]{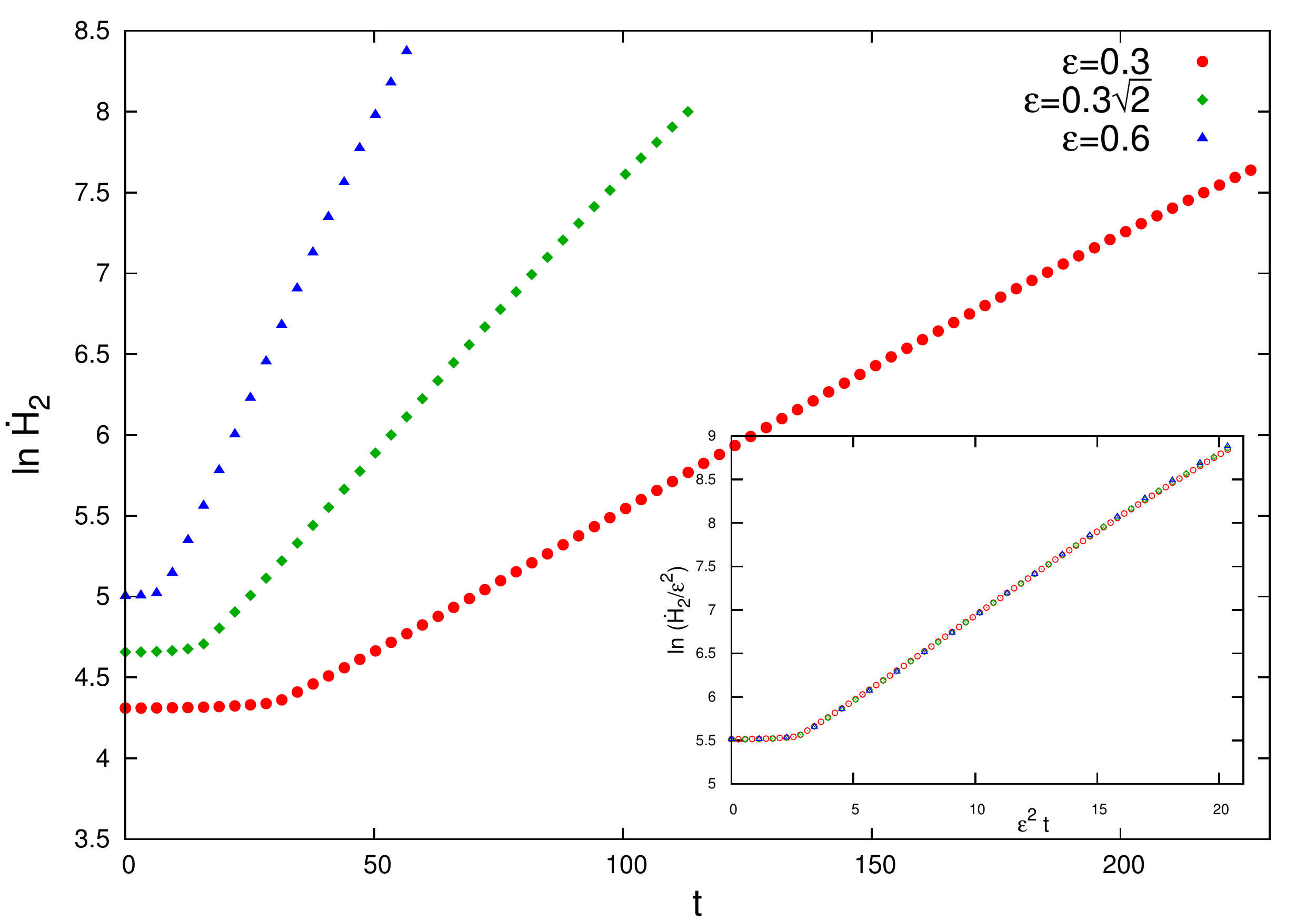}
   \caption[Analyticity radius and Sobolev norm during collaspe in AdS$_3$]{Left: analyticity radius $\rho$ of a Gaussian initial
      data in 3-dimensional \gls{ads} space-time. The decrease is exponential, but $\rho$ never hits zero. Right: Sobolev norm $H_2(t) =
      \lVert \phi''(t,x) \rVert_2$ of the scalar field for three different amplitudes of the Gaussian initial data. The inset
      shows the same curves in terms of the slow-time $\varepsilon^2 t$, so that it is clear that the blow up is $\propto
      \exp(O(\varepsilon^2 t))$. Credits: \cite{Bizon13}.}
   \label{sobolev}
\end{myfig}

The instability conjecture has thus a different flavour in three dimensions \cite{Bizon14}:

\begin{conjecture}[Anti-de Sitter instability in 3 dimensions]
Small smooth perturbations of 3-dimensional \gls{ads} remain smooth at all times but their radius of analyticity shrinks to
zero exponentially fast.
\end{conjecture}

Said differently, even if no black hole is formed for sufficiently small amplitude initial data, the turbulent cascade still
occurs as the perturbations do not remain small in any reasonable norm that captures the turbulent behaviour.

\subsection{Einstein-Gauss-Bonnet gravity}

The extension of the \gls{ekg} setup to 5-dimensional \gls{egb} gravity was performed in \cite{Deppe15a,Deppe16a}, adding the
following term to the Lagrangian of the theory
\begin{equation}
   \lambda(R^2 - 4R_{\mu\nu}R^{\mu\nu} + R_{\mu\nu\rho\sigma}R^{\mu\nu\rho\sigma}),
\end{equation}
where $\lambda$ is the \gls{gb} parameter. In this theory, a limiting black hole of radius $r_H\rightarrow 0$ has a
minimum mass $M_0 = \lambda/2$. This implies that, in analogy with the 3-dimensional \gls{ads} case, no black hole can form if the
initial data has mass lower than $M_0$.

The authors chose to study Gaussian initial data, for which the amplitude corresponding to $M_0$ is $\varepsilon = 21.86$. However, no black
hole formation was observed for amplitudes below $\varepsilon = 36$, whose last collapsing solution bounced 24 times. This means
that there was a whole range of amplitudes where black hole formation was theoretically possible but yet did not occur. Running a simulation with
an amplitude in this range but turning off the \gls{gb} term did indeed reignite collapse. The intuition was that the \gls{gb}
term contributes largely to defocusing mechanisms that resists black hole formation. The instability was thus suppressed more
severely than expected.  The comparison of collapses with and without the \gls{gb} term is clearly visible on
figure \ref{gaussbonnet}, where the structure of the instability seems much more chaotic in the \gls{gb} gravity.

\begin{myfig}
   \includegraphics[width = 0.49\textwidth]{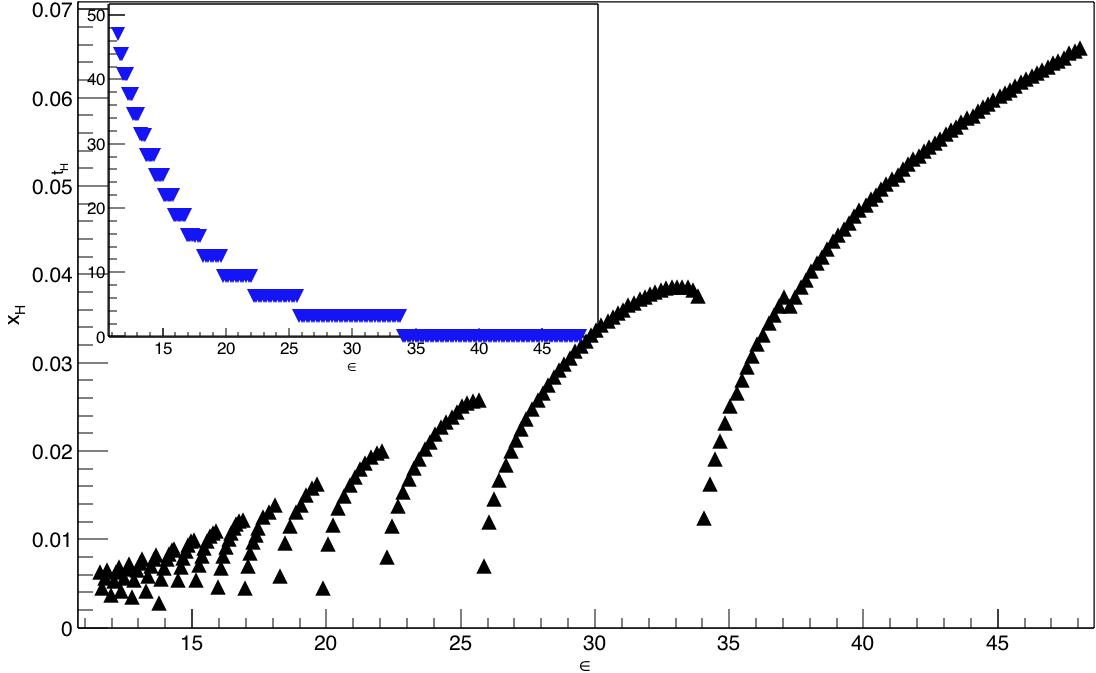}
   \includegraphics[width = 0.49\textwidth]{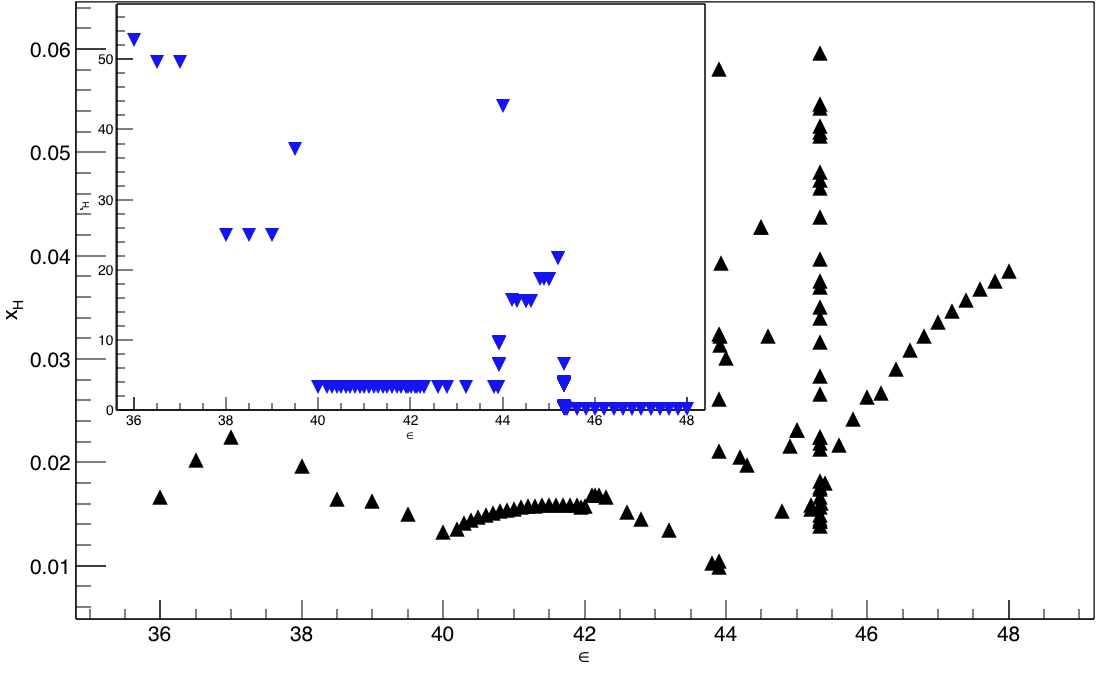}
   \caption[Anti-de Sitter Gauss-Bonnet instability]{Horizon radius $x_H$ as a function of the initial amplitude of the
      Gaussian scalar wave packet in Einstein's (left) and \gls{egb} gravity (right). Insets show the black hole
      formation times $t_H$. Credits: \cite{Deppe15a}.}
   \label{gaussbonnet}
\end{myfig}

\section{The CFT interpretation}
\label{cftinterp}

Of course, one of the main motivations to study the \gls{ads} instability is the \gls{ads}-\gls{cft} correspondence. What does the
instability means on the \gls{cft} side? And what are the quantum observables that may be impacted by the instability?

\subsection{Dual thermodynamics}

From a thermodynamical point of view, black hole formation is usually understood as thermalisation of the dual system in the
\gls{cft} side, and the weakly turbulent cascade came as no surprise for the \gls{ads}-\gls{cft} community who expected any
thermodynamical system to thermalise \cite{Hubeny15}. However, several authors cautioned that this was not always true
\cite{Abajo14,Silva15,Dimitrakopoulos15a}. Indeed, black holes obtained in numerical simulations are always classical. But if one
considers the Hawking radiation \cite{Bekenstein73,Hawking75}, these black holes formed after several bounces could be themselves
thermodynamically unstable and may evaporate, partially or totally, according the \gls{hp} phase transition (see chapter
\ref{adscft} section \ref{hawkingpage}).  The thermalised state could thus be described by a smaller black hole in equilibrium
with its Hawking radiation or by a thermal gas in \gls{ads} space-time. This means that the final classical black holes observed
in the simulations, if too small, are not thermalised states, but just pre-thermalisation steps in the dual theory, that only
achieve equilibrium on a longer timescale.  This is strikingly reminiscent of revivals and equilibrations that were observed
experimentally in some isolated Bose-Einstein condensates \cite{Gring12,Trotzky12,Kinoshita06}. Others discussions of \gls{cft}
duals in the context of the \gls{ads} instability can be found in \cite{Garfinkle11,Dias12a,Garfinkle12,Balasubramanian14}.

As first noticed in \cite{Dias12a,Buchel13}, the existence of non-linearly stable solutions is even more surprising, since they
are dual to systems that never thermalise in the \gls{cft} side, and do not exhibit black hole formation even if their mass is
above the \gls{hp} limit. A potential heuristic explanation is that such systems may form a meta-stable state that can survive by
continuously exchanging energy between its constitutive parts, as observed in
\cite{Pretorius00,Abajo14,Silva15,Deppe15a,Deppe15b,Deppe16a,Okawa15,Brito16b}, which is translated in a gravitational balance
between focusing and defocusing effects in the gravitational \gls{ads} dual (discussed in section \ref{competition}).

\subsection{Dual systems}

Several \gls{cft} interpretations are available in the literature. For example, the authors of \cite{Abajo14,Silva15} focused on
the \gls{cft} duals of direct and delayed (i.e.\ with bounces) collapses in \gls{ads} space-time. They computed the entanglement entropy of the
dual system and observed that this quantity was oscillating in the latter case. The authors also suggested that the
maximum of matter energy distribution in \gls{ads} was dual to the density of strongly correlated excitations in the field
theory.

In the hard wall model of \cite{Craps14c,Silva16}, the authors observed that non-collapsing solutions induced a modulated
oscillation of the boundary operator $\langle \mathcal{O} \rangle$, and the time scale of the modulation was precisely
$O(\varepsilon^{-2})$. The oscillations were interpreted in terms of conversions of a collection of glueballs into another
collection (and back).

Finally, from an \gls{ads}-\gls{cft} point of view and since a linearised graviton can be interpreted as a spin-2 excitation, gravitational
geons are equivalent to a Bose-Einstein condensate of spin-2 field excitations \cite{Dias12a}, namely glueballs excitations. This
is illustrated in figure \ref{dual}.

\begin{myfig}
   \includegraphics[width = 0.49\textwidth]{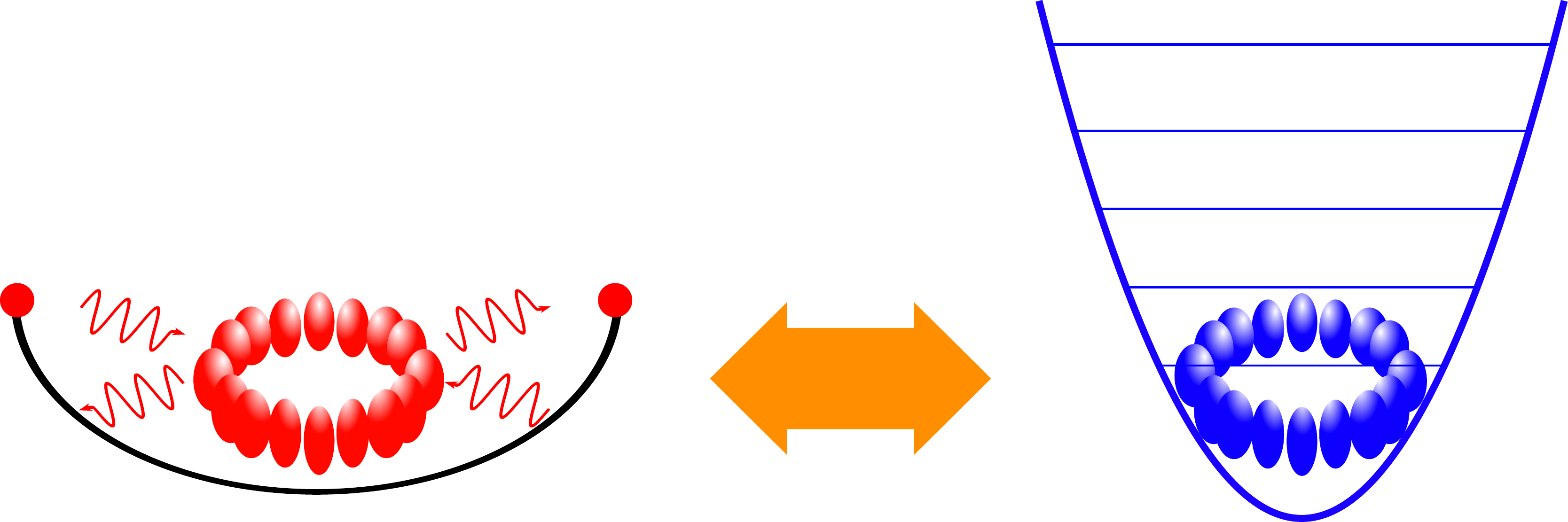}
   \caption[Dual system of geons]{Left: diagrammatic representation of a gravitational geon in an \gls{aads} space-time. Right:
   dual system living on the boundary. It is a Bose-Einstein condensate of a spin-2 field, namely glueball excitations. Credits: G. Martinon.}
   \label{dual}
\end{myfig}

Because of the holographic principle, the dual system lives always in a space-time that is one dimension less than the bulk
space-time. For example, a 4-dimensional geon is dual to a 3-dimensional Bose-Einstein condensate of glueball excitations. The
latter, however, can well be interpreted as a 4-dimensional system with one symmetry (or invariance), so much as to reduce the
\textit{effective} dimensionality to three.

\section{Summary and future prospects}

Since the original work of Bizo\'n and Rostworowski in 2011 \cite{Bizon11}, much progress has been made to understand the
\gls{ads} instability. In table \ref{adsrecap}, we have tried to recap all the different aspects of the problem. The statements
are mainly a short synopsis of the previous sections of this chapter.

\begin{mystab}
\begin{small}
\begin{tabular}{llllcc}
\hline
field               & initial data            & statement                                                                             & year & first reference           \\
\hline
gravitons           & -                       & linear stability of \gls{ads}                                                         & 1981 & \cite{Abbott82}           \\
gravitons           & -                       & \gls{ads} ``rigidity'' theorem                                                        & 2006 & \cite{Anderson06}         \\
massless scalar     & Gaussian                & non-linear instability of a Gaussian wave packet                                      & 2011 & \cite{Bizon11}            \\
massless scalar     & single-mode             & non-linear stability of a single-mode excitation                                      & 2011 & \cite{Bizon11}            \\
massless scalar     & two-mode                & non-linear instability of a two-mode excitation with turbulent cascade                & 2011 & \cite{Bizon11}            \\
massless scalar     & -                       & secular resonances on timescale $O(\varepsilon^{-2})$                                 & 2011 & \cite{Bizon11}            \\
massless scalar     & Gaussian                & instability due to the resonant spectrum in all dimensions                            & 2011 & \cite{Jalmuzna11}         \\
massless scalar     & Gaussian                & instability recovered in flat space-time enclosed in a cavity                         & 2011 & \cite{Maliborski12}       \\
massless scalar     & Gaussian                & critical phenomena in 3-dimensional \gls{ads} space-time                              & 2000 & \cite{Pretorius00}        \\
massless scalar     & Gaussian                & critical phenomena in 4-dimensional \gls{ads} space-time                              & 2003 & \cite{Husain03}           \\
massless scalar     & Gaussian                & critical phenomena for an arbitrary number of bounces                                 & 2016 & \cite{Olivan16a}          \\
perfect fluid       & -                       & singularity theorems in \gls{ads} space-time                                          & 2012 & \cite{Ishibashi12}        \\
null dust           & -                       & demonstration of the instability conjecture for radial Einstein-Vlasov                & 2017 & \cite{Moschidis17a}       \\
massless scalar     & -                       & first time-periodic solution                                                          & 2010 & \cite{Basu10}             \\
massless scalar     & time-periodic           & first numerical evolution of an island of stability                                   & 2013 & \cite{Maliborski13b}      \\
massless scalar     & Gaussian                & non-linear stability of a Gaussian profile of width $0.4 \lesssim \sigma \lesssim 8$  & 2013 & \cite{Buchel13}           \\
massless scalar     & exponential spectrum    & \gls{ttf} equations can probe the instability conjecture                              & 2014 & \cite{Balasubramanian14}  \\
massless scalar     & oscillatory spectrum    & \gls{ttf} can help generating non-linearly stable solutions                           & 2015 & \cite{Green15}            \\
perfect fluid       & thin shells             & fractal signature of collapse and chaotic islands of stability                        & 2016 & \cite{Brito16b}           \\
massless scalar     & Gaussian injection      & dichotomy of (in)stability in the hard wall model                                     & 2014 & \cite{Craps14c}           \\
massive scalar      & -                       & Boson stars in \gls{ads} space-time                                                   & 2003 & \cite{Astefanesei03}      \\
gravitons           & spherical harmonic seed & perturbative construction of gravitational geons in \gls{ads} space-time              & 2012 & \cite{Dias12a}            \\
gravitons           & spherical harmonic seed & numerical construction of gravitational geons in \gls{ads} space-time                 & 2015 & \cite{Horowitz15}         \\
photons             & spherical harmonic seed & Einstein-Maxwell spinning solitons in \gls{ads} space-time                            & 2012 & \cite{Herdeiro16a}        \\
massive Proca       & -                       & Proca stars in \gls{ads} space-time                                                   & 2016 & \cite{Duarte16}           \\
-                   & -                       & \gls{ads} instability obeys the ``mixed stability'' picture                           & 2015 & \cite{Dimitrakopoulos15b} \\
massless scalar     & -                       & competition between focusing and defocusing effects                                   & 2015 & \cite{Dimitrakopoulos15a} \\
massless scalar     & phase coherent          & phase coherence with power-law spectrum as a sufficient condition for collapse        & 2016 & \cite{Freivogel16}        \\
massive scalar      & Gaussian                & a resonant spectrum is a necessary condition for the instability                      & 2012 & \cite{Dias12b}            \\
massless scalar     & Gaussian                & instability in 3-dimensional \gls{ads} means blow-up of the Sobolev norm              & 2013 & \cite{Bizon13}            \\
massless scalar     & Gaussian                & instability less systematic in \gls{gb} gravity                                       & 2015 & \cite{Deppe15a}           \\
massless scalar     & Gaussian shell          & \gls{ads} instability as a pre-thermalisation process with \gls{hp} phase transition  & 2014 & \cite{Abajo14}            \\
massless scalar     & non-spherical           & first numerical evolution of the instability beyond spherical symmetry                & 2017 & \cite{Bantilan17}         \\
\hline
\end{tabular}
\end{small}
\caption[Summary of AdS instability]{Summary of the recent progress in the field of \gls{ads} instability. Credits: G. Martinon.}
\label{adsrecap}
\end{mystab}

The vast majority of the literature focused on spherical symmetry, very often with a massless scalar field and Gaussian initial
data. The few breakthroughs beyond spherical symmetry were carried out in the case of gravitational geons (section
\ref{beyondspher}). The remaining of this manuscript focuses on the techniques used to construct such geons and the results that
were obtained in \cite{Martinon17}. Our intuition is that the science of \gls{ads} instability beyond spherical symmetry will play
a prominent role in the near future.

\part{Numerical construction of gravitational geons}
\label{part2}
\chapter{Kodama-Ishibashi-Seto formalism}
\label{perturbations}
\citationChap{If I have seen further it is by standing on the shoulders of Giants.}{Isaac Newton}
\minitoc

From now on and in the chapters hereafter, we focus on 4-dimensional solutions. The goal of this chapter is to build linear
\gls{aads} gravitational geons, i.e.\ periodic solutions of the linearised Einstein's equation in vacuum with an \gls{ads}
background. At this end, we first rely on perturbative methods, that allow to obtain good approximations of solutions in the
low-amplitude limit. This perturbative approach is based upon the \gls{kis} formalism, that was first applied to cosmology in
\cite{Kodama00,Kodama03,Kodama04} and then adapted to \gls{aads} space-times in \cite{Ishibashi04}.

The main idea is to start from a spherically symmetric background, namely the pure vacuum \gls{ads} space-time. The background is
then treated with a $p+2$-decomposition (with $p = 2$ in our case of interest), i.e.\ a time-radial plane times the 
$p$-sphere. One then defines a metric $\widehat{g}_{ab}$ with $a,b$ restricted to the time-radial plane and another one
$\widetilde{\gamma}_{ij}$, with $i,j$ describing the $p$-sphere. The advantage of this split is that it separates the spherically
symmetric part of the metric from the time-radial components.

Once this background is settled, it is modified by a metric perturbation $h_{\alpha\beta}$. The main simplification of the
\gls{kis} formalism comes from the decomposition of the $h_{\alpha\beta}$ tensor. Indeed, the most general perturbation can be decomposed
into several pieces, each piece belonging to one of the following families: scalar-type, vector-type or tensor-type perturbations.
The denomination (scalar, vector or tensor) merely describes how the different components of the perturbation transform under
rotations on the $p$-sphere. Furthermore, each one of these families can be decomposed onto an adapted functional basis, namely
scalar, vector and tensor spherical harmonics, indexed by quantum numbers $(l,m)$.

It turns out that the background split combined with the perturbation decomposition onto adapted spherical harmonics always lead
to a system of decoupled equations, i.e.\ each type (scalar, vector, tensor) and each spherical harmonic $(l,m)$ can be treated
separately since they are not influencing each other. This greatly simplifies the perturbative scheme and allows to focus on each
type of perturbation independently and successively. On physical grounds however, we should keep in mind that any physical
perturbation is a priori a linear combination of all of these modes.

This formalism is relevant for the construction of geons. In this chapter, we focus on the first order only, since it
provides a precise enough seed for numerical simulations. Some complements about higher orders can be found in \cite{Martinon17}.

Our notations are the followings. Indices from the beginning of the alphabet $a,b,c,d$ are related to the time-radial plane while
indices $i,j,k,l$ are related to the unit 2-sphere angular directions. Hatted quantities always refer to the time-radial
background metric and quantities with a tilde to the unit 2-sphere metric. We also resort to the usual notations
\begin{equation}
   \widehat{\Box}f = \widehat{\nabla}_a \widehat{\nabla}^af, \quad (\widehat{\nabla}f)^2 = \widehat{\nabla}_a f \widehat{\nabla}^a
   f, \quad (H\cdot \widehat{\nabla}f) = H_a \widehat{\nabla}^a f.
\end{equation}

\section{Background decomposition}

The first step of the \gls{kis} formalism is to provide a decomposition adapted to the spherical symmetry of the background,
namely a $2+2$-decomposition, with 2 time-radial coordinates, and 2 angular coordinates. This allows to decompose not only the
4-dimensional metric but also all geometrical tensors and symbols of \gls{gr}, defined in appendix \ref{grd}.

\subsection{Decomposition of the metric}

We consider the \gls{ads} background, in an arbitrary coordinates system. We simply assume
that the background metric can be decomposed as
\begin{equation}
   \overline{g}_{\mu\nu}dx^\mu dx^\nu = \widehat{g}_{ab}(y)dy^a dy^b + r^2(y) \widetilde{\gamma}_{ij}dz^i dz^j,
   \label{bgsplit}
\end{equation}
where $y^a$ are the time-radial coordinates, and $z^i$ are angular coordinates (typically
$\theta$ and $\varphi$) describing the 2-sphere. The metric $\widehat{g}_{ab}$ is the 2-dimensional metric of the time-radial plane
and it is independent of the angular coordinates by spherical symmetry. The metric $\widetilde{\gamma}_{ij}$ is the one of the unit
2-sphere and is independent of $y^a$. The function $r(y)$ represents the radius of spheres in the coordinates
system. For instance, $r(y) = \gls{L} \sinh\rho$ in global coordinates (equation \eqref{adsglobal}) and $r(y) = \gls{L}\tan x$ in
conformal coordinates (equation \eqref{adsconformal}). In matrix-style representation, the split of the background reads
\begin{equation}
   \overline{g}_{\alpha\beta} = \left(
   \begin{array}{cc}
      \widehat{g}_{ab}(y) & 0 \\
      0 & r^2(y)\widetilde{\gamma}_{ij} \\
   \end{array}
   \right) \quad \tn{and} \quad \overline{g}^{\alpha\beta} = \left(
   \begin{array}{cc}
      \widehat{g}^{ab}(y) & 0 \\
      0 & \dfrac{1}{r^2(y)}\widetilde{\gamma}^{ij}
   \end{array}
   \right),
   \label{gsplit}
\end{equation}
where $\widehat{g}^{ab}$ and $\widetilde{\gamma}^{ij}$ are respectively the inverses of $\widehat{g}_{ab}$ and
$\widetilde{\gamma}_{ij}$.

\subsection{Decomposition of the Christoffel symbols}

This $2+2$-decomposition of the metric also holds for the Christoffel symbols and curvature tensors, which play an important role in
the perturbative scheme. Their computation directly follows from the definitions of appendix \ref{grd}. It can be shown that
the Christoffel symbols of the background metric are related to the ones of $\widehat{g}_{\alpha\beta}$ and
$\widetilde{\gamma}_{ij}$ by
\begin{subequations}
\begin{align}
   \overline{\Gamma}\indices{^c_{ab}} &= \widehat{\Gamma}\indices{^c_{ab}}, &\quad &\overline{\Gamma}\indices{^a_{ib}} = 0, &\quad &\overline{\Gamma}\indices{^a_{ij}} = -r \widehat{\nabla}^a r \widetilde{\gamma}_{ij},\\
   \quad \overline{\Gamma}\indices{^k_{ij}} &= \widetilde{\Gamma}\indices{^k_{ij}}, &\quad & \overline{\Gamma}\indices{^i_{ab}} = 0 &\quad&  \overline{\Gamma}\indices{^i_{aj}} = \frac{1}{r}\widehat{\nabla}_a r \delta \indices{^i_j},
\end{align}
\label{barGamma}
\end{subequations}
where $\widehat{\nabla}$ denotes the covariant derivative associated to the time-radial metric $\widehat{g}_{ab}$.

\subsection{Decomposition of the curvature tensors}

Since $\widehat{g}_{ab}$ is a 2-dimensional maximally symmetric negatively
curved metric and $\widetilde{\gamma}_{ij}$ a 2-dimensional maximally symmetric positively curved metric, their respective Ricci
tensors and scalars are (see equation \eqref{Rmaxsym2})
\begin{subequations}
\begin{align}
   \widehat{R}_{ab} &= -\frac{1}{\gls{L}^2}\widehat{g}_{ab}, &\quad \widetilde{R}_{ij} &= \widetilde{\gamma}_{ij},\\
   \widehat{R} &= -\frac{2}{\gls{L}},&\quad \widetilde{R} &= 2.
\end{align}
\end{subequations}
This allows to compute the background Riemann tensor
\begin{subequations}
\begin{align}
   \overline{R}\indices{^a_{bcd}} &= -\frac{1}{\gls{L}^2}(\delta \indices{^a_c}\widehat{g}_{bd} - \delta \indices{^a_d}\widehat{g}_{bc}), &\quad &\overline{R}\indices{^a_{ijb}} = r \widehat{\nabla}^a \widehat{\nabla}_b r \widetilde{\gamma}_{ij},\\
   \overline{R}\indices{^i_{jkl}} &= (1 - (\widehat{\nabla}r)^2)(\delta \indices{^i_k}\widetilde{\gamma}_{jl} - \delta
   \indices{^i_l}\widetilde{\gamma}_{jk}), &\quad &\overline{R}\indices{^i_{abj}} = \frac{1}{r}\widehat{\nabla}_a
   \widehat{\nabla}_b r \delta \indices{^i_j},
\end{align}
\label{barRiem}%
\end{subequations}
all the other components being zero. By contraction, the components of the background Ricci tensor are
\begin{equation}
   \overline{R}_{ab} = -\frac{1}{\gls{L}^2}\widehat{g}_{ab} - \frac{2}{r}\widehat{\nabla}_a \widehat{\nabla}_b r, \quad
   \overline{R}_{ai} = 0, \quad \overline{R}_{ij} = (1 - r \widehat{\Box}r - (\widehat{\nabla}r)^2)\widetilde{\gamma}_{ij}.
   \label{barRab}
\end{equation}
Contracting again, the background Ricci scalar is
\begin{equation}
   \overline{R} = -\frac{2}{\gls{L}^2} + \frac{2}{r^2}(1 - 2r \widehat{\Box}r - (\widehat{\nabla}r)^2).
   \label{barR}
\end{equation}
At this point we can combine \eqref{barRab} and \eqref{barR} with Einstein's equation in vacuum, namely
\begin{equation}
   \overline{R}_{\alpha\beta} = -\frac{3}{\gls{L}^2}\overline{g}_{\alpha\beta} \quad \tn{and} \quad \overline{R} = -\frac{12}{\gls{L}^2},
\end{equation}
and get
\begin{equation}
   \widehat{\nabla}_a r \widehat{\nabla}_b r = \frac{r}{\gls{L}^2}\widehat{g}_{ab}, \quad \widehat{\Box}r = \frac{2r}{\gls{L}^2}, \quad
   (\widehat{\nabla}r)^2 = 1 + \frac{r^2}{\gls{L}^2}.
   \label{backrel}
\end{equation}
These expressions make clear the link between the function $r(y)$ and the background geometry. This brings the following
definitive expressions of \eqref{barRiem}, \eqref{barRab} and \eqref{barR}
\begin{subequations}
\begin{align}
   \overline{R}\indices{^a_{bcd}} &= -\frac{1}{\gls{L}^2}(\delta \indices{^a_c}\widehat{g}_{bd} - \delta \indices{^a_d}\widehat{g}_{bc}), &\quad &\overline{R}\indices{^a_{ijb}} = \frac{r^2}{\gls{L}^2} \delta \indices{^a_b} \widetilde{\gamma}_{ij},\\
   \overline{R}\indices{^i_{jkl}} &= -\frac{r^2}{\gls{L}^2}(\delta \indices{^i_k}\widetilde{\gamma}_{jl} - \delta\indices{^i_l}\widetilde{\gamma}_{jk}), &\quad &\overline{R}\indices{^i_{abj}} = \frac{1}{\gls{L}^2}\widehat{g}_{ab} \delta \indices{^i_j},\\
   \overline{R}_{ab} &= -\frac{3}{\gls{L}^2}\widehat{g}_{ab}, &\quad &\overline{R}_{ij} = -\frac{3r^2}{\gls{L}^2}\widetilde{\gamma}_{ij}.
\end{align}
\label{Riemsplit}%
\end{subequations}

All these expressions of the usual geometrical tensors (and symbols) are thus mere consequences of the $2+2$ decomposition
\eqref{bgsplit}. We use them hereafter in order to simplify the perturbative scheme.

\section{Classification of the perturbations}

The cornerstone of the \gls{kis} formalism is the decomposition of the metric perturbation onto scalar, vector and tensor type
components. Each of these three components is in turn decomposed onto an adapted functional basis, namely the scalar, vector and
tensor spherical harmonics.

\subsection{Decomposition theorems}

In \cite{Ishibashi04}, two fundamental decomposition theorems were demonstrated.

\begin{theorem}[Vector decomposition]
For a compact Riemannian manifold equipped with the metric $\widetilde{\gamma}_{ij}$, any vector field $v^i$ can be uniquely decomposed as
\begin{equation}
   v^i = V^i + \widetilde{D}^i S, \quad \tn{where} \quad \widetilde{D}_i V^i = 0,
   \label{vectordecomposition}
\end{equation}
with $S$ a scalar field. $V^i$ and $S$ are referred respectively as the vector and scalar parts of $v^i$.
\label{theoV}
\end{theorem}

\begin{theorem}[Tensor decomposition]
For a compact Riemannian manifold equipped with the metric $\widetilde{\gamma}_{ij}$ and whose Ricci tensor obeys
$\widetilde{R}_{ij} = c \widetilde{\gamma}_{ij}$ for some constant $c$, any symmetric tensor $t_{ij}$ can be uniquely decomposed
as
\begin{equation}
   t_{ij} = T_{ij} + 2\widetilde{D}_{(i}V_{j)} + \frac{1}{2}\widetilde{\gamma}_{ij}t \indices{^m_m} + \left( \widetilde{D}_i \widetilde{D}_j - \frac{1}{2}\widetilde{\gamma}_{ij} \widetilde{D}_m
   \widetilde{D}^m \right)S,
   \label{tensordecomposition}
\end{equation}
where $V^i$ is a divergence-free vector and $T_{ij}$ is a transverse-traceless tensor, namely
\begin{equation}
   \widetilde{D}_i V^i = 0,\quad T \indices{^i_i} = 0,\quad \widetilde{D}_i T^{ij} = 0.
\end{equation}
$T_{ij}$, $V^i$ and $(S,t \indices{^m_m})$ are referred respectively as the tensor, vector and scalar parts of $t_{ij}$.
\label{theoT}
\end{theorem}

These theorems are fundamental for the metric perturbation decomposition. In particular, they readily apply to the 2-sphere, which
is relevant to the metric split \eqref{gsplit}, as well as the decomposition of the metric perturbation.

\subsection{Decomposition of the metric perturbation}

We consider the most general gravitational perturbations $h_{\alpha\beta}$. According to the background decomposition
\eqref{gsplit}, we can write it
\begin{equation}
   h_{\mu\nu}dx^\mu dx^\nu = h_{ab}dy^a dy^b + 2 h_{ai}dy^adz^i + h_{ij}dz^i dz^j.
\end{equation}
Under rotations on the 2-sphere, $h_{ab}$ transforms as a scalar field, $h_{ai}$ as a vector field and $h_{ij}$ as a tensor field.
We can thus make use of the two decomposition theorems \ref{theoV} and \ref{theoT}, to write
\begin{subequations}
\begin{align}
   h_{ab} &= h_{ab}^{(s)},\\
   h_{ai} &= h_{ai}^{(v)} + \widetilde{D}_i h_a^{(s)},\\
   h_{ij} &= h_{ij}^{(t)} + 2 \widetilde{D}_{(i}h_{j)}^{(v)} + h_L^{(s)} \widetilde{\gamma}_{ij} + \left( \widetilde{D}_i \widetilde{D}_j - \frac{1}{2}\widetilde{\gamma}_{ij} \widetilde{D}_m \widetilde{D}^m \right)h_T^{(s)},
\end{align}
\label{hdecomp}
\end{subequations}
where
\begin{equation}
   \widetilde{D}^i h_{ai}^{(v)} = 0, \quad \widetilde{D}^j h_{ij}^{(t)} = 0, \quad \widetilde{\gamma}^{ij}h_{ij}^{(t)} = 0, \quad \widetilde{D}^i h^{(v)}_i = 0.
\end{equation}
The $(s)$, $(v)$, and $(t)$ exponents denote respectively the scalar, vector and tensor parts of the metric perturbation. The
advantage of this decomposition is that all scalar (s), vector (v), and tensor (t) parts can be decomposed in turn onto scalar,
vector and tensor spherical harmonics respectively.

\subsection{Spherical harmonic decomposition}

On the one hand, the scalar spherical harmonics are
\begin{equation}
   \mathbb{S}_{lm} = \left\{
   \begin{array}{ll}
      (-1)^m \sqrt{\frac{2l+1}{2\pi}\frac{(l-m)!}{(l+m)!}}P^m_l (\cos\theta)\cos(m\varphi) & (m > 0),\\
      \sqrt{\frac{2l+1}{4\pi}}P_l(\cos\theta) & (m = 0),\\
      (-1)^{|m|}\sqrt{\frac{2l+1}{2\pi}\frac{(l-|m|)!}{(l+|m|)!}}P^{|m|}_l(\cos\theta)\sin(|m|\varphi) & (m < 0),
   \end{array}
   \right .
\end{equation}
where $l$ is a positive integer and $m$ is an integer obeying $|m| \leq l$. These harmonics satisfy
\begin{equation}
   \widetilde{D}_i \widetilde{D}^i \mathbb{S}_{lm} + l(l+1) \mathbb{S}_{lm} = 0,
   \label{stype}
\end{equation}
where $\widetilde{D}$ is the connection associated to $\widetilde{\gamma}_{ij}$. The following normalisation condition holds
\begin{equation}
   \int_0^\pi \int_0^{2\pi} \mathbb{S}_{lm} \mathbb{S}_{l'm'} \sin\theta d\theta d\varphi = \delta_{ll'}\delta_{mm'}.
\end{equation}
Any function of $(\theta,\varphi)$ can be decomposed onto the scalar spherical harmonics basis $\mathbb{S}_{lm}$.

On the other hand, any divergence-free 2-vector on the 2-sphere $V^i$ can be decomposed on the basis of the
vector spherical harmonics, defined by
\begin{equation}
   \mathbb{V}_{lm}^i = \frac{1}{\sqrt{l(l+1)}}\widetilde{\epsilon}^{ij}\widetilde{D}_j \mathbb{S}_{lm},
\end{equation}
where $\widetilde{\epsilon}^{ij}$ is the volume 2-form of the unit 2-sphere. They obey to
\begin{equation}
   \widetilde{D}_i \mathbb{V}_{lm}^i = 0, \quad \tn{and} \quad \widetilde{D}^j \widetilde{D}_j \mathbb{V}^i_{lm} + [l(l+1) -
   1]\mathbb{V}^i_{lm} = 0,
   \label{vtype}
\end{equation}
and are normalised according to
\begin{equation}
   \int_0^\pi \int_0^{2\pi} \widetilde{\gamma}_{ij}\mathbb{V}_{lm}^i \mathbb{V}_{l'm'}^j \sin \theta d\theta d\varphi = \delta_{ll'}\delta_{mm'}.
\end{equation}

Last but not least, in dimension $n = 4$, tensor spherical harmonics do not exist \cite{Higuchi87}. This implies that there is no
tensor degree of freedom for the perturbations. Namely, there is no tensor part in \eqref{hdecomp} and $h_{ij}^{(t)} = 0$. We thus
focus only on scalar and vector spherical harmonics and write, according to \eqref{hdecomp}
\begin{subequations}
\begin{align}
   h_{ab}^{(s)} &= \sum_{l,m} H_{ab}^{(l,m)}(y) \mathbb{S}^{lm},\\
   h_{ai}^{(v)} &= \sum_{l,m} H_a^{(v)(l,m)}(y) \mathbb{V}^{lm}_i,\\
   h_a^{(s)}    &= \sum_{l,m} H_a^{(s)(l,m)}(y) \mathbb{S}^{lm},\\
   h_i^{(v)}    &= \sum_{l,m} H_T^{(v)(l,m)}(y) \mathbb{V}^{lm}_i,\\
   h_L^{(s)}    &= \sum_{l,m} H_L^{(l,m)}(y) \mathbb{S}^{lm},\\
   h_T^{(s)}    &= \sum_{l,m} H_T^{(s)(l,m)}(y) \mathbb{S}^{lm}.
\end{align}
\label{hspherdecomp}%
\end{subequations}
Due to the rotational invariance of the background and of the mathematical operators involved in Einstein's equation, all
$(l,m)$ components of the scalar and vector-type perturbations decouple \cite{Ishibashi04}. This allows to simplify the argument by examining only
one type of perturbation (either scalar or vector) with one fixed $(l,m)$ independently and separately. In order to alleviate the
notations hereafter, we drop the $(l,m)$ indices since we can always study one $(l,m)$ component at a time.

\section{Vector-type perturbations}

We start by examining the family of vector-type perturbations, for fixed $(l,m)$. Following equations \eqref{hdecomp} and
\eqref{hspherdecomp}, they are defined by
\begin{equation}
   h_{ab} = 0, \quad h_{ai} = H_a \mathbb{V}_i, \quad h_{ij} = 2 H_T \widetilde{D}_{(i}\mathbb{V}_{j)},
   \label{hv}
\end{equation}
where $H_a$ and $H_T$ depends only on the time-radial coordinates $y^a$. Together, $H_{ab}$ and $H_T$ account for three degrees of
freedom (out of ten a priori) of the 4-dimensional metric perturbation. We can reduce further this number by choosing a suitable gauge.

\subsection{Gauge freedom}

In order to simplify the argument, we look for a gauge in which the computations are easier. We thus examine the first order change of coordinates
\begin{equation}
   x^{\alpha'} = x^\alpha + \chi^\alpha(x^\mu),
   \label{coorchangev}
\end{equation}
where $\chi^\alpha$ depends on the coordinates $(x^\alpha)$ and is of the same order as $h_{\alpha\beta}$.
Since we want to preserve the vector type of the metric perturbation, $\xi^\alpha$ is restricted to be of the form
\begin{equation}
   \chi_a = 0 \quad \tn{and} \quad \chi_i = \xi \mathbb{V}_i,
   \label{xiv}
\end{equation}
where $\xi$ depends only on the $y^a$ coordinates. The new perturbation is then (see equation \eqref{Ttrans} of appendix \ref{grd})
\begin{equation}
   h_{\alpha'\beta'} = h_{\alpha\beta} - \overline{\nabla}_\alpha \chi_\beta - \overline{\nabla}_\beta \chi_\alpha.
   \label{hgauge}
\end{equation}
With the expression \eqref{barGamma} of the Christoffel symbols, as well as \eqref{hv}, \eqref{xiv} and \eqref{hgauge}, the metric
perturbation in the new coordinates system can be expressed as
\begin{subequations}
\begin{align}
   H_{a'} &= H_a - r^2 \widehat{\nabla}_a \left( \frac{\xi}{r^2} \right),\\
   H_T'   &= H_T - \xi.
\end{align}
\label{HaHtprime}
\end{subequations}
In light of this results, there appears a natural gauge-invariant variable, namely
\begin{equation}
   Z_a \equiv H_a - r^2 \widehat{\nabla}_a \left( \frac{H_T}{r^2} \right),
   \label{Za}
\end{equation}
that remains unchanged under the transformation \eqref{xiv}. It turns out that it is possible to express the linearised Einstein's
equation in terms of $Z_a$, which has only two degrees of freedom.

\subsection{Linearised Einstein's equation}

In order to simplify the calculation, we choose a gauge in which
\begin{equation}
   H_T = 0.
   \label{HT0}
\end{equation}
This is possible according to \eqref{HaHtprime}. The components of the metric perturbation are now
\begin{equation}
   h_{ab} = 0, \quad h_{ai} = H_a \mathbb{V}_i, \quad h_{ij} = 0,
\end{equation}
and its trace $g^{\alpha\beta}h_{\alpha\beta}$ is zero. We also assume that $l \geq 2$. The case $l = 1$ is delayed to
section \ref{l1v}. As for the $l = 0$ case, since it is spherically symmetric, it cannot represent dynamical gravitational waves
by Birkhoff's theorem, and thus corresponds to a change of background.

Einstein's equation being a second order differential equation, we need the second order derivatives of $h_{\alpha\beta}$. In our
gauge \eqref{HT0}, the first order derivatives of the metric perturbation are
\begin{subequations}
\begin{align}
   \overline{\nabla}_c h_{ab} &= 0,& \quad \overline{\nabla}_a h_{ij} &= 0,\\
   \overline{\nabla}_b h_{ai} &= r \widehat{\nabla}_b\left( \frac{H_a}{r} \right)\mathbb{V}_i, &\quad \overline{\nabla}_i h_{aj} &= H_a \widetilde{D}_i \mathbb{V}_j,\\
   \overline{\nabla}_i h_{ab} &= -\frac{2}{r}H_{(a}\widehat{\nabla}_{b)}r \mathbb{V}_i, &\quad \overline{\nabla}_k h_{ij} &=
   2r(H\cdot \widehat{\nabla}r) \widetilde{\gamma}_{k(i}\mathbb{V}_{j)}.
\end{align}
\end{subequations}
The second order derivatives of the metric perturbation are
\begin{subequations}
\begin{align}
   \overline{\nabla}_a \overline{\nabla}_b h_{cd} &= 0,\\
   \overline{\nabla}_i \overline{\nabla}_j h_{ab} &= -\frac{4}{r}H_{(a}\widehat{\nabla}_{b)}r \widetilde{D}_{(i}\mathbb{V}_{j)},\\
   \overline{\nabla}_a \overline{\nabla}_b h_{ic} &= r \widehat{\nabla}_a \widehat{\nabla}_b\left( \frac{H_c}{r} \right)\mathbb{V}_i ,\\
   \overline{\nabla}_a \overline{\nabla}_i h_{bj} &= r^2 \widehat{\nabla}_a\left( \frac{H_b}{r^2} \right) \widetilde{D}_i \mathbb{V}_j,\\
   \overline{\nabla}_a \overline{\nabla}_i h_{bc} &= -2r \widehat{\nabla}_a\left( \frac{H_{(b}\widehat{\nabla}_{c)}r}{r^2} \right)\mathbb{V}_i,\\
   \overline{\nabla}_i \overline{\nabla}_a h_{bj} &= r^2 \widehat{\nabla}_a\left( \frac{H_b}{r^2} \right)\widetilde{D}_i \mathbb{V}_j,\\
   \overline{\nabla}_i \overline{\nabla}_a h_{bc} &= -2r \widehat{\nabla}_a \left( \frac{1}{r^2}H_{(b} \right)\widehat{\nabla}_{c)}r \mathbb{V}_i, \\
   \overline{\nabla}_a \overline{\nabla}_b h_{ij} &= 0,\\
   \overline{\nabla}_a \overline{\nabla}_k h_{ij} &= 2r^3 \widehat{\nabla}_a \left( \frac{(H\cdot \widehat{\nabla}r)}{r^2} \right)\widetilde{\gamma}_{k(i}\mathbb{V}_{j)},\\
   \overline{\nabla}_k \overline{\nabla}_a h_{ij} &= 2r^3 \widehat{\nabla}^b r \widehat{\nabla}_a \left( \frac{H_b}{r^2} \right)\widetilde{\gamma}_{k(i}\mathbb{V}_{j)},\\
\nonumber   \overline{\nabla}_i \overline{\nabla}_j h_{ak} &= H_a \widetilde{D}_i \widetilde{D}_j \mathbb{V}_k + r \widehat{\nabla}^b r \widehat{\nabla}_b H_a \widetilde{\gamma}_{ij}\mathbb{V}_k - 2 H_a (\widehat{\nabla}r)^2 \widetilde{\gamma}_{i(j}\mathbb{V}_{k)}\\
   &- (H\cdot \widehat{\nabla}r) \widehat{\nabla}_a r (\widetilde{\gamma}_{ik}\mathbb{V}_j + \widetilde{\gamma}_{jk}\mathbb{V}_i + \widetilde{\gamma}_{ij}\mathbb{V}_k),\\
   \overline{\nabla}_i \overline{\nabla}_j h_{kl} &= 4r (H\cdot \widehat{\nabla}r) \widetilde{\gamma}_{(k(i}\widetilde{D}_{j)}\mathbb{V}_{l)}.
\end{align}
\label{seconddev}%
\end{subequations}

We can now use these results for the linearised Einstein's equation. For a traceless perturbation, Einstein's equation in vacuum can be
computed readily with the results of appendix \ref{leastaction} (equations \eqref{variations}). At first
order it reads
\begin{equation}
   -\frac{1}{2}\overline{g}^{\mu\nu}\overline{\nabla}_\mu \overline{\nabla}_\nu h_{\alpha\beta} + \frac{1}{2}\overline{g}^{\mu\nu}
   \overline{\nabla}_\alpha \overline{\nabla}_\mu h_{\beta\nu} + \frac{1}{2}\overline{g}^{\mu\nu} \overline{\nabla}_\beta \overline{\nabla}_\mu
   h_{\alpha\nu} + \overline{R}_{\mu\alpha\beta\nu}h^{\mu\nu} = 0.
   \label{linein}
\end{equation}
This expression comes from Palatini's identity \eqref{palatini}, combined with Ricci's identity \eqref{ricci} and Einstein's
equation \eqref{eineq2} (see e.g.\ \cite{Flanagan05}). With the help of \eqref{gsplit}, \eqref{Riemsplit}, \eqref{vtype} and
\eqref{seconddev}, we can compute the components of \eqref{linein}. The $(a,b)$ components are trivially zero since they are
proportional to $D_i \mathbb{V}^i$ which vanishes for vector perturbations (see equation \eqref{vtype}). The $(i,j)$ components
are found to be
\begin{equation}
   \widehat{\nabla}^b H_b D_{(i}\mathbb{V}_{j)} = 0.
\end{equation}
As for the $(a,i)$ components, using the Ricci identity, namely
\begin{equation}
   \widehat{\nabla}_a \widehat{\nabla}^b H_b - \widehat{\nabla}^b \widehat{\nabla_a}H_b = \widehat{R}\indices{_a^b}H_b = -\frac{1}{\gls{L}^2}H_a,
   \label{ricciid}
\end{equation}
and \eqref{vtype}, it comes
\begin{equation}
   \frac{H_a}{2r^2}l(l+1) + \frac{2 H_a}{\gls{L}^2} - \frac{1}{2}\widehat{\Box}H_a + \frac{1}{2}\widehat{\nabla}^b \widehat{\nabla}_a H_b
   - \frac{\widehat{\nabla}_a r}{r^2}(H \cdot \widehat{\nabla}r) + \frac{\widehat{\nabla}^b r}{r}\widehat{\nabla}_a H_b -
   \frac{\widehat{\nabla}_ar}{r}\widehat{\nabla}^bH_b = 0.
   \label{einvai}
\end{equation}
Trading $H_a$ for $Z_a$ in our gauge (see equations \eqref{Za} and \eqref{HT0}), we have thus recovered that
\begin{subequations}
\begin{align}
   \label{eqV1}
   \widehat{\nabla}^a Z_a = 0,\\
   \label{eqV2}
   \widehat{\nabla}^b\left( r^4\left[ \widehat{\nabla}_b \left( \frac{Z_a}{r^2} \right)  - \widehat{\nabla}_a \left( \frac{Z_b}{r^2}\right) \right] \right) - [l(l+1) - 2] Z_a = 0,
\end{align}
\label{eqV}%
\end{subequations}
where \eqref{eqV2} is a more compact writing of \eqref{einvai}. These two equations are strictly equivalent to the linearised
Einstein's equation for vector-type perturbations. Furthermore, even if we have performed the computations in our gauge
\eqref{HT0}, the final expression is gauge-invariant since $Z_a$ itself is gauge-invariant (equation \eqref{Za}).

\subsection{Master equation}

It turns out that equations \eqref{eqV} can be simplified further and reduced to a single scalar equation. Indeed,
from \eqref{eqV1}, we deduce that $Z_a$ is a curl. Namely, there exist a scalar field $\phi(y)$ such that
\begin{equation}
   Z_a = \widehat{\epsilon}_{ab}\widehat{\nabla}^b \phi,
   \label{zaphi}
\end{equation}
where $\widehat{\epsilon}_{ab}$ is the volume 2-form (or Levi-Civita symbol) of the time-radial plane, satisfying
\begin{equation}
   \widehat{\epsilon}_{01} = \sqrt{-\widehat{g}}, \quad \widehat{\epsilon}^{01} = -\frac{1}{\sqrt{-\widehat{g}}}, \quad \widehat{\nabla}_c
   \widehat{\epsilon}_{ab} = 0.
   \label{eps2d}
\end{equation}
Equation \eqref{eqV2} thus becomes
\begin{equation}
   \widehat{\epsilon}_{ac}\widehat{\nabla}^b\left[ r^4 \widehat{\nabla}_b\left( \frac{\widehat{\nabla}^c \phi}{r^2} \right) \right] -
   \widehat{\epsilon}_{bc}\widehat{\nabla}^b\left[ r^4 \widehat{\nabla}_a \left( \frac{\widehat{\nabla}^c\phi}{r^2} \right) \right] -
   [l(l+1)-2]\widehat{\epsilon}_{ac}\widehat{\nabla}^c \phi = 0.
\end{equation}
Contracting with $\widehat{\epsilon}^{ad}$, using the properties of the Levi-Civita symbols \eqref{leviCrel}, and the background
properties \eqref{backrel}, we get
\begin{equation}
   r^2 \widehat{\Box} \widehat{\nabla}^a \phi - \left[ l(l+1) - 2 + \frac{r^2}{\gls{L}^2} \right]\widehat{\nabla}^a \phi +
   2r \widehat{\nabla}^a r \widehat{\Box} \phi - 2r \widehat{\nabla}^b r \widehat{\nabla}_b \widehat{\nabla}^a \phi - 2
   \widehat{\nabla}^a r (\widehat{\nabla} r\cdot \widehat{\nabla} \phi) = 0.
\end{equation}
This equation can put into a more compact form by using Ricci's identity \eqref{ricciid}
\begin{equation}
   \widehat{\nabla}^a\left[ r^4 \widehat{\nabla}_b \left( \frac{\widehat{\nabla}^b\phi}{r^2} \right) - [l(l+1) - 2]\phi \right] = 0.
\end{equation}
Since the hole left-hand side is a gradient, this means that the expression inside the brackets is a constant. However, since $Z_a$ is invariant by the transformation $\phi \to \phi + cst$ according to
\eqref{zaphi}, we have the freedom to choose $\phi$ so as to satisfy directly
\begin{equation}
   r^4 \widehat{\nabla}_b \left( \frac{\widehat{\nabla}^b\phi}{r^2} \right) - (l(l+1) - 2)\phi = 0.
   \label{premasterV}
\end{equation}
Now defining the master function
\begin{equation}
   \Phi_V \equiv \frac{\phi}{r},
\end{equation}
equation \eqref{premasterV} becomes a master equation for $\Phi_V$ that is
\begin{equation}
   \widehat{\Box}\Phi_V - \frac{l(l+1)}{r^2}\Phi_V = 0.
   \label{vtypemaster}
\end{equation}
This equation encodes fully the linearised Einstein's equation for vector-type perturbations and is by construction
gauge-invariant. Recall  that we started with three degrees of freedom ($H_{a}$ and $H_T$), that we have reduced to two with the
gauge-invariant variable $Z_a$, and finally, we have reduced it further to one with the scalar field $\Phi_V$.

\subsection{Case $l=1$}
\label{l1v}

If now we consider the special case of $l = 1$ perturbations, $H_T$ in equation \eqref{hv} is not defined any longer since $\widetilde{D}_i
\mathbb{V}_j = 0$. According to \eqref{HaHtprime}, the only gauge-invariant quantity that we can build is the antisymmetric
tensor
\begin{equation}
   Z_{ab} = r^2 \left[\widehat{\nabla}_a \left( \frac{H_b}{r^2} \right) - \widehat{\nabla}_b\left( \frac{H_a}{r^2} \right)\right].
\end{equation}
The linearised Einstein's equation \eqref{eqV2} then reduces to
\begin{equation}
   \widehat{\nabla}^b(r^2 Z_{ab}) = 0,
\end{equation}
whose antisymmetric solutions are
\begin{equation}
   Z_{ab} = \widehat{\epsilon}_{ab} \frac{C}{r^2},
\end{equation}
where $C$ is a constant of integration. On this solution, it is clear that $Z_{ab}$ is independent of time, so that the metric
perturbation is independent of time too. The $l=1$ case is then just a change of background and is not dynamical.

\section{Scalar-type perturbations}

We now consider the scalar-type perturbations, for fixed $(l,m)$. Following equations \eqref{hdecomp} and
\eqref{hspherdecomp}, they are defined by
\begin{equation}
   h_{ab} = H_{ab}\mathbb{S}, \quad h_{ai} = H_a \widetilde{D}_i \mathbb{S}, \quad h_{ij} = H_L \widetilde{\gamma}_{ij} \mathbb{S} + H_T
   \left( \widetilde{D}_i \widetilde{D}_j + \frac{l(l+1)}{2}\widetilde{\gamma}_{ij} \right)\mathbb{S},
   \label{hs}
\end{equation}
where $H_{ab}$, $H_a$, $H_L$ and $H_T$ depend only on the time-radial coordinates $y^a$. We have also used \eqref{stype} to
replace the 2-sphere Laplacian. The number of degrees of freedom carried by $H_{ab}$, $H_a$, $H_L$ and $H_T$ is seven (out of ten for
the full metric perturbation). Since the vector-type perturbations accounted for three degrees of freedom initially, we recover that vector
and scalar perturbations are sufficient, in the 4-dimensional case, to describe fully the whole metric perturbation.

\subsection{Gauge freedom}

Just like the previous vector-type case, we are looking for gauge-invariant variables. We thus perform the first order change of coordinates
\begin{equation}
   x^{\alpha'} = x^\alpha + \chi^\alpha(x^\mu).
\end{equation}
The scalar-type character of the metric perturbation is only preserved under the class of transformations
\begin{equation}
   \chi_a = \xi_a \mathbb{S} \quad \tn{and} \quad \chi_i = \xi \widetilde{D}_i \mathbb{S},
   \label{xis}
\end{equation}
where $\xi_a$ and $\xi$ depend only on the $y^a$ coordinates. The new metric perturbation is (equation \eqref{Ttrans}) then
\begin{equation}
   h_{\alpha'\beta'} = h_{\alpha\beta} - \overline{\nabla}_\alpha \chi_\beta - \overline{\nabla}_\beta \chi_\alpha.
   \label{hgauge2}
\end{equation}
Using the expression of the background Christoffel symbols \eqref{barGamma} as well as \eqref{hs}, \eqref{xis} and \eqref{hgauge2}, we obtain
\begin{subequations}
\begin{align}
   H_{a'b'} &= H_{ab} - \widehat{\nabla}_a \xi_b - \widehat{\nabla}_b \xi_a,\\
   H_{a'} &= H_a - \xi_a - r^2 \widehat{\nabla}_a \left( \frac{\xi}{r^2} \right),\\
   H_L' &= H_L + l(l+1)\xi - 2r \widehat{\nabla}^a r \xi_a,\\
   H_T' &= H_T - 2\xi.
\end{align}
\label{Hs}
\end{subequations}
Out of these results, we can construct two natural gauge-invariant variables, namely
\begin{subequations}
\begin{align}
   Z &= \frac{2}{r^2}\left(H_L + \frac{l(l+1)}{2}H_T + 2r \widehat{\nabla}^a r X_a\right),\\
   Z_{ab} &= H_{ab} + \widehat{\nabla}_a X_b + \widehat{\nabla}_b X_a + \frac{1}{2}Z \widehat{g}_{ab},
\end{align}
\label{ZZab}
\end{subequations}
where we have defined
\begin{equation}
   X_a = -H_a + \frac{r^2}{2}\widehat{\nabla}_a \left( \frac{H_T}{r^2} \right).
\end{equation}
Under the change of coordinates \eqref{xis}, $Z$ and $Z_{ab}$ remain unchanged while $X_a$ transforms like
\begin{equation}
   X_{a'} = X_a + \xi_a.
\end{equation}
In the following, we express the linearised Einstein's equation only in terms of these gauge-invariant variables.

\subsection{Linearised Einstein's equation}

In order to make computations simpler, we work in a gauge where
\begin{equation}
   H_a = 0 \quad \tn{and} \quad H_T = 0.
   \label{HaHT0}
\end{equation}
This is possible according to \eqref{Hs}. We also assume that the quantum number $l$ satisfies $l \geq 2$. The special case $l = 1$
is treated in section \ref{l1s} and the case $l = 0$ is not dynamical since it is spherically symmetric (Birkhoff's theorem). The metric components of the perturbation then read
\begin{equation}
   h_{ab} = H_{ab}\mathbb{S}, \quad h_{ai} = 0, \quad h_{ij} = H_L \widetilde{\gamma}_{ij}\mathbb{S}.
\end{equation}
We have thus reduced the number of degrees of freedom to four, contained in $H_{ab}$ and $H_L$. The trace of the metric perturbation is
\begin{equation}
   h = \left( H \indices{^a_a} + \frac{2H_L}{r^2} \right)\mathbb{S}.
\end{equation}

To prepare the computation of Einstein's equation at first order, we give the derivatives of the perturbation in our gauge
\eqref{HaHT0}. The first order derivatives of the metric are
\begin{subequations}
\begin{align}
   \overline{\nabla}_c h_{ab} &= \widehat{\nabla}_c H_{ab}\mathbb{S},& \quad \overline{\nabla}_a h_{ij} &= r^2 \widehat{\nabla}_a \left( \frac{H_L}{r^2} \right)\widetilde{\gamma}_{ij}\mathbb{S},\\
   \overline{\nabla}_b h_{ai} &= 0, &\quad \overline{\nabla}_i h_{aj} &= \left( r H_{ab}\widehat{\nabla}^b r - H_L \frac{\widehat{\nabla}_a r}{r} \right)\widetilde{\gamma}_{ij}\mathbb{S},\\
   \overline{\nabla}_i h_{ab} &= H_{ab}\widetilde{D}_i \mathbb{S}, &\quad \overline{\nabla}_k h_{ij} &= H_L \widetilde{\gamma}_{ij}\widetilde{D}_k \mathbb{S}.
\end{align}
\end{subequations}
The second order derivatives of the metric perturbation are
\begin{subequations}
\begin{align}
   \overline{\nabla}_a \overline{\nabla}_b h_{cd} &= \widehat{\nabla}_a \widehat{\nabla}_b H_{cd}\mathbb{S},\\
   \overline{\nabla}_i \overline{\nabla}_j h_{ab} &= H_{ab} \widetilde{D}_i \widetilde{D}_j \mathbb{S} + \left( r \widehat{\nabla}^cr \widehat{\nabla}_c H_{ab} - 2 \widehat{\nabla}_{(a}r H_{b)c}\widehat{\nabla}^c r + \frac{2H_L}{r^2}\widehat{\nabla}_a r \widehat{\nabla}_b r \right)\widetilde{\gamma}_{ij}\mathbb{S},\\
   \overline{\nabla}_a \overline{\nabla}_b h_{ic} &= 0,\\
   \overline{\nabla}_a \overline{\nabla}_i h_{bj} &= r^2 \widehat{\nabla}_a\left( \frac{H_{bc}\widehat{\nabla}^cr}{r} - \frac{H_L \widehat{\nabla}_b r}{r^3} \right)\widetilde{\gamma}_{ij} \mathbb{S},\\
   \overline{\nabla}_a \overline{\nabla}_i h_{bc} &= r \widehat{\nabla}_a \left( \frac{H_{bc}}{r} \right)\widetilde{D}_i \mathbb{S},\\
   \overline{\nabla}_i \overline{\nabla}_a h_{bj} &= r^2 \left( \widehat{\nabla}_a \left( \frac{H_{bc}}{r} \right)\widehat{\nabla}^c r - \widehat{\nabla}_a \left( \frac{H_L}{r^3} \right)\widehat{\nabla}_b r \right)\widetilde{\gamma}_{ij}\mathbb{S},\\
   \overline{\nabla}_i \overline{\nabla}_a h_{bc} &= r \widehat{\nabla}_a \left( \frac{H_{bc}}{r} \right)\widetilde{D}_i \mathbb{S}, \\
   \overline{\nabla}_a \overline{\nabla}_b h_{ij} &= r^2 \widehat{\nabla}_a \widehat{\nabla}_b \left( \frac{H_L}{r^2} \right)\widetilde{\gamma}_{ij}\mathbb{S},\\
   \overline{\nabla}_a \overline{\nabla}_k h_{ij} &= r^3 \widehat{\nabla}_a\left( \frac{H_L}{r^3} \right)\widetilde{\gamma}_{ij}\widetilde{D}_k \mathbb{S},\\
   \overline{\nabla}_k \overline{\nabla}_a h_{ij} &= r^3 \widehat{\nabla}_a \left( \frac{H_L}{r^3} \right)\widetilde{\gamma}_{ij}\widetilde{D}_k \mathbb{S},\\
   \overline{\nabla}_i \overline{\nabla}_j h_{ak} &= 2\left( r H_{ab}\widehat{\nabla}^b r - H_L \frac{\widehat{\nabla}_a r}{r} \right)\widetilde{\gamma}_{k(i}\widetilde{D}_{j)}\mathbb{S},\\
\nonumber   \overline{\nabla}_i \overline{\nabla}_j h_{kl} &= H_L \widetilde{\gamma}_{kl}\widetilde{D}_i \widetilde{D}_j \mathbb{S} + r^3 \widehat{\nabla}^a r \widehat{\nabla}_a \left( \frac{H_L}{r^2} \right)\widetilde{\gamma}_{ij}\widetilde{\gamma}_{kl}\mathbb{S}\\
   &+ (r^2 H_{ab}\widehat{\nabla}^a r \widehat{\nabla}^b r - H_L (\widehat{\nabla}r)^2)(\widetilde{\gamma}_{ik}\widetilde{\gamma}_{jl} + \widetilde{\gamma}_{il}\widetilde{\gamma}_{jk})\mathbb{S}.
\end{align}
\label{seconds}
\end{subequations}

Now, the first order Einstein's equation is (deduced from equation \eqref{variations} of the appendix)
\begin{equation}
   -\frac{1}{2}\overline{g}^{\mu\nu}\overline{\nabla}_\mu \overline{\nabla}_\nu h_{\alpha\beta} - \frac{1}{2}\overline{\nabla}_\alpha
   \overline{\nabla}_\beta h + \frac{1}{2}\overline{g}^{\mu\nu}
   \overline{\nabla}_\alpha \overline{\nabla}_\mu h_{\beta\nu} + \frac{1}{2}\overline{g}^{\mu\nu} \overline{\nabla}_\beta \overline{\nabla}_\mu
   h_{\alpha\nu} + \overline{R}_{\mu\alpha\beta\nu}h^{\mu\nu} = 0.
   \label{linein2}
\end{equation}
In light of \eqref{seconds}, the $(a,i)$ components of \eqref{linein2} boil down to
\begin{equation}
   -\frac{1}{2}\widehat{\nabla}_a \left( H \indices{^b_b} + \frac{2H_L}{r^2} \right) + \frac{1}{2}\widehat{\nabla}^b H_{ab} +
   \frac{1}{2} \widehat{\nabla}_a \left( \frac{H_L}{r^2} \right) + \frac{H \indices{^b_b}}{2r}\widehat{\nabla}_a r = 0.
   \label{aicomp}
\end{equation}
If now we look at the $(i,j)$ components of \eqref{linein2} with the help of \eqref{stype}, they reduce to
\begin{align}
\nonumber   0 &= -\frac{1}{2}H \indices{^a_a}\widetilde{D}_i \widetilde{D}_j \mathbb{S} + \widetilde{\gamma}_{ij}\mathbb{S}\bigg[
   -\frac{r^2}{2}\widehat{\Box} \widehat{\nabla} \left( \frac{H_L}{r^2} \right) + r^2 \widehat{\nabla}^a\left(
\frac{H_{ab}}{r} \right)\widehat{\nabla}^b r - r^2 \widehat{\nabla}_a \left( \frac{H_L}{r^3} \right)\widehat{\nabla}^a r\\
  &+ \frac{H_Ll(l+1)}{2r^2} + 2 H_{ab} \widehat{\nabla}^a r \widehat{\nabla}^b r - \frac{H_L}{r^2}(\widehat{\nabla}r)^2 -
  \frac{H_L}{r^2} + rH_{ab} \widehat{\nabla}^a \widehat{\nabla}^b r - \frac{r}{2}\widehat{\nabla}^a r \widehat{\nabla}_a \left( H
  \indices{^b_b} + \frac{2H_L}{r^2} \right)\bigg].
\label{ijcomp}
\end{align}
The right-hand side has a non-vanishing trace. The traceless part is simpler and reads
\begin{equation}
   H \indices{^a_a}\left( \widetilde{D}_i \widetilde{D}_j \mathbb{S} + \frac{l(l+1)}{2}\widetilde{\gamma}_{ij}\mathbb{S} \right) = 0 \underset{l \geq 2}{\iff} H \indices{^a_a} = 0.
   \label{ijcomp2}
\end{equation}
This equation combined with the background identities \eqref{backrel} and \eqref{stype} gives finally an equation for the trace of \eqref{ijcomp}, namely
\begin{equation}
   -\frac{r^2}{\gls{L}^2}\widehat{\Box} \left( \frac{H_L}{r^2} \right) - 2r \widehat{\nabla}_a \left(
   \frac{H_L}{r^2} \right)\widehat{\nabla}^a r + \frac{H_L}{2r^2}[l(l+1) - 2] + r \widehat{\nabla}^a H_{ab} \widehat{\nabla}^b r +
   H_{ab} \widehat{\nabla}^a r \widehat{\nabla}^b r = 0.
   \label{trijcomp}
\end{equation}
As for the $(a,b)$ components of the linearised Einstein's equation \eqref{linein2}, we can use \eqref{seconds} and \eqref{stype} and get
\begin{align}
\nonumber   0 &= -\frac{1}{2}\widehat{\Box} H_{ab} - \frac{1}{2}\widehat{\nabla}_a \widehat{\nabla}_b \left( H \indices{^c_c} + \frac{2 H_L}{r^2} \right) + \frac{1}{2}\widehat{\nabla}_a \widehat{\nabla}^c H_{bc} + \frac{1}{2} \widehat{\nabla}_b \widehat{\nabla}^c H_{ac} + \frac{H_{ab}}{2r^2}\left( l(l+1) + \frac{2r^2}{\gls{L}^2} \right)\\
   &- \frac{\widehat{\nabla}^c r}{r}\left( \widehat{\nabla}_c H_{ab} - \widehat{\nabla}_a H_{bc} - \widehat{\nabla}_b H_{ac} \right) - \frac{1}{r}\widehat{\nabla}_a \left( \frac{H_L}{r^2} \right)\widehat{\nabla}_b r - \frac{1}{r}\widehat{\nabla}_b \left( \frac{H_L}{r^2} \right)\widehat{\nabla}_a r + \frac{H \indices{^c_c}}{\gls{L}^2}\widehat{g}_{ab}.
\label{abcomp}
\end{align}
Now that we have at hand all components of Einstein's equation, we can make appear the gauge-invariant variables
\eqref{ZZab}, which in our gauge \eqref{HaHT0} read
\begin{equation}
   Z = \frac{2 H_L}{r}, \quad Z_{ab} = H_{ab} + \frac{H_L}{r^2}\widehat{g}_{ab}, \quad Z \indices{^a_a} = H \indices{^a_a} + \frac{2 H_L}{r^2}.
   \label{gaugevars}
\end{equation}
If we combine the $(a,i)$ and $(i,j)$ components (equations \eqref{aicomp} and \eqref{ijcomp2}) of the linearised Einstein's
equation, we find the two equations
\begin{equation}
   \widehat{\nabla}^b Z_{ab} = \widehat{\nabla}_a Z \indices{^b_b} \quad \tn{and} \quad Z \indices{^a_a} = Z.
   \label{gaugevareqs}
\end{equation}
Despite their simplicity, these two equations entirely summarise the $(a,i)$ and $(i,j)$ components of Einstein's equation at
first order for scalar-type perturbations. It can be shown that these equations imply the existence of a function $\phi$ such that \cite{Ishibashi04}
\begin{subequations}
\begin{align}
   \label{zabdef}
   Z_{ab} &= \widehat{\nabla}_a \widehat{\nabla}_b \phi - \frac{\phi}{\gls{L}^2}\widehat{g}_{ab},\\
   Z &= \widehat{\Box}\phi - \frac{2\phi}{\gls{L}^2}.
\end{align}
\label{eqgauges}
\end{subequations}
This is similar to the property ``a vector of vanishing divergence is necessary a curl'' encountered in the vector-type case.
The proof relies on the following arguments (readers ready to admit equations \eqref{eqgauges} can jump directly to section
\eqref{mastereq}).

Consider the unique function $\phi$ solution of
\begin{equation}
   \left( \widehat{\nabla}^a \widehat{\nabla}_a - \frac{2}{\gls{L}^2} \right)\phi = 0,
   \label{lapphi}
\end{equation}
with initial data satisfying
\begin{equation}
   \partial_t^b\left( \widehat{\nabla}_a \widehat{\nabla}_b - \frac{1}{\gls{L}^2}\widehat{g}_{ab} \right)\phi = \partial_t^b Z_{ab}, \quad
   \tn{on the Cauchy surface} \quad x = x_0.
   \label{initphi}
\end{equation}
Then the tensor defined by
\begin{equation}
   T_{ab} \equiv Z_{ab} - \left(\widehat{\nabla}_a \widehat{\nabla}_b \phi - \frac{\phi}{\gls{L}^2}\widehat{g}_{ab}\right),
\end{equation}
vanishes because
\begin{myenum}
   \item it is transverse $\widehat{\nabla}_b T \indices{^b_a} = 0$,
   \item it is traceless $\widehat{g}^{ab}T_{ab} = 0$,
   \item its contraction with $\partial_t^a$ is zero: $T_{ab}\partial_t^b = 0$.
\end{myenum}
These conditions are well sufficient to infer that $T_{ab} = 0$ (and hence \eqref{eqgauges}). Indeed, from condition 3, we deduce that
$T_{00}$ and $T_{01}$ are zero. Combined with the traceless condition 2, this ensures that $T_{11}$ vanishes too, so that all
$(a,b)$ components of this tensor are zero, ending the proof of \eqref{eqgauges}. Conditions 1 and 2 are direct consequences of
\eqref{lapphi} and of the 2-dimensional Ricci identity \eqref{ricciid}. As for condition 3, it is harder to prove. If we define
\begin{equation}
   v_a \equiv T_{ab}\partial_t^b,
\end{equation}
it results from
\begin{myenum2}
   \item $v_a$ is a curl. Since $\widehat{\nabla}_a v^a = 0$, there exists a scalar field $s$ such that $v^a =
      \widehat{\epsilon}^{ab}\widehat{\nabla}_b s$.
   \item The scalar $s$ satisfies $s = 0$ and $\frac{\partial s}{\partial x} = 0$ on the hypersurface $x = x_0$.
   \item The symmetry of $\widehat{\nabla}_a v_b$: $\widehat{\nabla}( \widehat{\epsilon}^{ab}v_b) = 0 \iff
      -\widehat{\nabla}_a \widehat{\nabla}^a s = 0$.
\end{myenum2}
These three conditions imply that $s = 0$ since the only solution of (c) with initial condition (b) is $s = 0$
and hence $v_a = 0$ by condition (a). Condition (a) is a consequence of condition 1 and of the antisymmetry of $\widehat{\nabla}^a
\partial_t^b$ (Killing equation for the background). Condition (b) holds because the initial condition \eqref{initphi} implies $v^a
= 0$ and thus $\widehat{\nabla}_a s = 0$ on the hypersurface $x = x_0$. Since $s$ is defined within a constant choice, we can
arbitrarily set $s$ to $0$ on the Cauchy surface for the initial data. Condition (c) is not trivial to establish, but we can remark that
\begin{equation}
   \widehat{\nabla}_a (\widehat{\epsilon}^{ab}v_b) = \widehat{\nabla}_a(\widehat{\epsilon}^{ab}T_{bc} \partial_t^c).
   \label{nabeps}
\end{equation}
Besides, $\widehat{\epsilon}_{ac}\widehat{\epsilon}^{ab}T \indices{_b^c} = T \indices{^b_b} = 0$ (thanks to condition 2 and
\eqref{leviCrel}), which means that $\widehat{\epsilon}^{ab}T \indices{^b_c}$ is symmetric. We can thus rewrite \eqref{nabeps} as
\begin{equation}
   \widehat{\nabla}_a (\widehat{\epsilon}^{ab}v_b) = \widehat{\epsilon}_{cb}T^{ba}\widehat{\nabla}_a \partial_t^c,
   \label{nabeps2}
\end{equation}
where we have used condition 1 and \eqref{eps2d}. Since $\widehat{\nabla}^a \partial_t^c$ is antisymmetric (Killing equation for the
background) and the time-radial plane is 2-dimensional, $\widehat{\nabla}^a \partial_t^c$ must be proportional to
$\widehat{\epsilon}^{ac}$. The proportionality factor is readily found by contracting with $\widehat{\epsilon}_{ac}$ and using
\eqref{leviCrel}:
\begin{equation}
   \widehat{\nabla}^a \partial_t^c = -\frac{1}{2}\widehat{\epsilon}_{de}\widehat{\nabla}^d \partial_t^e \widehat{\epsilon}^{ac}.
   \label{nabpartt}
\end{equation}
From \eqref{nabpartt}, \eqref{leviCrel}, and \eqref{nabeps2}, follows directly condition (c).

\subsection{Master equation}
\label{mastereq}

The important feature we have proved so far is that the $(a,i)$ and $(i,j)$ components of Einstein's equation \eqref{gaugevareqs} imply that the
gauge-invariant variables take the form \eqref{eqgauges}. Namely, we have reduced the number of degrees of freedom to 1, since
$\phi$ encodes all the scalar-type metric perturbation. In order to express all our results in terms of the $\phi$ function, we
notice that \eqref{gaugevars} and \eqref{eqgauges} imply
\begin{subequations}
\begin{align}
   H_{ab} = \widehat{\nabla}_a \widehat{\nabla}_b \phi - \frac{\widehat{\Box}\phi}{2}\widehat{g}_{ab},\\
   \frac{H_L}{r^2} = \frac{\widehat{\Box}\phi}{2} - \frac{\phi}{\gls{L}^2},\\
   \widehat{\nabla}^b H_{ab} = \widehat{\nabla}_a \left( \frac{H_L}{r^2} \right).
\end{align}
\end{subequations}
Using these expressions in \eqref{trijcomp} and \eqref{abcomp} yields
equations for $\phi$ only, namely
\begin{subequations}
\begin{align}
\nonumber   0 &= -\frac{\widehat{\Box}^2 \phi}{4} + \frac{\widehat{\Box} \phi}{4 r^2}[l(l+1) - 4] - \frac{\phi [l(l+1) - 2]}{2r^2 \gls{L}^2} + \frac{1}{r^2}\widehat{\nabla}_a \widehat{\nabla}_b \phi \widehat{\nabla}^a r \widehat{\nabla}^b r \\
     &- \frac{1}{2r}\widehat{\nabla}^a r \widehat{\nabla}_a \widehat{\Box}\phi + \frac{1}{r \gls{L}^2}\widehat{\nabla}^a r \widehat{\nabla}_a \phi,\\
\nonumber   0 &= -\frac{1}{2}\widehat{\nabla}_a \widehat{\nabla}_b \widehat{\Box} \phi + \frac{\widehat{\Box}^2 \phi}{4}\widehat{g}_{ab} + \frac{\widehat{\nabla}_a \widehat{\nabla}_b \phi}{2 r^2}\left[ l(l+1) + \frac{6r^2}{\gls{L}^2} \right] - \frac{\widehat{\Box}\phi}{4 r^2}\widehat{g}_{ab}\left[ l(l+1) + \frac{6r^2}{\gls{L}^2}\right]\\
\nonumber   &+ \frac{\widehat{\nabla}^c r}{r}\widehat{\nabla}_a \widehat{\nabla}_b \widehat{\nabla}_c \phi + \frac{\widehat{g}_{ab}}{2r}\widehat{\nabla}^c \widehat{\Box}\phi \widehat{\nabla}_c r - \frac{\widehat{\nabla}_b r}{r}\widehat{\nabla}_a \widehat{\Box}\phi - \frac{\widehat{\nabla}_ar}{r}\widehat{\nabla}_b \widehat{\Box}\phi + \frac{\widehat{\nabla}_a \phi \widehat{\nabla}_b r}{r \gls{L}^2} \\
 &+ \frac{2 \widehat{\nabla}_b\phi \widehat{\nabla}_a r}{r \gls{L}^2} - \frac{\widehat{g}_{ab}}{r \gls{L}^2}\widehat{\nabla}^c r \widehat{\nabla}_c \phi,
\end{align}
\end{subequations}
where we have extensively used the Ricci identity \eqref{ricciid}. These two equations can be combined and simplified further as
follows. Eliminating $\widehat{\Box}^2 \phi$ between these two equations, we get
\begin{align}
\nonumber   0 &=-\frac{1}{2}\widehat{\nabla}_a \widehat{\nabla}_b \widehat{\Box}\phi + \frac{\widehat{\nabla}^c r}{r}\widehat{\nabla}_a \widehat{\nabla}_b \widehat{\nabla}_c\phi - \frac{\widehat{\nabla}_b r}{r}\widehat{\nabla}_a \widehat{\Box}\phi - \frac{\widehat{\nabla}_a r}{r}\widehat{\nabla}_b \widehat{\Box}\phi + \frac{\widehat{g}_{ab}}{r^2}\widehat{\nabla}_c \widehat{\nabla}_d\phi \widehat{\nabla}^cr \widehat{\nabla}^d r\\
\nonumber   &- \frac{\widehat{g}_{ab}}{r^2}\widehat{\Box}\phi \left( 1 + \frac{3r^2}{2 \gls{L}^2} \right) + \frac{\widehat{\nabla}_a \widehat{\nabla}_b \phi}{2r^2}\left( l(l+1) + \frac{6r^2}{\gls{L}^2} \right) + \frac{\widehat{\nabla}_a \phi \widehat{\nabla}_b r}{r \gls{L}^2} + \frac{2 \widehat{\nabla}_b \phi \widehat{\nabla}_a r}{r \gls{L}^2}\\
&- \frac{\phi \widehat{g}_{ab}}{2 r^2 \gls{L}^2}[l(l+1) - 2] = 0.
\label{eqs1}
\end{align}
Finally, this equation admits the more compact form
\begin{equation}
   -\frac{1}{2r^2}\left( \widehat{\nabla}_a \widehat{\nabla}_b - \frac{\widehat{g}_{ab}}{\gls{L}^2} \right)(r^2 \widehat{\Box}\phi - 2r
   \widehat{\nabla}^c r \widehat{\nabla}_c \phi - [l(l+1)-2]\phi) = 0.
   \label{eqs2}
\end{equation}
Establishing the equivalence between \eqref{eqs1} and \eqref{eqs2} is not trivial at all. Indeed the difference between \eqref{eqs1} and
\eqref{eqs2} is
\begin{align}
\nonumber   A_{ab} &\equiv \frac{\widehat{\nabla}_a \widehat{\nabla}_b \phi}{r^2}\left( 1 + \frac{r^2}{\gls{L}^2} \right) -
\frac{\widehat{g}_{ab}\widehat{\Box}\phi}{r^2}\left( 1 + \frac{r^2}{\gls{L}^2} \right) + \frac{\widehat{g}_{ab}}{r^2}\widehat{\nabla}_c \widehat{\nabla}_d \phi \widehat{\nabla}^c r \widehat{\nabla}^d r + \frac{\widehat{\Box}\phi}{r^2}\widehat{\nabla}_ar \widehat{\nabla}_b r \\
&- \frac{1}{r^2}\widehat{\nabla}_b r \widehat{\nabla}^c r \widehat{\nabla}_a \widehat{\nabla}_c \phi - \frac{1}{r^2}\widehat{\nabla}_a r \widehat{\nabla}^c r \widehat{\nabla}_b \widehat{\nabla}_c \phi.
\end{align}
Proving that $A_{ab}$ is zero is not obvious at first sight. However, if exceptionally we choose to work in a given background frame, say the
one associated to static coordinates (equation \eqref{adsstatic} of chapter \ref{aads}), we have
\begin{equation}
   \widehat{g}_{ab} = \left(
   \begin{array}{cc}
      -f & 0 \\
      0 & \frac{1}{f} \\
   \end{array}
   \right), \quad \widehat{g}^{ab} = \left(
   \begin{array}{cc}
      -\frac{1}{f} & 0 \\
      0 & f
   \end{array}
   \right), \quad \partial_0 r = 0, \quad \partial^0 r = 0, \quad \partial_1 r = 1, \quad \partial^1 r = f,
\end{equation}
where $f(r) = 1 + r^2/\gls{L}^2$. It can then be shown readily that each component of $A_{ab}$ is zero, so that the tensor
$A_{ab}$ is identically zero. This proves the equivalence between \eqref{eqs1} and \eqref{eqs2}. 

We have thus recovered that
\begin{subequations}
\begin{align}
   \left( \widehat{\nabla}_a \widehat{\nabla}_b - \frac{\widehat{g}_{ab}}{\gls{L}^2} \right)E(\phi) &= 0\\
   \tn{with}\quad E(\phi) &= r^2 \widehat{\Box}\phi - 2r \widehat{\nabla}^a r \widehat{\nabla}_a \phi - [l(l+1) - 2] \phi.
\end{align}
\label{eqphimas}%
\end{subequations}
In a quest for further simplification, we notice that solving this equation is equivalent to solving $E(\phi) = 0$. Indeed, recall that
$Z_{ab}$ (equation \eqref{eqgauges}) is invariant under the transformation $\phi \to \phi + \phi_0$ provided
\begin{equation}
   \left(\widehat{\nabla}_a \widehat{\nabla}_b - \frac{\widehat{g}_{ab}}{\gls{L}^2}\right)\phi_0 = 0.
   \label{eqphi0}
\end{equation}
It turns out that the space of solutions of \eqref{eqphi0} is entirely generated by three functions $f_1$, $f_2$ and $f_3$
\cite{Ishibashi04}, so that any solution of \eqref{eqphi0} has the form
\begin{equation}
   \phi_0 = a_1 f_1 + a_2 f_2 + a_3 f_3,
\end{equation}
where $a_1$, $a_2$ and $a_3$ are three constants. Viewing \eqref{eqphimas} as a differential equation for $E$ (actually the same
as \eqref{eqphi0}), it comes similarly
\begin{equation}
   E(\phi) = b_1 f_1 + b_2 f_2 + b_3 f_3.
\end{equation}
This structure propagates such that
\begin{equation}
   E(\phi + \phi_0) = c_1 f_1 + c_2 f_2 + c_3 f_3,
\end{equation}
where $c_i$ are simple functions of $a_i$ and $b_i$. We can thus always choose $\phi_0$ such that $E(\phi + \phi_0) = 0$. Said
differently, solving \eqref{eqphimas} is equivalent to solve $E(\phi) = 0$, as announced above.

We end the discussion by defining the master variable
\begin{equation}
   \Phi_S \equiv \frac{\phi}{r},
\end{equation}
so that $E(\phi) = 0$ is equivalent to the master equation
\begin{equation}
   \widehat{\Box}\Phi_S - \frac{l(l+1)}{r^2}\Phi_S = 0.
\end{equation}
Notice that this equation is gauge-invariant by construction and is exactly the same as for vector-type perturbations, equation \eqref{vtypemaster}. This equation is
thus appropriately called a master equation. Notice also that out of the ten initial degrees of freedom contained in
$h_{\alpha\beta}$, we have reduced the number of degrees of freedom to two, one for the vector-type perturbation $\Phi_V$ and one
for the scalar-type perturbation $\Phi_S$. We thus have recovered the well-known result that first order gravitational waves in
four dimensions have only two degrees of freedom.

\subsection{Case $l=1$}
\label{l1s}

If $l=1$, $H_T$ in \eqref{hs} is not defined any more since $\widetilde{D}_i \widetilde{D}_j \mathbb{S} + \widetilde{\gamma}_{ij} \mathbb{S} = 0$.
Thus, $Z_{ab}$ and $Z$ are no more gauge-invariant. We recall that if $H_T = 0$, we have defined (equations \eqref{ZZab})
\begin{subequations}
\begin{align}
   Z_{ab} &= H_{ab} - \widehat{\nabla}_a H_b - \widehat{\nabla}_b H_a + \frac{Z}{2}\widehat{g}_{ab},\\
   Z &= \frac{2}{r^2}(H_L - 2r \widehat{\nabla}^a rH_a).
\end{align}
\end{subequations}
By an infinitesimal first order change of coordinates of the form \eqref{xis}, it comes
\begin{subequations}
\begin{align}
   Z' &= Z + \frac{4\xi}{r^2} + 4 r \widehat{\nabla}_a \left( \frac{\xi}{r^2} \right)\widehat{\nabla}^a r,\\
   Z_{ab}' &= Z_{ab} + \widehat{\nabla}_a \left( r^2 \widehat{\nabla}_b \left( \frac{\xi}{r^2} \right) \right) + \widehat{\nabla}_b \left(
   r^2 \widehat{\nabla}_a \left( \frac{\xi}{r^2} \right) \right) + \frac{2\xi}{r^2}\widehat{g}_{ab} + 2r \widehat{\nabla}^c r \widehat{\nabla}_c
   \left( \frac{\xi}{r^2} \right)\widehat{g}_{ab}.
\end{align}
\end{subequations}
These equations can be made more compact. Indeed, introducing
\begin{equation}
   \Xi \equiv -\frac{2\xi}{r},
\end{equation}
it comes
\begin{subequations}
\begin{align}
   Z' &= Z + \frac{2r}{\gls{L}^2}\Xi - 2 (\widehat{\nabla}r \cdot \widehat{\nabla}\Xi),\\
   \label{Zabprime}
   Z_{ab}' &= Z_{ab} - r \widehat{\nabla}_a \widehat{\nabla}_b \Xi + \frac{2r}{\gls{L}^2}\Xi \widehat{g}_{ab} - (\widehat{\nabla}r \cdot
   \widehat{\nabla}\Xi) \widehat{g}_{ab}.
\end{align}
\end{subequations}
In particular, we notice that
\begin{equation}
   (Z\indices{^a_a} - Z)' = Z \indices{^a_a} - Z - r\left( \widehat{\Box}\Xi - \frac{2\Xi}{\gls{L}^2} \right).
\end{equation}
the condition $Z \indices{^a_a} = Z$ (equation \eqref{gaugevareqs}) is thus not coming from Einstein's equation any more, but
can be interpreted as a gauge choice (recall for example that in our demonstration, \eqref{ijcomp2} is automatically satisfied if
$l = 1$ and consequently does not imply $H \indices{^a_a} = 0$).

Within this gauge, any first order change of coordinates satisfying $\widehat{\Box}\Xi = 2 \Xi/\gls{L}^2$ leaves
the gauge condition $Z \indices{^a_a} = Z$ unchanged. If we succeed, within this gauge class, to find $\Xi$ such that
$Z'_{ab} = 0$, it would imply $Z = 0$ and cascade to $h_{\alpha\beta} = 0$. Said differently, it would prove that $l=1$ scalar
perturbations are pure gauge modes. Thus, we would like to solve
\begin{subequations}
\begin{align}
   Z \indices{^a_a} = Z,\\
   Z'_{ab} = 0,\\
   E(\phi) = 0.\\
\end{align}
\end{subequations}
Keeping in mind \eqref{zabdef}, it is equivalent to
\begin{subequations}
\begin{align}
   \label{xi1}
   \widehat{\Box}\Xi - \frac{2}{\gls{L}^2}\Xi = 0,\\
   \label{xi2}
   -r \widehat{\nabla}_a \widehat{\nabla}_b \Xi + \frac{2r}{\gls{L}^2}\Xi \widehat{g}_{ab} - (\widehat{\nabla}r\cdot
   \widehat{\nabla}\Xi)
   \widehat{g}_{ab} + \widehat{\nabla}_a \widehat{\nabla}_b \phi - \frac{\phi}{\gls{L}^2} \widehat{g}_{ab} = 0,\\
   r^2 \widehat{\Box}\phi - 2r \widehat{\nabla}^a r \widehat{\nabla}_a \phi = 0.
\end{align}
\end{subequations}
In \cite{Ishibashi04}, it is shown that we can choose $\Xi$ as the solution of \eqref{xi1} with initial data satisfying
\eqref{xi2}. This implies that \eqref{xi2} remains zero at all times. In particular, according to \eqref{Zabprime}, it means
that $Z_{ab}$ remains zero at all times. Since we remained in the gauge class for which $Z \indices{^a_a} = Z$, $Z$ vanishes too.
This implies that the remaining degrees of freedom encoded in $H_a$ and $H_L$ also vanish (recall that $H_T = 0$ and $H_{ab} = 0$
are also gauge choices), and so does the whole perturbation $h_{\alpha\beta}$. The $l=1$ perturbation can thus be brought identically to
zero by a sophisticated gauge choice. They are thus no more than pure gauge modes.

\section{Linear geons}

In the previous sections, we have seen that the number of degrees of freedom of the metric
perturbation could be reduced to two: one degree of freedom $\Phi_V$ for vector-type perturbations, and one for scalar-type
perturbations $\Phi_S$. Furthermore, both master functions $\Phi_V$ and $\Phi_S$ obey the same gauge-invariant master equation, namely
\begin{equation}
   \widehat{\nabla}_a \widehat{\nabla}^a \Phi - \frac{l(l+1)}{r^2}\Phi = 0.
   \label{master}
\end{equation}
Our objective is now to solve this master equation and to find its general solutions. This allows to reconstruct the full
metric perturbation. The route of this reconstruction is however very different between vector-type and scalar-type
perturbations. Notably, the way to ensure that the space-time is \gls{aads} in the sense of chapter \ref{aads} put severe constraints on
the solutions of \eqref{master} that are clearly distinct for vector and scalar perturbations.

The reconstruction procedure depends in general on the gauge choice and on
the coordinates system. For the sake of concreteness, we fix the coordinates of the \gls{ads} background to be conformal (chapter
\ref{aads} equation \eqref{adsconformal}). The background metric then reads
\begin{equation}
   d\overline{s}^2 = \frac{\gls{L}^2}{\cos^2x}(-dt^2 + dx^2 + \sin^2x[d\theta^2 + \sin^2\theta d\varphi^2]),
   \label{adsconformal2}
\end{equation}
where $x \in [0,\pi/2[$, $x = 0$ being the origin and $x = \pi/2$ the \gls{ads} boundary (spatial infinity). Beware that we have
rescaled the time coordinate so that $t$ is dimensionless in this equation.

Our goal is to build linear \gls{aads} gravitational geons, i.e.\ periodic solutions of the linearised Einstein's equation in
vacuum with an \gls{ads} background (recall definition \ref{defgeon} of chapter \ref{geons} and definition \ref{aadsdefinition} of
chapter \ref{aads}). We thus focus on time-periodic solutions.

\subsection{Time-periodic solutions of the master equation}

As we have shown that $l=1$ modes are pure gauge modes in sections \ref{l1v} and \ref{l1s}, we consider only the case $l
\geq 2$. In the coordinate system \eqref{adsconformal2}, the master equation \eqref{master} is equivalent to
\begin{equation}
   -\partial_t^2 \Phi + \partial_x^2 \Phi - \frac{l(l+1)}{\sin^2x}\phi = 0.
   \label{masterx}
\end{equation}
We look for a time-periodic solution of the form
\begin{equation}
   \Phi(t,x) = P(x) \cos(\omega t + \delta).
   \label{ansatzPhi}
\end{equation}
Equation \eqref{masterx} thus becomes a differential equation for the function $P$, namely,
\begin{equation}
   P'' + \left( \omega^2 - \frac{l(l+1)}{\sin^2 x} \right)P = 0.
   \label{difP}
\end{equation}
This equation can be cast into the form of the hypergeometric differential equation. Indeed, let us perform the change of
variable
\begin{equation}
   z = \sin^2x, \quad \frac{dz}{dx} = 2 \sqrt{z(1-z)},
\end{equation}
which is a one-to-one transformation of $x\in [0,\pi/2[$ to $z\in [0,1[$. Equation \eqref{difP} yields
\begin{equation}
   4z(1-z)P'' + 2(1-2z)P' + \left( \omega^2 - \frac{l(l+1)}{z} \right)P = 0.
   \label{zppP}
\end{equation}
Trading $P$ for the rescaled function $Q$ defined by
\begin{equation}
   Q \equiv \frac{P}{z^{\frac{l+1}{2}}},
   \label{Qdef}
\end{equation}
the differential equation \eqref{zppP} becomes
\begin{equation}
   z(1-z)Q'' + \left[ l+ \frac{3}{2} - (l+2)z \right]Q' + \left[ \omega^2 - \frac{(l+1)^2}{4} \right]Q = 0.
\end{equation}
This equation is indeed the hypergeometric differential equation in its standard form (see \cite{nist} section 15.10)
\begin{equation}
   z(1-z)Q'' + \left[c -(a+b+1)z \right]Q' - abQ = 0,
   \label{hypergeomdiffeq}
\end{equation}
where $a$, $b$, and $c$ are the three constants
\begin{equation}
   a = \frac{1}{2}(l+1-\omega), \quad b = \frac{1}{2}(l+1+\omega), \quad c = l + \frac{3}{2}.
   \label{abc}
\end{equation}

The differential equation \eqref{hypergeomdiffeq} is closely linked to the hypergeometric function, which is defined by
\begin{equation}
   \tensor[_2]{F}{_1} (a,b;c;z) \equiv \sum_{n=0}^\infty \frac{(a)_n(b)_n}{(c)_n}\frac{z^n}{n!},
   \label{hypergeomdef}
\end{equation}
where we have introduced the Pochhammer's symbol
\begin{equation}
   (a)_n \equiv \left\{
   \begin{array}{ll}
      1 & (n = 0),\\
      a(a+1)\cdots(a+n-1) & (n > 0).
   \end{array}
   \right .
\end{equation}
It can be noticed that if $a$ is a negative integer, its Pochhammer's symbol $(a)_n$ is zero for $n \geq 1 - a$. Thus, if $a$ or
$b$ are negative integers, the hypergeometric function closes at finite order and reduces to an ordinary polynomial. Note also that
this function admits a smooth and finite limit when $z \to 1$.

The space of solutions of the hypergeometric differential equation \eqref{hypergeomdiffeq} is of dimension two, meaning that it is
sufficient to find two independent solutions to build the space of solutions. Furthermore, we are interested in solutions that show no
singularities at the two poles $z=0$ (origin) and $z=1$ (\gls{ads} boundary).

A pair of two independent solutions of \eqref{hypergeomdiffeq} that are numerically satisfactory (in the sense of \cite{nist}
section 2.7(iv)) near $z=0$ is
\begin{equation}
   Q_1 = \tensor[_2]{F}{_1}(a,b;c;z) \quad \tn{and} \quad Q_2 = z^{1-c}\tensor[_2]{F}{_1}(a-c+1,b-c+1;2-c;z),
   \label{q1q2}
\end{equation}
while a pair of two numerically satisfactory solutions near $z=1$ is
\begin{equation}
   Q_3 = \tensor[_2]{F}{_1}(a,b;a+b-c+2;1-z) \quad \tn{and} \quad Q_4 = z^{c-a-b}\tensor[_2]{F}{_1}(c-a,c-b;c-a-b+1;1-z).
   \label{q3q4}
\end{equation}
Of course, since the space of solutions is only of dimension two, the function $Q_1$, $Q_2$, $Q_3$ and $Q_4$ are not independent.
They are related to each other by
\begin{subequations}
\begin{align}
   \label{q1toq4}
   Q_1 &= \frac{\Gamma(c)\Gamma(c-a-b)}{\Gamma(c-a)\Gamma(c-b)}Q_3 + \frac{\Gamma(c)\Gamma(a+b-c)}{\Gamma(a)\Gamma(b)}Q_4,\\
   Q_2 &= \frac{\Gamma(2-c)\Gamma(c-a-b)}{\Gamma(1-a)\Gamma(1-b)}Q_3 + \frac{\Gamma(2-c)\Gamma(a+b-c)}{\Gamma(a-c+1)\Gamma(b-c+1)}Q_4,
\end{align}
\end{subequations}
where $\Gamma$ is Euler's Gamma function. At this point, we can dismiss $Q_2$ since with our set of $(a,b,c)$ parameters
(equation \eqref{abc}), the $z^{1-c}$ factor in \eqref{q1q2} is always singular at the origin $z = 0$ while the hypergeometric
function smoothly tends to 1. This implies that the master function $\Phi$ would diverge as $O(z^{-l/2})$ near the origin. Our regular
geon solution thus cannot be generated by $Q_2$. As for $Q_4$, it is not diverging near $z=1$ since the factor $c-a-b$
appearing in \eqref{q3q4} is $1/2$. We are thus left with three regular generating functions for geon solutions, $Q_1$, $Q_3$ and $Q_4$.

\subsection{Scalar-type linear geons}

Let us summarise the results of the scalar-type perturbative approach. The reconstruction of the metric perturbation in this case
proceeds as follows.
\begin{myenum}
   \item Construct a time-periodic master function $\Phi_S$ obeying the master equation \eqref{master}.
   \item Compute the gauge-invariant variable
      \begin{equation}
         Z_{ab} = \left(\widehat{\nabla}_a \widehat{\nabla}_b - \frac{1}{\gls{L}^2}\widehat{g}_{ab}\right)(r\Phi_S).
      \end{equation}
   \item The gauge-invariant variable $Z_{ab}$ being related to the metric perturbation by
      \begin{subequations}
      \begin{align}
         Z \indices{^a_a} &= \frac{2}{r^2}\left(H_L + \frac{l(l+1)}{2}H_T + 2r \widehat{\nabla}^a r X_a\right),\\
         Z_{ab} &= H_{ab} + \widehat{\nabla}_a X_b + \widehat{\nabla}_b X_a + \frac{1}{2}Z \widehat{g}_{ab},
      \end{align}
      \end{subequations}
      where
      \begin{equation}
         X_a = -H_a + \frac{r^2}{2}\widehat{\nabla}_a \left( \frac{H_T}{r^2} \right),
      \end{equation}
      choose a convenient gauge where these relations are easily invertible (e.g.\ a gauge in which $H_a = 0$ and $H_T = 0$).
   \item Reconstruct the metric perturbation with any $(l,m)$ angular dependence knowing that
      \begin{equation}
         h_{ab} = H_{ab}\mathbb{S}, \quad h_{ai} = H_a \widetilde{D}_i \mathbb{S}, \quad h_{ij} = H_L \widetilde{\gamma}_{ij} \mathbb{S} + H_T
         \left( \widetilde{D}_i \widetilde{D}_j + \frac{l(l+1)}{2}\widetilde{\gamma}_{ij} \right)\mathbb{S}.
      \end{equation}
   \item Check that the obtained linear geon is well \gls{aads} (in the sense of definition \ref{aadsdefinition}), in particular check that
      the pseudo-magnetic Weyl tensor $\widehat{B}_{\alpha\beta}$ (equation \eqref{defmagn}) vanishes at the \gls{ads} boundary:
      \begin{equation}
         \widehat{B}_{\alpha\beta} \mathrel{\widehat{=}} 0.
      \end{equation}
\end{myenum}
Step 1 is achieved by choosing $Q$ as a linear combination of $Q_1$, $Q_3$, and $Q_4$ and using relations \eqref{Qdef} and
\eqref{ansatzPhi}. It turns out that if we choose $Q = Q_4$ in step 1, the pseudo-magnetic Weyl tensor $\widehat{B}_{\alpha\beta}$ is not vanishing
at the \gls{ads} boundary $x = \pi/2$. This means that this kind of solutions is not \gls{aads} and must be discarded. The master
function can thus have a $Q_1$ or a $Q_3$ component but $Q_4$ is strictly forbidden. In particular, this means that in
\eqref{q1toq4}, the proportionality factor in front of $Q_4$ must vanish. This is possible only if $a$ or $b$ is a negative
integer\footnote{Recall this property of the Gamma function: $\forall n \in \mathbb{Z}^-, \dfrac{1}{\Gamma(n)} = 0$.}, or
equivalently (see equation \eqref{abc}), if
\begin{equation}
   \omega = \omega_S \equiv l + 1 + 2n, \quad n \in \mathbb{N},
   \label{omegaS}
\end{equation}
where we restricted the frequency $\omega$ to be positive. In this case, $Q_1$ is merely proportional to $Q_3$. Furthermore,
\eqref{omegaS} implies that $a = -n$ is necessarily a negative integer, so that the hypergeometric function in the expression
\eqref{q3q4} of $Q_3$ closes at finite order and reduces to a mere polynomial. In particular, we have
\begin{equation}
   \tensor[_2]{F}{_1}(-n,b;c;z) = \frac{n!}{(c)_n}P_n^{(c-1,b-c-n)}(1 - 2z),
   \label{Fmn}
\end{equation}
where $P_n^{(\alpha,\beta)}$ denotes Jacobi polynomials. The master function can then be rewritten with \eqref{Qdef} and \eqref{ansatzPhi} as
\begin{equation}
   \Phi_S = \alpha_S \sin^{l+1}x P_n^{\left( l+\frac{1}{2},-\frac{1}{2} \right)}(\cos(2x))\cos(\omega_S t + \delta),
   \label{PhiS}
\end{equation}
where $\alpha_S$ is an arbitrary (but small) amplitude factor. This expression can be readily employed in the five-step procedure
described above to construct any $(l,m,n)$ scalar-type linear geon.

\subsection{Vector-type linear geons}

In the vector-type case, the reconstruction of the metric perturbation proceeds as follows.
\begin{myenum}
   \item Construct a time-periodic master function $\Phi_V$ obeying the master equation \eqref{master}.
   \item Compute the gauge-invariant variable
      \begin{equation}
         Z_a = \widehat{\epsilon}_{ab}\widehat{\nabla}^b (r \Phi_V).
      \end{equation}
   \item The gauge-invariant variable $Z_a$ being related to the metric perturbation by
      \begin{equation}
         Z_a = H_a - r^2 \widehat{\nabla}_a\left(\frac{H_T}{r^2}\right),
      \end{equation}
      choose a convenient gauge where this relation is easily invertible (e.g.\ a gauge in which $H_T = 0$).
   \item Reconstruct the metric perturbation with any $(l,m)$ angular dependence knowing that
      \begin{equation}
         h_{ab} = 0, \quad h_{ai} = H_a \mathbb{V}_i, \quad h_{ij} = 2 H_T \widetilde{D}_{(i}\mathbb{V}_{j)}.
      \end{equation}
   \item Check that the obtained linear geon is well \gls{aads}, namely
      \begin{equation}
         \widehat{B}_{\alpha\beta} \mathrel{\widehat{=}} 0.
      \end{equation}
\end{myenum}
Similarly to the scalar-type reconstruction, step 1 is achieved by choosing $Q$ a linear combination of $Q_1$, $Q_3$, and $Q_4$ and using relations \eqref{Qdef} and
\eqref{ansatzPhi}. It turns out that if we choose $Q = Q_3$ in step 1, the pseudo-magnetic Weyl tensor $\widehat{B}_{\alpha\beta}$ is not vanishing
at the \gls{ads} boundary $x = \pi/2$. This means that this kind of solutions is not \gls{aads} and must be discarded. The master
function can thus display a $Q_1$ or a $Q_4$ component but $Q_3$ is strictly forbidden. In particular, this means that in
\eqref{q1toq4}, the proportionality factor in front of $Q_3$ must vanish. This is possible only if $c-a$ or $c-b$ is a negative
integer, or equivalently, if
\begin{equation}
   \omega = \omega_V \equiv l + 2 + 2n, \quad n \in \mathbb{N},
   \label{omegaV}
\end{equation}
where we restricted the frequency $\omega$ to be positive. In this case, $Q_1$ is merely proportional to $Q_4$. Furthermore,
\eqref{omegaV} implies that $c-b = -n$ is a negative integer, so that the hypergeometric function in the expression of $Q_4$
closes at finite order and reduces to a mere polynomial. Using \eqref{Fmn} and the symmetry property $P_n^{(\alpha,\beta)}(-z) =
(-1)^n P_n^{(\alpha,\beta)}(z)$, the master function can be rewritten
\begin{equation}
   \Phi_V = \alpha_V \sin^{l+1}x \cos x P_n^{\left( l+\frac{1}{2},\frac{1}{2} \right)}(\cos(2x))\cos(\omega_V t + \delta),
   \label{PhiV}
\end{equation}
where $\alpha_V$ is an arbitrary (but small) amplitude factor. This expression can be readily employed in the five-step procedure
described above to construct any $(l,m,n)$ vector-type linear geon.

\subsection{Helically symmetric solutions}
\label{helsymgeon}

The two procedures described above give us the power to build any linear geon solution. In particular, we are interested in
helically symmetric geons. They can be constructed by the following procedure.
\begin{myenum}
   \item Choose a set of quantum numbers $(l,m,n)$ and determine the type (scalar or vector) of the solution. This determines
      fully the frequency $\omega$.
   \item Pick one $\Phi$ with phase $\delta = 0$ and build the corresponding metric perturbation with $(l,m)$ angular dependence.
   \item Pick the same $\Phi$ but this time with $\delta = \pi/2$ and build the corresponding metric perturbation with $(l,-m)$ angular dependence.
   \item Add the two above metric perturbations.
   \item Perform the change of coordinate $\varphi \to \varphi - \omega t / m$, that impacts the $(t,\varphi)$ block of the
      metric as (see equation \eqref{Ttrans})
      \begin{equation}
         g_{tt} \to g_{tt} + \frac{\omega^2}{m^2}g_{\varphi\varphi}, \quad g_{t\varphi} \to \frac{\omega}{m}g_{t\varphi}, \quad
         g_{\varphi\varphi} \to g_{\varphi\varphi}.
      \end{equation}
\end{myenum}
The metric perturbation resulting from step 4 have components that are proportional to either $\cos(\omega t - m\varphi)$ or
$\sin({\omega t - m\varphi)}$, where $\omega$ is entirely determined by $n$ and $l$ (equations \eqref{omegaS} and \eqref{omegaV}). Hence, step 5 is
simply a change of coordinates toward the co-rotating frame in which the metric is time-independent. In particular, the solution
admits a helical Killing vector
\begin{equation}
   \partial_{t'}^\alpha \equiv \partial_t^\alpha + \frac{\omega}{m}\partial_\varphi^\alpha.
\end{equation}
This kind of solution is interesting since it is stationary in the co-rotating frame. This means that
the time dependence can be dropped out. It also suggests that fully non-linear geons belonging to this helically symmetric family do not
depend on time in this frame either.

From a numerical point of view, we can try to find helically symmetric solutions by solving the full Einstein's equation and
dropping all time derivatives in the co-rotating frame. Einstein's equation can then be considered as an elliptic (and thus
invertible) 3-dimensional problem, that is much faster to solve than a 4-dimensional one. The linear analytic solution constructed
above can serve as a good initial guess for an iterative scheme. Of course, the gauge choice is
crucial in the inversion of the non-linear Einstein's operator, and this problem is discussed in detail in chapter \ref{gaugefreedom}.

We can go further in the construction of helically symmetric geons by mixing modes of different type with different quantum
numbers $(l,m,n)$ provided they share a common frequency $\omega$. For example, equations \eqref{omegaS} and \eqref{omegaV} allow
us to mix the scalar-type $(l,m,n) = (2,2,1)$, $(4,2,0)$ modes and the vector-type $(3,2,0)$ mode, since they
all share the same frequency $\omega = 5$. Appropriately combining with the $-m$ counterparts (step 3 of the above procedure), we can construct radially excited
(since one component has $n = 1$) helically symmetric linear geons. The possibilities of construction of helically symmetric
linear geons are thus infinite.

\section{Beyond the linear approximation}
\label{beyondlin}

Beyond first order, the perturbative expansion can be pursued. The philosophy remains unchanged \cite{Martinon17}, namely we can build
gauge-invariant variables at each order that are simply related to the metric perturbation and whose evolutions are dictated by the
same master equation encountered in this chapter, equation \eqref{master}. The notable difference though is that at order $k$,
this master equation contains source terms instead of a zero on the right-hand side. These source terms are functions of the
less-than-$k$ orders.

These inhomogeneous terms contaminate the higher orders equations, until there appears a resonant term in the master equation, namely
a term in $\cos(\omega t)$ where $\omega$ is precisely the same as in the homogeneous part (left-hand side). This happens at third order and
is thus highly reminiscent of the massless scalar field case in the context of the \gls{ads} instability (chapter
\ref{adsinsta}). When such resonances appear and cannot be cured, it is a strong indication that the perturbative expansion will
not converge any more after a finite amount of time. In this regard, resonance implies non-existence of geons.
However, just like some time-periodic solutions in the scalar field case (section \ref{tpsol} of chapter \ref{adsinsta}),
gravitational geons can be cured from these secular resonances by simply using the Poincaré-Lindstedt method, i.e.\ promoting
$\omega$ to a function of the amplitude.

This procedure was successively applied for the first time in the special scalar $(l,m,n) = (2,2,0)$ case in \cite{Dias12a}. The
authors concluded that, at fixed quantum numbers $(l,m,n)$, the perturbative construction of this kind of geons could be theoretically feasible at arbitrary high
orders, suggesting the existence of fully non-linear geons. A consecutive study \cite{Dias16a,Dias17a} tried to generalise this result to
all types of geons but failed to do so. According to the authors, the number of secular resonances arising at third order highly
depended on the quantum numbers $(l,m,n)$ and they were unable to remove all secular resonances in all cases, thus concluding that
some families of geons (for example the scalar $(l,m,n) = (2,2,1)$ one) had no fully non-linear existence.

This claim was denied very soon after. Indeed, the author of \cite{Rostworowski16,Rostworowski17a,Rostworowski17b} stressed that there were indeed
irremovable secular resonances at third order for some single-mode families of geons (i.e.\ fixed $(l,m,n)$). However, even if more mathematically
convenient, there is no reason to consider exclusively single-mode families. As we stressed in the above section about helically
symmetric geons, it is perfectly right to build mixed-mode families by combining perturbations with different types and quantum
numbers provided they share the same frequency $\omega$ and azimuthal number $m$. In this case, the number of inhomogeneous terms appearing at third
order in the master equation increases, but the combination of different modes offers more freedom to fine-tune the amplitude parameters
and perform a successful Poincaré-Lindstedt regularisation. There are thus much more families of geons that can be constructed perturbatively
to arbitrary order. Thus, at fixed parameters $(\omega,m)$, all geons seem to admit fully non-linear extensions, as illustrated
in chapter \ref{adsinsta} in figure \ref{geoncontroversy}.

Actually, there are even clues that the number of fully non-linear geons at a given frequency $\omega$
matches exactly the multiplicity of $\omega$. For example, if we consider again the case $\omega = 5$ discussed
in section \ref{helsymgeon}, we have found three single-mode families that share this frequency (the scalar-type $(l,m,n) =
(2,2,1)$, $(4,2,0)$ modes and the vector-type $(3,2,0)$ mode). It turns out that there are exactly three
possible linear combinations of them that survive the perturbative expansion at third order via successful Poincaré-Lindstedt
regularisation. This property (coincidence?) suggests that the perturbative expansion has a refined structure that is yet to be understood.

The number of fully non-linear geons is thus much higher than originally observed in \cite{Dias16a,Dias17a}. In
\cite{Martinon17}, we have confirmed this expectation by numerically constructing all three fully non-linear and radially excited geons with $\omega =
5$ and $m=2$, using the appropriate linear combinations of single-mode excitations. The remaining of this manuscript is dedicated to the
theoretical and numerical implementations that allowed such numerical results.

\chapter{Harmonic gauges}
\label{gaugefreedom}
\citationChap{Everything must be made as simple as possible. But not simpler.}{Albert Einstein}
\minitoc

In fully non-linear \gls{gr}, perturbative methods are non longer valid when gravity is too strong. As a consequence we have to rely on the
fully non-linear Einstein's system of equations. It constitutes a system
of second order partial differential equations for the metric $g_{\alpha\beta}$. In order to build gravitational geons, we focus
on the 4-dimensional Einstein's equation in vacuum with \gls{aads} asymptotics. The unknowns are the ten metric
components, and Einstein's system provides the same number of equations. However, not all are independent: they
display six degrees of freedom only.

In order to get an invertible system of equations, we thus have to fix one way or another four degrees of freedom of the metric,
and solve the remaining six others via Einstein's equation. This is the so-called gauge freedom. Namely, the four degrees of
freedom to be fixed a priori are equivalent to a choice of a coordinate system. So there are infinitely many possibilities to fix
the gauge freedom. However, not all choices are well behaved mathematically, and not all choices make Einstein's system of
equations invertible. This is called ill-posedness and is due to the gauge invariance of the theory.
Actually, there are very few gauges which are known to give robust results in the field of numerical relativity. We refer to
standard textbooks \cite{Alcubierre08,Baumgarte10,Gourgoulhon07} for reviews of all known and successful gauges in this area.

In this chapter, we focus on a particular class of gauges, namely the harmonic and spatial harmonic gauge. The harmonic gauge was
notably used in 1952 by Choquet-Bruhat in the first proof of existence and uniqueness of solutions of Einstein's field
equations \cite{Bruhat52,Bruhat62}. The proof is based on a hyperbolic reduction of Einstein's equation \cite{Friedrich96}, i.e.\
a reformulation into a well-posed initial-value (or Cauchy) problem.

In mathematics, such a problem is defined as follows \cite{Sarbach12}.
\begin{definition}[Well-posed Cauchy problem]
The Cauchy problem
\begin{subequations}
\begin{align}
   \partial_t f(t,x) &= P(t,x,\partial_x)f(t,x), \quad x \in \mathbb{R}^n, \quad t \geq 0,\\
   f(0,x) &= g(x), \quad x \in \mathbb{R}^n,
\end{align}
\label{cauchy}
\end{subequations}
is well-posed if and only if any $g \in C_0^\infty(\mathbb{R}^n)$ gives rise to a unique $C^\infty$ solution $f(t,x)$ and if there
are constants $K \geq 1$ and $\alpha \in \mathbb{R}$ such that
\begin{equation}
   \lVert f(t,\cdot) \rVert \leq K e^{\alpha t} \lVert g \rVert,
   \label{estimate}
\end{equation}
for all $g \in C_0^\infty(\mathbb{R}^n)$ and all $t \geq 0$.
\label{cauchydef}
\end{definition}
In this definition, $f$ is a state vector encoding all the unknown functions of the problem, $C_0^\infty$ is the set of smooth
functions with compact support, and $P(t,x,\partial_x)$ is an operator with variable coefficients that involves spatial
derivatives. All norms refer to the $L^2$ norm. This definition does not state anything about growth of the solution with time other that this growth is bounded by
an exponential. In this sense, well-posedness is not a statement about the stability in time, but rather about stability with
respect to mode fluctuations (the so-called constraints violating modes in \gls{gr}). 

Let us notice that if $u_1$ and $u_2$ are
two solutions corresponding to initial data $g_1,g_2 \in C_0^\infty(\mathbb{R}^n)$, then the difference $f = f_2 - f_1$ satisfies
the Cauchy problem \eqref{cauchy} with $g = g_2 - g_1$ and the estimate \eqref{estimate} implies that
\begin{equation}
   \lVert f_2(t,\cdot) - f_1(t,\cdot)\rVert \leq K e^{\alpha t} \lVert g_2 - g_1 \rVert.
\end{equation}
In particular, this implies that $f_2 (t,\cdot)$ converges to $f_1(t,\cdot)$ if $g_2$ converges to $g_1$ in the $L^2$-sense. Said
differently, the solution depends continuously on the initial data. Furthermore, the estimate \eqref{estimate}, implies uniqueness
of the solution because for two solutions $f_1$ and $f_2$ with the same initial data $g_1 = g_2 \in C_0^\infty(\mathbb{R}^n)$, the
inequality \eqref{estimate} implies $f_1 = f_2$.

We retain that an initial value problem is well-posed when it admits a unique solution whose behaviour changes
continuously with initial conditions. We apply this vocabulary equally to boundary-value problems of the form
\begin{subequations}
\begin{align}
   0 = P(x,\partial_x)f(x), \quad x \in \mathcal{V},\\
   f(x) = g(x), \quad x \in \partial \mathcal{V},
\end{align}
\end{subequations}
where $\partial \mathcal{V}$ is the boundary of the domain $\mathcal{V}$.

The hyperbolic reduction of \cite{Friedrich96} consists in converting the geometrical initial-value problem for Einstein's field
equations into a well-posed Cauchy problem for hyperbolic differential systems. This procedure is precisely made possible in the harmonic gauge
because it allows the principal part of the Einstein operator to be reduced to a quasi-linear wave (or d'Alembertian) operator.

Moreover, this gauge played a very important role in the numerical relativity breakthrough of 2005 \cite{Sperhake15}, when Pretorius
\cite{Pretorius05a,Pretorius06} revealed the first successful numerical evolution of a black hole binary. His work was based on
the generalised harmonic formulation of Einstein's equation, first described in \cite{Garfinkle02,Szilagyi03}, and then tested and
reviewed in \cite{Pretorius05b,Babiuc07,Sorkin10}. The harmonic formulation is a very close cousin of the Z4 formulation,
initially studied in \cite{Bona03,Bona04a,Bona04b,Gundlach05,Lindblom06}. Such simulations were pursued with success in the
\gls{ads}-\gls{cft} context in \cite{Bantilan12,Bantilan15} in simulations of axisymmetric black hole collisions in \gls{aads}
space-times and for the first time in the \gls{ads} instability context in \cite{Bantilan17}.

Hereafter, we first present the harmonic gauge. We then discuss how to enforce the harmonic condition with the help of the De Turck
method in boundary-value problems. After an examination of the $3+1$ decomposition of the \gls{edt} system, we discuss a closely related
gauge, namely the \gls{am} gauge, which is more appropriate to $3+1$ formalism \cite{Andersson03,Martinon17}. We discuss the
well-posedness of the resulting system of equations and examine how the constraints propagate during the evolution of an
initial-value problem. Inspired by these results, we explain how the \gls{eam} system can be adapted to a boundary-value problem,
that is relevant to the numerical construction of helically symmetric geons.

Hereafter, we use the usual notations of $3+1$ formalism (appendix \ref{d+1}). We denote by Latin letters $i,j,k,l$ the spatial indices
and set $\gls{G} = \gls{c} = 1$.

\section{Harmonic gauge}

The harmonic gauge enforces four conditions on the coordinates. Namely, each coordinate has to be harmonic, i.e.\ solution of the harmonic
wave equation. These four conditions can be reformulated in terms of vector components $\xi^\alpha$, so that enforcing the gauge
is equivalent to bring this vector to zero.

\subsection{Harmonic coordinates}

The harmonic gauge is by definition a choice of harmonic coordinates satisfying the wave equation
\begin{equation}
   \nabla_\mu \nabla^\mu x^\alpha = 0.
   \label{harmonicgauge}
\end{equation}
An important feature of this equation is that $x^\alpha$, despite the notations, is not a vector. Instead each coordinate
$x^\alpha$ is considered independently as a scalar field. Consequently, equation \eqref{harmonicgauge} is not a vector equation but a set of
four mathematical equations fixing the four coordinates. We can rewrite it with the definition of the covariant derivative
\eqref{defcov} as follows:
\begin{equation}
   g^{\mu\nu}(\partial_\mu \partial_\nu x^\alpha - \Gamma \indices{^\rho_{\mu\nu}}\partial_\rho x^\alpha) = 0 \iff g^{\mu\nu}\Gamma \indices{^\alpha_{\mu\nu}} = 0,
\end{equation}
where we have used the property $\partial_\nu x^\alpha = \delta \indices{^\alpha_\nu}$. We thus define, in light of
\eqref{Gammadet},
\begin{equation}
   \Gamma^\alpha \equiv g^{\mu\nu}\Gamma \indices{^\alpha_{\mu\nu}} = -\frac{1}{\sqrt{-g}}\partial_\mu (\sqrt{-g}g^{\alpha\mu}),
   \label{gammaa}
\end{equation}
so that the harmonic gauge condition \eqref{harmonicgauge} reads
\begin{equation}
   \Gamma^\alpha = 0.
   \label{Gammaa}
\end{equation}
This equation is a mere rewriting of \eqref{harmonicgauge}. As such, it still cannot be considered a vector equation. This is most
easily seen with the transformation rule of the Christoffel symbols \eqref{Gammatrans}, which do not transform as tensors
(equation \eqref{Ttrans}). Equation \eqref{Gammaa} is thus a set of four coordinate-dependent conditions.

Interestingly, this harmonic criterion is not satisfied by Minkowski space-time in spherical coordinates. Indeed, considering the
corresponding length element
\begin{equation}
   ds^2 = -dt^2 + dr^2 + r^2(d\theta^2 + \sin^2 \theta d\varphi^2),
\end{equation}
a computation shows that
\begin{equation}
   \Gamma^t = 0, \quad \Gamma^r = -\frac{2}{r}, \quad \Gamma^\theta = -\frac{1}{r^2\tan\theta},\quad \Gamma^\varphi = 0.
\end{equation}
The condition \eqref{Gammaa} thus restricts the coordinates to be of Cartesian type and forbids the use of spherical ones.
In order to allow any type of coordinates, a natural extension of the harmonic condition is thus to enforce $\Gamma^\alpha$ to be
fixed to a reference quantity. Namely, let us define
\begin{equation}
   \xi^\alpha \equiv g^{\mu\nu}(\Gamma \indices{^\alpha_{\mu\nu}} - \overline{\Gamma}\indices{^\alpha_{\mu\nu}}),
   \label{xidef}
\end{equation}
where a bar refers to a predefined reference metric, and enforce
\begin{equation}
   \xi^\alpha = 0,
   \label{gammagauge}
\end{equation}
The choice of the reference metric cannot be arbitrary. In particular, it has to share the same asymptotics as
the physical metric in order for \eqref{gammagauge} to be satisfied far away from the gravitational source \cite{Dias16b}. In
practice, it is very often sufficient to choose Minkowski space-time in asymptotically flat settings and \gls{ads} space-time in
\gls{aads} configurations.

Condition \eqref{gammagauge} is quite reminiscent of the generalised harmonic gauge of Pretorius \cite{Pretorius05b}.
In the latter, the gauge condition is $\Gamma^\alpha = H^\alpha$, where $H^\alpha$ is a given function of the coordinates
determined locally. However, one major difference with \eqref{gammagauge} is that the choice of a reference metric is global while
$H^a$, as a function of the coordinates, is local. The choice of $H_\alpha$ is not covariant and might be fine-tuned depending on
the particular problem at hand. In contrast, equation \eqref{gammagauge} is covariant and its left-hand side transforms as a
vector under an arbitrary change of coordinates. This is because under a general coordinate transformation, the parts that do not transforms
as tensors in the Christoffel symbols (equation \eqref{Gammatrans}) cancel each other in \eqref{xidef}.

Let us mention that there are several equivalent formulas for $\xi^\alpha$. Thanks to the properties of the Christoffel symbols,
featured in equations \eqref{christoffel}, \eqref{varGamma}, \eqref{Gammadet} and \eqref{defcov}, it can be shown that
\cite{Friedrich96,Headrick10}
\begin{subequations}
\begin{align}
   \xi^\alpha &= g^{\mu\nu}(\Gamma \indices{^\alpha_{\mu\nu}} - \overline{\Gamma}\indices{^\alpha_{\mu\nu}}) \\
              &= -\overline{\nabla}_\mu g^{\alpha\mu} + \frac{1}{2}g^{\alpha\mu}g_{\rho\sigma}\overline{\nabla}_\mu g^{\rho\sigma}\\
              &= -\overline{g}^{\alpha\mu}\left( \nabla^\nu \overline{g}_{\mu\nu} - \frac{1}{2}g^{\rho\sigma}\nabla_\mu \overline{g}_{\rho\sigma} \right)\\
              &= -\sqrt{\frac{\overline{g}}{g}}\overline{\nabla}_\mu\left(\sqrt{\frac{g}{\overline{g}}}g^{\alpha\mu}\right).
\end{align}
\end{subequations}

\subsection{Linear level}
\label{linharmo}

Now that we have defined the harmonic gauge condition, our goal is to give a method that can bring any metric in this gauge.
We start by the linear level which is the simplest. We consider a small first order perturbation $h_{\alpha\beta}$ of a background
metric $\overline{g}_{\alpha\beta}$ in vacuum. This is relevant in particular for the case of \gls{aads} geons that we
already have built in chapter \ref{perturbations}. By construction, the background metric satisfies the gauge condition \eqref{gammagauge}, but this is
not the case of the whole perturbed metric a priori. We thus perform an infinitesimal change of coordinates
\begin{equation}
   x^{\alpha'} = x^\alpha + \chi^\alpha(x^\mu),
\end{equation}
where $\chi^\alpha$ depends on the coordinates $(x^\mu)$ and is of the same order as $h_{\alpha\beta}$. We aim at bringing the
new perturbation $h_{\alpha'\beta'}$ to the harmonic gauge while keeping the background fixed. Applying the transformation formulas (appendix \ref{grd}, equations
\eqref{Ttrans} and \eqref{Gammatrans}) at first order in $\chi^\alpha$, we get the metric and Christoffel symbols in the new
coordinate system $(x^{\alpha'})$:
\begin{subequations}
\begin{align}
   h_{\alpha'\beta'} &= h_{\alpha\beta} - \mathcal{L}_\chi h_{\alpha\beta},\\
   h^{\alpha'\beta'} &= h^{\alpha\beta} - \mathcal{L}_\chi h^{\alpha\beta},\\
   \Gamma \indices{^{\gamma'}_{\alpha'\beta'}} &= \Gamma \indices{^\gamma_{\alpha\beta}} - \mathcal{L}_\chi \Gamma \indices{^\gamma_{\alpha\beta}} - \partial_\alpha\partial_\beta \chi^\gamma,
\end{align}
\label{hapbp}
\end{subequations}
where $\mathcal{L}_\chi$ indicates the Lie derivative (equation \eqref{lie}) in the $\xi^\alpha$ direction. The background quantities are simply left
untouched. Applying these results to the gauge vector $\xi^\alpha$, equation \eqref{xidef} becomes, in the new coordinate system,
\begin{equation}
   \xi^{\alpha'} = \xi^\alpha - g^{\mu\nu}(\mathcal{L}_\chi \Gamma \indices{^\alpha_{\mu\nu}} + \partial_\mu \partial_\nu \chi^\alpha) -
   (\Gamma \indices{^\alpha_{\mu\nu}} - \overline{\Gamma}\indices{^\alpha_{\mu\nu}})\mathcal{L}_\chi g^{\mu\nu},
\end{equation}
where we have conserved only first order terms in $\chi^\alpha$. Expanding the Lie derivatives with \eqref{lie} and forcing the appearance of
$\nabla^\mu \nabla_\mu \xi^\alpha$ with the definition of the covariant derivative \eqref{defcov}, we obtain
\begin{equation}
   \xi^{\alpha'} = \xi^\alpha -\nabla^\mu \nabla_\mu \chi^\alpha - R \indices{^\alpha_\mu}\chi^\mu + 2 (\Gamma \indices{^\alpha_{\mu\nu}} - \overline{\Gamma} \indices{^\alpha_{\mu\nu}})\nabla^\mu \chi^\nu.
   \label{xiprime}
\end{equation}
Since a priori $(\Gamma \indices{^\alpha_{\mu\nu}} - \overline{\Gamma} \indices{^\alpha_{\mu\nu}})$ is first order in the
perturbation, just like $\chi^\alpha$, we could even drop the last term in this equation since it is second order. What this
relation tells us is that we can enforce the harmonic gauge $\xi^{\alpha'} = 0$ by simply solving a linear wave
equation for $\chi^\alpha$ \cite{Headrick10,Figueras11,Adam12,Dias16b}. Once a solution $\chi^\alpha$ has been found, the metric
in the new coordinate system can be reconstructed via \eqref{hapbp}, and is by construction in the harmonic gauge. The linear
problem is thus not of great difficulty as far as the harmonic condition is concerned. However, this does not apply any more to
the non-linear level.

\subsection{Principal part of the Ricci tensor}

In classical electrodynamics, it is well-known that the field equations can be reduced to a wave equation in the Lorentz gauge, in
which the vector potential vanishes, namely $A^\alpha = 0$. In \gls{gr}, the same reduction can be performed with the harmonic gauge. The
main idea is to reduce the principal part of the Ricci tensor, i.e.\ its highest order derivative terms, to a pure wave
operator in curved space-time. This gives rise to an invertible system of equations whose solutions are
unique under an appropriate choice of initial data and boundary conditions \cite{Sarbach12}. In order to grasp how this reduction
is provided by the harmonic gauge, it is instructive to scrutinise the structure of the Ricci tensor.

So we start with the following rewriting of the Riemann tensor,
\begin{equation}
   R_{\alpha\beta\gamma\delta} = \frac{1}{2}(\partial_\beta \partial_\gamma g_{\alpha\delta} + \partial_\alpha \partial_\delta
   g_{\beta\gamma} - \partial_\alpha\partial_\gamma g_{\beta\delta} - \partial_\beta\partial_\delta g_{\alpha\gamma}) +
   g_{\mu\nu}(\Gamma \indices{^\mu_{\alpha\delta}}\Gamma \indices{^\nu_{\beta\gamma}} - \Gamma \indices{^\mu_{\alpha\gamma}}\Gamma
   \indices{^\nu_{\beta\delta}}).
\end{equation}
This equation follows directly from the definition \eqref{defriem}. By contraction of the first and third indices, we obtain an
expression for the Ricci tensor that is
\begin{equation}
   R_{\alpha\beta} = -\frac{1}{2}g^{\mu\nu}(\partial_\mu\partial_\nu g_{\alpha\beta} + \partial_\alpha \partial_\beta g_{\mu\nu} -
   \partial_\alpha \partial_\mu g_{\beta\nu} - \partial_\beta\partial_\mu g_{\alpha\nu}) + g^{\mu\nu}g_{\rho\sigma}(\Gamma
   \indices{^\rho_{\mu\alpha}}\Gamma \indices{^\sigma_{\nu\beta}} - \Gamma \indices{^\rho_{\alpha\beta}} \Gamma
   \indices{^\sigma_{\mu\nu}}).
   \label{ricciparts}
\end{equation}
On this equation, we find that the principal part of the Ricci tensor is
\begin{equation}
   PP[R_{\alpha\beta}] \equiv -\frac{1}{2}g^{\mu\nu}(\underbrace{\partial_\mu\partial_\nu g_{\alpha\beta}}_{\tn{wave operator}} +
   \underbrace{\partial_\alpha \partial_\beta g_{\mu\nu} - \partial_\alpha \partial_\mu g_{\beta\nu} - \partial_\beta\partial_\mu
   g_{\alpha\nu}}_{\tn{ill-posed operator}}).
   \label{PP}
\end{equation}
The first term is nothing but a non-linear wave operator for the metric. On the
other hand, the three other terms are responsible for the ill-posedness of Einstein's system of equation, since they are not
uniquely invertible \cite{Sarbach12}. Moreover, there are an infinity of solutions satisfying $PP[R_{\alpha\beta}] = 0$ \cite{Sarbach12}. This is a direct
consequence of the gauge invariance of \gls{gr} \cite{Dias16b}: this operator cannot be inverted without prescribing the
coordinates.

However, it turns out that these ill-posed terms are precisely encoded in the harmonic gauge vector $\xi^\alpha$ \eqref{xidef}.
Venturing to compute its covariant derivative (defined in \eqref{defcov}), we find
\begin{equation}
   \nabla_\alpha \xi_\beta = \partial_\alpha \xi_{\beta} - \Gamma \indices{^\mu_{\alpha\beta}}\xi_\mu.
\end{equation}
If we rewrite the first term with the help of $\xi^\alpha$ (equation \eqref{xidef}), the Christoffel symbols (equation
\eqref{christoffel}) and trade the derivatives of $g^{\alpha\beta}$ for those of $g_{\alpha\beta}$ with \eqref{commute}, we obtain
\begin{align}
\nonumber   \nabla_\alpha \xi_\beta &= (\Gamma \indices{^\rho_{\mu\nu}} -
   \overline{\Gamma}\indices{^\rho_{\mu\nu}})\partial_\beta(g_{\alpha\rho}g^{\mu\nu}) -g^{\mu\nu}\Gamma
   \indices{^\rho_{\mu\nu}} \partial_\beta g_{\alpha\rho} + g^{\mu\nu} \left(\partial_\beta\partial_\mu g_{\alpha\nu} -
   \frac{1}{2}\partial_\alpha \partial_\beta g_{\mu\nu}\right) \\
   &- g_{\alpha\rho}g^{\mu\nu}\partial_\beta \overline{\Gamma}\indices{^\rho_{\mu\nu}} - \Gamma
   \indices{^\sigma_{\alpha\beta}}g_{\rho\sigma}g^{\mu\nu}(\Gamma
   \indices{^\rho_{\mu\nu}} - \overline{\Gamma}\indices{^\rho_{\mu\nu}}).
\end{align}
Even if not appealing at first sight, this expression needs only to be symmetrised to reveal its connection to the Ricci tensor.
The computation gives
\begin{align}
\nonumber   \nabla_{(\alpha}\xi_{\beta)} &= \frac{1}{2}g^{\mu\nu}\overbrace{(\partial_\mu \partial_\beta g_{\alpha\nu} +
\partial_\alpha\partial_\mu g_{\beta\nu} - \partial_\alpha\partial_\beta g_{\mu\nu})}^{\tn{ill-posed operator}} + (\Gamma \indices{^\rho_{\mu\nu}} - \overline{\Gamma}\indices{^\rho_{\mu\nu}})\partial_{(\alpha}(g_{\beta)\rho}g^{\mu\nu}) \\
                                         & - g^{\mu\nu}\Gamma \indices{^\rho_{\mu\nu}}\partial_{(\alpha} g_{\beta)\rho} -
                                         g^{\mu\nu}g_{\rho(\alpha}\partial_{\beta)} \overline{\Gamma}\indices{^\rho_{\mu\nu}} -
                                         \Gamma \indices{^\sigma_{\alpha\beta}}g_{\rho\sigma}g^{\mu\nu}(\Gamma
                                         \indices{^\rho_{\mu\nu}} - \overline{\Gamma}\indices{^\rho_{\mu\nu}}).
   \label{dvterm}
\end{align}
It turns out that the principal part of this expression (first parenthesis), match
precisely the ill-posed part of the Ricci tensor in \eqref{PP}. Before substituting this result into \eqref{ricciparts}, it is wiser to
reintroduce covariance by trading partial derivatives with covariant derivatives (equation \eqref{defcov}). The wave operator in \eqref{ricciparts} can thus
be rewritten accordingly
\begin{equation}
   \frac{1}{2}g^{\mu\nu}\overline{\nabla}_\mu\overline{\nabla}_\nu g_{\alpha\beta} = \frac{1}{2}g^{\mu\nu}(\partial_\mu
   \partial_\nu g_{\alpha\beta} - g_{\alpha\rho}\partial_\mu \overline{\Gamma}\indices{^\rho_{\beta\nu}} -
   g_{\beta\rho}\partial_\mu \overline{\Gamma}\indices{^\rho_{\alpha\nu}}) + \tn{ terms }\propto \overline{\Gamma}.
   \label{ricciterm}
\end{equation}
The epitome of the demonstration lies in the combination of \eqref{ricciparts}, \eqref{dvterm} and \eqref{ricciterm}. The computation is most easily
done in a frame where\footnote{But beware that
   $\partial_\delta\overline{\Gamma}\indices{^\gamma_{\alpha\beta}} \neq 0$ and contributes to background Riemann tensor terms.}
$\overline{\Gamma}\indices{^\gamma_{\alpha\beta}} = 0$ and $\overline{\nabla} = \partial$. The first order derivative part can
also be simplified by expanding all Christoffel symbols with \eqref{christoffel}. We finally obtain \cite{Andersson03}
\begin{align}
\nonumber   R_{\alpha\beta} &= -\frac{1}{2}g^{\mu\nu} \overline{\nabla}_\mu\overline{\nabla}_\nu g_{\alpha\beta} +
   \nabla_{(\alpha}\xi_{\beta)} + g^{\mu\nu}g_{\rho(\alpha}\overline{R}\indices{^\rho_{\mu\beta)\nu}} \\
   &+ \frac{1}{2} g^{\mu\nu}g^{\rho\sigma}\left(\overline{\nabla}_\alpha
   g_{\mu\rho} \overline{\nabla}_\nu g_{\beta\nu} + \overline{\nabla}_\beta g_{\mu\rho} \overline{\nabla}_\nu g_{\alpha\sigma} -
   \frac{1}{2} \overline{\nabla}_\alpha g_{\mu\rho} \overline{\nabla}_\beta g_{\nu\sigma} + \overline{\nabla}_\mu g_{\alpha\rho}
   \overline{\nabla}_\nu g_{\beta\sigma} - \overline{\nabla}_\mu g_{\alpha\rho} \overline{\nabla}_\sigma g_{\beta\nu}\right).
   \label{riccifinal}
\end{align}
Even if we have performed the computation in a given frame where $\overline{\Gamma}\indices{^\gamma_{\alpha\beta}} = 0$, this
equation is a tensorial equation and is thus valid in any frame. What is remarkable is that the principal part \eqref{PP} of the Ricci tensor is entirely
encoded into the first two terms of the first line, while all the other terms are at most first order derivatives of the metric.
This equation thus makes clear that a sufficient condition to bring the Ricci tensor to a well-posed form is to enforce
$\xi^\alpha = 0$, in which case its principal part reduce to a wave operator. The harmonic gauge is thus nothing but a convenient
way of suppressing the ill-posed part of Einstein's equation.

\subsection{Einstein-De Turck system}
\label{edtsyst}

In numerical relativity, a usual problem consists in evolving in time some initial data. This is the so-called initial-value or Cauchy
problem of definition \ref{cauchydef}. However, the initial data has to be constructed and to solve in turn Einstein's equation,
with suitable boundary conditions. The construction of initial data (or equivalently stationary solutions) is called a
boundary-value problem \cite{Sarbach12}.

At first, the generalised harmonic gauge with source term $\Gamma^\alpha = H^\alpha$ and its declinations
\cite{Garfinkle02,Szilagyi03,Pretorius05a,Pretorius05b,Pretorius06,Sorkin10,Babiuc07,Bona03,Bona04a,Bona04b,Gundlach05,Lindblom06}
where originally dedicated to initial-value problems, i.e.\ for time evolution of initial data. Indeed, Einstein's equation with ill-posed terms
replaced by $\nabla_{(\alpha}H_{\beta)}$ admits a hyperbolic reduction \cite{Friedrich96,Sarbach12}, and can be
formulated as a well-posed Cauchy problem. In particular, this means that initial data satisfying (i) Einstein's equation and (ii) the gauge
condition can be uniquely evolved in time without spoiling (i) or (ii).

More recently, and notably, in the context of the numerical construction of helically symmetric geons in \gls{aads} space-times, the system of
equation has been formulated as a boundary-value problem in a frame that is co-rotating with the solution. We thus legitimately
ask: how can we solve Einstein's equation and fix the gauge simultaneously?

An answer is precisely provided by the
increasingly popular De Turck method. It was employed in
\cite{DeTurck83,Headrick10,Figueras11,Figueras13,Fischetti13,Adam12,Horowitz15,Dias15,Herdeiro16c,Figueras17} exclusively for boundary-value
problems involving stationary solutions. These kinds of solutions were generally obtained with iterative methods such as the
Newton-Raphson algorithm. Originally formulated by De Turck in 1983 \cite{DeTurck83} and resurrected in 2010 by \cite{Headrick10},
this method consists in imposing the harmonic gauge by solving the so-called \gls{edt} system of equations. In this
framework, a term is added to the Einstein tensor so as to make the operator invertible. Once the system is solved, it is checked
a posteriori that the additional term is indeed zero, ensuring that the solution of the \gls{edt} system is also a solution
of Einstein's system.

The De Turck method was successfully used in Kaluza-Klein theory in \cite{Headrick10},
within the \gls{ads}-\gls{cft} correspondence in \cite{Figueras11,Figueras13,Fischetti13}, and in combination with the Ricci flow
method in \cite{Adam12}. Let us also mention that the \gls{aads} multipolar black holes \cite{Herdeiro16c} presented in chapter
\ref{adsinsta} figure \ref{bhmultipole} as well as the first gravitational geons obtained in \cite{Horowitz15} and the black
resonators of \cite{Dias15} all relied on this method. It is extensively reviewed in \cite{Dias16b}.

Suppose we want to solve Einstein's equation \eqref{eineq2} in vacuum
\begin{equation}
   R_{\alpha\beta} = \gls{Lambda}g_{\alpha\beta}.
   \label{einequation}
\end{equation}
Since we know that this system of equations is ill-posed, we consider instead the equation
\begin{equation}
   R_{\alpha\beta} - \nabla_{(\alpha}\xi_{\beta)} = \gls{Lambda}g_{\alpha\beta}.
   \label{edt}
\end{equation}
This is the so-called \gls{edt} system of equations. According to our previous analysis (see equation \eqref{riccifinal}), the
principal part of this equation is merely a wave operator, and is uniquely invertible once suitable boundary conditions are
chosen. This system is not the one of \gls{gr} though: if we find a solution $g_{\alpha\beta}$ solving \eqref{edt}, it is not
necessarily a solution of Einstein's equation \eqref{einequation}, unless this solution exhibits the particular property $\xi^\alpha = 0$. This is
precisely the protocol of the \gls{edt} method: first solve the \gls{edt} equation \eqref{edt}, and second check a posteriori that
this solution is indeed solution of Einstein's equation, i.e.\ satisfies the harmonic condition $\xi^\alpha = 0$. This method is
not guaranteed success though: there exist solutions of the \gls{edt} system that do not solve Einstein's system and exhibit a
non-vanishing $\xi^\alpha$ vector. They are called Ricci solitons.

\subsection{Ricci solitons}
\label{riccisol}

The \gls{edt} system can be reformulated as a wave equation for the vector $\xi^\alpha$. Indeed, let us start from
\eqref{edt}
\begin{subequations}
\begin{align}
   R_{\alpha\beta} - \nabla_{(\alpha}\xi_{\beta)} - \gls{Lambda}g_{\alpha\beta} &= 0,\\
   R - \nabla_\mu \xi^\mu - 4 \gls{Lambda} &= 0.
\end{align}
\end{subequations}
Taking one covariant derivative yields
\begin{subequations}
\begin{align}
   \label{eindt1}
   \nabla^\mu R_{\alpha\mu} - \frac{1}{2}\nabla^\mu\nabla_\alpha\xi_\mu - \frac{1}{2}\nabla^\mu \nabla_\mu \xi_\alpha &= 0,\\
   \label{eindt2}
   \nabla_\alpha R - \nabla_\alpha \nabla_\mu \xi^\mu &= 0.
\end{align}
\end{subequations}
Subtracting half of \eqref{eindt2} to \eqref{eindt1} yields
\begin{equation}
   \nabla^\mu \nabla_\mu \xi_\alpha + R \indices{^\mu_\alpha}\xi_\mu = 0,
   \label{boxxi}
\end{equation}
where we have used the vanishing divergence of the Einstein tensor \eqref{divG} and the Ricci identity \eqref{ricci}. This is the
equivalent of the contracted Bianchi identity \eqref{divG} for the \gls{edt} system. Equation \eqref{boxxi} admits non-trivial
solutions, called Ricci solitons. The corresponding metric, despite being solutions of the \gls{edt} system, do not solve
Einstein's system. It is indeed expected that the space of solutions of the \gls{edt} system is larger than the one of Einstein's system, since
there are ten independent components for the former and only six for the latter. The equivalence between the two is only met for
the restricted set of \gls{edt} solutions displaying a vanishing $\xi^\alpha$.

There is a variety of situations described in \cite{Headrick10,Figueras11,Dias16b,Figueras17} where it can be demonstrated that Ricci
solitons do not exist and that Einstein and \gls{edt} systems are equivalent. However it is not the case in general. Nonetheless,
there are local uniqueness theorems than ensures that solutions with $\xi^\alpha = 0$ are distinguishable from those with
$\xi^\alpha \neq 0$ \cite{Dias16b}. In other words, Ricci solitons cannot be arbitrarily close to Einstein solutions, except for a measure zero
set in moduli space. In practice, this means that Ricci solitons can be ruled out on a case by case basis by verifying if
$\xi^\alpha = 0$ to a sufficient numerical precision. In numerical relativity, it can thus be checked that $\xi^\alpha$ tends to zero
in accordance with increasing numerical resolution. If so, it is clear that the solution is not a Ricci soliton but a true
Einstein solution.

\subsection{$3+1$ decomposition}
\label{3+1harmo}

Numerical \gls{gr} is most often formulated in terms of $3+1$ formalism (see appendix \ref{d+1}). Now that the philosophy of the
De Turck method has been reviewed, it is natural to examine its $3+1$ decomposition. In particular, we are looking for a link
between the 4-dimensional vector $\xi^\alpha$ of equation \eqref{xidef} and its 3-dimensional counterpart
\begin{equation}
   V^i \equiv \gamma^{kl} (\digamma \indices{^i_{kl}} - \overline{\digamma} \indices{^i_{kl}}),
   \label{defV}
\end{equation}
where $\digamma \indices{^i_{kl}}$ denotes the Christoffel symbols of the 3-metric $\gamma_{ij}$ induced on the $t = cst$
hypersurfaces $\Sigma_t$. Recall that in our notation, a bar indicates a background quantity (that is pure \gls{ads} space-time
in our case).

We first consider the gauge vector $\xi^\alpha$ defined in \eqref{xidef}. We can $3+1$ decompose this vector into
\begin{equation}
   \xi^\alpha = \zeta u^\alpha + \chi^\alpha, \quad \tn{with} \quad u_{\alpha}\chi^\alpha = 0,
\end{equation}
where $u^\alpha$ (equation \eqref{ua} and \eqref{ua}) is the unit normal vector to $\Sigma_t$ and $\chi^\alpha$ is the spatial projection of
$\xi^\alpha$. We can then compute $\zeta$ according to $\zeta = -u_\alpha
\xi^\alpha = N\xi^t$, which in light of  \eqref{xidef}, the $3+1$ decomposition of the metric \eqref{metricd+1} and the one of
the Christoffel symbols \eqref{d+1christo}, \eqref{d+1christo2}, gives
\begin{equation}
   \zeta = -K -\frac{1}{N^2}\mathcal{L}_m N - \frac{2}{N}\beta^i\partial_i \ln \overline{N},
   \label{zeta}
\end{equation}
where $N$ is the lapse function, $\beta^i$ is the shift vector and $K$ is the mean extrinsic curvature of $\Sigma_t$. As for the
spatial components of $\chi^\alpha$ it comes $\chi^i = \xi^i - \zeta u^i = \xi^i + \beta^i \xi^t$, or equivalently
\begin{align}
   \nonumber   \chi^i &= V^i - \frac{1}{N^2}\mathcal{L}_m\beta^i - \gamma^{ij}\partial_j \ln N +
   \frac{\beta^j\beta^k}{N^2}\overline{\digamma}\indices{^i_{jk}} + \frac{1}{N^2}(\overline{N}\overline{\gamma}^{ij}\partial_j \overline{N} + \overline{\beta}^j\overline{D}_j \overline{\beta}^i)\\
&- \frac{2\beta^j}{N^2}((\beta^j- \overline{\beta}^j)\partial_j\ln \overline{N} + \overline{D}_j \overline{\beta}^i ),
\label{chii}
\end{align}
where $D$ is the covariant derivative associated to the 3-metric $\gamma_{ij}$ and $\mathcal{L}_m = \partial_t -
\mathcal{L}_{\beta}$ the evolution operator. What is remarkable in \eqref{zeta} and
\eqref{chii} is that
\begin{equation}
   (\zeta,\chi^i) = (-K,V^i) + \tn{terms in } \partial N, \partial \beta^i.
   \label{remarkable}
\end{equation}
Namely, since $N$ and $\beta^i$ are non-dynamical quantities, the $3+1$ decomposition of the 4-dimensional vector $\xi^\alpha$
catches essentially the mean extrinsic curvature $K$ of
$\Sigma_t$ as well as its 3-dimensional counterpart $V^i$. In particular, $V^i$ encodes the ill-posed part of the 3-dimensional Ricci
tensor $\mathcal{R}_{ij}$ in much the same way as $\xi^\alpha$ does in four dimensions (equation \eqref{riccifinal}), namely
\begin{align}
\nonumber   \mathcal{R}_{ij} &= -\frac{1}{2}\gamma^{kl} \overline{D}_k\overline{D}_l \gamma_{ij} + D_{(i}\xi_{j)} + \gamma^{kl}\gamma_{m(i}\overline{\mathcal{R}}\indices{^m_{kj)l}} \\
   &+ \frac{1}{2} \gamma^{kl}\gamma^{mn}\left(\overline{D}_i \gamma_{km} \overline{D}_l \gamma_{jl} + \overline{D}_j \gamma_{km}
   \overline{D}_l \gamma_{in} - \frac{1}{2} \overline{D}_i \gamma_{km} \overline{D}_j \gamma_{ln} + \overline{D}_k \gamma_{im}
   \overline{D}_l \gamma_{jn} - \overline{D}_k \gamma_{im} \overline{D}_n \gamma_{jl}\right).
   \label{riccifinal3d}
\end{align}
The missing degree of freedom between the three of $V^i$ and the four of $\xi^\alpha$ is encoded in $K$. Thus, in terms of 3+1
formalism, the harmonic gauge $\xi^\alpha = 0$ is strikingly close to a combined maximal slicing $K = 0$ and
spatial harmonic $V^i = 0$ gauge, even if not equivalent.

If now we come back to the \gls{edt} equation \eqref{edt}, we can $3+1$ decompose in turn the De Turck term. Namely, we can compute
the various projections of $\nabla_{(\alpha} \xi_{\beta)}$ as follows \cite{Martinon17}
\begin{subequations}
\begin{align}
   \gamma^\alpha_{\mu}\gamma^\beta_{\nu}\nabla_{(\alpha}\xi_{\beta)} &= -\zeta K_{\mu\nu} + D_{(\mu}\chi_{\nu)},\\
   \gamma^\alpha_{\mu}u^\beta\nabla_{(\alpha}\xi_{\beta)} &= -\frac{1}{2}D_{\mu}\zeta + \frac{1}{2}\zeta D_{\mu}\ln N + K_{\mu\nu}\chi^\nu + \frac{1}{2N}\mathcal{L}_m\chi_\mu,\\
   u^\alpha u^\beta\nabla_{(\alpha}\xi_{\beta)} &= -\frac{1}{N}\mathcal{L}_m \zeta - \chi^\mu D_\mu \ln N,\\
   \nabla_{\alpha} \xi^\alpha &= \frac{1}{N}\mathcal{L}_m \zeta - \zeta K + D_{\mu}\chi^\mu + \chi^\mu D_{\mu}\ln N.
\end{align}
\end{subequations}
These results can be substituted into the projections of the \gls{edt} system \eqref{edt} in much the same way as appendix
\ref{d+1} section \ref{d+1ein}. This gives
\begin{align}
   \label{hamDT}
   \mathcal{R}_{ij} + K^2 - K_{ij}K^{ij} - 2\gls{Lambda} + \frac{1}{N}\mathcal{L}_m \zeta + \zeta K - D_i \chi^i + \chi^i D_i \ln N = 0,\\
   D_j K^j_i - D_i K - \frac{1}{2}D_i \zeta + \frac{1}{2}\zeta D_i \ln N + K_{ij}\chi^j + \frac{1}{2N}\mathcal{L}_m \chi_i = 0, \\
\label{evoDT}
\mathcal{L}_m K_{ij} = -D_iD_jN + N( \mathcal{R}_{ij} + K K_{ij} - 2 K_{ij}K^k_j - \gls{Lambda} \gamma_{ij} + \zeta K_{ij} - D_{(i}\chi_{j)}) = 0,
\end{align}
where we recognise the Hamiltonian and momentum constraints as well as the evolution equation, with corrections due to the De
Turck term. Combining the trace of $\eqref{evoDT}$ with the evolution equation \eqref{evolgamma} and the Hamiltonian constraint \eqref{hamDT}, we also get
\begin{equation}
   \mathcal{L}_m K = - D_i D^i N + N (K_{ij}K^{ij} - \chi^i D_i \ln N - \gls{Lambda}) - \mathcal{L}_m \zeta.
\end{equation}
Paying a careful attention to the Hamiltonian constraint \eqref{hamDT} and to \eqref{remarkable}, we notice that compared to the
usual constraint \eqref{ham}, the term in $K^2$ vanishes and the term $-D_i V^i$ is added. Similarly, comparing the evolution equation
\eqref{evoDT} with \eqref{lmKij1}, the term $NKK_{ij}$ vanishes while a term $-D_{(i}V_{j)}$ is added. These features are highly reminiscent of the
so-called \gls{eam} formulation that we describe below. Even if it went unnoticed in the literature, it thus seems that the 3+1
decomposition of the \gls{edt} system is very close to the \gls{eam} system \cite{Martinon17}.

\section{Andersson-Moncrief gauge}

The \gls{am} gauge was described in 2003 by Andersson and Moncrief \cite{Andersson03}. The idea is to enforce
\begin{equation}
   K = 0 \quad \tn{and} \quad V^i = 0.
   \label{amgauge}
\end{equation}
In this regard, this is very close (but not equivalent) to the standard 4-dimensional harmonic gauge. The first condition is
the so-called maximal slicing condition \cite{Gourgoulhon07}, while the second one is nothing but a spatial
harmonic gauge condition. This is why the \gls{am} gauge is sometimes referred as a combined maximal slicing-spatial harmonic gauge.

It was originally designed for initial-value problems. In particular, the authors of
\cite{Andersson03} proved that the so-called \gls{eam} system (presented in section \ref{eamsystem}) was strongly well-posed. If initial data
satisfying the \gls{am} gauge \eqref{amgauge} as well as the Hamiltonian and momentum constraints is evolved with the \gls{eam}
system of equations, then both the gauge and the constraints are preserved during evolution, ensuring that the data remains
solution of Einstein's equation at all times. However, with the tremendous successes of the generalised harmonic and
\gls{bssn} formulations \cite{Sperhake15}, the \gls{eam} formulation went unnoticed in numerical relativity. On theoretical grounds it has as
many good properties as the generalised harmonic and \gls{bssn} formulations. It provides indeed a well-posed hyperbolic reduction
of the $3+1$ Einstein's equation. We aim at illustrating these arguments in the subsequent sections.

Even if to the best of our knowledge, the \gls{am} gauge was never used for boundary-value problems, we succeeded in
\cite{Martinon17} to obtain helically symmetric geons within this gauge, by simply adapting the De Turck method.

\subsection{Linear level}

Since we have already seen in chapter \ref{perturbations} how to build linear geons, it is natural to ask how to bring linear solutions to
the \gls{am} gauge and make them satisfy \eqref{amgauge}. Since there are two distinct conditions to satisfy in \eqref{amgauge},
we proceed in two steps.

On the one hand, we perform a change of the time coordinate alone (i.e.\ a change of foliation)
\begin{equation}
   t' = t + \alpha(x^i),
\end{equation}
where $\alpha$ is a function of the spatial coordinates only, the latter being unchanged. Since we are interested in linear
solutions, we assume that $\alpha$ is of first order in amplitude. The transformation rules \eqref{Ttrans} for $g_{\alpha\beta}$ combined
with the $3+1$ matrix representation \eqref{metricd+1} yield, at first order in $\alpha$
\begin{subequations}
\begin{align}
   N' &= N (1 + \beta^i\partial_i \alpha),\\
   \beta^{i'} &= \beta^i + (N^2 \gamma^{ij} + \beta^i\beta^j)\partial_j\alpha, \\
   \beta_{i'} &= \beta_i + (N^2 - \beta_k\beta^k)\partial_i\alpha, \\
   \gamma_{i'j'} &= \gamma_{ij} - \beta_i \partial_j\alpha - \beta_j \partial_i\alpha,\\
   \gamma^{i'j'} &= \gamma^{ij} + (\gamma^{ik}\beta^j + \gamma^{jk}\beta^i)\partial_k \alpha,
\end{align}
\label{primealpha}%
\end{subequations}
where primes indicates geometrical quantities in the new coordinate system. The trace of the evolution equation \eqref{evolgamma}
yields
\begin{equation}
   -2NK = \beta^k \gamma^{ij}\partial_k \gamma_{ij} + 2 \partial_k \beta^k.
   \label{2NK}
\end{equation}
Evaluating this expression in the new coordinate system with \eqref{primealpha} gives rise to
\begin{equation}
   2N'K' =  2NK(1 + \beta^iD_i\alpha) - 2N^2 D^iD_i \alpha - 4N D^iN D_i\alpha.
   \label{eqK}
\end{equation}
If the background geometry exhibits a vanishing extrinsic curvature tensor (this is the case e.g.\ of \gls{ads} space-time), then $K$ is first order in the
metric perturbation and the term $2NKD_i\alpha$, being
second order, can be dropped out at the linear level. Enforcing $K' = 0$ gives nothing but an elliptic equation for the first
order function $\alpha$. Such an equation admits a unique solution provided suitable boundary conditions are chosen for
$\alpha$. Once a solution $\alpha$ has been found, the metric in the new coordinate system can be reconstructed via
\eqref{primealpha}, and is by construction in the maximal slicing gauge.

On the other hand, we perform a coordinate transformation of the spatial coordinates
\begin{equation}
   x^{i'} = x^i + \chi^i(x^j),
   \label{xp}
\end{equation}
where the time is left unchanged and the first order vector $\chi^i$ is independent of the time coordinate.
The transformation rules \eqref{Ttrans} for $g_{\alpha\beta}$ combined with the $3+1$ matrix representation \eqref{metricd+1} yield,
at first order in $\chi^i$
\begin{subequations}
\begin{align}
   N'& = N - \mathcal{L}_{\chi} N,\\
   \beta^{i'} &= \beta^i - \mathcal{L}_{\chi}\beta^i,\\
   \beta_{i'} &= \beta_i - \mathcal{L}_{\chi}\beta_i,\\
   \gamma_{i'j'} &= \gamma_{ij} - \mathcal{L}_{\chi}\gamma_{ij},\\
   \gamma^{i'j'} &= \gamma^{ij} - \mathcal{L}_{\chi}\gamma^{ij},\\
   \Gamma\indices{^{i'}_{k'l'}} &= \Gamma\indices{^i_{kl}} - \mathcal{L}_\chi \Gamma\indices{^i_{kl}} - \partial_k \partial_l \chi^i,
\end{align}
\label{primechi}%
\end{subequations}
where all quantities are evaluated at $x'$, and $\mathcal{L}_{\chi}$ stands for the Lie derivative along $\chi$. Since $K$ is a scalar field for
the 3-dimensional space-time, it obeys $K'(x') = K(x)$. Thus the vanishing of the mean extrinsic curvature $K$ is preserved under
\eqref{xp}. Now repeating the arguments already discussed for the harmonic gauge in section \ref{linharmo}, we obtain the
3-dimensional counterpart of equation \eqref{xiprime}, namely
\begin{equation}
   V^{i'} = V^i - D^jD_j \chi^i - \mathcal{R} \indices{^i_j}\chi^j + 2 (\digamma \indices{^i_{jk}} - \overline{\digamma} \indices{^i_{jk}})D^j \chi^k.
   \label{eqVV}
\end{equation}
Rigorously speaking, the last term in this equation is second order and can be dropped out.
Enforcing $V^{i'} = 0$ thus amounts to solving an elliptic equation for the first order
vector $\chi^i$, that admits a unique solution under suitable boundary conditions. The metric in the new coordinates
system can then be reconstructed with \eqref{primechi}, and is by construction in the spatial harmonic gauge.

Finally, with these two successive coordinate transformations, any first order Einstein's solution, in particular the linear geons of
chapter \ref{perturbations}, can be brought in the \gls{am} gauge by simply solving two linear and elliptic equations.

\subsection{Einstein-Andersson-Moncrief system}
\label{eamsystem}

Since we want to impose $K=0$ and $V^i = 0$, (equation \eqref{amgauge}), it is natural to determine how $K$ and $V^i$ evolve in
time. This is of crucial importance in initial-value problems. For this purpose, we recall some results of the $3+1$
decomposition, among them the evolution equation of the metric \eqref{evolgamma}, which is equivalent to
\begin{subequations}
\begin{align}
   \dot{\gamma}_{ij} &= -2NK_{ij} + D_i\beta_j + D_j \beta_i,\\
   \dot{\gamma}^{ij} &= 2NK^{ij} - D^i\beta^j - D^j \beta^i,
\end{align}
   \label{dotgamma}%
\end{subequations}
where a dot indicates partial derivative with respect to time. We also recall the definition of the evolution operator (section
\ref{evolop} of appendix \ref{d+1})
\begin{equation}
   \mathcal{L}_m = \partial_t - \mathcal{L}_\beta,
   \label{deflm}
\end{equation}
where $\mathcal{L}$ denotes the Lie derivative and $\beta^i$ the shift vector.

In order to compute $\mathcal{L}_m V^i$, we start by examining its time derivative. Namely, from its definition \eqref{defV}, it comes
\begin{equation}
   \dot{V}^i = \dot{\gamma}^{kl}(\digamma \indices{^i_{kl}} - \overline{\digamma}\indices{^i_{kl}}) + \gamma^{kl}\dot{\digamma}\indices{^i_{kl}},
\end{equation}
where we have assumed the background metric to be time-independent. With the variations of Christoffel's symbols \eqref{varGamma}, the
time derivative of the metric \eqref{dotgamma} and the Ricci identity \eqref{ricci}, this expression can be rewritten in
\begin{equation}
   \dot{V}^i = (2NK^{kl} - 2D^k\beta^l)(\digamma \indices{^i_{kl}} - \overline{\digamma}\indices{^i_{kl}}) + D_j D^j \beta^i +
   \mathcal{R}^{ij}\beta_j + D^i(NK) - 2D_j (NK^{ij}).
   \label{dotV}
\end{equation}
Since we already have an evolution equation for $K$ (see appendix \ref{d+1} equation \eqref{lmK}), we thus have recovered
\cite{Andersson03}
\begin{subequations}
\begin{align}
   \mathcal{L}_m K &= -D_i D^i N + NK_{ij}K^{ij} -N \gls{Lambda},\\
\nonumber   \mathcal{L}_m V^i &= 2(NK^{kl} - D^k\beta^l)(\digamma \indices{^i_{kl}} - \overline{\digamma}\indices{^i_{kl}}) + D_j D^j \beta^i + \mathcal{R}^{ij}\beta_j \\
\label{lmV}
   &+ D^i (NK) - 2D_j (NK^{ij}) - \mathcal{L}_\beta V^i.
\end{align}
\end{subequations}
These are the sought-after evolution equations for $K$ and $V^i$.

Now, let us recall the standard $3+1$ system of equations. It reads (see appendix \ref{d+1}, equations \eqref{evolgamma},\eqref{traceevol},
\eqref{mom}, \eqref{lmKij1})
\begin{subequations}
\begin{align}
   \label{ein3+1evol}
   \mathcal{L}_m \gamma_{ij} + 2NK_{ij} &= 0,\\
   \label{ein3+1ham}
   -D_i D^i N + NK_{ij}K^{ij} - N \gls{Lambda} - \mathcal{L}_m K &= 0,\\
   \label{ein3+1mom}
   D_j K \indices{^j_i} - D_iK &= 0,\\
   \label{ein3+1evo}
   \mathcal{L}_m K_{ij} + D_iD_jN - N(\mathcal{R}_{ij} + KK_{ij} - 2K_{ik}K \indices{^k_j} - \gls{Lambda}\gamma_{ij}) &= 0.
\end{align}
\label{ein3+1}%
\end{subequations}
Of course, this system is ill-posed since the gauge freedom is not addressed. In \cite{Andersson03}, the authors proposed to solve
instead the so-called \gls{eam} system of equations, namely
\begin{subequations}
\begin{align}
   \label{systgamma}
   \mathcal{L}_m \gamma_{ij} + 2NK_{ij} &= 0,\\
   \label{systK}
   -D_i D^i N + NK_{ij}K^{ij} -N \gls{Lambda} &= 0,\\
   \label{systV}
   D_j D^j \beta^i + \mathcal{R}^{ij}\beta_j - 2K^{ij}D_j N + 2(NK^{kl} - D^k\beta^l)(\digamma \indices{^i_{kl}} - \overline{\digamma}\indices{^i_{kl}}) - \mathcal{L}_\beta V^i&= 0,\\
   \label{systE}
   \mathcal{L}_m K_{ij} + D_iD_jN - N(\mathcal{R}_{ij} - D_{(i}V_{j)} - 2K_{ik}K \indices{^k_j} - \gls{Lambda}\gamma_{ij}) &= 0.
\end{align}
\label{syst}%
\end{subequations}
The system \eqref{syst} is not the $3+1$ Einstein's system \eqref{ein3+1}. However, it is equivalent to it provided both the gauge
\eqref{amgauge} ($K = 0$, $V^i = 0$) and the momentum constraint \eqref{ein3+1mom} ($D_j K\indices{^i_j} = D^i K = 0$) are satisfied.
Indeed, equations \eqref{ein3+1evol} and \eqref{systgamma} are identical, equations \eqref{ein3+1ham} and \eqref{systK} are
equivalent when $K=0$, equation \eqref{systV} is a rewriting of \eqref{lmV} in combination with \eqref{ein3+1mom} and $K=0$, and
equation \eqref{systE} is equivalent to \eqref{ein3+1evo} when $K=0$ and $V^i = 0$.

The upper indices counterpart of \eqref{systgamma} is
\begin{equation}
   \mathcal{L}_m \gamma^{ij} - 2NK^{ij} = 0.
   \label{uppersystgamma}
\end{equation}
Combined with \eqref{systE}, this equation gives the alternative formulations of the \gls{eam} evolution equation
\begin{subequations}
\begin{align}
   \mathcal{L}_m K_{ij} + D_iD_jN - N(\mathcal{R}_{ij} - D_{(i}V_{j)} - 2K_{ij}K \indices{^k_j} - \gls{Lambda}\gamma_{ij}) &= 0,\\
   \mathcal{L}_m K^{ij} + D^i D^j N - N (\mathcal{R}^{ij} - D^{(i}V^{j)} + 2 K^{ik} K \indices{_k^j} - \gls{Lambda}\gamma^{ij}) &= 0,\\
   \label{evolKijud}
   \mathcal{L}_m K \indices{^i_j} + D^i D_j N - N\left(\mathcal{R}\indices{^i_j} - \frac{1}{2}D^iV_j - \frac{1}{2}D_j V^i - \gls{Lambda}\delta\indices{^i_j}\right) &=0.
\end{align}
\label{evolKall}
\end{subequations}

In \cite{Andersson03}, it was shown that it \textit{suffices} to build initial data satisfying the gauge condition and Einstein's
constraints to get the whole solution in gauge in initial-value problems. Said differently, if $K = 0$, $V^i = 0$ and the
constraints are satisfied at $t = 0$, then they remain zero during the whole evolution under \eqref{syst}. This essential property
is a direct consequence of the equations of propagation for the constraints.

\subsection{Propagation of constraints}

Any initial-value problem within the \gls{am} gauge has to fulfil the gauge conditions $K = 0$, $V^i = 0$ as well as the
Hamiltonian and momentum constraints $H = 0$ and $M_i = 0$, where we have defined (equations \eqref{ham}, \eqref{mom})
\begin{subequations}
\begin{align}
   \label{Hdef}
   H &\equiv \mathcal{R} - D_i V^i + K^2 - K_{ij}K^{ij} - 2 \gls{Lambda},\\
   \label{Mdef}
   M_i &\equiv D_i K - D_j K \indices{^j_i}.
\end{align}
\end{subequations}
Our goal is to get evolution equations for $K$, $V^i$, $H$, $M_i$,
assuming that the \gls{eam} system \eqref{syst} is satisfied at all times. In particular, we would like to compute $\mathcal{L}_m
K$, $\mathcal{L}_m V^i$, $\mathcal{L}_m H$, $\mathcal{L}_m M_i$. To get these evolution equations, we start by examining all the
time derivatives of the constraints. It turns out that all these calculations naturally reveal $\mathcal{L}_\beta$ lie
derivatives, leading to the sought-after equations (according to \eqref{deflm}).

Before addressing these equations, we need to derive some preliminary results. From the variations of the Christoffel symbols
\eqref{varGamma} and the time derivative of the 3-metric \eqref{dotgamma}, we get
\begin{align}
\nonumber   \dot{\digamma}\indices{^i_{kl}} &= -K \indices{^i_l}D_k N - K \indices{^i_k}D_l N + K_{kl}D^i N - N D_k K \indices{^i_l} - N D_l
   K \indices{^i_k} + N D^i K_{kl}\\
   &- \frac{1}{2}\gamma^{ij}\mathcal{R}\indices{^m_{lkj}}\beta_m -
   \frac{1}{2}\gamma^{ij}\mathcal{R}\indices{^m_{klj}}\beta_m + D_{(k}D_{l)}\beta^i.
   \label{dotdigamma1}
\end{align}
Contracting with $\gamma^{kl}$ and using the symmetries of the Riemann tensor \eqref{symmetries} yields
\begin{equation}
   \gamma^{kl}\dot{\digamma}\indices{^i_{kl}} = -2 K^{ij}D_j N - 2N D_j K^{ij} + D^i (NK) + \mathcal{R}_{ij}\beta^j + D_k D^k \beta^i,
\end{equation}
as well as
\begin{equation}
   \dot{\digamma}\indices{^j_{ij}} = - D_i (NK) - \frac{1}{2}\mathcal{R}_{ij}\beta^j + \frac{1}{2}D_i D_j \beta^j +
   \frac{1}{2}D_j D_i \beta^j.
   \label{dotdigamma2}
\end{equation}
For the Ricci scalar, with the help of the variations \eqref{dR}, the evolution of the 3-metric \eqref{dotgamma}, the Ricci identity \eqref{ricci},
and the zero-divergence of the 3-dimensional Einstein tensor \eqref{divG}, we obtain
\begin{equation}
   \dot{\mathcal{R}} = \beta^i D_i \mathcal{R} + 2D_iD^i (NK) - 2D^iD^j(NK_{ij}) + 2N \mathcal{R}_{ij}K^{ij}.
   \label{dotR}
\end{equation}

Fortified by these results, we can now focus on the evolution of the mean extrinsic curvature $K$. Since we assume that the \gls{eam}
system is solved at any time, we can use the trace of the evolution equation \eqref{evolKijud}, which gives
\begin{equation}
   \mathcal{L}_m K = - D_i D^i N + N(\mathcal{R} - D_i V^i + K^2 - 3 \gls{Lambda}).
\end{equation}
Since \eqref{systK} is supposed satisfied too, it comes with the definition \eqref{Hdef}
\begin{equation}
   \mathcal{L}_m K = NH.
   \label{lmKfinal}
\end{equation}
This is how $K$ propagates under the \gls{eam} system \eqref{syst}.

Now we address the evolution of $V^i$. We thus start from the general result \eqref{lmV} and, since we assume that equation
\eqref{systV} is satisfied, we get
\begin{equation}
   \mathcal{L}_m V^i = D^i(NK) - 2N D_j K^{ij}.
   \label{lmVfinal}
\end{equation}
This is how $V^i$ propagates under the \gls{eam} system \eqref{syst}.

In order to obtain an evolution equation for the constraint $H$ (equation \eqref{Hdef}), we need to address the evolution of its constitutive parts. For
its first term, we have already shown with our preliminary result \eqref{dotR} that
\begin{equation}
   \mathcal{L}_m \mathcal{R} = 2 D_i D^i (NK) - 2 D^i D^j (N K_{ij}) + 2 N \mathcal{R}_{ij}K^{ij}.
   \label{lmR}
\end{equation}
For the $K^2$ term, it follows directly from the evolution \eqref{lmKfinal} of $K$ that
\begin{equation}
   \mathcal{L}_m K^2 = 2NKH.
   \label{lmK2}
\end{equation}
For $K_{ij}K^{ij}$, we infer directly from \eqref{evolKall} that
\begin{equation}
   \mathcal{L}_m (K_{ij}K^{ij}) = -2 K^{ij}D_i D_j N + 2N(K_{ij}\mathcal{R}^{ij} - K^{ij}D_i V_j + KK_{ij}K^{ij} - \gls{Lambda}K).
   \label{lmKijKij}
\end{equation}
As for $D_i V^i$, we have
\begin{equation}
   \partial_t (D_i V^i) = D_i \dot{V}^i + \dot{\digamma}\indices{^i_{ij}}V^j.
\end{equation}
Using our preliminary results \eqref{dotdigamma2}, \eqref{dotV} and with the Ricci identity \eqref{ricci}, it comes
\begin{equation}
   \mathcal{L}_m D_i V^i = D_i D^i (NK) - 2 D_i N D_j K^{ij} - 2 ND_i D_j K^{ij} - V^i D_i (NK).
   \label{lmDiVi}
\end{equation}
Finally, combining \eqref{Hdef}, \eqref{lmR}, \eqref{lmK2}, \eqref{lmKijKij} and \eqref{lmDiVi}, we get the following propagation equation
\begin{align}
\nonumber   \mathcal{L}_m H &= D_i D^i (NK) - 2 D_i N D_j K^{ij} - 2 N K K_{ij} K^{ij} + 2NK^{ij}D_i V_j\\
& + V^i D_i (NK) + 2NK(H+2 \gls{Lambda}).
\label{lmHfinal}
\end{align}
This is how the Hamiltonian constraint $H$ propagates under the \gls{eam} system \eqref{syst}.

At this point, only remains the case of the momentum constraints. Its first term is $D_i K$, whose time derivative is
\begin{equation}
   \partial_t D_i K = D_i \dot{K}.
\end{equation}
Hence from the evolution of $K$ \eqref{lmKfinal}
\begin{equation}
   \mathcal{L}_m D_i K = D_i \beta^j D_j K + H D_i N + N( D_i \mathcal{R} - 2 K_{jk}D_i K^{jk} + 2K D_iK - D_i D_j V^j ) .
   \label{lmDiK}
\end{equation}
On the other hand, the term $D_j K \indices{^j_i}$ obeys
\begin{equation}
   \partial_t D_j K \indices{^j_i} = D_j \dot{K}\indices{^j_i} + \dot{\digamma}\indices{^j_{kj}} K \indices{^k_i} -
   \dot{\digamma}\indices{^k_{ij}} K \indices{^j_k}.
\end{equation}
With the evolution equation \eqref{evolKijud}, the time derivatives of the Christoffel symbols \eqref{dotdigamma1},
\eqref{dotdigamma2}, the divergence of the 3-dimensional Einstein tensor \eqref{divG}, the Hamiltonian constraint of the
\gls{eam} system \eqref{systK} as well as the Ricci identity \eqref{ricci} and the symmetries of the Riemann tensor
\eqref{symmetries}, a slightly involved computation shows
\begin{equation}
   \mathcal{L}_m D_j K \indices{^j_i} = - D_{(i}V_{j)}D^j N - N\left(\frac{1}{2}D_i \mathcal{R} + K_{jk}D_i K^{jk} + KD_j K \indices{^j_i} - \frac{1}{2}D_j D_i V_j - \frac{1}{2} D_j D^j V_i \right).
   \label{lmDjKji}
\end{equation}
Finally, combining \eqref{lmDiK} and \eqref{lmDjKji} with the Ricci identity \eqref{ricci}, we have recovered that
\begin{align}
   \nonumber   \mathcal{L}_m M_i &= N\left(\frac{1}{2}D_i \mathcal{R} - K_{jk}D_i K^{jk} + 2K D_i K - K D_j K \indices{^j_i} - \frac{1}{2}D_i D_j V^j + \frac{1}{2}D_j D^j V_i + \frac{1}{2} \mathcal{R}_{ij} V^j\right)\\
                                 &+ D_i \beta^j D_j K + HD_i N + D_{(i} V_{j)} D^j N.
   \label{lmMfinal}
\end{align}
This is how the momentum constraint $M_i$ propagates under the \gls{eam} system \eqref{syst}.

Thus, under the system \eqref{syst}, equations \eqref{lmKfinal}, \eqref{lmVfinal}, \eqref{lmHfinal} and \eqref{lmMfinal} show that
the constraints propagate as follows
\begin{subequations}
\begin{align}
   \mathcal{L}_m K &= NH,\\
   \mathcal{L}_m V^i &= K D^i N + N (M^i - D_j K \indices{^j_i}),\\
\nonumber   \mathcal{L}_m H &= 2NK(H - K_{ij}K^{ij} + 2 \gls{Lambda}) + D_i D^i (NK) - 2 D_i N D_j K^{ij}\\
   & + 2NK^{ij}D_i V_j + V^i D_i (NK),\\
   \nonumber   \mathcal{L}_m M_i &= H D_i N + NK(M_i + D_i K) + \frac{N}{2}D_i (H + K^2) + D_i \beta^j D_j K\\
\label{lmMi}
   &+ D_{(i} V_{j)} D^j N + \frac{N}{2} (D_j D^j V_i + \mathcal{R}_{ij} V^j).
\end{align}
\end{subequations}
As a consequence, if we start with initial data satisfying the constraints and the gauge, namely $H = 0$, $K = 0$, $M_i = 0$, $V^i = 0$ at $t =
0$ (which is equivalent to provide an initial data solution of Einstein's system in the \gls{am} gauge), the equations above ensure that this
remains so during all the evolution. Furthermore, since the \gls{eam} system is strongly well-posed, this ensures that there are
no constraint violating modes that could lead to a blow up of numerical errors in numerical simulations.

\subsection{3+1 De Turck method}

So far, we have only discussed the \gls{eam} system as an initial-value problem, since this was in this context that it was
studied originally \cite{Andersson03}. Regarding boundary-value problems, the De Turck method was adapted to the \gls{am} gauge
for the first time in \cite{Martinon17}. In this paper, we were able to build helically symmetric geons with unprecedented
precision and high amplitudes. Our results and boundary conditions are discussed in chapter \ref{simulations}.

For now, we focus on the corresponding \gls{eam} system of equation. Recall the standard $3+1$ system of equations. It reads (see
appendix \ref{d+1} equations \eqref{evolgamma}, \eqref{ham}, \eqref{mom}, \eqref{lmKij1})
\begin{subequations}
\begin{align}
   \mathcal{L}_m \gamma_{ij} + 2NK_{ij} &= 0,\\
   \mathcal{R} + K^2 - K_{ij}K^{ij} - 2 \gls{Lambda} &= 0,\\
   D_j K \indices{^j_i} - D_iK &= 0,\\
   \mathcal{L}_m K_{ij} + D_iD_jN - N(\mathcal{R}_{ij} + KK_{ij} - 2K_{ik}K \indices{^k_j} - \gls{Lambda}\gamma_{ij}) &= 0.
\end{align}
\label{3+1syst}%
\end{subequations}
This system is not invertible. However, we know that our solution must satisfy $K = 0$ and $V^i = 0$. As is customary in
numerical relativity with maximal slicing foliations, we thus suppress all occurrences of $K$ in \eqref{3+1syst}. In addition,
since we know that the Ricci tensor is an ill-posed operator for the metric (equation \eqref{riccifinal3d}), we suppress also the
ill-posed term by replacing all occurrences of $R_{ij}$ by $R_{ij} - D_{(i}V_{j)}$, leading to
\begin{subequations}
\begin{align}
   \mathcal{L}_m \gamma_{ij} + 2NK_{ij} &= 0,\\
   \mathcal{R} - D_i V^i - K_{ij}K^{ij} - 2 \gls{Lambda} &= 0,\\
   D_j K \indices{^j_i} &= 0,\\
   \mathcal{L}_m K_{ij} + D_iD_jN - N(\mathcal{R}_{ij} - D_{(i}V_{j)} - 2K_{ik}K \indices{^k_j} - \gls{Lambda}\gamma_{ij}) &= 0.
\end{align}
\label{3+1syst2}%
\end{subequations}
By construction, this system of equation is invertible as a boundary-value problem. This is the system
of equation that was inverted in \cite{Martinon17}. This kind of arguments is highly reminiscent of the ones used in the Dirac
gauge which involves a conformal decomposition of the spatial metric \cite{Bonazzola04}.

However, just like the 4-dimensional harmonic gauge, it is expected that solutions of the system \eqref{3+1syst2} form a larger
class of solutions than that of \eqref{3+1syst}, including some kinds of Ricci solitons. One thus needs to check, a posteriori, that the
obtained solution satisfy the gauge conditions $K=0$ and $V^i=0$. In this regard, this is an adapted De Turck method, that turns
out to be successful in the construction of helically symmetric geons \cite{Martinon17}.

\section{Gauge freedom and geons}

In the context of geons construction, helically symmetric ones have the characteristic feature of being stationary in a co-rotating frame.
They are thus solutions of a boundary-value problem in \gls{aads} space-time. The first construction of geons \cite{Horowitz15}, dealing with the
particular $(l,m,n) = (2,2,0)$ case, solved this problem via the standard De Turck method. In \cite{Martinon17} however, motivated by a 3+1
formulation and the well-posedness of the \gls{eam} system, we have constructed several families of helically symmetric geons
within the \gls{am} gauge. Our technique adapted the 4-dimensional De Turck method to the $3+1$ formalism, borrowing ideas from
the maximal slicing and the Dirac gauge. The resulting procedure is thus half-way between a De Turck method and the \gls{eam}
procedure that was originally designed for an initial-value problem. In particular, we still have to check a posteriori that the
gauge conditions $K = 0$, $V^i = 0$ are satisfied. Of course, any boundary-value problem, as its name indicates, has to be
supplied with correct boundary conditions, i.e.\ boundary conditions that preserve the good asymptotics. As discussed in detail in
chapter \ref{aads}, there are several criteria that have to be met to preserve the \gls{aads} character of space-time. We return
to this point in chapter \ref{simulations}.

Nonetheless, the material of this chapter makes straightforward the transition between the boundary-value problem of the
construction of geon and the initial-value problem of their evolution. The time evolution of geons appears as a very interesting but challenging problem. Recall that, even if there
are some indications of their non-linear stability (see chapter \ref{adsinsta} section \ref{beyondspher}), no definitive numerical
of this statement proof was ever given. This is because numerical evolution in \gls{aads} space-times with only one Killing vector
is a daunting task and is quite a stammering research area for now. However, the expertise that was developed in asymptotically
flat space-times for astrophysical purposes are very likely to be reconcilable with \gls{ads} space-time. In particular, it is
very encouraging to observe that Bantilan and Pretorius \cite{Bantilan12,Bantilan15,Bantilan17} employed with great success the generalised
harmonic scheme to evolve in time axially symmetric black holes collisions in 5-dimensional \gls{ads} space-times. This is in compliance with
this philosophy that we have tried to give a second life to the \gls{eam} system within the problem of geon construction. Next
chapter is entirely dedicated to this boundary-value problem.

\chapter{Simulations of gravitational geons}
\label{simulations}
\addcontentsline{lof}{chapter}{\nameref{simulations}}
\addcontentsline{lot}{chapter}{\nameref{simulations}}
\citationChap{Hei! Aa-shanta 'nygh! You are off! Send back earth's gods to their haunts of unknown Kadath, and pray to all space that you may never meet me in my thousand other forms. Farewell, Randolph Carter, and beware; for I am Nyarlathotep, the Crawling Chaos.}{Howard Philips Lovecraft}
\minitoc

In the previous chapters, we have seen how to build linear geons with the \gls{kis} formalism (chapter
\ref{perturbations}) and how to fix the gauge by solving the \gls{eam} system (chapter \ref{gaugefreedom}). In this chapter, we
use these results to build numerically fully non-linear \gls{aads} geons with helical symmetry. Starting with a linear geon as initial guess, we solve the
\gls{eam} system of equations iteratively, and check afterwards that the gauge conditions are
satisfied. This ensures that our solutions are indeed geons and not Ricci solitons.

The discretisation of the \gls{eam} system is performed with spectral methods \cite{Grandclement09}, and the system is solved by a Newton-Raphson
algorithm. These two crucial steps are accomplished with KADATH \cite{kadath}, a general C++ library dedicated to the numerical
resolution of
\gls{gr} equations \cite{Grandclement10}. This library was inspired by the LORENE library \cite{lorene}, which was developed in
the 90's and aimed at solving the structure of neutrons stars, among other things. KADATH is a general purpose library that can solve very different
problems, among them vortons, critical collapse, binary black holes initial data \cite{Grandclement10,Uryu12}, oscillatons
\cite{Grandclement11}, \gls{aads} scalar breathers \cite{Fodor14,Fodor15}, boson stars
\cite{Grandclement14,Meliani15,Vincent16a,Meliani16,Grandclement17} and \gls{aads} geons \cite{Martinon17}.

There remain two conceptual issues that we have not discussed yet. First, because the \gls{ads} metric components are notoriously diverging
near the \gls{ads} boundary, we need to rely on a regularisation procedure that allows us to represent the whole space-time in
a computer, as was done for example in \cite{Bantilan12,Bantilan15,Bantilan17}. Basically, this implies to work with the conformal metric introduced in
chapter \ref{aads}. Second, since the construction of stationary helically symmetric geons is a boundary-value problem, we need to
impose boundary conditions. It turns out that for this particular problem, Dirichlet conditions for the conformal metric leads to
a well-posed problem.

Once geon solutions are computed, there are several validations that are necessary to handle. For instance, a solution of the
\gls{eam} system is a solution of Einstein's system if and only if the gauge conditions $K = 0$ and $V^i = 0$ are satisfied
(chapter \ref{gaugefreedom}), or equivalently, if the solution computed has a vanishing Einstein tensor. Numerically speaking, and
in particular for spectral methods, this translates the exponential decrease of these residual quantities with increasing
resolution. Furthermore, solving Einstein's equation with a negative cosmological constant is not sufficient to ensure that the
solution displays correct \gls{aads} asymptotics. We thus have to check that the Weyl tensor, its pseudo-magnetic part, as well as
the quasi-local stress tensor do
indeed vanish on the \gls{ads} boundary (chapter \ref{aads}). We also have to check that the \gls{amd} and \gls{bk} charges
converge to each other since we are in 4-dimensional space-times. Again, all these tests are expected to converge exponentially to
zero with increasing resolution when using spectral discretisation.

Our numerical geons thus obtained belong to five families presented in \cite{Martinon17}, namely the $(l,m,n) = (2,2,0), (4,4,0)$
single-mode and the three degenerated $\omega = 5$ radially excited ones. The $(l,m,n) = (2,2,0)$ ones constitute an independent
construction from the previous work \cite{Horowitz15}. Our result disagree to some extent with these results, but are strongly supported by
convergent and independent numerical and analytical arguments. Furthermore, as discussed in chapter \ref{adsinsta} section
\ref{beyondspher}, this first-time
numerical construction of the three radially excited families $\omega = 5$ constitutes a definitive argument in favour of their fully
non-linear existence, that was questioned in \cite{Dias16a,Dias17a,Rostworowski16,Rostworowski17a,Rostworowski17b}.

Our notations and conventions are the same as in the previous chapter.

\section{Spectral methods}

One important technical aspect of the construction of geons as presented in \cite{Martinon17} lies in spectral methods. Hereafter,
we give a brief overview of this topic. We refer the interested reader to the review \cite{Grandclement09} and to standard textbooks
\cite{Gottlieb78,Fornberg98,Boyd01,Hesthaven07,Canuto07,Canuto10,Canuto12}.

\subsection{Spectral interpolant}

Let us consider the interval $[-1,1]$. We introduce a scalar product for functions defined on this interval
\begin{equation}
   (f,g)_{w} = \int_{-1}^1 f(x)g(x)w(x)dx,
   \label{scalarprod}
\end{equation}
where $w(x)$ is a positive function on $[-1,1]$ and is called the measure of the scalar product. A usual approximation of the
function $f$ relies on its projection on a basis of orthogonal polynomials $(p_n)_{n\in \mathbb{N}}$. Its truncated projection of order $N$ is then
\begin{equation}
   P_Nf(x) = \sum_{n=0}^N \widehat{f}_n p_n(x) \quad \tn{with} \quad \widehat{f}_n = \frac{(f,p_n)}{(p_n,p_n)}.
   \label{PN}
\end{equation}
A discrete version of this formula is given by the theorem of Gauss quadratures. Namely, there exist $N+1$ positive real numbers
$w_n$ and $N+1$ real numbers $x_n \in [-1,1]$ such that
\begin{equation}
   \forall f \in \mathbb{P}_{2N+\delta}, \quad \int_{-1}^1 f(x) w(x) dx = \sum_{n=0}^N f(x_n) w_n.
   \label{Gaussquad}
\end{equation}
In this expression $\mathbb{P}_{2N+\delta}$ is the set of polynomials of degree inferior to $2N + \delta$, and the $x_n$ are
called the collocation points. The parameter $\delta$ can take values in $\{-1,0,1\}$.
\begin{myitem}
   \item If $\delta = 1$, this is the standard Gauss quadrature.
   \item If $\delta = 0$, this is the Gauss-Radau quadrature, and $x_0$ is fixed to the lower bound of the interval $x_0 = -1$.
   \item If $\delta = -1$, this is the Gauss-Lobatto quadrature, and $x_0$ and $x_N$ are the fixed to the lower and upper bound $x_0 = -1$ and $x_N = 1$.
\end{myitem}
In the standard case $\delta = -1$, the Gauss quadrature theorem states that the collocation points $(x_n)_{n\in \llbracket 0,N\rrbracket}$ are precisely
the roots of the order $N+1$ polynomial $p_{N+1}(x)$. As for the weight coefficients, they can be simply obtained by inverting the system of
equations
\begin{equation}
   \left(
   \begin{array}{ccc}
      p_0(x_0) & \cdots & p_0(x_N)\\
      p_1(x_0) & \cdots & p_1(x_N)\\
      \vdots   &        & \vdots\\
      p_N(x_0) & \cdots & p_N(x_N)\\
   \end{array}
   \right)
   \left(
   \begin{array}{c}
      w_0\\
      w_1\\
      \vdots\\
      w_N
   \end{array}
   \right) = \left(
   \begin{array}{c}
      \int_{-1}^1 p_0(x)w(x)dx\\
      0\\
      \vdots\\
      0
   \end{array}
   \right),
\end{equation}
which follows directly from \eqref{Gaussquad} applied to the basis polynomials. The zeros on the right-hand side simply follows
from the orthogonality of $(p_n)_{n\in \llbracket 1, N \rrbracket}$ with $p_0$ that is a constant function in general. Such a formula can be
condensed into \cite{Press07}
\begin{equation}
   w_i = \frac{(p_N,p_N)}{p_{N-1}(x_i) p'_{N+1}(x_i)}.
\end{equation}
Similar formulas for the $\delta = 0,1$ cases can be found in e.g.\ \cite{Canuto10}. Of these formulas, we only retain that the
collocation points $x_n$ and the weight coefficients are utterly determined by the choice of the polynomial basis. Even the
weight function $w(x)$ follows from the choice of the family $(p_n)_{n\in \mathbb{N}}$ since the polynomials have to be
orthogonal with respect to the scalar product \eqref{scalarprod}. Usually in spectral methods, polynomials are chosen to be
Chebychev, Legendre or Laguerre polynomials.

The main idea behind spectral methods is to discretise the scalar products of \eqref{PN} with Gauss quadratures
\eqref{Gaussquad}. The latter is only valid for polynomial functions $f$, but we use it for smooth functions all the same. The projection
$P_Nf$ is thus traded for its interpolant
\begin{equation}
   I_Nf(x) = \sum_{n=0}^N f_n p_n(x)  \quad \tn{with} \quad f_n = \frac{\displaystyle{\sum_{i=0}^N
   f(x_i) p_n(x_i)w_i}}{\displaystyle{\sum_{i=0}^N p_n^2(x_i)w_i}}.
   \label{IN}
\end{equation}
The name interpolant comes from the property
\begin{equation}
   \forall i \in \llbracket 0,N\rrbracket,\quad I_Nf(x_i) = f(x_i),
   \label{interpolant}
\end{equation}
at all collocation points. It is the relevant approximation of the function $f$ for numerical computations. It can be
noticed that $I_Nf$ is entirely determined by either
\begin{myitem}
   \item its coefficients $(f_n)_{n\in\llbracket 0,N \rrbracket}$, this is the so-called coefficient space.
   \item its evaluations $(f(x_n))_{n\in\llbracket 0,N \rrbracket}$ at the collocation points, this is the so-called configuration space.
\end{myitem}
Indeed, from \eqref{IN} and \eqref{interpolant}, we get
\begin{equation}
   \forall i \in \llbracket 0, N \rrbracket, \quad f(x_i) = \sum_{n=0}^N f_n p_n(x_i).
\end{equation}
This is nothing but a system of equations that translates the bijection between coefficient and configuration space. This
equation allows to compute the configuration space knowing the coefficient space while the other way around is simply given by the
definition of $f_n$ in \eqref{IN}.

The main advantage of spectral methods is that for smooth functions, the interpolant $I_Nf$ converges to $f$ faster than any power
law with increasing $N$ in the $L^2$-sense. In practice, it is very often equivalent to an exponential convergence. This property
is highly valuable in terms of computational cost, and can be used as a validation of spectral codes. This very fast convergence is
illlustrated on figure \ref{speconv}.

\begin{myfig}
   \includegraphics[width = 0.49\textwidth]{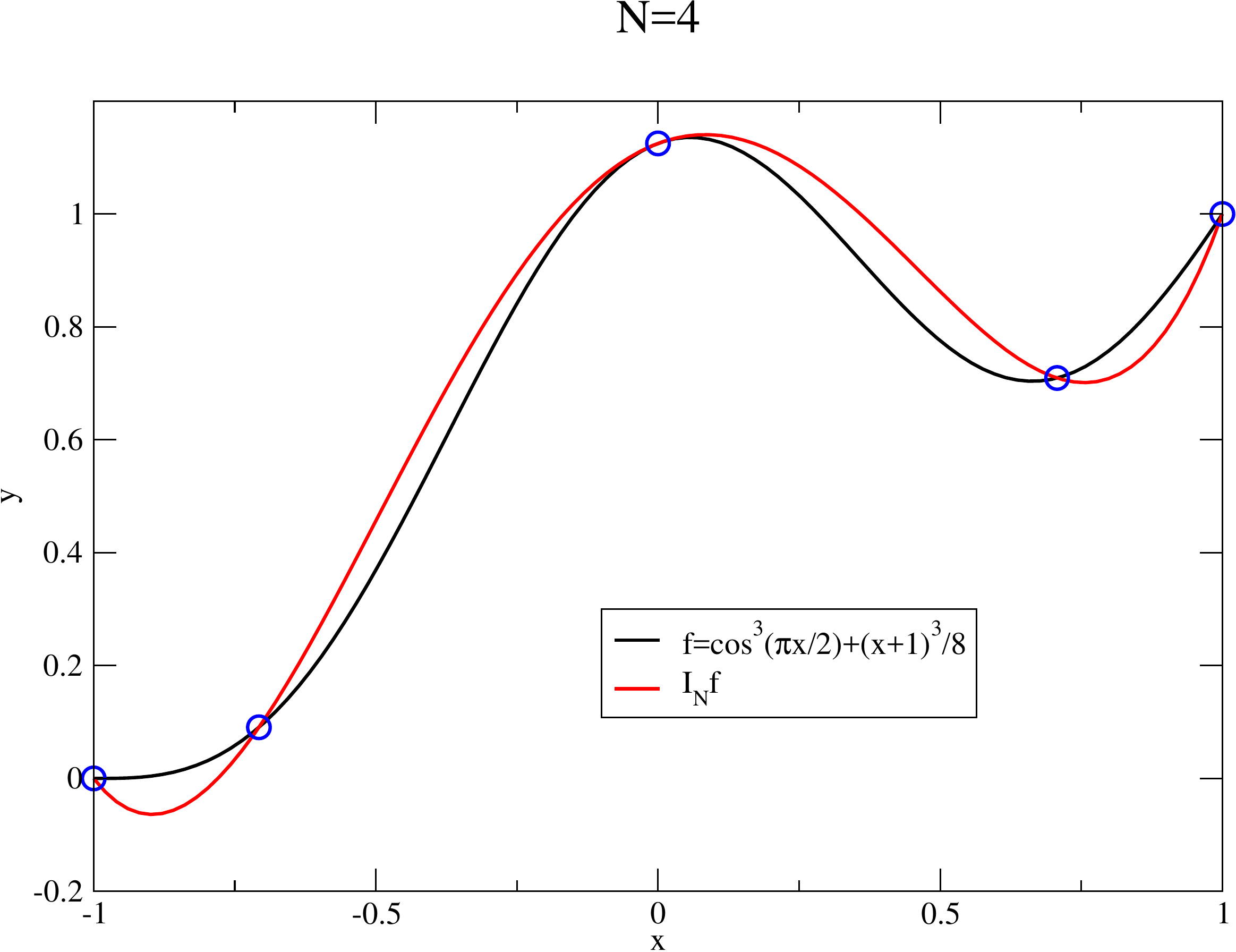}
   \includegraphics[width = 0.49\textwidth]{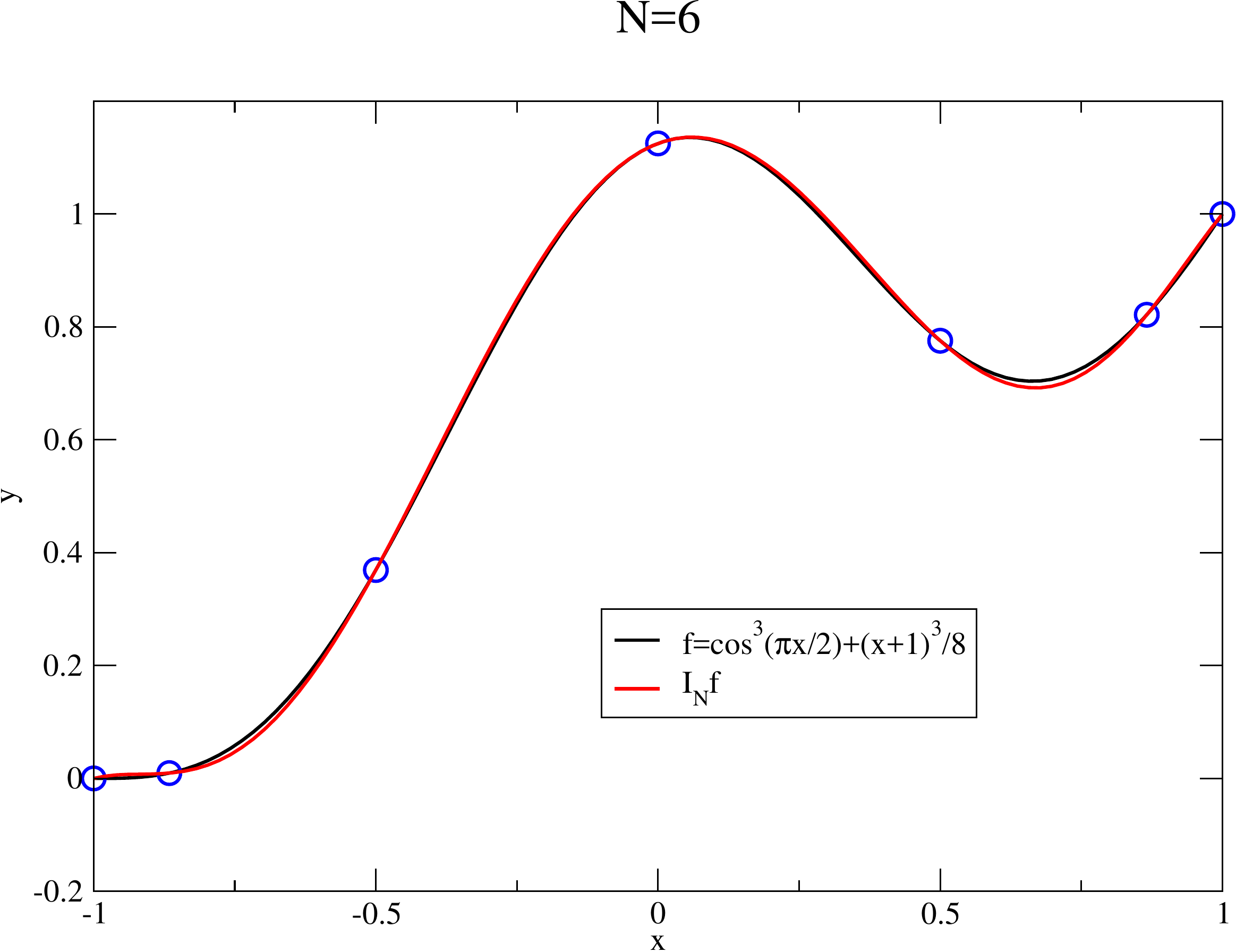}
   \caption[Spectral convergence]{The function $f(x) = \cos^3(\pi x/2) + (x+1)^3/8$, which is neither periodic nor polynomial,
      is approximated by its spectral interpolant $I_N f$ on $[-1,1]$. On the left panel, the truncation number is $N=4$, while on the
      right panel it is $N=6$. In this latter case, the difference between the function and its interpolant is hardly visible at
      naked eye. This illustrates that a handfull of coefficients is often enough to represent complicated functions with a good
      precision. Credits: \cite{Grandclement09}.}
   \label{speconv}
\end{myfig}

\subsection{Discretisation of equations}

Suppose now that we want to solve a differential equation of the form
\begin{equation}
   Lf(x) = S(x),
   \label{Lf}
\end{equation}
where $L$ is a differential operator and $S$ a source term. Our goal is to perform the spectral decomposition of $Lf(x)$ and
$S(x)$. For concreteness, let us work with Chebychev polynomials. Thanks to their properties, and notably several recursion
relations, simple operators admit the following decomposition \cite{Gottlieb78}
\begin{subequations}
\begin{align}
   \label{diff1}
   f'(x) = \sum_{n=0}^N a_n p_n(x) \quad \tn{with} \quad a_n = \frac{1}{1 + \delta_{0n}}\sum_{\substack{p=n+1\\n+p\ \tn{odd}}}^N 2pf_p,\\
   f''(x) = \sum_{n=0}^N b_n p_n(x) \quad \tn{with} \quad b_n = \frac{1}{1 + \delta_{0n}}\sum_{\substack{p=n+2\\n+p\ \tn{even}}}^N p(p^2 - n^2)f_p,\\
   \label{div1mx2}
   \frac{f(x)}{1-x^2} = \sum_{n=0}^N c_n p_n(x) \quad \tn{with} \quad c_n = -\frac{2}{1 + \delta_{0n}} \sum_{\substack{p=n+2\\n+p\ \tn{even}}}^N (p-n)f_p,
\end{align}
\label{specop}
\end{subequations}
the symbol $\delta$ denoting Kronecker's delta symbol. On the one hand, some operations are most easily done in configuration space. For
instance, to compute $f(x)g(x)$, it suffices to compute the product $f(x_i)g(x_i)$ at the collocation points and to deduce the
coefficients of the result with equation \eqref{IN}. On the other hand, some operations can only be performed in coefficient space. This is
the case of $f(x)/(1-x^2)$ for functions obeying $f(\pm 1) = 0$. Since the denominator diverges at the boundary of the interval $[-1,1]$, the division is forbidden in
configuration space. However, the result admits a suitable representation in coefficient space that is given by \eqref{div1mx2}. This kind
of arguments is highly relevant to \gls{ads} computations where some quantities has to be regularised by a factor that is
precisely zero on the space-time boundary.

Let us also mention that these formulas encode how numerical errors propagate with spectral operations. For example, suppose that
there is a small numerical error $(\delta f_n)_{n\in\llbracket 0,N\rrbracket}$ on the coefficients of the function $f$. Given the
spectral representation of the derivative \eqref{diff1}, the errors on $f'$ are $\sim N$ times larger than the errors on
$f$. Since usually, $N \sim 10-100$ in most numerical computations, errors are roughly multiplied by a factor $\sim 100$ at each
derivative step. This is why a function determined at machine precision $\sim 10^{-14}$ (with double arithmetics) has a second derivative determined
only at the $\sim 10^{-10}$ level. In general, any operation that takes place in coefficient space multiplies the numerical errors by a
non-negligible amount. This is also an advantage in some cases, since this makes spectral methods very sensitive. In particular,
this facilitates greatly the detection of numerical mistakes.

Coming back to equation \eqref{Lf}, it can be rewritten
\begin{equation}
   R \equiv Lf - S = \sum_{n=0}^N (l_n - s_n) p_n(x) = 0,
   \label{eqresidual}
\end{equation}
where $R$ is called the residual, $(l_n)_{n\in\llbracket 0,N \rrbracket}$ are the coefficients of $Lf(x)$ (computed with equations
\eqref{specop}), and $(s_n)_{n\in\llbracket 0,N \rrbracket}$ the ones of the source term $S$. The equation can be considered
solved if the residual is as small as possible, in a sense that has yet to be defined. In the weighted residual method, one
considers a set of $N+1$ test functions $(\xi_n)_{n\in\llbracket 0,N \rrbracket}$ defined on $[-1,1]$ and requires
\begin{equation}
   \forall i \in \llbracket 0,N \rrbracket, \quad (R,\xi_i) = 0.
   \label{systeN}
\end{equation}
Depending on the choice of the test functions, we obtain different spectral methods. For example,
\begin{myitem}
   \item the tau method, in which $\xi_n(x) = p_n(x)$ and \eqref{systeN} amounts to (by orthogonality of the polynomials)
      \begin{equation}
         \forall i \in \llbracket 0,N \rrbracket, \quad l_i - s_i = 0,
         \label{taueq}
      \end{equation}
   \item the collocation method, in which $\xi_i$ are the Lagrange polynomials defined by $\xi_i(x_j) = \delta_{ij}$ and
      \eqref{systeN} amounts to (recall the Gauss quadrature formula \eqref{Gaussquad})
      \begin{equation}
         \forall i \in \llbracket 0,N \rrbracket, \quad \sum_{n=0}^N (l_n - s_n) p_n(x_i) = 0.
         \label{coleq}
      \end{equation}
\end{myitem}
Both equations \eqref{taueq} and \eqref{coleq} constitute a system of $N+1$ equations for the $N+1$ unknown coefficients $(f_n)_{n\in
\llbracket 0,N \rrbracket}$.  If the operator $L$ is linear, each $l_n$ is a mere linear combination of the
$f_n$, and either equation can be solved by a single matrix inversion. 

Note that without specification of boundary conditions, the system \eqref{systeN} is generally not
invertible. For second order differential equations, two boundary conditions have to be specified.  A usual method is to replace
the last two equations in \eqref{systeN} by the two boundary conditions. If the underlying problem is well-posed, this ensures
that the matrix inversion gives rise to a unique solution. Other methods are reviewed in \cite{Grandclement09}. As far as KADATH
is concerned, we work with the tau method.

In the most general case, the operator $L$ is non-linear though, and each $l_n$ is then a non-linear function of the $(f_n)_{n\in\llbracket 0,N \rrbracket}$. The
system \eqref{systeN}, supplied with boundary conditions, can therefore be rewritten
\begin{equation}
   \vec{L}(\vec{f}) = 0,
   \label{vecL}
\end{equation}
where $\vec{L}$ is a vector of $N+1$ components $(l_n)_{n\in\llbracket 0,N \rrbracket}$, each one of them being a non-linear function of all
the coefficients $(f_n)_{n\in\llbracket 0,N \rrbracket}$, collected in turn in $\vec{f}$. This problem is nothing but a
multi-dimensional root finding problem. It can be solved with an initial guess and a root-finding algorithm, such as
Newton-Raphson, that inverts iteratively the Jacobian of $\vec{L}$. 

\subsection{Application to Einstein's equation}

Even if we have only presented 1-dimensional functions $f(x)$ in the interval $[-1,1]$, these methods generalise readily to
multi-dimensional problems in arbitrary intervals. Any change of variables realising a bijection from $[-1,1]$ to the interval of
interest allows to use the spectral representation. If $f$ depends on two variables $x$ and $y$, we can write
\begin{equation}
   f(x,y) = \sum_{i=0}^N \sum_{j=0}^M f_{ij}p_i(x)p_j(y).
\end{equation}
The coefficients are thus stored in a 2-dimensional array $f_{ij}$, but the very nature of the problem
\eqref{vecL} is unchanged.

As far as Einstein's equation is concerned, the order of magnitude of the size $m$ of the Jacobian matrix can be estimated to be
\begin{equation}
   m \propto N_{fields} \times N_{domains} \times N^d,
\end{equation}
where $N_{fields}$ is the number of fields, $N_{domains}$ the number of spectral domains, $N$ the average number of collocation
points in each direction, and $d$ the dimensionality of the problem. For 4-dimensional helically symmetric geons, the stationarity
allows to suppress the time dependence, so that $d =3$. Furthermore, the unknowns are the ten metric components, which means
$N_{fields} = 10$. And a typical number of domains is $N_{domains} = 2$. Requiring a resolution of $N = 20$ points in all three
space directions, means that the number of columns of the Jacobian is $m \sim 10^5$. In practice, such matrices need a large
amount of memory to be stored, and require a significant amount of time to be inverted. Parallel programming is then of particular
relevance. More details can be found in \cite{Grandclement10}.

Let us briefly mention the case of higher-dimensional geons. Adding a supplementary dimension means enlarging
the size of the Jacobian matrix by a factor $N = 20$. This means that the computation of the Jacobian which is $O(m^2)$ would take $\sim
400$ times longer. As for the matrix inversion which is $O(m^{2.3})$ at best, this operation would take $\sim 1000$ times longer.
This is why we worked in 4-dimensional space-times, even if, from an \gls{ads}-\gls{cft} point of view, the 5-dimensional case is
more appropriate. On practical grounds, no supplementary dimension can be added unless an additional symmetry (or invariance) is assumed. The
existence of geons is believed to share the same properties in 4 and 5-dimensional space-times though.

In our numerical setup, we decompose the radial direction onto Chebychev polynomials, while angular directions are decomposed with
discrete Fourier transforms (which is a particular case of a spectral decomposition). In the following, since we focus on geons that are even
in their $(l,m)$ quantum numbers, they all display an octant symmetry. This allows to reduce the computation time by restricting
the angular spectral expansion onto a subset of trigonometrical functions. Accordingly, when quoting a resolution of, say,
$(37,9,9)$, it means that we have used 37 coefficients in the radial direction, and 9 coefficients in each angular direction
$(\theta,\varphi)$. This is equivalent to 9 collocation points \textit{per octant} in each angular direction.

\section{Regularisation}
\label{reg}

We now have to suitably represent the geometry numerically, and this cannot be done without regularisation. This points was
already noticed in \cite{Bantilan12,Bantilan15,Bantilan17}. For the sake of concreteness, we chose to work in the so-called isotropic
Cartesian coordinates, in which the background \gls{ads} length element takes the form \eqref{adsisotropic}
\begin{equation}
   \overline{ds}^2 = \overline{g}_{\alpha\beta}dx^\alpha dx^\beta =  -\left(\frac{1+\rho^2}{1 - \rho^2}\right)^2dt^2 + \frac{4}{(1-\rho^2)^2}(dx^2+dy^2+dz^2),
\label{confcoor}
\end{equation}
where $\rho = r/\gls{L}$ and $r = \sqrt{x^2 + y^2 + z^2} \in [0,\gls{L} [$. A bar indicates, as usual, a background \gls{ads}
quantity. As already discussed in chapter \ref{aads} section \ref{adsboundary}, the metric components are all diverging like
$O((L-r)^{-2})$ in the neighbourhood of the \gls{ads} boundary, which makes this length element inadequate for numerical computations.
This unwanted behaviour propagates to all geometrical tensors and symbols. We thus need to perform a regularisation that ensures
all quantities to be finite within the whole numerical manifold.

\subsection{Regularised $3+1$ quantities}
\label{3+1quant}

We start by regularising all the geometrical quantities. Accordingly, let us define the following conformal factor
\begin{equation}
   \Omega = \frac{1-\rho^2}{1+\rho^2}.
\end{equation}
It is clear that the \gls{ads} metric components diverge at the boundary $r=\gls{L}$ like $O(\Omega^{-2})$ (this is even implied
by the definition \ref{aadsdefinition} of \gls{aads} space-times given in chapter \ref{aads}). A regularised metric is then
\begin{equation}
   \widetilde{\overline{ds}}^2 \equiv \Omega^2 \overline{ds}^2 = \widetilde{\overline{g}}_{\alpha\beta}dx^\alpha dx^\beta =  -dt^2 + \frac{4}{(1+\rho^2)^2}(dx^2+dy^2+dz^2),
   \label{confads}
\end{equation}
which is regular and flat at the \gls{ads} boundary $r = \gls{L}$. Hereafter, we use a tilde to denote all geometrical quantities that we
regularise using some power of $\Omega$, such that all tilded quantities are regular at the boundary. These are not to be confused with conformal
quantities of chapter \ref{aads}, that are denoted by a hat. It turns out that for the \gls{ads} metric background,
$\widetilde{\overline{g}}_{\alpha\beta} = \widehat{\overline{g}}_{\alpha\beta}$, but this does not remain the case for all the
other quantities. In table \ref{quantities}, we summarise the behaviour of various geometric tensors and symbols near the
\gls{ads} boundary as well as their regularisations. Notations are those of $3+1$ formalism (appendix \ref{d+1}), and we have used
extensively the matricial representation of $3+1$ decomposition \eqref{metricd+1}.

\begin{mytab}
\begin{tabular}{lclc}
\hline
Quantity                    & behaviour at $r=\gls{L}$ & Regularisation                                                                 & behaviour at $r=\gls{L}$\\
\hline
$N$                         & $O(\Omega^{-1})$         & $\widetilde{N} \equiv \Omega N$                                                & $O(1)$  \\
$\beta^i$                   & $O(1)$                   & $\widetilde{\beta}^i \equiv \beta^i$                                           & $O(1)$  \\
$\beta_i$                   & $O(\Omega^{-2})$         & $\widetilde{\beta}_i \equiv \Omega^2\beta_i$                                   & $O(1)$  \\
$\gamma_{ij}$               & $O(\Omega^{-2})$         & $\widetilde{\gamma}_{ij} \equiv \Omega^2 \gamma_{ij}$                          & $O(1)$  \\
$\gamma^{ij}$               & $O(\Omega^{2})$          & $\widetilde{\gamma}^{ij} \equiv \gamma^{ij}/\Omega^2$                          & $O(1)$  \\
$\digamma\indices{^k_{ij}}$ & $O(\Omega^{-1})$         & $\widetilde{\digamma}\indices{^k_{ij}} \equiv \Omega\digamma\indices{^k_{ij}}$ & $O(1)$  \\
$K_{ij}$                    & $O(\Omega^{-2})$         & $\widetilde{K}_{ij} \equiv \Omega^2K_{ij}$                                     & $O(1)$  \\
$K\indices{^j_i}$           & $O(1)$                   & $\widetilde{K}^j_i \equiv K^j_i$                                               & $O(1)$  \\
$K$                         & $O(1)$                   & $\widetilde{K} \equiv K$                                                       & $O(1)$  \\
$K^{ij}$                    & $O(\Omega^{2})$          & $\widetilde{K}^{ij} \equiv K^{ij}/\Omega^2$                                    & $O(1)$  \\
$\mathcal{R}_{ij}$          & $O(\Omega^{-2})$         & $\widetilde{\mathcal{R}}_{ij} \equiv \Omega^2 \mathcal{R}_{ij}$                & $O(1)$  \\
$\mathcal{R}$               & $O(1)$                   & $\widetilde{\mathcal{R}} \equiv \mathcal{R}$                                   & $O(1)$  \\
$V^i$                       & $O(\Omega)$              & $\widetilde{V}^i \equiv V^i/\Omega$                                            & $O(1)$  \\
$V_i$                       & $O(\Omega^{-1})$         & $\widetilde{V}_i \equiv \Omega V_i$                                            & $O(1)$  \\
\hline
\end{tabular}
\caption[Regularisation of 3+1 quantities]{Behaviour of $3+1$ geometrical quantities and their regularisations near the \gls{ads} boundary expressed in powers of
the conformal factor $\Omega$. Notations are those of $3+1$ formalism, appendix \ref{d+1}. Credits: \cite{Martinon17}.}
\label{quantities}
\end{mytab}

More precisely, denoting
\begin{equation}
   \Omega_i = \partial_i \Omega,
\end{equation}
the regularisations of table \ref{quantities} are performed according to
\begin{subequations}
\begin{align}
   \label{reggamma}
   \widetilde{\digamma}\indices{^k_{ij}} &= \frac{1}{2}\Omega\widetilde{\gamma}^{kl}(\partial_i \widetilde{\gamma}_{jl} + \partial_j \widetilde{\gamma}_{il} - \partial_l \widetilde{\gamma}_{ij}) - \widetilde{\gamma}^{kl}(\widetilde{\gamma}_{il}\Omega_j + \widetilde{\gamma}_{jl}\Omega_j - \widetilde{\gamma}_{ij}\Omega_l),\\
   \widetilde{K}_{ij} &= -\frac{1}{2 \widetilde{N}}(\Omega \mathcal{L}_m \widetilde{\gamma}_{ij} + 2 \widetilde{\gamma}_{ij}\widetilde{\beta}^k\Omega_k),\\
   \widetilde{K}\indices{^j_i} &= \widetilde{\gamma}^{jk}\widetilde{K}_{ki},\\
   \widetilde{K} &= \widetilde{\gamma}^{ij} K_{ij},\\
   \widetilde{K}^{ij} &= \widetilde{\gamma}^{ik}\widetilde{\gamma}^{jl}\widetilde{K}_{kl},\\
   \widetilde{\mathcal{R}}_{ij} &= \Omega (\partial_k \widetilde{\digamma}\indices{^k_{ij}} - \partial_i \widetilde{\digamma}\indices{^k_{jk}}) - (\widetilde{\digamma}\indices{^k_{ij}}\Omega_k - \widetilde{\digamma}\indices{^k_{jk}}\Omega_i) + \widetilde{\digamma}\indices{^k_{ij}} \widetilde{\digamma}\indices{^l_{kl}} - \widetilde{\digamma}\indices{^l_{ik}} \widetilde{\digamma}\indices{^k_{jl}},\\
   \widetilde{\mathcal{R}} &= \widetilde{\gamma}^{ij} \widetilde{\mathcal{R}}_{ij},\\
   \widetilde{V}^i &= \widetilde{\gamma}^{kl}(\widetilde{\digamma}\indices{^i_{kl}} - \widetilde{\overline{\digamma}}\indices{^i_{kl}}),\\
   \widetilde{V}_i &= \widetilde{\gamma}_{ij}\widetilde{V}^j,
\end{align}
\label{regquant}%
\end{subequations}
where we have used the definition of the Christoffel symbols \eqref{christoffel}, the evolution equation \eqref{evolgamma}, the
definition of the Ricci tensor \eqref{riccidef} and the definition of the spatial harmonic gauge vector \eqref{defV}. The
evolution operator is $\mathcal{L}_m = \partial_t - \mathcal{L}_{\beta}$, where $\mathcal{L}$ stands for the Lie derivative.
These expressions can be readily checked to be regular and $O(1)$ near the \gls{ads} boundary $\Omega = 0$.

\subsection{Regularisation of the first order gauge equations}
\label{reglingauge}

One of the first step in the construction of a geon is to bring a linear solution in the \gls{am} gauge. Recall that for helically
symmetric geons, these solutions are stationary in the co-rotating frame. As discussed in chapter
\ref{gaugefreedom}, we need to solve in turn equations \eqref{eqK} and \eqref{eqVV} at first order, namely
\begin{subequations}
\begin{align}
   \label{2nk}
   2N^2 D^iD_i \alpha + 4N D^iN D_i\alpha &= 2NK,\\
   \label{ddchi}
   D^jD_j \chi^i + \mathcal{R} \indices{^i_j}\chi^j &= V^i,
\end{align}
\end{subequations}
for the first order unknown scalar $\alpha(x^i)$ and 3-vector $\chi^i(x^j)$. Then we reconstruct the metric with \eqref{primealpha} and
\eqref{primechi}, that can be condensed into
\begin{subequations}
\begin{align}
   N' &= N (1 + \beta^i\partial_i \alpha) - \mathcal{L}_{\chi} N,\\
   \beta^{i'} &= \beta^i + (N^2 \gamma^{ij} + \beta^i\beta^j)\partial_j\alpha - \mathcal{L}_{\chi}\beta^i, \\
   \beta_{i'} &= \beta_i + (N^2 - \beta_j\beta^j)\partial_i\alpha - \mathcal{L}_{\chi}\beta_i, \\
   \gamma_{i'j'} &= \gamma_{ij} - \beta_i \partial_j\alpha - \beta_j \partial_i\alpha - \mathcal{L}_{\chi}\gamma_{ij},\\
   \gamma^{i'j'} &= \gamma^{ij} + (\gamma^{ik}\beta^j + \gamma^{jk}\beta^i)\partial_k \alpha - \mathcal{L}_{\chi}\gamma^{ij}.
\end{align}
\label{recons}%
\end{subequations}
Since the right-hand side of \eqref{2nk} is $O(\Omega^{-1})$ near the \gls{ads} boundary according to table \ref{quantities}, we
multiply it by $\Omega$ to get regular quantities. Using the covariant derivative definition \eqref{defcov} and substituting the
equations of table \ref{quantities}, we thus obtain
\begin{subequations}
\begin{align}
   2 \widetilde{N}(\Omega^2 \partial_i \partial_j \alpha - \widetilde{\gamma}^{ij}\widetilde{\Gamma}\indices{^k_{ij}}\partial_k \alpha) + 4 \widetilde{N}\widetilde{\gamma}^{ij}(\Omega \partial_i \widetilde{N} - \Omega_i) \partial_j \alpha &= 2\widetilde{N}\widetilde{K},\\
   \label{regxi}
   \tilde{\gamma}^{kl}(\Omega^2 \partial_k \partial_l \chi^i + \Omega \mathcal{L}_\chi \widetilde{\Gamma}\indices{^i_{kl}} - \widetilde{\Gamma}\indices{^i_{kl}}\chi^m\Omega_m) &= \Omega \widetilde{V}^i.
\end{align}
\label{reggauge}%
\end{subequations}
Equation \eqref{regxi} is most easily derived from \eqref{defV} and \eqref{primechi} than with \eqref{ddchi}.
Note that, rigorously speaking, any term containing an $\alpha$ or a $\chi^i$ is first order only if it is multiplied by background
quantities. Thus, all the geometrical quantities on the left-hand sides of equations \eqref{reggauge} can be computed with the pure
\gls{ads} metric \eqref{confads}. The simplest boundary conditions for $\alpha$ and $\chi^i$ are
\begin{equation}
   \alpha \mathrel{\widehat{=}} 0 \quad \tn{and} \quad \chi^i \mathrel{\widehat{=}} 0,
   \label{bcgauge}
\end{equation}
where $\widehat{=}$ means equality restricted to the \gls{ads} boundary. They give rise to invertible boundary-value problems
and equations \eqref{reggauge} can be solved by the iterative scheme of KADATH with trivial initial guesses. Once the solutions
are obtained, the regularised metric can be reconstructed from the regularised version of \eqref{recons} and table \ref{quantities} as
\begin{subequations}
\begin{align}
   \widetilde{N}' &= \widetilde{N} (1 + \widetilde{\beta}^i\partial_i \alpha) - \mathcal{L}_{\chi} \widetilde{N} + \frac{1}{\Omega}\widetilde{N}\chi^i \Omega_i,\\
   \widetilde{\beta}^{i'} &= \widetilde{\beta}^i + (\widetilde{N}^2 \widetilde{\gamma}^{ij} + \widetilde{\beta}^i\widetilde{\beta}^j)\partial_j\alpha - \mathcal{L}_{\chi}\widetilde{\beta}^i, \\
   \widetilde{\beta}_{i'} &= \widetilde{\beta}_i + (\widetilde{N}^2 - \widetilde{\beta}_j\widetilde{\beta}^j)\partial_i\alpha - \mathcal{L}_{\chi}\widetilde{\beta}_i + \frac{2}{\Omega}\widetilde{\beta}_i \chi^j \Omega_j, \\
   \widetilde{\gamma}_{i'j'} &= \widetilde{\gamma}_{ij} - \widetilde{\beta}_i \partial_j\alpha - \widetilde{\beta}_j \partial_i\alpha - \mathcal{L}_{\chi}\widetilde{\gamma}_{ij} + \frac{2}{\Omega}\widetilde{\gamma}_{ij}\chi^k \Omega_k,\\
   \widetilde{\gamma}^{i'j'} &= \widetilde{\gamma}^{ij} + (\widetilde{\gamma}^{ik}\widetilde{\beta}^j + \widetilde{\gamma}^{jk}\widetilde{\beta}^i)\partial_k \alpha - \mathcal{L}_{\chi}\widetilde{\gamma}^{ij} - \frac{2}{\Omega}\widetilde{\gamma}^{ij} \chi^k \Omega_k.
\end{align}
\end{subequations}
Again, on the right-hand sides, only first order terms are needed, so that the background metric derived from \eqref{confads} can
be used extensively. It is important to note that there are terms in $\chi^i/\Omega$ that appears in this reconstruction. These
terms are regular beyond doubt since $\chi^i = O(\Omega)$ at $r = \gls{L}$ according to the boundary condition \eqref{bcgauge}. Numerically
speaking, we can take advantage of the spectral representation of $\chi^i$ to perform the division in coefficient space with the
formula \eqref{div1mx2}. In some sense, we use the non-local nature of the spectral decomposition.

Equipped with the linear helically symmetric geons of chapter \ref{perturbations} and this procedure to bring them in the \gls{am} gauge,
we are now able to build numerically any of these linear solutions in terms of regularised quantities.

\subsection{Regularisation of the EAM system}

Now that we are able to build, in practice, suitable initial guesses, we have to solve the full \gls{eam} system of equation
\eqref{3+1syst2}, that we reproduce here for convenience:
\begin{subequations}
\begin{align}
   \mathcal{L}_m K_{ij} + 2N \gamma_{ij} &= 0,\\
   \mathcal{R} - D_i V^i - K_{ij}K^{ij} - 2 \gls{Lambda} &= 0,\\
   D_j K \indices{^j_i} &= 0,\\
   \mathcal{L}_m K_{ij} + D_iD_jN - N(\mathcal{R}_{ij} - D_{(i}V_{j)} - 2K_{ik}K \indices{^k_j} - \gls{Lambda}\gamma_{ij}) &= 0.
\end{align}
\label{eam}%
\end{subequations}
Since we work in a frame co-rotating with the geon, there are no time derivative and the evolution operator is just $\mathcal{L}_m
= -\mathcal{L}_\beta$. With the results of table \ref{quantities}, it can be shown that the left-hand sides of these equations
behave respectively like $O(\Omega^{-3})$, $O(1)$, $O(\Omega^{-1})$ and $O(\Omega^{-3})$ near the \gls{ads} boundary. We then introduce a
regularised $O(1)$ version of them with the help of table \ref{quantities} and equations \eqref{regquant}
\begin{subequations}
\begin{align}
   2 \widetilde{N} \widetilde{K}_{ij} + \Omega \mathcal{L}_m \widetilde{\gamma}_{ij} + 2 \widetilde{\gamma}_{ij}\widetilde{\beta}^k\Omega_k &= 0,\\
   \label{hamreg}
   \widetilde{\mathcal{R}} - \partial_i (\Omega \widetilde{V}^i) - \widetilde{\digamma}\indices{^i_{ji}} \widetilde{V}^j -
   \widetilde{K}_{ij}\widetilde{K}^{ij} - 2\gls{Lambda} &= 0,\\
   \label{momreg}
   \Omega \partial_j \widetilde{K}\indices{^j_i} + \widetilde{\digamma}\indices{^j_{kj}} \widetilde{K}\indices{^k_i} -
   \widetilde{\digamma}\indices{^k_{ij}} \widetilde{K}\indices{^j_k} &= 0, \\
   \label{evoreg}
\nonumber   -\Omega\mathcal{L}_{m} \widetilde{K}_{ij} - 2 \widetilde{K}_{ij}\widetilde{\beta}^k \Omega_k - \delta \widetilde{N}_{ij}&\\
   + \widetilde{N}(\widetilde{\mathcal{R}}_{ij} - \Omega \partial_{(i}\widetilde{V}_{j)} + \widetilde{V}_{(i}\Omega_{j)}
   + \widetilde{\digamma}\indices{^k_{ij}} \widetilde{V}_k - 2\widetilde{K}_{ik}\widetilde{K}\indices{^k_j} - \gls{Lambda} \widetilde{\gamma}_{ij}) &= 0,
\end{align}
\label{eamreg}
\end{subequations}
where we have used the shortcut notation
\begin{align}
   \delta \widetilde{N}_{ij} &\equiv \Omega^3 D_i D_j \left( \frac{\widetilde{N}}{\Omega} \right) \\
\nonumber   &= \Omega^2 \partial_i\partial_j \widetilde{N} - \Omega(\Omega_i \partial_j \widetilde{N} + \Omega_j \partial_i \widetilde{N} + \widetilde{N} \partial_i \partial_j \Omega + \widetilde{\digamma}\indices{^k_{ij}} \partial_k \widetilde{N}) +
2\widetilde{N}\Omega_i\Omega_j + \widetilde{N}\widetilde{\digamma}\indices{^k_{ij}} \Omega_k.
\end{align}
The system of equations \eqref{eamreg} is then a regularised \gls{eam} system with negative cosmological constant. This is the
system of relevance for our numerical computations.

\subsection{Regularisation of the quasi-local stress tensor}
\label{tcftreg}

Once a non-linear solution is obtained, we need to compute its mass and angular momentum. In order to compute the \gls{bk} charges
of chapter \ref{aads} section \ref{globcharge}, we need to regularise the quasi-local stress tensor (equation
\eqref{quasilocalst})
\begin{equation}
   \tau_{\alpha\beta} = \frac{1}{8\pi}\left(\Theta_{\alpha\beta} - \Theta q_{\alpha\beta} - \frac{2}{\gls{L}}q_{\alpha\beta} +
   \gls{L} \mathfrak{G}_{\alpha\beta}\right),
   \label{eqtau}
\end{equation}
where $\mathfrak{G}_{\alpha\beta}$ is the Einstein tensor of the metric $q_{\alpha\beta}$ induced on $r=cst$ hypersurfaces
(denoted $\Sigma_r$). As in chapter \ref{aads}, we introduce $r_{\alpha}$ the unit normal to $\Sigma_r$, its acceleration
$a_{\alpha}$, $q_{\alpha\beta}$ the metric induced by $g_{\alpha\beta}$ on $\Sigma_r$ and $\Theta_{\alpha\beta}$ the corresponding
extrinsic curvature tensor. They are defined by (see appendix \ref{d+1})
\begin{subequations}
\begin{align}
   r_\alpha &= \frac{\partial_{\alpha}r}{\sqrt{g^{\mu\nu}\partial_{\mu}r \partial_\nu r}},\\
   a_\alpha &= r^\mu \nabla_{\mu}r_{\alpha},\\
   q_{\alpha\beta} &= g_{\alpha\beta} - r_{\alpha}r_{\beta},\\
   \Theta_{\alpha\beta} &= -\nabla_{\beta}r_{\alpha} + a_{\alpha}r_\beta.
\end{align}
\end{subequations}
This quantities can be regularised in the same way as the ones of section \ref{3+1quant}. In table \ref{quantities2}, we
specify their behaviours and regularisations near the \gls{ads} boundary $\Omega = 0$. In particular, the regularisation can be made
more precise with
\begin{subequations}
\begin{align}
   \widetilde{r}_{\alpha} &= \frac{\partial_{\alpha}r}{\sqrt{\widetilde{g}^{\mu\nu}\partial_{\mu}r \partial_\nu r}},\\
   \widetilde{a}_{\alpha} &= r^\mu [\partial_{\mu}(\Omega \widetilde{r}_{\alpha}) - \widetilde{\Gamma}\indices{^\nu_{\mu\alpha}} \widetilde{r}_{\nu} - 2 \widetilde{r}_{\alpha}\Omega_\mu],\\
   \widetilde{q}_{\alpha\beta} &= \widetilde{g}_{\alpha\beta} - \widetilde{r}_{\alpha}\widetilde{r}_{\beta},\\
   \widetilde{\Theta}_{\alpha\beta} &= -\partial_{\beta}(\Omega \widetilde{r}_{\alpha}) + \widetilde{\Gamma}\indices{^\mu_{\alpha\beta}} \widetilde{r}_\mu + 2 \widetilde{r}_{\alpha} \Omega_\beta + \widetilde{a}_\alpha \widetilde{r}_\beta,
\end{align}
\end{subequations}
where $\widetilde{\Gamma}_{\alpha\beta}^\mu = \Omega \Gamma_{\alpha\beta}^\mu$ can be regularised in the same spirit as equation
\eqref{reggamma}.

\begin{mytab}
\begin{tabular}{lclc}
\hline
Quantity                                & behaviour at $r=\gls{L}$ & Regularisation                                                                        & behaviour at $r=\gls{L}$\\
\hline
$q_{\alpha\beta}$                       & $O(\Omega^{-2})$         & $\widetilde{q}_{\alpha\beta} \equiv \Omega^2 q_{\alpha\beta}$                         & $O(1)$ \\
$q\indices{^\alpha_\beta}$              & $O(1)$                   & $\widetilde{q}\indices{^\alpha_\beta} \equiv q\indices{^\alpha_\beta}$                & $O(1)$ \\
$r_\alpha$                              & $O(\Omega^{-1})$         & $\widetilde{r}_\alpha \equiv \Omega r_\alpha$                                         & $O(1)$ \\
$a_\alpha$                              & $O(\Omega^{-1})$         & $\widetilde{a}_\alpha \equiv \Omega a_\alpha$                                         & $O(1)$ \\
$\Theta_{\alpha\beta}$                  & $O(\Omega^{-2})$         & $\widetilde{\Theta}_{\alpha\beta} \equiv \Omega^2 \Theta_{\alpha\beta}$               & $O(1)$ \\
$R_{\alpha\beta\gamma\delta}$           & $O(\Omega^{-4})$         & $\widetilde{R}_{\alpha\beta\gamma\delta} \equiv \Omega^4 R_{\alpha\beta\gamma\delta}$ & $O(1)$ \\
$u^\alpha$                              & $O(\Omega)$              & $\widetilde{u}^\alpha \equiv u^\alpha/\Omega$                                         & $O(1)$ \\
$\sigma_{\alpha\beta}$                  & $O(\Omega^{-2})$         & $\widetilde{\sigma}_{\alpha\beta} \equiv \Omega^2 \sigma_{\alpha\beta}$               & $O(1)$ \\
\hline
\end{tabular}
\caption[Regularisation of the quasi-local stress tensor]{Behaviour of geometrical quantities involved in the computation of the quasi-local stress tensor $\tau_{\alpha\beta}$.
All behaviours are intended at the \gls{ads} boundary. Credits: G. Martinon.}
\label{quantities2}
\end{mytab}

In order to obtain the regularised Einstein tensor $\mathfrak{G}_{\alpha\beta}$ of the induced metric $q_{\alpha\beta}$, we first determine its
regularised Riemann tensor\footnote{Not to be confounded with $R_{\alpha\beta\gamma\delta}$ the Riemann tensor of
$g_{\alpha\beta}$ or with $\mathcal{R}_{\alpha\beta\gamma\delta}$ the one of
$\gamma_{\alpha\beta}$.} $\mathfrak{R}_{\alpha\beta\gamma\delta}$. It can be obtained via the 4-dimensional regularised
$\widetilde{R}_{\alpha\beta\gamma\delta}$ which admits the following regularisation
\begin{equation}
   \widetilde{R}_{\alpha\beta\mu\nu} = \widetilde{g}_{\alpha\rho} (\Omega\partial_\mu \widetilde{\Gamma}\indices{^\rho_{\beta\nu}}
   - \Omega\partial_\nu \widetilde{\Gamma}\indices{^\rho_{\beta\mu}} - \widetilde{\Gamma}\indices{^\rho_{\beta\nu}} \Omega_\mu +
   \widetilde{\Gamma}\indices{^\rho_{\beta\mu}} \Omega_\nu +
   \widetilde{\Gamma}\indices{^\rho_{\sigma\mu}} \widetilde{\Gamma}\indices{^\sigma_{\beta\nu}} -
   \widetilde{\Gamma}\indices{^\rho_{\sigma\nu}} \widetilde{\Gamma}\indices{^\sigma_{\beta\mu}}).
\end{equation}
The link between the 4 and 3-dimensional Riemann tensors is given by the Gauss relation (appendix \ref{d+1} equation
\eqref{gauss}). After multiplication by $\Omega^4$, it reads
\begin{equation}
   \widetilde{\mathfrak{R}}_{\alpha\beta\gamma\delta} = \widetilde{q}\indices{^\mu_\alpha} \widetilde{q}\indices{^\nu_\beta}
   \widetilde{q}\indices{^\rho_\gamma} \widetilde{q}\indices{^\sigma_\delta} \widetilde{R}_{\mu\nu\rho\sigma} + \widetilde{\Theta}_{\alpha\gamma}\widetilde{\Theta}_{\beta\delta} - \widetilde{\Theta}_{\alpha\delta}\widetilde{\Theta}_{\beta\gamma}.
\end{equation}
From this, it is straightforward to get the regularised Einstein tensor of $q_{\alpha\beta}$, $\widetilde{\mathfrak{G}}_{\alpha\beta} =
\Omega^2 \mathfrak{G}_{\alpha\beta}$ with
\begin{equation}
   \widetilde{\mathfrak{G}}_{\alpha\beta} = \widetilde{\mathfrak{R}}_{\alpha\beta} -
   \frac{\widetilde{\mathfrak{R}}}{2}\widetilde{q}_{\alpha\beta}, \quad \widetilde{\mathfrak{R}}_{\alpha\beta} =
   \widetilde{q}^{\mu\nu}\widetilde{\mathfrak{R}}_{\mu\alpha\nu\beta}, \quad \widetilde{\mathfrak{R}} = \widetilde{q}^{\mu\nu}
   \widetilde{\mathfrak{R}}_{\mu\nu}.
\end{equation}
The regularised quasi-local stress tensor \eqref{eqtau} is then given by
\begin{equation}
   \tau_{\alpha\beta} = \frac{1}{8\pi\Omega^2}\left(\widetilde{\Theta}_{\alpha\beta} -
   \widetilde{\Theta}\widetilde{q}_{\alpha\beta} - \frac{2}{\gls{L}}\widetilde{q}_{\alpha\beta} + \gls{L} \widetilde{\mathfrak{G}}_{\alpha\beta}\right),
   \label{taupar}
\end{equation}
where $\widetilde{\Theta} = \widetilde{q}^{\mu\nu}\widetilde{\Theta}_{\mu\nu}$. As discussed in chapter \ref{aads} section
\ref{bkcharge}, and even if not obvious at first sight, The parenthesis is $O(\Omega^3)$ near the \gls{ads}
boundary, while its constitutive parts are all $O(1)$. Furthermore, the quasi-local stress tensor is involved in the computation
of charges according to \eqref{QBK}:
\begin{equation}
   Q^{BK}_\xi [\Sigma] = \oint_\Sigma \tau_{\mu\nu}u^\mu \xi^\nu \sqrt{\sigma}d^2z,
\end{equation}
where $u^\alpha$ is the unit normal vector to $\Sigma_t$, $\xi^\alpha$ is a Killing vector of the \gls{ads} boundary, and
$\sigma_{\alpha\beta}$ is the induced metric on $\Sigma$, a $t=cst$ slice of the boundary. As already reviewed in chapter \ref{aads} and
summarised in table \ref{quantities2}, the regularisations of $u^\alpha$ and $\sigma_{\alpha\beta}$ imply that
\begin{equation}
   Q^{BK}_\xi [\Sigma] = \oint_\Sigma \frac{1}{\Omega}\tau_{\mu\nu}\widetilde{u}^\mu \xi^\nu \sqrt{\widetilde{\sigma}}d^2z.
   \label{QBKreg}
\end{equation}
Thus, what needs to be calculated is the regular quantity $\tau_{\alpha\beta}/\Omega$. Consequently, to compute this formula numerically, we
need to first compute the parenthesis of \eqref{taupar} and check that it is zero to machine precision at $r=\gls{L}$. We then compute
its division by $\Omega^3$, i.e.\ we take advantage of the spectral representation provided by the KADATH library and perform the
division in coefficient space with \eqref{div1mx2}. In essence one uses the non-local nature of the spectral representation. This operation brings some
numerical errors that can be monitored (see section \ref{res} below).

\section{Numerical tests}

Before presenting the results of geon computations, we discuss several tests arguing in favour of the validity of the code. In
particular, we check that our initial guess solutions are indeed first order solutions of Einstein's equation and are successfully
brought to the \gls{am} gauge. This validates partially our regularisation of the system of equations. A complementary check of
this regularisation is to compute the global charges of a known analytical solution, the Kerr-\gls{ads} black hole.

\subsection{Linear geons in the AM gauge}

We start with our analytical solutions of helically symmetric geons obtained in chapter \ref{perturbations}. We derived them
in conformal coordinates \eqref{adsconformal}, but they can be readily brought to isotropic coordinates \eqref{adsisotropic} as
explained in section \ref{coordinates} of chapter \ref{aads}.

Analytically again, we can perform the $3+1$ decomposition of the perturbed metric (chapter \ref{d+1} equation \eqref{metricd+1}) and compute the lapse function $N$, the shift
vector $\beta^i$ and the 3-metric $\gamma_{ij}$ at first order in amplitude. These expressions can be regularised according to table
\ref{quantities}. Then, and only then, can we use the obtained analytical expressions to build their numerical and spectral
representations.

A strong validation of this procedure consists in verifying that the numerical output is indeed an
\gls{aads} first order solution of the regularised Einstein's system. Denoting by
$\alpha$ the amplitude of the linear geon\footnote{This can be the $\alpha_S$ or $\alpha_V$ defined in chapter \ref{perturbations},
equations \eqref{PhiS} and \eqref{PhiV}.}, we thus expect that the followings quantities behave as $O(\alpha^2)$ (see chapter
\ref{aads} for definitions):
\begin{myitem}
   \item the left-hand side of Einstein's equation $\widetilde{R}_{\alpha\beta} - \gls{Lambda}\widetilde{g}_{\alpha\beta}$,
   \item the boundary conformal Weyl tensor $\widehat{C}_{\alpha\beta\gamma\delta}(r = \gls{L})$,
   \item the boundary pseudo-magnetic Weyl tensor $\widehat{B}_{\alpha\beta}(r = \gls{L})$,
   \item the boundary quasi-local stress tensors $\tau_{\alpha\beta}(r = \gls{L})$.
\end{myitem}
Note that conformal quantities denoted by a hat are not diverging: they directly follow from $\widehat{g}_{\alpha\beta} =
\Omega^2 g_{\alpha\beta}$ and thus are not subject to regularisation. As for the quasi-local stress tensor, it should be zero at
first order at the \gls{ads} boundary as well, as discussed in section \ref{tcftreg}.

If any of this quantity is linear in the amplitude, this means that the solution is not an \gls{aads} Einstein's solution, in the
sense of chapter \ref{aads}. On the left panel of figure \ref{testlin}, we show these quantities for an $(l,m,n) = (2,2,0)$
helically symmetric geon seed. It is clear that the $O(\alpha^2)$ behaviour is met for the above quantities, thus validating both our linear
geon construction and regularisation procedure. Two quantities are only $O(\alpha)$ though: the mean extrinsic curvature $\widetilde{K}$ and the spatial harmonic vector $\widetilde{V}^i$.
Indeed, these quantities are first order in amplitude in general.

We can now bring the linear geon to the \gls{am} gauge, as
explained in section \ref{reglingauge}. At the end of the metric reconstruction procedure, we compute again the above Einstein
and \gls{aads} residuals. The results is shown of the right panel of figure \ref{testlin}. This time, not only the gauge is
satisfied at first order, i.e.\ $\widetilde{K}$ and $\widetilde{V}^i$ are second order in amplitude, but we have not broken the
$O(\alpha^2)$ dependence of the other quantities.

This clearly validates that our initial guess are first order \gls{aads} Einstein's solutions in the \gls{am} gauge.

\begin{myfig}
   \includegraphics[width = 0.49\textwidth]{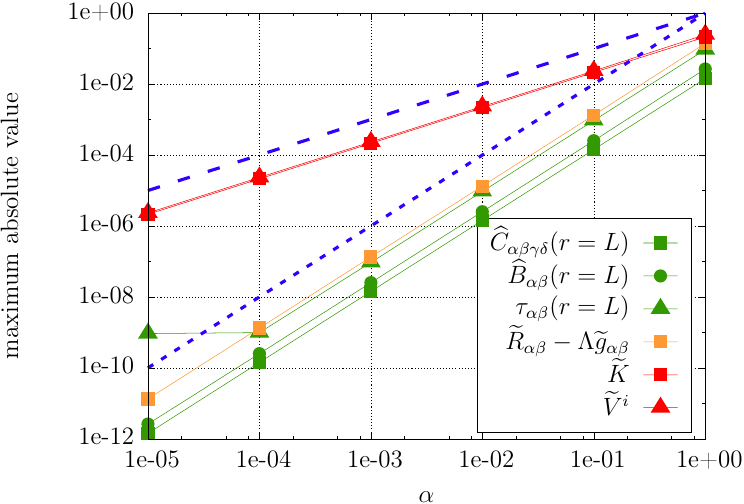}
   \includegraphics[width = 0.49\textwidth]{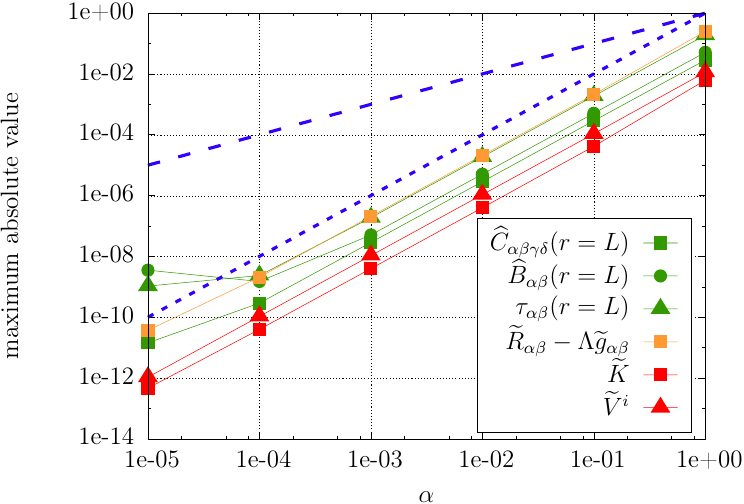}
   \caption[Gauge enforcement on linear geons]{Several residual are represented here: the Einstein's equation residual
      $\widetilde{R}_{\alpha\beta} - \gls{Lambda}\widetilde{g}_{\alpha\beta}$, the boundary conformal Weyl residual
      $\widehat{C}_{\alpha\beta\gamma\delta}(r = \gls{L})$, the boundary conformal pseudo-magnetic Weyl residual
      $\widehat{B}_{\alpha\beta}(r = \gls{L})$ and the boundary quasi-local stress tensor residual $\tau_{\alpha\beta}(r =
      \gls{L})$. By maximum absolute value we intend the maximum value at the collocation points, for all components, in the whole
      numerical volume or only at the \gls{ads} boundary $r = \gls{L}$ when so specified. Left panel: linear $(l,m,n) = (2,2,0)$
      helically symmetric geon as given by the perturbative approach. Right panel: same geon after bringing it into the \gls{am}
      gauge with the procedure of section \ref{reglingauge}. These plots were obtained at a fixed resolution of $(21,5,5)$. In
      dotted lines, we have also represented the curve $f(\alpha) = \alpha$ and $g(\alpha) = \alpha^2$, in order to ease the reading
      of slopes and power-law behaviours. Credits: G. Martinon.}
   \label{testlin}
\end{myfig}

\subsection{Mass and angular momentum tests}
\label{kerr}

Another valuable and independent test consists in comparing charges computed numerically and analytically. Since linear geons have
no mass nor angular momentum, it is relevant to use the analytical Kerr-\gls{ads} metric, uncovered for the first time in
\cite{Carter68} (see however \cite{Gibbons05} for a more immediate reading of the metric components). This allows us to probe our absolute
numerical precision on mass and angular momentum for a large number of different resolutions.

The Kerr-\gls{ads} metric expressed in our isotropic coordinates \eqref{confcoor} reads \cite{Martinon17}
\begin{subequations}
\begin{align}
 ds^2 &= -\frac{\Delta - (1-\rho^2)^4\Delta_{\theta}a^2\sin^2\theta}{(1-\rho^2)^2 \Sigma}dt^2 +
   \frac{4}{(1-\rho^2)^2}\frac{(1+\rho^2)^2\Sigma}{\Delta}dr^2\\
\nonumber   &+ \frac{\Sigma}{(1-\rho^2)^2\Delta_\theta}d\theta^2
   + \frac{\Delta_\theta(4r^2 + a^2(1-\rho^2)^2)^2 - \Delta a^2\sin^2\theta}{(1-\rho^2)^2 \Sigma\Xi^2}d\varphi^2 \\
\nonumber   &- \frac{\Delta_\theta(1-\rho^2)^2(4r^2 + a^2(1-\rho^2)^2) - \Delta}{(1-\rho^2)^2\Sigma\Xi}2a\sin^2\theta dt d\varphi,
\end{align}
\end{subequations}
where we have defined
\begin{subequations}
\begin{align}
   \Delta &= (4r^2 + a^2(1-\rho^2)^2)(1+\rho^2)^2 - 4mr(1-\rho^2)^3,\\
   \Sigma &= 4r^2 + (1-\rho^2)^2 a^2 \cos^2\theta,\\
   \Delta_\theta &= 1 - \frac{a^2}{L^2}\cos^2\theta,\\
   \Xi &= 1 - \frac{a^2}{L^2}.
\end{align}
\end{subequations}
In these expressions, we assumed without loss of generality that the two parameters $m$ and $a$ were real and positive. The
horizon of the black hole lies at $\Delta = 0$.

In such axisymmetric space-times for which $\partial_\varphi^\alpha$ is a Killing vector, a key concept is the one of
\glspl{zamo}. By definition, a \gls{zamo} has a 4-velocity $u^\alpha$ obeying
\begin{equation}
   g_{\mu\nu}u^\mu\partial_\varphi^\nu = 0 \iff u^\alpha = A(\partial_t^\alpha + \omega\partial_\varphi^\alpha),
\end{equation}
where $A$ is a normalisation factor that can be determined with the time-like condition $u_\mu u^\mu = -1$. The $\omega$
parameter translates the frame-dragging effect, and is nothing but the angular velocity of the \gls{zamo}. In Kerr-\gls{ads}
space-times, a boundary \gls{zamo} has actually an angular velocity $\omega = -g_{t\varphi}/g_{\varphi\varphi} \underset{r =
   \gls{L}}{=} -a/L^2$. This has many consequences on the global charges.

As already noticed in \cite{Caldarelli00,Gibbons05}, the parameters $m$ and $a$ are not the thermodynamical mass $M$ and angular
momentum $J$ of the black hole, but are related to them via
\begin{equation}
   M = \frac{m}{\Xi^2} \quad \tn{and} \quad J = -\frac{am}{\Xi^2}.
   \label{ja}
\end{equation}
However, applying either the \gls{amd} or \gls{bk} definitions of charge (see equations \eqref{QAMD} and \eqref{QBK} of chapter \ref{aads}),
a naive analytical computation gives
\begin{equation}
   Q_{\partial_t}[\Sigma] = \frac{m}{\Xi} \quad \tn{and} \quad Q_{\partial_\varphi}[\Sigma] = J.
\end{equation}
The reason why $M \neq Q_{\partial_t}[\Sigma]$ is that the observer whose worldline is attached to $\partial_t^\alpha$ at the \gls{ads}
boundary is not a \gls{zamo}. This implies that charges computed with respect to this worldline take
into account its non-vanishing angular momentum. The correct mass is thus given by the charge attached
to the \gls{zamo} worldline
\begin{equation}
   M = Q_{\partial_t + \omega\partial_\varphi}[\Sigma_t] = Q_{\partial_t}[\Sigma_t] + J\omega(r = \gls{L}).
   \label{zamo}
\end{equation}
On top of that, there is another counter-intuitive but physical effect in Kerr-\gls{ads} space-times: $J$ and\footnote{Recall that
   $a$ is directly proportional to the angular velocity of the horizon, the latter being defined as the angular velocity of a
   \gls{zamo} on the horizon $\Delta = 0$.} $a$ have opposite signs\footnote{For geons,
we also observed that $J$ and $\Omega$ had opposite signs, however for the results presented in this manuscript we changed the
sign of $J$, as it seems common in the literature.} in equation \eqref{ja}. This is because the frame-dragging function $\omega$ is positive near the horizon but
negative and finite at the \gls{ads} boundary, incidentally impacting the sign $J$. As a point of comparison, in asymptotically flat
space-times, $\omega$ is positive everywhere and goes to zero at infinity.

These considerations taken into account, we can test our charge computation numerically by selecting a particular Kerr-\gls{ads}
configuration with, say, $m/\gls{L} = 1$ and $a/\gls{L}^2 = 0.5$. The analytical charges are then $M^0/\gls{L} = 16/9$ and
$J^0/\gls{L}^2 = -8/9$. We then compute numerically both \gls{amd} and \gls{bk} charges taking into account \eqref{zamo}. For this
particular configuration, the horizon lies at $\sim 0.37\gls{L}$, so we use only one domain describing $r\in[0.5,1]\gls{L}$ to avoid the
coordinate singularity at the horizon.

In figure \ref{kerrtest}, we show the relative difference between analytical and numerical charges as a function of radial resolution.
We fixed the angular resolution to nine points per octant. It is clear on this plot that both
\gls{amd} and \gls{bk} charges converge exponentially to the analytical value up to $N_r = 37$-$41$, after which rounding errors start to
increase. It can also be noticed that, at fixed resolution, \gls{bk} charges are less precise than \gls{amd} ones, because of a more involved regularisation
procedure (section \ref{tcftreg}). Furthermore, our precision saturates at $\sim10^{-6}\%$, so that our absolute precision is around $\sim10^{-8}$ at a
resolution of $(37,9,9)$. This is quite large for an analytical metric, but we can't do much better in double precision
arithmetics, since we need to perform many spectral operations like second order derivatives, divisions in coefficient space as
well as surface integration.

\begin{myfig}
   \includegraphics[width = 0.49\textwidth]{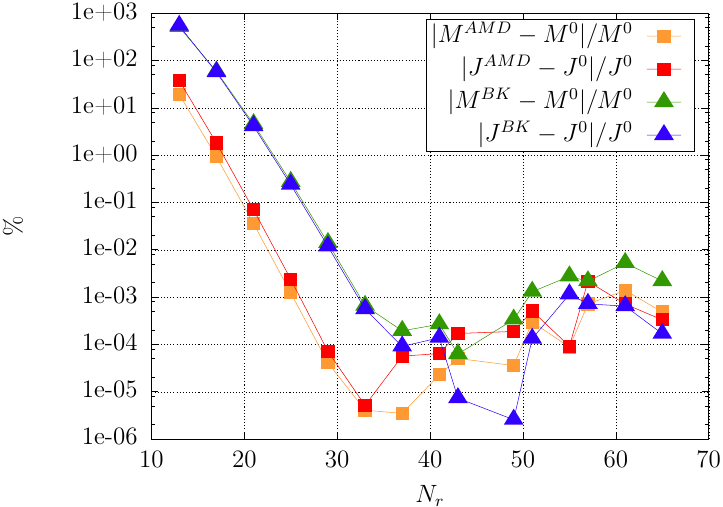}
   \caption[Numerical tests with the Kerr-AdS metric]{Relative differences between numerical \gls{amd}, \gls{bk} and analytical charges for the
   Kerr-\gls{ads} metric ($m/\gls{L} = 1$, $a/\gls{L}^2 = 1/2$) as a function of radial resolution. Angular resolution is fixed to
   $N_\theta = N_\varphi = 9$ assuming an octant symmetry. Credits: \cite{Martinon17}.}
   \label{kerrtest}
\end{myfig}

These results are nonetheless a strong validation of our regularisation procedure, notably for the quasi-local stress tensor
involved in the \gls{bk} mass (section \ref{tcftreg}) as well as of our \gls{amd} charge determination.

\section{Numerical setup}
\label{numproc}

Now that we have tested our numerical code, we can start our construction of fully non-linear geons. The idea is to
start from a first order initial guess and then to compute a full sequence by increasing step by step an amplitude parameter
called wiggliness. Additionally, there are a number of a posteriori validations of the solutions that have to be met.

\subsection{Numerical algorithm}

In order to solve systems of differential equations, we use the open source KADATH library \cite{kadath,Grandclement10}, which provides
a C++ interface for solving relativistic systems with multi-domain spectral methods. The vast range of problems this
library is able to solve (e.g.\ vortons, critical collapse, binary black holes initial data \cite{Grandclement10,Uryu12}, oscillatons
\cite{Grandclement11}, \gls{aads} scalar breathers \cite{Fodor14,Fodor15}, boson stars
\cite{Grandclement14,Meliani15,Vincent16a,Meliani16,Grandclement17} and \gls{aads} geons \cite{Martinon17}) make it reliable and
robust for our purposes. In order to solve problems in \gls{aads} space-times, we had to develop in particular the regularisation
scheme of section \ref{reg}. The library manages non-linear systems with a Newton-Raphson scheme coupled to automatic
differentiation \cite{Grandclement10}, which makes it flexible to address many different problems.

In order to construct non-linear numerical geons, we proceed as follows.
\begin{myenum}
   \item We analytically construct a helically symmetric first order geon with the results of chapter \ref{perturbations},
      and express the perturbative solution in the co-rotating frame in isotropic coordinates. The linear geon then admits a helical
      Killing vector that is simply the generator of our time coordinate $t'$ in this frame. Namely, its expression is
      \begin{equation}
         \partial_{t'}^\alpha = \partial_t^\alpha + \frac{\omega}{m}\partial_{\varphi}^\alpha,
         \label{killingvector}
      \end{equation}
      such that $\partial_{t'} g_{\alpha\beta} = 0$ and $\mathcal{L}_m = -\mathcal{L}_{\widetilde{\beta}}$ in \eqref{eamreg}.
   \item After having chosen a suitably small amplitude, the linearised first order geon can be brought to the \gls{am} gauge with
      the procedure described in section \ref{reglingauge}.
   \item The resulting first order geon in the \gls{am} gauge is then used as an initial guess for the full $3+1$
      regularised \gls{eam} system of ten equations \eqref{eamreg} whose ten unknowns are
      $N$,$\beta^i$,$\gamma_{ij}$. As suitable boundary conditions at the \gls{ads} boundary, we choose to enforce Dirichlet
      conditions
      \begin{equation}
         \widetilde{N} \mathrel{\widehat{=}} \widetilde{\overline{N}},\quad \widetilde{\beta}^i \mathrel{\widehat{=}} \frac{\omega}{m}\partial_{\phi}^i, \quad \widetilde{\gamma}_{ij} \mathrel{\widehat{=}} \widetilde{\overline{\gamma}}_{ij}.
         \label{BC}
      \end{equation}
      The condition on the shift merely translates that we work in a frame that is co-rotating with the
      geon. Since $\omega$ is expected to change with the geon amplitude, it is treated as an additional unknown of the system,
      solved together with the metric components in the numerical solver. Meanwhile we provide an additional equation that forces
      a marching parameter, or geon wiggliness $w$, to take a user-defined value. This wiggliness parameter translates the idea of
      amplitude of non-linear solutions. It can be for example the value of some spectral coefficient. The Newton-Raphson
      algorithm of the KADATH library is then in charge of finding the nearest fully non-linear solution. The iterative procedure
      is stopped when the error, measured as the highest coefficient of the Einstein equation residuals, reaches about $10^{-8}$.
   \item Once a numerical and non-linear solution is obtained, it is used as an initial guess for the \gls{eam} system with a
      $w$ fixed to a slightly incremented value. The system then converges to the nearby solution with this
      wiggliness requirement. Repeating the process allows us to build sequences of geons parametrised by larger and larger $w$.
\end{myenum}
This methodology is illustrated in figure \ref{methodo}.

\begin{myfig}
      \includegraphics[width = 0.49\textwidth]{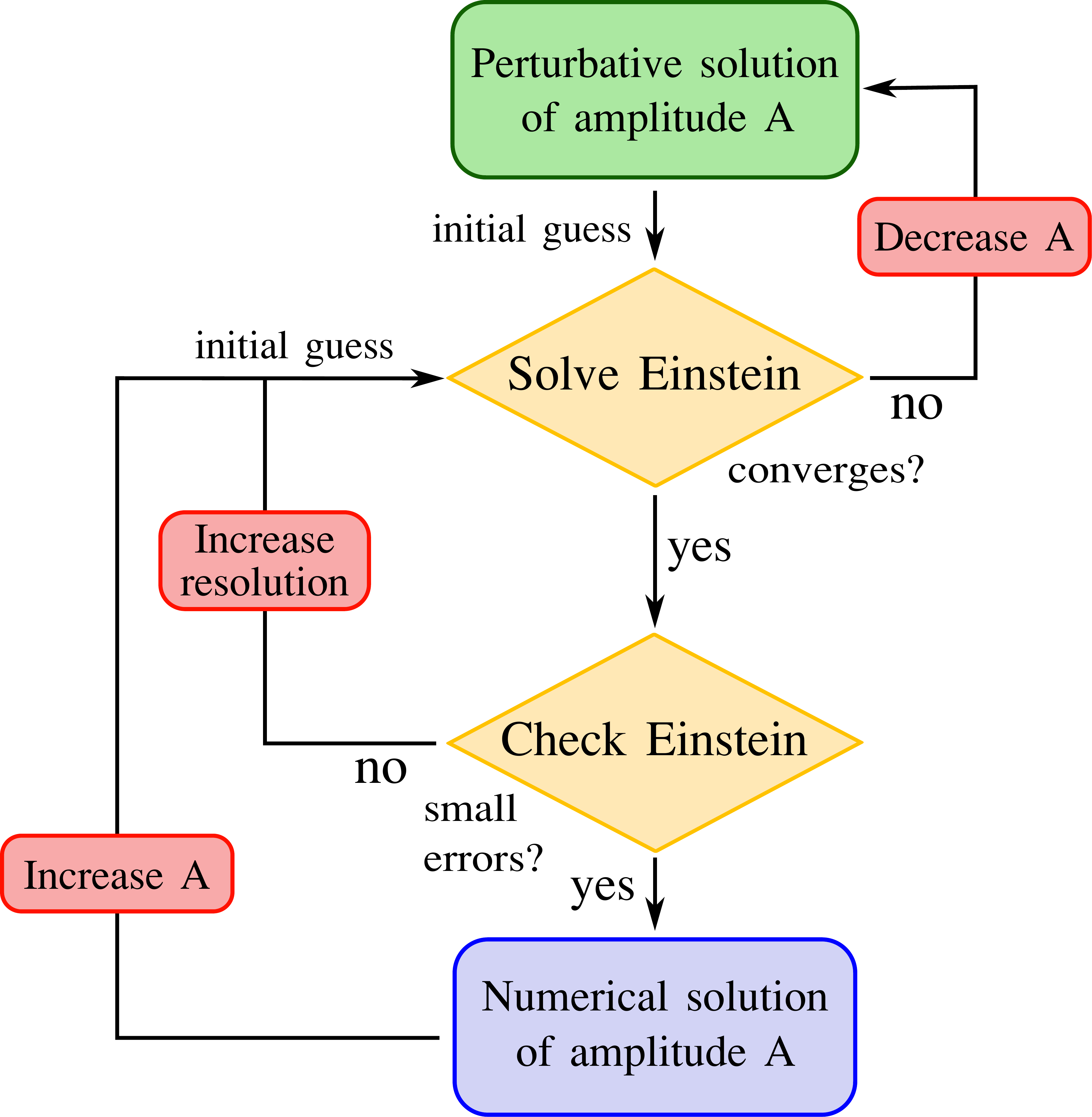}
      \caption[Numerical methodology]{We start with a first order in-gauge initial guess (green box). If the amplitude is too
      large, it could be that we are out of the basin of attraction of the Newton-Raphson algorithm (upper yellow diamond). Else,
      if the initial guess is small
   enough and admits a non-linear extension, the code converges to a fully non-linear solution of the same amplitude
(wiggliness). The frequency $\omega$ is solved together with the metric components to match this amplitude requirement. The errors can then be decreased by increasing the resolution, down to sufficiently small values. Once a non-linear
solution of fixed amplitude with small errors is obtained (blue box), we can use it as an initial guess for the near solution featuring a
slightly increased amplitude requirement. This is how a full sequence is built step by step. Credits: G.  Martinon.}
   \label{methodo}
\end{myfig}

\subsection{Precision monitoring}
\label{prec}

In numerical physics, working at only one resolution gives few information on the solution. We thus build several sequences of geons at different
resolutions. In particular, if a physical quantity should be zero in theory, in practice this means that this quantity is
\textit{tending} to zero when the numerical resolution is increased. Within spectral methods in particular, this convergence should
be exponential with the number of coefficients involved. Thus, in order to monitor the precision of our numerical results, various
tests can be performed.
\begin{myenum}
   \item \textbf{Spectral convergence:} if the metric components are well described by the spectral expansion, the
      coefficients should decrease exponentially. With double precision arithmetics and a second order differential system of
      equations, the saturation level is expected to be around $10^{-10}$.
   \item \textbf{Gauge residual:} $\widetilde{K}$ and $\widetilde{V}^i$ should be as low as possible (but are expected to saturate at a $10^{-10}$
      level). Their infinity norm should decrease exponentially with increasing numerical resolution. An other complementary check
      in the \gls{eam} framework consists in observing a similar convergence for the components of $\widetilde{R}_{\alpha\beta} -
      \gls{Lambda} \widetilde{g}_{\alpha\beta}$ which should be zero for any solution of Einstein's equation in vacuum.
   \item \textbf{\gls{aads} space-times:} we chose to enforce Dirichlet boundary conditions on the system, however this might
      not be enough to ensure the right asymptotics. According to chapter \ref{aads}, we can check that $(i)$
      $\widehat{C}_{\alpha\beta\mu\nu}$, $\widehat{B}_{\alpha\beta}$ and $\tau_{\alpha\beta}$ have boundary values
      decreasing exponentially to zero with increasing resolution, and $(ii)$ that the \gls{amd} and \gls{bk} charges
      converge to each other exponentially with increasing resolution.
   \item \textbf{Agreement with perturbative approach:} any numerical sequence of geon should coincide with perturbative
      results for low enough amplitudes. In particular, the global charges of our numerical results should match the perturbative
      predictions for small enough marching parameters.
\end{myenum}

\section{Results}
\label{res}

We are now able to present our results, described in \cite{Martinon17}. We build five different families of geons,
including radially excited ones. Some of our results are compared to higher-than-first order perturbative computations, that are
summarised in \cite{Martinon17}.

\subsection{Geons with $(l,m,n) = (2,2,0)$}

Following the procedure of section \ref{numproc}, we first consider geons with excitation numbers $(l,m,n) = (2,2,0)$, i.e.\
helically symmetric geons in their fundamental state. We are able to reach unprecedented amplitudes, exhibiting deviations from
third order perturbative expansion as large as $50\%$. Furthermore, we use this family of geons as a testbed for our numerics. 

Let us first introduce the difference between the geon metric and the \gls{ads} background
\begin{equation}
   h_{\alpha\beta} \equiv g_{\alpha\beta} - \overline{g}_{\alpha\beta} \quad \tn{and} \quad \widetilde{h}_{\alpha\beta} \equiv \Omega^2 h_{\alpha\beta}.
\end{equation}
It turns out that the perturbed metric component $\widetilde{h}_{xx}$ has a egg-shape in the \gls{am} gauge, as pictured in figure
\ref{hxxgallery}. We thus choose our marching parameter, or wiggliness, to be
\begin{equation}
   w \equiv \widetilde{h}_{xx}(r=0).
   \label{w220}
\end{equation}
Note that in this equation, since $\Omega = 1$ at the origin, we have $\widetilde{h}_{xx}(r=0) = h_{xx}(r=0)$. A sequence
typically starts at a small amplitude, and hence small wiggliness $w = 0.1$, and finishes at $w \sim 13$. As a point of comparison,
the \gls{ads} background metric displays a background wiggliness of $\widetilde{\overline{g}}_{xx}(0) = 4$.

Since the extrinsic curvature tensor $\widetilde{\Theta}_{\alpha\beta}$ of $\Sigma_r$ (involved in the \gls{bk} charge
computation \eqref{QBKreg}) diverges
like $O(r^{-1})$ near the origin, and is only needed at the \gls{ads} boundary $r = \gls{L}$, we use two spectral domains: one
nucleus describing $r\in[0,0.5]L$ and one shell describing $r\in[0.5,1]L$. This allows us to compute the boundary quasi-local stress tensor
$\tau_{\alpha\beta}$ with equation \eqref{taupar} only in the shell domain. We checked that the results were almost insensitive to the
position of the domain separation, as expected for the global representation of smooth fields in spectral methods.

Furthermore, since the quantum numbers $l$ and $m$ are both even, there is an octant symmetry. Accordingly, quoting a resolution
of, say $(37,9,9)$, means that, in each domain, one uses $N_r = 37$ points in the $r$ coordinate and $N_\theta = N_\varphi = 9$
points per octant in the $\theta$ and $\varphi$ coordinates.

\begin{myfig}
      \includegraphics[width = 0.49\textwidth]{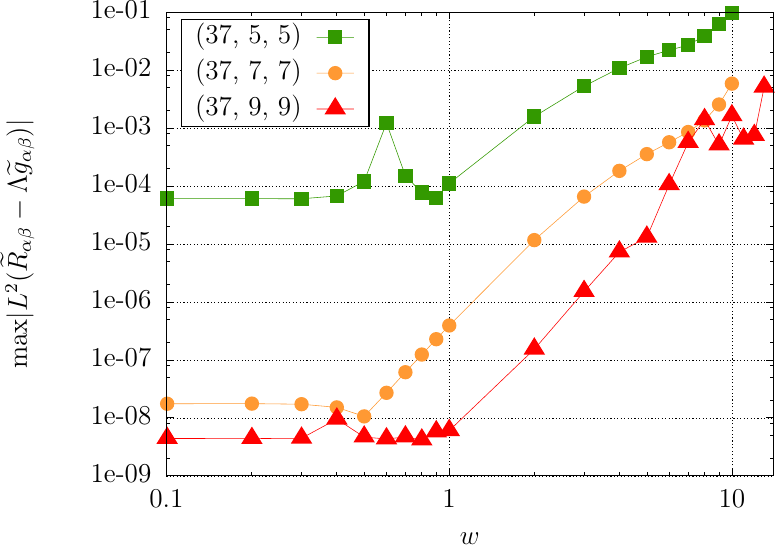}
      \includegraphics[width = 0.49\textwidth]{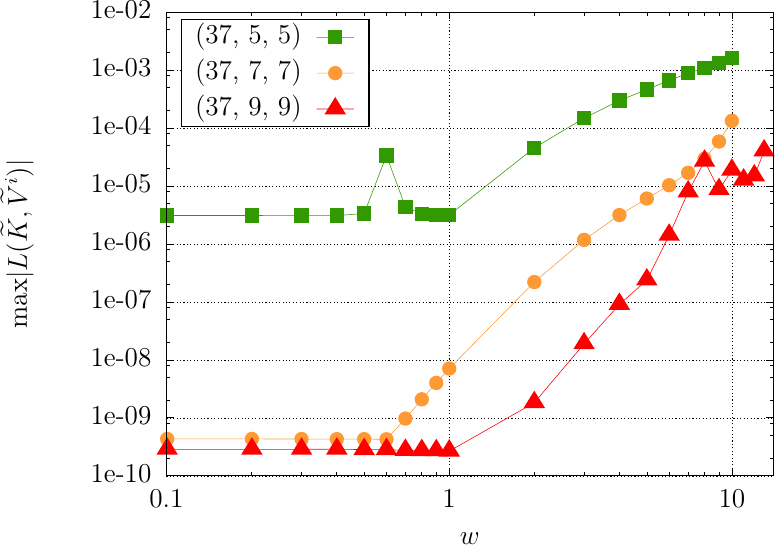}\\
      \includegraphics[width = 0.49\textwidth]{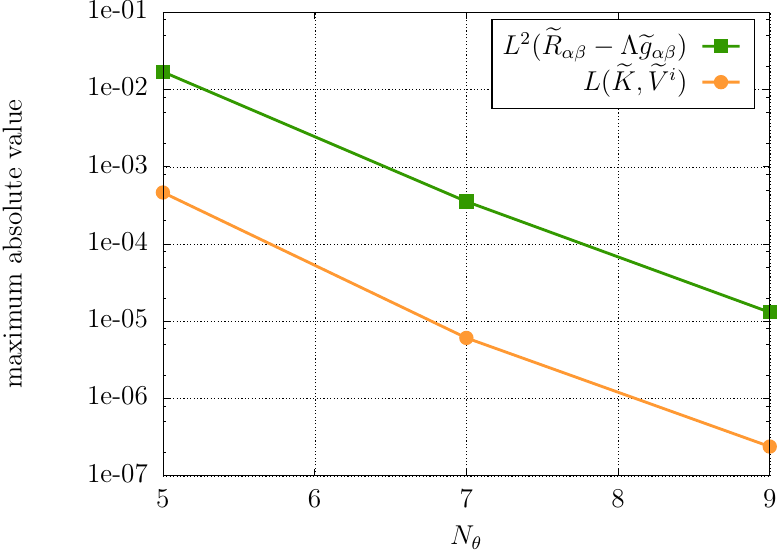}
      \caption[Einstein residuals for $(l,m,n)=(2,2,0)$ geons]{Top left panel: regularised Einstein residual
      $\widetilde{R}_{\alpha\beta} - \gls{Lambda} \widetilde{g}_{\alpha\beta}$. Top right panel: regularised
      gauge vector residual $(\widetilde{K},\widetilde{V}^i)$. Bottom panel: spectral convergence of the Einstein and gauge
      residuals as a function of the angular resolution at fixed wiggliness $w = 5$. Each residual is the maximum value in the
      whole configuration space, i.e.\ at collocation points. These plot describe the $(l,m,n) = (2,2,0)$ geons. Credits: G.
   Martinon.}
   \label{eingauge22}
\end{myfig}

As a first validation of our solutions, we verify that the computed geons well satisfy the \gls{am} gauge. In figure \ref{eingauge22}, we
show how the Einstein and gauge residuals vary with amplitude and numerical resolution.
The data points being almost insensitive to radial resolution in the range $N_r \in [29,37]$, we only show the angular resolution
dependence.

The residuals are increasing with the amplitude of the geon, but decrease exponentially by several orders of magnitude
with increasing angular resolution, indicating spectral convergence. This confirms that our solutions not only respect the
\gls{am} gauge but also obey the full Einstein's system. At a resolution of $(37,9,9)$, we are able to lower the Einstein residual
down to $\sim 10^{-3}$ for our most massive configuration. This is to be compared to the largest metric coefficient which can
reach $\sim 15$ at the origin.

\begin{myfig}
   \includegraphics[width = 0.49\textwidth]{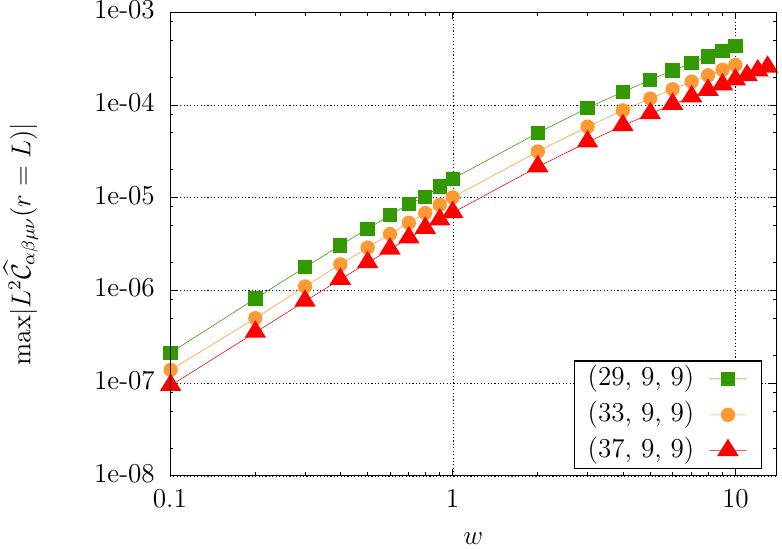}
   \includegraphics[width = 0.49\textwidth]{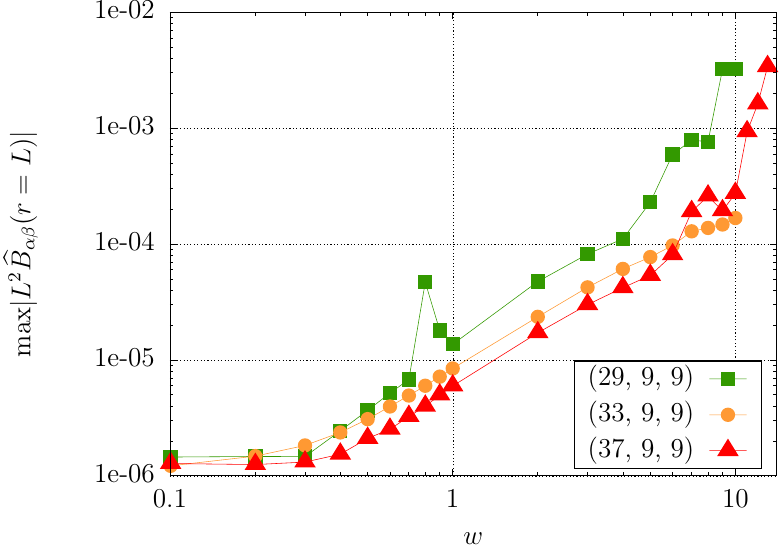}
   \includegraphics[width = 0.49\textwidth]{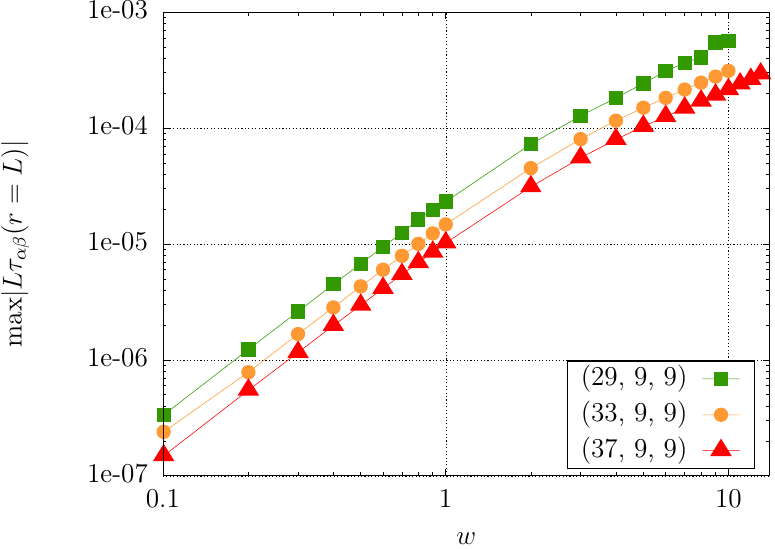}
   \includegraphics[width = 0.49\textwidth]{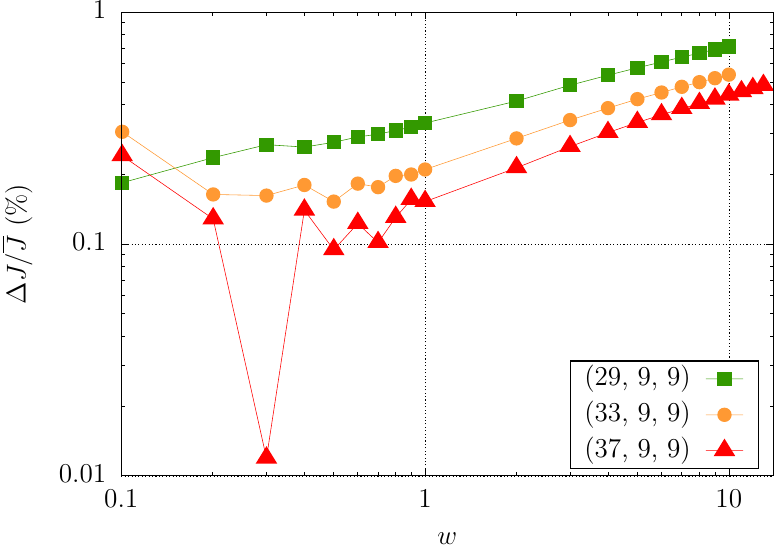}
   \caption[AAdS residuals for $(l,m,n) = (2,2,0)$ geons]{\gls{aads} residuals for the $(l,m,n) = (2,2,0)$ geons. Upper left
   panel: boundary conformal Weyl tensor residual $\widehat{\mathcal{C}}_{\alpha\beta\mu\nu}$. Upper right panel: boundary conformal
   pseudo-Weyl tensor residual $\widehat{B}_{\alpha\beta}$. Bottom left panel: boundary quasi-local stress tensor
   $\tau_{\alpha\beta}$. Bottom right panel: relative difference in percent between the \gls{amd} and \gls{bk} angular momenta,
   with $\Delta J = J^{AMD} - J^{BK}$ and $\overline{J} = (J^{AMD} + J^{BK})/2$. By boundary residual we intend the maximum value
   of the field restricted to collocation points of the \gls{ads} boundary. Credits: \cite{Martinon17}.}
   \label{aads22}
\end{myfig}

A second validation of the solutions consists in verifying the \gls{aads} asymptotics. In particular, and as discussed in section
\ref{prec}, the boundary residual of the Weyl and pseudo-magnetic Weyl tensors have to vanish on the boundary, as well as the
quasi-local stress tensor. We should also observe a vanishing of the difference between \gls{amd} and \gls{bk} charges.

In figure \ref{aads22}, we show how these quantities vary with amplitude and resolution. The angular resolution
has essentially no effect in the range $N_{\theta,\varphi}\in[5,9]$, so we only show the radial resolution declination. The higher the
radial resolution, the closer to zero they are, which shows that our solutions are well \gls{aads}. We have checked
that the decrease, even if slow, was nevertheless exponential. The bottom right panel
shows that our \gls{amd} and \gls{bk} charges agree with each other at a $\sim 0.5\%$ level that also decreases exponentially with
resolution. This is an extra confirmation that our solutions display indeed \gls{aads} asymptotics.

The convergence rates of the \gls{aads} residuals seems to be quite slower than those of the Einstein and gauge residuals of
figure \ref{eingauge22}. However, this is to be expected: these indicators are quite demanding in terms of precision and undergo
an involved regularisation procedure (section \ref{reg}). Notably, several expensive (in terms of accuracy) operations are at
play: second order derivatives, divisions in coefficient space and evaluation and integration at the \gls{ads} boundary.

\begin{myfig}
   \includegraphics[width = 0.49\textwidth]{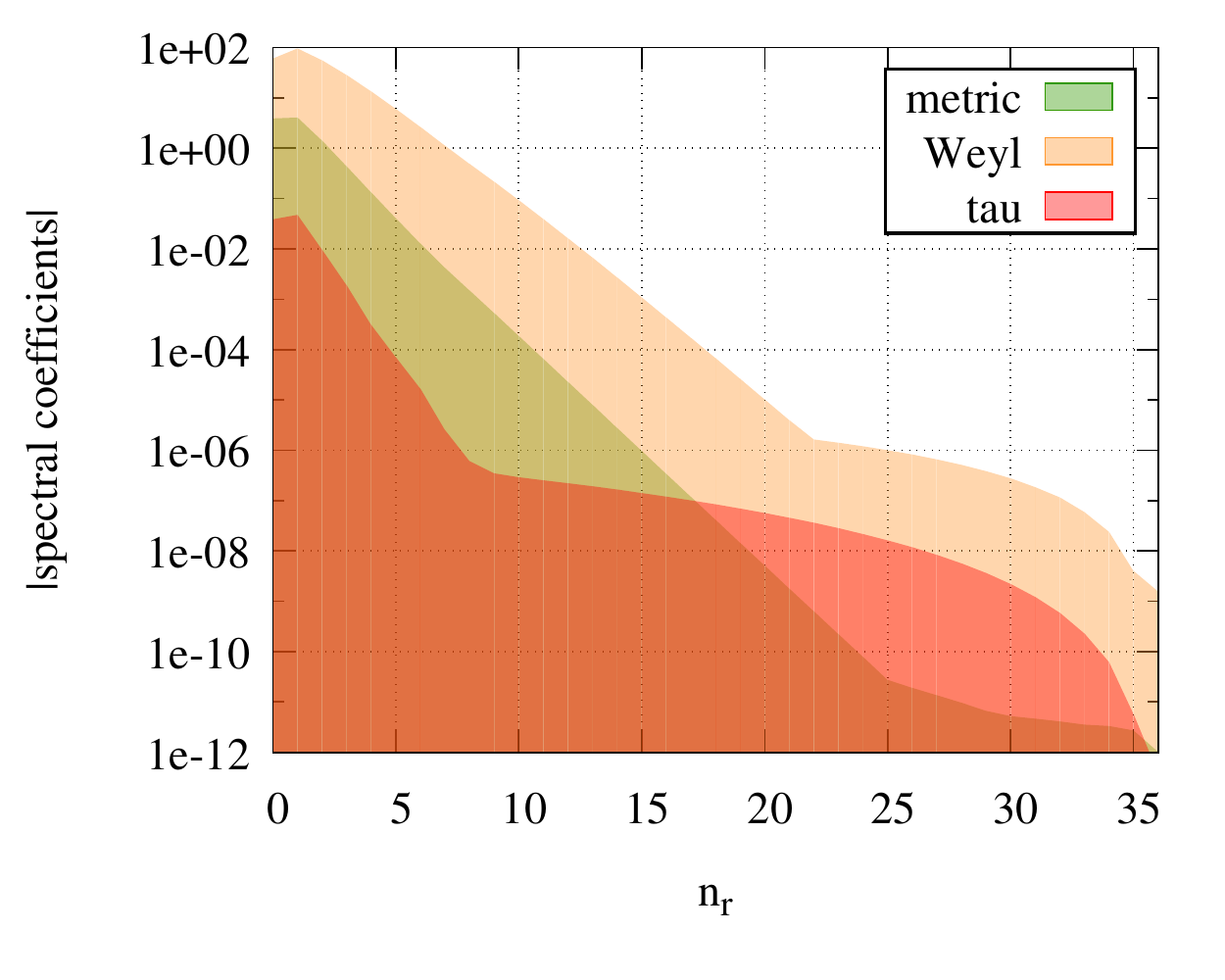}
   \includegraphics[width = 0.49\textwidth]{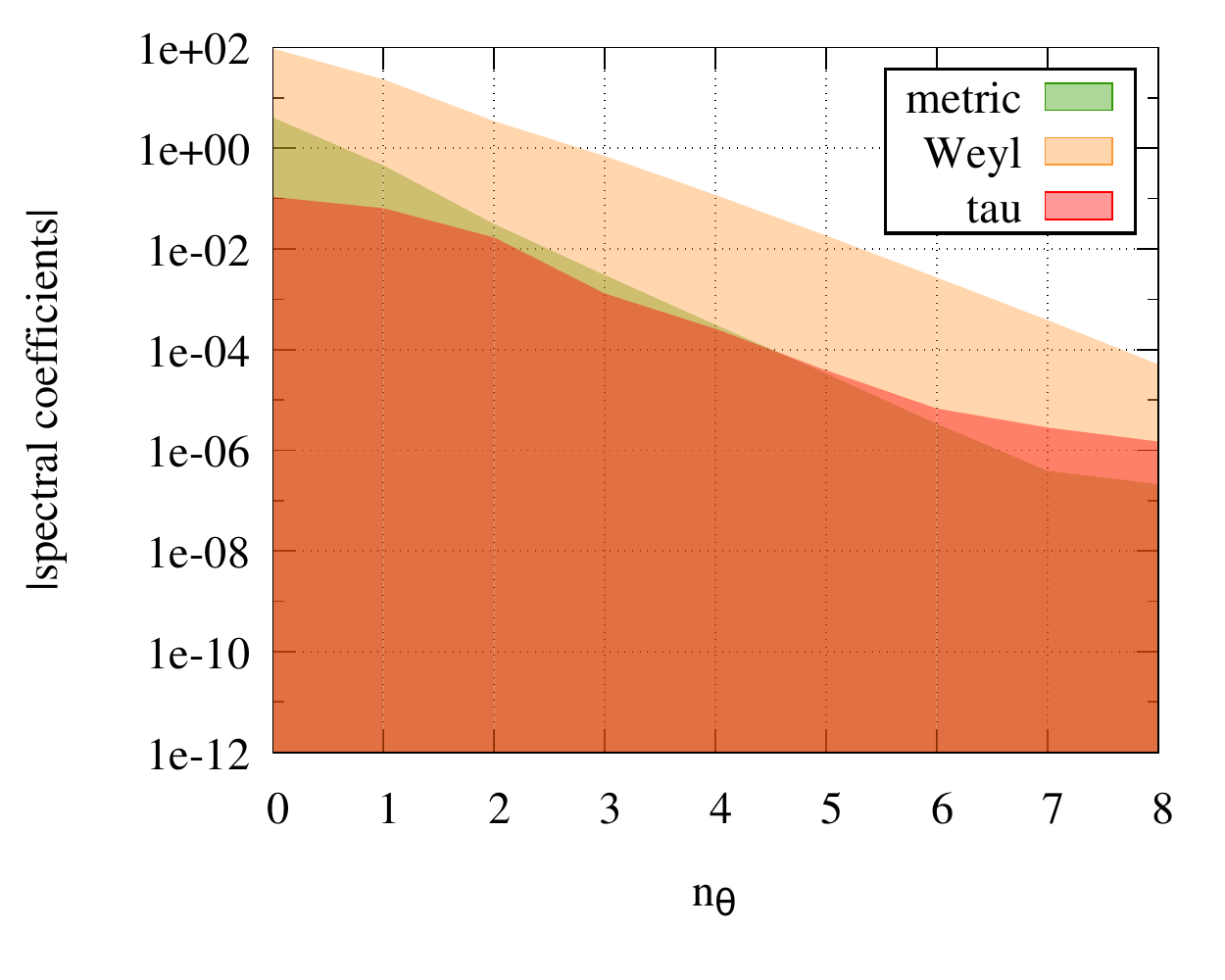}
   \caption[Spectral coefficients of $(l,m,n)=(2,2,0)$ geons]{Spectral coefficients of the regularised metric $\widetilde{g}_{\alpha\beta}$ in green,
   of the conformal Weyl tensor $\widehat{\mathcal{C}}_{\alpha\beta\mu\nu}$ in yellow and of the quasi-local stress tensor
   $\tau_{\alpha\beta}$ in red for the $(l,m,n) = (2,2,0)$ geons. The resolution is fixed at $(37,9,9)$ and the amplitude at $w = 10$. Our
   coefficients collection is actually a three-dimensional array indexed by three integers $n_r \in \{0,\cdots,N_r\}$, $n_\theta
   \in \{0,\cdots,N_\theta\}$ and $n_\varphi \in \{0,\cdots,N_\varphi\}$. Left panel: coefficients versus $n_r$ for arbitrary
   $n_\theta$ and $n_\varphi$. Right panel: coefficients versus $n_\theta$ for arbitrary $n_r$ and $n_\varphi$. For the sake of
   clarity, only the upper envelope of the coefficients collection is shown. Credits: \cite{Martinon17}.}
   \label{coefs22}
\end{myfig}

As a third validation of our solutions, we check that each field features a spectral (i.e.\ exponential) decrease of its
coefficients. In figure \ref{coefs22}, we show the coefficients of our solution with large amplitude $w=10$ and highest resolution
$(37,9,9)$ for the three tensors $\widetilde{g}_{\alpha\beta}$, $\widehat{C}_{\alpha\beta\gamma\delta}$ and $\tau_{\alpha\beta}$.
In particular, the last two are involved in the \gls{amd} and \gls{bk} charges respectively.

The sought-after spectral convergence is clearly visible both radially and angularly until a threshold is reached. This saturation
threshold is larger for the Weyl tensor $\widehat{\mathcal{C}}_{\alpha\beta\mu\nu}$ and the quasi-local stress tensor
$\tau_{\alpha\beta}$ because they involve second order derivatives of the metric and, for the latter, a regularisation procedure
(section \ref{tcftreg}) that increases numerical errors and hence the noise level.

If it were not too computationally demanding, we could increase the angular resolution to describe better the coefficient tail.
It would probably decrease the errors on Einstein's equation and gauge residuals, according to figure \ref{eingauge22}.

\begin{myfig}
   \includegraphics[width = 0.49\textwidth]{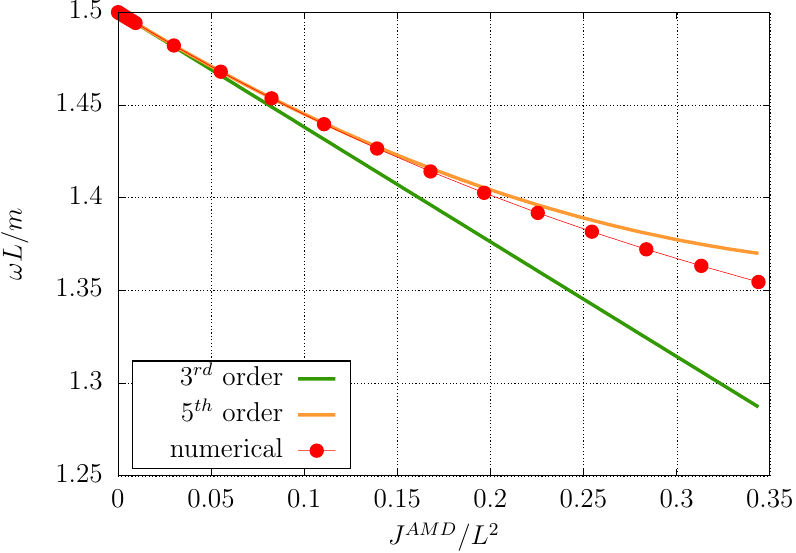}
   \includegraphics[width = 0.49\textwidth]{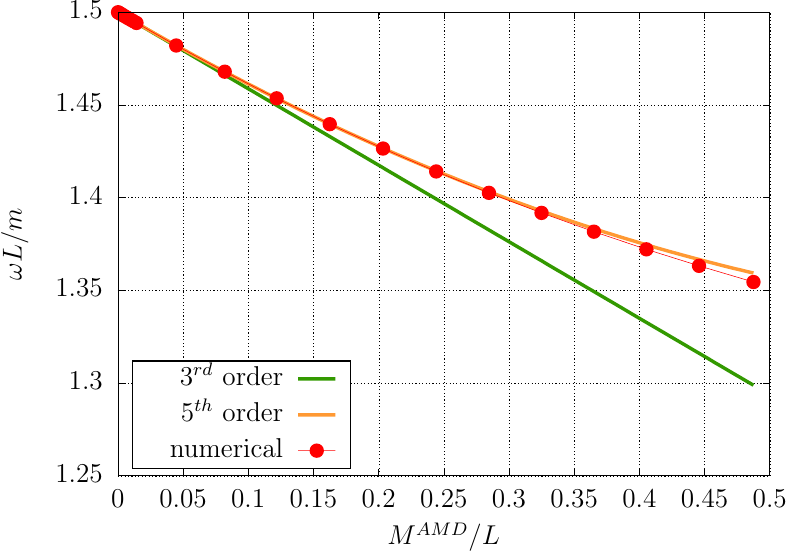}\\
   \includegraphics[width = 0.49\textwidth]{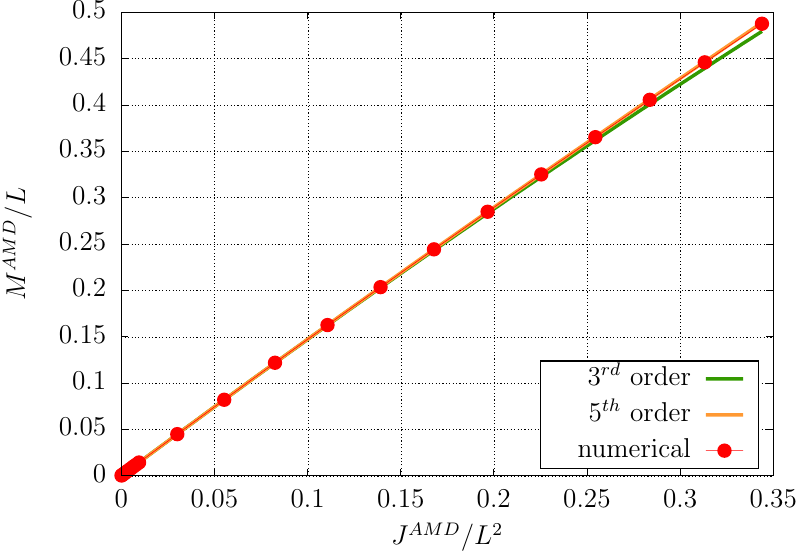}
   \includegraphics[width = 0.49\textwidth]{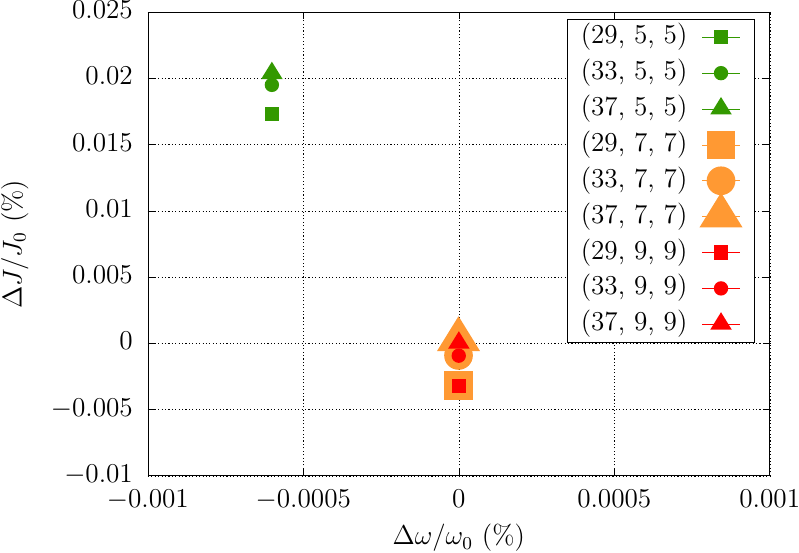}
   \caption[Global charges of $(l,m,n) = (2,2,0)$ geons]{The $\omega$-$M$-$J$ planes for our numerical sequences $(l,m,n) =
      (2,2,0)$ at a resolution of $(37,9,9)$. We only show the \gls{amd} definitions of charges. The successive perturbative
      orders are shown in comparison with our numerical results. At the bottom right panel, we have chosen a typical
      point of the $(37,9,9)$ sequence featuring a wiggliness of $w_0 = 5$, an angular frequency of $\omega_0 L/m = 1.44$ and an angular
      momentum of $J_0 = 0.11$. The plot shows the relative difference $\Delta \omega = \omega - \omega_0$ and $\Delta J = J -
      J_0$ in percentage at several radial and angular resolutions. The point corresponding to the highest resolution $(37,9,9)$
   being taken as a reference, it naturally lies at the origin of the plot. Credits: \cite{Martinon17}.}
   \label{amd22}
\end{myfig}

When it comes to computing global charges, it turns out that the precision on the mass (be it \gls{amd} or \gls{bk}) is less than that on
the angular momentum. In particular, even if the angular momentum $J$ is in a very strong agreement with the perturbative approach in the low amplitude
limit, the mass $M$ is usually overestimated by $\sim5$-$10\%$ depending on the radial resolution. Increasing the number of points improves
the match but very slowly. Our guess is that we lose precision on $M$ because of the numerous steps of regularisation and
spectral operations that all contribute to the accumulation of numerical errors. The reason of such an asymmetry between $M$ and $J$ is unclear, but
probably the terms involved in the computation of $J$ are simpler than those involved in the computation of $M$. We think that, if
affordable, quadruple precision could improve this point.

Thus, in order to give reliable results on masses within reasonable computing times, we compute $M$ using the first law of geon dynamics
\begin{equation}
   \delta M - \frac{\omega}{m} \delta J = 0,
   \label{firstlaw}
\end{equation}
which relates the variations of mass with those of angular momentum. This relation has been
demonstrated in the asymptotically flat case for helically symmetric systems in \cite{Friedman02}. In this case, the charges were the
\gls{adm} ones. A similar result holds for Kerr-\gls{ads} with an additional entropy
term \cite{Gibbons05}. Even if a sketch of a proof of this first law was presented in \cite{Horowitz15}, a rigorous
mathematical demonstration in the general \gls{aads} helically symmetric case is still missing, to the best of our knowledge.
Nevertheless the first law is widely accepted and actually confirmed by perturbative results up to sixth order
\cite{Martinon17,Dias12a,Horowitz15}.

In practice, we write the integral version of \eqref{firstlaw}
\begin{equation}
   M = \int_0^J \frac{\omega(J')}{m}dJ',
\end{equation}
where the function $\omega(J)$ is obtained by a polynomial fit (reduced $\chi^2 < 10^{-13}$). This ensures that $M$ is computed
with as much precision as $\omega$ and $J$ are.

In figure \ref{amd22} we show the three global quantities of importance: the angular velocity $\omega/m$ (with $m=2$), the
\gls{amd} mass $M$ and the \gls{amd} angular momentum\footnote{Recall that the \gls{amd} charges are in general easier to compute
and more precise (section \ref{kerr}). Furthermore, figure \ref{aads22} already demonstrated that the \gls{amd} and \gls{bk}
charges were equivalent to the $\sim 0.5\%$ level.} $J$. The numerical results are given at our best resolution $(37,9,9)$. They are compared to
analytical ones which were obtained with a fifth order perturbative approach, whose results are also presented in
\cite{Martinon17}. On these plots, it appears unambiguously that successive orders of perturbations are closer and closer to our
numerical solutions, which is again a strong validation of our fully non-linear results.

In order to estimate the numerical error bars, we examine a typical point of the sequence with fixed wiggliness $w = 5$,
and compute the solution at a dozen of different resolutions. Taking as reference values the highest one $(37,9,9)$, we can
observe how the global quantities vary with resolution. The differences on $\omega$ and $J$ are pictured on the bottom right panel
of figure \ref{amd22}.

First, it is clear that, when the numerical resolution is increased, global quantities are converging to the highest resolution
results, which lies at the origin of the plot. Second, this allows us to read off error bars on $J$ and $\omega$, namely $\Delta
J \sim 0.02 \%$ and $\Delta \omega \sim 0.0006 \%$ between our worst and best resolutions. Restricting the angular resolutions
between seven and nine angular points (hardly distinguishable on the figure) gives $\Delta J \sim 0.003 \%$ and $\Delta \omega
\sim 0.000001 \%$. Furthermore, we observed that these values remained approximately constants along the entire sequence. Such
magnitudes are obviously invisible at naked eye on the three other panels of figure \ref{amd22}.

A potential explanation of these very small error bars is the following. The initial geon seed is a scalar spherical harmonic
$\mathbb{S}_{22}$. At first order, because of the periodic $\varphi$ dependence of this harmonic, there is no contribution to the
surface integrals involved in the computation of mass and angular momentum. At second order though, there are terms proportional to
$(\mathbb{S}_{22})^2$. With simple trigonometric multiplication formulas, it can be shown that this generates terms
\footnote{This is highly reminiscent of the combination of spins in quantum mechanics.} with $l=0$ and $l=1$ harmonics. At even
higher order, even more harmonics are generated just by multiplication of the precedent ones. However, it turns out that only the
$l=0$ harmonics display a non-vanishing mass integral, and only the $l=1$ harmonics ensure a non-vanishing angular momentum
integral. This is so because all the others harmonics give rise to sinusoidal integrands which vanish after integration. In terms of
spectral representation, this means that global charges are utterly determined by the very first coefficients of the non-linear
solution. When the coefficients are decreasing exponentially, as is our case (figure \ref{coefs22}), it is expected that higher
coefficients contribute marginally to the global charges. This may explain the very minor influence of resolution on mass and angular momentum.

\begin{myfig}
   \includegraphics[width = 0.49\textwidth]{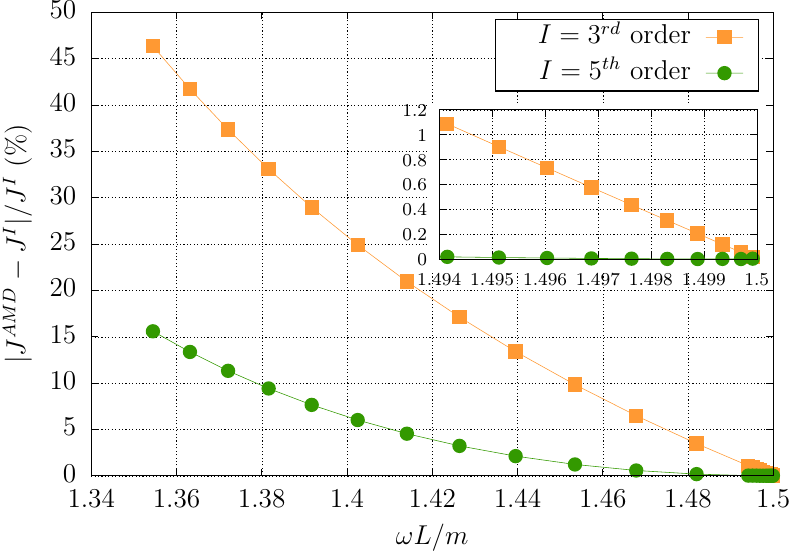}
   \caption[Comparison between numerical and perturbative geons]{Difference in percentage between the numerical angular
      momentum $J^{AMD}$ for $(l,m,n) = (2,2,0)$ geons at resolution $(37,9,9)$ and perturbative results at $I^{th}$ order, $I$ being 3 (yellow squares) or 5
      (green circles). Inset: zoom on the low amplitude limit $\omega \gls{L}/m \to 3/2$. Credits: \cite{Martinon17}.}
   \label{4order22}
\end{myfig}

Now that we have estimated the error bars on the numerical global charges, we can perform the comparison with the perturbative
approach more easily. In figure \ref{4order22}, we show the difference between the numerical results and perturbative predictions.
In particular, this plot clearly demonstrates that our numerical configurations agree with the small expansion procedure in the
zero-amplitude limit. The agreement is even better with fifth order expansion than third order. At the highest amplitudes, our
non-linear solutions deviate from $3^{rd}$ order by at most $50\%$ and from $5^{th}$ order by at most $15\%$, emphasising that we
are probing amplitudes that are much beyond reach of perturbative techniques.

The results of figure \ref{amd22} suggest to write $\omega$ as a function of $J$. If we consider the expansion $\omega \gls{L}/m = f(J)$
\begin{equation}
   \frac{\omega \gls{L}}{m} = a_0 + a_1 \frac{J}{\gls{L}^2} + a_2 \frac{J^2}{\gls{L}^4} + a_3 \frac{J^3}{\gls{L}^6} + \cdots,
   \label{omjexpansion2}
\end{equation}
the $a_i$ coefficients can be computed by a polynomial fit of the curves of our figure \ref{amd22}. In table \ref{tabcoefs22} we show
the obtained coefficients either with perturbative arguments or by fitting our numerical results. Again, some details about
the high order perturbative results are available in \cite{Martinon17}.

\begin{mytab}
\begin{tabular}{lcccc}
\hline
              & $a_0$           & $a_1$           & $a_2$           & $a_3$           \\
\hline
perturbative  & $1.5000000$     & $-0.619062$     & $0.70049$       &  -              \\
numerical     & $1.5000000$     & $-0.619064$     & $0.70031$       & $-0.345$        \\
fit precision & $\pm 6.10^{-9}$ & $\pm 2.10^{-6}$ & $\pm 7.10^{-5}$ & $\pm 1.10^{-3}$ \\
\hline
\end{tabular}
\caption[$\omega-J$ expansion for $(l,m,n) = (2,2,0)$ geons]{Coefficients in the polynomial expansion $\omega \gls{L}/m = f(J)$
(equation \eqref{omjexpansion2}) for both perturbative and numerical results at resolution $(37,9,9)$ for $(l,m,n)=(2,2,0)$ geons.
Error bars on the coefficients $a_i$ are given by the Levenberg-Marquardt fit algorithm. Credits: \cite{Martinon17}.}
\label{tabcoefs22}
\end{mytab}

In this table, it is clear that our results match very well the perturbative expansion. For instance the relative
difference in $a_2$ is of order $\sim 0.01 \%$. We can even predict the sign and value of $a_3$, that was never obtained analytically.
The sign of $a_3$ can be readily estimated on figure \ref{amd22} since the numerical points are below the fifth order curve.

At this point, let us mention that our results are in disagreement with those of \cite{Horowitz15}, whose authors were the first (and
single) to propose a numerical construction of the $(l,m,n) = (2,2,0)$ geons. In figure \ref{compare}, we reproduce their
figure (1.b) next to the top left panel of figure \ref{amd22}. It is clear that they found $a_2$ to be negative since their numerical points are below the
third order curve. According to our results listed in table \ref{coefs22}, we find, however, that $a_2\sim +0.700$. We are very
confident in this result since our perturbative computations carried out to sixth order (see \cite{Fodor17} for details) agree very well with our precise numerical
measurements. We thus provide two independent arguments in favour of the positivity of $a_2$, and are unable to find any reason
why this coefficient should be negative in \cite{Horowitz15}. Additional and independent future derivations of these results
would provide a very welcome clarification of this point.

\begin{myfig}
   \includegraphics[width = 0.98\textwidth]{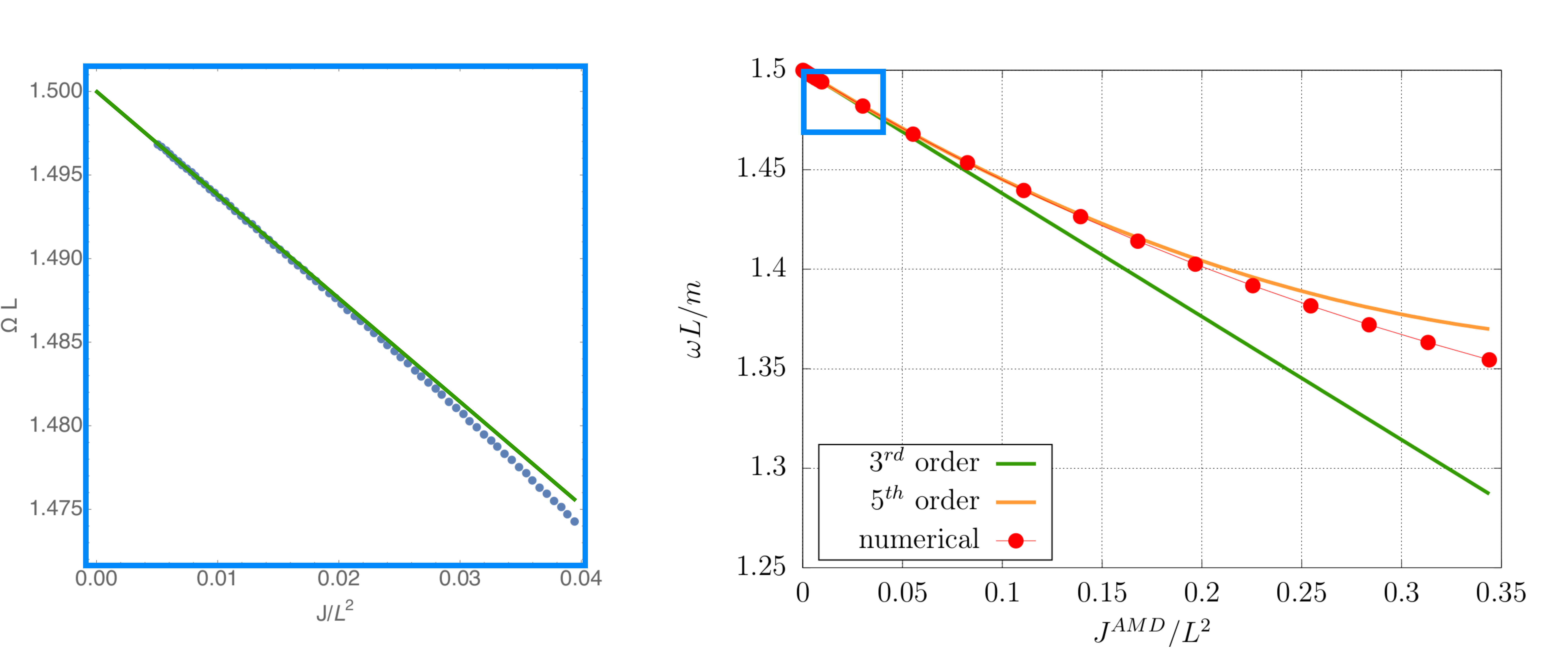}
   \caption[Comparison with previous works]{Left panel: $\omega-J$ plane taken from the numerical simulations of
      \cite{Horowitz15}. Blue points are numerical results while the solid green curve denotes $3^{rd}$ order results. Right panel:
      $\omega-J$ plane computed in \cite{Martinon17}. The range of amplitudes probed on the left panel is reported on the right
      panel with the help of a light blue frame at scale. Credits: \cite{Horowitz15,Martinon17}.}
   \label{compare}
\end{myfig}

\subsection{Geons with $(l,m,n) = (4,4,0)$}

Now that we have extensively tested our numerical procedure and studied the simplest $(l,m,n)=(2,2,0)$ geons, we are ready to
tackle more complicated configurations. Increasing the angular excitation number, we can construct geons with $(l,m,n) =
(4,4,0)$. This time, we choose for the wiggliness parameter the largest positive spectral coefficient of the first order geon in the
\gls{am} gauge. Namely:
\begin{equation}
   w \equiv \tn{coefficient } (n_r = 0,n_\theta = 1,n_\varphi = 1) \tn{ of } \widetilde{h}_{yy}.
\end{equation}
We are able to build sequences at several resolutions and the Einstein residuals are shown in figure \ref{err44}. The exponential
decrease (right panel) in the errors with increasing number of collocation points demonstrates that our solutions do indeed solve
Einstein's equation. We have checked that all the other residuals shared the same features. This comes as no surprise since only
the initial guess and the wiggliness have been changed compared to the previous $(l,m,n)=(2,2,0)$ case.

\begin{myfig}
   \includegraphics[width = 0.49\textwidth]{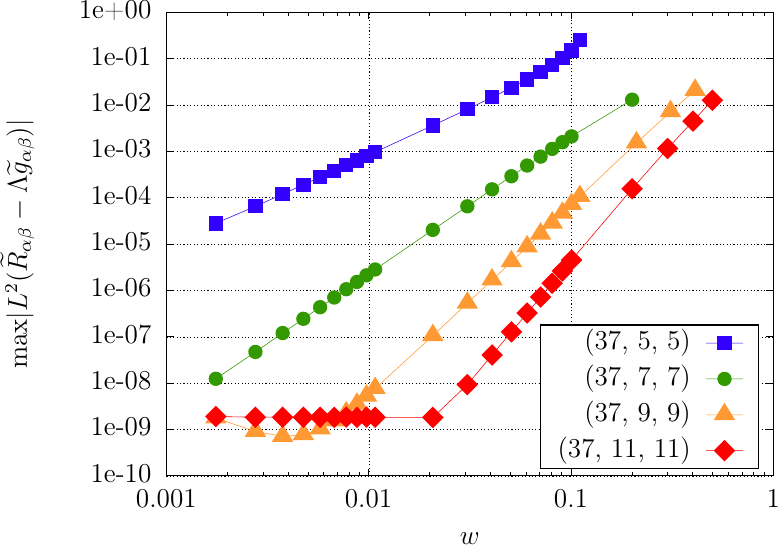}
   \includegraphics[width = 0.49\textwidth]{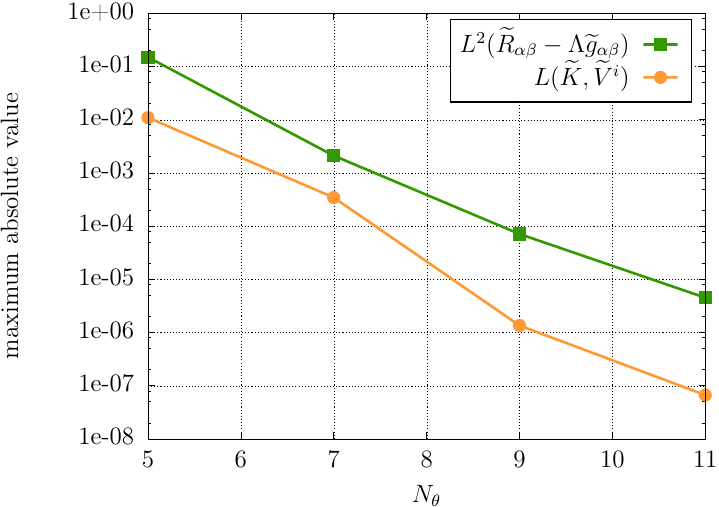}
   \caption[Einstein residual for $(l,m,n)=(4,4,0)$ geons]{Left panel: residual of the regularised Einstein's equation
      $\widetilde{R}_{\alpha\beta} - \gls{Lambda} \widetilde{g}_{\alpha\beta}$ for $(l,m,n) = (4,4,0)$ geons at four angular
   resolutions. The wiggliness $w$ translates the amplitude of the non-linear geon. Right panel: spectral convergence of the Einstein and gauge
      residuals as a function of the angular resolution at fixed wiggliness $w = 0.1$. Credits: \cite{Martinon17}.}
   \label{err44}
\end{myfig}

Global \gls{amd} quantities are shown on figure \ref{amd44}. We again observe a good match between perturbative and numerical results.
We can reach masses of order $\sim 0.5\gls{L}$. In this regard, this can be considered a high amplitude.
Note also that this is the very first fully non-linear construction of this family of geons \cite{Martinon17}.

\begin{myfig}
   \includegraphics[width = 0.49\textwidth]{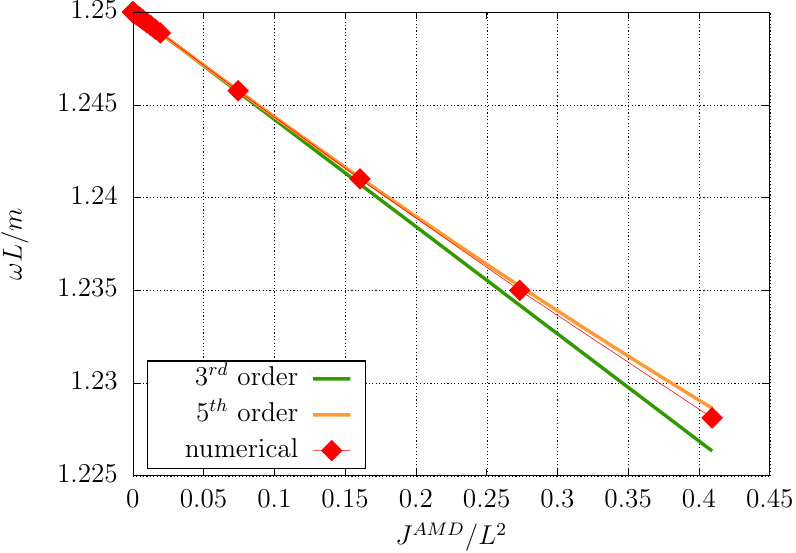}
   \includegraphics[width = 0.49\textwidth]{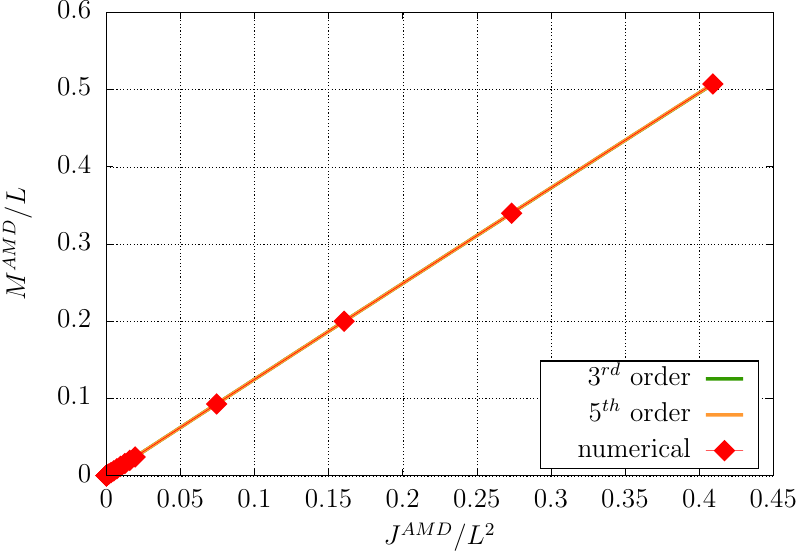}
   \caption[Global charges of $(l,m,n) = (4,4,0)$ geons]{The $\omega$-$M$-$J$ planes for our numerical sequences $(l,m,n) =
      (4,4,0)$ at resolution $(37,11,11)$ using \gls{amd} definitions. On the right panel, the curves are indistinguishable from
   perturbative results at naked eye. Credits: \cite{Martinon17}.}
   \label{amd44}
\end{myfig}

By fitting our numerical data, we infer the numerical values of the coefficients in the $\omega \gls{L}/m = f(J)$ expansion
\eqref{omjexpansion2}. They are presented in table \ref{coefs44} in comparison with fifth order perturbative results, detailed in
\cite{Martinon17}. There is a discrepancy of $\sim20\%$ between our numerical and perturbative results for the coefficient $a_2$.
We think that this is mainly due to the small number of points involved in the fit. Indeed, as can be seen in figure \ref{amd44}, our
numerical points are almost on a straight line. This severely impacts the precision of the fit when it involves terms in $J^4$.
With higher amplitudes points, the numerical points would exhibit a clearer departure from the straight line that would make the
fit more accurate. As for the coefficient $a_3$, it should be taken as a rough guess and the error bars provided by the
Levenberg-Marquardt fit algorithm are questionable. Nevertheless, the negative sign of $a_3$ is attested by figure \ref{amd44}
since the numerical points are below the fifth order curve. This is to be expected since it is also a feature of the very similar
$(l,m,n)=(2,2,0)$ case.

\begin{mytab}
\begin{tabular}{lcccc}
\hline
              & $a_0$            & $a_1$            & $a_2$           & $a_3$           \\
\hline
perturbative  & $1.2500000$      & $-0.05781990$    & $0.0137851$     &  -              \\
numerical     & $1.2500000$      & $-0.05782040$    & $0.0111142$     & $-0.0011196$    \\
fit precision & $\pm 4.10^{-10}$ & $\pm 3.10^{-8}$  & $\pm 2.10^{-7}$ & $\pm 4.10^{-7}$ \\
\hline
\end{tabular}
\caption[$\omega-J$ expansion for $(l,m,n) = (4,4,0)$ geons]{Coefficients in the polynomial expansion $\omega \gls{L}/m = f(J)$
(equation \eqref{omjexpansion2}) for both perturbative and numerical results of $(l,m,n) = (4,4,0)$ geons at a resolution of
$(37,11,11)$. Error bars are given by the Levenberg-Marquardt fit algorithm. Credits: \cite{Martinon17}.}
\label{coefs44}
\end{mytab}

\subsection{Geons with one radial node}

We now address the problem of radially excited geons, whose perturbative approach is more complicated. Initially in
\cite{Dias16a,Dias17a}, the authors claimed that the single $(l,m,n) = (2,2,1)$ linear mode could not seed a stable non-linear family of
geons, because some secular resonances could not be suppressed by Poincaré-Lindstedt regularisation. However, recently a paper and a
comment came out suggesting that a linear combination of several seeds sharing the same $\omega$ could indeed survive at arbitrary
order \cite{Rostworowski16,Rostworowski17a,Rostworowski17b}. We thus tried to build these linear combinations of single-mode
excitations.

As discussed in chapter \ref{perturbations} section \ref{beyondlin}, and illustrated in figure \ref{geoncontroversy}, the angular
frequency belonging to the helical extension of the $(2,2,1)$ scalar mode is $\omega \gls{L}/m=5/2$, and is the same  of the
scalar mode $(4,2,0)$ and the vector mode $(3,2,0)$.  Taking a linear combination of the helically symmetric metric perturbation
generated by these three modes, we obtain a two-parameter seed for the perturbative formalism, namely
\begin{equation}
   h_{\mu\nu} = \alpha h_{\mu\nu}^{(2,2,1)(s)} + \beta h_{\mu\nu}^{(4,2,0)(s)} + \gamma h_{\mu\nu}^{(3,2,0)(v)},
\end{equation}
with two independent parameters being the ratios of amplitudes $\beta/\alpha$ and $\gamma/\alpha$. Performing a small amplitude
expansion in $\alpha$ up to third order gives rise to several secular resonances. However, a computation shows that they can be removed via
Poincaré-Lindstedt regularisation for a very restricted set of parameters $\beta/\alpha$ and $\gamma/\alpha$. Actually, there are
only three possibilities for these ratios, displayed in table \ref{threefamtable}. This gives rise to exactly three families of
excited geons with one radial nodes. Any other value of the ratios $\beta/\alpha$ and
$\gamma/\alpha$ suffers form irremovable secular resonances that forbid the fully non-linear construction of the corresponding
geons. In particular, the case $\beta = \gamma = 0$ cannot seed a non-liner geon, as was first noticed in \cite{Dias16a,Dias17a}.

\begin{mytab}
\begin{tabular}{lcccc}
\hline
            & $\beta/\alpha$ & $\gamma/\alpha$ \\
\hline
family I    & $-0.00286074$  & $0.154618$ \\
family II   & $0.0367439$    & $-1.67172$ \\
family III  & $1.07086$      & $1.39907$  \\
\hline
\end{tabular}
\caption[Allowed coefficients for radially excited geons]{Numerical values of the relative amplitude parameters for the three
   degenerated families of geons with angular frequency $\omega \gls{L}/m=5/2$ in the zero-amplitude limit. Credits:
   \cite{Martinon17}.}
\label{threefamtable}
\end{mytab}

There are thus three different families of excited geons with a frequency obeying $\omega \gls{L}/m=5/2$ in the zero-amplitude limit. In
order to compute radially excited geons, we start with one of the linear combinations given in Table \ref{threefamtable}. Our
marching parameters, determined on the basis of the largest spectral coefficient, are
\begin{eqnarray}
   w &\equiv& \tn{coefficient } (n_r = 1,n_\theta = 0,n_\varphi = 0) \tn{ of } \widetilde{h}_{yy} \tn{ for family I},\\
   w &\equiv& \tn{coefficient } (n_r = 0,n_\theta = 1,n_\varphi = 0) \tn{ of } \widetilde{h}_{xx} \tn{ for family II},\\
   w &\equiv& \tn{coefficient } (n_r = 0,n_\theta = 1,n_\varphi = 0) \tn{ of } \widetilde{h}_{xz} \tn{ for family III}.
\end{eqnarray}
We have also tried to naively start with a single-mode $(l,m,n) = (2,2,1)$ first order seed, and observed that the code was
spontaneously converging to family I's branch of solutions. This is to be expected since this is the one family with highest
contribution from this seed. Since we succeeded in building numerically all three families of excited geons, this confirms the results of
\cite{Dias16a,Dias17a,Rostworowski16,Rostworowski17a,Rostworowski17b}: on the one hand, there is no non-linear geon belonging to
the single scalar mode $(2,2,1)$ class, and on the other hand there exist three families with the same linear frequency $\omega
\gls{L}/m = 5/2$ (in the zero-amplitude limit) and with radial nodes that are constituted by the linear combinations of table \ref{threefamtable}.

\begin{myfig}
   \includegraphics[width = 0.49\textwidth]{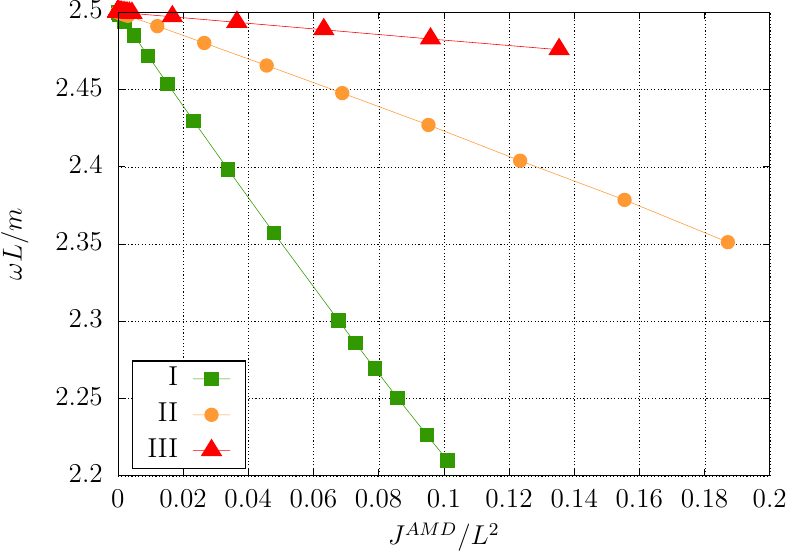}
   \includegraphics[width = 0.49\textwidth]{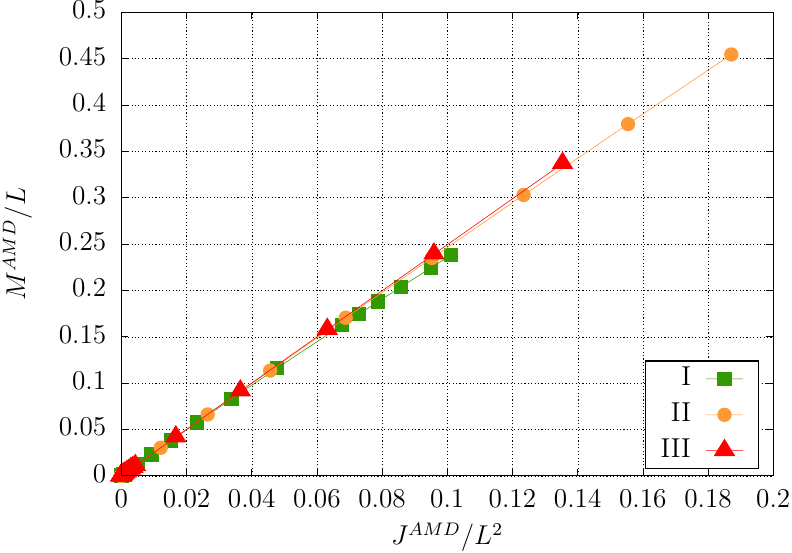}
   \caption[Global charges of radially excited geons]{The $\omega$-$M$-$J$ planes for the three families I-III of radially excited geons with one radial node and $m
      = 2$, using \gls{amd} definitions. All curves are computed at a common resolution of $(37,9,9)$. Credits: \cite{Martinon17}.}
   \label{amd1node}
\end{myfig}

Global quantities of these three families of excited geons are displayed in figure \ref{amd1node}. We can reach masses of order
$\sim 0.5 \gls{L}$, depending on the family under study.

\begin{mytab}
\begin{tabular}{llccccc}
\hline
                           &               & $a_0$           & $a_1$           & $a_2$           & $a_3$           \\
\hline
                           & perturbative  & $2.5000000$     & $-3.058658$     &     -           &    -            \\
                           & numerical     & $2.5000000$     & $-3.058670$     & $1.239$         & $8.2$           \\
\multirow{-3}*{family I}   & fit precision & $\pm 2.10^{-9}$ & $\pm 7.10^{-6}$ & $\pm 2.10^{-3}$ & $\pm 3.10^{-1}$ \\
\hline
                           & perturbative  & $2.5000000$     & $-0.747498$     &     -           &    -            \\
                           & numerical     & $2.5000000$     & $-0.747533$     & $-0.1723$       & $-0.39$         \\
\multirow{-3}*{family II}  & fit precision & $\pm 3.10^{-9}$ & $\pm 3.10^{-6}$ & $\pm 3.10^{-4}$ & $\pm 2.10^{-2}$ \\
\hline
                           & perturbative  & $2.5000000$     & $-0.180638$     &     -           &    -            \\
                           & numerical     & $2.5000000$     & $-0.180628$     & $-0.0072$       &    -            \\
\multirow{-3}*{family III} & fit precision & $\pm 5.10^{-9}$ & $\pm 5.10^{-6}$ & $\pm 7.10^{-4}$ &    -            \\
\hline
\end{tabular}
\caption[$\omega-J$ expansion for radially excited geons]{Coefficients in the polynomial expansion $\omega \gls{L}/m = f(J)$ (equation
\eqref{omjexpansion2}) for both perturbative and numerical results of the three geon families I-III at a resolution of $(37,9,9)$. Error bars are given by the
Levenberg-Marquardt fit algorithm. Credits: \cite{Martinon17}.}
\label{coefs1nodeFamily1}
\end{mytab}

In table \ref{coefs1nodeFamily1}, we show the fit of our numerical results on global charges in the $\omega \gls{L}/m=f(J)$ expansion of
equation \eqref{omjexpansion2}. These values can be regarded as predictions we can make on the $a_i$ coefficients.
We expect that higher amplitudes sequences at higher resolutions could allow us to refine these values. Since we are in perfect
agreement with perturbative results, notably for the coefficients $a_1$, our results clearly identify the three distinct families of
table \ref{threefamtable}. This definitely confirms the arguments of \cite{Rostworowski16,Rostworowski17a,Rostworowski17b}
according to which the number of excited geon families precisely matches the multiplicity of the frequency $\omega \gls{L}/m =
5/2$.

\section{Visualisation of geons}

All five families of geons we were able to construct are pictured in figures \ref{hxxgallery} and \ref{killinggallery}. In figure
\ref{hxxgallery}, we show the shape of the geon, namely the map of the component $\widetilde{h}_{xx}$. In
particular, the $(l,m,n) = (2,2,0)$ geon exhibit an egg-shape, centred on the origin. This motivated our definition of wiggliness
\eqref{w220}, which was basically the maximum value of this component. Such a definition cannot be extended to other families of
geons though, since some of them always display a vanishing $\widetilde{h}_{xx}(r = 0)$.

\begin{myfig}
   \includegraphics[width = 0.49\textwidth]{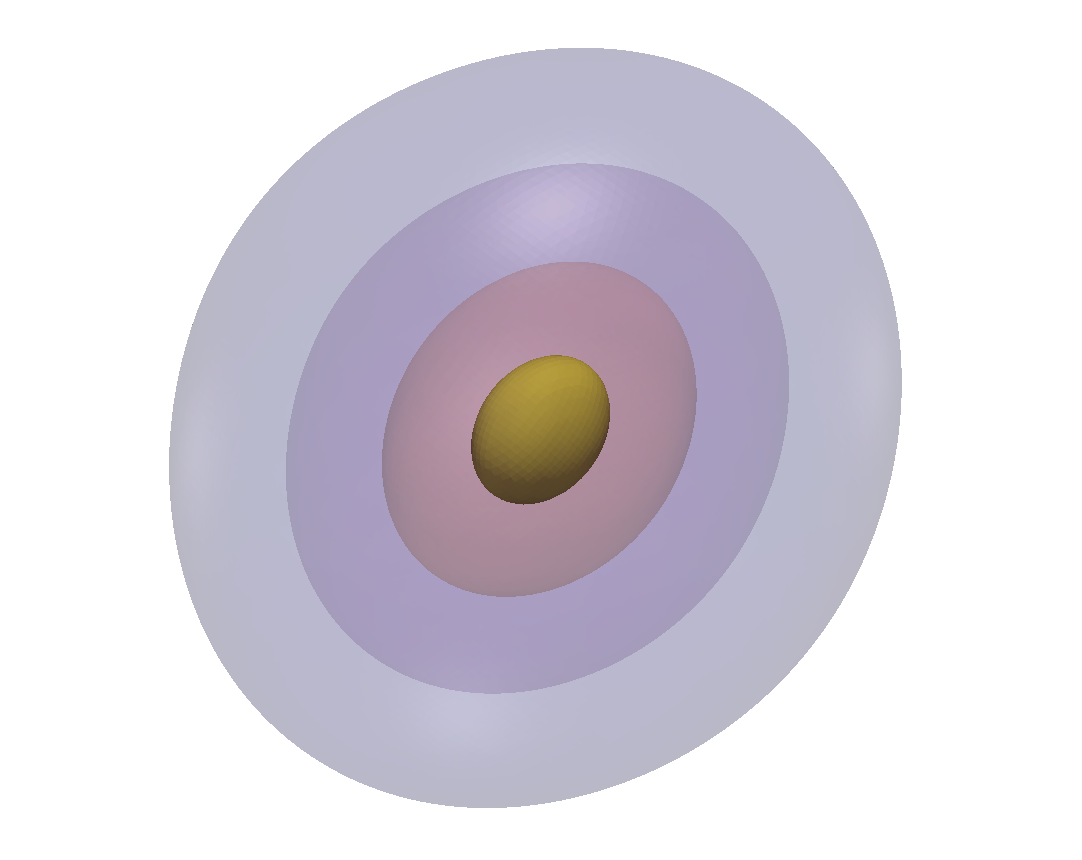}
   \includegraphics[width = 0.49\textwidth]{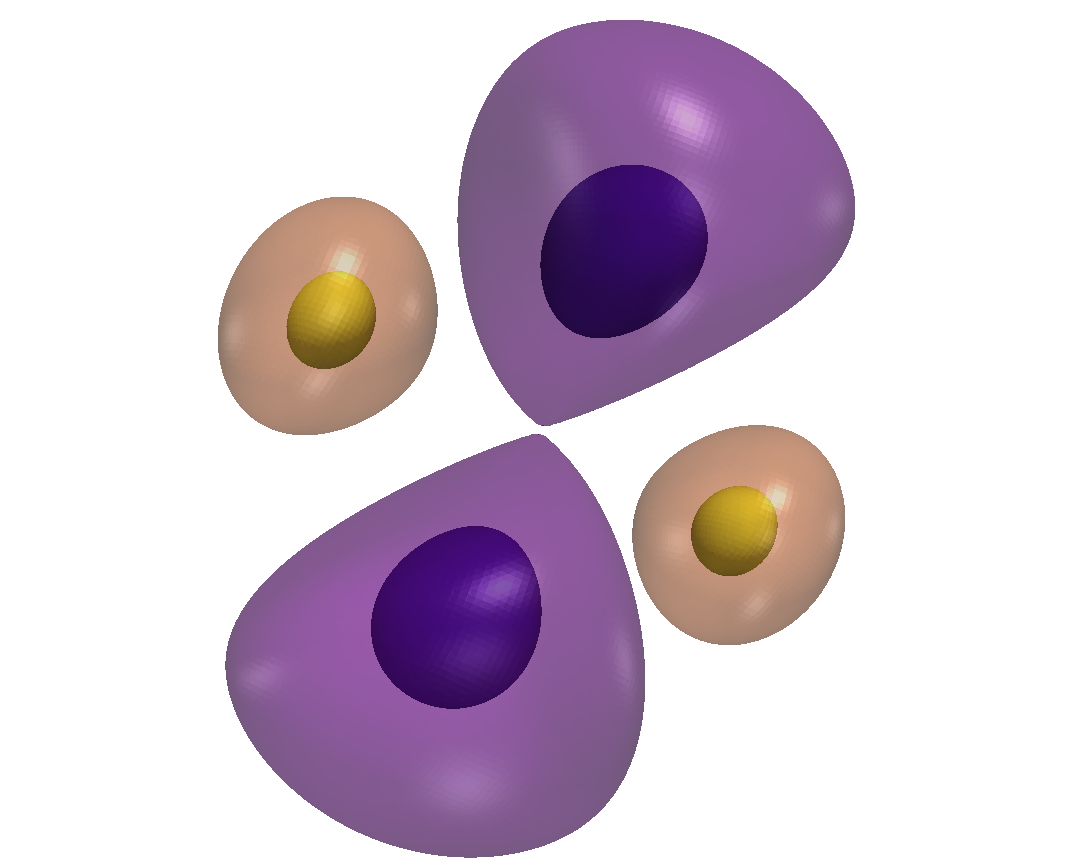}\\
   \includegraphics[width = 0.49\textwidth]{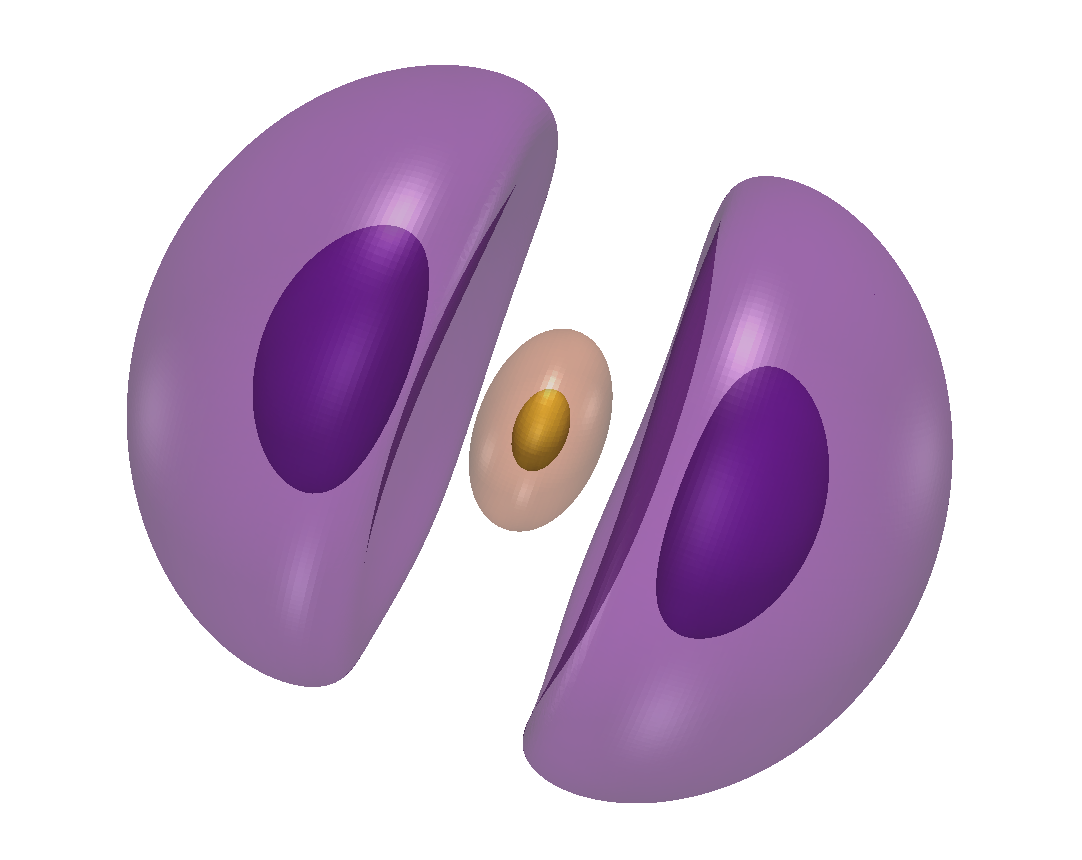}\\
   \includegraphics[width = 0.49\textwidth]{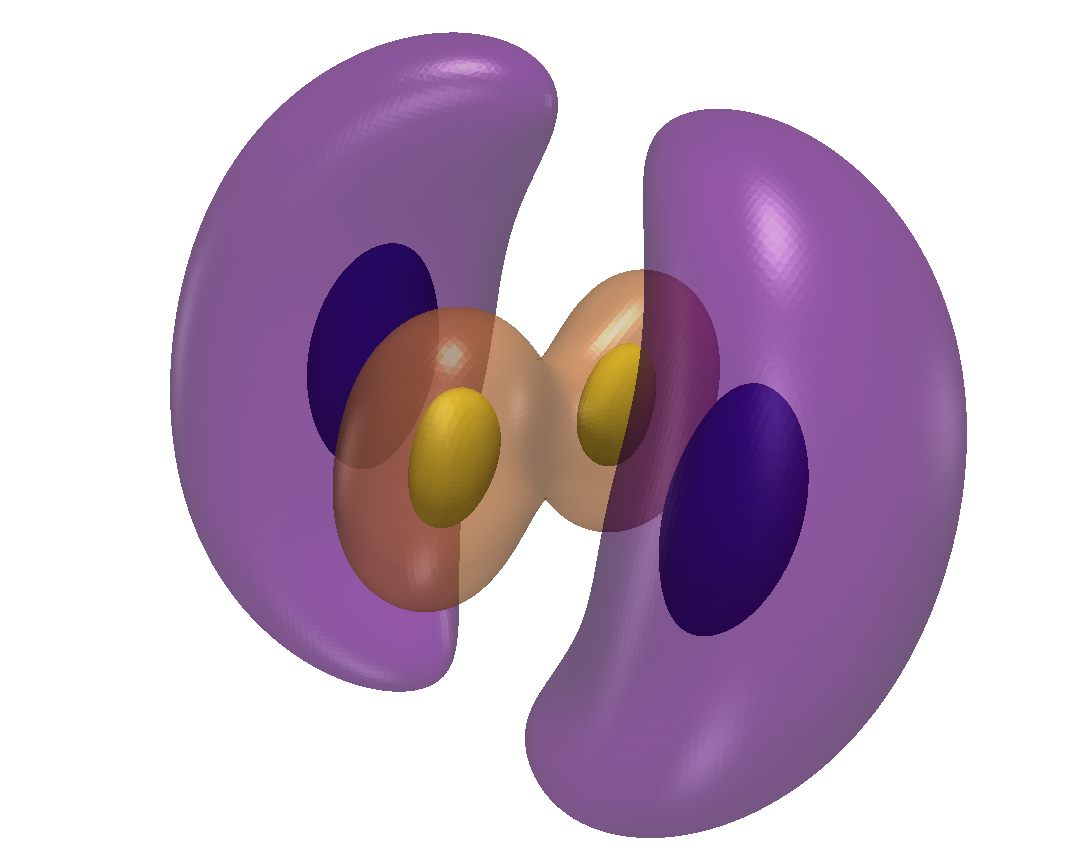}
   \includegraphics[width = 0.49\textwidth]{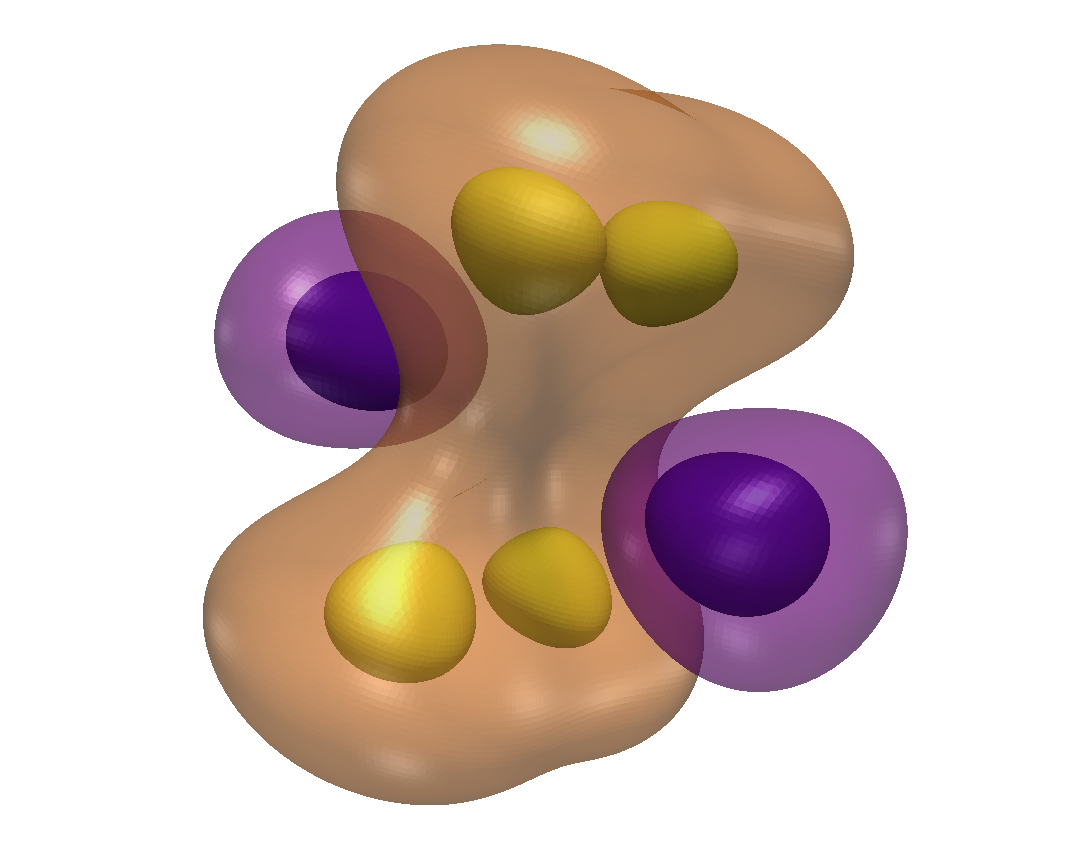}
   \caption[Metric components of geons]{Component $\widetilde{h}_{xx}$ regularised metric
   perturbation in the \gls{am} gauge for all five families of geons numerically computed, namely the $(l,m,n) = (2,2,0)$,
   $(4,4,0)$ and families I-III in this order. The masses of the pictured solutions are
   around $\sim 0.2 \gls{L}$, and they were computed at our best numerical resolutions. Credits: G. Martinon.}
   \label{hxxgallery}
\end{myfig}

In figure \ref{killinggallery}, we show the $\widetilde{h}_{t't'}$ component, that is directly related to the regularised square norm of the
Killing vector $\partial_{t'}^\alpha$ of equation \eqref{killingvector}. Indeed, we have
\begin{equation}
   \Omega^2\partial_{t'}\cdot \partial_{t'} = \widetilde{g}_{\mu\nu}\partial_{t'}^\mu \partial_{t'}^\nu = \widetilde{g}_{t't'}.
\end{equation}
After background subtraction (the \gls{ads} background being common to all solutions), the contribution of the geon to this norm
is directly pictured in figure \ref{killinggallery}. The shape of the isocontours increases in complexity for excited geons.

\begin{myfig}
   \includegraphics[width = 0.49\textwidth]{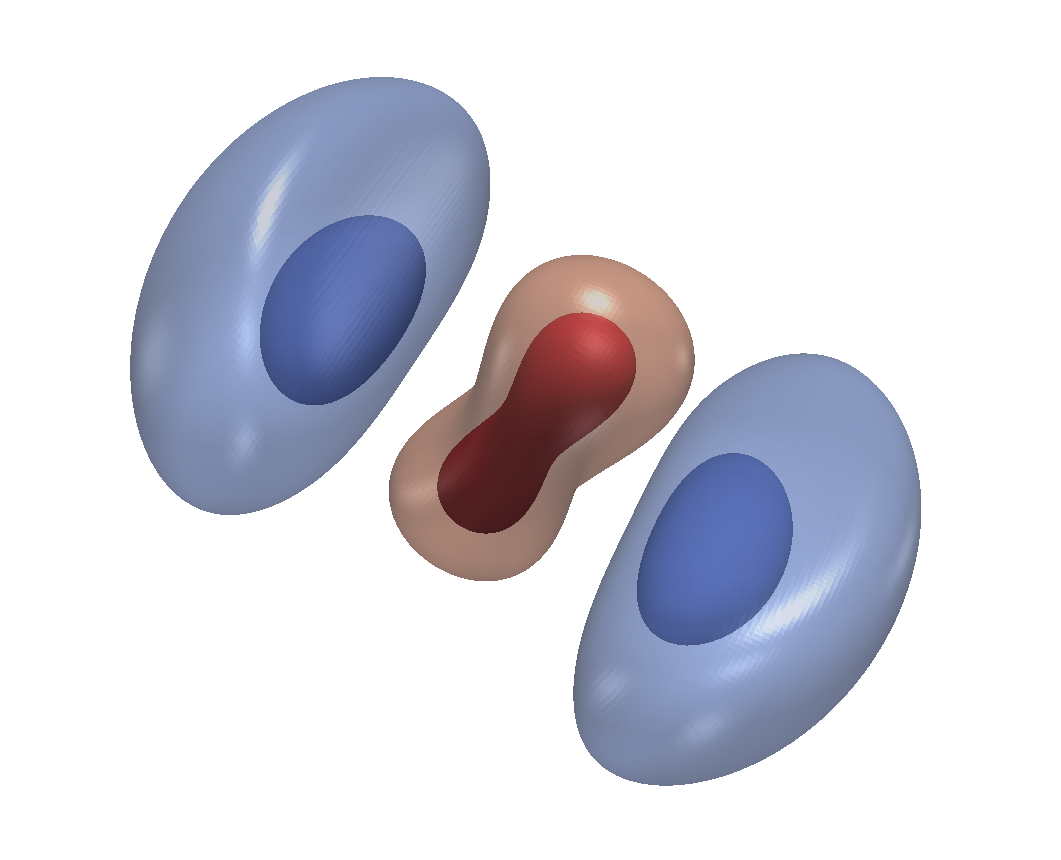}
   \includegraphics[width = 0.49\textwidth]{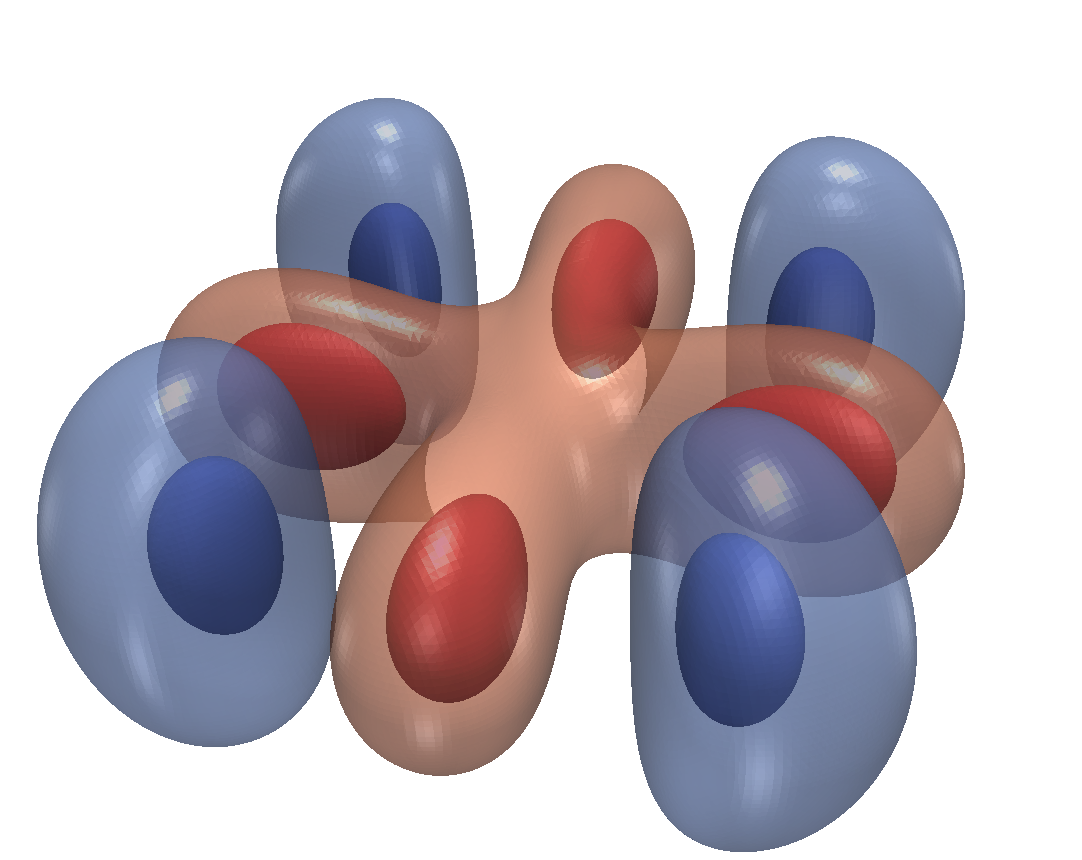}\\
   \includegraphics[width = 0.49\textwidth]{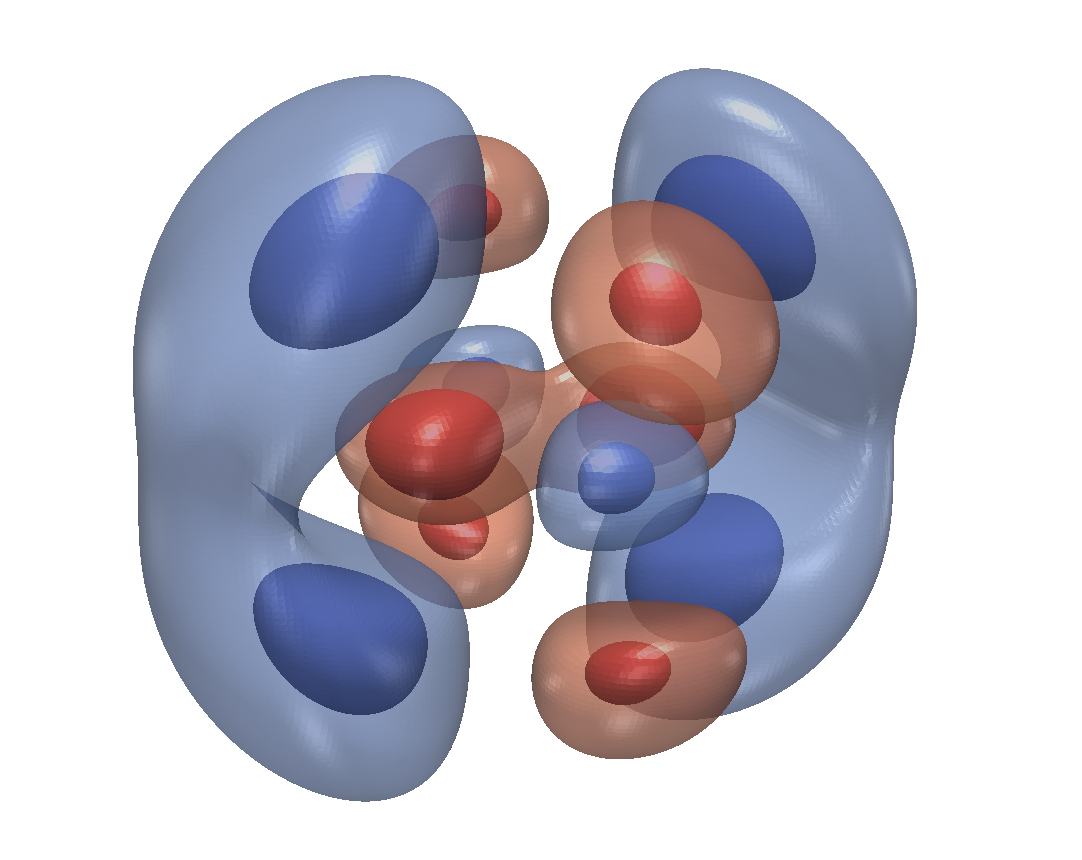}\\
   \includegraphics[width = 0.49\textwidth]{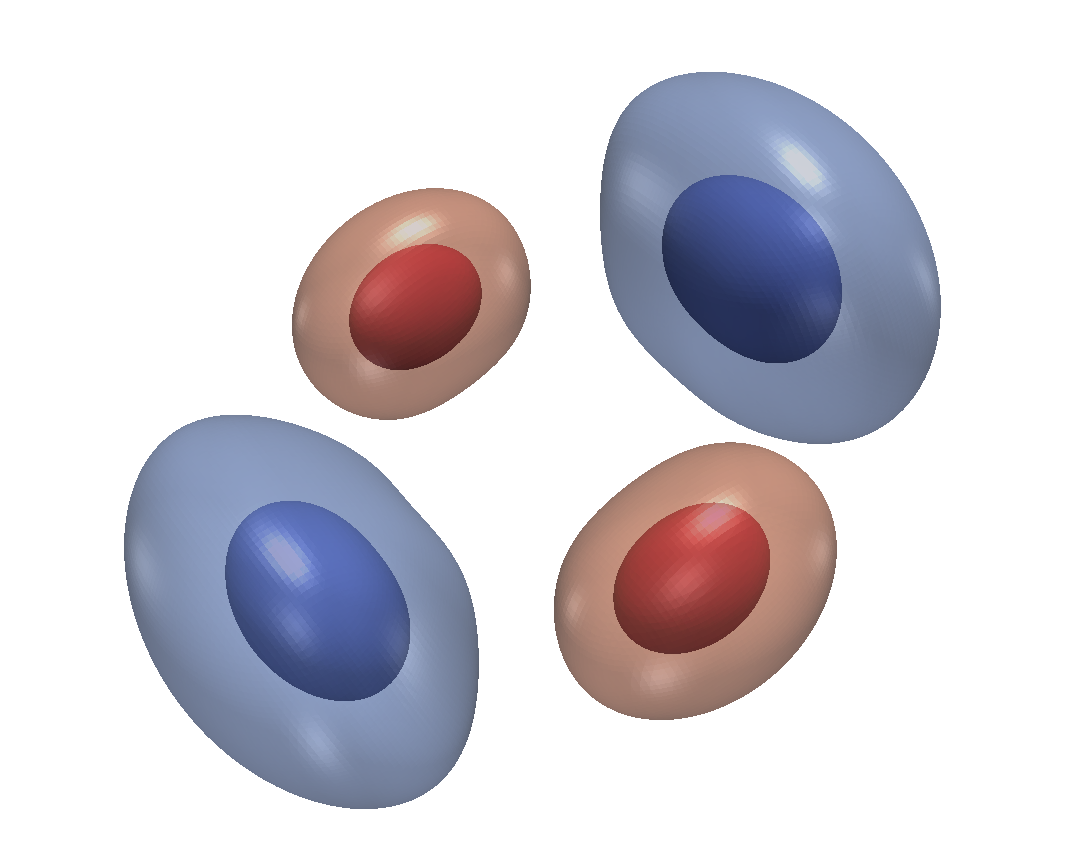}
   \includegraphics[width = 0.49\textwidth]{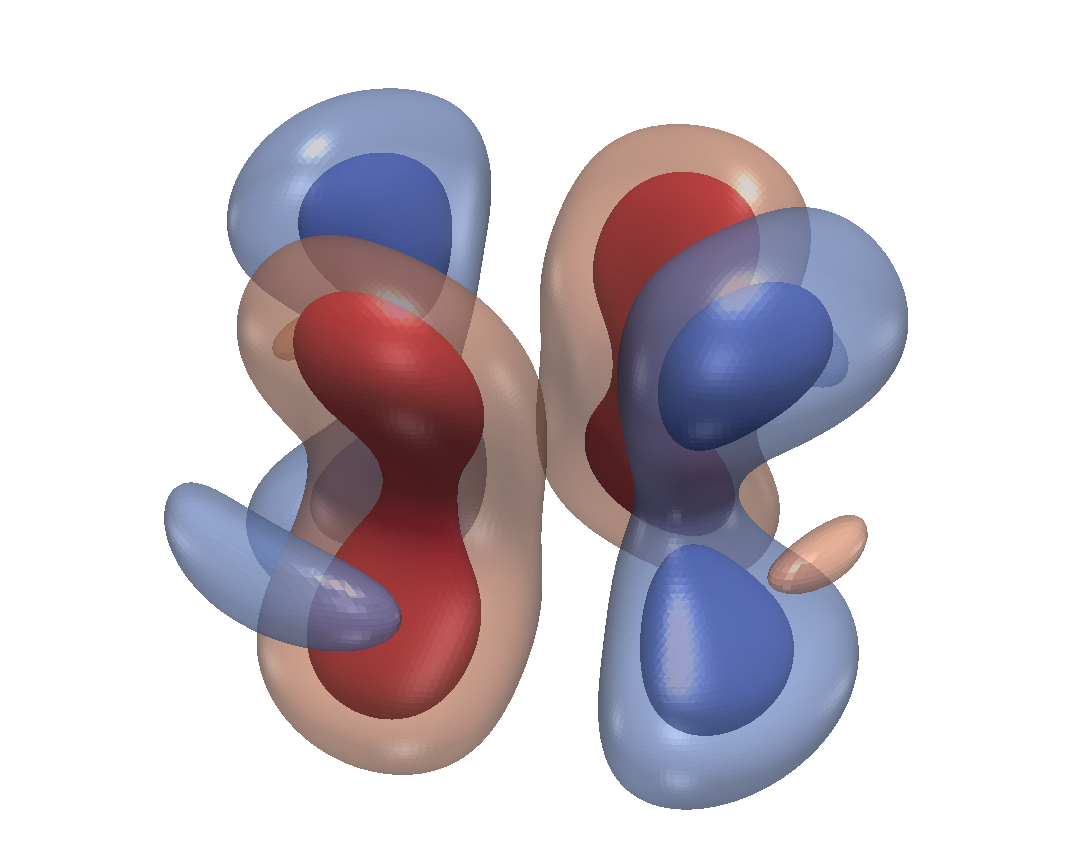}
   \caption[Regularised norm of the helical Killing vector in geons]{Isocontours of $\widetilde{h}_{t't'}$. They correspond to
   the regularised norm of the helical Killing vector $\partial_{t'}$ in the co-rotating frame of geons, after background
   subtraction. All five families of geons are presented, namely the scalar $(l,m,n)=(2,2,0),(4,4,0)$ modes and the three excited families
   I-III, in this order. Red isocontours denote positive contributions to the norm. Credits: G.  Martinon.}
   \label{killinggallery}
\end{myfig}

Of the five families of geons we have built in \cite{Martinon17}, all but the $(l,m,n)=(2,2,0)$ were first-time numerical
constructions. Without a doubt, many other families will be constructed in the future, with other symmetries than these helically
symmetric ones. A straightforward extension of this work would be to construct $m=0$ axisymmetric geons, that have no $\varphi$
dependence, but are periodic in time. In terms of our numerical representation, this is equivalent to our helically symmetric geons
but this time there is no co-rotating frame. We can thus simply switch the spectral representations of time $t'$ and azimuthal angle $\varphi$. Other families of
geons could be constructed too, e.g.\ scalar modes with odd quantum numbers $l$ and $m$. These geons, however, have less
symmetries than the even-even ones, and are thus potentially more greedy in terms of numerical resolution.

We hope that our work \cite{Martinon17} has given a general method to build non-linear geons. With these tools at
end, these fundamental gravitational excitations of \gls{ads} space-time will, without a doubt, shed new light on the
\gls{ads} instability problem. Notably, the very particular property according to which the number of non-linear families of
geons precisely matches the multiplicity of a given frequency is a strong hint that there is an underlying structure in geons that
has yet to be unveiled.

\chapter*{Conclusion}
\addstarredchapter{Conclusion}
\citationChap{An expert is a person who has made all the mistakes that can be made in a very narrow field.}{Niels Bohr}

In this manuscript, we have studied in details the numerical construction of geons in \gls{aads} space-times. The concept of geons
was defined in chapter \ref{geons} as well as their link to dark matter candidates. The notion of \gls{aads} space-times was
outlined in chapter \ref{aads} and we have sketched in chapter \ref{adscft} why such space-times were of physical interest, in
particular within the context of the \gls{ads}-\gls{cft} correspondence. The geons originally studied by Wheeler were notably
revived in the tremendous discovery of the so-called \gls{ads} instability conjecture, reviewed in chapter \ref{adsinsta}. In this
regard, geons can be interpreted as the fundamental gravitational excitations of \gls{ads} space-times, and are believed to form
islands of stability. Because of their intricate geometry, they have very few symmetries if at all. This enters in deep contrast
with the vast majority of spherically symmetric configurations studied in the literature. The second part of this manuscript was
precisely dedicated to the numerical construction of \gls{aads} geons. From a perturbative point of view first, in chapter
\ref{perturbations}, we have detailed the construction of linear geons with the covariant and gauge-invariant \gls{kis} formalism.
In chapter \ref{gaugefreedom}, we have discussed how to choose a gauge and have explained the motivations of the harmonic as well
as the \gls{am} gauges. Finally in chapter \ref{simulations}, we have described numerical and general techniques to build geons,
and presented the results of five different families of helically symmetric configurations \cite{Martinon17}.

Recall that geons came out in the \gls{aads} literature in 2012 with the work of \cite{Dias12a,Dias12b}. The authors applied for
the first time the \gls{kis} formalism to the perturbative construction of geons and argued that they constituted stable islands
of stability. It was only in 2015 that one family of geon, the $(l,m,n)=(2,2,0)$ one, was constructed numerically for the first
time, in \cite{Horowitz15}. The construction then relied on the 4-dimensional De Turck method, with a rather involved reference
metric.  In a subsequent work, the authors of \cite{Dias16a,Dias17a} studied further the perturbative approach and focused mainly
on single-mode geon excitations. This lead them to conclude that the instability of \gls{ads} space-time was densely populated
with secular resonances in the gravitational sector, so much as to forbid large sets of spherical harmonic geon seeds. Among them,
the authors unveiled that radially excited geons seeded by a scalar $(l,m,n)=(2,2,1)$ mode were subject to irremovable secular
resonances. The scalar $(2,2,1)$ geons were thus sentenced to non-existence. It thus came quite as a big surprise when we observed
numerically that our own code was converging to a well-behaved solution when seeded with this scalar $(2,2,1)$ initial guess. In
the meantime, \cite{Rostworowski16} came out and argued that even if indeed one single scalar mode was subject to robust secular
resonances, it was closely connected to linear combinations (or families) of geons that were indeed free of any resonance. At this
time, we thus understood that our code had ``jumped'' spontaneously to one of these families, thus confirming that (i) the single
scalar mode $(2,2,1)$ was not a valid non-linear solution of Einstein's equation and (ii) that there were nonetheless nearby
solutions that were indeed valid.  These arguments were given more weight afterwards in \cite{Rostworowski17a,Rostworowski17b},
while we decided in the meantime to construct all three possible families of excited geons. Our results were finally presented in
\cite{Martinon17}. Our numerical and fully non-linear construction of geons thus provided definitive arguments in this lively
debate, and we confirmed the previous perturbative works \cite{Dias16a,Dias17a,Rostworowski16,Rostworowski17a,Rostworowski17b}.
Furthermore, our numerical results allowed us to probe amplitudes much beyond reach of perturbative techniques. In particular, we
have not detected any maximum mass despite our high amplitudes. Do geon have a maximum mass? This question is still unanswered,
but we can recall that the answer is negative in asymptotically flat space-times (chapter \ref{geons}). We hope that the numerical
construction of geons will accelerate in a near future. Our precision monitoring employed many techniques of spectral methods and
took advantage of several relevant concepts in \gls{aads} space-times, in particular the \gls{aads} asymptotics.

As mentioned at the end of chapter \ref{simulations}, a straightforward extension of our work would be to compute axially
symmetric $m=0$ geons. These configurations are not rotating and are independent of the $\varphi$ coordinate. However, they oscillate in
time. On numerical grounds, this is equivalent to switch the time $t$ and angle $\varphi$ spectral representation of helically
symmetric geons, with minimal changes in the evolution operator $\mathcal{L}_m$ and the boundary conditions. In the same spirit,
other helically symmetric geons with less symmetries could be constructed, notably ones with odd quantum numbers $l$ and $m$. This
would imply to modify the spectral basis of decomposition since the octant symmetry would be lost. In the same vein of
\cite{Dias16a,Dias17a} that undertook the huge task of classifying all geon seeds in perturbative approach, we expect
numerical constructions of geons to be more and more exhaustive in the future. We endeavoured to give a general and covariant
framework to do so. We also expect numerical approaches to be able to discover new families of geons that can sometimes be missed
by perturbative techniques, as we have felt so for the radially excited helically symmetric configurations.

Geons can be viewed as gravitational and non-spherically symmetric generalisations of the scalar time-periodic solutions, reviewed
in chapter \ref{adsinsta}. A natural extension of our work thus lies in the time evolution of our non-linear geons. Indeed, we
have proved beyond doubt the existence of five different families of geons, but a still unanswered question is: are these
solutions stable attractors? Even if there are perturbative arguments in favour of a positive answer \cite{Dias12b}, to this day
there is no definitive proof of the non-linear stability of geons. In some sense, our numerical solutions are willing to become
initial data for time evolution codes in \gls{aads} space-times. Such simulations would be a huge improvement of the current
spherically symmetric ones, and could largely probe the \gls{ads} instability beyond spherical symmetry. Given how the instability
is understood up to now, it is expected that initial geon data, even if numerically perturbed, would evolve in time in a
quasi-periodic manner for arbitrary long times, as was observed for spherically symmetric time-periodic stable attractors in the
scalar sector. But such simulations are yet to be implemented.

Evolution codes in \gls{gr} are a very technical issue. It suffices to count the decades during which relativists
struggled to obtain their first stable evolutions to understand how a daunting task it is. However, we believe
that a large part of the know-how acquired in asymptotically flat space-times context extends readily to \gls{aads} space-times. The
\gls{aads} asymptotics bring however their own difficulties, and cannot be considered without regularisation techniques and
appropriate gauge choices. This was the philosophy of our work: borrowing ideas and techniques form relativistic and numerical
astrophysics to apply them to \gls{aads} space-times. In this regard, Bantilan and Pretorius, who was the first to successfully
evolve a numerical black hole binary within the generalised harmonic gauge, had made important progresses in \gls{aads}
space-times, notably in \cite{Bantilan12,Bantilan15}. In these works, the authors evolved axisymmetric frontal
collisions of non-spinning \gls{aads} black holes. These methods were also re-employed in \cite{Bantilan17} with the first ever
evolutions of a massless scalar field beyond spherical symmetry in \gls{ads} space-time.

Our intuition is that time evolution of geons in \gls{aads} space-times is as much
difficult as promising. Had this project had more time, no doubt that we would have tried hard to obtain numerical evolutions of
our first results. This notably motivated our choice of the \gls{am} gauge, which was initially designed for initial-value problems.
Indeed, the link between the time evolution scheme of \cite{Andersson03} and our results is straightforward, since we already have
geons in the \gls{am} gauge. Our regularisation procedure readily applies to time evolution, so that on theoretical grounds,
we are ready to tackle such problems. However, a large part of the difficulty lies in several numerical technicalities, like e.g.\
outer trapped surface detection and \gls{amr} scheme for precise resolution of black hole formation, not to mention the
essential parallel computing. Implementing these features is a slow and not so straightforward process.

Finally, there is another topic indirectly linked to geons: non-coalescing black hole binaries. In \gls{aads} space-times, we
could perfectly imagine a configuration where two black holes are orbiting each other in a periodic manner. This would be
forbidden in asymptotically flat space-times, but in \gls{aads} ones, due to the reflective boundary conditions on the boundary, such a
configuration would be in equilibrium with its own radiation. Even if the space-time is not smooth and contains singularities,
such non-coalescing binaries could well incarnate a different flavour of the concept of island of stability, since no collision
ever happens. Furthermore, they also enjoy helical symmetry and hence co-rotating stationarity. However, non-coalescing binaries
would be much more computationally demanding than geons. This stresses how subtle, how intricate and how rich the notion of
stability can be in \gls{aads} space-times.

\part{Appendices}
\label{part3}
\begin{appendix}
\chapter{General relativity in dimension $n$}
\label{grd}
\citationChap{Any fool can know. The point is to understand.}{Albert Einstein}
\minitoc

This appendix is a condensed memo of \gls{gr} in $n$-dimensional space-times. We denote by $g_{\alpha\beta}$ the physical
metric. We also use the symmetric and antisymmetric indices defined by
\begin{equation}
   A_{(\alpha\beta)} \equiv \frac{1}{2}(A_{\alpha\beta} + A_{\beta\alpha}) \quad \tn{and} \quad A_{[\alpha\beta]} \equiv
   \frac{1}{2}(A_{\alpha\beta} - A_{\beta\alpha}).
\end{equation}
The main occurrence of the dimension $n$ of the space-time is through the trace of the metric tensor, namely
\begin{equation}
   g^{\mu\nu}g_{\mu\nu} = n.
\end{equation}
This $n$ factor propagates into several other tensors. We carefully keep track of all the dimensional dependence throughout all
calculations.

\section{Geometrical tensors}

In this section, we recap the different tensors involved in \gls{gr}. Readers are referred to standard textbooks for details
\cite{Misner73,Poisson04,Gourgoulhon07}.

\subsection{Definitions and properties}
\label{defpropgeom}

\gls{gr} works with a few key geometrical tensors (or symbols), that can be constructed solely from the metric $g_{\alpha\beta}$. They are the
followings.
\begin{myitem}
   \item The Christoffel symbols
      \begin{equation}
         \Gamma\indices{^\gamma_{\alpha\beta}} \equiv \frac{1}{2}g^{\gamma\mu}(\partial_\alpha g_{\beta\mu} + \partial_\beta g_{\alpha\mu} - \partial_\mu g_{\alpha\beta}).
         \label{christoffel}
      \end{equation}
   \item The Riemann tensor
      \begin{equation}
         R\indices{^\alpha_{\beta\gamma\delta}} \equiv \partial_\gamma \Gamma\indices{^\alpha_{\beta\delta}} - \partial_\delta
         \Gamma\indices{^\alpha_{\beta\gamma}} + \Gamma\indices{^\alpha_{\gamma\mu}}\Gamma\indices{^\mu_{\beta\delta}} -
         \Gamma\indices{^\alpha_{\delta\mu}}\Gamma\indices{^\mu_{\beta\gamma}}.
         \label{defriem}
      \end{equation}
   \item The Ricci tensor
      \begin{equation}
         R_{\alpha\beta} \equiv R\indices{^\mu_{\alpha\mu\beta}} =  \partial_\mu \Gamma\indices{^\mu_{\alpha\beta}} -
         \partial_\beta \Gamma\indices{^\mu_{\alpha\mu}} + \Gamma\indices{^\mu_{\alpha\beta}}\Gamma\indices{^\nu_{\mu\nu}} -
         \Gamma\indices{^\mu_{\alpha\nu}}\Gamma\indices{^\nu_{\beta\mu}}.
         \label{riccidef}
      \end{equation}
   \item The Ricci scalar
      \begin{equation}
         R \equiv g^{\mu\nu}R_{\mu\nu}.
      \end{equation}
   \item The Einstein tensor
      \begin{equation}
         G_{\alpha\beta} \equiv R_{\alpha\beta} - \frac{1}{2}R g_{\alpha\beta}.
      \end{equation}
   \item The trace of the Einstein tensor
      \begin{equation}
         G \equiv g^{\mu\nu}G_{\mu\nu} = -\frac{n-2}{2}R.
      \end{equation}
   \item The Schouten tensor
      \begin{equation}
         S_{\alpha\beta} \equiv R_{\alpha\beta} - \frac{R}{2(n-1)}g_{\alpha\beta}.
         \label{schoutendef}
      \end{equation}
   \item The trace of the Schouten tensor
      \begin{equation}
         S \equiv g^{\mu\nu}S_{\mu\nu} = \frac{n-2}{2(n-1)}R.
         \label{trschoutendef}
      \end{equation}
   \item The Cotton tensor
      \begin{equation}
         C_{\alpha\beta\gamma} \equiv \nabla_{[\alpha}S_{\beta]\gamma}.
         \label{cottondef}
      \end{equation}
   \item The Weyl tensor
      \begin{equation}
         C_{\alpha\beta\gamma\delta} \equiv R_{\alpha\beta\gamma\delta} - \frac{2}{n-2}(g_{\alpha[\gamma}S_{\delta]\beta} - g_{\beta[\gamma}S_{\delta]\alpha}).
         \label{weyldef}
      \end{equation}
   \item The Kronecker delta symbol
      \begin{equation}
         \delta\indices{^\alpha_\beta} \equiv \left\{
         \begin{array}{cl}
            1 & \tn{if } \alpha = \beta,\\
            0 & \tn{otherwise}.
         \end{array}
         \right.
      \end{equation}
   \item The generalised Kronecker delta symbol
      \begin{equation}
         \delta_{\alpha_1\ldots\alpha_k}^{\beta_1\ldots\beta_k} \equiv \left\{
         \begin{array}{cl}
            +1 & \tn{if } (\alpha_1,\ldots,\alpha_k) \tn{ is an even permutation of } (\beta_1,\ldots,\beta_k),\\
            -1 & \tn{if } (\alpha_1,\ldots,\alpha_k) \tn{ is an odd permutation of } (\beta_1,\ldots,\beta_k),\\
             0 & \tn{otherwise}.
         \end{array}
         \right.
      \end{equation}
   \item The Levi-Civita tensor
      \begin{subequations}
      \begin{align}
         \varepsilon_{\alpha_1\ldots\alpha_n} &\equiv \left\{
         \begin{array}{cl}
            +\sqrt{-g} & \tn{if } (\alpha_1,\ldots,\alpha_n) \tn{ is an even permutation of } (0,\ldots,n-1),\\
            -\sqrt{-g} & \tn{if } (\alpha_1,\ldots,\alpha_n) \tn{ is an odd permutation of }  (0,\ldots,n-1),\\
            0 & \tn{otherwise},
      \end{array}
      \right.\\
         \varepsilon^{\alpha_1\ldots\alpha_n} &\equiv \left\{
         \begin{array}{cl}
            -\dfrac{1}{\sqrt{-g}} & \tn{if } (\alpha_1,\ldots,\alpha_n) \tn{ is an even permutation of } (0,\ldots,n-1),\\
            +\dfrac{1}{\sqrt{-g}}& \tn{if } (\alpha_1,\ldots,\alpha_n) \tn{ is an odd permutation of }  (0,\ldots,n-1),\\
            0 & \tn{otherwise}.
      \end{array}
      \right.
      \end{align}
      \label{leviCdef}
      \end{subequations}
\end{myitem}

From the definition of the Christoffel symbols \eqref{christoffel} and the general property of matrices that $\delta \ln|\det A| =
\Tr(A^{-1}\delta A)$, it comes
\begin{equation}
   \Gamma \indices{^\mu_{\alpha\mu}} = \frac{1}{2}g^{\mu\nu}\partial_\alpha g_{\mu\nu} = \frac{1}{\sqrt{-g}}\partial_\alpha \sqrt{-g}.
   \label{Gammadet}
\end{equation}

We also define the covariant derivative of a tensor $T \indices{^\alpha_\beta}$ as
\begin{equation}
   \nabla_\gamma T\indices{^\alpha_\beta} \equiv \partial_\gamma T\indices{^\alpha_\beta} + \Gamma\indices{^\alpha_{\gamma\mu}} T\indices{^\mu_\beta} - \Gamma\indices{^\mu_{\gamma\beta}} T\indices{^\alpha_\mu}.
   \label{defcov}
\end{equation}
The generalisation to tensors of arbitrary valence is straightforward. Furthermore, it can be shown that
\begin{equation}
   \nabla_\gamma g_{\alpha\beta} = 0 \quad \tn{and} \quad \nabla_\gamma g^{\alpha\beta} = 0.
   \label{divmetric}
\end{equation}
With \eqref{Gammadet}, we find that the divergence of a vector is given by
\begin{equation}
   \nabla_\mu v^\mu = \frac{1}{\sqrt{-g}}\partial_\mu (\sqrt{-g}v^\mu).
   \label{divv}
\end{equation}

Another derivative that is relevant is the Lie derivative along a given vector $v^\alpha$. It is defined by
\begin{equation}
   \mathcal{L}_v T \indices{^\alpha_\beta} \equiv v^\mu \nabla_\mu T \indices{^\alpha_\beta} - T \indices{^\mu_\beta}\nabla_\mu
   v^\alpha + T \indices{^\alpha_\mu} \nabla_\beta v^\mu.
   \label{lie}
\end{equation}
It is remarkable that in this expression, all the implicit Christoffel symbols cancel off each other, so that the expression could
be just as well written with $\partial$ only.

Finally, under a coordinate change
\begin{equation}
   x^{\alpha'} = f^\alpha(\xi^\mu) \iff x^\alpha = (f^{-1})^\alpha(x^{\mu'}),
   \label{xap}
\end{equation}
a tensor transforms like
\begin{equation}
   T \indices{^{\alpha'}_{\beta'}}(x^{\rho'}) = \frac{\partial x^{\alpha'}}{\partial x^\mu}\frac{\partial x^\nu}{\partial x^{\beta'}} T \indices{^\mu_\nu} (x^\rho),
   \label{Ttrans}
\end{equation}
where on the right-hand side, $x^\rho$ is considered a function of $x^{\rho'}$ (and conversely) via \eqref{xap}. This formula generalises readily to
tensors of arbitrary valence. Note that the Christoffel symbols, as their name indicate, are not tensors and do not obey
\eqref{Ttrans}. Instead, it can be shown that \eqref{christoffel} combined with \eqref{Ttrans} for the metric gives
\begin{equation}
   \Gamma \indices{^{\gamma'}_{\alpha'\beta'}}(x^{\sigma'}) = \frac{\partial x^{\gamma'}}{\partial x^\rho}\frac{\partial x^\mu}{\partial
      x^{\alpha'}}\frac{\partial x^\nu}{\partial x^{\beta'}}\Gamma \indices{^\rho_{\mu\nu}}(x^\sigma) + \frac{\partial^2 x^\mu}{\partial
         x^{\alpha'} \partial x^{\beta'}} \frac{\partial x^{\gamma'}}{\partial x^\mu}.
         \label{Gammatrans}
\end{equation}

\subsection{Symmetries}
\label{symm}

From the above definitions, the following symmetries hold
\begin{subequations}
\begin{align}
   \Gamma\indices{^\gamma_{[\alpha\beta]}} &= 0,\\
   R_{\alpha\beta\gamma\delta} &= R_{\gamma\delta\alpha\beta},\\
   \label{sym1R}
   R_{(\alpha\beta)\gamma\delta} &= 0,\\
   \label{sym2R}
   R_{\alpha\beta(\gamma\delta)} &= 0,\\
   R_{[\alpha\beta]} &= 0,\\
   G_{[\alpha\beta]} &= 0,\\
   S_{[\alpha\beta]} &= 0,\\
   C_{(\alpha\beta)\gamma} &= 0,\\
   \label{sym1C}
   C_{\alpha\beta\gamma\delta} &= C_{\gamma\delta\alpha\beta},\\
   \label{sym2C}
   C_{(\alpha\beta)\gamma\delta} &= 0,\\
   \label{sym3C}
   C_{\alpha\beta(\gamma\delta)} &= 0.
\end{align}
\label{symmetries}
\end{subequations}

The contractions of the Levi-Civita tensor are
\begin{subequations}
\begin{align}
   \varepsilon_{\alpha_1\ldots\alpha_n}\varepsilon^{\beta_1\ldots\beta_n} &= -\delta^{\beta_1\ldots\beta_n}_{\alpha_1\ldots\alpha_n},\\
   \label{leviC2}
   \varepsilon_{\mu_1\ldots\mu_{k}\alpha_{k+1}\ldots\alpha_n}\varepsilon^{\mu_1\ldots\mu_k\beta_{k+1}\ldots\beta_n} &= -k!\delta^{\beta_{k+1}\ldots\beta_n}_{\alpha_{k+1}\ldots\alpha_n},\\
   \varepsilon_{\mu_1\ldots\mu_n}^{\mu_1\ldots\mu_n} &= -n!.
\end{align}
\label{leviCrel}
\end{subequations}

Last but not least, the Weyl tensor is traceless:
\begin{equation}
   C\indices{^\mu_{\mu\alpha\beta}} = C\indices{^\mu_{\alpha\mu\beta}} = C\indices{^\mu_{\alpha\beta\mu}} = 0.
   \label{traceC}
\end{equation}

\subsection{Identities}
\label{ide}

From the definitions of section \ref{defpropgeom}, it is possible to demonstrate the first Bianchi identity
\begin{equation}
   R\indices{^\alpha_{\beta\gamma\delta}} + R\indices{^\alpha_{\delta\beta\gamma}} + R\indices{^\alpha_{\gamma\delta\beta}} = 0,
\end{equation}
which is also valid for the Weyl tensor
\begin{equation}
   C\indices{^\alpha_{\beta\gamma\delta}} + C\indices{^\alpha_{\delta\beta\gamma}} + C\indices{^\alpha_{\gamma\delta\beta}} = 0.
   \label{bianchi12}
\end{equation}
The definitions of the geometrical tensors also imply the second Bianchi identity
\begin{equation}
   \nabla_\rho R\indices{^\alpha_{\beta\gamma\delta}} + \nabla_\delta R\indices{^\alpha_{\beta\rho\gamma}} + \nabla_\gamma
   R\indices{^\alpha_{\beta\delta\rho}} = 0.
   \label{bianchi2}
\end{equation}
The contraction on $\alpha$ and $\gamma$ of this Bianchi identity yields
\begin{equation}
   \nabla^\mu G_{\alpha\mu} = 0,
   \label{divG}
\end{equation}
which states that the Einstein tensor is divergence-free. Trading the Riemann tensor for the Weyl tensor with \eqref{weyldef}, the
second Bianchi identity \eqref{bianchi2} can be rewritten as
\begin{equation}
   \nabla_\mu C\indices{^\mu_{\alpha\beta\gamma}} = 2\frac{n-3}{n-2}C_{\beta\gamma\alpha}.
   \label{bianchi22}
\end{equation}
The Riemann tensor is directly linked to the non-commutativity of covariant derivatives. Indeed, the Ricci identity is valid for any
vector $v^\alpha$ or 1-form $v_\alpha$ and reads
\begin{subequations}
\begin{align}
   \nabla_\beta\nabla_\gamma v^\alpha - \nabla_\gamma \nabla_\beta v^\alpha &= R\indices{^\alpha_{\mu\beta\gamma}}v^\mu,\\
   \nabla_\beta\nabla_\gamma v_\alpha - \nabla_\gamma \nabla_\beta v_\alpha &= -R\indices{^\mu_{\alpha\beta\gamma}}v_\mu,
\end{align}
\label{ricci}
\end{subequations}
where we have used equation \eqref{sym1R}.

Finally, given the symmetries of \ref{symm} and the above identities, it can be shown that the numbers of independent components
for the Riemann, Weyl and Cotton tensor are \cite{Garcia04,Alcubierre08}
\begin{subequations}
\begin{align}
   \frac{1}{12}n^2(n^2-1) &\quad \tn{(Riemann)},\\
   \label{weylcompind}
   \frac{1}{12}n(n+1)(n+2)(n-3) &\quad \tn{(Weyl)},\\
   \frac{1}{3}n(n^2-4) &\quad \tn{(Cotton)}.
\end{align}
\end{subequations}
From this result, we deduce that in 3-dimensional space-times, the Weyl tensor is identically zero.

\subsection{Einstein's equations}

Einstein's equation in its standard form is (in geometrical units $\gls{G} = \gls{c} = 1$)
\begin{equation}
   R_{\alpha\beta} - \frac{1}{2}R g_{\alpha\beta} + \gls{Lambda}g_{\alpha\beta} = 8\pi T_{\alpha\beta},
   \label{eineq}
\end{equation}
and its trace is
\begin{equation}
   -\frac{n-2}{2}R + \gls{Lambda} n = 8\pi T,
   \label{treineq}
\end{equation}
where $T_{\alpha\beta}$ is the energy-momentum tensor and $T$ its trace. Combining \eqref{eineq} and \eqref{treineq}, a
second form of Einstein's equation is
\begin{equation}
   R_{\alpha\beta} = \frac{2 \gls{Lambda}}{n-2}g_{\alpha\beta} + 8\pi\left( T_{\alpha\beta} - \frac{T}{n-2}g_{\alpha\beta} \right).
   \label{eineq2}
\end{equation}
Trading the Ricci tensor and its trace by the Schouten tensor and its trace in \eqref{eineq} with the definitions
\eqref{schoutendef} and \eqref{trschoutendef}, it comes a third form of Einstein's equation
\begin{equation}
   S_{\alpha\beta} - Sg_{\alpha\beta} + \gls{Lambda}g_{\alpha\beta} = 8\pi T_{\alpha\beta},
   \label{eineq3}
\end{equation}
whose trace is
\begin{equation}
   -(n-1)S + \gls{Lambda}n = 8\pi T.
   \label{treineq3}
\end{equation}
Combining \eqref{eineq3} and \eqref{treineq3}, a fourth form of Einstein's equation is
\begin{equation}
   S_{\alpha\beta} = \frac{\gls{Lambda}}{n-1}g_{\alpha\beta} + 8\pi\left( T_{\alpha\beta} - \frac{T}{n-1}g_{\alpha\beta} \right).
   \label{eineq4}
\end{equation}
Finally, combining the Bianchi identity \eqref{bianchi22} with Einstein's equation \eqref{eineq4}, it
comes
\begin{equation}
   \nabla_\mu C\indices{^\mu_{\alpha\beta\gamma}} = 16\pi\frac{n-3}{n-2}\left( \nabla_{[\beta} T_{\gamma]\alpha} + \frac{1}{n-1}g_{\alpha[\beta}\nabla_{\gamma]} T\right).
   \label{einbianchi}
\end{equation}

\section{Conformal transformations}
\label{conftransform}

Let us denote by $\widehat{g}_{\alpha\beta}$ the conformal transformation of the metric:
\begin{equation}
   \widehat{g}_{\alpha\beta} \equiv \Omega^2 g_{\alpha\beta} \quad \tn{and} \quad \widehat{g}^{\alpha\beta} \equiv \frac{1}{\Omega^2} g^{\alpha\beta},
   \label{defconfg}
\end{equation}
where $\Omega$ is the so-called conformal factor and can be any function of the coordinates. Hereafter, we denote by
\begin{equation}
   \varphi \equiv \ln \Omega,
\end{equation}
its logarithm in order to make the formulas more concise.

\subsection{Conformal tensors}

The geometrical tensors of the conformal metric $\widehat{g}_{\alpha\beta}$ are linked to the one of the original metric
$g_{\alpha\beta}$. Hereafter, hatted tensors (or symbols) are derived from the conformal metric. It can be shown that
\begin{subequations}
\begin{align}
   \label{confGamma}
   \widehat{\Gamma}\indices{^\gamma_{\alpha\beta}} &= \Gamma\indices{^\gamma_{\alpha\beta}} + \delta\indices{^\gamma_\alpha}
   \nabla_\beta \varphi + \delta\indices{^\gamma_\beta} \nabla_\alpha\varphi - g_{\alpha\beta}g^{\gamma\mu}\nabla_\mu \varphi,\\
 \nonumber \widehat{R}\indices{^\alpha_{\beta\gamma\delta}} &= R\indices{^\alpha_{\beta\gamma\delta}} +
 \delta\indices{^\alpha_\delta} \nabla_\gamma\nabla_\beta \varphi - \delta\indices{^\alpha_\gamma}\nabla_\delta\nabla_\beta\varphi + g_{\beta\gamma} \nabla^\alpha\nabla_\delta\varphi - g_{\beta\delta}\nabla^\alpha\nabla_\gamma\varphi \\
 \nonumber                                    &+ \delta\indices{^\alpha_\gamma} \nabla_\beta\varphi \nabla_\delta\varphi -
 \delta\indices{^\alpha_\delta}\nabla_\beta\varphi\nabla_\gamma\varphi \\
 &+ (\delta\indices{^\alpha_\delta} g_{\beta\gamma} - \delta\indices{^\alpha_\gamma} g_{\beta\delta})\nabla^\mu\varphi\nabla_\mu\varphi + (g_{\beta\delta}\nabla_\gamma\varphi - g_{\beta\gamma}\nabla_\delta\varphi)\nabla^\alpha\varphi,\\
\nonumber \widehat{R}_{\alpha\beta} &= R_{\alpha\beta} - (n-2)\nabla_\alpha\nabla_\beta\varphi - g_{\alpha\beta}\nabla^\mu\nabla_\mu\varphi \\
                                             &+ (n-2)\nabla_\alpha\varphi\nabla_\beta\varphi - (n-2)g_{\alpha\beta}\nabla^\mu\varphi\nabla_\mu\varphi,\\
          \widehat{R} &= \frac{1}{\Omega^2}[R - 2(n-1)\nabla^\mu\nabla_\mu\varphi - (n-1)(n-2)\nabla^\mu\varphi\nabla_\mu\varphi],\\
          \widehat{S}_{\alpha\beta} &= S_{\alpha\beta} - (n-2)\nabla_\alpha\nabla_\beta\varphi + (n-2)\nabla_\alpha\varphi\nabla_\beta\varphi - \frac{n-2}{2}g_{\alpha\beta}\nabla^\mu\varphi\nabla_\mu\varphi,\\
          \label{weylconf}
          \widehat{C}_{\alpha\beta\gamma\delta} &= \Omega^2 C_{\alpha\beta\gamma\delta},\\
          \label{cottonconf}
          \widehat{C}_{\alpha\beta\gamma} &= C_{\alpha\beta\gamma} - \frac{n-2}{2}C_{\alpha\beta\gamma\mu}\nabla^\mu\varphi,
\end{align}
\end{subequations}
where $\nabla$ is the covariant derivative associated to the physical metric
$g_{\alpha\beta}$. Let us mention that the computation of the conformal Riemann tensor is more easily done in an astutely chosen frame
where $\Gamma\indices{^\gamma_{\alpha\beta}} = 0$ (but $\partial_\delta \Gamma\indices{^\gamma_{\alpha\beta}} \neq 0$) and $\partial_\alpha =
\nabla_\alpha$. Let us also mention that equation \eqref{weylconf} is equivalent to
\begin{equation}
   \widehat{C}\indices{^\alpha_{\beta\gamma\delta}} = C\indices{^\alpha_{\beta\gamma\delta}},
   \label{weylconf2}
\end{equation}
which expresses the standard conformal invariance of the Weyl tensor. Note also the particular case of $n=3$ for which the Weyl
tensor is identically zero (see equation \eqref{weylcompind}) and the role of the conformally invariant tensor is taken over by
the Cotton tensor (according to \eqref{cottonconf}).

\subsection{Conformal Killing vectors}

Suppose that $\xi^\alpha$ is a Killing vector for the metric $g_{\alpha\beta}$. It obeys the Killing equation
\begin{equation}
   \nabla^{(\alpha}\xi^{\beta)} = 0.
\end{equation}
In terms of hatted quantities, trading the $\nabla$ for $\widehat{\nabla}$ with the help of \eqref{defcov}, \eqref{defconfg} and
\eqref{confGamma} it comes
\begin{equation}
   \nabla^{(\alpha}\xi^{\beta)} = \Omega^2 \widehat{\nabla}^{(\alpha}\xi^{\beta)} - \Omega\widehat{g}^{\alpha\beta} \xi^\mu
   \widehat{\nabla}_\mu \Omega = 0.
   \label{killing1}
\end{equation}
The trace of this equation is
\begin{equation}
   \Omega \widehat{\nabla}_\mu\xi^\mu = n\xi^\mu\widehat{\nabla}_\mu \Omega .
   \label{trkilling1}
\end{equation}
Eliminating $\xi^\mu\widehat{\nabla}_\mu \Omega$ between \eqref{killing1} and \eqref{trkilling1}, we recover that
\begin{equation}
   \nabla^{(\alpha}\xi^{\beta)} = 0 \iff \widehat{\nabla}^{(\alpha}\xi^{\beta)} = \frac{1}{n}\widehat{g}^{\alpha\beta}\widehat{\nabla}_\mu
   \xi^\mu
   \label{killingconfkilling}
\end{equation}
The second equation is the conformal Killing equation. This means that a Killing vector for $g_{\alpha\beta}$
is a conformal Killing vector for $\widehat{g}_{\alpha\beta}$.

\chapter{$d+1$ formalism}
\label{d+1}
\citationChap{Research is what I'm doing when I don't know what I'm doing.}{Wernher Von Braun}
\minitoc

In order to get a more intuitive interpretation of the system of partial differential equations provided by Einstein's equation,
the $d+1$ formalism separates distinguishes evolution and constraint equations. This setup is based on projections on
a family of hypersurfaces and allows to reformulate the system of equations in a way that is more appropriate to numerical
computations. Furthermore, the formalism makes the discussion of gauge freedom much easier than in the original formulation.
The formalism is also fundamental to the Hamiltonian formulation of \gls{gr}, discussed in detail in \cite{Poisson04}.

Some of the arguments are too lengthy to be exposed in this appendix and the reader is thus referred to standard textbooks like
\cite{Poisson04,Alcubierre08,Baumgarte10,Gourgoulhon07}. Throughout all this chapter, the dimension of the space-time is $d+1$.
We thus particularise one direction in space-time, and hypersurfaces orthogonal to this direction are of dimension $d$. We use the
geometrical units in which $\gls{G} = \gls{c} = 1$.

\section{$d+1$ decomposition}

The main idea behind the standard $d+1$ formalism is to study the time evolution of $d$-dimensional geometrical quantities, which is somewhat
reminiscent of Newton's approach. However, unlike Newton's theory, time is not absolute and hypersurfaces of constant time
coordinate are not flat a priori. This split of space-time into a time direction and a family of space-like hypersurfaces is called
the space-time foliation. However, in this chapter, we consider foliations of both space-like and time-like hypersurfaces in
$d+1$-dimensional space-times. In other words, we particularise one space-time direction that is not necessarily the time. All
geometrical and tensorial quantities can then be projected either onto the family of hypersurfaces or along the normal direction.
These $d$-dimensional hypersurfaces, embedded in the $d+1$-dimensional space-time, have both an intrinsic and an extrinsic
curvature, that determine the overall geometry. Hereafter, we use Latin indices $i,j,k,l$ instead of Greek letters
$\alpha,\beta,\gamma,\delta$ to denote indices restricted to coordinates adapted to the hypersurfaces. Furthermore, we introduce
\begin{equation}
   \epsilon = \left\{
   \begin{array}{cc}
      -1 & \tn{(space-like foliation)},\\
      +1 & \tn{(time-like foliation)},
   \end{array}
   \right .
\end{equation}
which allows us to treat both cases simultaneously.

\subsection{Space-time foliation}

Let us consider a foliation of space-time with hypersurfaces of constant $x$ coordinate. In general, $x$ is either the time
coordinate $t$ or the radial coordinate $r$. Namely,
\begin{equation}
   \mathcal{M} = \bigcup_{x} \Sigma_x,
\end{equation}
where $\mathcal{M}$ is the space-time manifold, and each $\Sigma_x$ corresponds to constant $x$ slices. We denote by $u_\alpha$
the unit normal to $\Sigma_x$. Since it is normal to constant $x$ hypersurfaces, it must be proportional to
$\nabla_\alpha x$. Thus
\begin{equation}
   u_{\alpha} = \epsilon N \nabla_\alpha x \quad \tn{with} \quad N = \frac{1}{\sqrt{\epsilon g^{\mu\nu}\nabla_\mu x \nabla_\nu x}},
   \label{udef}
\end{equation}
where $N$ is the so-called lapse function and ensures the normalisation
\begin{equation}
   g^{\mu\nu}u_\mu u_\nu = \epsilon.
   \label{unormal}
\end{equation}
We have chosen $u_\alpha$ to be oriented toward the future if $x$ is a time coordinate and toward the increasing radii if $x$ is a
radial coordinate. The normalisation condition can also be read $u^\mu u_\mu = \epsilon$, so that $u^\alpha$ (with upper index)
can be written
\begin{equation}
   u^\alpha = \frac{1}{N}(\partial_x^\alpha - \beta^\alpha) \quad \tn{with} \quad u_\mu \beta^\mu = 0,
   \label{ua}
\end{equation}
where the components $\beta^\alpha$ are those of a $d$-dimensional vector living on $\Sigma_x$, the so-called shift vector. This equation can be interpreted as a
definition of the latter. The lapse function and the shift vector are not imposed by the equations of motion and can be
interpreted as the four gauge degrees of freedom. Choosing $N$ and $\beta^\alpha$ is thus, in some sense, a choice of coordinates.

\subsection{Projection onto hypersurfaces}

We define the projection operator onto $\Sigma_x$ by
\begin{equation}
   \gamma\indices{^\alpha_\beta} \equiv \delta\indices{^\alpha_\beta} - \epsilon u^\alpha u_\beta.
   \label{projdef}
\end{equation}
It can be shown that $\gamma \indices{^\alpha_\beta}$ is a projector since
\begin{equation}
   \gamma \indices{^\mu_\alpha}u_\mu = 0 \quad \tn{and} \quad \forall v^\alpha, v^\mu u_\mu = 0, \gamma \indices{^\alpha_\mu}v^\mu = v^\alpha.
   \label{projeq}
\end{equation}
If we lower the first index, it comes
\begin{equation}
   \gamma_{\alpha\beta} = g_{\alpha\beta} - \epsilon u_\alpha u_\beta.
\end{equation}
Since $u_i = 0$ (equation \eqref{udef}), we have $\gamma_{ij} = g_{ij}$. The metric $\gamma_{ij}$ is thus called the $d$-metric, or
induced metric on $\Sigma_x$. It measures distances between points restricted to $\Sigma_x$. Besides, its trace is
\begin{equation}
   \gamma^{\mu\nu}\gamma_{\mu\nu} = d.
\end{equation}
This metric has an associated covariant derivative, which is unique and can be shown to be \cite{Gourgoulhon07}
\begin{equation}
   D_\gamma T \indices{^\alpha_\beta} = \gamma \indices{^\rho_\gamma} \gamma \indices{^\alpha_\mu} \gamma \indices{^\nu_\beta} \nabla_\rho T \indices{^\mu_\nu}.
   \label{covdefd+1}
\end{equation}
The generalisation to tensors of arbitrary valence is straightforward. It is compatible with the $d$-metric, namely
\begin{equation}
   D_{\gamma}\gamma_{\alpha\beta} = 0.
\end{equation}
Let us also denote by $\gamma(v)$ the projection of a vector $v$:
\begin{equation}
   \gamma(v)^\alpha \equiv \gamma \indices{^\alpha_\mu}v^\mu = v^\alpha - \epsilon u_\mu v^\mu u^\alpha.
\end{equation}
As for the generalisation to tensor of arbitrary valence, each index has to be contracted once with the projection
tensor.

In order to get the $d+1$ metric as a function of $N$, $\beta^\alpha$ and $\gamma_{\alpha\beta}$, we notice that in coordinates
adapted to the foliation we get, with the help of \eqref{ua}
\begin{subequations}
\begin{align}
   g_{xx} &= \partial_x \cdot \partial_x = (Nu_\mu + \beta_\mu)(Nu^\mu + \beta^\mu) = \epsilon N^2 + \beta_\mu \beta^\mu,\\
   g_{xi} &= \partial_x \cdot \partial_i = (Nu_\mu + \beta_\mu)\partial_i^\mu = \beta_i,\\ g_{ij} &= \gamma_{ij}.
\end{align}
\end{subequations}
Inverting the matrix $g_{\alpha\beta}$ gives finally
\begin{subequations}
\begin{align}
   g_{xx} &= \epsilon N^2 + \beta_\mu \beta^\mu, &\quad g_{xi} &= \beta_i, &\quad g_{ij} &= \gamma_{ij},\\
   g^{xx} &= \frac{\epsilon}{N^2}, &\quad g^{xi} &= -\frac{\epsilon\beta^i}{N^2}, &\quad g^{ij} &= \gamma^{ij} + \frac{\epsilon \beta^i \beta^j}{N^2}.
\end{align}
\label{metricd+1}
\end{subequations}

The $d$-metric $\gamma_{\alpha\beta}$ is, however, not sufficient to describe the whole geometry, since it contains much less information than the
full metric $g_{\alpha\beta}$. The missing degrees of freedom are contained in the extrinsic curvature tensor.

\subsection{Extrinsic curvature}

Let us define the extrinsic curvature bilinear form\footnote{Recall that the covariant derivative along a vector is $(\nabla_w
v)^\alpha = w^\mu \nabla_\mu v^\alpha$.} by \cite{Gourgoulhon07}
\begin{equation}
   \forall (v,w), \quad K(v,w) \equiv -\gamma(v)\cdot \nabla_{\gamma(w)}u,
\end{equation}
where $v$ and $w$ are two arbitrary vectors. If we introduce the acceleration of an Eulerian observer of world-line $u^\alpha$
\begin{equation}
   a_{\alpha} \equiv (\nabla_u u)_\alpha = u^\mu \nabla_\mu u_{\alpha},
   \label{adef}
\end{equation}
we can demonstrate with \eqref{unormal} that
\begin{equation}
   a_\mu u^\mu = \frac{1}{2}u^\mu\nabla_\mu(u^\nu u_\nu) = \frac{1}{2}u^\mu \nabla_\mu (\epsilon) = 0.
\end{equation}
This implies that the extrinsic curvature tensor obeys
\begin{equation}
   K(v,w) = -[v - \epsilon (v\cdot u)u]\cdot [\nabla_w u -\epsilon (u\cdot w) a] = -v\cdot \nabla_w u + \epsilon(u\cdot w)(a\cdot v).
\end{equation}
This equation can be translated into components of the extrinsic curvature tensor by choosing $v$ and $w$ to be the different
coordinates generators. It thus comes
\begin{equation}
   K_{\alpha\beta} = -\nabla_\beta u_\alpha + \epsilon a_{\alpha}u_\beta.
   \label{Kdef}
\end{equation}
Furthermore, since $K_{\alpha\beta}$ is built only with projected vectors, it is invariant by projection and is orthogonal to
$u^\alpha$. Namely
\begin{equation}
   \gamma \indices{^\mu_\alpha}\gamma \indices{^\nu_\beta} K_{\mu\nu} = K_{\alpha\beta} \quad \tn{and} \quad K_{\alpha\mu}u^\mu =
   0.
\end{equation}
Since $\gamma \indices{^\mu_\alpha}u_\mu = 0$ (equation \eqref{projeq}), it comes also
\begin{equation}
   K_{\alpha\beta} = -\gamma \indices{^\mu_\alpha}\gamma \indices{^\nu_\beta}\nabla_\mu u_\nu.
   \label{Kdef2}
\end{equation}
The trace of the extrinsic curvature tensor is called the mean extrinsic curvature, and is given by
\begin{equation}
   K \equiv \gamma^{\mu\nu}K_{\mu\nu} = -\nabla_\mu u^\mu.
\end{equation}
The extrinsic curvature tensor encodes part of the degrees of freedom of the $d+1$-dimensional metric. The tensor $\gamma_{ij}$ and
$K_{ij}$ are sometimes called the first and second fundamental forms of the space-time. Actually, Einstein's equations can be
formulated as evolutions equations for both of them.

\subsection{Evolution operator}
\label{evolop}

The acceleration of an observer of worldline $u^\alpha$ can be rewritten by combining \eqref{udef} and \eqref{adef} as
\begin{equation}
   a_{\alpha} = -\epsilon D_{\alpha}\ln N.
\end{equation}
We also introduce the vector (compare with \eqref{ua})
\begin{equation}
   m^\alpha \equiv N u^\alpha = \partial_x^\alpha - \beta^\alpha.
   \label{mdef}
\end{equation}
With these quantities at hand, it comes
\begin{equation}
   \mathcal{L}_m \gamma_{\alpha\beta} = -2NK_{\alpha\beta}, \quad \mathcal{L}_m \gamma \indices{^\alpha_\beta} = 0, \quad
   \mathcal{L}_m \gamma^{\alpha\beta} = 2NK^{\alpha\beta},
   \label{evolgamma}
\end{equation}
where $\mathcal{L}_m$ denotes the Lie derivative along the vector $m$ (see equation \eqref{lie}). These relations are purely geometrical and are not
equations of motions. Instead they are just a consequence of the $d+1$ decomposition of the metric. The operator $\mathcal{L}_m$
being roughly a evolution operator along $x$, we can thus interpret the extrinsic curvature tensor as the ``speed'' of the $d$-metric along the
$x$ coordinate.

\subsection{$d+1$ decomposition of Christoffel symbols}

The Christoffel symbols (equation \eqref{christoffel}) can also be expressed in terms of the $d+1$ geometrical quantities
\cite{Alcubierre08}. Combining \eqref{metricd+1},\eqref{mdef} and \eqref{evolgamma} it comes
\begin{subequations}
\begin{align}
   \Gamma\indices{^x_{xx}} &= \frac{1}{N}(\partial_x N + \beta^k\partial_k N + \epsilon \beta^k\beta^l K_{kl}),\\
   \Gamma\indices{^x_{xi}} &= \frac{1}{N}(\partial_i N + \epsilon \beta^k K_{ik}),\\
   \Gamma\indices{^x_{ij}} &= \frac{\epsilon K_{ij}}{N},\\
   \Gamma\indices{^i_{xx}} &= -\epsilon N\partial^i N - 2N\beta^k K \indices{^i_k} + \partial_x \beta^i + \beta^k D_k \beta^i - \frac{\beta^i}{N}(\partial_x N + \beta^k \partial_kN + \epsilon \beta^k\beta^l K_{kl}),\\
   \Gamma\indices{^i_{jx}} &= -\frac{\beta^i}{N}(\partial_j N + \epsilon \beta^k K_{jk} - N K \indices{^i_j} + D_j \beta^i),\\
   \Gamma\indices{^k_{ij}} &= \digamma\indices{^k_{ij}} - \frac{\epsilon \beta^k K_{ij}}{N},
\end{align}
\label{d+1christo}
\end{subequations}
where $\digamma \indices{^k_{ij}}$ are the Christoffel symbols of the $d$-metric $\gamma_{ij}$. Defining also
\begin{equation}
   \Gamma^\alpha \equiv g^{\mu\nu}\Gamma \indices{^\alpha_{\mu\nu}} \quad \tn{and} \quad \digamma^\alpha \equiv
   \gamma^{\mu\nu}\digamma \indices{^\alpha_{\mu\nu}},
\end{equation}
it comes
\begin{subequations}
\begin{align}
   \Gamma^x &= \frac{\epsilon}{N}(\partial_x N - \beta^k \partial_k N + N^2 K),\\
   \Gamma^i &= \digamma^i - \beta^i \Gamma^x + \frac{\epsilon}{N}(\partial_x \beta^i - \beta^k \partial_k \beta^i - \epsilon N \partial^i N).
\end{align}
\label{d+1christo2}
\end{subequations}
The extrinsic curvature tensor is thus tightly linked to the first order derivatives of the metric and the Christoffel symbols.

\section{The Gauss-Codazzi relations}

In the previous section, we have defined the fundamental geometrical quantities of the $d+1$ decomposition. However, Einstein's theory
is all about curvature. How to relate the (intrinsic) curvature $R$ of the $d+1$-metric $g_{\alpha\beta}$ to the intrinsic curvature
$\mathcal{R}$ of the $d$-metric $\gamma_{ij}$ and its extrinsic curvature tensor $K_{ij}$? The answer lies in the so-called Gauss-Codazzi
relations. Their demonstrations is mainly based on the projections of the Ricci identity \eqref{ricci} and can be found in
\cite{Poisson04,Gourgoulhon07}.

\subsection{The Gauss relations}

Starting from the Ricci identity for the $d$-metric
\begin{equation}
   \forall v \in \Sigma_x, D_\alpha D_\beta v^\gamma - D_\beta D_\alpha v^\gamma = \mathcal{R}
   \indices{^\gamma_{\mu\alpha\beta}}v^\mu,
\end{equation}
and combining with the $d+1$-dimensional Ricci identity \eqref{ricci} as well as \eqref{projeq}, \eqref{covdefd+1} and
\eqref{Kdef2}, the Gauss relation is the following \cite{Gourgoulhon07}
\begin{equation}
   \gamma \indices{^\mu_\alpha}\gamma \indices{^\nu_\beta}\gamma \indices{^\rho_\gamma}\gamma
   \indices{^\sigma_\delta}R_{\mu\nu\rho\sigma} = \mathcal{R}_{\alpha\beta\gamma\delta} -\epsilon( K_{\alpha\gamma}K_{\beta\delta}
   - K_{\beta\gamma}K_{\alpha\delta}).
   \label{gauss}
\end{equation}
Note that with all the projections involved, the demonstration not only holds for all $v \in \Sigma_x$, but for all $v$ in the manifold
$\mathcal{M}$. This relation thus directly relates the projection of the $d+1$-dimensional Riemann tensor to the $d$-dimensional one
and the extrinsic curvature tensor. Contracting the indices $\alpha$ and $\gamma$, and playing with the symmetries of the Riemann
tensor (equation \eqref{symmetries}), it comes
\begin{equation}
   \gamma \indices{^\mu_\alpha}\gamma \indices{^\nu_\beta} R_{\mu\nu} -\epsilon \gamma \indices{^\mu_\alpha}\gamma
   \indices{^\rho_\beta}u^\nu u^\sigma R_{\mu\nu\rho\sigma} = \mathcal{R}_{\alpha\beta} - \epsilon( K K_{\alpha\beta} -
   K_{\alpha\mu}K \indices{^\mu_\beta}).
   \label{gausscontract}
\end{equation}
This is the contracted Gauss relation and relates the $d+1$ and $d$-dimensional Ricci tensors to each other. Contracting again this
relation with $\gamma^{\alpha\beta}$, it comes
\begin{equation}
   R - 2\epsilon R_{\mu\nu}u^\mu u^\nu = \mathcal{R} -\epsilon( K^2 - K_{\mu\nu}K^{\mu\nu}).
   \label{gaussscalar}
\end{equation}
This is the scalar Gauss equation and relates the $d+1$ and $d$-dimensional Ricci scalars to each other.

\subsection{The Codazzi relations}

Unlike the Gauss relations, we start from the $d+1$-dimensional Ricci identity \eqref{ricci} applied to the unit normal vector
$u^\alpha$:
\begin{equation}
   \nabla_\alpha \nabla_\beta u^\gamma - \nabla_\beta \nabla_\alpha u^\gamma = R \indices{^\gamma_{\mu\alpha\beta}}u^\mu.
   \label{ricciu}
\end{equation}
Projecting with $\gamma \indices{^\mu_\alpha}\gamma \indices{^\nu_\beta} \gamma \indices{^\gamma_{\rho}}$ and using
\eqref{covdefd+1} and \eqref{Kdef}, it comes
\begin{equation}
   \gamma \indices{^\mu_\alpha}\gamma \indices{^\nu_\beta}\gamma \indices{^\rho_\gamma}u^\sigma R_{\mu\nu\rho\sigma} =
   D_\beta K\indices{^\gamma_\alpha} - D_\alpha K \indices{^\gamma_\beta}.
   \label{ggguR}
\end{equation}
This is the Codazzi relation. Unlike its Gauss counterpart, the Riemann tensor is projected three times onto $\Sigma_x$ and one time
onto $u^\alpha$. Contracting the $\alpha$ and $\gamma$, playing with the symmetries \eqref{symmetries} and using the projector
definition \eqref{projdef}, it comes
\begin{equation}
   \gamma \indices{^\mu_\alpha}u^\nu R_{\mu\nu} = D_{\alpha} K - D_\mu K \indices{^\mu_\alpha}.
   \label{guR}
\end{equation}
This is the contracted Codazzi relation, and it involves the Ricci tensor contracted once onto $\Sigma_x$ and once onto $u^\alpha$.

\subsection{Last non-trivial projection of the Riemann tensor}

The Gauss-Codazzi relations above do not explore all possible projections of the Riemann tensor. Projecting it three times on
$u^\alpha$ is trivial by the antisymmetry \eqref{sym2R}. However, a double projection on both $\Sigma_x$ and $u^\alpha$ is not
trivial. Starting again with the Ricci identity applied to the unit normal vector $u^\alpha$ (equation \eqref{ricciu}), using the
$d$-covariant derivative \eqref{covdefd+1} and projecting with $\gamma_{\gamma\mu}\gamma \indices{^\nu_\alpha}u^\beta$, it comes
\begin{equation}
   \gamma \indices{^\mu_\alpha} \gamma \indices{^\nu_\beta}u^\rho u^\sigma R_{\mu\rho\nu\sigma} = \frac{1}{N}\mathcal{L}_m
   K_{\alpha\beta} - \frac{\epsilon}{N} D_\alpha D_\beta N + K_{\alpha\mu}K\indices{^\mu_\beta}.
   \label{gguuR}
\end{equation}
This is the last non-trivial projection of the Riemann tensor. If we combine it with \eqref{gausscontract}, we can isolate the projection
of the Ricci tensor. It is
\begin{equation}
   \gamma \indices{^\mu_\alpha}\gamma \indices{^\nu_\beta} R_{\mu\nu} = \frac{\epsilon}{N}\mathcal{L}_m K_{\alpha\beta} -
   \frac{1}{N} D_\alpha D_\beta N + \mathcal{R}_{\alpha\beta} - \epsilon(K K_{\alpha\beta} - 2 K_{\alpha\mu}K
   \indices{^\mu_\beta}).
   \label{ggR}
\end{equation}
Contracting with $\gamma^{\alpha\beta}$ and combining with \eqref{evolgamma}, it comes
\begin{equation}
   R -\epsilon R_{\mu\nu}u^\mu u^\nu = \mathcal{R} - \epsilon K^2 + \frac{\epsilon}{N} \mathcal{L}_m K - \frac{1}{N}D_\mu D^\mu N.
   \label{R1}
\end{equation}
This equation is independent and reminiscent of the scalar Gauss relation. Indeed, if we combine with \eqref{gaussscalar}, we can
isolate the Ricci scalar as the following
\begin{equation}
   R = \mathcal{R} - \epsilon(K^2 + K_{\mu\nu}K^{\mu\nu}) + \frac{2\epsilon}{N}\mathcal{L}_m K - \frac{2}{N}D_\mu D^\mu N.
   \label{R2}
\end{equation}
Eliminating $R$ between \eqref{R1} and \eqref{R2} gives the expression of the double projection of the Ricci tensor onto $u^\alpha$:
\begin{equation}
   R_{\mu\nu}u^\mu u^\nu = -K_{\mu\nu}K^{\mu\nu} + \frac{1}{N}\mathcal{L}_m K - \frac{\epsilon}{N}D_\mu D^\mu N.
   \label{Ruu}
\end{equation}
All the Gauss-Codazzi relations and the relations above are the fundamental consequences of the $d+1$ decomposition. They always
relate one projection of a $d+1$-dimensional curvature tensor to the $d$-dimensional counterpart, the extrinsic curvature tensor and the
lapse function and shift vector.

\section{$d+1$ decomposition of Einstein's equation}
\label{d+1ein}

With the Gauss-Codazzi relations above, we can now tackle the problem of converting the $d+1$-dimensional Einstein's equation in a
set of evolution and constraint equations for the $d$-dimensional quantities.

\subsection{Decomposition of the energy-momentum tensor}

Let us denote by $T_{\alpha\beta}$ the energy-momentum tensor. Its projections are defined as follows:
\begin{equation}
   \rho \equiv T_{\mu\nu}u^\mu u^\nu, \quad p_\alpha \equiv -\gamma \indices{^\mu_\alpha}u^\nu T_{\mu\nu},\quad S_{\alpha\beta}
   \equiv \gamma \indices{^\mu_\alpha} \gamma \indices{^\nu_\beta}T_{\mu\nu}.
\end{equation}
They correspond respectively to the energy density, the momentum and the stress tensor of the matter field.
Making these equations more explicit with \eqref{projdef}, it comes
\begin{equation}
   T_{\alpha\beta} = S_{\alpha\beta} - \epsilon(u_\alpha p_\beta + u_\beta p_\alpha) + \rho u_\alpha u_\beta.
\end{equation}
This is actually the general $d+1$ decomposition of a symmetric bilinear form. Denoting by $S$ the trace of $S_{\alpha\beta}$ and
by $T$ the trace of $T_{\alpha\beta}$, we get
\begin{equation}
   T = S + \epsilon\rho.
   \label{tsr}
\end{equation}
The $d+1$ Einstein's equations thus simply result from the Gauss-Codazzi relations and the decomposition of the energy-momentum tensor.

\subsection{Decomposition of Einstein's equation}

We start with the standard form of Einstein's equation \eqref{eineq}
\begin{equation}
   R_{\alpha\beta} - \frac{1}{2}R g_{\alpha\beta} + \gls{Lambda}g_{\alpha\beta} = 8\pi T_{\alpha\beta},
   \label{eini}
\end{equation}
and projecting onto $u^\alpha u^\beta$, it comes
\begin{equation}
   -\epsilon\mathcal{R} + K^2 - K_{\mu\nu}K^{\mu\nu} + 2\epsilon \gls{Lambda} = 16\pi \rho.
   \label{ham}
\end{equation}
This is the so-called Hamiltonian constraint.

The second projection of Einstein's equation in its standard form \eqref{eini} is onto $\gamma
\indices{^\alpha_\mu}u^\beta$, and gives
\begin{equation}
   D_\mu K \indices{^\mu_\alpha} - D_\alpha K = 8\pi p_\alpha.
   \label{mom}
\end{equation}
This is the so-called momentum constraint.

Now if we consider Einstein's equation in the alternative form \eqref{eineq2}
\begin{equation}
   R_{\alpha\beta} = \frac{2\gls{Lambda}}{d-1}g_{\alpha\beta} + 8\pi\left(T_{\alpha\beta} - \frac{T}{d-1}g_{\alpha\beta}\right),
\end{equation}
and project it with $\gamma \indices{^\alpha_\mu} \gamma \indices{^\beta_\nu}$, it comes
\begin{subequations}
\begin{align}
\label{lmKij1}
   \mathcal{L}_m K_{\alpha\beta} &= \epsilon D_\alpha D_\beta N + N \left[-\epsilon\mathcal{R}_{\alpha\beta} + K K_{\alpha\beta} - 2 K_{\alpha\mu} K \indices{^\mu_\beta} + \frac{2\epsilon\gls{Lambda}}{d-1}\gamma_{\alpha\beta} + 8\pi\epsilon\left(S_{\alpha\beta} - \frac{S+\epsilon\rho}{d-1}\gamma_{\alpha\beta} \right)\right],\\
\label{lmKij2}
   \mathcal{L}_m K^{\alpha\beta} &= \epsilon D^\alpha D^\beta N + N \left[-\epsilon\mathcal{R}^{\alpha\beta} + K K^{\alpha\beta} + 2 K^{\alpha\mu} K \indices{_\mu^\beta} + \frac{2\epsilon\gls{Lambda}}{d-1}\gamma^{\alpha\beta} + 8\pi\epsilon\left(S^{\alpha\beta} - \frac{S+\epsilon\rho}{d-1}\gamma^{\alpha\beta} \right)\right],\\
\label{lmKij3}
   \mathcal{L}_m K \indices{^\alpha_\beta} &= \epsilon D^\alpha D_\beta N + N \left[-\epsilon\mathcal{R}\indices{^\alpha_\beta} + K K\indices{^\alpha_\beta} + \frac{2\epsilon\gls{Lambda}}{d-1}\delta\indices{^\alpha_\beta} + 8\pi\epsilon\left(S\indices{^\alpha_\beta} - \frac{S+\epsilon\rho}{d-1}\delta\indices{^\alpha_\beta} \right)\right],
\end{align}
\label{lmKij}%
\end{subequations}
where \eqref{lmKij2} and \eqref{lmKij3} follow from \eqref{lmKij1} and \eqref{evolgamma}.
The trace of equation \eqref{lmKij1}, with the help of \eqref{evolgamma}, is
\begin{equation}
   \mathcal{L}_m K = \epsilon D_\mu D^\mu N + N\left[-\epsilon R + K^2 + \frac{2\epsilon\gls{Lambda}d}{d-1} -
   \frac{8\pi}{d-1}(\epsilon S + d\rho)\right],
   \label{traceevol}
\end{equation}
or equivalently using \eqref{ham}
\begin{equation}
   \mathcal{L}_m K = \epsilon D_\mu D^\mu N + N\left[K_{\mu\nu}K^{\mu\nu} + \frac{2\epsilon\gls{Lambda}}{d-1} -\frac{8\pi}{d-1}(\epsilon S -
   (d-2)\rho)\right].
   \label{lmK}
\end{equation}
Let us stress that equation \eqref{evolgamma} is a first order evolution equation for the $d$-metric and that \eqref{lmKij} is a
first order evolution equation for the extrinsic curvature tensor. We thus have, in some sense, reduced the second order Einstein
equation to a system of two first order equations. However, besides evolution equations, the projections of the $d+1$-dimensional
Einstein's equation also brought in the constraint equations \eqref{ham} and \eqref{mom}.

At this point, let us notice that we have four degrees of freedom (the lapse function $N$ and the three components of the shift
vector $\beta^i$) as well as four constraints (the Hamiltonian constraint \eqref{ham} and the three components of the momentum
constraint \eqref{mom}). This suggests that we could enforce the gauge freedom with the help of the constraint equations. This is
indeed the case and we refer to \cite{Gourgoulhon07} for a detailed discussion of the gauge freedom in the context of numerical
simulations within the $d+1$ formalism.

\section{3+1 decomposition of the Weyl tensor}

The Weyl tensor has the following expression (see equations \eqref{weyldef} and \eqref{schoutendef})
\begin{equation}
   C_{\alpha\beta\gamma\delta} = R_{\alpha\beta\gamma\delta} -\frac{1}{d-1}(g_{\alpha\gamma}R_{\beta\delta} -
   g_{\alpha\delta}R_{\beta\gamma} - g_{\beta\gamma}R_{\alpha\delta} + g_{\gamma\delta}R_{\alpha\gamma}) +
   \frac{R}{d(d-1)}(g_{\alpha\gamma}g_{\beta\delta} - g_{\alpha\delta}g_{\beta\gamma}).
   \label{weyldef2}
\end{equation}
Even if not directly involved in Einstein's equation, it is useful to give its $3+1$ decomposition as it is notably involved in
the computation of global charges in \gls{ads} space-time. This decomposition mainly relies on the split of the Weyl tensor into
an electric and a magnetic part.

\subsection{Electric part}
\label{elecpartweyl}

The electric part of the Weyl tensor is defined by
\begin{equation}
   E_{\alpha\beta} \equiv C_{\alpha\mu\beta\nu}u^\mu u^\nu.
   \label{defE}
\end{equation}
From this definition, we can infer several properties of this tensor.
\begin{myitem}
   \item It is symmetric: $E_{[\alpha\beta]} = 0$. This is a direct consequence of the symmetry of the Weyl tensor (equation
      \eqref{sym1C}).
   \item It is traceless: $g^{\mu\nu}E_{\mu\nu} = 0$. This comes from the traceless character of the Weyl tensor (equation
      \eqref{traceC}).
   \item It is transverse: $u^\mu E_{\alpha\mu} = 0$. This results from the antisymmetry of the Weyl tensor (equation
      \eqref{sym2C}).
\end{myitem}
The three properties above reduce its number of independent components to $\frac{d(d+1)}{2}-1$, i.e.\ five if $d = 3$.

The $d+1$ decomposition of $E_{\alpha\beta}$ proceeds as follows. Transforming \eqref{defE} with \eqref{weyldef2} gives
\begin{equation}
   E_{\alpha\beta} = R_{\alpha\mu\beta\nu}u^\mu u^\nu - \frac{1}{d-1}(g_{\alpha\beta}R_{\mu\nu}u^\mu u^\nu -
   u_{\alpha}R_{\beta\mu}u^\mu - u_\beta R_{\alpha\mu}u^\mu + \epsilon R_{\alpha\beta}) +\frac{\epsilon R}{d(d-1)}\gamma_{\alpha\beta}.
\end{equation}
Now contracting with $\gamma\indices{^\alpha_\mu} \gamma\indices{^\beta_\nu}$, and combining with \eqref{gguuR}, \eqref{Ruu}, \eqref{ggR},
\eqref{lmKij}, \eqref{lmK}, \eqref{treineq} and \eqref{tsr} it comes
\begin{equation}
   \gamma(E)_{\alpha\beta} = -\epsilon\mathcal{R}_{\alpha\beta} + KK_{\alpha\beta} - K_{\alpha\mu}K^\mu_\beta +
   \frac{2\epsilon\gls{Lambda}}{d}\gamma_{\alpha\beta} + \frac{8\pi\epsilon(d-2)}{d(d-1)}\left(
   dS_{\alpha\beta} - \left[S + \frac{2\epsilon(d-1)\rho}{d-2}\right]\gamma_{\alpha\beta}\right).
   \label{Ed+1}
\end{equation}
This equation thus relates directly the electric part of the Weyl tensor to the $d+1$ geometrical quantities. Any other projection
or mixed projection onto $u^\alpha$ is trivial due to the antisymmetry of the Weyl tensor \eqref{symmetries}.

\subsection{Magnetic part}

From now on, we restrict our computations to $d = 3$. The magnetic part of the Weyl tensor is then defined by
\begin{equation}
   B_{\alpha\beta} \equiv \star C_{\alpha\mu\beta\nu} u^\mu u^\nu = \frac{1}{2}\varepsilon\indices{^{\rho\sigma}_{\beta\nu}}C_{\alpha\mu\rho\sigma}u^\mu u^\nu.
   \label{defB}
\end{equation}
From this definition, we can infer several properties of this tensor.
\begin{myitem}
   \item It is symmetric: $B_{[\alpha\beta]} = 0$. Indeed, this is equivalent to
      $\varepsilon\indices{^{\mu\nu}_{\alpha\beta}}B_{\mu\nu} = 0$, and with the help of \eqref{leviC2}, it can be shown that it
      is a direct consequence of the traceless (equation \ref{traceC}) and antisymmetric (equation \ref{sym3C}) characters of the
      Weyl tensor.
   \item It is traceless: $g^{\mu\nu}B_{\mu\nu} = 0$. Indeed, this is equivalent to $g^{\mu\nu}\star
      C_{\mu\alpha\nu\beta} = \frac{1}{2}\varepsilon\indices{^{\mu\nu\tau}_\beta} C_{\tau\alpha\mu\nu} = 0$. Playing with the
      Bianchi identity \eqref{bianchi12}, the symmetries of the Weyl tensor \eqref{sym1C} and \eqref{sym3C}, and the antisymmetry
      of the Levi-Civita tensor \eqref{leviCdef}, it can be shown that $\frac{1}{2}\varepsilon\indices{^{\mu\nu\tau}_\beta}
      C_{\tau\alpha\mu\nu} = -\varepsilon\indices{^{\mu\nu\tau}_\beta} C_{\tau\alpha\mu\nu}$, which can only be true if both left
      and right-hand sides are zero.
   \item It is transverse: $u^\mu B_{\alpha\mu} = 0$.  This is a direct consequence of the anti-symmetry of the Weyl tensor
      (equation \ref{sym2C}).
\end{myitem}
The three properties above reduce its number of independent components to five.

The $3+1$ decomposition of $B_{\alpha\beta}$ proceeds as follows. We define
\begin{equation}
   \varepsilon_{\alpha\beta\gamma} \equiv u^\mu\varepsilon_{\mu\alpha\beta\gamma}.
\end{equation}
Notice that due to the antisymmetric character of the Levi-Civita tensor (equation \eqref{leviCdef}),
$\varepsilon_{\alpha\beta\mu}u^\mu = 0$. Then, transforming \eqref{defB} with \eqref{weyldef2} gives
\begin{equation}
   B_{\alpha\beta} = -\frac{1}{2}\varepsilon\indices{^{\rho\nu}_\beta}R_{\alpha\mu\rho\nu}u^\mu - \frac{1}{2}\varepsilon\indices{^\nu_{\alpha\beta}}R_{\mu\nu}u^\mu.
\end{equation}
Now contracting with $\gamma^\alpha_\mu \gamma^\beta_\nu$, and combining with \eqref{ggguR}, \eqref{guR} and \eqref{mom}, it comes
\begin{equation}
   \gamma(B)_{\alpha\beta} = \varepsilon\indices{^{\mu\nu}_\beta}(D_\mu K_{\alpha\nu} - 4\pi \gamma_{\alpha \mu}p_\nu).
\end{equation}
This equation thus relates directly the magnetic part of the Weyl tensor to the $3+1$ geometrical quantities.

\subsection{Weyl decomposition}

The tensors $E_{\alpha\beta}$ and $B_{\alpha\beta}$ are called electric and magnetic because when rewriting the Einstein-Bianchi
identity \eqref{einbianchi} in terms of these tensors, the equations have the same structure as Maxwell's equations (see e.g.\
\cite{Friedrich96}). We have seen that $E_{\alpha\beta}$ and $B_{\alpha\beta}$ are independent and have five independent
components each. Since the Weyl tensor has precisely ten independent components in 4-dimensional space-times (see equation
\eqref{weylcompind}), it means that the information carried by the Weyl tensor is entirely encoded in the electric and magnetic
parts. The Weyl tensor should then be expressible as a function of these two tensors. Indeed, the following relation holds
\begin{equation}
   C_{\alpha\beta\gamma\delta} = -2\epsilon(l_{\alpha[\gamma}E_{\delta]\beta} - l_{\beta[\gamma}E_{\delta]\alpha} + \epsilon
   u_{[\gamma}B_{\delta]\mu}\varepsilon\indices{^\mu_{\alpha\beta}} + \epsilon u_{[\alpha}B_{\beta]\mu}\varepsilon\indices{^\mu_{\gamma\delta}}),
   \label{d+1C}
\end{equation}
where
\begin{equation}
   l_{\alpha\beta} \equiv g_{\alpha\beta} - 2\epsilon u_\alpha u_\beta.
\end{equation}
The demonstration proceeds as follows. From the definitions \eqref{defE} and \eqref{defB}, it is possible to show that the right-hand
side of \eqref{d+1C} is
\begin{align}
\nonumber &g_{\alpha\gamma}C_{\beta\mu\delta\nu}u^\mu u^\nu - g_{\alpha\delta}C_{\beta\mu\gamma\nu}u^\mu u^\nu - g_{\beta\gamma}C_{\alpha\mu\delta\nu}u^\mu u^\nu + g_{\beta\delta}C_{\alpha\mu\gamma\nu}u^\mu u^\nu \\
+& C_{\delta\mu\alpha\beta}u_\gamma u^\mu - C_{\gamma\mu\alpha\beta}u_\delta u^\mu - C_{\alpha\mu\gamma\delta}u_\beta u^\mu +
C_{\beta\mu\gamma\delta}u_\alpha u^\mu.
\end{align}
It is then possible to show that the electric and magnetic part of this expression are precisely $E_{\alpha\beta}$ and
$B_{\alpha\beta}$. Since $E_{\alpha\beta}$ and $B_{\alpha\beta}$ encode all the information about the Weyl tensor, this ends the
proof.

\chapter{Principle of least action in general relativity}
\label{leastaction}
\citationChap{For every complex problem, there is an answer that is clear, simple, and wrong.}{Henry Louis Mencken}
\minitoc

In this appendix, we derive the equations of motion of \gls{gr} with a principle of least action. Namely, we start with the
standard \gls{eh} action with Dirichlet boundary conditions and discuss its fundamental flaws. We then provide several corrections
to bring the action to a well-defined functional form, notably with the addition of the \gls{ghy} term. Even if this term
regularises the action, it contributes to give infinite global quantities (like mass and angular momentum), so this term needs a
correction too. After introducing the counter-term solution, we relax the standard Dirichlet conditions in order to unveil the
quasi-local stress tensor. In the context of the \gls{ads}-\gls{cft} correspondence, it is interpreted as the energy-momentum
tensor of the dual \gls{cft} living on the boundary of \gls{aads} space-times.

We consider that the space-time has one time dimension and $d$ spatial dimensions. We use geometrical units in which
$\gls{G} = \gls{c} = 1$.

\section{The Einstein-Hilbert action}

The \gls{eh} action was the first formulation of \gls{gr} with a principle of least action \cite{Hilbert15}. In this
section, we derive its standard variational calculus and discuss why it is insufficient on mathematical grounds.

\subsection{Variations of the geometrical tensors}

We denote by $g_{\alpha\beta}$ the metric, $g$ its determinant, $\Gamma\indices{^\gamma_{\alpha\beta}}$ the Christoffel symbols,
$R_{\alpha\beta\gamma\delta}$ the Riemann tensor, $R_{\alpha\beta}$ the Ricci tensor and $R$ the scalar curvature. Denoting
$\delta$ the first order variations of these quantities with respect to a fixed background, it is a standard exercise to show that
\cite{Landau75,Misner73}:
\begin{subequations}
\begin{align}
   \label{commute}
   \delta g^{\alpha\beta} &= -g^{\alpha\mu}g^{\beta\nu}\delta g_{\mu\nu},\\
   \label{deltadetg}
   \delta \sqrt{-g} &= -\frac{\sqrt{-g}}{2}g_{\mu\nu}\delta g^{\mu\nu},\\
   \label{varGamma}
   \delta \Gamma\indices{^\gamma_{\alpha\beta}} &= \frac{g^{\gamma\mu}}{2}(\nabla_\alpha \delta g_{\beta\mu} + \nabla_\beta \delta g_{\alpha\mu} - \nabla_\mu \delta g_{\alpha\beta}),\\
   \label{palatini}
   \delta R_{\alpha\beta} &= \nabla_\mu \delta\Gamma\indices{^\mu_{\alpha\beta}} - \nabla_\beta \delta\Gamma\indices{^\mu_{\alpha\mu}}\\
   \nonumber &= \frac{1}{2}g^{\mu\nu}(\nabla_\mu\nabla_\alpha \delta g_{\beta\nu} + \nabla_\mu \nabla_\beta \delta g_{\alpha\nu} - \nabla_\mu \nabla_\nu \delta g_{\alpha\beta} - \nabla_\beta\nabla_\alpha \delta g_{\mu\nu}),\\
   \label{dR}
   \delta R &= R_{\mu\nu}\delta g^{\mu\nu} + g^{\mu\nu}(\nabla_\rho \delta \Gamma\indices{^\rho_{\mu\nu}} - \nabla_\nu \delta \Gamma\indices{^\rho_{\mu\rho}})\\
   \nonumber &= R_{\mu\nu}\delta g^{\mu\nu} - g^{\mu\nu}\Box \delta g_{\mu\nu} + \nabla^\mu\nabla^\nu \delta g_{\mu\nu},
\end{align}
\label{variations}%
\end{subequations}
where $\Box = \nabla^\mu \nabla_\mu$. Equation \eqref{palatini} is called Palatini's identity.

\subsection{Variations of the action}

Let us consider the standard \gls{eh} action with a matter field in $d+1$ dimensions:

\begin{equation}
   S_{EH} = \int_\mathcal{M} \left(\frac{1}{16\pi}[R - 2 \gls{Lambda}] + \mathcal{L}\right) \sqrt{-g}d^{d+1}x,
   \label{EH}
\end{equation}
where $\mathcal{L} = \mathcal{L}(\mathcal{\phi,\nabla \phi})$ is the Lagrangian of some generic matter field $\phi$,
$\gls{Lambda}$ is the cosmological constant and $\mathcal{M}$ is the domain of integration covering the entire space-time manifold.
Hereafter, we also use the notation
\begin{equation}
   \mathcal{S} = \int_\mathcal{M} \mathcal{L}\sqrt{-g} d^{d+1}x,
\end{equation}
to refer to the matter part of the action. We suppose that the metric and the matter field can vary and we impose Dirichlet
conditions
\begin{equation}
   \delta g^{\mu\nu} \underset{\partial \mathcal{M}}{=}  0 \quad \tn{and} \quad \delta \phi \underset{\partial \mathcal{M}}{=} 0,
   \label{bcvar}
\end{equation}
on the boundary $\partial \mathcal{M}$ of the integration domain. For the variations with respect to the matter field $\phi$, the argument is the
following:
\begin{equation}
   \delta_\phi S_{EH} = \int_\mathcal{M}\left( \frac{\partial \mathcal{L}}{\partial \phi} \delta \phi + \frac{\partial \mathcal{L}}{\partial(\nabla_\mu \phi)}\delta \nabla_\mu \phi \right) \sqrt{-g}d^{d+1}x.
\end{equation}
By commuting the $\nabla$ and $\delta$ in the second term and using Leibniz product rule (or integration by parts), it comes
\begin{equation}
   \delta_\phi S_{EH} = \int_\mathcal{M} \left( \frac{\partial \mathcal{L}}{\partial \phi} \delta \phi + \nabla_\mu \left[\frac{\partial \mathcal{L}}{\partial(\nabla_\mu \phi)}\delta \phi\right] - \nabla_\mu\left[ \frac{\partial \mathcal{L}}{\partial(\nabla_\mu\phi)} \right]\delta \phi \right)\sqrt{-g}d^{d+1}x.
   \label{divmatter}
\end{equation}
The second term in the parenthesis is the volume integral of a perfect divergence. With the help of Stokes' theorem
\cite{Poisson04,Gourgoulhon13}, it can be reduced to a boundary integral of a term proportional to $\delta \phi$ which is zero
according to \eqref{bcvar}.

Thus, the whole variations of the action with respect to both $g_{\alpha\beta}$ and $\phi$ are straightforward
\cite{Landau75,Misner73}. With the results \eqref{variations}, it comes
\begin{align}
\nonumber \delta S_{EH} &= \frac{1}{16\pi}\int_\mathcal{M}\left( R_{\mu\nu} - \frac{R}{2}g_{\mu\nu} + \gls{Lambda} g_{\mu\nu} + \frac{16\pi}{\sqrt{-g}} \frac{\partial(\sqrt{-g}\mathcal{L})}{\partial g^{\mu\nu}} \right)\delta g^{\mu\nu}\sqrt{-g} d^{d+1}x\\
\nonumber               &+ \frac{1}{16\pi}\int_\mathcal{M} g^{\mu\nu}(\nabla_\rho \delta \Gamma\indices{^\rho_{\mu\nu}} - \nabla_\nu \delta \Gamma\indices{^\rho_{\mu\rho}})\sqrt{-g}d^{d+1}x\\
                        &+ \int_\mathcal{M} \left( \frac{\partial \mathcal{L}}{\partial \phi} - \nabla_\mu\left[ \frac{\partial \mathcal{L}}{\partial (\nabla_\mu \phi)} \right] \right) \delta \phi \sqrt{-g} d^{d+1}x.
\end{align}
Assuming that the laws of physics are those for which the action is stationary\footnote{See \cite{Bekenstein15} for an argumentative
criticism of this paradigm. We also stress that $\delta S = 0$ does not implies that the action is extremal, as is often claimed
in the literature. For example, $\delta S$ can be zero for a saddle point. The proper vocabulary is that of stationarity.}, we should bring
$\delta S_{EH}$ to zero for every variations $\delta \phi$ and $\delta g_{\alpha\beta}$ around this stationary point. This requires at
least

\begin{subequations}
\begin{align}
   R_{\alpha\beta} - \frac{R}{2}g_{\alpha\beta} + \gls{Lambda} g_{\alpha\beta} &= 8\pi T_{\alpha\beta},\\
   \frac{\partial \mathcal{L}}{\partial \phi} - \nabla_\mu\left[ \frac{\partial \mathcal{L}}{\partial (\nabla_\mu \phi)}\right] &= 0,
\end{align}
\label{motion}%
\end{subequations}
where we have recovered the standard Euler-Lagrange equation for the matter field as well as Einstein's equation for the
gravitational field, and have defined the energy-momentum tensor
\begin{equation}
   T_{\alpha\beta} \equiv -\frac{2}{\sqrt{-g}} \frac{\partial(\sqrt{-g}\mathcal{L})}{\partial g^{\alpha\beta}},
\end{equation}
or equivalently using \eqref{commute}
\begin{equation}
   T_{\alpha\beta} = -\frac{2}{\sqrt{-g}} \frac{\delta \mathcal{S}}{\delta g^{\alpha\beta}} \quad \tn{or} \quad
   T^{\alpha\beta} = +\frac{2}{\sqrt{-g}} \frac{\delta \mathcal{S}}{\delta g_{\alpha\beta}}.
   \label{Tdef}
\end{equation}

However, even with these requirements, we have not set exactly $\delta S$ to zero. The on-shell residual is

\begin{equation}
   \delta S_{EH}^\star = \frac{1}{16\pi}\int_\mathcal{M} g^{\mu\nu}(\nabla_\rho \delta \Gamma\indices{^\rho_{\mu\nu}} - \nabla_\nu
   \delta \Gamma\indices{^\rho_{\mu\rho}})\sqrt{-g}d^{d+1}x.
   \label{remain}
\end{equation}

The temptation is great at this point to invoke Stokes' theorem to get a boundary integral, as the integrand is a pure divergence. Nevertheless, it would imply an
integral on the domain $\partial \mathcal{M}$ of terms in $\delta \Gamma$, that are \textit{not} zero: our boundary conditions
\eqref{bcvar} do not imply at all that these terms should vanish. This means that the action is ill-defined and not truly
differentiable in the sense of variational calculus. In other words, the solutions of the equations of motion \eqref{motion} are
not truly making the action stationary. The reason is that the \gls{eh} Lagrangian $\mathcal{L} = R - 2 \gls{Lambda}$ is second
order in the derivatives, which brings first order derivative boundary terms after integration by parts. Next section illustrates this
point on a simple example.

\subsection{The 1-dimensional second order Lagrangian example}
\label{1daction}

Consider the illustrative example of a 1-dimensional second order theory described by the action
\begin{equation}
   S = \int_{-\infty}^{+\infty} \mathcal{L}(\phi,\phi',\phi'')dx,
\end{equation}
where primes indicate derivatives with respect to $x$. The variations of this action can be computed as
\begin{align*}
   \delta S &= \int_{-\infty}^{+\infty}\left( \frac{\partial \mathcal{L}}{\partial \phi}\delta \phi + \frac{\partial \mathcal{L}}{\partial \phi'}\delta \phi' +  \frac{\partial \mathcal{L}}{\partial \phi''}\delta \phi''\right)dx\\
            &= \int_{-\infty}^{+\infty}\Bigg[ \frac{\partial \mathcal{L}}{\partial \phi}\delta \phi + \left(\frac{\partial \mathcal{L}}{\partial \phi'}\delta \phi\right)' - \left(\frac{\partial \mathcal{L}}{\partial \phi'}\right)'\delta \phi\\
            &+ \left(\frac{\partial \mathcal{L}}{\partial \phi''}\delta \phi'\right)' - \left(\left(\frac{\partial \mathcal{L}}{\partial \phi''}\right)'\delta \phi\right)' + \left(\frac{\partial \mathcal{L}}{\partial \phi''}\right)''\delta\phi\Bigg]dx,
\end{align*}
where we have intensively used the Leibniz product rule. The total derivative terms can be integrated out to give boundary terms, such that
\begin{equation}
   \delta S = \int_{-\infty}^{+\infty}\left[ \frac{\partial \mathcal{L}}{\partial \phi} - \left(\frac{\partial \mathcal{L}}{\partial \phi'}\right)' + \left(\frac{\partial \mathcal{L}}{\partial \phi''}\right)''\right]\delta \phi dx
   + \left[\left(\frac{\partial \mathcal{L}}{\partial \phi'} - \left(\frac{\partial \mathcal{L}}{\partial \phi''}\right)'\right)\delta\phi +
   \frac{\partial \mathcal{L}}{\partial \phi''}\delta\phi'\right]_{-\infty}^{+\infty}.
\end{equation}
The equation of motion can be read in the bulk integrand: this is nothing but the generalisation of the Euler-Lagrange equation
for a second order Lagrangian. The boundary term cannot vanish unless \textit{both} $\delta \phi$ \textit{and} $\delta \phi'$ be zero at
the boundary. This is a too strong condition for the variational principle, as far as physics is involved.

\section{Corrections to the action}

In order to get rid of this kind of spurious boundary terms without destroying the equations of motion, the action needs to be
regularised. The simplest way of doing this is to add a corrective boundary term to the action.

\subsection{On-shell residual as a boundary term}

Let us first try and identify the problematic boundary term in \eqref{remain}. In light of \eqref{varGamma}, let us define
\begin{equation}
   V^\alpha = g^{\mu\nu} \delta \Gamma\indices{^\alpha_{\mu\nu}} - g^{\alpha\mu}\delta \Gamma\indices{^\nu_{\mu\nu}} = (g^{\alpha\mu}g^{\nu\rho} - g^{\alpha\rho}g^{\mu\nu})\nabla_\rho \delta g_{\mu\nu},
   \label{eqVa}
\end{equation}
so that the residual of the on-shell action reads
\begin{equation}
   \delta S_{EH}^\star = \frac{1}{16\pi} \int_\mathcal{M} \nabla_\mu V^\mu \sqrt{-g}d^{d+1}x.
\end{equation}
We denote by $u^\alpha$ the normal vector to the boundary $\partial \mathcal{M}$ and
\begin{equation}
   \epsilon = g^{\mu\nu}u_\mu u_\nu = \left\{
      \begin{array}{l}
         -1 \quad \tn{(timelike vector)},\\
         +1 \quad \tn{(spacelike vector)},\\
      \end{array}
      \right.
      \label{eps}
\end{equation}
its square. Via Stokes' theorem, one can get
\begin{equation}
   \delta S_{EH}^\star = \frac{1}{16\pi}\oint_{\partial \mathcal{M}} \epsilon V^\mu u_\mu \sqrt{|q|}d^dy,
   \label{sehstar}
\end{equation}
where $\sqrt{|q|}d^dy$ is the elementary boundary volume\footnote{As $q_{\alpha\beta}$ has $(+++\cdots)$ signature on space-like hypersurfaces and
   $(-++\cdots)$ signature on time-like hypersurfaces, we use the unambiguous notation $\sqrt{|q|}$ for the square-root of its
   determinant. In practice, $\partial \mathcal{M}$ has to be divided in one time-like hypersurface and two space-like
hypersurfaces, like a space-time cylinder, to perform the integrations properly.} with
\begin{equation}
   q_{\alpha\beta} \equiv g_{\alpha\beta} - \epsilon u_\alpha u_\beta,
   \label{inducedmetric}
\end{equation}
the induced metric on $\partial \mathcal{M}$. Incidentally, from \eqref{eps} and \eqref{inducedmetric} follows directly the orthogonality condition
\begin{equation}
   q^{\alpha\mu}u_\mu = 0.
   \label{ortho}
\end{equation}
Using \eqref{eqVa}, \eqref{sehstar} becomes
\begin{equation}
   \delta S_{EH}^\star = \frac{1}{16\pi}\oint_{\partial \mathcal{M}} \epsilon (u^\mu \underbrace{q^{\nu\rho}\nabla_\rho \delta g_{\mu\nu}}_{\tn{tangential}} - q^{\mu\nu}\underbrace{u^\rho \nabla_\rho \delta g_{\mu\nu}}_{\tn{normal}}) \sqrt{|q|}d^dy.
   \label{discussion}
\end{equation}
Inside the parenthesis, the first term involves the derivative of $\delta g_{\alpha\beta}$ tangential to $\partial \mathcal{M}$.
Indeed, the $\nabla$ can be replaced by $\partial$ on the boundary because of the Dirichlet conditions \eqref{bcvar}. Moreover, in coordinates adapted to $\partial \mathcal{M}$, the
derivative is only along the $\partial \mathcal{M}$ generators, which preserve \eqref{bcvar}. This first term is thus zero.
On the other hand, the second term of the parenthesis involves the normal derivative of $\delta g_{\alpha\beta}$ that is not zero.
Thus we have recovered that the on-shell \gls{eh} action reads \cite{Blau16}
\begin{equation}
   \delta S_{EH}^\star = -\frac{1}{16\pi}\oint_{\partial \mathcal{M}} \epsilon q^{\mu\nu}u^\rho \nabla_\rho \delta g_{\mu\nu} \sqrt{|q|}d^dy.
   \label{deltasehstar}
\end{equation}

\subsection{The Gibbons-Hawking-York term}
\label{ghy}

This on-shell residual can be eliminated by the \gls{ghy} correction \cite{York72,Gibbons77}. Let us define the acceleration $a_\alpha$,
the extrinsic curvature tensor $\Theta_{\alpha\beta}$ and the mean extrinsic curvature $\Theta$ of $\partial \mathcal{M}$ as
\begin{subequations}
\begin{align}
   a_\alpha &\equiv u^\mu \nabla_\mu u_\alpha,\\
   \label{thetaab}
   \Theta_{\alpha\beta} &\equiv -\nabla_\beta u_\alpha + \epsilon a_\alpha u_\beta ,\\
   \Theta &\equiv q^{\mu\nu}\Theta_{\mu\nu}.
\end{align}
\end{subequations}
Because of the orthogonality condition \eqref{ortho} and the normalisation condition \eqref{eps}, it comes
\begin{equation}
   a_\nu u^\nu =  u^\mu u^\nu \nabla_\mu u_\nu = \frac{1}{2}u^\mu \nabla_\mu(u_\nu u^\nu) = \frac{1}{2}u^\mu \nabla_\mu \epsilon = 0,
\end{equation}
and
\begin{equation}
   \Theta = -q^{\mu\nu}\nabla_\mu u_\nu = -q^{\mu\nu}(\partial_\mu u_\nu - \Gamma\indices{^\rho_{\mu\nu} }u_\rho).
   \label{theta}
\end{equation}
Let us now consider the amended action
\begin{equation}
   S_{GHY} = \int_\mathcal{M} \left(\frac{1}{16\pi}[R - 2 \gls{Lambda}] + \mathcal{L}\right) \sqrt{-g}d^{d+1}x - \frac{1}{8\pi}\oint_{\partial \mathcal{M}} \epsilon \Theta \sqrt{|q|}d^dy.
   \label{sghy}
\end{equation}
The only difference with the \gls{eh} action in \eqref{EH} is the addition of a boundary term. Being a surface term, it leaves
untouched the equations of motion. Notice that the \gls{eh} action involves the intrinsic curvature $R$ of the bulk while the
correction involves the mean extrinsic curvature of the boundary. The on-shell variations are now (compare with \eqref{deltasehstar})
\begin{equation}
   \delta S_{GHY}^\star = -\frac{1}{16\pi}\oint_{\partial \mathcal{M}} \epsilon (q^{\mu\nu}u^\rho \nabla_\rho \delta g_{\mu\nu} + 2\delta \Theta)\sqrt{|q|}d^dy.
   \label{sghyos}
\end{equation}
In order to derive an expression for $\delta \Theta$, let us first consider the variations of $u_\alpha$. The hypersurface $\partial
\mathcal{M}$ is characterised by an equation of the form
\begin{equation}
   F(x^\mu) = 0,
\end{equation}
where $F$ is a real function of the coordinates. By definition, $u_\alpha$ is the normal vector to $\partial \mathcal{M}$ and is
thus proportional to the gradient of $F$ \cite{Wald84}:
\begin{equation}
   u_{\alpha} = A \partial_\alpha F,
\end{equation}
where $A$ is a normalisation factor. The function $F$ and the coordinates are independent of the metric variations so that the
variations of $u_\alpha$ are
\begin{equation}
   \delta u_\alpha = \delta A\partial_\alpha F = \frac{\delta A}{A}u_\alpha.
   \label{dna}
\end{equation}
On the other hand, differentiating \eqref{eps} yields
\begin{equation}
   \delta g^{\mu\nu}u_\mu u_\nu + 2 g^{\mu\nu}\delta u_\mu u_\nu = 0,
\end{equation}
which can be combined with \eqref{dna} and \eqref{eps} again to give
\begin{equation}
   \delta u_\alpha = -\frac{\epsilon}{2}\delta g^{\mu\nu}u_\mu u_\nu u_\alpha.
   \label{dn}
\end{equation}
In particular, as a consequence of the Dirichlet boundary conditions \eqref{bcvar}, \eqref{commute} and \eqref{inducedmetric} we get
\begin{equation}
   \delta u_\alpha \underset{\partial \mathcal{M}}{=} 0 \quad \tn{and} \quad \delta q_{\alpha\beta} \underset{\partial
      \mathcal{M}}{=} 0,
   \label{dndv}
\end{equation}
at the boundary $\partial \mathcal{M}$.

We can now come back to the variations of $\Theta$. Combining \eqref{theta}, \eqref{dndv} and \eqref{varGamma}, and
discarding tangential derivatives on $\partial \mathcal{M}$, we find
\begin{equation}
   \delta \Theta|_{\partial \mathcal{M}} = \underbrace{-q^{\mu\nu}\partial_\mu \delta u_\nu|_{\partial
      \mathcal{M}}}_{\tn{tangential} = 0} +  q^{\mu\nu}u_\rho\delta \Gamma\indices{^\rho_{\mu\nu}}|_{\partial \mathcal{M}} = \frac{1}{2}q^{\mu\nu}u_\rho g^{\rho\sigma}(\underbrace{\nabla_\mu \delta g_{\nu\sigma} + \nabla_\nu \delta g_{\mu\sigma}}_{\tn{tangential} = 0} - \underbrace{\nabla_\sigma \delta g_{\mu\nu}}_{\tn{normal}})|_{\partial \mathcal{M}},
\end{equation}
where the index $|_{\partial \mathcal{M}}$ indicates restriction to the boundary of the integration volume.
Repeating the discussion below \eqref{discussion}, we finally get
\begin{equation}
   \delta \Theta \underset{\partial \mathcal{M}}{=} -\frac{1}{2}q^{\mu\nu} u^\rho \nabla_\rho \delta g_{\mu\nu}.
\end{equation}
In view of \eqref{sghyos}, this allows us to conclude that the on-shell residual of the \gls{ghy} action is zero, namely
\begin{equation}
   \delta S_{GHY}^\star = 0.
\end{equation}
The \gls{ghy} action is thus differentiable, and any solution of the equations of motion \eqref{motion} satisfies $\delta S_{GHY} = 0$, i.e.\
truly makes the action stationary.

\subsection{The background term}

The on-shell action is often involved in the computation of global quantities like mass and angular momentum. Unfortunately, the
\gls{ghy} action is quite awkward in this regard, as it is notoriously\ldots\ infinite! Indeed, consider $\partial
\mathcal{M}$ the sphere $\mathcal{S}$ of radius $R$ in Minkowski space-time between two dates $t_1$ and $t_2$. Using Cartesian
coordinates $(t,x,y,z)$ and $r = \sqrt{x^2 + y^2 + z^2}$, it can be shown that
\begin{subequations}
\begin{align}
   u_\alpha &= (1,0,0,0)\quad &\tn{and} \quad \Theta &= 0 \quad &\tn{on the hypersurfaces}\quad t = cst,\\
   u_\alpha &= \dfrac{1}{r}(0,x,y,z)\quad &\tn{and} \quad \Theta &= -\dfrac{2}{r} \quad &\tn{on the hypersurfaces}\quad r = cst.
\end{align}
\end{subequations}
The cut-off at radius $R$ of the on-shell \gls{ghy} action then reduces to
\begin{equation}
   S_{GHY}^R = -\frac{1}{8\pi}\int_{t_1}^{t_2}\oint_{\mathcal{S}} \left( -\frac{2}{R} \right)R^2\sin\theta dtd\theta d\varphi = R (t_2 - t_1).
\end{equation}
However, the true on-shell \gls{ghy} action is
\begin{equation}
   S_{GHY} = \lim_{\substack{t_1\rightarrow-\infty\\t_2\rightarrow+\infty\\R\rightarrow+\infty}} S_{GHY}^R = +\infty.
\end{equation}
Solutions of the equations of motion are thus making the action stationary, but this stationary point is infinite! This is definitely a serious
problem. A usual cure to this divergence is to add a background term (BT) in the action, and to write
\begin{equation}
   S_{GHY}^{BT} = \int_\mathcal{M} \left(\frac{1}{16\pi}[R - 2 \gls{Lambda}] + \mathcal{L}\right) \sqrt{-g}d^{d+1}x - \frac{1}{8\pi}\oint_{\partial \mathcal{M}} \epsilon (\Theta - \Theta_0) \sqrt{|q|}d^dy,
\end{equation}
where $\Theta_0$ is the mean extrinsic curvature of $\partial \mathcal{M}$ computed with a background metric. Very often, it is
chosen to be the Minkowski metric is asymptotically flat space-times \cite{Gourgoulhon07} or the \gls{ads} metric in \gls{aads}
space-times \cite{Abbott82}. This is sufficient to make the action finite.

\subsection{The counter-term}

However, one can feel ill-at-ease with this kind of argument as the background metric seems to be an arbitrary and unphysical additional
ingredient. In \cite{Balasubramanian99}, the authors tried and found the simplest counter-term (CT) in \gls{aads} space-times that
cancels off the divergence of the on-shell action and that is covariant. They wrote
\begin{equation}
   S_{GHY}^{CT} = \int_\mathcal{M} \left(\frac{1}{16\pi}[R - 2 \gls{Lambda}] + \mathcal{L}\right) \sqrt{-g}d^{d+1}x - \frac{1}{8\pi}\oint_{\partial \mathcal{M}} \epsilon\Theta \sqrt{|q|}d^dy + \frac{1}{8\pi}\mathcal{S}_{CT},
   \label{ghyct}
\end{equation}
where
\begin{equation}
   \mathcal{S}_{CT} = \oint_{\partial \mathcal{M}} \mathcal{L}_{CT}\sqrt{|q|}d^dy.
\end{equation}

In \gls{aads} space-times having dimensions $d+1 = 3,4,5$, the authors of \cite{Balasubramanian99} showed that a suitable
counter-term Lagrangian was
\begin{equation}
   \mathcal{L}_{CT} = -\frac{1}{\gls{L}}\left( d-1 + \frac{\gls{L}^2}{2(d-2)}\mathcal{R} \right),
   \label{lctd}
\end{equation}
where $\gls{L}$ is the \gls{ads} radius and $\mathcal{R}$ is the Ricci scalar\footnote{we also use the letter $\mathcal{R}$ for
the corresponding Riemann and Ricci tensors.} of the induced metric $q_{\alpha\beta}$. Note that the second term in the
parenthesis is only present for $d > 2$. This time, the additional term is an intrinsic curvature terms whereas the \gls{ghy}
correction involved a mean extrinsic curvature term. With the help of \eqref{variations}, we can compute the variations of the
counter-term action\footnote{The astute reader may have noticed that we have disregarded the last two terms of \eqref{dR}. Indeed,
they are pure divergences. With Stokes' theorem, they can be reduced to an integral on the boundary of $\partial \mathcal{M}$,
namely $\partial \partial \mathcal{M}$. However, the boundary of a boundary is zero \cite{Misner73}.}
\begin{equation}
   \delta \mathcal{S}_{CT} = \oint_{\partial \mathcal{M}} \left[ \frac{(d-1)}{2\gls{L}}q_{\mu\nu} - \frac{\gls{L}}{2(d-2)}\mathcal{G}_{\mu\nu} \right]\delta q^{\mu\nu}\sqrt{|q|}d^3y,
   \label{dsctd}
\end{equation}
where $\mathcal{G}_{\mu\nu} = \mathcal{R}_{\mu\nu} - \frac{\mathcal{R}}{2}q_{\mu\nu}$ is Einstein's tensor of the induced
metric. The advantage of the counter-term is that not only does it make finite the on-shell action, but also it does not spoil its
differentiability. Namely, the on-shell variations are still zero when Dirichlet boundary conditions \eqref{bcvar} are imposed, as
can be read on \eqref{dsctd}. For dimensions $d+1 > 5$, other terms in $\mathcal{R}^2$, $\mathcal{R}_{\mu\nu}\mathcal{R}^{\mu\nu}$
etc. may appear in \eqref{lctd}

\section{Generalisation of the least action principle}

Hitherto, we have always considered the standard variational principle with Dirichlet boundary conditions \eqref{bcvar}. However,
relaxing this condition allows to define a quasi-local stress tensor, that was first derived in \cite{Brown93}, and used in
\cite{Balasubramanian99} in the context of the \gls{ads}-\gls{cft} correspondence.

\subsection{Dropping the Dirichlet assumption}

In \cite{Brown93}, the authors examined the consequences of relaxing the Dirichlet boundary conditions. So let us consider the action
\eqref{ghyct}. Under general perturbations, its variations are
\begin{align}
\nonumber \delta S_{GHY}^{CT} &= \frac{1}{16\pi}\int_\mathcal{M}(G_{\mu\nu} + \gls{Lambda} g_{\mu\nu} - 8\pi T_{\mu\nu})\delta g^{\mu\nu}\sqrt{-g}d^{d+1}x \\
\nonumber                     &+ \int_\mathcal{M} \left( \frac{\partial \mathcal{L}}{\partial \phi} - \nabla_\mu\left[ \frac{\partial \mathcal{L}}{\partial (\nabla_\mu \phi)} \right] \right) \delta \phi \sqrt{-g} d^{d+1}x\\
\nonumber                     &+ \frac{1}{16\pi}\oint_{\partial \mathcal{M}} \epsilon (u^\mu q^{\nu\rho}\nabla_\rho \delta g_{\mu\nu} - q^{\mu\nu}u^\rho \nabla_\rho \delta g_{\mu\nu})\sqrt{|q|}d^dy + \oint_{\partial \mathcal{M}}\epsilon \frac{\partial \mathcal{L}}{\partial(\nabla_\mu \phi)}u_\mu \delta \phi\sqrt{|q|} d^dy \\
                              &- \frac{1}{8\pi}\oint_{\partial \mathcal{M}} \epsilon (\delta \Theta \sqrt{|q|} + \Theta \delta \sqrt{|q|})d^dy + \frac{1}{8\pi} \oint_{\partial \mathcal{M}} \frac{\delta \mathcal{S}_{CT}}{\delta q_{\mu\nu}}\delta q_{\mu\nu}d^dy.
\end{align}
The first two lines deal with bulk integrals on which can be read the equations of motion \eqref{motion}. The third line contains the
boundary terms obtained by Stokes' theorem for both the gravitational field \eqref{discussion} and the matter field
\eqref{divmatter}. On the fourth line, the first term is the variations of the \gls{ghy} correction and the second term is the
variations of the counter-term, both present in \eqref{ghyct}. If we now consider on-shell variations, i.e.\ variations around a
solution of the equations of motion, the first two lines disappear. If we enforce Dirichlet boundary conditions for the matter
field variations $\delta\phi \underset{\partial \mathcal{M}}{=} 0$, the second term of the third line disappears too. Under these
assumptions, we are left with
\begin{align}
\nonumber \delta S_{GHY}^{CT\star} &= \frac{1}{16\pi}\oint_{\partial \mathcal{M}} \epsilon (u^\mu q^{\nu\rho}\nabla_\rho \delta g_{\mu\nu} - q^{\mu\nu}u^\rho \nabla_\rho \delta g_{\mu\nu})\sqrt{|q|}d^dy\\
                               &- \frac{1}{8\pi}\oint_{\partial \mathcal{M}} \epsilon (\delta \Theta \sqrt{|q|} + \Theta \delta \sqrt{|q|})d^dy + \frac{1}{8\pi} \oint_{\partial \mathcal{M}} \frac{\delta \mathcal{S}_{CT}}{\delta q_{\mu\nu}}\delta q_{\mu\nu}d^dy.
\label{onshellds}
\end{align}
This time, we have to compute the variations without assuming the Dirichlet boundary conditions for $\delta
g_{\alpha\beta}$. However, we can still take advantage of the orthogonality condition \eqref{ortho}, whose variations are
\begin{equation}
   \delta q^{\alpha\mu}u_\mu + q^{\alpha\mu}\delta u_\mu = 0.
\end{equation}
The second term is zero because of \eqref{dn} and \eqref{ortho}, so that
\begin{equation}
   \delta q^{\alpha\mu}u_\mu = 0 \quad \tn{and} \quad q^{\alpha\mu}\delta u_\mu = 0.
   \label{dqn}
\end{equation}
These equations are useful for deriving the variations of the different terms in the action.

\subsection{Variations of boundary terms}

The variations $\delta \Theta$ are less easy to handle compared to section \ref{ghy}. In full generality, according to
\eqref{varGamma} and the definition \eqref{theta}, we get
\begin{equation}
   \delta \Theta = -\delta q^{\mu\nu}\nabla_\mu u_\nu - q^{\mu\nu}\nabla_\mu\delta u_\nu + \frac{1}{2}q^{\mu\nu}u^\rho(\nabla_\mu \delta g_{\nu\rho} + \nabla_\nu \delta g_{\mu\rho} - \nabla_\rho \delta g_{\mu\nu}).
   \label{dtheta}
\end{equation}
In light of \eqref{thetaab} and \eqref{dqn}, the first term is $\Theta_{\mu\nu}\delta q^{\mu\nu}$. The second term is
\begin{equation}
   - q^{\mu\nu}\nabla_\mu\delta u_\nu = -q^{\mu\nu}\nabla_\mu \left(-\frac{\epsilon}{2}\delta g^{\rho\sigma}u_\rho u_\sigma u_\nu\right) =
   \frac{\epsilon}{2}q^{\mu\nu}\delta g^{\rho\sigma}u_\rho u_\sigma \nabla_\mu u_\nu = -\frac{\epsilon}{2}\Theta \delta g^{\rho\sigma}u_\rho
   u_\sigma,
\end{equation}
where we used successively \eqref{dn}, the Leibniz product rule in combination with the orthogonality condition \eqref{ortho}, and
\eqref{theta}. The third term of \eqref{dtheta} is
\begin{align*}
   \frac{1}{2}q^{\mu\nu}u^\rho\nabla_\mu \delta g_{\nu\rho} &= \frac{1}{2}q^{\mu\nu}u^\rho\nabla_\mu\delta q_{\nu\rho} + \frac{1}{2}q^{\mu\nu}u^\rho\nabla_\mu(\epsilon u_\rho \delta u_\nu + \epsilon u_\nu \delta u_\rho)\\
   &= -\frac{1}{2}q^{\mu\nu}\delta q_{\nu\rho}\nabla_\mu u^\rho - \frac{1}{2}q^{\mu\nu}u^\rho\nabla_\mu(\delta g^{\sigma \lambda}u_\sigma u_\lambda u_\rho u_\nu)\\
   &= -\frac{1}{2}g^{\mu\nu}\delta q_{\nu\rho}\nabla_\mu u^\rho - \frac{1}{2}q^{\mu\nu}u^\rho\delta g^{\sigma\lambda} u_\sigma u_\lambda u_\rho \nabla_\mu u_\nu\\
   &= \underbrace{\frac{1}{2}\Theta^{\mu\nu}\delta q_{\mu\nu}}_{-\frac{1}{2}\Theta_{\mu\nu}\delta q^{\mu\nu}} + \frac{\epsilon}{2}\Theta\delta g^{\sigma\lambda}u_\sigma u_\lambda,
\end{align*}
where we have intensively used the Leibniz product rule and equations \eqref{eps}, \eqref{inducedmetric}, \eqref{ortho}, \eqref{thetaab},
\eqref{theta}, \eqref{dn}, \eqref{dqn}. Thus
\begin{equation}
   \delta \Theta = \frac{1}{2}\Theta_{\mu\nu}\delta q^{\mu\nu} + \frac{1}{2}q^{\mu\nu}u^\rho(\nabla_\nu \delta g_{\mu\rho} - \nabla_\rho\delta g_{\mu\nu}).
   \label{dtheta2}
\end{equation}
The last two terms \eqref{dtheta2} cancel exactly the first line of \eqref{onshellds}, so that finally the variations of the on-shell action
are
\begin{equation}
   \delta S_{GHY}^{CT\star} = \frac{1}{16\pi}\oint_{\partial \mathcal{M}} \left(\epsilon[\Theta^{\mu\nu} - \Theta q^{\mu\nu}] + \frac{2}{\sqrt{|q|}}\frac{\delta \mathcal{S}_{CT}}{\delta q_{\mu\nu}}\right)\delta q_{\mu\nu}\sqrt{|q|}d^dy,
\end{equation}
where we have used the 3-dimensional counterpart of \eqref{deltadetg}. The tensor inside the brackets is sometimes called the
conjugate momentum, and is defined by
\begin{equation}
   \pi^{\mu\nu} \equiv \Theta^{\mu\nu} - \Theta q^{\mu\nu}.
   \label{pi}
\end{equation}

\subsection{The quasi-local stress tensor}

For the sake of completeness, the full action is finally
\begin{equation}
   S_{GHY}^{CT} = \int_\mathcal{M} \left(\frac{1}{16\pi}[R - 2 \gls{Lambda}] + \mathcal{L}\right) \sqrt{-g}d^{d+1}x -
   \frac{1}{8\pi}\oint_{\partial \mathcal{M}} \epsilon\Theta \sqrt{|q|}d^dy + \frac{1}{8\pi}\mathcal{S}_{CT},
\end{equation}
and its variations are
\begin{align}
\nonumber \delta S_{GHY}^{CT} &= \frac{1}{16\pi}\int_\mathcal{M}(G_{\mu\nu} + \gls{Lambda} g_{\mu\nu} - 8\pi T_{\mu\nu})\delta g^{\mu\nu}\sqrt{-g}d^{d+1}x \\
\nonumber                     &+ \int_\mathcal{M} \left( \frac{\partial \mathcal{L}}{\partial \phi} - \nabla_\mu\left[ \frac{\partial \mathcal{L}}{\partial (\nabla_\mu \phi)} \right] \right) \delta \phi \sqrt{-g} d^{d+1}x\\
                              &+ \oint_{\partial \mathcal{M}}\frac{\partial \mathcal{L}}{\partial(\nabla_\mu \phi)}\epsilon u_\mu
                              \delta \phi\sqrt{|q|} d^dy + \frac{1}{16\pi}\oint_{\partial \mathcal{M}}\epsilon \pi^{\mu\nu}\delta q_{\mu\nu}\sqrt{|q|}d^dy + \frac{1}{8\pi} \oint_{\partial \mathcal{M}} \frac{\delta \mathcal{S}_{CT}}{\delta q_{\mu\nu}}\delta q_{\mu\nu}d^dy.
\label{deltasfinal}
\end{align}
The first two lines of \eqref{deltasfinal} describe the equations of motion while the third line collects all boundary terms that
vanish only with Dirichlet boundary conditions.

The Hamilton-Jacobi approach of \cite{Brown93} allows us to define a quasi-local stress tensor similarly to \eqref{Tdef}
\begin{equation}
   \tau^{\alpha\beta} \equiv \frac{2}{\sqrt{|q|}}\frac{\delta S_{GHY}^{CT}}{\delta q_{\alpha\beta}} \quad \iff \quad \tau_{\alpha\beta} \equiv -\frac{2}{\sqrt{|q|}}\frac{\delta S_{GHY}^{CT}}{\delta q^{\alpha\beta}}.
\end{equation}
Combined with \eqref{dsctd}, \eqref{pi}, \eqref{deltasfinal}, we get in \gls{aads} space-times, for $d+1 = 3,4,5$,
\begin{equation}
   \tau^{\alpha\beta} = \frac{1}{8\pi}\left(\Theta^{\alpha\beta} - \Theta q^{\alpha\beta} - \frac{d-1}{\gls{L}}q^{\alpha\beta} +
   \frac{\gls{L}}{d-2} \mathcal{G}^{\alpha\beta}\right),\\
   \label{taudef}
\end{equation}
where $\mathcal{G}_{\alpha\beta}$ is the Einstein tensor of the induced metric $q_{\alpha\beta}$. The tensor $\tau^{\alpha\beta}$ is closely related
to the energy-momentum tensor of the \gls{cft} dual system living on the \gls{ads} boundary.

\chapter{The perfect quantum gas}
\label{quantumgas}
\citationChap{When I hear of Schrödinger's cat, I reach for my gun.}{Stephen William Hawking}
\minitoc

We derive standard calculations in quantum mechanics with Bose-Einstein and Fermi-Dirac statistics.
We set $\gls{hbar} = \gls{c} = \gls{kb} = 1$.

\section{Partition function}

Consider a quantum perfect gas. We denote by $\omega_i$ the energy levels and $n_i$ the occupation
numbers, i.e.\ the number of particles in the $i^{th}$ level of energy. One micro-state is entirely determined by the $n_i$
numbers, and the total energy is simply
\begin{equation}
   E = \sum_{i=1}^\infty n_i \omega_i.
\end{equation}
We can thus write the partition function as the sum over all possible micro-states of the Boltzmann factor $e^{-\beta E}$, where
$\beta = 1/T$ is the inverse of the gas temperature:
\begin{equation}
   Z = \sum_{n_1}\sum_{n_2}\cdots e^{-\beta(n_1 \omega_1 + n_2\omega_2 + \ldots)}.
\end{equation}
From the partition function, all thermodynamical quantities can be determined. For example, the energy density $e$ and the entropy
density $s$ are
\begin{equation}
   e = -\frac{1}{V}\frac{\partial \ln Z}{\partial \beta} \quad \tn{and} \quad s = -\frac{1}{V}\ln Z + \beta e,
   \label{thermo}
\end{equation}
where $V$ denotes the total volume of the gas. Hereafter, we compute these quantities for both the massless bosonic and the massless fermionic case.

\section{Bosonic case}

For bosons, there can be an arbitrary number of particles in the same energy level, so that $n_i \in \llbracket 0,\infty
\llbracket$ and
\begin{equation}
   Z_B = \left( \sum_{n_1 = 0}^\infty e^{-\beta n_1 \omega_1} \right)\times\left( \sum_{n_2 = 0}^\infty e^{-\beta n_2 \omega_2} \right)\times \cdots = \prod_{i=1}^\infty \frac{1}{1 - e^{-\beta \omega_i}}.
\end{equation}
Taking the logarithm of this expression yields
\begin{equation}
   \ln Z_B = -\sum_{i=1}^\infty \ln(1 - e^{-\beta \omega_i}).
\end{equation}
If now we consider massless particles, the energy is just the momentum $\omega = |\vec q|$. Taking the continuum limit, we replace the
infinite sum by an integral over phase space $\int d^3x d^3q/(2\pi)^3$. It thus comes after angular integration
\begin{equation}
   \ln Z_B = -\frac{V}{(2\pi)^3}\int_0^{\infty} \ln(1 - e^{-\beta q})4\pi q^2 dq.
\end{equation}
If now we perform a change of variable $x = \beta q$, it comes
\begin{equation}
   \ln Z_B = -\frac{V}{2\pi^2}T^3 \underbrace{\int_0^\infty x^2 \ln(1 - e^{-x})dx}_{-\frac{\pi^4}{45}}.
\end{equation}
Finally, the logarithm of the partition function is
\begin{equation}
   \ln Z_B = \frac{\pi^2}{90}V T^3.
\end{equation}
Using \eqref{thermo}, we infer that the energy density and the entropy density of the bosonic gas are
\begin{equation}
   e_B =  \frac{\pi^2}{30}T^4 \quad \tn{and} \quad s_B = \frac{2\pi^2}{45}T^3.
   \label{eB}
\end{equation}
Note that in \eqref{eB}, we get half of the usual black-body result. Indeed in the case of photons, summation
over the right and left polarisations brings a factor two.

\section{Fermionic case}

For fermions, according to Pauli's exclusion principle, each energy level can contain at most one particle, so that $n_i \in
\llbracket 0,1\rrbracket$ and
\begin{equation}
   Z_F = \prod_{i=1}^\infty (1 + e^{-\beta\omega_i}).
\end{equation}
Taking the logarithm and the continuum limit, it comes, for massless fermions obeying the dispersion relation $\omega = |\vec{q}|$
\begin{equation}
   \ln Z_F = \sum_{i=1}^\infty \ln(1 + e^{-\beta \omega_i}) = \frac{V}{(2\pi)^3}\int_0^\infty 4\pi q^2 \ln(1 + e^{-\beta q})dq.
\end{equation}
Setting $x = \beta q$ yields
\begin{equation}
   \ln Z_F = \frac{V}{2\pi^2}T^3 \underbrace{\int_0^\infty x^2\ln(1 + e^{-x})dx}_{\frac{7\pi^4}{360}},
\end{equation}
so that finally
\begin{equation}
   \ln Z_F = \frac{7}{8}\ln Z_B, \quad e_F = \frac{7}{8} e_B, \quad s_F = \frac{7}{8} s_B.
\label{78}
\end{equation}
We thus recover the well-known result according to which fermions contribute for $(7/8)^{th}$ of the energy density compared to bosons.

\end{appendix}

\pagenumbering{Roman}
\bibliographystyle{unsrt}
\bibliography{biblio}

\clearevenpage
\thispagestyle{empty}
\noindent\textbf{Titre :} Systèmes gravitationnels en espace-temps asymptotiquement anti-de Sitter\\

\noindent\textbf{Résumé :} Ingrédient clé de la correspondance \gls{ads}-\gls{cft}, l'espace-temps \gls{ads} est soupçonné depuis 2011
d'être non-linéairement instable. Si on y ajoute une perturbation d'amplitude arbitrairement petite, la formation d'une
singularité en un temps fini est quasi-systématique. Cependant, un certain nombre de perturbations pourraient résister à
l'instabilité, et osciller de manière régulière et quasi-périodique. Ce sont les îlots de stabilité. Parmi eux, on distingue les
geons, qui sont des excitations purement gravitationnelles d'\gls{ads}.

Dans cette thèse, nous nous proposons de construire numériquement pour la première fois plusieurs familles de geons
asymptotiquement \gls{ads}. Nous avons été capables de démontrer de manière
définitive l'existence de geons dits excités, qui faisaient l'objet d'une controverse dans la littérature. Nous détaillons
également de manière détaillée comment nombre de propriétés propres à cet espace-temps peuvent être utilisées numériquement pour
servir de diagnostics et valider les résultats. Enfin, nous soulignons comment les geons éclairent le problème délicat de
l'instabilité \gls{ads}.\\

\noindent\textbf{Mots-clés :} geons, anti-de Sitter, AdS/CFT, relativité générale\\

\vspace{1.0cm}

\noindent\textbf{Title:} Gravitational systems in asymptotically anti-de Sitter space-times\\

\noindent\textbf{Abstract:} Being a key ingredient of the \gls{ads}-\gls{cft} correspondence, \gls{ads} space-time is suspected
to be non-linearly unstable since 2011. Even with arbitrarily small initial data, a singularity almost invariably emerges. However, some
configurations allow for perfectly regular and quasi-periodic solutions. These are the so-called islands of stability. Among
them, geons play the role of purely gravitational excitations of \gls{ads} space-time.

In this manuscript, we tackle the problem of the numerical construction several families of asymptotically \gls{ads} geons for the first
time. We are able to definitely demonstrate the existence of the so-called excited geons, a point that was the subject to a lively
debate in the literature. We also detail how several features of such space-times can be used as precision monitors and help us to
validate numerics. Finally, we discuss in detail how geons shed new light on the \gls{ads} instability problem and advocate
for a renewed prudence in some analytical arguments.\\

\noindent\textbf{Keywords:} geons, anti-de Sitter, AdS/CFT, general relativity

\end{document}